\documentclass[10pt,a4paper,oneside,openright]{report}

\usepackage[square,comma]{natbib}                    
\usepackage{style/aas_macros}          
\usepackage{style/noindent}            
\usepackage{graphicx}                  
\usepackage{fancyhdr}                  
\usepackage{longtable}                 
\usepackage{style/import }             
\usepackage{times}                     
\usepackage[scaled=.92]{helvet}	       
\usepackage{bigstrut}                  
\usepackage{pict2e}                    
\usepackage[margin=10pt,font=footnotesize,labelfont=bf]{caption} 
\usepackage{multicol}                  

\input{texfiles/mn2e_compat.tex}

\makeatletter
\renewcommand{\@schapter}[1]{
                    \chaptermark{#1}%
                    \addtocontents{lof}{\protect\addvspace{10\p@}}%
                    \addtocontents{lot}{\protect\addvspace{10\p@}}%
                    \if@twocolumn
                      \@topnewpage[\@makeschapterhead{#1}]%
                    \else
                      \@makeschapterhead{#1}%
                      \@afterheading
                    \fi}

\renewenvironment{thebibliography}[1]
     {\chapter*{\bibname
        \@mkboth{{\bf \bibname}}{{\bf \bibname}}}%
      \list{\@biblabel{\@arabic\c@enumiv}}%
           {\settowidth\labelwidth{\@biblabel{#1}}%
            \leftmargin\labelwidth
            \advance\leftmargin\labelsep
            \@openbib@code
            \usecounter{enumiv}%
            \let\p@enumiv\@empty
            \renewcommand\theenumiv{\@arabic\c@enumiv}}%
      \sloppy
      \clubpenalty4000
      \@clubpenalty \clubpenalty
      \widowpenalty4000%
      \sfcode`\.\@m}
     {\def\@noitemerr
       {\@latex@warning{Empty `thebibliography' environment}}%
      \endlist}
\renewcommand\newblock{\hskip .11em\@plus.33em\@minus.07em}
\let\@openbib@code\@empty

\renewcommand\tableofcontents{%
    \if@twocolumn
      \@restonecoltrue\onecolumn
    \else
      \@restonecolfalse
    \fi
    \chapter*{\contentsname
        \@mkboth{%
           {\bf \contentsname}}{{\bf \contentsname}}}%
    \@starttoc{toc}%
    \if@restonecol\twocolumn\fi
    }

\renewcommand\listoffigures{%
     \if@twocolumn
       \@restonecoltrue\onecolumn
     \else
       \@restonecolfalse
     \fi
     \chapter*{List of Figures
         \@mkboth{%
            {\bf List of Figures}}{{\bf List of Figures}}}%
     \@starttoc{lof}%
     \if@restonecol\twocolumn\fi
     }

\makeatother

\renewcommand{\chaptermark}[1]{\markboth{#1}{#1}} 
\renewcommand{\sectionmark}[1]{} 
\fancypagestyle{plain}{
\fancyhead{} 
}

\newcommand{\ssp}{\renewcommand{\baselinestretch}{1}\normalsize}
\newcommand{\msp}{\renewcommand{\baselinestretch}{1.3}\normalsize} 
\newcommand{\dsp}{\renewcommand{\baselinestretch}{1.5}\normalsize} 

\makeatletter
\renewenvironment{table}
                {\ssp\@float{table}}
                {\end@float\dsp}
\renewenvironment{table*}		
                {\ssp\@dblfloat{table}}
                {\end@dblfloat\dsp}
\renewenvironment{figure}
                {\ssp\@float{figure}}
                {\end@float\dsp}
\renewenvironment{figure*}		
                {\ssp\@dblfloat{figure}}
                {\end@dblfloat\dsp}
\makeatother

\newcommand{\ie}{\emph{i.e.}}
\newcommand{\eg}{\emph{e.g.}}
\newcommand{\etal}{\emph{et al.}}

\newcommand{\be}{\begin{eqnarray}}
\newcommand{\ee}{\end{eqnarray}}

\def\simless{\mathbin{\lower 3pt\hbox
  {$\rlap{\raise 5pt\hbox{$\char'074$}}\mathchar"7218$}}}   
\def\simgreat{\mathbin{\lower 3pt\hbox  
   {$\rlap{\raise 5pt\hbox{$\char'076$}}\mathchar"7218$}}}  

\def \min{\,{\rm min}}
\def \max{\,{\rm max}}

\newcommand{\grumpy}				
{\mbox{\begin{picture}(20,20)(0,7)
\put(10,10){\circle{15}}
\put(12.5,12.5){\circle*{2}}
\put(7.5,12.5){\circle*{2}}
\put(6,1){\shortstack[l]{$\mbox{}^{\frown}$}}
\end{picture}}}

\newcommand{\happy}				
{\mbox{\begin{picture}(20,20)(0,7)
\put(10,10){\circle{15}}
\put(12.5,12.5){\circle*{2}}
\put(7.5,12.5){\circle*{2}}
\put(6,1){\shortstack[l]{$\mbox{}^{\smile}$}}
\end{picture}}}

\usepackage{amssymb}
\usepackage{amsmath}
\usepackage{mathrsfs}
\usepackage{graphicx}
\usepackage{color}

\usepackage{subfigure,epsfig,psfrag}
\usepackage[section]{placeins}

\newcommand{\nc}{\newcommand}
\nc{\renc}{\renewcommand} \nc{\nt}{\newtheorem}
\nc{\nb}{\nonumber}
\nc{\beq}[1]{\begin{equation} \mbox{$\label{#1}$}}
\nc{\eeq}{\vspace{\undereqskip}\end{equation}}
\nc{\bea}[1]{\begin{eqnarray} \mbox{$\label{#1}$}}
\nc{\eea}{\vspace{\undereqskip}\end{eqnarray}}
\nc{\state}[1]{\left<#1\right>} \nc{\set}[1]{\left\{#1\right\}}
\nc{\bset}[1]{\left(#1\right)} \nc{\sbset}[1]{\left[#1\right]}
\nc{\tol}{\rightarrow} \nc{\To}{\Rightarrow}
\nc{\lr}{\Leftrightarrow} \nc{\bo}{\textbf} \nc{\p}{\prime}
\nc{\delbar}{\delta\mkern-9mu\mathchar'26} \nc{\ho}{\hbar\omega}
\nc{\mz}{\mathcal{Z}} \nc{\md}{\mathcal{D}} \nc{\ml}{\mathcal{L}} \nc{\mS}{\mathcal{S}} \nc{\mU}{\mathcal{U}}
\nc{\mx}{\mathbf{x}} \nc{\mh}{\mathcal{H}} \nc{\mxi}{\mathbf{\xi}}
\nc{\mM}{\mathcal{M}}\nc{\mV}{\mathcal{V}}\nc{\mP}{\mathbf{P}}\nc{\mQ}{\mathbf{Q}}
\nc{\mR}{\mathbf{R}}
\nc{\mk}{\mathbf{k}} \nc{\order}[1]{\mathcal{O}(#1)}
\nc{\oi}{\omega_i} \nc{\frtwo}{\frac{1}{\rtwo}} \nc{\Eo}{E_\omega}
\nc{\So}{S_\omega} \nc{\eo}{\epsilon_\omega} \nc{\mo}{m_\omega}
\nc{\Uo}{U_\omega} \nc{\bsig}{\bar{\sigma}}
\nc{\wt}[1]{\tilde{#1}} \nc{\bx}{\bar{x}}
\nc{\wtPhi}{\wt{\Phi}} \nc{\wtMG}{\wt{M_G}} 
\nc{\dDx}{\int d^{D+1}x\sqrt{-g}} \nc{\sqg}{\sqrt{\abs{g}}}
\nc{\bq}{\bar{q}}
\def\half{\frac{1}{2}}
\renc{\thefootnote}{\arabic{footnote}} \nc{\capt}[1]{{\bf Figure.}{\small\sl #1}}
\nc{\eqss}[3]{\mbox{Eqs.~(\ref{#1}, \ref{#2}, \ref{#3})}}
\nc{\eqs}[2]{\mbox{Eqs.~(\ref{#1}, \ref{#2})}}
\nc{\eqm}[2]{\mbox{Eqs.~(\ref{#1}-\ref{#2})}}
\nc{\eq}[1]{\mbox{Eq.~(\ref{#1})}}
\nc{\figs}[2]{\mbox{Figs.~\ref{#1} and \ref{#2}}}
\nc{\fig}[1]{\mbox{Fig.~\ref{#1}}}
\nc{\tbl}[1]{\mbox{Table ~\ref{#1}}}
\nc{\tbls}[2]{\mbox{Tables ~\ref{#1} and \ref{#2}}}

\nc{\spc}{\hspace*{10pt}}
\newlength{\overeqskip}
\newlength{\undereqskip}
\nc{\arreq}{&\!=\!&} \nc{\arrmi}{&\!-\!&} \nc{\arrpl}{&\!+\!&}
\nc{\arrap}{&\!\!\!\approx\!\!\!&} \nc{\non}{\nonumber\\*}

\def\lsim{\; \raise0.3ex\hbox{$<$\kern-0.75em
       \raise-1.1ex\hbox{$\sim$}}\; }
\def\gsim{\; \raise0.3ex\hbox{$>$\kern-0.75em
       \raise-1.1ex\hbox{$\sim$}}\; }
\nc{\slask}{\!\!\!/} \nc{\bis}{{\prime\prime}} \nc{\pa}{\partial}
\nc{\na}{\nabla} \nc{\ra}{\rangle} 
\nc{\goto}{\rightarrow} \nc{\swap}{\leftrightarrow}
\nc{\abs}[1]{\left|#1\right|}
\nc{\at}[2]{\left.#1\right|_{#2}} \nc{\norm}[1]{\|#1\|}
\nc{\vek}[1]{{\rm\bf #1}} \nc{\integral}[2]{\int\limits_{#1}^{#2}}
\nc{\inv}[1]{\frac{1}{#1}} \nc{\dd}[2]{{\frac{\partial #1}{\partial
#2}}} \nc{\ddd}[2]{{\frac{{\partial}^2 #1}{\partial {#2}^2}}}
\nc{\al}{\alpha} 
\nc{\Del}{\Delta}
\nc{\eps}{\epsilon} \nc{\lam}{\lambda} \nc{\om}{\omega}
\nc{\Om}{\Omega} \nc{\ve}{\varepsilon} \nc{\mn}{{\mu\nu}}
\nc{\tim}{\tilde{m}}
\nc{\tiom}{\tilde{\omega}}
\nc{\tisig}{\tilde{\sigma}}
\nc{\tir}{\tilde{r}}
\nc{\alom}{\kappa_\omega}
\nc{\lamom}{\lambda_\omega}
\nc{\vp}{\varphi}
\nc{\hatsig}{\hat{\sigma}}
\nc{\tirhoE}{\tilde{\rho}_E}
\nc{\dsig}{\delta\sigma}
\nc{\dtheta}{\delta\theta}
\nc{\tirhoQ}{\tilde{\rho}_Q}
\nc{\tiu}{\tilde{u}}

\nc{\dbar}{d\mkern-6mu\mathchar'26}
\nc{\ka}{\frac{\mathbf{k}}{a}}
\nc{\spt}{(t,\mathbf{x})}
\nc{\ok}{\omega_k}
\nc{\frie}{\int \dbar^3k}
\nc{\friew}{\int \frac{\dbar^3k}{2\omega_k}}
\nc{\friewpi}{\int \dbar^3k \sqrt{\frac{\omega_k}{2}}}
\nc{\ladd}{a^\dag_{\bf{k}}}
\nc{\laddm}{a^\dag_{-\bf{k}}}
\nc{\lad}{a_{\bf{k}}}
\nc{\ladm}{a_{-\bf{k}}}
\nc{\laddb}{b^\dag_{\bf{k}}}
\nc{\laddmb}{b^\dag_{-\bf{k}}}
\nc{\ladb}{b_{\bf{k}}}
\nc{\ladbm}{b_{-\bf{k}}}
\nc{\uk}{u_k}
\nc{\vk}{v_k}
\nc{\uka}{u^\ast_k}
\nc{\vka}{v^\ast_k}
\nc{\alk}{\alpha_k}
\nc{\bek}{\beta_k}
\nc{\kx}{\mathbf{k}\cdot\mathbf{x}}
\nc{\norder}[1]{\left:#1\right:}
\nc{\rtwo}{\sqrt{2}}
\nc{\cf}{{\it cf. }}
\renc{\figurename}{\textbf{FIG.}}
\renc{\tablename}{\textbf{TABLE}}
\renewcommand{\baselinestretch}{1.1}
\newcommand{\captionfonts}{\small}
\makeatletter  
\long\def\@makecaption#1#2{%
  \vskip\abovecaptionskip
  \sbox\@tempboxa{{\captionfonts #1: #2}}%
  \ifdim \wd\@tempboxa >\hsize
    {\captionfonts #1: #2\par}
  \else
    \hbox to\hsize{\hfil\box\@tempboxa\hfil}%
  \fi
  \vskip\belowcaptionskip}
\nc{\tblcaption}[1]{\def\@captype{table}\caption{\captionfonts{#1}}}
\makeatother   

\begin{document}
%
%

\newcommand{\HRule}{\rule{\linewidth}{2mm}}
\begin{titlepage}

\begin{center}

\HRule

\vspace*{\stretch{3}}

\Huge{\textbf{The Physics of $Q$-balls}}

\vspace{\stretch{1}}

\Large{\textbf{MITSUO\hspace*{5pt} TSUMAGARI}}

\vspace{\stretch{4}}

\includegraphics[width=0.5\textwidth]{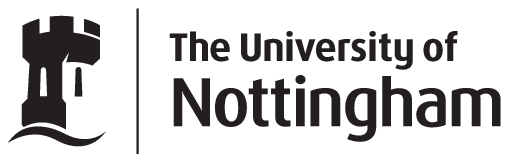}

\vspace{\stretch{1}}

	Thesis submitted to the University of Nottingham\\
	for the degree of Doctor of Philosophy\\ 
	\vspace*{10pt}
	OCTOBER 2009\\

\vspace{\stretch{2}}

\HRule
\end{center}

\end{titlepage}

%
%

\clearpage
\thispagestyle{empty}

\begin{center}

\vspace*{\stretch{6}}

\emph{``The career of a young theoretical physicist consists of treating the harmonic oscillator in ever-increasing levels of abstraction.''}
\vspace{\stretch{1}}
\begin{flushright}
{\bf -- Sidney Coleman. }
\end{flushright}
\vspace{\stretch{10}}
\begin{flushright}
\begin{minipage}[c]{6cm}
\begin{tabular}{cl}
{\bf Supervisor:} & Prof. Edmund Copeland\\
~ & ~ \\
{\bf Examiners:} & Dr. Anne Green\\
~ & Dr. Anupam Mazumdar\\
\end{tabular}
\end{minipage}
\end{flushright}
\vspace{\stretch{4}}
\end{center}

\clearpage

\msp
%
%

\chapter*{Abstract}

In this thesis we investigate the stationary properties and formation process of a class of nontopological solitons, namely $Q$-balls. We explore both the quantum-mechanical and classical stability of $Q$-balls that appear in polynomial, gravity-mediated and gauge-mediated potentials. By presenting our detailed analytic and numerical results, we show that absolutely stable non-thermal $Q$-balls may exist in any kinds of the above potentials. The latter two types of potentials are motivated by Affleck-Dine baryogenesis, which is one of the best candidate theories to solve the present baryon asymmetry. By including quantum corrections in the scalar potentials, a naturally formed condensate in a post-inflationary era can be classically unstable and fragment into $Q$-balls that can be long-lived or decay into the usual baryons/leptons as well as the lightest supersymmeric particles. This scenario naturally provides the baryon asymmetry and the similarity of the energy density between baryons and dark matter in the Universe. Introducing detailed lattice simulations, we argue that the formation, thermalisation and stability of these $Q$-balls depend on the properties of models involved with supersymmetry breaking.
\clearpage



\chapter*{List of Papers}

This thesis consists of the following three papers.
\begin{enumerate}
 \item   M.~I.~Tsumagari, E.~J.~Copeland and P.~M.~Saffin,\\
  \emph{Some stationary properties of a $Q$-ball in arbitrary space dimensions}\\
  Phys.\ Rev.\  D {\bf 78} (2008) 065021
 \item E.~J.~Copeland and M.~I.~Tsumagari,\\
  \emph{$Q$-balls in flat potentials}\\
  Phys.\ Rev.\  D {\bf 80} (2009) 025016

 \item M.~I.~Tsumagari,\\
  \emph{Affleck-Dine dynamics, $Q$-ball formation and thermalisation}\\
  Phys.\ Rev.\  D {\bf 80} (2009) 085010

\vspace*{20pt}

Ref. 1 is described in chapter 2 and chapter 3.\\
Ref. 2 is described in chapter 4.\\
Ref. 3 is described in chapter 5.

\end{enumerate}

\newpage

%
%

\chapter*{{\Huge Acknowledgements}}

I would like to thank my supervisor, Prof. Ed Copeland, for his kindness and assistance in numerous ways. His deep physical insight helped my understanding of a wide variety of branches within physics, and his generous supervision gave me freedom to pursue my own interests. I am also grateful to Drs. Paul Saffin and Osamu Seto for all the help and interesting discussions, and to all the members in the Particle Theory Group and Astronomy Group at the University of Nottingham. In particular, I would like to thank Duncan Buck and Amandeep Josan for their encouraging comments, Shuntaro Mizuno for his valueable suggestions, and all of my office mates for having been nice around me.

\vspace*{5pt}

I would like to express my deep appreciation for the financial support from the University of Nottingham, which initiated and funded this PhD project. This thesis would have not been possible without access to the supercomputers at both the University of Nottingham and University of Cambridge. In particular, I acknowledge the Nottingham HPC facility and UK National Cosmology Supercomputer, Cosmos, respectively. The system administrators in both of the institutes, Colin Bannister (Nottingham), Andrey Kaliazin (Cambridge) and Victor Travieso (Cambridge), helped to optimise and visualise my codes. I am also very happy to thank Drs. Mark Hindmarsh and Neil Bevis for having produced sophisticated parallel codes, LATfield, and Dr. Anders Tranberg and Prof. Noriaki Shibazaki for initiating my present physics career. Finally, I would like to thank my family, Teruyuki, Nobuko, Taro, Yukio, Onticha and Chion Tsumagari, my grand parents, Teruko Fuchibe and Iwato Tsumagari, and my relatives for their continuous encouragement and financial support, Maya Hirabara for her patience and forgiveness, and my friends for being interested in my tales of cosmology.

\clearpage

\ssp                                            
\pagenumbering{roman}                           
\tableofcontents                                
\newpage                                        %
\addcontentsline{toc}{chapter}{List of Figures} 
\listoffigures                                  
\newpage                                        %
\addcontentsline{toc}{chapter}{List of Tables}  
\listoftables                                   
\newpage


\pagenumbering{arabic}                          
\dsp                                            
%
%

\chapter*{{\Huge The Physics of $Q$-balls}}
\addtocontents{toc}{\bigskip \bf The Physics of $Q$-balls \normalfont \endgraf}

\vspace{\stretch{10}}

\begin{figure}[!ht]
  \begin{center}
	\includegraphics[scale=0.5]{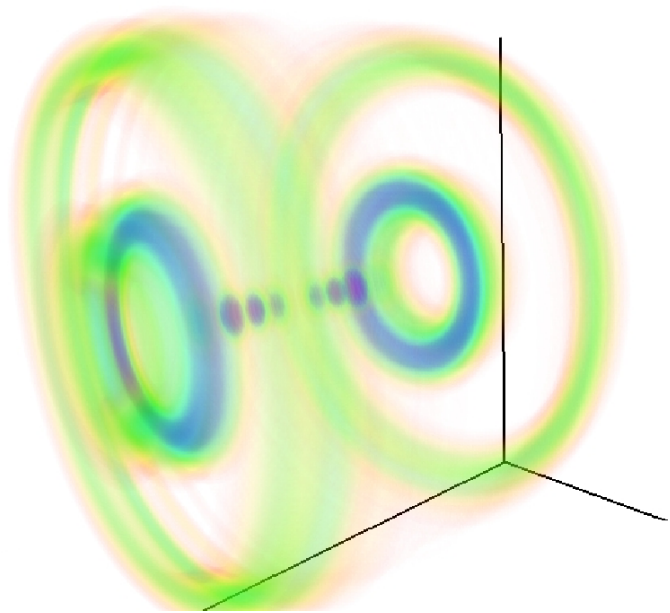}
  \end{center}
  \caption{Ring formation after the collision of a pair of $Q$-balls \cite{mit-web-page}.}
  \label{fig:qring}
\end{figure}

\emph{``What one man can invent, another can discover.''}
\vspace{\stretch{1}}
\begin{flushright}
{\bf -- Sherlock Holmes. }
\end{flushright}
\vspace{\stretch{10}}

\clearpage

\newpage

\chapter[Introduction]{Introduction}\label{ch:chapter1}

\section{The Standard Model to the theory of supersymmetry}

All of the complexities around us come from the mixtures of many simple events. With this belief, theoretical particle physicists have developed our understanding on the extremely small scales of physics (around the size of atoms and even smaller scales). Almost all of the theoretical predictions have been supported well by the experimental results to date, and now we know that the Standard Model (SM) is the fundamental theory to understand the dynamics of elementary particles, like photons and protons. In the SM, the simplicities are symmetries, such that a right-handed person sees the left-handed person in a mirror. The SM in particle physics consists of the mixtures of the three independent groups of the symmetries, each of which can describe one of the four fundamental forces (electromagnetic, weak and strong forces) except gravity. The most familiar one among them might be the electromagnetic force generated through the exchange of photons; for instance, electronically positive charged objects repel each other where photons (massless gauge bosons) are known as the mediators to generate the repulsion force. Similarly, the two other forces come from the exchanges of the corresponding mediators, \ie\ $W^{\pm}$ and $Z$ (massive gauge) bosons and gluons. These mediators including photons are named as gauge bosons, each of which has its own strength (coupling). While the strong force tightly binds protons and neutrons together, the weak force is involved in radioactive decay. Both forces work only in an atom scale, but the strength of the weak force is smaller than the one of the strong force by a factor of $10^6$; in fact, the weak force is $10^{32}$ times stronger than the gravitational force, which works in an infinite distance range, like the electromagnetic force.

Under the conjecture that the physics above some energy scale should be described by one symmetry group parameter, Glashow \etal\ successfully unified the two symmetry groups for electromagnetic and weak forces by introducing the higher (electroweak) symmetry group \cite{Glashow:1970gm, Weinberg:1979bn}. The energy scale of this unified theory (hereafter, the electroweak theory) is around a few $10^2$ GeV, below which the unified electroweak symmetry inevitably breaks down. This symmetry-breaking mechanism requires hypothetical objects, Higgs bosons \cite{Higgs:1964ia}, which give rise to the masses of both weak gauge bosons and other elementary particles, namely quarks and leptons. These theoretical accomplishment in the SM have been in good agreement with the independent experiments, LEP, HERA, and Tevatron run-I \& -II; typically, the different massive weak gauge bosons were well verified in high precision.

Despite the agreement with these extensive experiments, the SM has still several shortcomings. Higgs bosons have not yet been found, and the detection of these particles is still an active research field. Without Higgs bosons, there are no appealing explanations why the weak gauge bosons have nonzero masses. Moreover, the quantum corrections to the Higgs mass become quadratically divergent unless the divergence is canceled out (renormalisation). This problem is known as the \emph{hierarchy problem}. Furthermore, the SM contains far too many parameters to be consistent with the observations in the sense of beauty. Theoretical high energy physicists believed that the physics above the electroweak scale should also unify the strong force. It implies that we could merge the strong force and electroweak interactions into the one theory known as the Grand Unified Theories (GUTs).  Independently, Einstein attempted to unify the electromagnetic force with gravity ultimately.

Can we actually unify the theory of the strong force and the electroweak theory at the high energy scale? The three different gauge strengths for electromagnetic, weak, and strong forces are determined by the ways of the divergence cancellations. Unfortunately, all of the strengths do not meet each other precisely at the GUTs energy scale ($\sim 10^{15}$ GeV). How about the unification to gravity? This problem is related to the hierarchy problem. The SM cannot include the theory of gravity since the quantum effects on gravity give unavoidable infinities, \ie\ nonrenormalisabilitites. The energy scale, at which both quantum and gravitational effects are equally significant, is expected to be around the Planck scale ($\sim 10^{19}$ GeV), and it is is far beyond the electroweak scale. The failures of the unifications are understandable, given that the Planck (or GUTs) scale corresponds to the earliest period of the thermal history of the Universe from zero to $10^{-43}$ (or $10^{-34}$) seconds just after the ``Big Bang''. Hence, we ended up failing to unify the three different forces as well as to solve the hierarchy problem.

One of the particularly exciting solutions for these problems is the addition of an exotic symmetry, supersymmetry (SUSY), to the SM gauge groups. The energy scale of the theory of SUSY lies in between the electroweak scale and the GUTs scale; therefore, SUSY solves the hierarchy problem. More nicely, the theory predicts the same matching point for the three gauge strengths at the GUTs scale. Thus, the discovery of SUSY would be one of the biggest successes in the 21st theoretical physics, and may solve other cosmological problems as we will discuss shortly.

\section{The Big Bang theory to cosmic inflation models}

The recent developments of observational equipments reveal the detailed thermal history of our Universe, starting from Big Bang to present (13.7 billion years).  The Big Bang Theory or BBT for short is based on both general relativity and the cosmological principles in which the energy density of our Universe was uniformly distributed over the large scale and the space-time topology of the Universe has been flat for a long time. In other words, there are no special regions in the Universe, where some small special regions play no important roles of the history and topology of the Universe. To describe such a simple profile over the largest observable scale of the Universe, modern cosmologists often use the following technical words: homogeneity and isotropy. The small scales of the Universe, on the other hand, consist of the inhomogeneous regions which are stars, galaxies, and clusters of galaxies. In 1922, Friedmann \etal\ solved the Einstein equation with the cosmological principles, and proposed that the Universe should be expanding. In the two years later, Hubble measured the distances and the receding speeds of 18 galaxies; he then concluded that each galaxies was indeed receding from us with the linear relation between the distance and the speed, known as the Hubble expansion.  
As a smoking gun of the cosmological principles, Penzias and Wilson discovered the isotropic cosmic microwave background radiation (CMB) which has a black body spectrum with the low temperature, $2.73$ Kelvin \cite{Komatsu:2008hk}.

The present individual observations, Wilkinson Microwave Anisotropy Probe (WMAP) \cite{wmap-web-page}, Sloan Digital Sky Survey \cite{sdss-web-page} and Type Ia supernova (SNIe), determine the precise magnitude of the Hubble expansion rate. They also back up a homogeneous and isotropic profile of the Universe on scales larger than $\sim 100$ Mpc. The history of the Universe is now well understood from the first few minutes after the Big Bang. At the few minutes cosmic time, nucleosynthesis took place, creating light nuclei, \eg\ hydrogen, helium, and lithium, while carbon and the heavier elements were rarely produced in the interior of stars far more later. The observations of the abundances of those light elements are in excellent agreement with the recent theoretical predictions. In fact, the $^4$He abundances are correctly calculated within $1-2\%$ \cite{Bernstein:1988ad}, and the semi-analytic estimations on the abundances of deuterium, $^3$He, and $^7$Li are accurate within a factor of $2-3$ \cite{Esmailzadeh}. These great successes of Big Bang Nucleosynthesis (BBN) and the detection of the Hubble expansion have firmly built up the BBT.

The BBT predicts that the early Universe was extremely hot and dense due to the fact of the Hubble expansion. In such small and high-energy environment, the quantum effects are not negligible; indeed, there are a number of issues of the BBT. We present the five principle problems from now on. First, no information can travel faster than the speed of light according to the standard BBT. Therefore, the Universe should consist of patches of the causally connected regions. In this sense, each of the disconnected regions should be uncorrelated with those neighbors. However, the actual temperature distribution of the CMB is almost isotropic over a large scale which is much larger than the predicted scale, only about 2 degrees on the sky, from the BBT, \ie\ \emph{the horizon problem}. The second issue is a fine-tuning problem on the space-time topology, \emph{the flatness problem}. The Friedmann equations give the three possibilities of the topology, depending on the total dimensionless energy density $\Omega$ of the Universe. The value of $\Omega$ has been extremely close to unity for billions of years, where $\Omega=1$ corresponds to the flat space-time geometry. It implies that the ``God'' must fine-tune the value of $\Omega$ to remain to be unity for the extremely long history of the Universe. This is because any small departure from the flat space-time leads to the other two kinds of topologies obtained by the Friedmann equations. The third problem is the production of \emph{magnetic monopoles}, which may naturally exist in many extensions to the SM. A magnetic monopole is a theoretical object, but it has not yet been detected in our Universe. In fact, we cannot obtain a single side of the magnet (either the north or south pole) even when cut in half. The origin of the \emph{imbalance between matter (baryons) and anti-matter (anti-baryons)} is another controversial puzzle. In BBN, the amount of ordinary matter density $n_b$ relative to the number density of radiation $n_\gamma$, namely the baryon-to-photon ratio $n_b/n_\gamma$, can explain the light element abundances, but it says nothing about the origin of the ratio. The physics within the BBT suggests that both baryons and anti-baryons should be equally created, conserving their charges. It implies that all of the elements (atoms, galaxies, and even human beings) should not exist now since the annihilations between matter and anti-matter take place instantly. The final question of the BBT is the existence of the non-luminous massive matter, \emph{dark matter}. According to the luminosity distribution of a given galaxy, the analytically predicted rotation velocity of the galaxy at large radius is slower than the observed value. It implies that a large amount of invisible massive matter must exist in the galaxy.

How can we solve these problems of the BBT? First of all, we have to modify the \emph{very} early epoch of the Big Bang cosmology. The widely accepted solutions of the first three problems, \emph{the horizon problem, the flatness problem}, and \emph{magnetic monopole problem}, require that a rapid space-time expansion should take place in the very short era just after the Big Bang. This idea, called cosmic inflation or just inflation, is compatible with many observational results.  The fast expansion of inflation gives the reasons why the temperature of the CMB is almost same for any directions and how the causally disconnected regions are correlated due to the past explosive expansion. Additionally, inflation stretches out the past curved space-time and dilutes the primordial inhomogeneity, anisotropy, and the density of the exotic particles, such as magnetic monopoles. The other two problems, \emph{the asymmetry between baryons and anti-baryons} and \emph{dark matter}, will be discussed in the following sections.

Is inflation alternative to the BBT? The inflation models compensate the weaknesses of the BBT, such as an origin of the cosmological principles and the generation of of the large scale structure of the Universe through quantum fluctuations. These density fluctuations deviated from the homogeneous and isotropic values are expected to be nearly scale-invariant and Gaussian, which impressively agree with the WMAP data. Inflation itself is not a complete theory; rather, it is a modification model of the successful BBT. Inflation has however several problems, too. Although the inflation energy scale should be around the GUTs scale, we do not know what the origin of the inflation is. Further, the temperature of the very early Universe was proposed to be nearly zero during inflation, but the early era of the Big Bang Universe should be hot. This discrepancy implies that we need a dynamical mechanism to heat up the cold Universe after inflation, jointing to the onset of the Big Bang cosmology. In a typical scenario of reheating the very early Universe, the dominated energy account for the rapid expansion was released to create the usual SM particles, and in principle the Universe was thermalised by the random motions and scatterings of the created particles. 

Let us itemize the two problems that we did not answer yet: \emph{baryon asymmetry} and \emph{dark matter}. It will turn out that these two problems are related each other, and the plausible solutions could be made by the use of the inflation theory and the new theory of particle physics, namely SUSY.

\section{The two quantities in particle cosmology \\ \hspace*{30pt} -- baryon asymmetry and dark matter}

The present asymmetry between baryons and anti-baryons in the Universe is one of the most mysterious problems in cosmology and particle physics. Indeed, no concentration of anti-baryons has been detected in our observable Universe. From the current several observations (CMB anisotropy measurements, SNIe data, and BAO peak measurements) \cite{Komatsu:2008hk}, the energy density of baryons is only about 4.6\% of the total energy density of the Universe. The remainder of the energy components consist of both dark matter (23.3\%), and dark energy (72.1\%). The baryon-to-photon ratio is also given in \cite{Fields:2008zz},
\be\label{basym}
\frac{n_b}{n_\gamma} \simeq (4.7-6.5) \times 10^{-10}.
\ee
The ratio of the dimensionless energy density between dark matter and baryons is independently obtained in \cite{Spergel:2006hy}
\be\label{dm/b}
\frac{\Omega_{DM}}{\Omega_{b}}=5.65\pm0.58.
\ee
This thesis deals with a number of issues related to the origin of the above two quantities.

\vspace*{10pt}

The first quantity, \eq{basym}, is larger by a factor of $10^9$ than that predicted within the conventional BBT where the quantity was assumed to be zero. In 1985, within the SM of particle physics, Shaposhnikov \etal\ \cite{Kuzmin:1985mm} considered a model based on electroweak physics to explain the origin of this baryon abundance, the so-called electroweak baryogenesis. It satisfies the well-known Sakharov's conditions required for successful baryogenesis \cite{Sakharov:1967dj}, namely baryon number production, the violation of discrete symmetries [charge conjugation (C) and charge parity (CP)], and departure from thermal equilibrium. The magnitude of the CP violation of the SM is, however, far too small to produce the present observed baryon asymmetry. To solve these problems within both BBT and the electroweak baryogenesis,
we require SUSY in addition to the usual gauge symmetry group of the SM. In the minimal super-symmetric extension of the SM (MSSM), Affleck \etal\ \cite{Affleck:1984fy} and Dine \etal\ \cite{Dine:1995kz} proposed a more successful baryogenesis scenario, known as Affleck-Dine (AD) baryogenesis. It can solve a number of severe cosmological problems, such as gravitino and moduli overproduction, which are harmful for successful BBN. More strikingly, AD baryogenesis may also provide the mechanism to obtain the second quantity, \eq{dm/b}, which implies that the baryonic matter and dark matter could share the same origin.

\vspace*{10pt}

How does AD baryogenesis naturally provide the quantities in \eqs{basym}{dm/b} ? Let us now look at the original AD baryogenesis scenario in the MSSM in more detail (for a review see \cite{Dine:2003ax}). The MSSM has nearly 300 flat directions, some of which are uplifted by SUSY breaking effects arising from nonrenormalisable terms, and we can parametrise one of the flat directions in terms of a complex scalar field known as an AD field, which consists of a combination of squarks and/or sleptons (supersymmetric partners of quarks and leptons). During an inflationary epoch in the very early Universe, the AD field evolves to a large field expectation value, and squarks and sleptons form homogeneous condensate. After inflation, the motion of the AD field can be kicked along the phase direction due to the A-terms arising from the nonrenormalisable terms, which are essential for the baryon generation. Through thermal scattering, the AD condensate decays into the usual baryons/leptons and the lightest SUSY particles (which are candidates for dark matter), the AD baryogenesis then becomes complete. By including radiative and/or thermal corrections to the mass term of the scalar potentials, it alters the above standard AD baryogenesis scenario and gives a rich variety of cosmological implications \cite{Enqvist:2003gh}. In this alternative scenario, the AD condensate can be classically unstable against spatial perturbations, and fragment to bubble-like objects, eventually evolving into a stable nontopological soliton, the SUSY $Q$-ball \cite{Kusenko:1997ad}, which is a candidate for self-interacting cold dark matter. The fraction of the $Q$-balls could also contribute to the number density of baryons. With a low-energy SUSY breaking scale $M_S \sim 1-10$ TeV and a plausible charge $Q \sim 10^{26}$ (baryon number) of the SUSY $Q$-balls, Laine \etal\ \cite{Laine:1998rg} found 
\be\label{qbasym}
\frac{n_b}{n_\gamma} \sim 10^{-10}\bset{\frac{M_S}{\rm TeV}}^{-2}\bset{\frac{Q}{10^{26}}}^{-1/2}, \hspace*{30pt} \frac{\Omega_{DM}}{\Omega_b}\sim 10,
\ee
which are the correct orders of magnitude required in \eqs{basym}{dm/b}.

\section{$Q$-ball and its stability}

What exactly is a soliton and $Q$-ball? A soliton is a nonlinear and nondissipative solution which appears in a large variety of both classical and quantum field theory. The energy density of this solution is smooth, and compacted in a finite region space, and solitons themselves behave as the usual elementary particles of the SM. Because of the origin of their stability, there exist two types of solitons, \ie\ topological solitons and nontopological solitons. A conserved Noether charge stabilises nontopological solitons, unlike the case of topological solitons whose stability is ensured by the presence of conserved topological charges. In a pioneering work by Freidberg, Lee, and Sirlin \cite{Friedberg:1976me}, nontopological solitons were introduced in a successful quantum chromodynamics (hadron) model. Later, Coleman \cite{Coleman:1985ki} proposed that it was possible for a new class of non-topological solitons to exist within a self-interacting scalar field theory by introducing the notion of a $Q$-ball. His model had a continuous unbroken global U(1) charge $Q$, which corresponds to an angular motion with angular velocity $\omega$ in the U(1) internal space. Once formed, a $Q$-ball is absolutely stable if five conditions are satisfied: (1) \textit{existence condition} - its potential should grow slower than the quadratic mass term, and this can be realised through a number of routes such as the inclusion of radiative or finite temperature corrections to a bare mass, or  nonlinear terms in a polynomial potential, (2) \textit{absolute stability condition} - the energy $E_Q$ (or mass) of a $Q$-ball must be lower than the corresponding energy that the collection of the lightest possible scalar particle quanta could have, (3) \textit{classical stability condition} - the $Q$-ball should be stable to linear fluctuations; with the threshold of the stability being located at the saddle point of the Euclidean action, (4) \textit{fission condition} - the energy of a single $Q$-ball must be less than the total energy of the smaller $Q$-balls that it could in principle fragment into, (5) \emph{decays into fermions}  - a $Q$-ball should not couple with fermions strongly. If coupling with light/massless fermions, the $Q$-ball evaporates via the surface area. For the first condition to be satisfied, we require
\bea{four}
\omega_-\leq \abs{\omega}<\omega_+,
\eea
where $\omega_\mp$ are the lower and upper limits of $\om$ that the $Q$-ball can have. The lower limit $\om \simeq \om_-$ can define thin-wall $Q$-balls, whilst the upper limit $\om\simeq \om_+$ can define ``thick-wall'' $Q$-balls. Although the thin-wall $Q$-ball can actually have a thin-wall thickness, the ``thick-wall'' limit does not imply that the ``thick-wall'' $Q$-ball has to have a large thickness which is comparable to the size of the core size. In chapter \ref{ch:chapter2} of this thesis, we review the fundamental properties of $Q$-balls with a complete classical stability analysis given in Appendix \ref{ap1}, following the original work in \cite{Friedberg:1976me}. As a first nontrivial example of standard $Q$-balls, in chapter \ref{ch:qpots} we inspect both analytically and numerically the stationary properties of a single $Q$-ball in an arbitrary number of the spatial dimensions with a general polynomial potential, working in the both thin- and thick-wall limits. We discover the connection of the analyses between the virial relation and the thin- and thick-wall approximations, giving an important quantity $\gamma$ defined by 
\be\label{eqpro}
E_Q\propto Q^{1/\gamma}.
\ee

\section{Supersymmetric $Q$-balls}

From a phenomenological point of view, the most interesting examples are the SUSY $Q$-balls arising within the MSSM. Since they suffer from evaporation, diffusion, dissociation and decay into light fermions \cite{Cohen:1986ct}, SUSY $Q$-balls are generally not stable but long-lived. The stability and cosmological consequences [such as \eqs{basym}{dm/b}] of these $Q$-balls depend on how SUSY is broken in the hidden sector, transmitting to the observable sector through so-called messengers. In the gravity-mediated \cite{de Gouvea:1997tn} or gauge-mediated scenarios \cite{Dvali:1997qv}, the messengers correspond respectively either to supergravity fields or to some heavy particles charged under the gauge group of the SM. So far, no reliable stability analyses on these SUSY $Q$-balls have been performed analytically as well as numerically. In chapter \ref{ch:qbflt}, we, therefore, present a thorough stability analysis of the SUSY $Q$-balls with flat potentials arising in both gravity-mediated and gauge-mediated models. We show that the associated $Q$-matter formed in gravity-mediated potentials can be stable against decays into their own free-particles as long as the coupling constant of the nonrenormalisable term is small, and that all of the possible three-dimensional $Q$-ball configurations are classically stable. Three-dimensional gauge-mediated $Q$-balls can be absolutely stable in the ``thin-wall'' limit, but are completely unstable in the ``thick-wall'' limit. In both of the above models, we also obtain the values of $\gamma$, \eg\ $1/\gamma=3/4$ for ``thin-wall'' $Q$-balls in gauge-mediated potentials. This example turns out that these $Q$-balls are the most energetically compact state given a sufficiently large charge, recalling \eq{eqpro}.

\section{Observational limits on $Q$-balls}

\begin{figure}[!ht]
  \begin{center}
	\includegraphics[scale=0.6]{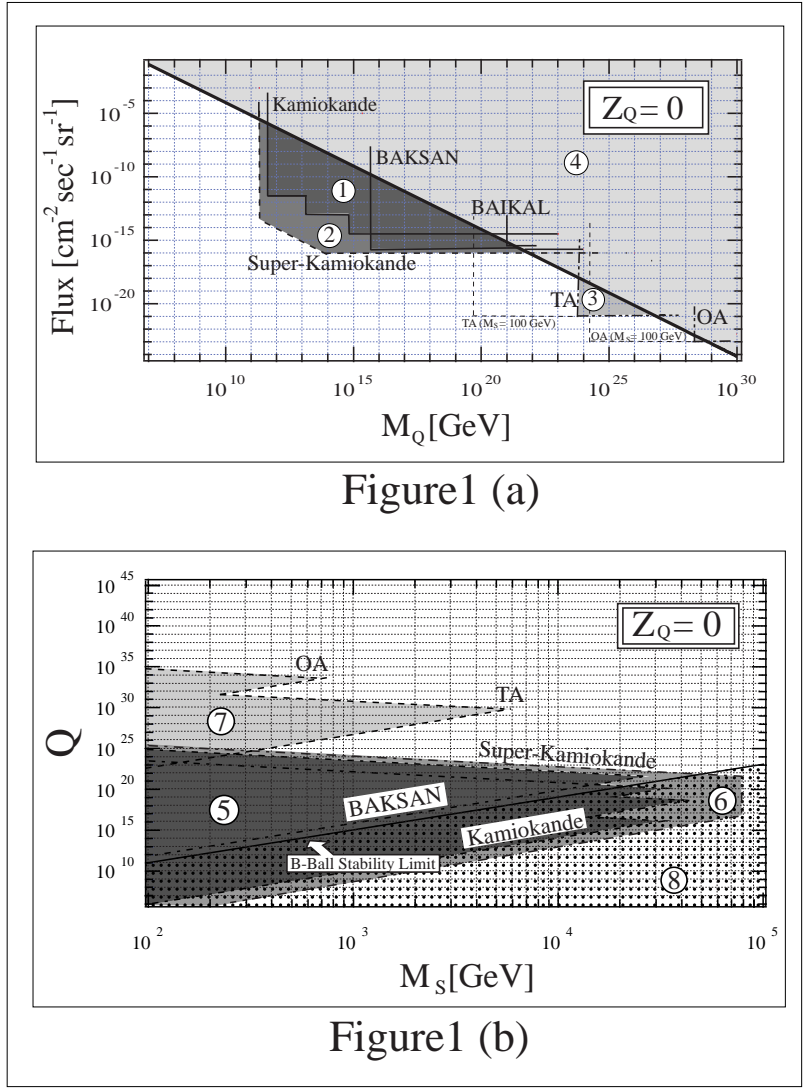}
  \end{center}
  \caption{Figure 1 (a) cited from \cite{Arafune:2000yv} plots the several bounds on flux and mass for SENS ($Z_Q=0$) where $Z_Q$ is an electric charge and $M_Q$ (or equivalently $E_Q$) is the mass of SENS. The diagonal line corresponds to the flux estimated under the assumption that dark matter in the galaxy ($\sim 0.3 {\rm GeV} /\mbox{cm}^3$) is mainly from SENS. Therefore, the regions 4, which is above this diagonal line, is ruled out. The region 1 is also experimentally banned by {\it
Gyrlyanda} \cite{Belolaptikov:1998mn}, {\it BAKSAN} \cite{Alekseev:1982yr}, and {\it Kamiokande} \cite{Kajita:1986nb}; similarly, the region 2 is also excluded by the {\it Super-Kamiokande} experiments. The future experiments, such as {\it TA} \cite{Ta} and {\it OA} \cite{Owl}, will clarify the region 3. Figure 1 (b) shows the bounds for the charge $Q$ and the SUSY breaking scale $M_S$ for SENS in the regions 5,6, and 7 where each regions was obtained by the same experiments as the regions 1, 2 and 3, respectively. Below the 'B-Ball Stability Limit' line, the region 8 is excluded so that the allowed region is  $Q>10^{22}$ where SENS can be dark matter or part of it.}
  \label{fig:oblimqb1}
\end{figure}

\begin{figure}[!ht]
  \begin{center}
	\includegraphics[scale=0.6]{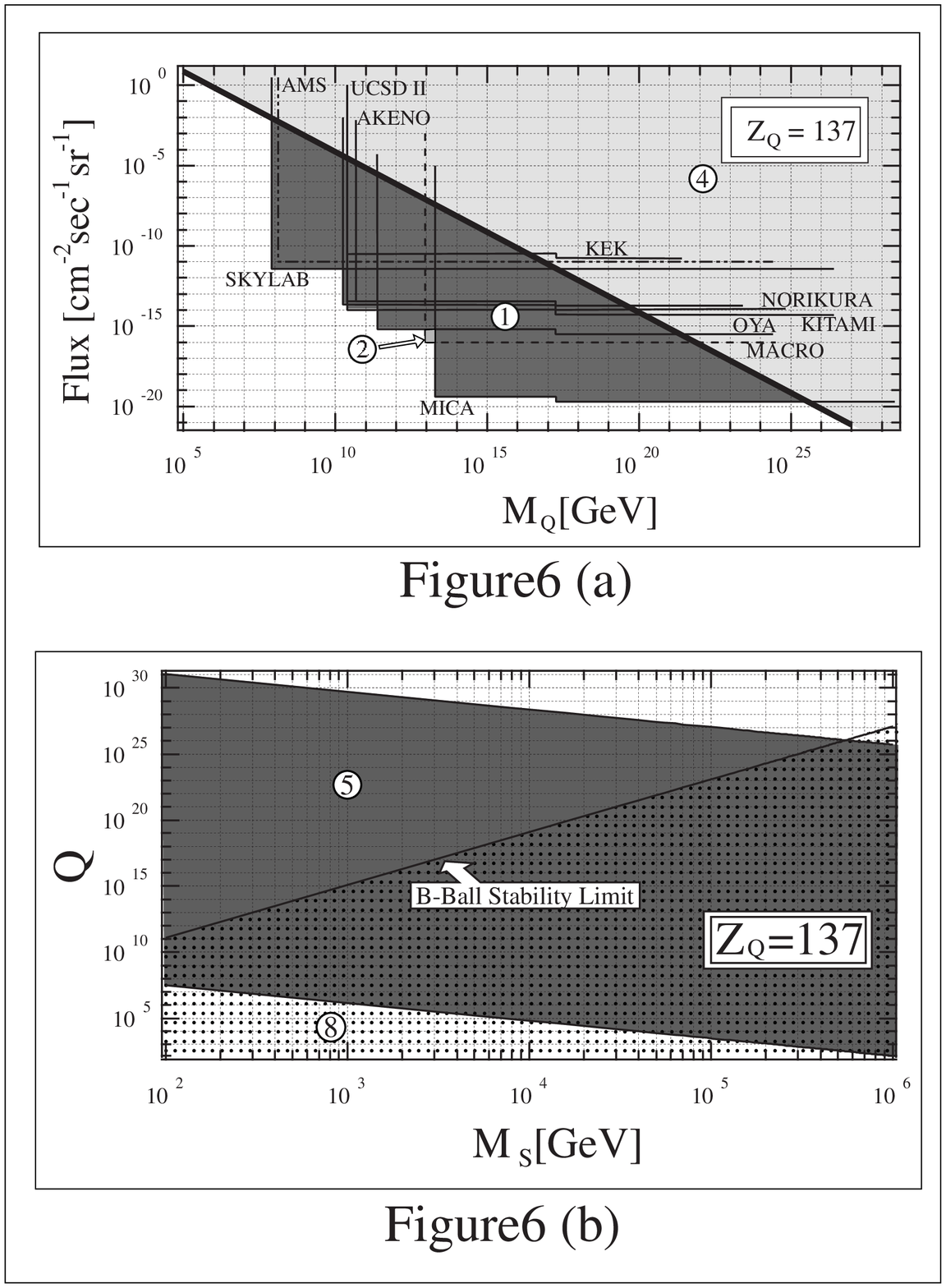}
  \end{center}
\caption{Figure 6 (a) also cited from \cite{Arafune:2000yv} plots the the flux and mass for SECS with 
$Z_Q=137$ instead of SENS. The region 4 above the diagonal line is also excluded as before. We can also exclude the regions 1 and 2 by the present and past experiments such as {\it KEK} \cite{Nakamura:1985rg}, {\it AKENO} \cite{Hara:1985bs}, {\it UCSD}$I\!I$ \cite{Buckland:1990ux}, {\it MACRO} \cite{Bakari:2000dq, Cei:1998qq, Ambrosio:1999gj}, {\it OYA} \cite{Orito:1990ny}, {\it NORIKURA} \cite{Nakamura:1990js}, {\it SKYLAB} \cite{Sky},
{\it KITAMI} \cite{Doke:1983if}, and {\it MICA} \cite{Price:1986ky}, and by the future experiments, \eg\
{\it AMS} \cite{Alcaraz:2000ss, Derkaoui:1998uv}. Similarly, Figure 6 (b) is plotted as Figure 1 (b).
}
  \label{fig:oblimqb2}
\end{figure}

Can we detect the signals of $Q$-balls through observations? The current experiments in the search for SUSY $Q$-balls are sensitive to electrically neutral $Q$-balls (SENS) \cite{Takenaga:2006nr} and electrically charged $Q$-balls (SECS) \cite{Cecchini:2008su} where the present and past experiment data and the estimations from the future experiments are summarised in \fig{fig:oblimqb1} for  SENS and \fig{fig:oblimqb2} for SECS. The core of a SENS has a large vacuum expectation value of squark, slepton, and/or Higgs fields, where the symmetry of the strong force (colour confinement, which is the binding of mesons and baryons, composed of two and three quarks ($q$), respectively) is broken. If a nucleon enters into this deconfinement region, it dissociates into three quarks, some of which may be absorbed by the SENS. This implies that the following reaction may happen, $qq\to \tilde{q}\tilde{q}$, releasing the energy, $\sim 1$ GeV/nucleon, where $\tilde{q}$ is the anti-quark of a quark ($q$). Moreover, a similar process to proton decay may take place around the thin-wall region of the SENS. From the Japan-US large underground water Cherenkov detector, Super-Kamiokande \cite{Arafune:2000yv}, an upper bound on the SENS flux has been obtained, which is equivalent to giving the lower bound on the mass of the SENS, \ie\ 
\be{}
E_Q > [4.0 \times 10^{11},\; 1.2\times 10^{13},\; 5.6\times 10^{13}] \hspace*{4pt} \bset{\frac{M_S}{\textrm{TeV}}}^4 {\rm GeV},
\ee
with the respective cross sections $[0.1,\; 1,\; 10]$ mb, where $M_S$ is a typical SUSY breaking scale appeared in \eq{qbasym}. On the other hand, for a sufficiently large charge, a SECS whose effective radius is $\sim 1$ \AA\ comparable to an atom size accompanied with electron clouds loses the energy due to the interaction with nuclei and electrons of the traversed medium. In the SLIM and MACRO experiments \cite{Cecchini:2008su}, which are designed to be sensitive to SECS, it also gives the upper bound on the SECS flux and equivalently the lower bound on the SECS mass with the electrical charge $Z_Q=137$, \ie\,
\be{}
E_Q > 2.5 \times 10^7 \hspace*{4pt} \bset{\frac{\rho L}{\rm gr/cm^2}} {\rm GeV}, 
\ee
where $\rho$ and $L$ are the density and length of the electronic medium. The present best experimental limit from Super-Kamiokande II \cite{Takenaga:2006nr} is 
\be{}
Q\gtrsim 10^{24},
\ee
\cf\  \eq{qbasym}; the future Cherenkov detectors are expected to tighten these limits further.

\section{$Q$-ball formation}

The final big question, which is the main goal of this thesis, might be 'How do $Q$-balls form and interact with each other in the very early Universe ?'. The dynamics and formation of $Q$-balls involve nonlinear, nonperturbative, and out-of-equilibrium processes, which generally require numerical simulations. With different relative phases and initial velocities between two $Q$-balls in a polynomial potential, we found rings form after the collision of a pair of the $Q$-balls \cite{mit-web-page}, \eg\ see \fig{fig:qring}. It has been found \cite{Radu:2008pp} that similar ring-like solutions are responsible for the excited states from the ground state ($Q$-ball) by introducing extra degrees of freedom, \ie\ spatial spins and twists. Further, the main $Q$-ball formation process has been examined in gauge-mediated and gravity-mediated models \cite{Kasuya:1999wu, Kasuya:2000wx}; however, those previous analyses used initial conditions, which were chosen under a simple assumption, and the lattice simulations were too small and short to reproduce satisfactory results. With more generic initial conditions and much larger and longer lattice simulations, we present, in chapter \ref{ch:adqbform}, both analytically and numerically the consistent analysis from the AD dynamics to the subsequent semiclassical evolution, \ie\ $Q$-ball formation, in both gravity-mediated and gauge-mediated models. We obtain analytically the elliptic motions in the AD dynamics as the analogy of the well-known planetary motions (\ie\ Kepler-problem). By solving the equations of motion in a $3+1$ (and $2+1$)-dimensional lattice with $512^3$ (and $512^2$) lattice units, we find that the $Q$-ball formation goes through three distinct stages as a model of reheating process in the very early Universe after inflation: \emph{pre-thermalisation}, \emph{bubble collisions} and main \emph{thermalisation}. The second stage of the $Q$-ball formation lasts rather long compared to the first stage, and the main thermalisation process is unique due to the presence of ``thermal thin-wall $Q$-balls''.

\vspace*{10pt}

\section{Outline of the thesis}

This thesis is organised as follows. In chapter \ref{ch:chapter2}, we introduce the fundamental aspects of $Q$-balls. We then show the detailed stability analysis and stationary properties of both thin- and thick-wall $Q$-balls in a general polynomial potential in chapter \ref{ch:qpots} \cite{Tsumagari:2008bv}. Following this analysis, we study both the classical and quantum-mechanical stability of $Q$-balls in the MSSM flat potentials in chapter \ref{ch:qbflt} \cite{Copeland:2009as}. With numerical lattice simulations, we investigate how those SUSY $Q$-balls form in chapter \ref{ch:adqbform} \cite{Tsumagari:2009na}. Finally, we summarise our main results and discuss possible future work in chapter \ref{ch:concl}. Six appendices are included. We present the complete classical stability analysis of $Q$-balls in Appendix \ref{ap1}. For the analysis of gravity-mediated $Q$-balls, we find an exact solution in Appendix \ref{exactsol}, and show the classical stability of the $Q$-balls in the thick-wall limit with a Gaussian ansatz in Appendix \ref{appxthick}. We find the equations of motion for multi-scalar fields in Appendix \ref{MULTI}. In Appendix \ref{App2}, we obtain elliptic forms for the orbits of AD fields. In Appendix \ref{BERT} we prove Bertrand's theorem that there are only two potential forms allowed to be closed ``planetary'' orbits.


\chapter{Foundations}\label{ch:chapter2}

\section{Introduction}

In a pioneering paper published in 1985 \cite{Coleman:1985ki}, Sidney Coleman showed that it was possible for a new  class of non-topological solitons \cite{Friedberg:1976me} to exist within a self-interacting system by introducing the notion of a $Q$-ball (for reviews see \cite{Dine:2003ax, Enqvist:2003gh, Lee:1991ax, raj}). Once formed, a $Q$-ball is absolutely stable if the five conditions, one of which is shown in \eq{four}, are satisfied. The lower limit, $\om \simeq \om_-$, of the existence condition in \eq{four} can define thin-wall $Q$-balls, either without \cite{Coleman:1985ki} or with \cite{Spector:1987ag, Shiromizu:1998rt} the wall thickness being taken into account, while the upper limit, $\om\simeq \om_+$, can define thick-wall $Q$-balls in \cite{Kusenko:1997ad} which may be approximated by a simple Gaussian ansatz \cite{Gleiser:2005iq}.

There is a vast literature on nontopological solitons, including $Q$-balls. They have been seen to be solutions in Abelian gauge theories \cite{Lee:1988ag, Levi:2001aw, Anagnostopoulos:2001dh, Li:2001he, Shiromizu:1998eh}, in non-Abelian theories \cite{Safian:1987pr, Safian:1988cz, Axenides:1998fc}, in non-Abelian gauge theories \cite{Friedberg:1976az, Friedberg:1976ay, Kusenko:1997vi}, in self-dual (Maxwell-) Chern-Simons theory \cite{Jackiw:1990pr, Jackiw:1990aw, Hong:1990yh, DeshaiesJacques:2006ae}, in noncommutative complex scalar field theory \cite{Kiem:2001ny}, in (nonlinear) sigma models \cite{Leese:1990hd, Abraham:1991ki}, and in hadron models which include fermionic interactions \cite{Lee:1988ag, Friedberg:1976eg, Friedberg:1977xf}, as well as in the presence of gravity \cite{Lee:1986tr, Lynn:1988rb, Mielke:1997re, Schunck:2003kk}. $Q$-balls themselves have been quantized either by canonical \cite{Friedberg:1976me} or by path integral schemes \cite{Benson:1989id, Rajaraman:1975qr}. With thermal effects, it has been shown that $Q$-balls coupled to light/massless fermions are able to non-perturbatively and semi-classically evaporate away on their surface \cite{Cohen:1986ct, Clark:2005zc, Clark:2006if}; however, at sufficiently low temperatures they become stable, indeed they then tend to grow \cite{Laine:1998rg, Benson:1991nj}. The authors in \cite{Friedberg:1976me, Volkov:2002aj} have discussed and analysed the spatially excited states of $Q$-balls, including radial modes or spatially dependent phase excitations. A more general mathematical argument concerning the stability of solitary waves can be found in \cite{Shatah:1985vp, Blanchard:87, St}. Standard $Q$-balls exist in an arbitrary number of space dimensions $D$ and are able to avoid the restriction arising from  Derrick's theorem  \cite{Derrick:1964ww} because they are time-dependent solutions. A related class of objects to $Q$-balls are known as oscillons \cite{oscillon1, Gleiser:1993pt, Copeland:1995fq} or as I-balls \cite{Kasuya:2002zs}, and recent attention has turned to the dynamics of these time-dependent, nonlinear, and metastable configurations \cite{Saffin:2006yk, Gleiser:2007ts, Hindmarsh:2007jb}.

In this chapter we review the important stationary properties of a standard $Q$-ball in an arbitrary number of spatial dimensions $D$. By introducing a $Q$-ball ansatz in Sec. \ref{sec:qans}, we obtain powerful tools, Legendre relations  and characteristic slopes, in Sec. \ref{sec:leg} and Sec. \ref{sec:ch}. In Sec. \ref{sec:qbeq}, we then obtain a $Q$-ball equation and the existence condition, which requires certain restrictions on the allowed potentials. In Sec. \ref{sec:stb} we obtain four types of $Q$-ball stability conditions. By scaling a $Q$-ball solution, we find the characteristic slopes, depending on the ratio between the surface energy and potential energy, in Sec. \ref{sec:viri}. In Appendix \ref{ap1}, we show a general classical stability analysis of $Q$-balls, following \cite{Lee:1991ax}. This chapter contains work that is published in \cite{Tsumagari:2008bv}.

\section{\textit{Q}-ball ansatz}\label{sec:qans}

We consider a complex scalar field $\phi$ in Minkowski spacetime of arbitrary spatial dimensions $D$ with a U(1) potential bounded by $U(\abs{\phi})\ge 0$ for any values of $\phi$:
\bea{act}
S&=& \int d^{D+1}x \sqrt{-g}\; \ml,\\
\textrm{where} \hspace{10pt} \ml&=& -\half g^{\mn}\pa_\mu \phi^{\dag} \pa_\nu \phi - U(|\phi|).
\eea
The metric is $ds^2=g_{\mn}dx^\mu dx^\nu=-dt^2+h_{ij}dx^idx^j$ and $g$ is the determinant of $g_{\mn}$, where $\mu,\, \nu$ run from $0$ to $D$, and $i,\, j$ denote spatial indices running from $1$ to $D$. Now, using the standard decomposition of $\phi$ in terms of two real scalar fields $\phi=\sigma e^{i \theta}$, the energy momentum tensor $T_{\mn}\equiv -\frac{2}{\sqrt{-g}}\frac{\delta S}{\delta g_{\mn}} + (symmetrising\; factors)$ and the conserved U(1) global current $j_{\mu,U(1)}$ via the Noether theorem, we obtain
\bea{}
T_{\mn}&=&(\pa_\mu \sigma \pa_\nu \sigma + \sigma^2 \pa_\mu \theta \pa_\nu \theta)+g_{\mn} \ml,\\
j_{\mu,U(1)}&=&\sigma^2 \pa_\mu \theta.
\eea
Using a basis of vectors $\{n^\mu_{(a)}\}$ where $n^\mu_{(t)}$ is time-like and
  $n^\mu_{(i)}$ are space-like unit vectors oriented along the spatial $i$-direction, the above currents give the definitions of energy density $\rho_E$, charge density $\rho_Q$, momentum flux $\hat{P}_i$ and pressure $p$:
\bea{defeq}
\rho_E \equiv T_{\mn} n^\mu_{(t)} n^\nu_{(t)};\hspace{10pt} \rho_Q\equiv j_\mu n^\mu_{(t)};\hspace{10pt} \hat{P}_i \equiv T_{\mn} n^\mu_{(t)} n^\mu_{(i)};\hspace{10pt} p\equiv T_{\mn} n^\mu_{(i)} n^\nu_{(i)}.
\eea
Defining the $D$ dimensional volume $V_D$ bounded by a $(D-1)$-sphere, the Noether charges (energy, momenta, and U(1) charge) become
\beq{epq}
E=\int_{V_D}\rho_E,\hspace{15pt} P_i=\int_{V_D} \hat{P}_i,\hspace{15pt} Q=\int_{V_D}\rho_Q,
\eeq
where $\int_{V_D}\equiv \int d^Dx \sqrt{h}$. Minimising an energy with a fixed charge $Q$ for any degrees of freedom, we find the $Q$-ball (lowest) energy $E_Q$ by introducing a Lagrange multiplier $\omega$ and setting $n^\mu_t=(-1,0,0,\dots,0)$:
\bea{EQ}
E_Q&=&E+\omega\bset{Q-\int_{V_D} \rho_Q},\\
\label{EQ2} &=& \omega Q + \int_{V_D} \bset{\half \set{ \dot{\sigma}^2+\sigma^2(\dot{\theta}-\omega)^2 +(\nabla \sigma)^2 + \sigma^2 (\nabla \theta)^2} + \Uo }, \\
\label{legtrns}&=& \omega Q + \So,
\eea
where $\Uo=U-\half \omega^2 \sigma^2$, $\dot{\sigma} \equiv \frac{d\sigma}{dt}$ etc... and $\omega$ will turn out to be the rotation frequency in the U(1) internal space. The presence of the positive definite terms in \eq{EQ2} suggests that the lowest energy solution is obtained by setting
$\dot{\sigma}=0=\dot{\theta}-\omega=\nabla \theta$. The Euclidean action $\So$ and the effective potential $U_\omega$ in \eqs{EQ2}{legtrns} are finally given by
\beq{Uo}
\So=\int_{V_D} \half (\nabla \sigma)^2 + U_\omega,\hspace{10pt} U_\omega \equiv U -\half \omega^2 \sigma^2.
\eeq
The second term in $U_\omega$ comes from the internal spin of the complex field. Following Friedberg et. al \cite{Friedberg:1976me}, it is useful to define the functional
\beq{legtrns2}
G_I\equiv \int_{V_D} \half (\nabla \sigma)^2 + U
=  E_Q - \bset{\half \omega^2} I = \So + \bset{\half \omega^2} I,
\eeq
where $\half \omega^2$ is the corresponding Lagrange multiplier and $I\equiv \int_{V_D} \sigma^2$.

Given that the spherically symmetric profile is the minimum energy configuration  \cite{spherical}, we are lead to the standard stationary $Q$-ball ansatz at zero-temperature
\beq{qansatz}
\phi=\sigma(r)e^{i\omega t}.
\eeq
Substituting \eq{qansatz} into \eq{defeq}, we find
\bea{rho_Q}
\label{cheg}    \rho_E&=& \half \sigma^{\p 2} + U+\half \sigma^2 \om^2 ,\hspace*{10pt} \rho_Q=\omega \sigma^2,\\
\label{radialp}  p_r&=&\half \sigma^{\p 2}-\Uo, \hspace*{10pt} P_i=0,
\eea
where $\sigma^{\p} \equiv \frac{d\sigma}{dr}$ and $p_r$ is a radial pressure given in terms of the radially oriented unit vector $n^\mu_s=(0,1,0,\dots,0)$.
Without loss of generality, we set both $\om$ and $Q$ as positive.
\section{Legendre relations}\label{sec:leg}
It is sometimes difficult to compute $E_Q$ directly, but using Legendre relations often helps \cite{Friedberg:1976me}. In our case, from \eq{legtrns} and \eq{legtrns2} we find
\beq{legendre}
\left.\frac{d E_Q}{d Q}\right|_{\So} =\omega,\hspace{10pt} \left.\frac{d \So}{d\omega}\right|_{E_Q}=-Q,\hspace{10pt} \left.\frac{d G_I}{dI}\right|_{\So}=\half \omega^2
\eeq
because $Q$-ball solutions give the extrema of $E_Q,\; \So$, and $G_I$ with respect to  $Q,\; \om,$ and $I$, respectively.
These variables match the corresponding ``thermodynamic'' ones:
$E_Q,\; \omega,\; Q,\; \So$, and $G_I$ correspond to the internal energy, chemical potential, particle number,
and ``thermodynamic'' potentials \cite{Laine:1998rg}. After computing $\So$ or $G_I$,
one can calculate $Q$ or $\half \om^2$ using the second or third relation in \eq{legendre},
and can compute $E_Q$ using \eq{legtrns} or \eq{legtrns2}, \ie\ 
\beq{easycalc}
\So \to Q=-\frac{d \So}{d \omega} \to E_Q=\omega Q + \So,
\eeq
or similarly $G_I \to \half \om^2=\frac{d G_I}{dI}\to E_Q=G_I+\bset{\half \om^2} I,\; \So =G_I -\bset{\half  \om^2} I.$ We shall make use of this powerful technique later.

\section{The characteristic slope}\label{sec:ch}
Let us define
\beq{ch}
\gamma(\omega)\equiv \frac{E_Q}{\omega Q}.
\eeq
If $\gamma$ is not a function of $\omega$, we can obtain the following proportional relation using the first expression of \eq{legendre}
\beq{leg2}
E_Q \propto Q^{1/\gamma}.
\eeq
\section{\textit{Q}-ball equation and existence condition}\label{sec:qbeq}

Let us consider the action
$S=-\int dt  \So$ in \eq{act} with our ansatz \eq{qansatz} and the following boundary condition on a $(D-1)$-sphere which represents spatial infinity
\beq{nontbdry}
\sigma^\p|=0\; \textrm{on the ($D-1$)-sphere}.
\eeq
Varying $\So$ with respect to $\sigma$, we obtain the $Q$-ball equation:
\bea{QBeq}
\frac{d^2\sigma}{dr^2}+\frac{D-1}{r}\frac{d\sigma}{dr}-\frac{dU_\omega}{d\sigma}&=&0,\\
\label{dampsig}\lr \frac{d}{dr}\bset{\half \bset{\frac{d\sigma}{dr}}^2-U_\omega}&=&-\frac{D-1}{r}\bset{\frac{d\sigma}{dr}}^2 \le 0.
\eea
There is a well known mechanical analogy for describing the $Q$-ball solution of \eq{QBeq} \cite{Coleman:1985ki}, and that comes from viewing \eq{QBeq} in terms of the Newtonian dynamics of an unit-mass particle with position $\sigma$, moving in potential $-U_\omega$ with a friction $\frac{D-1}{r}$, where $r$ is interpreted as a time co-ordinate. Moreover, $\rho_Q=\omega \sigma^2$ can be considered as the angular momentum \footnote{$I$ is realised as an inertia moment in this mechanical analogy \cite{Coleman:1985ki, Friedberg:1976me}.}. Note that the friction term is proportional to $\frac{D-1}{r}$, and hence becomes significant for high $D$ and/or small $r$. According to \eq{dampsig}, the ``total energy'', $\half \bset{\frac{d\sigma}{dr}}^2-U_\omega$, is conserved for $D=1$ and/or $r \tol \infty$, implying that in that limit the $Q$-balls have no radial pressure, see the first expression of \eq{radialp}. Of course these are really field theory objects and consequently more restrictions apply: 

$\bullet$ no symmetry breaking, in other words $\sigma(r \to large)=0;\ U^{\p\p}(\sigma=0)\equiv m^2>0$ with an effective mass $m$, 

$\bullet$ regularity condition: $\sigma^\p(r=0)=0$,

$\bullet$ reflection symmetry under $\sigma \to -\sigma$.

\begin{figure}[ht]
  \begin{center}
   \subfigure{\label{fig:deg}\includegraphics[angle=-90,scale=0.3]{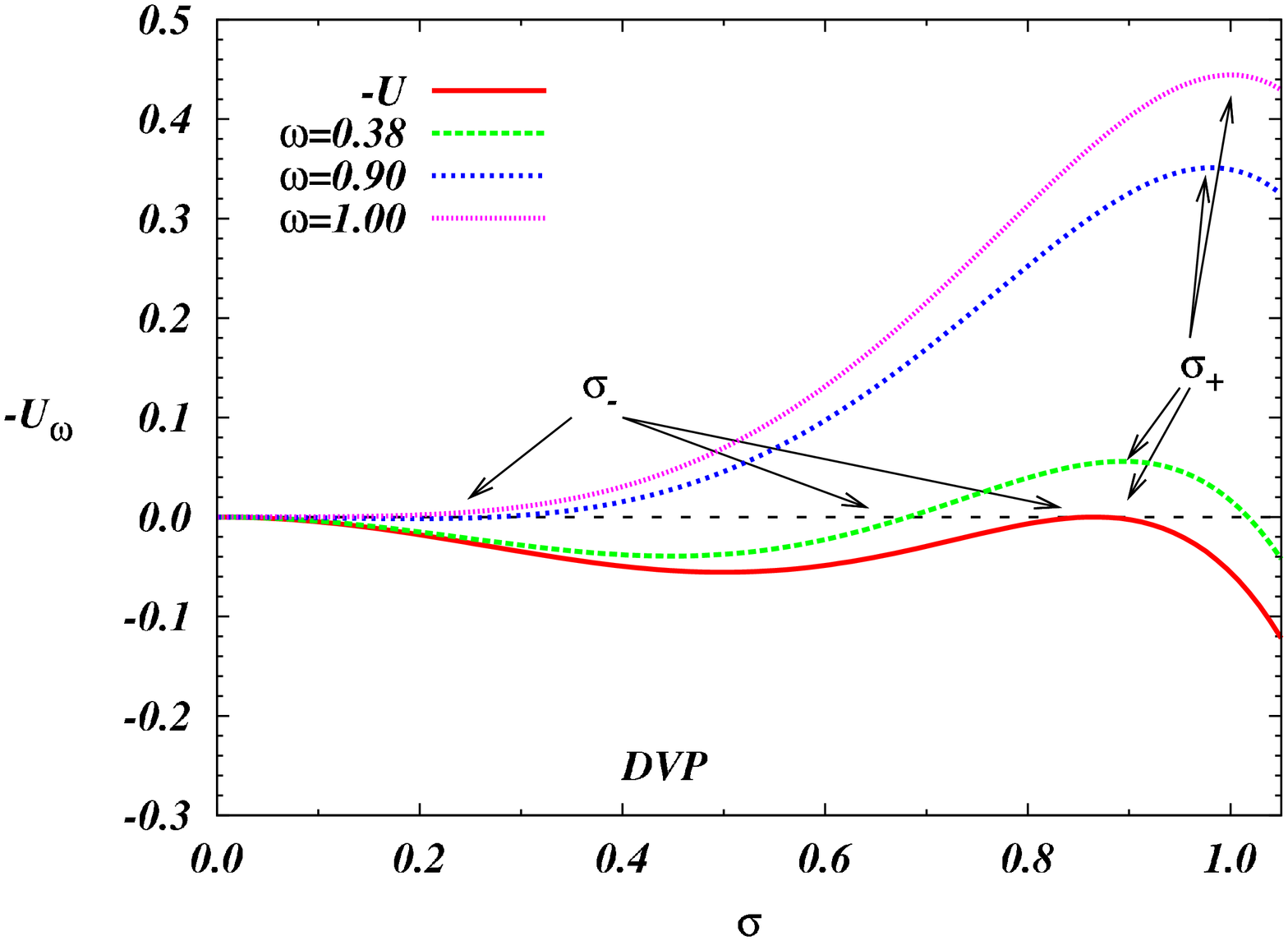}}
   \subfigure{\label{fig:nondeg} \includegraphics[angle=-90, scale=0.3]{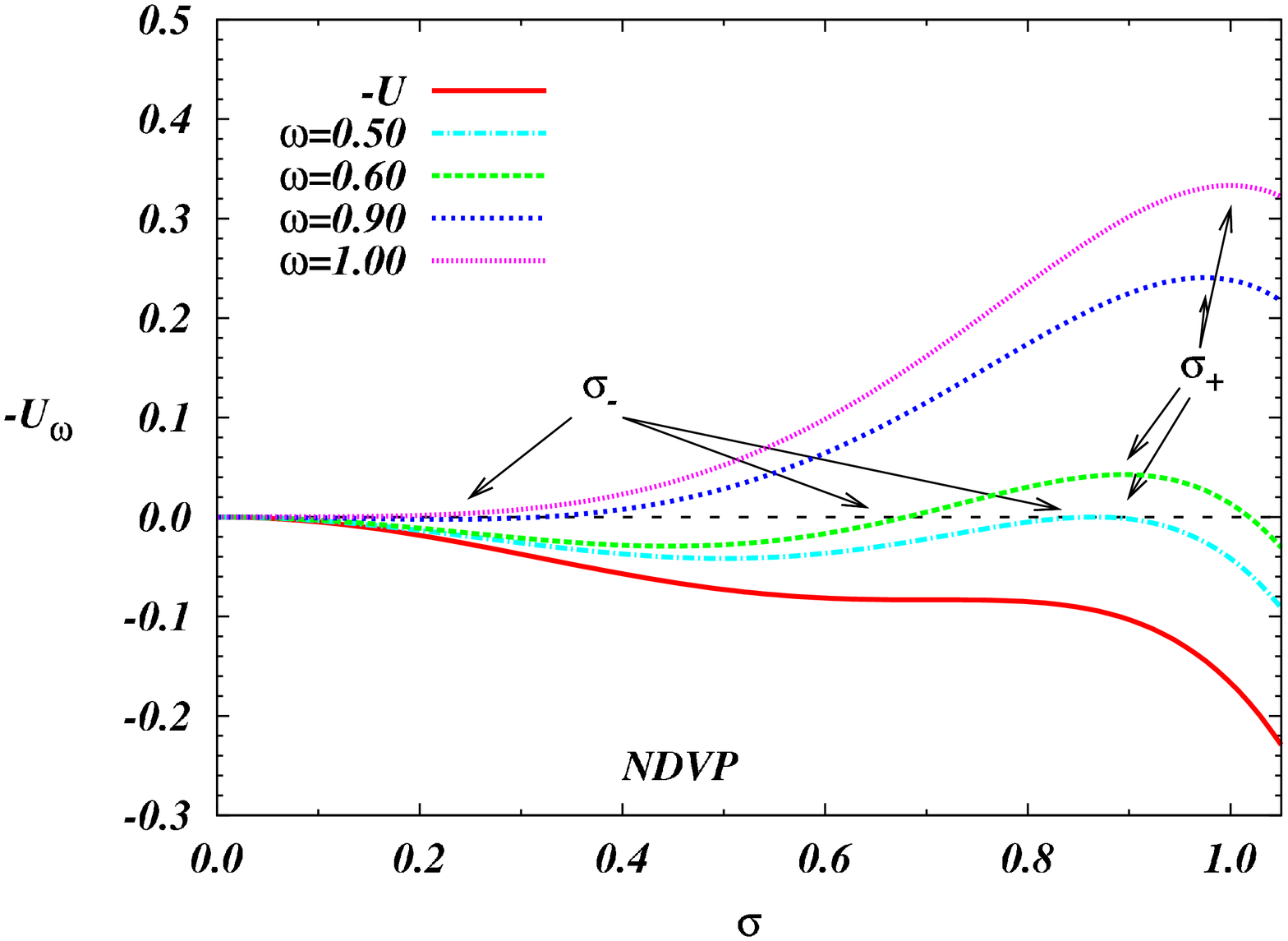}}
  \end{center}
  \caption{Parameters $\sigma_\pm(\om)$ in two typical
  potentials $U(\sigma)=\half \sigma^2-A \sigma^4+ B \sigma^6$ where $\om_+=m=1$ and the effective potentials $-\Uo$ are plotted for various values of $\om$: degenerate vacua potential (DVP) with $A=\frac43,\; B=\frac89$ on the left and non-degenerate vacua potential (NDVP) with $A=1,\; B=\frac23$ on the right. The DVP has degenerate vacua in the original potential $-U$ (red solid line) where we set $\om_-=0$. The NDVP does not have degenerate vacua, but with $\om=\om_-=0.5$ (sky-blue dot-dashed line) the effective potential $-\Uo$ does have degenerate vacua. The two lines in the lower limit $\om=\om_-$ show that $\sigma_-(\om)\to \sigma_+(\om)$ where we have defined the maximum of the effective potential to be at $\sigma_+(\om)$ and $\Uo(\sigma_-(\om))=0$ for $\sigma_-(\om)\neq 0$. The purple dotted-dashed lines show $\sigma_-(\om) \to 0$ with the thick-wall limit $\om=\om_+$. With some values of $\om$ (green dotted lines) satisfying the existence condition \eq{QBeq}, both potentials show the values of $\sigma_\mp(\om)$ clearly.}
\label{fig:twomdl}
\end{figure}

Note that \eq{QBeq} coupled with the boundary condition \eq{nontbdry} implies  $\sigma(r)$ is a monotonically decreasing function, \ie\  $\sigma^\p<0$ when the solution is nodeless. In fact, according to \eqs{nontbdry}{QBeq} and the above conditions, our mechanical analogy implies that a particle with an unit mass initially at rest should be released somewhere on its potential, eventually reaching the origin at large (but finite) time and stopping there due to the presence of a position- and $D$- dependent friction. It implies that the initial ``energy'' of the particle will monotonically decrease due to the friction term, and eventually lose all of the energy when it will reach at the origin, $\sigma=0$. These requirements constrain the allowed forms of the U(1) potentials: for example if the local maximum of the effective potential $-\Uo$ is less than $0$, the ``particle'' can not reach the origin, a process known as $undershooting$. To avoid undershooting we require
\beq{LEFT}
\max(-U_\omega) \geq 0 \lr \min\bset{\frac{2U}{\sigma^2}}\leq \omega^2.
\eeq
If $-U_\omega$ is convex at $\sigma=0$, the ``particle'' cannot stop at the origin, a situation  termed $overshooting$ such that
\beq{RIGHT}
\left.\frac{d^2 U_\omega}{d\sigma^2}\right|_{\sigma=0}<0 \lr \omega^2 < \left.\frac{d^2U}{d\sigma^2}\right|_{\sigma=0}.
\eeq
Combining \eqs{LEFT}{RIGHT}, we find the condition on $\omega$ for the existence of a single $Q$-ball at zero-temperature:
\beq{EXIST}
 \omega_-\leq \abs{\omega} < \omega_+,
\eeq
where we have defined the lower limit $\omega^2_-\equiv \min\bset{\frac{2U}{\sigma^2}}=\left.\frac{2U}{\sigma^2}\right|_{\sigma_+(\omega_-)} \ge 0$, $\sigma_+(\omega)$ is the nonzero value of $\sigma$ where $U_{\omega}(\sigma_+(\omega))$ is minimised (see \figs{fig:twomdl}{fig:gravpot}), and the upper limit $\omega^2_+ \equiv \left.\frac{d^2 U}{d\sigma^2}\right|_{\sigma=0}$. Here, we defined the maximum of the effective potential to be at $\sigma_+(\omega)$ (i.e. $\left.\frac{dU_\om}{d\sigma}\right|_{\sigma_+(\omega)}=0)$; thus, $\om^2_-=\left.\frac{2U}{\sigma^2}\right|_{\sigma_+}$ and $U_{\omega_-}(\sigma_+)=0$ where $\sigma_+\equiv \sigma_+(\omega_-)$. Moreover,  $\sigma_-(\omega)$ satisfies $\Uo(\sigma_-(\omega))=0$ for $\sigma_-(\omega)\neq 0$. Notice $\sigma_-(\omega)\simeq \sigma_+(\omega)$ when $\om \simeq \om_-$. The case, $\om_-=0$, corresponds to degenerate vacua potentials (DVPs), while the other case, $\om_-\neq 0$, does not have degenerate vacua (NDVPs). In \figs{fig:twomdl}{fig:gravpot}, we indicate the above introduced parameters, $\sigma_\pm(\omega)$ and $\om_-$, using typical original and effective potentials for both DVP (left) and NDVP (right), which we will use later. 

The \textit{existence condition} in \eq{EXIST} restricts the allowed form of the potential $U$, which implies that the potential should grow less quickly than the quadratic term (i.e. mass term) for small values of $\sigma$; hence, U(1) potentials must have a nonlinear interaction and $U_\omega$ is weakly attractive \cite{Lee:1991ax}. In chapter \ref{ch:qpots} we examine the case of polynomial potentials and restrict ourselves to the case of $\omega^2_+=m^2$, where $m$ is a bare mass in the potentials. In chapter \ref{ch:qbflt} we extend our analysis allowing us to investigate the case, $\omega^2_+ \gg m^2$, needed since the potentials include one-loop radiative corrections to the bare mass $m$. Here, the potential which we will consider in the gravity-mediated models is $U=U_{grav}+U_{NR}$, where $U_{NR}$ is a nonrenormalisation term (to be discussed in Sec. \ref{gravity-mediated}), and 
\beq{Ugrav}
U_{grav}\equiv \half m^2 \sigma^2\bset{1+K\ln\bset{\frac{\sigma^2}{M^2}}}.
\eeq
Here, $K$ is a constant factor arising from the one-loop correction and  $M$ is the renormalisation scale. When the sign of $K$ is negative, $Q$-balls may exist subject to the coupling constant in $U_{NR}$.

\section{Thin- and ``thick-wall'' $Q$-balls}\label{thinckqb}
\subsection{Definitions}
To proceed with analytical arguments, we consider the two limiting values of $\om$ or $\sigma_0(\omega)\equiv \sigma(0)$ that describe
\beq{QBdef}
\begin{cases}
  \bullet \hspace*{5pt} \textrm{thin- wall $Q$-balls when}\; \omega\simeq \omega_-\; \textrm{or equivalently}\; \sigma_0(\om) \sim \sigma_+(\omega),\\
  \bullet \hspace*{5pt} \textrm{``thick-wall'' $Q$-balls when}\; \omega\simeq \omega_+\; \textrm{or equivalently}\; \sigma_0(\om) \simeq \sigma_-(\omega).
\end{cases}
\eeq
Note, the ``thick-wall'' limit does not imply that the ``thick-wall'' $Q$-ball has to have a large thickness which is comparable to the size of the core size. For the extreme thin-wall limit, $\omega = \omega_-$, thin-wall $Q$-balls satisfy, $\frac{E_Q}{Q}=\gamma(\omega_-) \omega_-$, see \eq{ch}. In particular Coleman demonstrated that a step-like profile for $Q$-balls, which generally exist for $\omega_-\neq 0$, satisfies $\gamma=1$, which implies that the charge $Q$ and energy $E_Q$ are proportional to the volume [see \eq{leg2}], and he called this $Q$-matter \cite{Coleman:1985ki}. We will not be considering $Q$-ball solutions that exist in a false vacua where $\om^2_-<0$ \cite{Spector:1987ag}. When it comes to obtaining $Q$-ball profiles numerically, we will adopt a standard shooting method which fine-tunes the ``initial positions'' $\sigma_0(\om)$ subject to $\sigma_-(\om) \le \sigma_0(\om)<\sigma_+(\om)$, in order to avoid both \emph{undershooting} and \emph{overshooting}.

\subsection{The infinitesimal variables: $\eo$ and $\mo$}
For later convenience, we define two positive definite quantities, $\eo$ and $\mo$,
\begin{eqnarray}
\nonumber \eo&\equiv& -U_\omega(\sigma_+(\omega))=\half \omega^2 \sigma^2_+(\omega)- U(\sigma_+(\omega)),\\
\label{infqt} &\simeq& \half \bset{\omega^2-\omega^2_-}\sigma^2_+,\\ 
\label{mo} m^2_\omega&\equiv& m^2-\omega^2,
\end{eqnarray}
which can be infinitesimally small for either thin- or thick-wall limits. By assuming $\sigma_+(\omega)\simeq \sigma_+(\omega_-) \equiv \sigma_+$ in the thin-wall limit, we immediately obtain the second line in \eq{infqt}. Let us remark upon important exceptions, which we will discuss in chapter \ref{ch:qbflt}, such that the above assumption is fine for the polynomial and gravity-mediated cases, while for gauge-mediated potentials which are extremely flat, the assumption, $\sigma_+(\omega)\simeq\sigma_+$, can not hold because $\sigma_+(\omega)$ does not exist. Therefore, we will not use the variable $\epsilon_\omega$ for the case of the gauge-mediated potentials. Indeed, it will turn out that the ``thin-wall'' $Q$-balls in the gauge-mediated cases do not have a thin-wall thickness. Further, the variable $m^2_\omega$ can not be infinitesimally small when we consider the gravity-mediated cases: $\omega^2_+ \not\sim m^2$.

\section{Four kinds of stability}\label{sec:stb}

\subsection{Absolute stability}
When the volume $V_D$ approaches infinity \cite{Lee:1991ax} and/or $\om$ is outside the limits of  \eq{EXIST}, then plane wave solutions may exist around the vacua of $U(|\phi|)$. The equation of motion for $\phi$ becomes a free Klein-Gordon equation whose solution can be written as $\phi=N e^{i(\textbf{k}\cdot \textbf{x}-\omega_k t)}$, where $\omega_k=\sqrt{m^2+\mathbf{k}^2}$ and the normalisation factor $N=\sqrt{\frac{Q}{2\omega_k V_D}}$ has been calculated from $Q$. Then, the energy of the plane wave solution is proportional to $\omega_k$ and $Q$ linearly: $E_{free}=\omega_k Q \tol E_{free} \simeq mQ$ where we have taken the infrared limit, to obtain the second relation. The energy $E_{free}$ can be interpreted as the energy of
a collection of $Q$ free particle quanta with the rest masses $m$. Furthermore, one might expect that the $Q$-ball energy approaches $E_{free}$ in the ``thick-wall'' limit, $\om \simeq \om_+$, since the $Q$-ball profiles approach zero exponentially at infinity \cite{Lee:1991ax}:
\beq{freeengy}
E_Q(\om=\om_+) \simeq E_{free}\simeq mQ.
\eeq
Hence, the $absolute\; stability$ condition for a $Q$-ball becomes
\beq{ABSCOND}
E_Q(\om)<E_{free} \lr \frac{E_Q}{Q}<m.
\eeq
We would expect \eq{ABSCOND} to be the strongest condition which a $Q$-ball solution has to satisfy. If the $Q$-ball has decay channels into other fundamental scalar particles which have the lowest mass $m_{min}$, we need to replace $m$ by $m_{min}$ in \eq{ABSCOND}.

\subsection{Classical stability}
The \textit{classical stability} \cite{Friedberg:1976me, Lee:1991ax} can be
defined in terms of the mass-squared of the fluctuations around a $Q$-ball
solution. For zero mass fluctuations this corresponds to a zero mode, \ie\  translation and phase transformation of the $Q$-ball solution. Using collective coordinates and \eq{QBeq} which extremises $\So$, such a mode should be treated with special efforts. Since a detailed analysis can be found in Appendix \ref{ap1} and the literature \cite{Friedberg:1976me, Lee:1991ax}, we simply state the final result which implies the \emph{classical stability} condition is
\beq{CLS}
\frac{\omega}{Q}\frac{dQ}{d\omega}\leq 0 \lr \frac{d^2\So}{d\om^2}\ge 0,
\eeq
where we have used \eq{legendre} in the second relation of \eq{CLS}. Since $\omega$ and $Q$ have the same sign, the sign of $\frac{dQ}{d\om}$ signals whether the solution is classically stable. The first relation of \eq{CLS} indicates the presence of an extreme charge in the parameter space of $\om$, (we will later see that the extreme charge at some critical value $\om=\om_c$ turns out to be the minimum allowed). Let us remark on the characteristic slope of $E_Q/Q$ as a function of $\om$:
\beq{chslope}
\frac{d}{d\om} \bset{\frac{E_Q}{Q}}=-\frac{\So}{Q^2}\frac{dQ}{d\om} \ge 0,
\eeq
where we have used \eq{legtrns} and \eq{legendre}. Since $\So$ is positive definite for $D\ge 2$ as we will see, the classically stable $Q$-balls should satisfy $\frac{d}{d\om} \bset{\frac{E_Q}{Q}}\ge 0$. The conditions from both \eq{CLS} and \eq{chslope} must be same.
\subsection{Stability against fission}
Suppose that the total energy of two $Q$-balls is less than the energy of a single $Q$-ball carrying the same total charge. The single $Q$-ball naturally decays into two or more with some release of energy. As shown in \cite{Lee:1991ax}, the stability condition against \emph{fission} for a $Q$-ball is given by
\beq{SAF}
\frac{d^2E_Q}{dQ^2} \leq 0 \lr \frac{d\omega}{dQ} \leq 0,
\eeq
where we have used \eq{legendre}, going from the first expression to the next expression in \eq{SAF}. Note that this is the same condition as we found above in \eqs{CLS}{chslope}, so the condition for classical stability is identical to that of stability against fission.

\vspace{10pt}
Trying to summarise the stability so far, we can categorise three types of a $Q$-ball: \ie\  absolutely stable, meta-stable, or unstable $Q$-balls. Absolutely stable $Q$-balls are stable quantum mechanically as well as classically; meta-stable $Q$-balls decay into free particle quanta, but are stable under small fluctuations; whereas completely unstable $Q$-balls sometimes called $Q$-clouds \cite{Alford:1987vs} decay into lower energy $Q$-balls or free particle quanta.

\subsection{Stability against fermions}

If coupling with light/massless fermions, a $Q$-ball decays through the surface area $A$ of the object. This decay rate is suppressed by Pauli blocking, and the authors in \cite{Cohen:1986ct} computed the upper bound on the rate per unit surface area for $Q$-matter
\be\label{SAF}
\frac{dQ}{dtdA} \le \frac{\om^3_-}{192\pi^2}.
\ee
For a small Yukawa coupling limit, they also obtained the decay rate for general $Q$-ball profile cases. The rate in \eq{SAF} can be used to compute the life-time of the $Q$-ball.

\section{Virial theorem}\label{sec:viri}
Derrick's theorem restricts the existence of static non-trivial scalar field solutions in terms of the number of spatial dimensions. For example in a real scalar field theory, non-trivial solutions exist only in one-dimension, \eg\  Klein-Gordon kink.
$Q$-balls (or any nontopological solitons), however, avoid this constraint because they are time-dependent (stationary) solutions \cite{Kusenko:1997ad, Gleiser:2005iq}. We can easily show this and in doing so obtain  useful information about the scaling properties of the $Q$-balls as a function of dimensionality as well as the ratio between their surface and potential energies. Following \cite{Gleiser:2005iq}, we begin by scaling the $Q$-ball ansatz, \eq{qansatz}, using a one-parameter family $r \to \alpha r$, whilst keeping $Q$ fixed. Defining a surface energy $\mS\equiv \int_{V_D} \half \sigma^{\p 2}$, a potential energy $\mU\equiv \int_{V_D} U$, and recalling that the charge satisfies $Q=I\omega$, we see that the energy of the $Q$-ball, \eq{legtrns}, becomes
\bea{}
E_Q &=& \mS+\mU + \frac{Q^2}{2I}.
\eea
Now, under the scaling $r \to \alpha r$, then $E_Q \to E_Q^\p$ where $\left.\dd{E_Q^\p}{\alpha}\right|_{\alpha=1}=0$ because the $Q$-ball solutions are the extrema (minima) of $E_Q$. Evaluating this, we obtain the virial relation relating $\mU$ and $\mS$
\beq{virial}
 D\ \mU = -(D-2)\mS +D\frac{Q^2}{2I} \geq 0
\eeq
where we have used our earlier notation, $U\ge 0$, for any values of $\sigma$. The case of $Q=0$ recovers Derrick's theorem, showing no time-independent solutions for $D\geq2$ \cite{Gleiser:2005iq}.

Using $\mS=\frac{D Q^2}{2I}\bset{D-2+D\; \frac{\mU}{\mS}}^{-1}$ from \eq{virial}, the characteristic slope \eq{ch} is
\beq{linear}
\gamma(\om)= \frac{E_Q}{\om Q}=1+\bset{D-2+D\; \frac{\mU}{\mS}}^{-1}.
\eeq
For $D\ge2$, we can see $\gamma(\omega)\ge 1$ because $\mS,\; \mU \ge 0$, which implies that $\So$ is positive definite for $D\ge2$, see \eq{Uo}, while $\So$ is positive for $D=1$ only when $\mU \ge \mS$.

Let us consider three cases for $D\ge2$: (i)$\,\mS \ll \mU$, (ii)$\,\mS \sim \mU$, and (iii)$\,\mS \gg \mU$. They lead to predictions for $\om$-independent characteristic slopes $\gamma$:
\beq{virieq}
    \gamma \simeq 
  \left\{
    \begin{array}{ll}
	1             \hspace*{87pt}  \textrm{for (i)}\hspace*{9pt}  \mS \ll \mU,\\
	(2D-1)/2(D-1) \hspace*{10pt} \textrm{for (ii)}\hspace*{7pt} \mS \sim \mU,\\
	(D-1)/(D-2)   \hspace*{20pt} \textrm{for (iii)}\hspace*{4pt} \mS \gg \mU.
    \end{array}
    \right.
\eeq

All of the $Q$-balls in the range of $\om$ are classically stable because the terms, $E_Q/Q$, monotonically increase as a function of $\om$, see \eqs{chslope}{virieq}.
The first case (i) corresponds to the extreme thin- and thick-wall limits $\omega \simeq \omega_\mp$ as will see. In the second case (ii), the potential energy is of the same order as the surface energy which means $\mS$ and $\mU$ have equally virialised. This case will turn out to be that of the thin-wall limit for DVPs when the surface effects are included. At present it is not known what kind of $Q$-ball potentials correspond to the third case; however, we will shortly find a duality relation between this case and the second case. Notice that in the case $\mS \gg \mU$ for $D=2$, we obtain the characteristic slope, $\gamma\gg 1$, from \eq{linear}. Similarly for $D=1$, the characteristic slopes are obtained, i.e. $\gamma\simeq 1,\; \gg 1,\; \simeq 0$, respectively for (i), (ii), and (iii). We will use these $1D$ analytic results to interpret numerical results of one-dimensional $Q$-balls in the thin-wall limit. We note a nice duality which appears in \eqs{linear}{virieq} between the two cases $ \mS \sim \mU$ and $ \mS \gg \mU$. In particular for $ \mS \sim \mU$ in $D$ dimensions, the same result for $\gamma$ is obtained (to leading order) in $2\times D$ dimensions when  $ \mS \gg \mU$.

Suppose $\mS/\mU=const.$ over a large range of $\om$ within the existence condition \eq{EXIST} except $\om\simeq \om_+$ where $E_Q/\om_+ Q\simeq 1$.  We can find an approximate threshold value $\om_a$ for a $Q$-ball to be absolutely stable using \eqs{freeengy}{virieq}:
\beq{viriwa}
\frac{\om_a}{m} \simeq
    \left\{
    \begin{array}{ll}
    1 &\; \textrm{for (i)}, \\
    \frac{2(D-1)}{2D-1} &\; \textrm{for (ii)},\\
    \frac{D-2}{D-1} &\;	 \textrm{for (iii)}.
    \end{array}
    \right.
\eeq
Roughly speaking, $Q$-balls are classically and absolutely stable if $\om< \om_a$ because of \eqs{ABSCOND}{chslope} and \eq{virieq}. These approximations can and will be justified by our numerical results in polynomial potentials in chapter \ref{ch:qpots}, however they will not hold in other models introduced in chapter \ref{ch:qbflt}. We will find that the virial relation is a powerful tool enabling us to find appropriate values of $\om_a$ as opposed to the rather complicated computations we will have to perform in the following two chapters by introducing detailed $Q$-ball profiles and specific potential forms. We should point out a caveat in this argument, the assumption we are making here, that most of the $Q$-balls have an identical energy ratio $\mS/\mU$ over a range of $\om$, does of course rely on the specific form of the potential. We have to remind the readers that the virial relation \eq{virial} gives only the relation between $\mS$ and $\mU$ if the system allows the time-dependent solutions, $Q$-balls, in \eq{qansatz} to exist.

To sum up, the virial theorem induces the characteristic slopes \eq{virieq} with the time-dependent nonlinear solutions in the system, and gives the approximate critical values for $\om_a$ in \eq{viriwa} without requiring a knowledge of the detailed profiles and potential forms.

\chapter{$Q$-balls in polynomial potentials}\label{ch:qpots}

\section{Introduction}

Standard $Q$-balls exist in an arbitrary number of space dimensions $D$ and are able to avoid the restriction arising from  Derrick's theorem  \cite{Derrick:1964ww} because they are time-dependent solutions. A number of examples include polynomial models both for $D=3$ \cite{Axenides:1999hs, Multamaki:1999an} and for arbitrary $D$ \cite{Gleiser:2005iq}, Sine-Gordon models \cite{Arodz:2008jk}, parabolic-type models \cite{Theodorakis:2000bz}, confinement models \cite{Morris:1978ns, Simonov:1979rd, Mathieu:1987mr, Werle:1977hs}, two-field models \cite{Friedberg:1976me, McDonald:2001iv}, and flat models with supersymmetry broken by gravity mediation  \cite{Multamaki:1999an}, and by gauge mediation \cite{Laine:1998rg, Multamaki:2000ey, Asko:2002phd}. Returning to the case of $D=3$, phenomenologically, it turns out that the $Q$-balls present in models with gravity-mediated supersymmetry breaking are quasi-stable but long-lived, allowing in principle for these $Q$-balls to be the source of both the baryons as well as the  lightest supersymmetric particle dark matter \cite{Enqvist:1997si}. On the other hand, $Q$-balls in models of gauge-mediated supersymmetry breaking can be a dark matter candidate as they can be absolutely stable \cite{Enqvist:2003gh}. Both types of $Q$-balls have been shown to be able to provide the observed baryon-to-photon ratio \cite{Laine:1998rg}.

The dynamics and formation of $Q$-balls involve solving complicated non-linear systems, which generally require numerical simulations. The dynamics of two $Q$-balls in flat Minkowski space-time depends on parameters, such as the relative phases between them, and the relative initial velocities \cite{Multamaki:2000ey, Battye:2000qj, Multamaki:2000qb}. In addition, the main formation process through the Affleck-Dine mechanism \cite{Affleck:1984fy} has been extensively examined in both gauge-mediated  \cite{Kasuya:1999wu}, gravity-mediated \cite{Kasuya:2000wx, Enqvist:2000cq, Multamaki:2002hv}, and
running inflaton mass models \cite{Enqvist:2002si}. As analysing individual $Q$-balls is difficult in its own right, it is extremely challenging to deal with multiple $Q$-balls. A number of analytical approaches to address that issue have been made over the past few years, \eg\  \cite{Lee:1994qb, Koutvitsky:2006mp, Griest:1989cb, Elphick:1996ux, Brihaye:2007tn, Mackay03lectureson}. Multiple thermal $Q$-balls have been described in a statistical sense in \cite{Enqvist:2000cq, Christ:1975wt}.

In this chapter, we aim to analytically address stationary properties of a single $Q$-ball with polynomial potentials in an arbitrary number of spatial dimensions $D$. The work will draw on earlier work of Correia and Schmidt \cite{Paccetti:2001uh} who derived analytic properties for the thin- and ``thick-wall'' limits of $Q$-balls in $D=3$. Recently, Gleiser and Thorarinson \cite{Gleiser:2005iq} proved the absolute stability for thin-wall $Q$-balls using the virial theorem. We generalise the main results of \cite{Gleiser:2005iq, Paccetti:2001uh} to the case of arbitrary spatial dimensions, and in doing so both analytically predict and numerically confirm the unique values of the angular velocity $\om_a$ in \eq{viriwa} for the absolute stability of the $Q$-balls via the thin-wall $Q$-ball approximations. Moreover, we obtain the classical stability conditions for the thin- and ``thick-wall'' approximations, and discover the connections between the virial relation and thin- or ``thick-wall'' approximation for the characteristic slopes $E_Q/\om Q$.

This chapter is organised as follows. By introducing a number of different ans\"{a}tze, we present a detailed analysis of the solutions in the thin-wall limit in Sec. \ref{sect:thin-pol} and in the ``thick-wall'' limit in Sec. \ref{sect:thick-pol}. In order to obtain minimise the numerical errors, we obtain a general asymptotic profile in Sec. \ref{sec:polasym}. We then demonstrate the advantages of using two particular modified ans\"{a}tze in Sec. \ref{numerical}, where we present detailed numerical results for the case of both degenerate and non-degenerate underlying potentials. Finally, we conclude in Sec. \ref{ch3:conc}. This chapter is partially published in \cite{Tsumagari:2008bv}.

\section{Thin- and thick-wall approximations}

In this section we obtain approximate solutions for $Q$-balls in $D$-dimensions based on the known thin- and thick-wall approximations for the radial profiles $\sigma(r)$ of the fields. Moreover, we show how we can then use these results to verify the solutions we obtained in the previous chapter for $\gamma(\om)$ in \eq{virieq}. Further, we are able to test the solutions against detailed numerical solutions in the next section, Sec. \ref{numerical}. We start with two simple ans\"{a}tze for the radial profiles, a step-like function for the thin-wall case $\om\simeq \om_-$ and a Gaussian function for the ``thick-wall'' case $\om\simeq \om_+$. In both cases, we evaluate $\So,\; Q,\; E_Q$, as well as the conditions for  classical and absolute stability before modifying the ans\"{a}tze. Following that, we repeat the same calculations using our more physically motivated ans\"{a}tze via the Legendre transformation technique described in \eq{easycalc}. Let us comment briefly on the form of the potential. We see that in the thin-wall limit, $\sigma_0(\om)\simeq \sigma_+(\om)$, with our modified ansatz, although in principle we do not have to restrict ourselves to particular potentials, we are not be able to investigate cases where the effective potential is extremely flat; hence, we have to limit our investigation to situations. We will consider flat potential cases in chapter \ref{ch:qbflt}. In the ``thick-wall'' limit, $\om\simeq \om_+$, we have to restrict our analysis to the case of polynomial potentials of the form:
\beq{thckpot}
U(\sigma)=\half m^2 \sigma^2-A\sigma^n+B \sigma^{p} \hspace*{10pt} \textrm{for}\hspace*{10pt} p>n,
\eeq
where $n\geq 3$, with the nonlinear couplings  $A>0$ and $B>0$. To ensure the existence of $Q$-ball solutions, we will restrict $A,\; B,\; n$ and $p$ later. We expect the thin-wall approximation to be valid for general $Q$-ball potentials in which the $Q$-ball contains a lot of charge, with $\omega^2 \simeq \omega^2_- \ge 0$. In this limit, we can define a positive infinitesimal parameter, $\eo$ in \eq{infqt}, and the effective mass around $\sigma_+(\om)$ is given by, $\mu^2(\om) \equiv \frac{d^2\Uo}{d\sigma^2}|_{\sigma_+(\om)}$. The other extreme case corresponds to the ``thick-wall'' limit which is valid for $Q$-balls containing a small amount of charge, and it satisfies  $\omega^2 \simeq \omega^2_+=m^2$. For later convenience, in this limit, we use a positive infinitesimal parameter, $m^2_\om$, defined in \eq{mo}.

\subsection{Thin-wall $Q$-ball}\label{sect:thin-pol}

\subsubsection{Step-like ansatz $\om = \om_-$}
At a first step, we review the standard results in the thin-wall approximation originally obtained by Coleman \cite{Coleman:1985ki}. Adopting a step-like ansatz for the profile we write
\beq{equation}
  \sigma(r)=\left\{
    \begin{array}{ll}
    \sigma_0 &\ \ \textrm{for $r<R_Q$}, \\
    0 &\ \ \textrm{for $R_Q \leq r$},
    \end{array}
    \right.
\eeq
where $R_Q$ and $\sigma_0$ will be defined in terms of the underlying parameters, by minimising the $Q$-ball energy $E_Q$. We can easily calculate $\So,\; Q,$ and $E_Q$:
\beq{QEQEGYTHIN}
  \So=\bset{U_0-\half \omega^2 \sigma^2_0}V_D,\ \ Q=\omega \sigma^2_0 V_D, \ \ E_Q = \half \frac{Q^2}{\sigma^2_0
  V_D} + U_0 V_D,
\eeq
where $U_0 \equiv U(\sigma_0)$ and $V_D=V_D(r=R_Q)$. Note that \eq{QEQEGYTHIN} satisfies the Legendre transformation results, \eq{easycalc}, as we would have hoped. Since the ansatz, \eq{equation}, neglects the surface effects, we are working in the regime $\,\mU \gg \mS$ in \eq{virieq}. Therefore, we should be able to reproduce the result, $\gamma=\frac{E_Q}{\om Q} \simeq 1$, with this solution. To see this, we note that the two terms in $E_Q$ are the contributions from the charge and potential energies.. These two contributions are virialised in that $E_Q$ is extremised with respect to $V_D$ for a fixed charge $Q$, \ie\ $\partial E_Q/ \partial V_D|_Q=0$; hence, $V_D=Q\sqrt{1/(2\sigma^2_0 U_0)}$. This then fixes $R_Q$ because we know for a $(D-1)$-sphere, $V_D=\frac{R_Q^D}{D} \Omega_{D-1}$, where $\Omega_{D-1} \equiv \int d\Omega_{D-1}=\frac{2\pi^{D/2}}{\Gamma(D/2)}$. Here $\Gamma$ is gamma function. Substituting $V_D$ into $E_Q$ [the third expression in \eq{QEQEGYTHIN} ] and minimising it with respect to $\sigma_0$, we obtain
\beq{QEGYTHIN}
  E_Q=Q \cdot min \bset{\sqrt{\frac{2U_0}{\sigma^2_0}}}= Q\omega_-=\om^2_-\sigma^2_+V_D,
\eeq
where we have used \eq{EXIST} in which $\omega^2_-=min \left.\bset{\frac{2U_0}{\sigma^2_0}}\right|_{\sigma_0=\sigma_+}$. Thus, we recover \eq{virieq} in the limit $\mU \gg \mS$. Finally, we remind the reader that we have obtained the minimised energy, $E_Q$, with respect to $V_D (R_Q)$ and $\sigma_0$ in the extreme limit, $\om =\om_-$, where we find
\beq{sig0}
\sigma_0 = \sigma_+.
\eeq
\eq{sig0} implies that the ``particle'' spends a lot of ``time'' around
$\sigma_+$ because the effective potential $-\Uo$ around $\sigma_+$ is ``flat''. Note that $Q$ and $E_Q$ are proportional to the volume $V_D$ in \eqs{QEQEGYTHIN}{QEGYTHIN} just as they are for ordinary matter, in this case Coleman  called it $Q$-matter \cite{Coleman:1985ki}.

\vspace*{5pt}

%
%
\subsubsection{The modified ansatz $\sigma_0\simeq \sigma_+$}

Having seen the effect of an infinitely thin-wall, it is natural to ask what happens if we allow for a more realistic case where the wall has a thickness associated with it?  Modifying the previous step-like ansatz to include this possibility \cite{Paccetti:2001uh, Coleman:1977py} will allow us to include surface effects \cite{Coleman:1985ki, Spector:1987ag, Shiromizu:1998rt} and is applicable for a wider range of $\om$ than in the step-like case $\om=\om_-$. Using the results, we will examine the two different types of potentials, DVPs and NDVPs, which lead to the different cases of \eq{virieq}.

Following \cite{Paccetti:2001uh}, the modified ansatz is written as
\bea{thindanstz}
\sigma(r)= \begin{cases}
	\sigma_+ -s(r)\; & $for$\; r<R_Q,\\
	\bsig (r)\;  & $for$\; R_Q\leq r \leq R_Q + \delta,\\
    0\; & $for$\; R_Q + \delta < r,
 \end{cases}
\eea
where as before the core size $R_Q$, the wall thickness $\delta$, the core profile $s(r)$, and the shell profile $\bsig (r)$ will be obtained in terms of the underlying parameters by extremising $\So$ in terms of a degree of freedom $R_Q$. Continuity of the solution demands that we smoothly continue the profile at $r=R_Q$, namely $\sigma_+-s (R_Q) =\bar{\sigma}(R_Q)$ and $-s^\p(R_Q)=\bar{\sigma}^\p(R_Q)$.

We expand $\Uo$ to leading order around $\sigma _+$, to give  $\Uo(\sigma)\sim -\eo +\half  \mu^2 s^2$ where $s(r)=\sigma_+ - \sigma(r)$. In terms of our mechanical analogy, the ``particle'' will stay around $\sigma_+$ for a long ``time''. Once it begins to roll off the top of the potential hill, the damping due to friction ($\propto (D-1)/r$) becomes negligible and the ``particle'' quickly reaches the origin. Therefore, we can naturally assume
\beq{coreQ}
R_Q \gg \delta,
\eeq
where $\delta$ is the wall thickness. We know that $\sigma^\p(0)=-s^\p(0)=0,\; s^\p(R_Q) \neq 0,$ and $s^\p(r)>0$. Using \eq{QBeq}, the core profile $s(r)$ for $r< R_Q$ satisfies the Laplace equation:
\beq{eoms}
s^{\p\p}+\frac{D-1}{r}s^\p-\mu^2 s=0
\eeq
whose solution is
\beq{seq}
s(r)=r^{(1-\frac{D}{2})} \bset{C_1  I_{\frac{D}{2} -1}(\mu r) + C_2 K_{\frac{D}{2} -1}(\mu r)}
\eeq
where $I$ and $K$ are, respectively, growing and decaying Bessel functions, $C_1$ and $C_2$ are constants. Since $s(0)$ is finite and $s^\p(r)>0$, it implies that $C_2:=0$. Since $I_\nu(z)\sim z^\nu/2\Gamma(\nu+1)$ for small $z=\mu r$ and $\nu\neq -1,-2,-3 \dots$; thus, $s(0)$ is finite:
\beq{s0}
s(0)\sim C_1 \frac{\mu^{D/2-1}}{2\Gamma(D/2)}=\sigma_+-\sigma_0
\eeq
which gives a relation between $C_1$ and $\sigma_0$. In addition, the analytic solution is regular at $r=0$: $s^\p(0) \simeq 0$.  For large $r\sim R_Q$, \eq{seq} leads to
\beq{sslope}
\frac{s^\p}{s} \simeq \mu - \frac{D-2}{r} \to \mu,
\eeq
where we are assuming
\beq{app2}
\mu \gg 1/R_Q,
\eeq
and have used the approximation $I_\nu(z) \sim  \frac{e^{z}}{\sqrt{2 \pi z}}$ for large $z\equiv \mu r$. As already mentioned, we note that this result is not strictly valid for extremely flat potentials, \ie\ $\mu \simeq 1/R_Q$, because the expansion is only valid for $z \equiv \mu r \gg 1$. We will therefore only be applying it to the cases where the effective potential is not very flat.

Turning our attention to the shell regime $R_Q\le r \le R_Q + \delta$. Considering the ``friction'' term in \eq{QBeq}, we see that it becomes less important for large $r$ compared to the first and third terms in \eq{eoms}, because
\beq{frict}
\abs{\frac{D-1}{R_Q}s^\p (R_Q)} \simeq \abs{\frac{D-1}{\mu R_Q} \mu^2 s(R_Q)} \ll \mu^2 s(R_Q) \simeq s^{\p\p}(R_Q) \simeq \abs{\frac{d\Uo}{d s}}_{r=R_Q}
\eeq
where we have made use of \eqs{sslope}{app2}. Imposing continuity conditions, namely $\sigma_+-s(R_Q)=\bar{\sigma}(R_Q)$, $-s^\p(R_Q)=\bar{\sigma}^\p(R_Q)$, \eq{QBeq} without the ``friction'' term becomes
\beq{barsig}
\frac{d^2\bar{\sigma}}{dr^2}- \left.\frac{d\Uo}{d\sigma}\right|_{\bar{\sigma}}=0,
\eeq
where $\bar{\sigma}(r)$ is defined as being the solution to \eq{barsig}. With
the condition $\bar{\sigma}(R_Q)=\sigma_+-s(R_Q)$ and \eq{s0}, we find $\bar{\sigma}(R_Q)\sim \sigma_+$ in the thin-wall limit. Therefore,
\beq{modthsig}
\bar{\sigma}(R_Q) \gg s(R_Q).
\eeq
 Although \eq{nontbdry} does not hold exactly, the ``total energy'', $\half \bset{\frac{d\bsig}{dr}}^2-\Uo \sim 0$ with \eq{nontbdry}, is effectively conserved with the radial pressure $p_r$ vanishing outside the $Q$-ball core, see \eq{barsig}. This fact implies that the surface and effective potential energies virialise with equal contributions, $\mS_{shell} \simeq \mU_{shell}-\half \om Q_{shell}$, where we have introduced shell and core regimes defined by $X_{core}=\Omega_{D-1}\int^{R_Q}_0 dr r^{D-1} F(r, \dots)$ and $X_{shell}=\Omega_{D-1} \int^{R_Q+\delta}_{R_Q} dr r^{D-1} F(r, \dots)$ for some quantity $X$ and a function $F(r,\dots)$. Using $\sigma^\p<0$  and the condition $\bar{\sigma}(R_Q+\delta)=0$, the thickness of the $Q$-ball can be written as $\delta(\om)=\int^{\bar{\sigma}(R_Q)}_{0}\frac{d\sigma}{\sqrt{2\Uo}}$. Since $\delta$ is real and positive, we have to impose
\beq{const}
\bar{\sigma}(R_Q) < \sigma_-,
\eeq
recalling $\Uo(\sigma_-)=0$ for $\sigma_-\neq 0$.

With the use of \eq{easycalc}, we turn our attention to extremising the Euclidean action $\So$ in \eq{Uo} for the degree of freedom $R_Q$. Using the obtained value $R_Q$, we will differentiate $\So$ with respect to  $\om$ to obtain $Q$ as in \eq{legendre} which leads us to the $Q$-ball energy $E_Q$ as in \eq{legtrns} and the characteristic slope $E_Q/\om Q$. For convenience we split $\So$ into the core part $\So^{core}$ for $r<R_Q$ and the shell part $\So^{shell}$ for $R_Q \leq r \leq R_Q + \delta$ using \eq{thindanstz}. Using $V_D=\frac{R_Q^D}{D} \Omega_{D-1} \gg \partial V_D \equiv R_Q^{D-1}\Omega_{D-1} \gg \partial^2 V_D \equiv R_Q^{D-2}\Omega_{D-2}$ and \eqs{eoms}{sslope}, we find,
\beq{coresw}
\So^{core}=-V_D \cdot \eo + \pa V_D \cdot \bset{\half \mu s^2(R_Q)}  - \pa^2 V_D \cdot \bset{ \frac{\Omega_{D-1}}{\Omega_{D-2}} \frac{(D-2)}{\mu} \half \mu s^2(R_Q)},
\eeq
where the first term, $\eo$, in \eq{coresw} comes from the effective potential energy, while the second and third terms arise from the surface energy. Since $\eo$ is an infinitesimal parameter in the other thin-wall limit $\om\simeq \om_-$, it gives
\beq{ratcore}
\mU_{core}\simeq \half \om Q_{core}.
\eeq
The effective potential energy balances the surface energy in the shell [see \eq{barsig}], therefore by introducing the definition $T\equiv \int^{\bsig(R_Q)}_0 d\sigma \sqrt{2\Uo}$, we see
\bea{shellsw}
\So^{shell} &=&\Omega_{D-1}\int^{\bsig(R_Q)}_0 d\sigma r^{D-1}\sqrt{2\Uo(\sigma)} \lsim \Omega_{D-1} (R_Q+\delta)^{D-1} T, \\
\label{shllsw2} &\to& \pa V_D \cdot T + \pa^2 V_D \cdot \bset{\frac{\Omega_{D-1} }{\Omega_{D-2}}(D-1)\delta \cdot T} + \order{R^{D-1}_Q,\frac{\delta^2}{R^2_Q}} \cdot T,
\eea
where we have used the fact that the integrand has a peak at $r=R_Q+\delta$ in the second relation of \eq{shellsw} \cite{Hong:1987ur} and Taylor-expanded $ (R_Q+\delta)^{D-1}$ in going from \eq{shellsw} to \eq{shllsw2} because of our approximation \eq{coreQ}. Combining both expressions \eqs{coresw}{shllsw2}, we obtain
\bea{}
 \So &=& \So^{core} + \So^{shell}, \\
\label{swall} &\simeq & - \eo \cdot V_D  + \tau \cdot \pa V_D  +h\cdot \partial^2 V_D,
\eea
where $\tau \equiv T+ \half \mu s^2(R_Q)$. Note that while $T$ in $\tau$ contains the equally virialised surface and effective potential energies from the shell, the second term $\half \mu s^2(R_Q)$ contains a surface energy term from the core. Moreover, we have defined $h\equiv \frac{\Omega_{D-1}}{\Omega_{D-2}}\sbset{ (D-1) \delta \cdot T - \frac{(D-2)}{\mu} \half \mu s^2(R_Q)}$ which is negligible compared to $\tau$ because of the assumptions, \eqs{coreQ}{app2}. Therefore, we will take into account only the first two terms in $\So$, \eq{swall}. It is also important to realise that
\beq{tens}
\tau = \int^{\bsig(R_Q)}_0 d\sigma \sqrt{2\Uo} + \int^{\sigma_+}_{\bsig(R_Q)} d\sigma \sqrt{2 U_{\om_-}}\to \int^{\sigma_+}_0 d\sigma \sqrt{2U_{\om_-}}=const
\eeq
which is independent of $\om$ and $D$ in the limit of $\om\to \om_-$, where we have used the extreme thin-wall limit $\om= \om_-$ explicitly. Our modified ansatz is not only valid in the extreme limit $\om=\om_-$ but also in the limit $\om\sim \om_-$ as long as $\tau$ depends on $\om$ ``weakly''. Note that the condition of \eq{const} also ensures that $\tau$ is positive and real. In addition, the second term in the first expression of \eq{tens} is negligible compared to the first term, \ie\ 
\beq{ratshell}
\mS_{shell} \simeq \mU_{shell} - \half \om Q_{shell} \gg  \mS_{core}
\eeq
because of $\sigma_+ \sim \bar{\sigma}(R_Q)$, see \eq{modthsig}.

We can make progress by using the Legendre transformation  of \eq{easycalc}, which implies that we need to find the extrema of $\So$ with fixed $\omega$, \ie\  $\frac{\partial \So}{\partial R_Q}=0$. This is equivalent to the virialsation between $\eo$ and $\tau$. Then one can compute the core radius,
\beq{RQ}
	R_Q=(D-1)\frac{\tau}{\eo}.
\eeq
Note that this implies that one-dimensional thin-wall $Q$-balls do not exist due to the positivity of $R_Q$ and one of our assumptions $R_Q \gg \delta$. By using \eqs{swall}{RQ} and \eq{easycalc}, we can compute the desired quantities to compare with the results we obtained using the step-like ansatz, in particular \eqs{QEQEGYTHIN}{QEGYTHIN}, and we can confirm that the classical stability condition \eq{CLS} is satisfied:
\bea{}
\label{soq1} \So&\simeq & \frac{\tau}{D}\; \partial V_D =
\frac{\eo}{D-1}\; V_D>0,\ \ Q(\omega)\simeq \omega \sigma^2_+ V_D, \\
\label{linearthin} E_Q &\simeq& \omega^2 \sigma^2_+ V_D  + \frac{\tau}{D}\; \partial V_D,\\
\label{surfthin} &\simeq& \omega Q \sbset{\frac{2D-1}{2(D-1)}-\frac{\om^2_-}{2(D-1)\om^2}}, \\
\label{q1class}\frac{\omega}{Q}\frac{dQ}{d\omega}&\simeq&1- \frac{D\omega^2 \sigma^2_+}{\eo}  \simeq - \frac{D\omega^2 \sigma^2_+}{\eo} <0.
\eea
We can see the virialisation between $\tau$ and $\eo$ for the second and third terms in \eq{soq1}. As in \eq{QEGYTHIN}, the first term of $E_Q$, in \eq{linearthin}, is a combination of an energy from the charge and potential energy from the core throughout the volume, while the new second term $\frac{\tau}{D}$, called the surface tension,
represents the equally virialised surface and effective potential energies from the shell as in \eq{ratshell}. In the limit $\om\simeq \om_-$, $\eo$ becomes zero which implies \eq{ratcore}. We have also seen $\mS_{shell}\gg \mS_{core}$. Using $\mU=\mU_{core}+ \mU_{shell}$, $\mS=\mS_{core}+\mS_{shell} \sim \mS_{shell}$, and \eqs{ratcore}{ratshell}, we obtain
\beq
{UoverS}
\mU \sim \mS + \om_- Q
\eeq
which we will use shortly. Since the characteristic function, $E_Q/Q$, increases monotonically as a function of $\om$ and $\So>0$, \ie\  $\frac{d}{d\om}\bset{\frac{E_Q}{Q}}>0$ or we found \eq{q1class}, the classical stability condition \eqs{CLS}{chslope} is satisfied without specifying any detailed potential forms. However, the physical properties of the finite thickness thin-wall $Q$-balls do depend on the vacuum structures of the underlying potential. To demonstrate this we consider two cases of non-degenerate vacuum potentials (NDVPs) with $\om_- \neq 0$ and degenerate vacuum potentials (DVPs) with $\om_-=0$ (see red solid lines in \fig{fig:twomdl}). Suppose that the thin-wall $Q$-balls have identical features over a large range of $\om$, we can find the approximate threshold frequency $\om_a$ using \eqs{freeengy}{virieq} as we assumed when we obtained \eq{viriwa}.

\paragraph*{\underline{NDVPs:}}

This type of potential reproduces the results we obtained in \eq{QEGYTHIN} corresponding to the regime $\mU\gg \mS$ which corresponds to the existence of $Q$-matter in that the charge and energy is proportional to the volume $V_D$ due to the negligible surface tension in \eq{linearthin}. Hence, the modified ansatz \eq{thindanstz} can be simplified into the original step-like ansatz \eq{QEQEGYTHIN} with negligible surface effects in the extreme limit $\om=\om_-$. To see that, we need to recall the definition of $\om_-$ in \eq{LEFT}.
We can realise that $\mu$ is the same order as $\om_-$ except the case of $\om_-=0$. Using $\mu\sim \om_-$, we can show that $\half \om Q \gg \mS_{core}\sim \half \mu s^2(R_Q) \partial V_D$ where we have used \eqs{app2}{modthsig}. Using \eqs{ratcore}{ratshell} and $\half \om Q \gg \mS_{core}$ which we just showed, we can obtain the desired result $\mU\gg \mS$. Similarly \eq{surfthin} in the limit $\om \simeq \om_-$ simplifies to give $\frac{E_Q}{\om Q} \sim 1$ which is the result of \eq{virieq} with the case $\mU\gg \mS$. Using \eq{surfthin} and \eqs{freeengy}{virieq},
we can also find the critical value $\om_a$ for absolute stability
\beq{ndvpwa}
\frac{\om_a}{m}=\frac{D-1}{2D-1}
\bset{1+\sqrt{1+\frac{(2D-1)}{(D-1)^2}\frac{\om^2_-}{m^2} }}.
\eeq
Finally, thin-wall $Q$-balls in NDVPs are classically stable without the need for the detailed potential forms; however, the absolute stability condition for $\om \sim \om_-$ depends on the spatial dimensions $D$ and on the mass $m$.

\paragraph*{\underline{DVPs:}}
For the case of the presence of degenerate minima where $\om_-=0$, since  $\eo=\half \omega^2 \sigma^2$, we immediately see from \eq{surfthin} that
\beq{eqqdeg}
  \frac{E_Q}{\om Q}=\gamma\simeq \frac{2D-1}{2(D-1)}
\eeq
which reproduces \eq{virieq} for the case of $\mS\sim \mU$. As in NDVPs, we know \eq{UoverS} in the limit $\om \simeq \om_-$, but the second term $\om_- Q$ becomes zero in the present potentials. It follows that $\mU_{core} \simeq 0$ and $\mU_{shell} \simeq \mS_{shell}\gg \mS_{core}$ from \eq{ratshell}; hence, $\mS \sim \mU$. In other words, most of the $Q$-ball energy is concentrated within the shell. In addition, the charge $Q$ and energy $E_Q$ are not scaled by the volume, which implies the modified ansatz does not recover the simple ansatz as opposed to NDVPs. Using \eqs{eqqdeg}{leg2}, it implies $E_Q\propto Q^{2(D-1)/(2D-1)}$, which reproduces the three dimensional results obtained in \cite{Paccetti:2001uh}.

Finally, let us recap, the key approximations and conditions we have made in this modified ansatz. They are Eqs. (\ref{coreQ}, \ref{app2}, \ref{const}, \ref{tens}) for $D\ge 2$. The estimates we have arrived at for the thin-wall $Q$-balls are valid as long as the core size is much larger than the wall thickness, the effective potential is not too flat around $\sigma_+$, the core thickness $\delta$ and surface tension $\tau/D$ are positive and real, and $\tau$ is insensitive to both $\om$ and $D$.  With the extreme limit $\om\to \om_-$, the $Q$-balls in DVPs recover the simple step-like ansatz, while the ones in NDVPs do not. One-dimensional $Q$-balls do not support thin-wall approximation due to the absence of the friction term in \eq{QBeq}.

\subsection{``Thick-wall'' $Q$-ball}\label{sect:thick-pol}

\subsubsection{Gaussian ansatz}

As we have started with the simple step-like ansatz in the thin-wall approximation, a Gaussian function is a simple approximate profile to describe the ``thick-wall'' $Q$-balls in the limit $\om\simeq \om_+$ \cite{Gleiser:2005iq}. Using a Gaussian ansatz
\beq{gaussansatz}
\sigma(r)=\sigma_0(\om)\exp\bset{-\frac{r^2}{R^2}},
\eeq
we will extremise $\So$ with respect to $\sigma_0(\om)$ and $R$ with fixed $\om$, instead of minimising $E_Q$ with fixed $Q$. Notice that the slope $-\sigma^\p/\sigma$ becomes $2r/R^2$ which is linearly proportional to $r$ and the solution is regular at $r=0$: $\sigma^\p(0)=0$. By neglecting higher order term $B$ in \eq{thckpot} with \eq{gaussansatz}. which we will justify shortly, one can obtain straightforwardly
\bea{THCKQ}
    Q&=&\bset{\frac{\pi}{2}}^{D/2} \omega \sigma^2_0(\om) R^D,\\ 
\label{TCHKQ2}   \So&\simeq& \bset{\half \mo^2+\frac{D}{R^2}-A\sigma^{n-2}_0(\om)\bset{\frac{2}{n}}^{D/2}}\frac{Q}{\omega},\\
\label{eqthick1}    E_Q&\simeq&\sbset{\half \bset{m^2 + \omega^2} +
    \frac{D}{R^2}- A\sigma^{n-2}_0(\om) \bset{\frac{2}{n}}^{D/2}} \frac{Q}{\omega}.
\eea
\eq{easycalc} can be easily checked in \eqs{THCKQ}{TCHKQ2}, and \eq{eqthick1}. The first ($\half \frac{m^2Q}{\om}$) and last terms in \eq{eqthick1} are the potential energy terms; the second term, $\half \om Q$,
comes from the charge energy, and the surface energy term appears in the third
term, $\frac{DQ}{R^2\om}$. By finding the extrema of $\So$ with respect to $\sigma_0(\om)$ with $\dd{\So}{\sigma_0(\om)}=0$, it defines the underlying parameter $\sigma_0(\om)$ as
\beq{apprxsigma0}
\sigma_0(\om)=\sbset{\bset{m^2_\omega+\frac{2D}{R^2}}\frac{1}{nA}\bset{\frac{n}{2}}^{D/2}}^{1/n-2}
\tol \bset{\frac{\mo^2}{2A}}^{1/n-2}\sim \sigma_-(\om)
\eeq
where we have neglected the surface term and used the approximation $D/2\simeq \order{1}$ in the second relation of \eq{apprxsigma0}. We are then able to check the Gaussian ansatz naturally satisfies the other ``thick-wall'' limit $\sigma_0(\om)\simeq \sigma_-(\om)\to 0$ since $\mo$ is a positive infinitesimal parameter in the limit, $\om\simeq \om_+$, and justify the fact that we have neglected the higher order term $B$ in \eq{thckpot}. Using the first relation of \eq{apprxsigma0}, one needs to extremise $\So$ with respect to  another degree of freedom $R$ with $\dd{\So}{R}=0$ which determines $R$:
\beq{gausscore}
R=\sqrt{\frac{2(2-D)}{\mo^2}}\ge 0.
\eeq
The reality condition on $R$ implies that the Gaussian ansatz is valid only for $D=1$. The width of the gaussian function $R$ in \eq{gausscore} becomes very large in the ``thick-wall'' limit $\mo \to 0$; thus, we can justify that the surface terms in \eqs{eqthick1}{apprxsigma0} are negligible. Therefore, we are looking at the regime $\mU \gg \mS$ which should lead us to $\gamma \simeq 1$ as in the first case of \eq{virieq}. To do this for $D=1$ we substitute \eq{apprxsigma0} into $Q,\; E_Q,\; \So$:
\bea{}
Q&=&\sqrt{\frac{\pi}{2}}\omega \sigma^2_0(\om) R,\hspace{10pt} \So=\bset{\half- \frac{1}{n}}\frac{2\mo^2 Q}{\omega}>0,\\
\label{gausseq}\frac{E_Q}{\om Q}&=&\bset{\half+\frac{1}{n}} +\bset{\half-\frac{1}{n}}\bset{\frac{2m^2}{\om^2}-1}\tol 1,
\eea
where we have considered the ``thick-wall'' limit $\omega\simeq m$ in the second relation of \eq{gausseq}. We can check \eq{virieq} and the analytic continuation \eq{freeengy}. In the same limit, the Euclidian action becomes an infinitesimally small positive value: $\So \to 0^+$.

Using the second relation $\sigma_0(\om)$ in \eq{apprxsigma0} and \eq{gausscore}, one can find
\beq{gausscls}
\frac{\om}{Q}\frac{dQ}{d\om} \simeq 1-\frac{\omega^2}{\mo^2}\bset{\frac{4}{n-2}-1}
\to  -\frac{\omega^2}{\mo^2}\bset{\frac{4}{n-2}-1}\le 0,
\eeq
where we have used the fact that $\mo$ is a positive infinitesimal parameter in the limit, $\om\simeq \om_+$ going from the first relation to the second one. \eq{gausscls} shows that the classical stability condition clearly depends on the non-linear power $n$ in the potential \eq{thckpot}: $n \le 6$. This is contradictory because \eq{gausseq} gives $\frac{d}{d\om}\bset{\frac{E_Q}{Q}}\to -1+\frac{4}{n}$ which implies $n \le 4$ for the other classical stability condition using \eq{chslope}. We will shortly see that this contradiction between \eq{CLS} and \eq{chslope} is an artefact of the Gaussian ansatz. Moreover, our conclusion should state that the Gaussian approximation is approximately valid only for $D=1$. These awkward consequences are improved with the following physically motivated ansatz.
\subsubsection{The modified ansatz}
Having considered the case of the simple Gaussian ansatz following the spirit of \cite{Gleiser:2005iq}, we found some problems for the classical stability. To fix these, we need a more realistic ansatz \cite{Kusenko:1997ad, Shatah:1985vp, Blanchard:87, Multamaki:1999an, Paccetti:2001uh}. To do this we drop an explicit detailed profile to describe ``thick-wall'' $Q$-balls and rescale the field profile so as to work in dimensionless units whilst extracting out the explicit dependence on $\omega$ from $\So$. As in the thin-wall approximation with the modified ansatz, we will again make use of the technique \eq{easycalc} to obtain other physical quantities from $\So$.

We begin by defining  $\sigma=a\wt{\sigma}$ and $r=b\wt{r}$ with $a$ and $b$ which will depend on $\om$. Substituting them into \eq{Uo} with the potential \eq{thckpot} we obtain:
\bea{}
\nonumber \So &=& b^D \Omega_{D-1} \int d\tilde{r} \tilde{r}^{D-1}
\set{\half \bset{\frac{a}{b}}^2 \wt{\sigma}^{\p 2}+\half a^2 \mo^2 \tilde{\sigma}^2-A a^n \tilde{\sigma}^n +
	B a^{p}\wt{\sigma}^{p}}, \\
\nonumber &=& b^D\bset{\frac{a}{b}}^2 \Omega_{D-1} \int d\tilde{r} \tilde{r}^{D-1} \half \set{\wt{\sigma}^{\p 2}+\tilde{\sigma}^2-\tilde{\sigma}^n + 2B b^2 a^{p-2}\wt{\sigma}^{p}},\\
\label{so3}&\simeq&  \mo^{4/(n-2) -D+2} A^{2/(2-n)}\Omega_{D-1} S_n
\eea
with the rescaled action $S_n=\int d\tilde{r} \tilde{r}^{D-1} \bset{\half \wt{\sigma}^{\p 2}+\wt{U}}$ with $\wt{U}=\half \wt{\sigma}^2-\half \wt{\sigma}^n$, and we have neglected the higher order term involving $B$, which will be justified shortly.
In going from the first line to the second one in \eq{so3}, we have set the coefficients of the first three terms in the brackets to be unity in order to explicitly remove the $\om$ dependence from the integral in $\So$. In other words we have set
$\half \bset{\frac{a}{b}}^2=\half a^2 \mo^2=Aa^n$. This implies, $a=\bset{\frac{\mo^2}{2A}}^{1/(n-2)}=\sigma_-(\om)$ and $b=\mo^{-1}$. Then we can justify that the higher order term involved with $B$ is negligible due to $\sigma_-(\om)\to 0$ in the ``thick-wall'' limit. Crucially $S_n$ is independent of $\omega$, and is positive definite \cite{Kusenko:1997ad, Multamaki:1999an, Paccetti:2001uh}. Adopting the powerful approach developed in \eq{easycalc}, given $S_\om$ we can differentiate it to obtain $Q$ and then use the Legendre transformation to obtain $E_Q$. This is straightforward and yields
\bea{}
\nb Q(\omega)&=& \om m^{4/(n-2)-D}_\omega \bset{\frac{4}{n-2}-D+2}A^{-2/(n-2)}\Omega_{D-1} S_n,\\ 
\label{qthck} &\propto& \mo^{4/(n-2)-D},\\
\nb E_Q      &=& m^{4/(n-2)-D}_\omega \sbset{\mo^2+  \omega^2\bset{\frac{4}{n-2}-D+2}}A^{-2/n-2}\Omega_{D-1} S_n,\\
\label{modslope}          &=& \om Q \sbset{1+\frac{\mo^2}{\om^2}\bset{\frac{4}{n-2}-D+2}^{-1}  }\to \om Q.
\eea
The first term involving $\mo^2$ in the first line \eq{modslope} is the energy contributed by the charge, while the second term is dominated by the effective potential energy; hence, $\mU\gg \mS$. Therefore, we can also recover the result $\gamma \simeq 1$ in the ``thick-wall'' limit $\om\simeq \om_+$ as we would expect from \eq{virieq} when $\mU\gg \mS$. Since $Q$ and $E_Q$ should be positive definite, it places the constraint \cite{Multamaki:1999an}
\beq{validthck}
D < \frac{4}{n-2}+2.
\eeq
With the condition \eq{validthck}, it is easy to see that $\So \to 0^+$ in the ``thick-wall'' limit, $\om\simeq \om_+$ where $\mo^2 \to 0^+$. There is another constraint emerging from the need for the solution to be classically stable:
\bea{moddiffchg}
\frac{\om}{Q}\frac{dQ}{d\om} \simeq 1-\frac{\omega^2}{\mo^2}\bset{\frac{4}{n-2}-D}
&\to&  -\frac{\omega^2}{\mo^2}\bset{\frac{4}{n-2}-D}\le 0, \\
\label{modcls} &\lr& D \le \frac{4}{n-2}
\eea
which coincides with \eq{gausscls} in the case of $D=1$. Notice that the modified ansatz is valid not only for $D=1$ but also $D<\frac{4}{n-2}+2$ in \eq{validthck}. For $D=3$ this result matches that of \cite{Paccetti:2001uh}. The classical stability condition, \eq{modcls}, is consistent with the need for $Q$ and $E_Q$ to be finite. \eq{modcls} is more restrictive than that given in \eq{validthck}. Furthermore, we should check the relation \eq{chslope} for the characteristic function $E_Q/Q$ in terms of $\om$. It follows that $\frac{d}{d\om}\bset{\frac{E_Q}{Q}} \simeq 1- 2\bset{\frac{4}{n-2}-D+2 }^{-1} \ge 0$, which requires the same condition as \eq{modcls}. With this fact and \eq{modslope}, it implies that the ``thick-wall'' $Q$-balls with condition \eq{modcls} are both classically and absolutely stable. The fact reproduces the previous results for the case of $D=2$ and $n=4,\; p=6$ (6-th order potential) in \cite{Kim:1992mm} using the Hoelder inequality. Unlike the Gaussian ansatz \eq{gaussansatz}, our modified ansatz now shows consistent results between \eq{CLS} and \eq{chslope}.

Let us remark on the validity of our analysis following \cite{Multamaki:1999an}. In this section we have used a modified ansatz which has involved a re-scaling of $\sigma$ and $r$ in such a way as to leave us with a dimensionless action $S_n$. There are restrictions on our ability to do this as first pointed out in  \cite{Multamaki:1999an} for the case of $D=3$. We can generalise this to our $D$ dimensional case. Given that the $Q$-ball solutions extremise $S_n$, we may rescale $r$ or $\sigma$ introducing a one-parameter rescaling,  $r \to \al r$ or $\sigma \to \lambda \sigma$ which will deform the original solution. Defining $X(\al)\equiv S_n[\al r,\sigma(\al r)]$ and $Y(\lambda)\equiv S_n[\lambda \sigma(r)]$, we impose the condition that the action $S_n$ is extremised when $\al=\lambda=1$, which implies $\frac{dX}{d\al} |_{\al=1}=0=\frac{dY}{d\lambda}|_{\lambda=1}$. It is possible to show that these conditions imply that consistent solutions require the same condition as \eq{validthck}. The three dimensional case leads to the result, $n<6$, as originally obtained in \cite{Paccetti:2001uh}. The particular choice  of $n=4$ which we will investigate shortly implies $D < 4$ for the validity of our ``thick-wall'' approximation with the modified ansatz. Moreover, ``thick-wall'' $Q$-balls become classically unstable for $D \geq 3$ as can be seen from \eq{modcls}.

What have we learnt from extending the ansatz beyond the Gaussian one? We have seen that they have lead to different results. For instance, the Gaussian ansatz essentially is valid only for $D=1$ and has a contradiction, whereas the solutions based on the modified ansatz are valid for $D$ which satisfies \eq{validthck} and give consistent results \eq{modcls} for classical stability.

\subsection{Asymptotic profile}\label{sec:polasym}

The generic asymptotic profiles for large $r$ in polynomial potentials can be obtained by naively ignoring the higher order terms in the polynomial potentials \eq{thckpot} and linearising the $Q$-ball \eq{QBeq}:
\beq{LQBeq}
\sigma^{\p\p}+\frac{D-1}{r}\sigma^\p-m^2_\omega\ \sigma=0.
\eeq
We then obtain the analytic solution
\beq{slope}
\sigma(r)\sim E \sqrt{\frac{\pi}{2m_\omega}}\ r^{-\frac{D-1}{2}}e^{-m_\omega r} \hspace{10pt} \lr \hspace{10pt} -\frac{\sigma^\p}{\sigma}\sim\frac{D-1}{2r}+m_\om,
\eeq
where $E$ is a constant which is determined later. Note that we have used the fact that the modified Bessel function of the second kind has the relation $K_\mu(r)\simeq \sqrt{\frac{\pi}{2r}}e^{-r}$ for large $r$ and any real number $\mu$. The second expressions in \eq{slope} gives a condition to smoothly continue our numerical solutions to the asymptotic profiles at some large radius $r=R_{ana}$.

As we will see in the next section, our numerical results in which we obtain the full $Q$-ball solution support the modified ans\"{a}tze introduced in the previous section for both thin- and ``thick-wall'' cases.

\section{Numerical results}\label{numerical}
In this section we obtain numerical solutions for $Q$-balls using the polynomial potential in \eq{thckpot}, where $A>0,\,B>0,\ p>n>2$. We shall confirm the results obtained analytically using the modified ans\"{a}tze for both the thin- and ``thick-wall'' $Q$-balls. Recall that $U_\om(\sigma)=U(\sigma) - \frac12 \om^2\sigma^2$, with $U_\om(\sigma_-)=0$ and $\sigma_+(\om)$ marks the maximum of the effective potential $-U_\om$ where $\sigma_+(\om)\neq 0$. For a particular case, $p=2(n-1)$, we find
\bea{sig-}
\sigma_-(\om)&=&\bset{\frac{A-\sqrt{A^2-2 B m^2_\omega}}{2B}}^{1/(n-2)},\\
\label{sig+}\sigma_+(\om)&=&\bset{\frac{An+\sqrt{(An)^2-4Bp m^2_\omega}}{2Bp}}^{1/(n-2)}.
\eea
Also, for convenience, we set
\beq{set}
\omega_+=m=1,\hspace{5pt} \omega_-=\sqrt{1-\frac{A^2}{2B}}\ge 0 \lr A \le \sqrt{2B},
\eeq
where we recall the definitions of $\om_+$ and $\om_-$ are that $\om_+^2 \equiv \frac{d^2U}{d\sigma^2}|_{\sigma=0} = m^2$ and $U_{\om_-} (\sigma_+) \equiv 0$.
Setting  $\omega_-=0$ in  \eq{set} implies that $U(\sigma)$ in \eq{thckpot} has degenerate vacua at $\sigma=0,\; \pm \sigma_+$, whilst the original potential $U$ with $\om_-\neq 0$ does not have degenerate vacua. In this section, we shall consider two examples of the potential $U$,  which can be seen as the red solid lines in \fig{fig:twomdl}. The degenerate vacua potential (DVP) on the left has $\omega_-=0$ ($A=\sqrt{2B}$) and the non-degenerate vacua potential (NDVP) on the right has $\omega_-=0.5$ ($A=\sqrt{3B/2}$).  In order to determine actual values for $A$ and $B$, we define $\sigma_+(\om_+) =1$ and set $n=4,\; p=6$ for both cases; hence, $A=\frac43,\,B=\frac89$ in DVP and $A=1,\,B=\frac23$ in NDVP. Figure \ref{fig:twomdl} also includes plots of the effective potentials for various values of $\om$.
\paragraph*{\underline{\bf{{Numerical techniques and parameters}}}}

To obtain the $Q$-ball profile we need to know the initial ``position'' $\sigma_0(\omega)=\sigma(r=0)$. This is done using a shooting method, whereby we initially guess at a value of $\sigma_0(\omega)$, then solve \eq{QBeq} for the $Q$-ball profile, and depending on whether we overshoot or undershoot the required final value of $\sigma$, we modify our guess for $\sigma_0(\omega)$ and try again. Throughout our simulations, we need to specify the following three small parameters, $\epsilon,\; \xi,\; \eta$ which, respectively, determine our simulation size $r_{max}$, the radius $R_{ana}$ at which we can match the analytic and numerical solutions, and the core size $R_Q$. The smoothly continued profile is computed up to $r=R_{max}$.

\paragraph*{\underline{\bf{{Shooting method}}}}

Let us consider an effective potential $-\Uo$ which satisfies the $Q$-ball \textit{existence condition}, \eq{EXIST}. We have to initially guess $\sigma_0$ subject to it be being in the appropriate region $\sigma_-(\om) \le \sigma_0(\om) <\sigma_+(\om)$. For example it might be $\sigma^0_G=\frac{\sigma_+ + \sigma_-}{2}$. There are then three possibilities, the particle could overshoot, undershoot, or shoot properly. The last case is unlikely unless we are really ``lucky''. If it overshoots then we would find $\sigma(r_O)<0$ at some  ``time'' $r_O$. If that were to happen we could update $\sigma^0_G$ to  $\sigma^1_G = \frac{\sigma^0_G + \sigma_-}{2}$ as our next  guess. On the other hand if it undershoots, the ``velocity'' of the ``particle'' might be positive at some ``time'' $r_U$, $\sigma^\p(r_U)>0$. If that were to happen we might update  $\sigma^0_G$ to  $\sigma^1_G=\frac{\sigma_+ + \sigma^0_G}{2}$ as our next guess. After repeating the same procedures say $N$ times, we obtain the finely-tuned initial ``position'' $\sigma_0(\om)\simeq \sigma^N_G$ as our true value. To be compatible with numerical errors, our numerical simulation should be stopped with an appropriate accuracy parameterised by $\epsilon$:
\beq{eps}
 \epsilon > \sigma(r_U=r_{max}) > 0,
\eeq
where $r_{max}$ is the size of our simulations, and $\epsilon$ measures the numerical accuracy where a small value of $\epsilon$ corresponds to good numerical accuracy. Unfortunately the final profiles still have small numerical errors for large $r$. To compensate for these errors, the profiles should continue to the analytical ones smoothly at some point $r=R_{ana}$ using the following technique.

\paragraph*{\underline{\bf{{Matching analytic and numerical solutions at $R_{ana}$}}}}
In order to smoothly continue to the asymptotic profile which satisfies the second relation in \eq{slope} at the continuing point $R_{ana}$, the following condition is required:
\beq{xi}
\abs{\frac{D-1}{2r}+m_\om+\frac{\sigma^\p_{num}}{\sigma_{num}}}<\xi,
\eeq
where the parameter $\xi$ should be relatively small. Hence, we can find the appropriate profile in the whole space
\beq{mixedprof}
    \sigma(r)=
    \left\{
    \begin{array}{ll}
    \sigma_{num}(r)&\ \ \textrm{for $r<R_{ana}$}, \\
    \sigma_{num}(R_{ana})\bset{\frac{R_{ana}}{r}}^{(D-1)/2}e^{-m_{\omega}(r-R_{ana})} &\ \ \textrm{for $R_{ana} \le r \le R_{max}$},
    \end{array}
    \right.
\eeq
where we have computed $E$ using \eq{slope} and our simulations are carried out up to $r=R_{max}$.
\paragraph*{\underline{\bf{Core size and wall thickness of thin-wall $Q$-ball}}}
Using \eq{sslope}, we can define the core size $r=R_Q$ and the numerical wall thickness $\delta_{num}(\om)$ by the slope $-\sigma^\p/\sigma$ with the following condition
\bea{core}
\abs{\bset{\frac{D-2}{r}-\mu}\bset{\frac{\sigma_+-\sigma}{\sigma}} +\frac{\sigma^\p_{num}}{\sigma_{num}}}&<&\eta,\\
\delta_{num} \equiv R_{ana}-R_Q. & &
\eea
Notice that the definition of $\delta_{num}(\om)$ is different from the definition in \eq{coreQ} where $\delta(\om)=\int^{\bar{\sigma}(R_Q)}_0 \frac{d\sigma}{\sqrt{2\Uo}}$.
\paragraph*{\underline{\bf{{Numerical parameters}}}}

We have run our code in two different regimes of $\om$ for both DVP and NDVP because the profiles for large $\om$ are needed to look into larger simulation size $r_{max}$ compared to the ones for small $\om$. Because of numerical complications, we do not conduct our simulations near the extreme thin-wall limit, \ie\ $\om \simeq \om_-$. However, by solving close to the thin-wall limit, our numerical results for $\sigma_0(\om)\simeq \sigma_+(\om)$ and $R_Q \gg \delta_{num}$ allow us to recover the expected properties of thin-wall $Q$-balls with the modified ansatz \eq{thindanstz}. Finally, our results presented here correspond to the particular sets of parameters summarised in \tbl{tbl:nondeg}.
\begin{figure}[!ht]
  \def\@captype{table}
    \begin{center}
      \begin{tabular}{|c||c|c|c|c|c|}
	\hline
	\multicolumn{6}{|c|}{DVP}\\
	\hline
	$\omega$ & $\epsilon$ & $r_{max}$ & $R_{max}$ & $\xi$ & $\eta$ \\
    	\hline
0.38-0.73 & 4.0 $\times 10^{-2}$ & 30 & 200 & 8.0 $\times 10^{-3}$ & 1.0 $\times 10^{-1}$ \\
0.73-0.99999 & 1.0$\times 10^{-5}$ & 40 & 200 & 8.0 $\times 10^{-3}$ & 1.0 $\times 10^{-1}$ \\
\hline
      \end{tabular}
    \end{center}
  \hfill
    \begin{center}
      \begin{tabular}{|c||c|c|c|c|c|}
	\hline
	\multicolumn{6}{|c|}{NDVP}\\
	\hline
	$\omega$  & $\epsilon$ & $r_{max}$ & $R_{max}$ & $\xi$ & $\eta$ \\
    	\hline
0.60-0.85 & 3.0 $\times 10^{-3}$ & 30 & 200 & 8.0 $\times 10^{-3}$ & 1.0 $\times 10^{-1}$ \\
0.85-0.99999 & 1.0$\times 10^{-5}$ & 50 & 200 & 8.0 $\times 10^{-3}$ & 1.0 $\times 10^{-1}$ \\
\hline
      \end{tabular}
	\end{center}    \tblcaption{The numerical parameters in DVP (top) and in NDVP (bottom).}
    \label{tbl:nondeg}
\end{figure}
\subsection{Stationary properties in DVP and NDVP}

We devote a large part of this section to justifying the previously obtained analytical results in the thin- and ``thick-wall'' approximations by obtaining the appropriate numerical solutions.

\paragraph*{\underline{\bf{{Profiles with our numerical algorithm}}}}

\begin{figure}[!ht]
  \begin{center}
   \subfigure{\label{fig:deggrad}
	\includegraphics[angle=-90, scale=0.27]{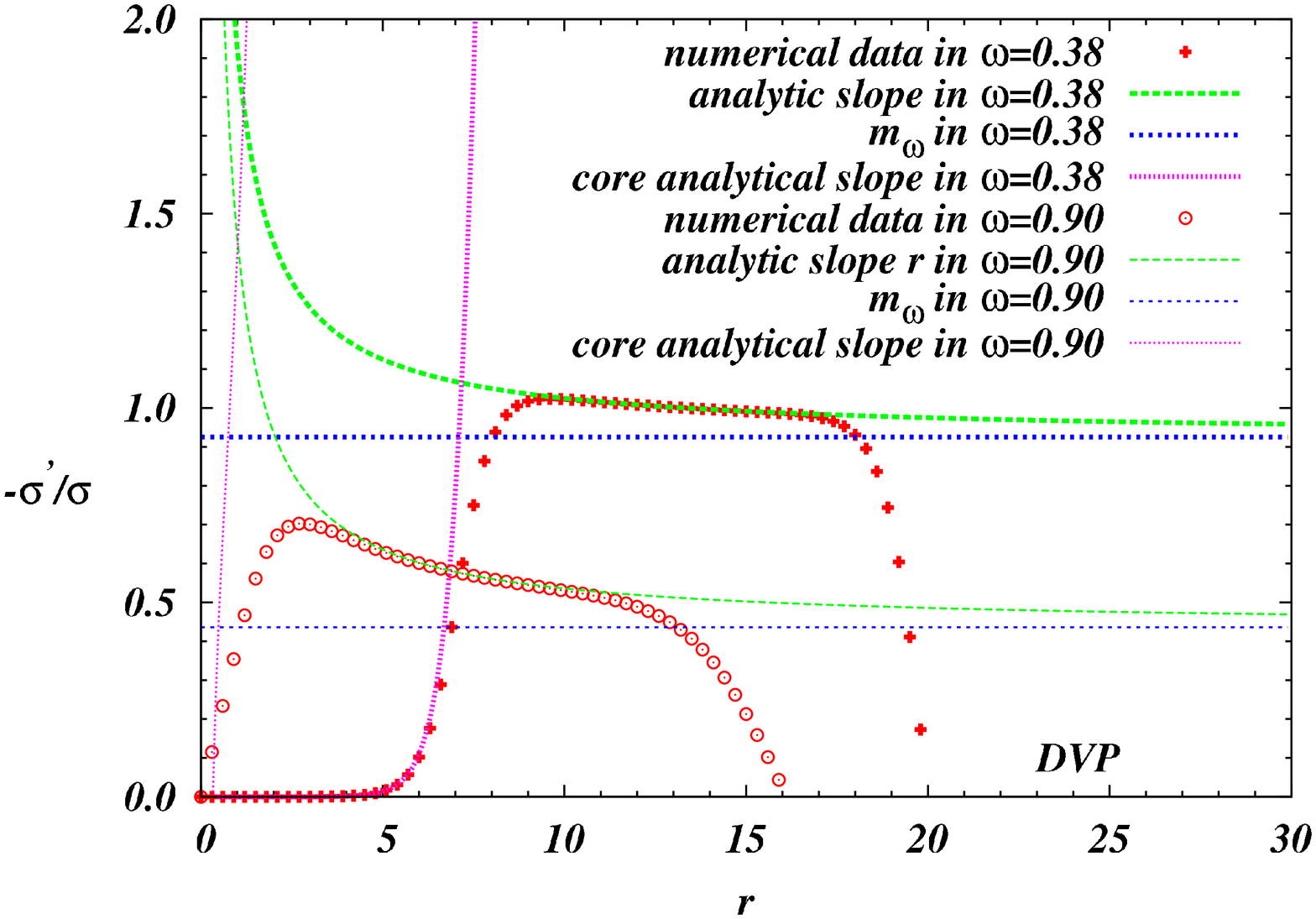}}
   \subfigure{\label{fig:nondeggrad}
	\includegraphics[angle=-90, scale=0.27]{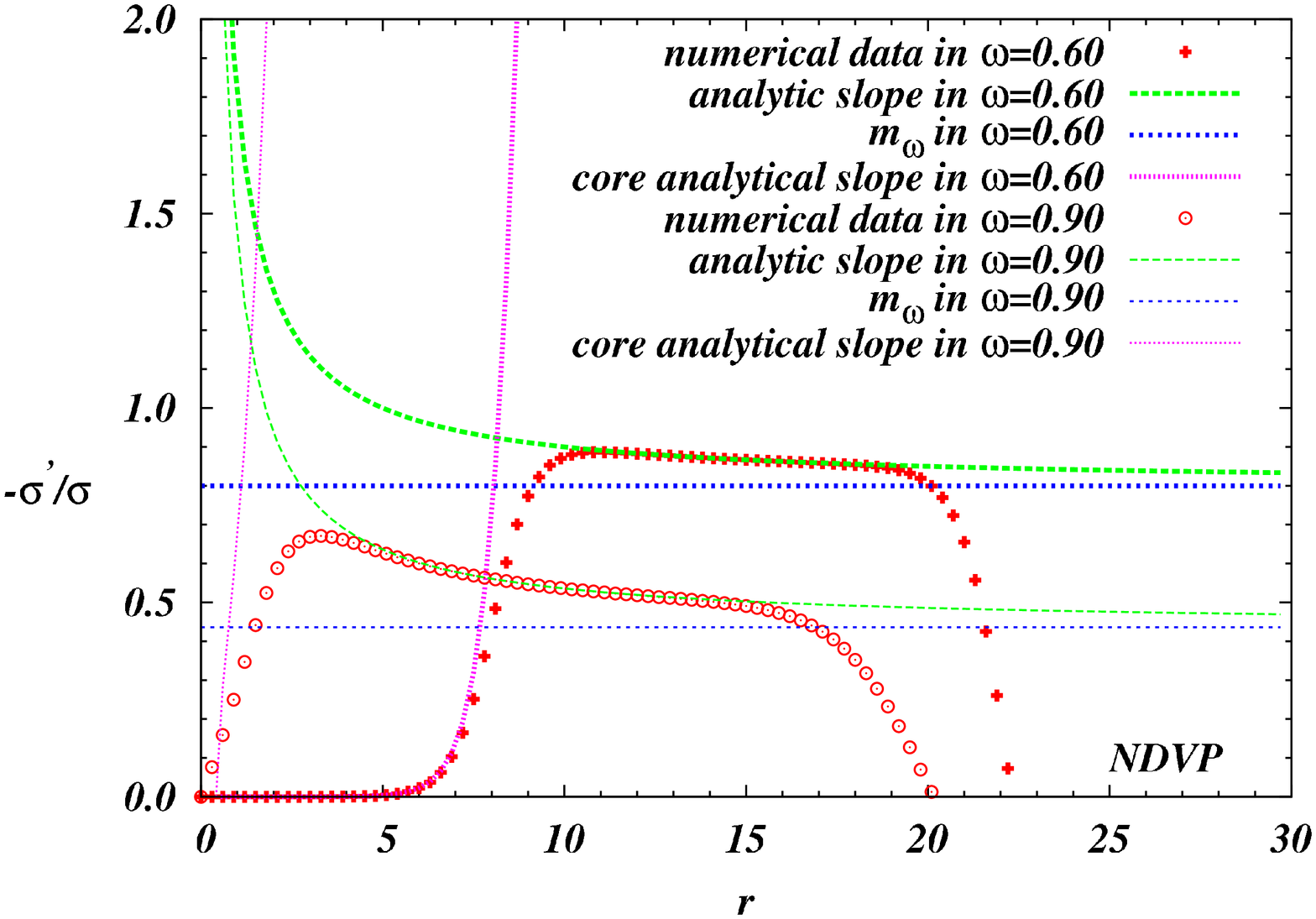}}\\
  \subfigure{\label{fig:degprof}  \includegraphics[angle=-90,scale=0.27]{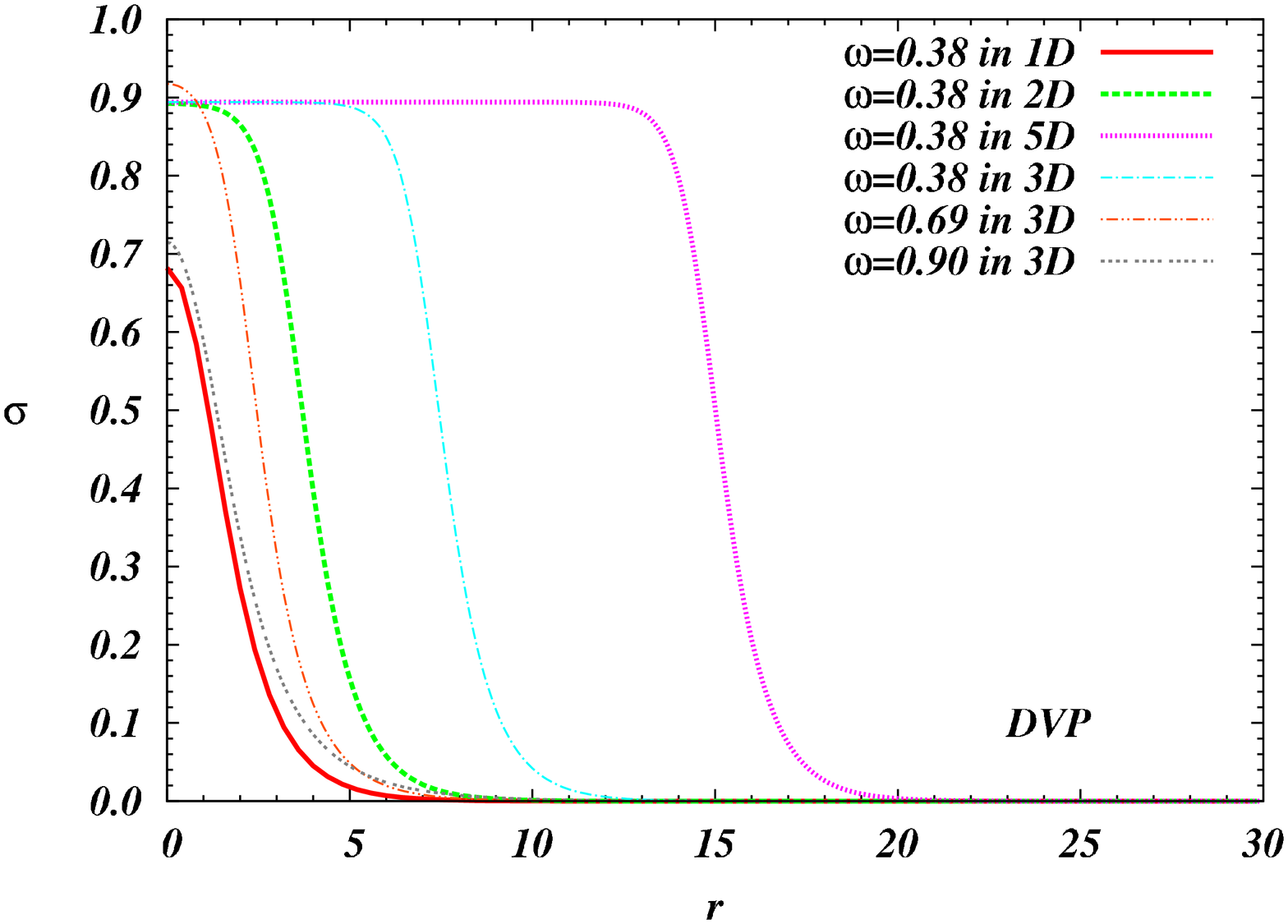}}
  \subfigure{\label{fig:nondegprof}\includegraphics[angle=-90,scale=0.27]{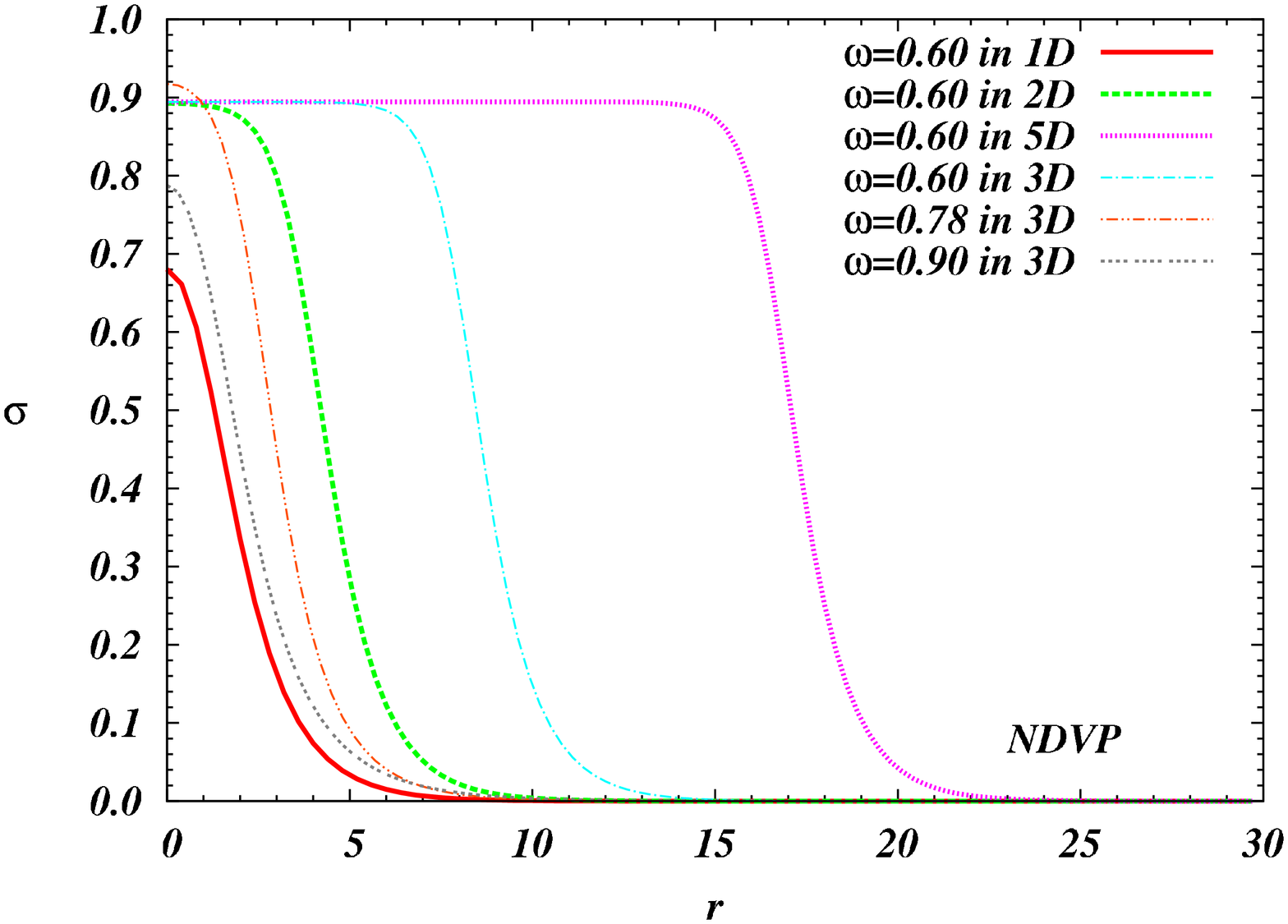}}
  \end{center}
  \caption{The top two panels show the numerical slopes $-\sigma^\p/\sigma$ for the case of $D=3$ for two values of $\om$ for both DVP (left) and NDVP (right). The red (one-dotted and one with circles) lines show the numerical slopes and the green dotted lines with two different widths the corresponding analytic solutions. The purple dot-dashed lines with two different widths show the analytic fits for the core profiles. The bottom two panels show the full $Q$-ball profile as described in \eq{mixedprof} for a number of values of $\om$ and $D$. Note how the core size increases with $D$.
}
  \label{fig:degexp}
\end{figure}

In the top two panels of \fig{fig:degexp} the two red lines (one dotted and one with circles) show the numerical slopes $-\sigma^\p/\sigma$ for the case of $D=3$ for two values of $\om$. These are then matched to the analytic profiles (green dotted lines) in order to achieve the full profile as given in  \eq{mixedprof}. Recall that we expect in general for all values of $\om$, the analytic fits to be accurate for large $r$, the numerical fits to be most accurate for small $r$ and there to be an overlap region where they are both consistent with each other as seen in \fig{fig:degexp}.  We have also plotted in dot-dashed purple lines our analytic fits,  \eq{core}, for the slopes of the thin-wall cores from $r=0.5$. We should remind the reader that this fit only really works for the case of small $\om$ because we are dealing with thin-wall $Q$-balls.  Notice, it is clear from the purple dot-dashed lines that the core sizes cannot be determined by this technique for the case $\om=0.9\simeq \om_+$.

\begin{figure}[!ht]
  \begin{center}
   \subfigure{\label{fig:degini}\includegraphics[angle=-90, scale=0.27]{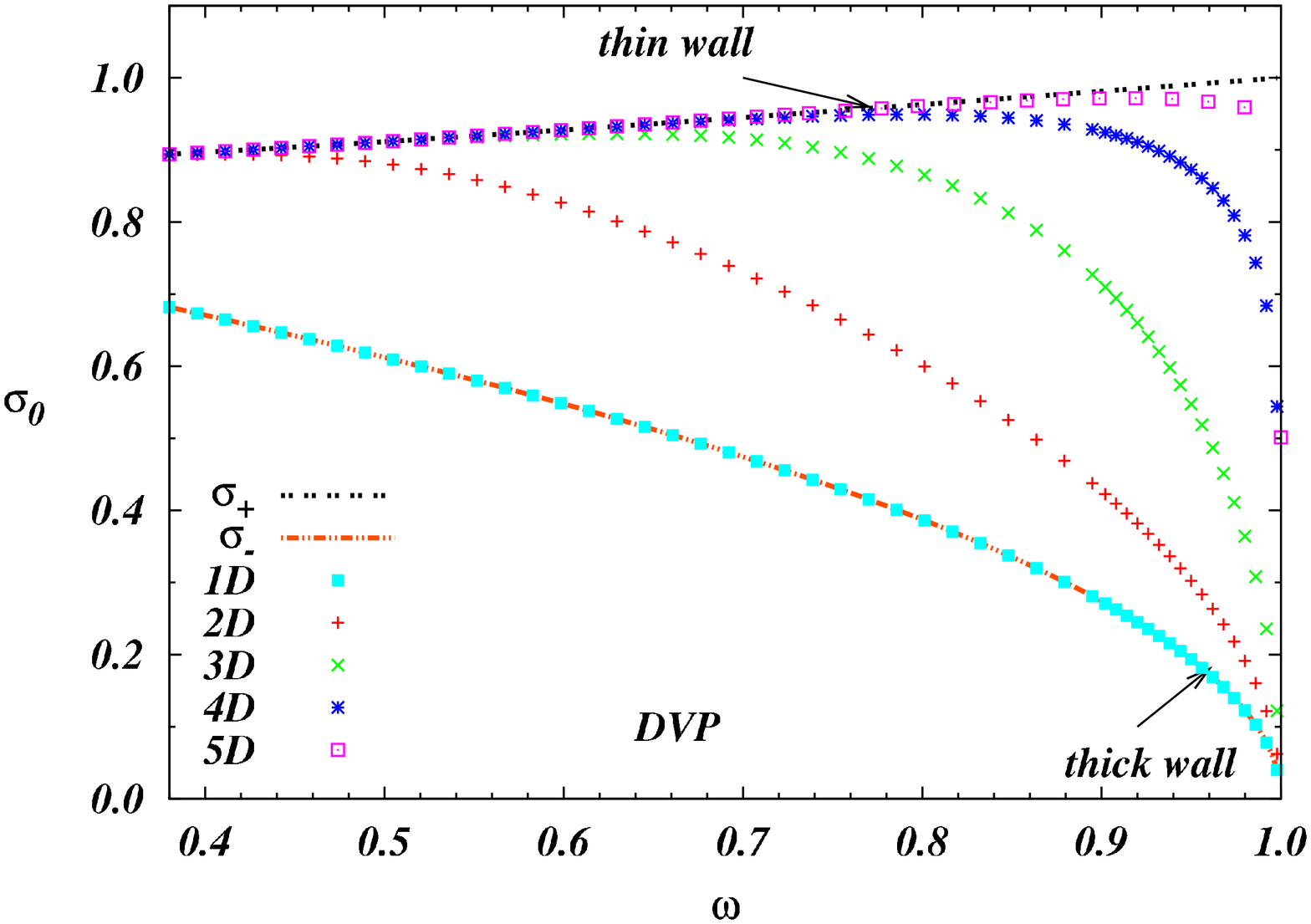}}
   \subfigure{\label{fig:nondegini}\includegraphics[angle=-90, scale=0.27]{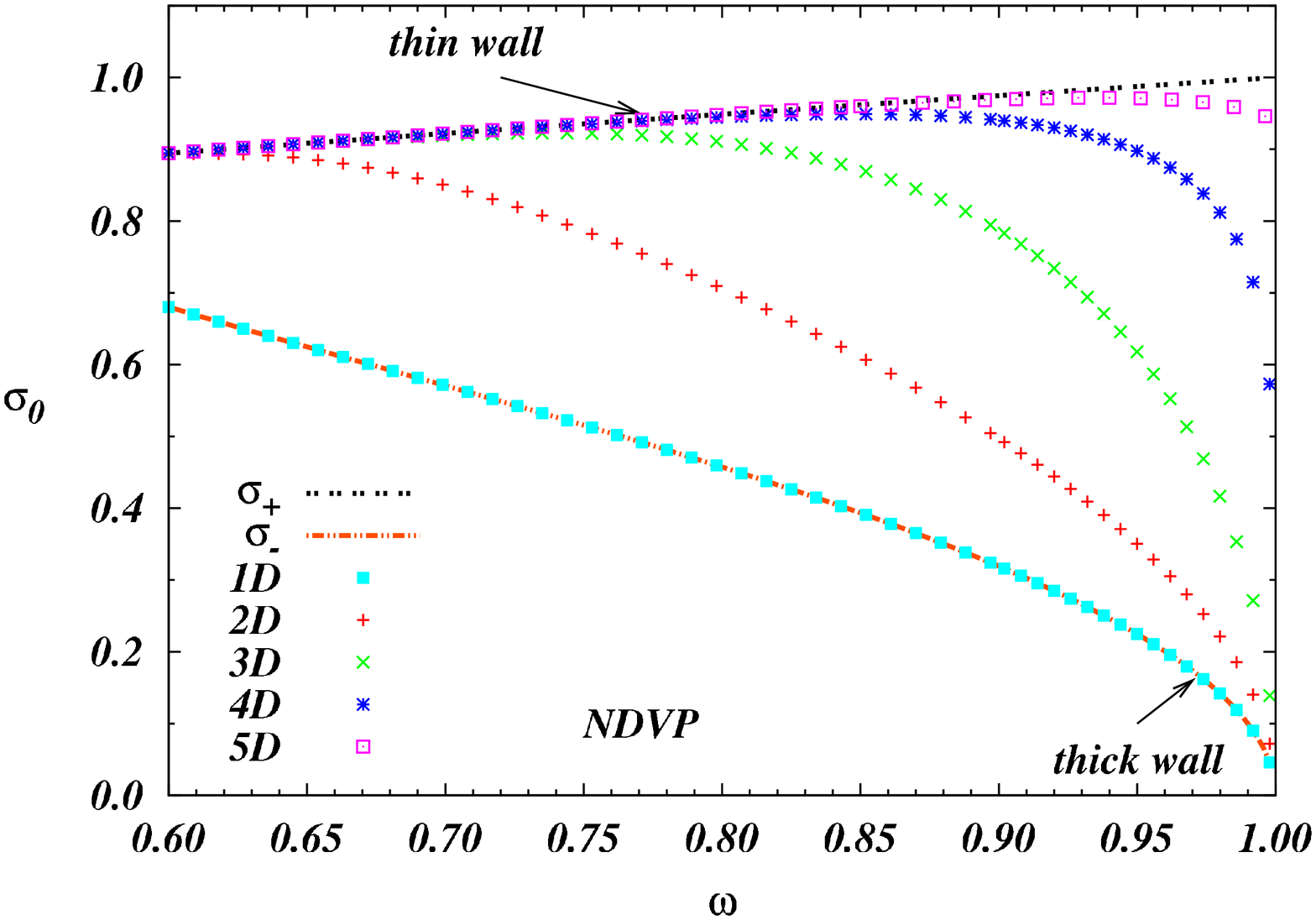}}\\
   \subfigure{\label{fig:dgthck}\includegraphics[angle=-90, scale=0.27]{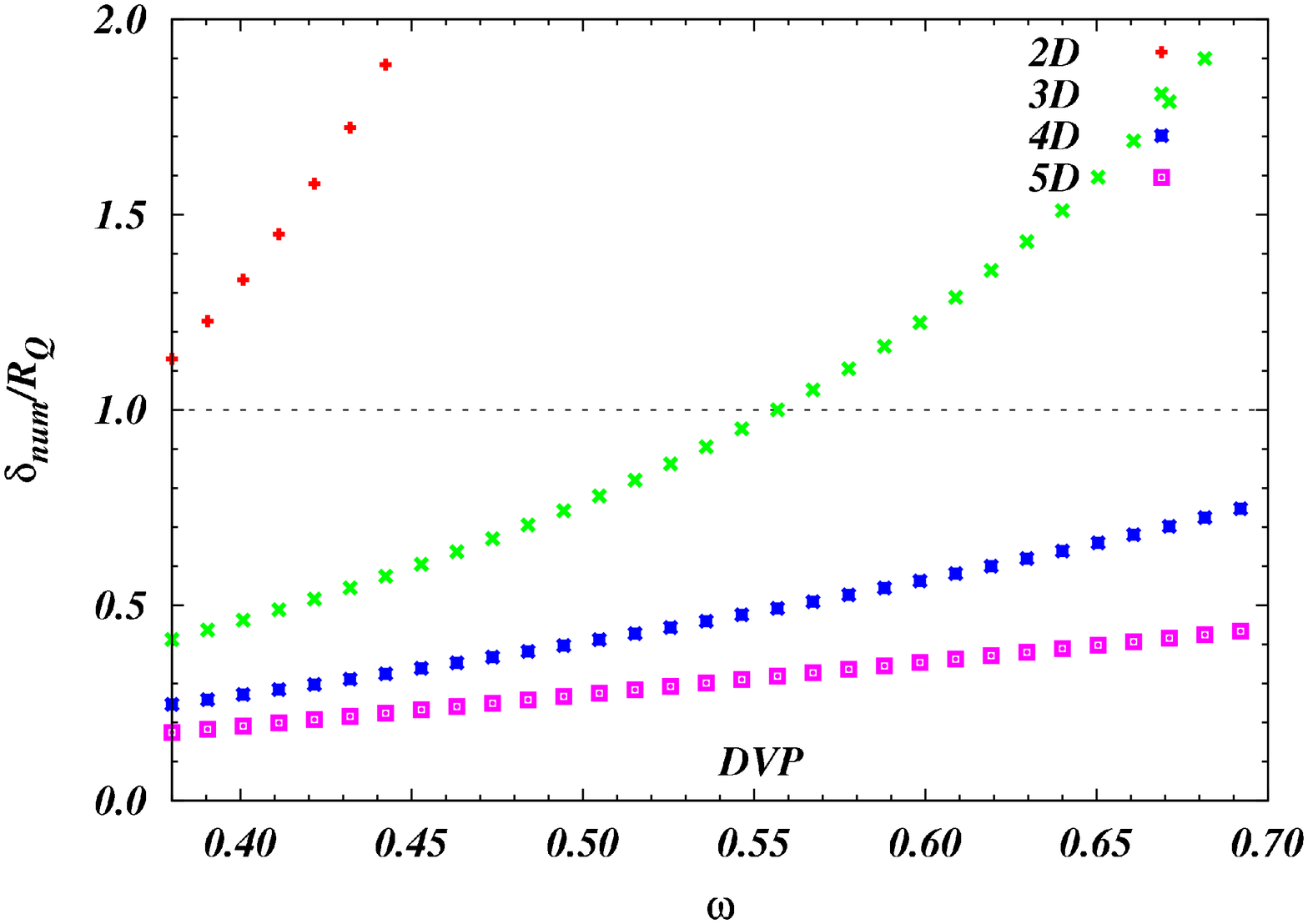}}
   \subfigure{\label{fig:ndgthck}\includegraphics[angle=-90, scale=0.27]{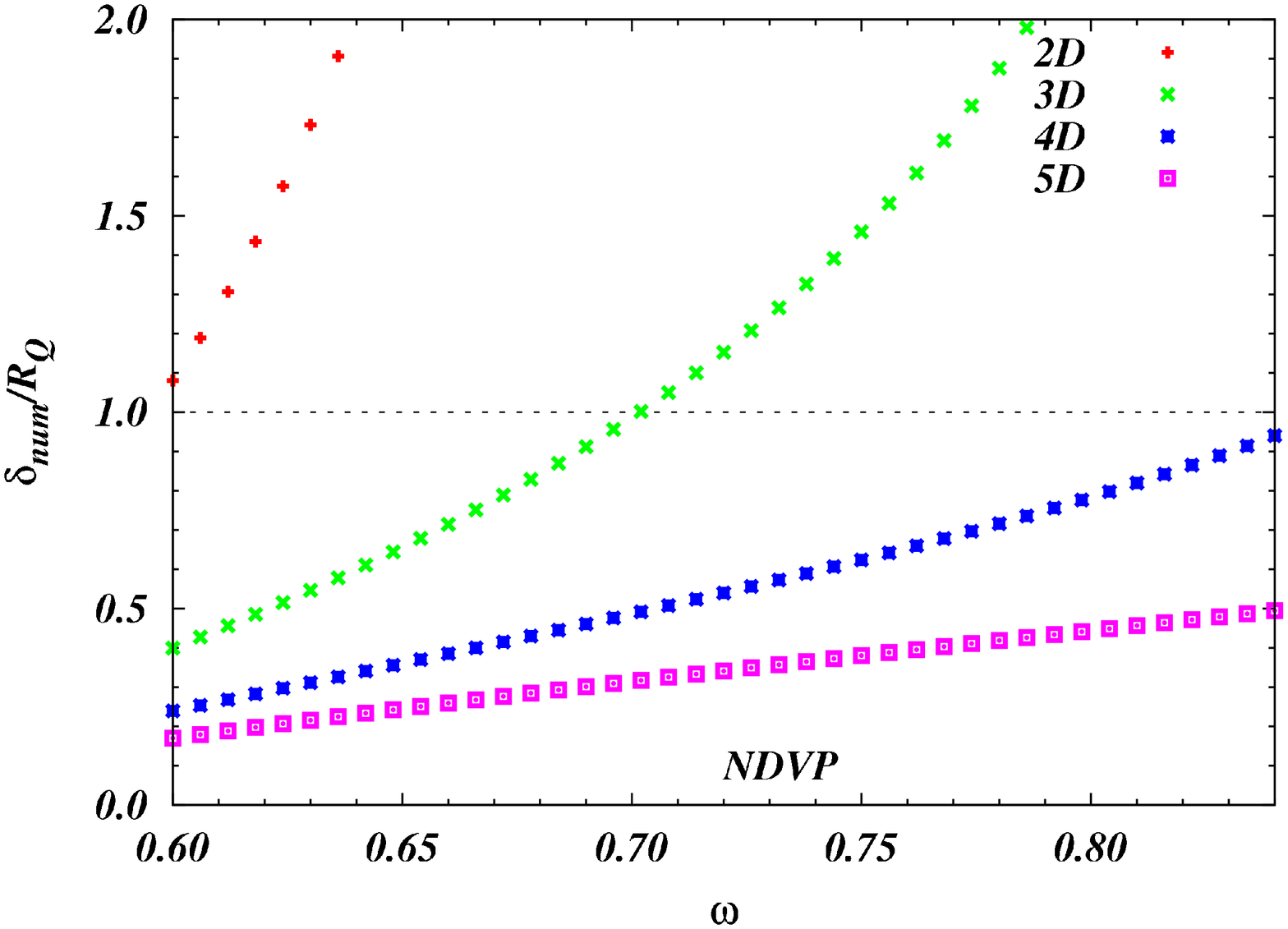}}\\
   \subfigure{\label{fig:degrq}\includegraphics[angle=-90, scale=0.27]{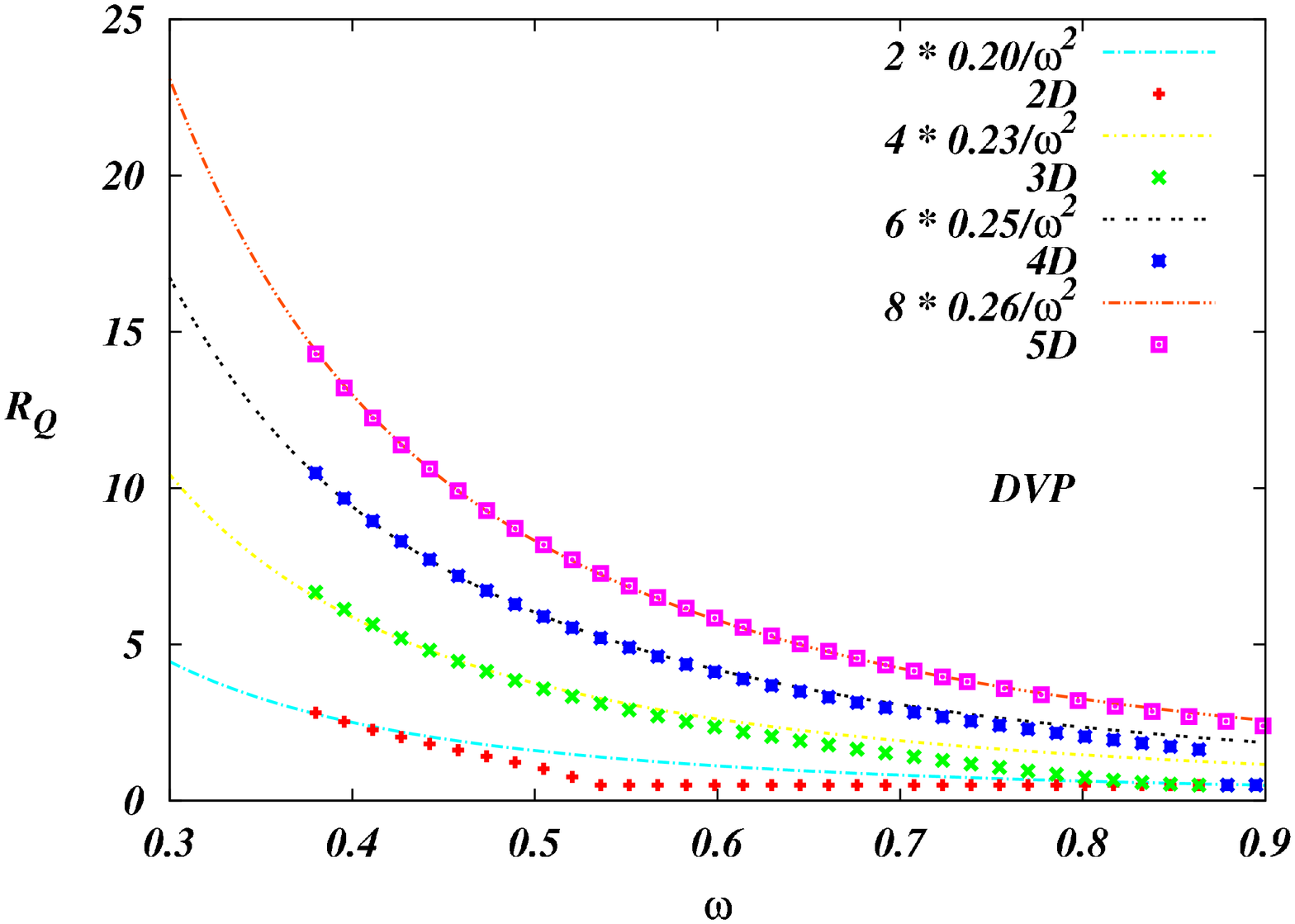}}
   \subfigure{\label{fig:nondegrq}\includegraphics[angle=-90, scale=0.27]{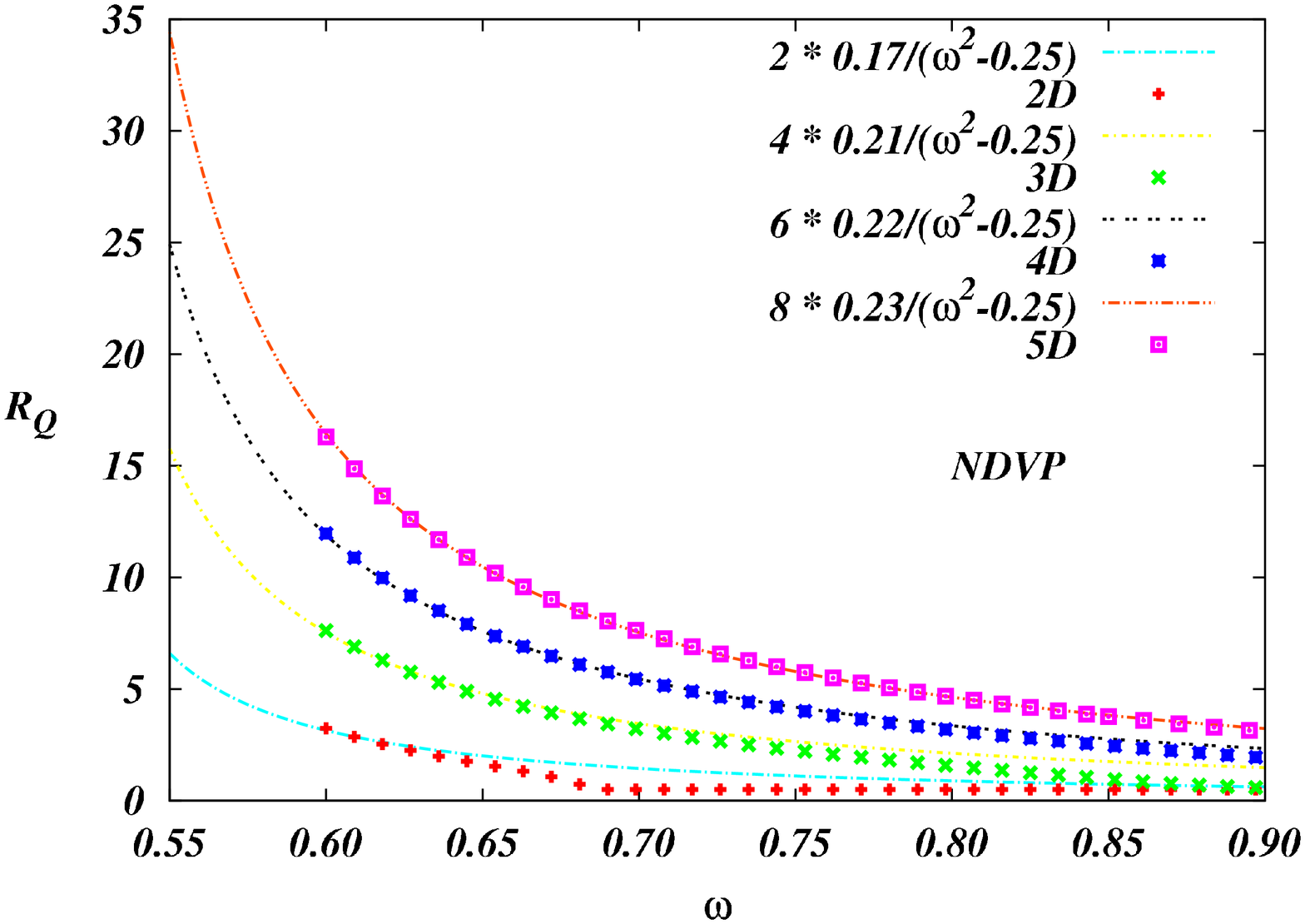}}
  \end{center}
  \caption{The initial ``positions'' $\sigma_0(\om)$ (top), $\delta_{num}/R_Q$ (middle), and the core sizes $R_Q(\om)$ (bottom). The top panels show $\sigma_\pm(\om)$, \eqs{sig-}{sig+} as black and orange dot-dashed lines respectively. The middle panels show the range of values of $\om$ for a given value of $D$ in which the core thickness is smaller than the core size, a crucial assumption we have to make. In the bottom panel, the analytical core sizes in \eq{RQ-DVP} are plotted with the numerical ones for the following $\om$ ranges: $[0.38-0.40],\ [0.38-0.55],\ [0.38-0.60],\ [0.38-0.70]$ in DVP, and $[0.60-0.62],\ [0.60-0.65],\ [0.60-0.75],\ [0.60-0.85]$ in NDVP and for $D=2,\ 3,\ 4,\ 5$, respectively.  As can be seen, the fits are excellent. The range of $\om$ values chosen have been based on the results shown in the top two panels and correspond to that range where the thin-wall $Q$-balls are solutions (except for $D=2$).}
  \label{fig:sig}
\end{figure}

The bottom two panels show the full profiles satisfying  \eq{mixedprof} for arbitrary $D$ up to $D=5$.  We have been able to obtain the $Q$-ball profiles in the whole parameter space $\om$ except for the extreme thin-wall region $\om\simeq \om_-$. Both DVP and NDVP $Q$-balls have profiles with similar behaviours in that as the spatial dimension increases, so does their core size.
\paragraph*{\underline{\bf{{Criteria for the existence of a thin-wall $Q$-ball with core size $R_Q$}}}}
The top and middle panels of \fig{fig:sig} show the numerical results for $\sigma_0(\om)$ and $\delta_{num}/R_Q$ against $\om$  for a number of spatial dimensions $D$. For the case of $D \geq 3$ it is clear from the top panels that the $Q$-balls are well described by the thin-wall result \eq{sig-} for most values of $\om$, with the range increasing as $D$ increases. The case of $D=2$ is less clear, it appears to asymptote onto the line. We believe there is a solution that exists for that case for small values of $\om$. An important point is that for the approximation to be valid we are working in the regime  $\delta_{num}/R_Q <1$ which can be seen to be true from the middle panels (again we believe the case of $D=2$ is heading below the line  $\delta_{num}/R_Q =1$ for small $\om$.

These results are consistent with our analytic solutions for finite thick-walled $Q$-balls given by \eq{thindanstz}, subject to the criteria $\sigma_0(\om)\simeq \sigma_+(\om)$ and $R_Q \gtrsim \delta_{num}$, even though $\om \sim \om_+$.

For $D=1$ we see in the top panels that  $\sigma_0(\om)$ exactly matches $\sigma_-(\om)$, (the orange dot-dashed lines). The bottom two panels in \fig{fig:sig} show the core sizes $R_Q$ of thin-wall $Q$-balls which satisfy our criterion \eq{core}. Recall that $R_Q$ in \eq{RQ} is a function of $\om$ assuming $\tau$ depends on $\om$ weakly; thus, we plot the numerical core sizes comparing them with our analytical approximation for DVP and NDVP, respectively
\beq{RQ-DVP}
R^{DVP}_Q \simeq \frac{2(D-1) \tau_{num}}{\omega^2};\hspace*{10pt} R^{NDVP}_Q \simeq \frac{2(D-1)\tau_{num}}{(\omega^2-\om^2_-)}
\eeq
where the parameter $\tau_{num}$ is
computed numerically (see \tbl{tbl:ndrq}). The presented numerical core sizes match excellently with the
analytical fittings over a wide range of $\om$. Some numerical errors
appear around $\om\simeq \om_+$ since we cannot determine the ``thin-wall'' cores with this technique, see the top two panels in \fig{fig:degexp}. \tbl{tbl:ndrq} shows analytical and numerical values of $\tau$ using \eq{tens} and the above fitting technique. We confirm that the values of $\tau$ (a part of the surface tension $\tau/D$ in \eq{linearthin}) are nearly constant, depending slightly on $D$. Therefore, the assumptions we made for thin-wall $Q$-balls are valid as long as $\sigma_0(\om)\simeq \sigma_+(\om)$ and $R_Q \gtrsim \delta_{num}$. 

\begin{figure}[!ht]
  \def\@captype{table}
  \begin{minipage}[t]{\textwidth}
    \begin{center}
      \begin{tabular}{|c||c|c|c|c|c|}
	\hline
	$\tau$ & $\tau_{ana}$ & $2D$ & $3D$ & $4D$  & $5D$ \\
    	\hline
		DVP & 0.19 & 0.20 & 0.23 & 0.25 & 0.26 \\
		NDVP & 0.16 & 0.17 & 0.21 & 0.22 & 0.23 \\
\hline
      \end{tabular}
	\end{center}
	\tblcaption{The values of $\tau_{ana}$ and $\tau_{num}$ in terms of $D$ in DVP and NDVP, see \eqs{tens}{RQ}.}
    \label{tbl:ndrq}
  \end{minipage}
\end{figure}

\paragraph*{\underline{\bf{{Configurations}}}}

\begin{figure}[htp]
    \begin{center}
    \subfigure{\label{fig:degrhoq}\includegraphics[angle=-90,scale=0.27]{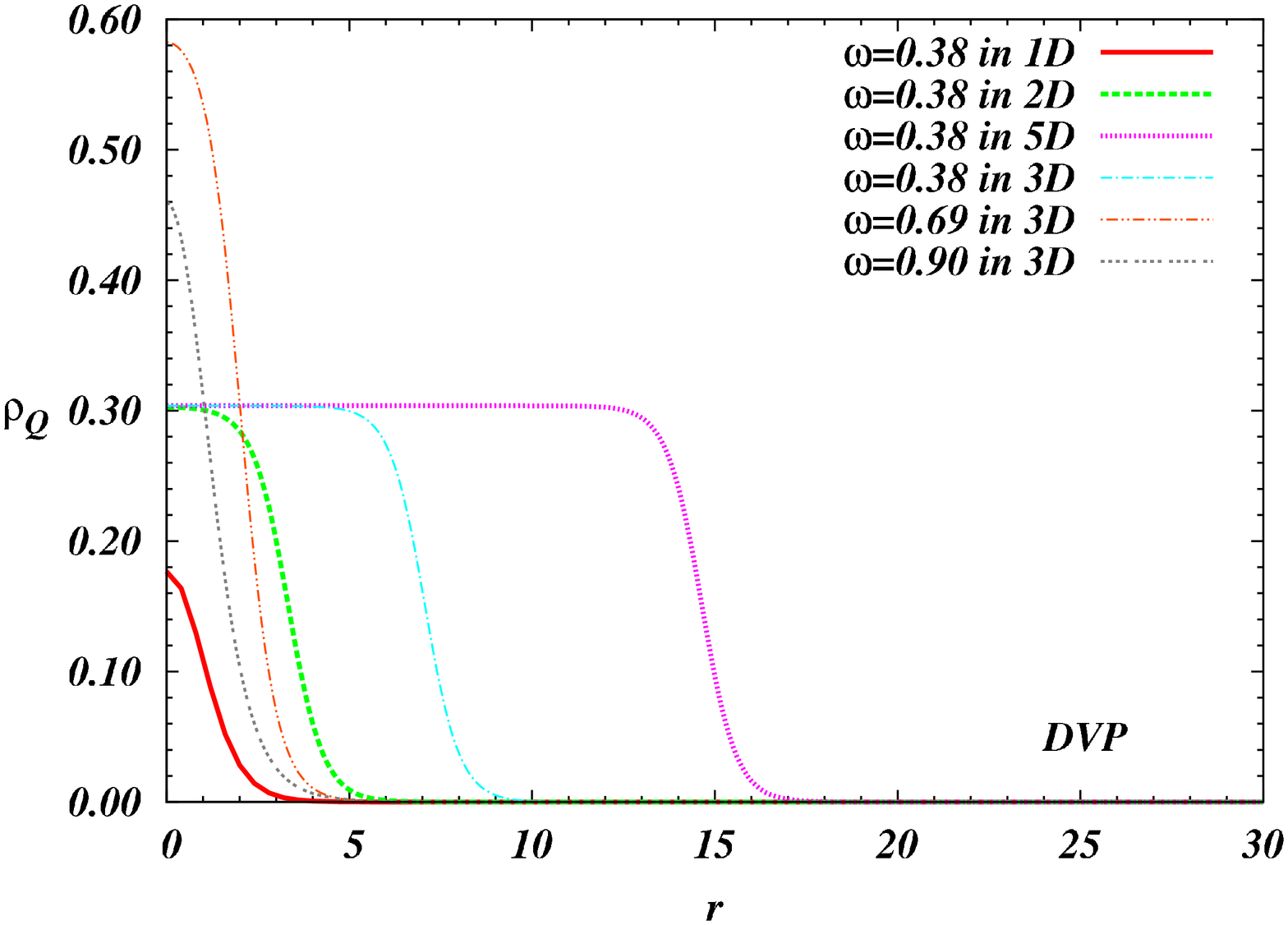}}
    \subfigure{\label{fig:nondegrhoq}\includegraphics[angle=-90,scale=0.27]{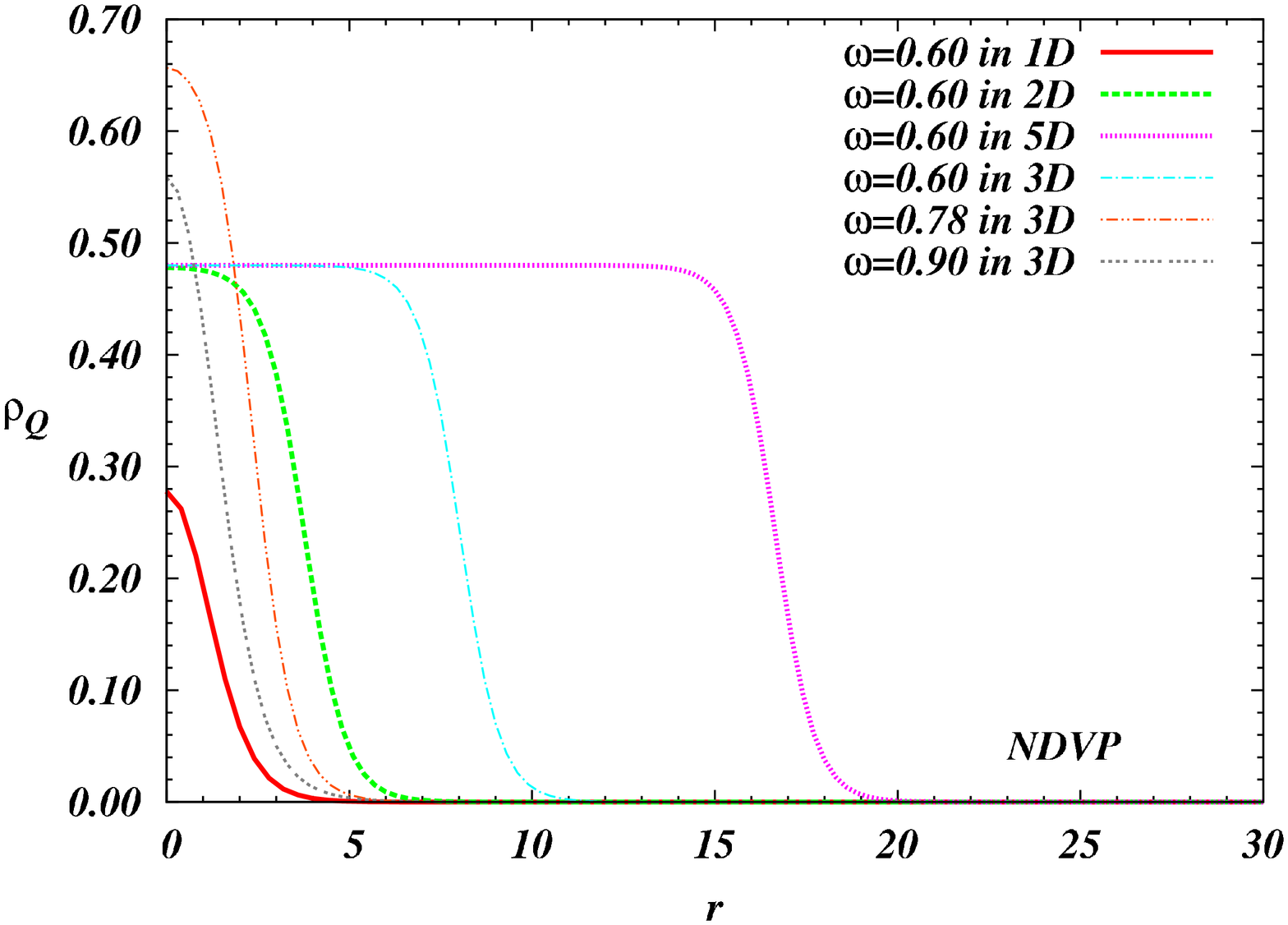}}\\
        \subfigure{\label{fig:degrhoe}\includegraphics[angle=-90,scale=0.27]{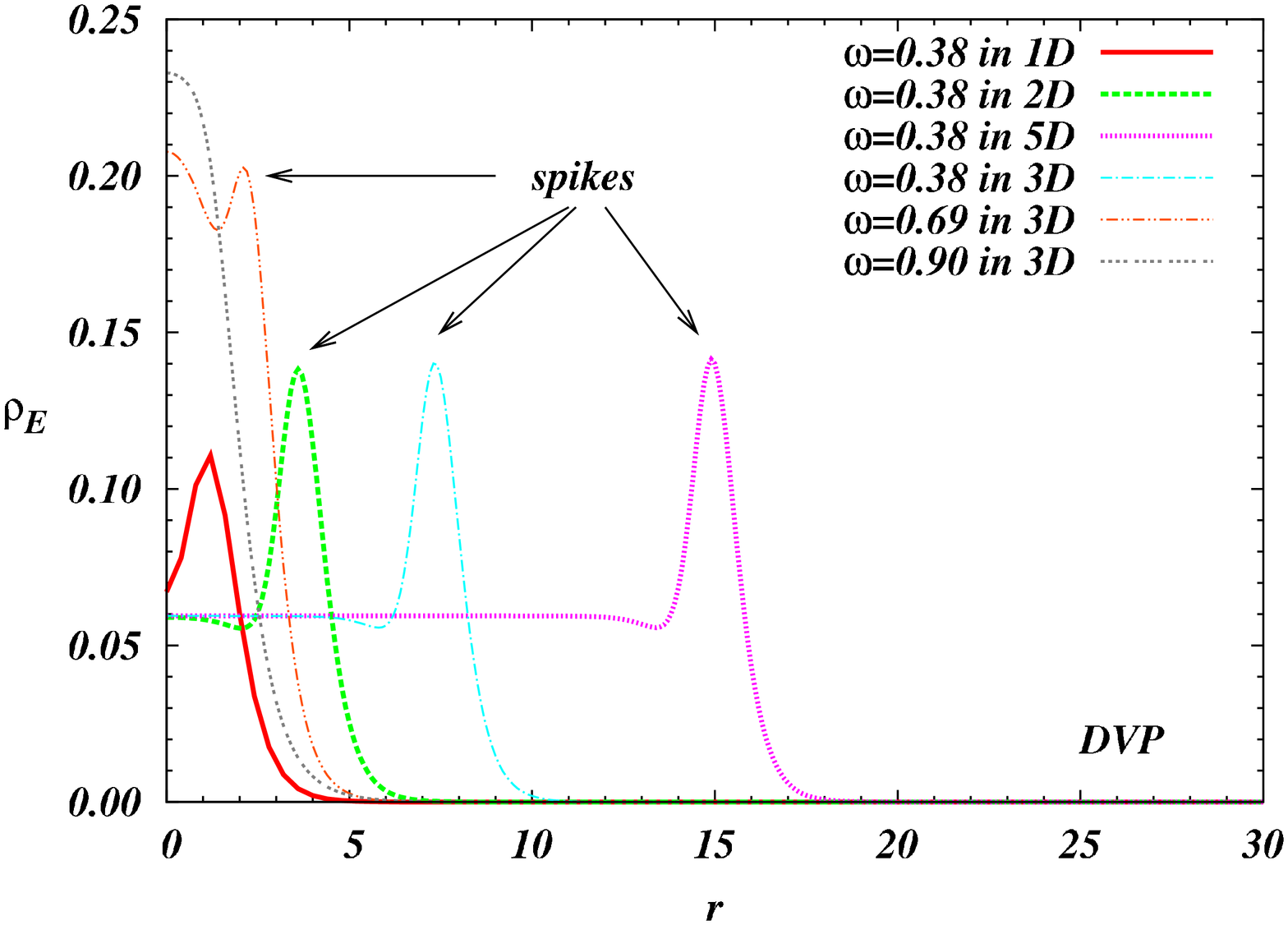}}
    \subfigure{\label{fig:nondegrhoe}\includegraphics[angle=-90,scale=0.27]{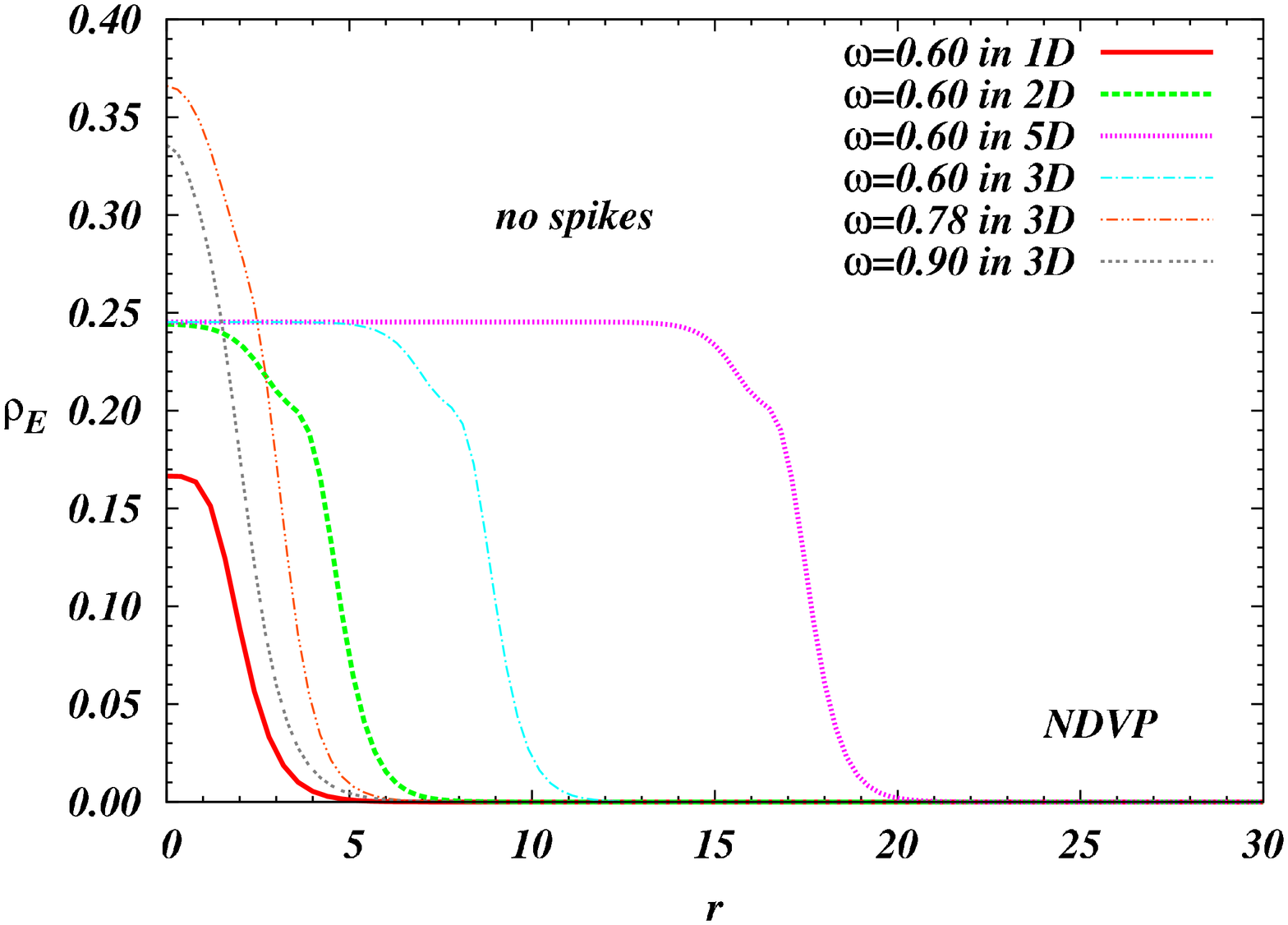}}
  \end{center}
  \caption{The configurations for charge density $\rho_Q$ (top) and energy density $\rho_E$ (bottom) computed using \eq{rho_Q} for both DVP (left) and NDVP (right). The presence of spikes of $\rho_E$ in DVPs contributes to their increased surface energies.}
  \label{fig:conf}
\end{figure}

\fig{fig:conf} illustrates the configurations of charge density $\rho_Q$ (top) and energy density $\rho_E$ (bottom), in both DVP (left) and NDVP (right). Each of the DVP energy densities around $\om\sim \om_-$ has a spike within the shells, while those spikes are not present in NDVP. The presence of spikes can contribute to the increase in surface energy $\mS$, which accounts for the different observed ratio for  $\mS/\mU$ in the two cases, where $\mU$ is the potential energies. Otherwise, DVP and NDVP models have similar profiles in \fig{fig:degexp}. Moreover, we have numerically checked that $Q$-balls for $D\ge 2$ generally have positive radial pressures, whereas the $1D$ radial pressures are always zero, \ie\ $\half \sigma^{\p2}=\Uo$ due to \eq{radialp}.

\paragraph*{\underline{\bf{{Virialisation and characteristic slope $E_Q/\om Q$}}}}

\begin{figure}[htp]
\begin{center}
    \subfigure{\label{fig:degsu}\includegraphics[angle=-90,scale=0.27]{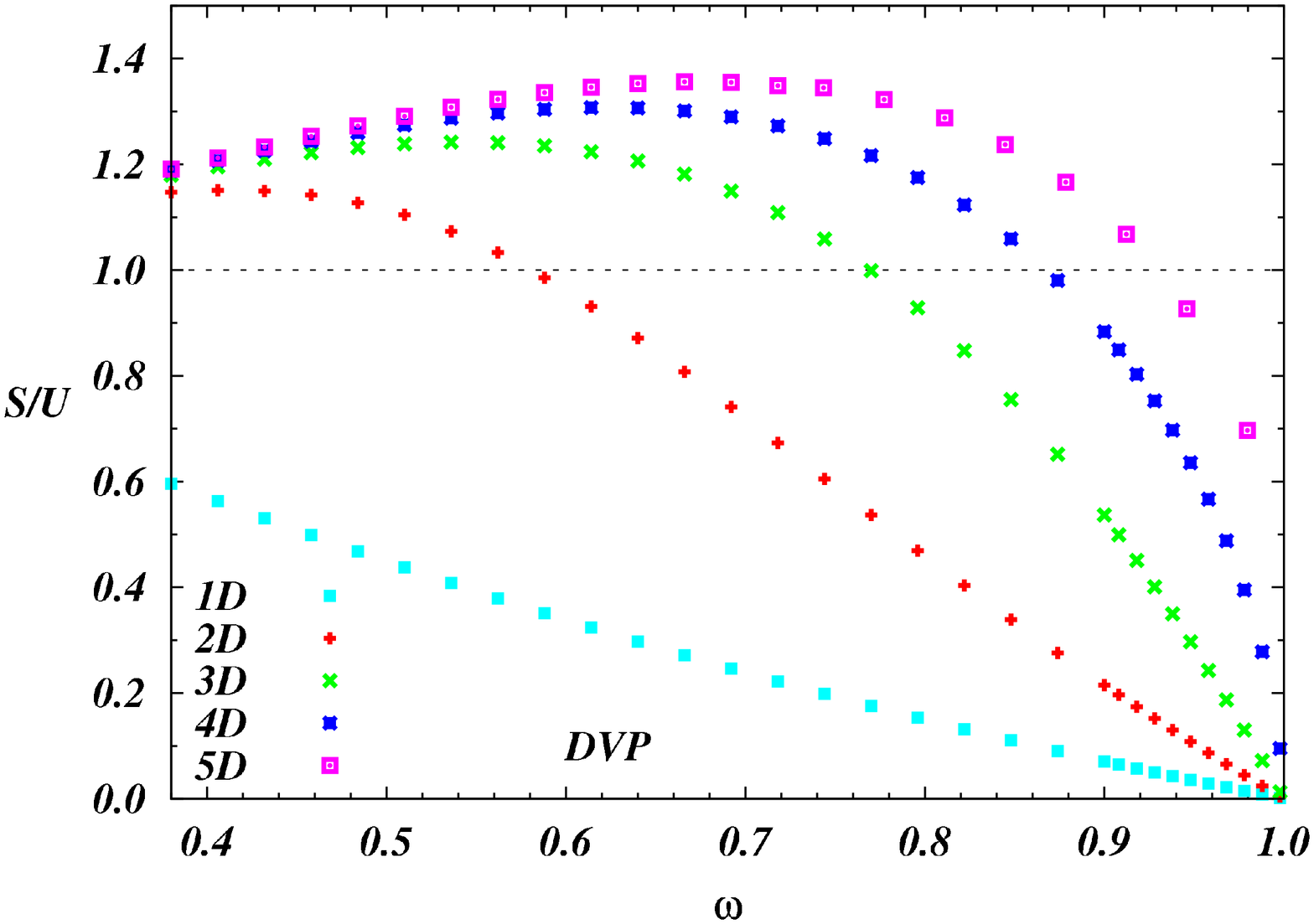}}
    \subfigure{\label{fig:nondegsu}\includegraphics[angle=-90,scale=0.27]{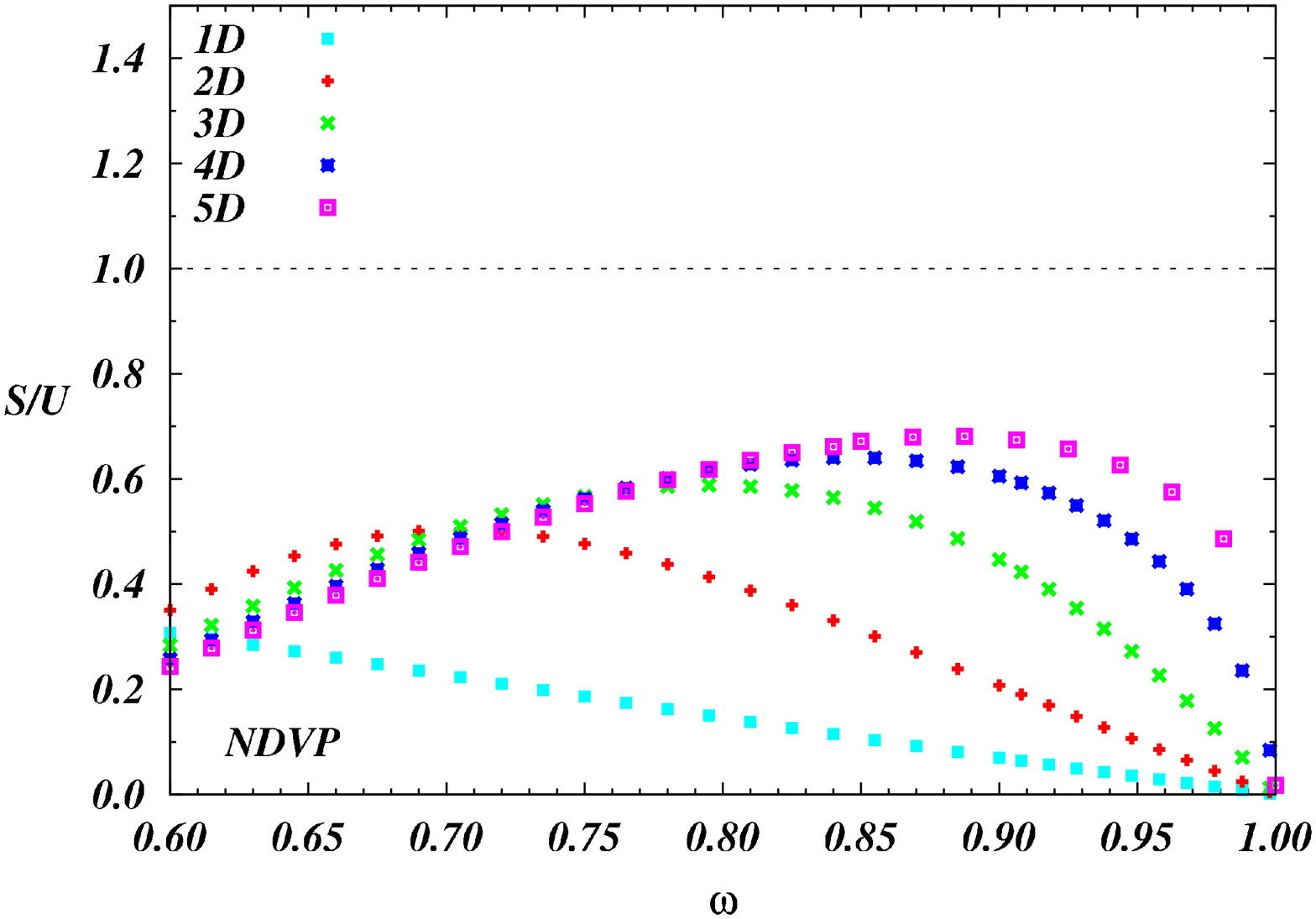}}\\
    \subfigure{\label{fig:degslope}\includegraphics[angle=-90,scale=0.27]{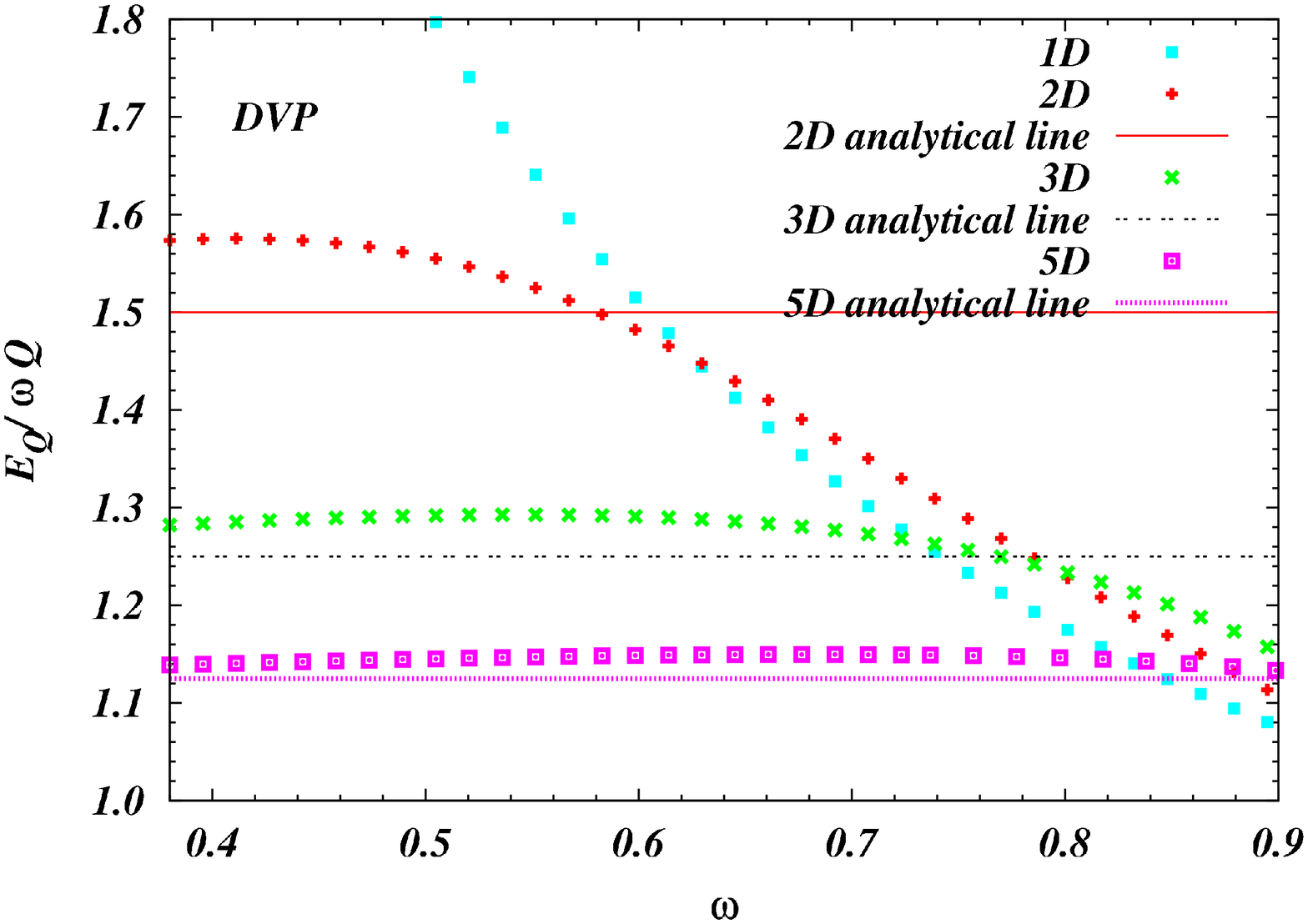}}
    \subfigure{\label{fig:nondegslope}\includegraphics[angle=-90,scale=0.27]{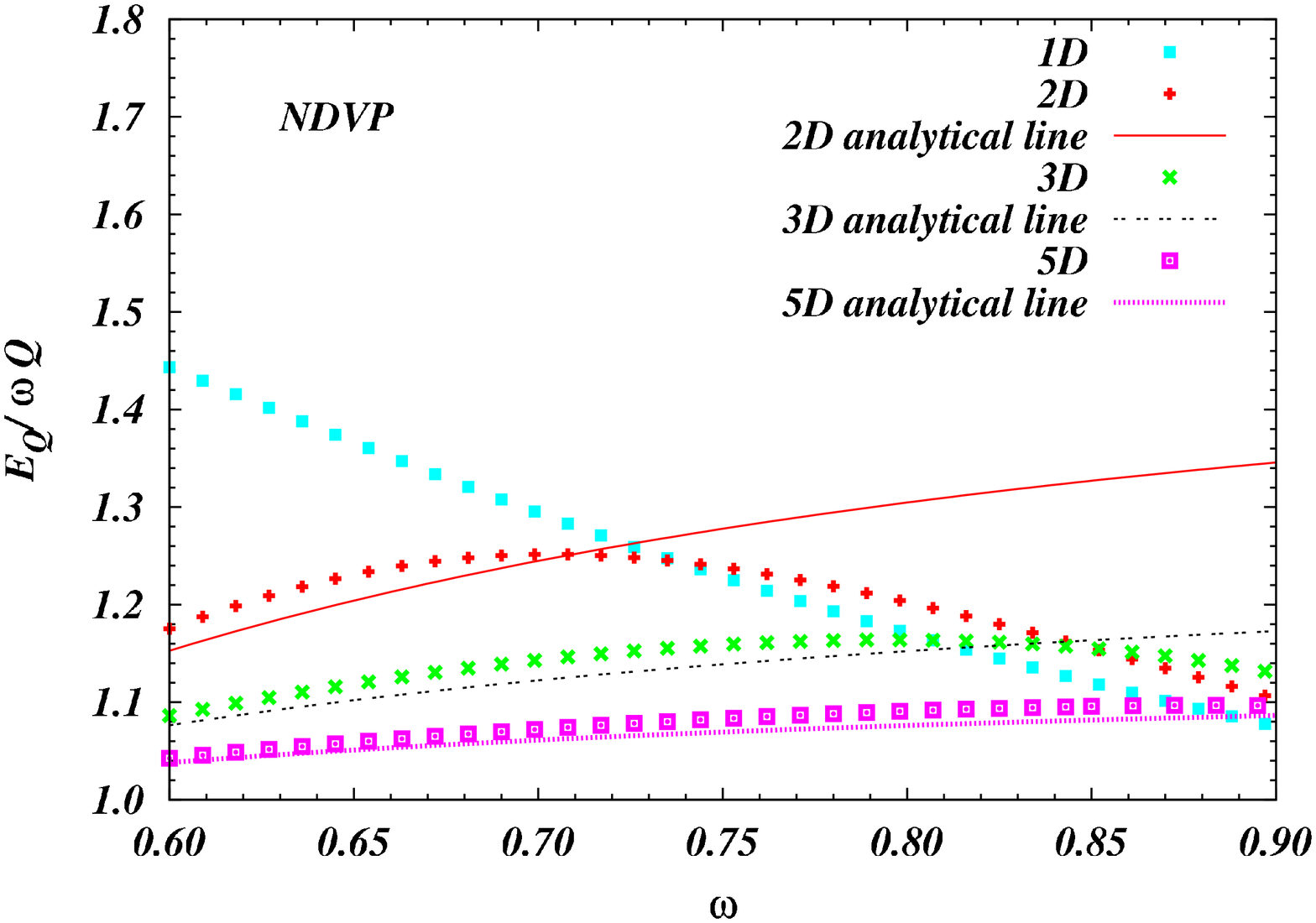}}\\
    \subfigure{\label{fig:degslopethck}\includegraphics[angle=-90,scale=0.27]{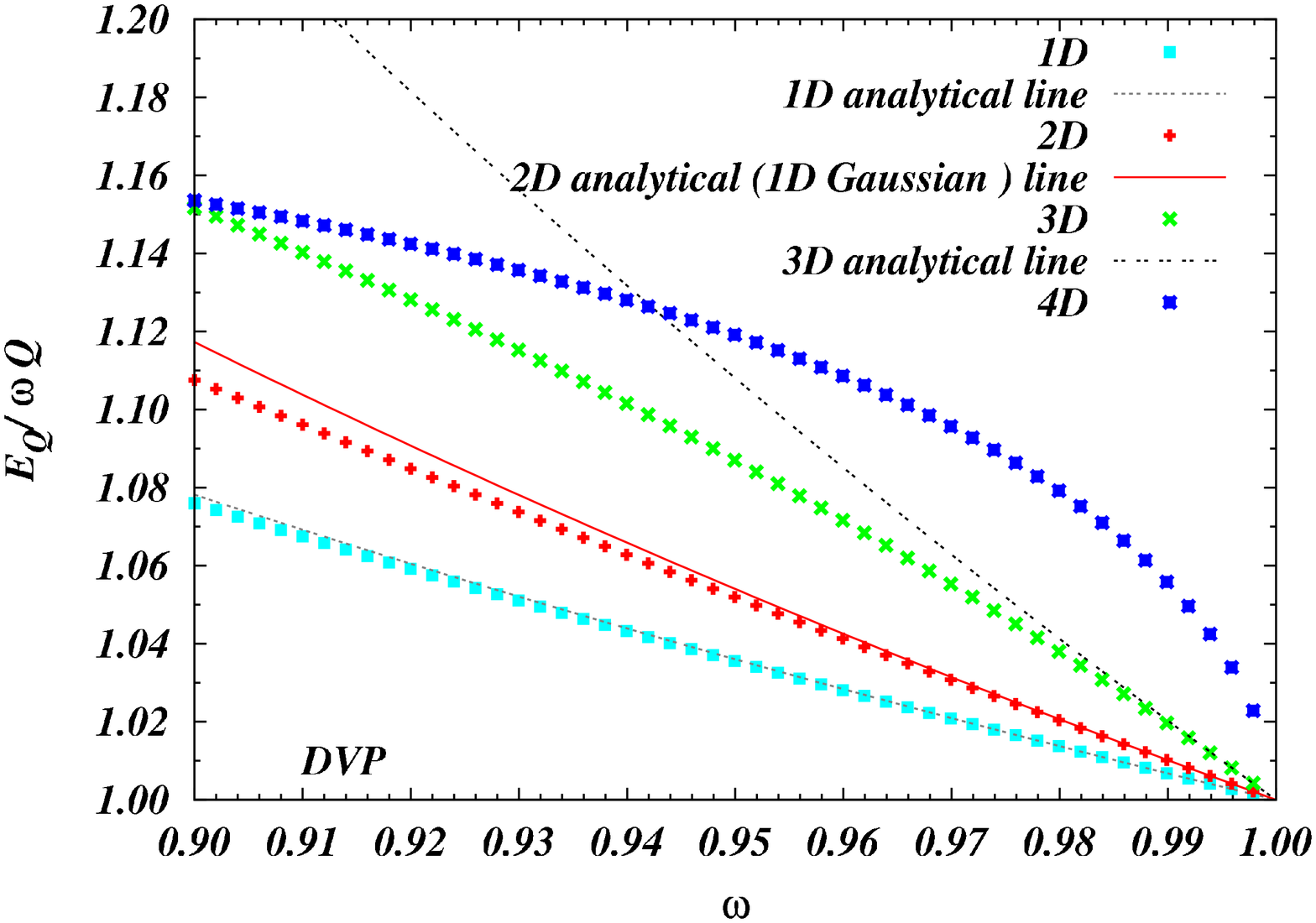}}
    \subfigure{\label{fig:nondegslopethck}\includegraphics[angle=-90,scale=0.27]{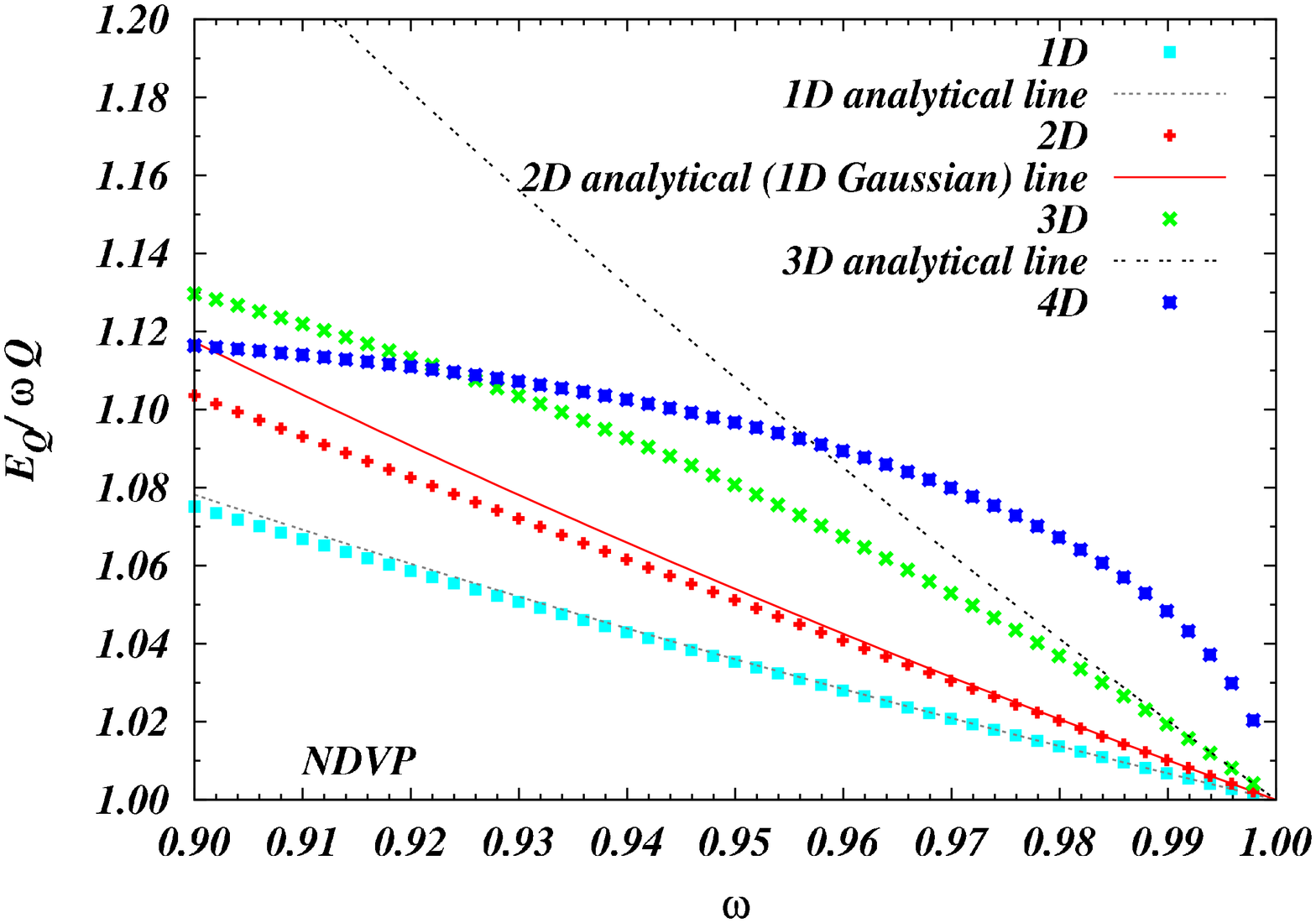}}\\
    \end{center}
  \caption{The ratio of $\mathcal{S/U}$ where $\mS$ and $\mU$ are surface and potential energies (top panels), the characteristic slope $\gamma(\om)\equiv E_Q/\om Q$ in the thin-wall-like limit, $\om\sim \om_-$, with the analytic lines \eq{surfthin} (middle panels),
and in the ``thick-wall-like'' limit, $\om \simeq \om_+$, (bottom panels), with the analytic lines \eqs{gausseq}{modslope}.}
  \label{fig:egy}
\end{figure}

The top panels in \fig{fig:egy} illustrate the ratios $\mS/\mU$ and the four bottom ones show the characteristic slopes of $E_Q/\om Q$ against $\om$ in both the thin-wall (middle-panels) and ``thick-wall'' (bottom-panels) limits. According to our analytic arguments \eq{UoverS}, we expect $\mS/\mU \simeq 1$ in the extreme limit $\om \simeq \om_-=0$ in DVP. Similarly, we expect
$\mS/\mU \sim 0$ in the same extreme thin-wall  limit $\om=\om_-=0.5$ for NDVP. The latter case corresponds to the existence of $Q$-matter with the simple step-like ansatz \eq{equation}. Although we are unable to probe these precise regimes, we believe the slopes of the curves indicate they are heading in the right direction.
The characteristic slopes $\gamma(\om)=E_Q/\om Q$ in the thin-wall limit in the two middle panels lie nearby the analytical ones, \eqs{surfthin}{eqqdeg}, as long as $\sigma_0(\om)\simeq\sigma_+(\om)$ (see \fig{fig:sig}) except for the $2D$ cases because for $D\le 2$ the profiles are not well fitted by thin-wall predictions.  Similarly, the characteristic slopes $E_Q/\om Q$ in the ``thick-wall'' limit in the bottom two panels agree with our analytical predictions \eq{modslope} using the modified ansatz rather than with \eq{gausseq} using the simple Gaussian ansatz. We have confirmed that the analytic characteristic slopes with \eq{modslope} can not apply to higher dimensions $D \ge 4$ in the ``thick-wall'' limit, see \eq{validthck}. Around the ``thick-wall'' limit $\om\simeq \om_+$, the behaviours in both potentials are $\mS \ll \mU$ (see top panel), which implies $E_Q \simeq \om Q$ as predicted in \eqs{gausseq}{modslope}; hence we can verify that the solutions are continued to the free particle solutions, see \eq{freeengy}. Our physically motivated modified ans\"{a}tze in both the thin- and ``thick-wall'' limits, therefore, have clear advantages over the simple ans\"{a}tze in \eqs{equation}{gaussansatz}.
\paragraph*{\underline{\bf{{$Q$-ball stability}}}}

\begin{figure}[htp]
  \begin{center}
    \subfigure{\label{fig:dcls}\includegraphics[angle=-90,scale=0.27]{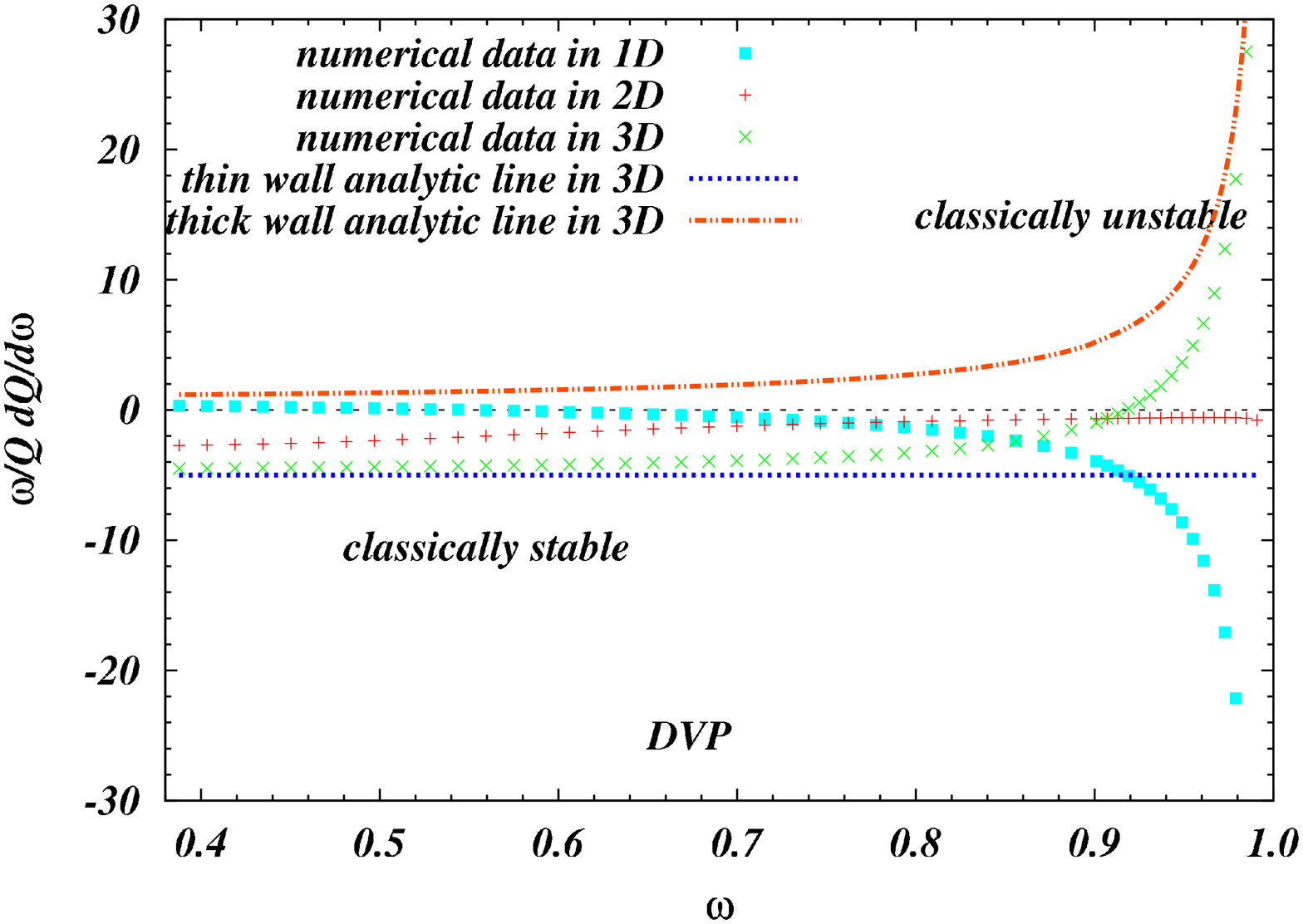}}
    \subfigure{\label{fig:ndcls}\includegraphics[angle=-90,scale=0.27]{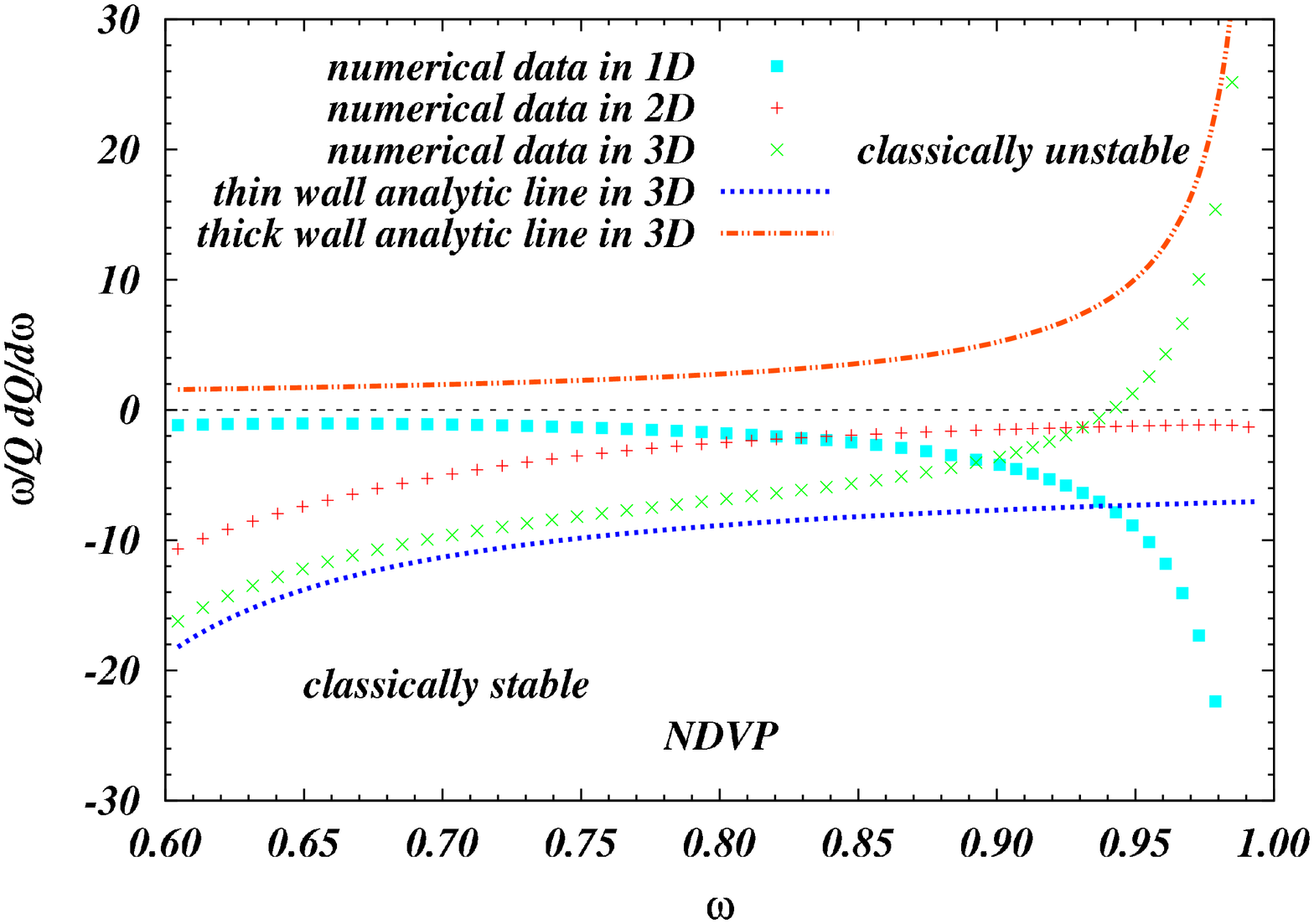}}\\
    \subfigure{\label{fig:degabs}\includegraphics[angle=-90,scale=0.27]{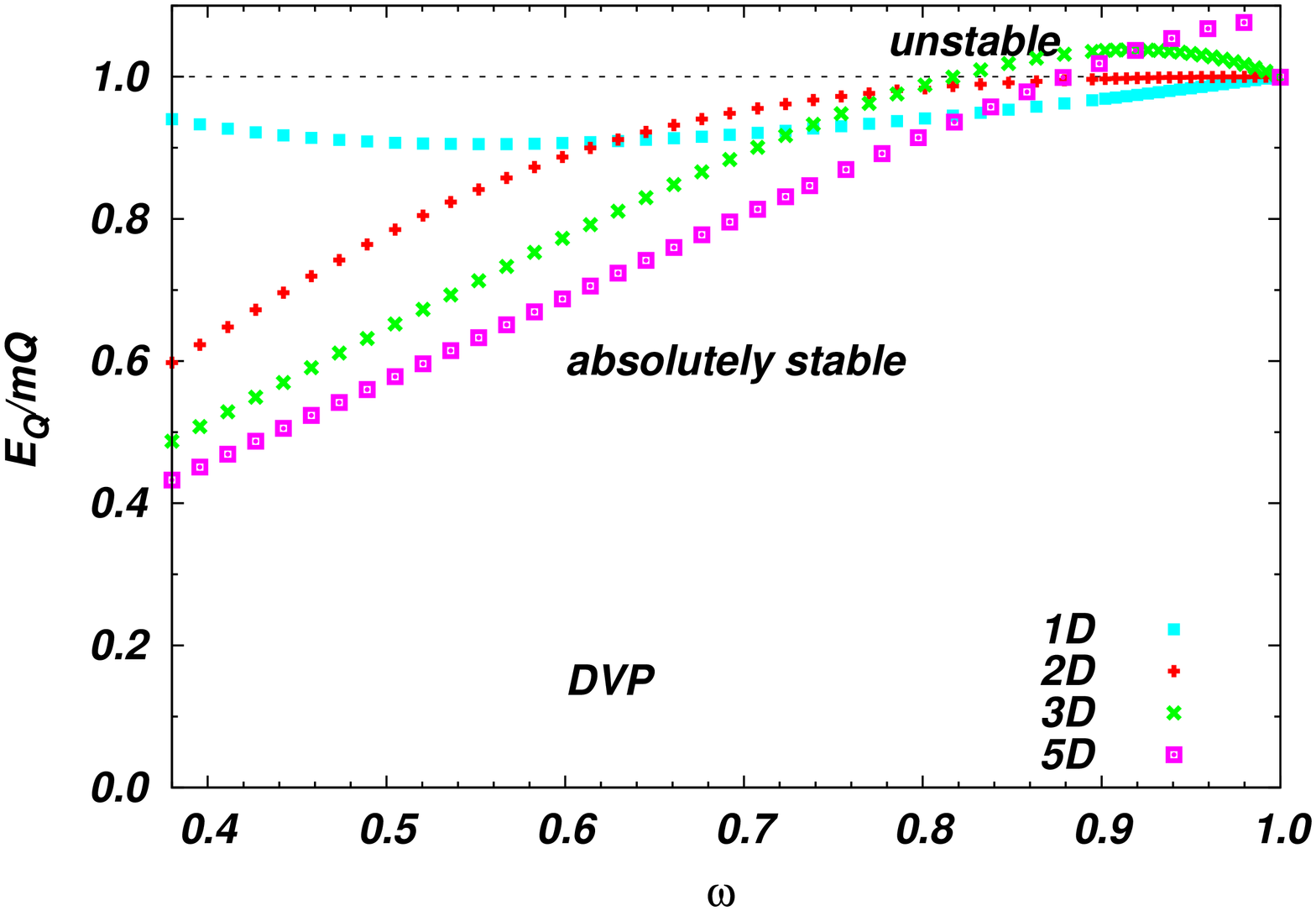}}
    \subfigure{\label{fig:nondegabs}\includegraphics[angle=-90,scale=0.27]{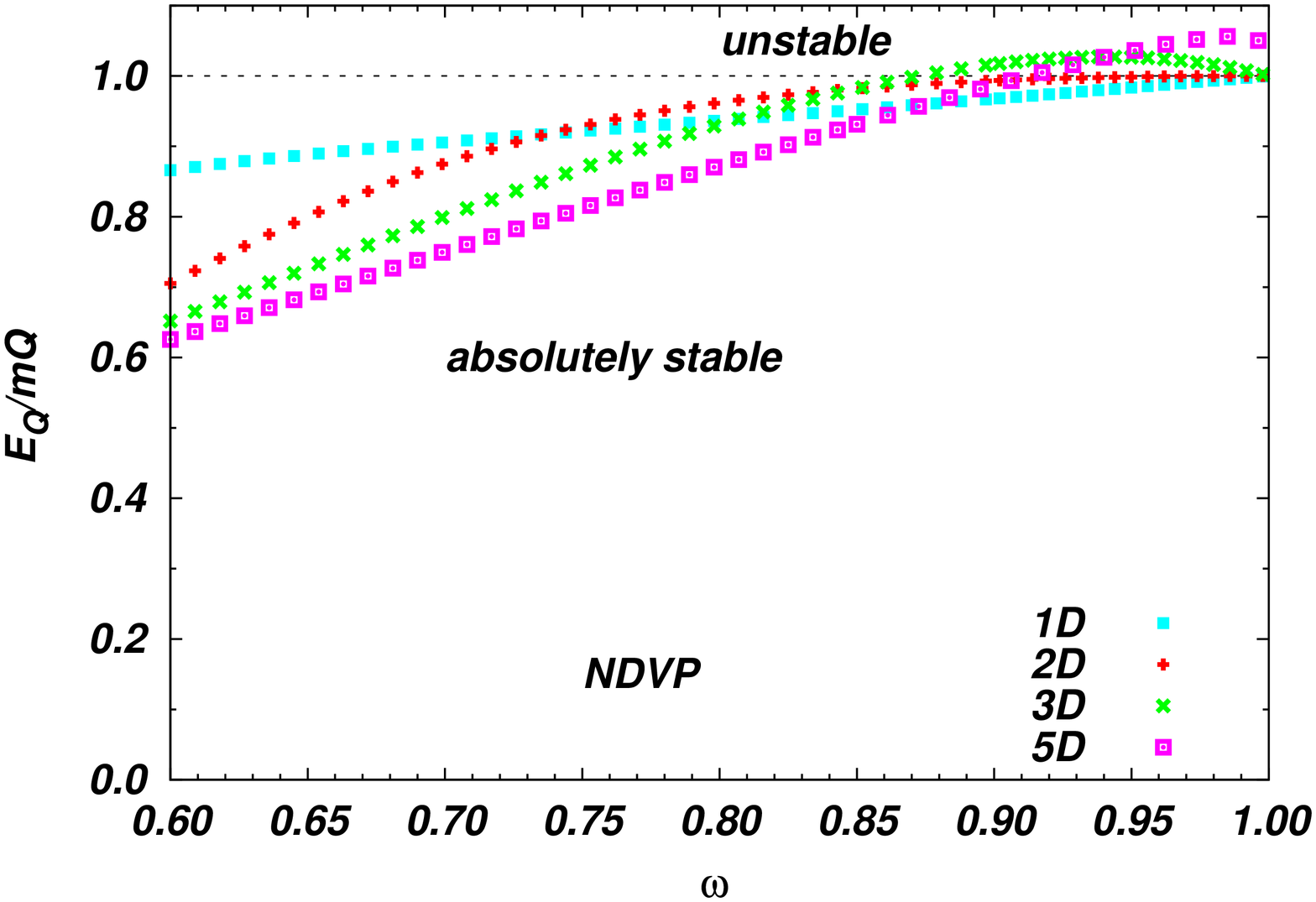}}
  \caption{Classical stability using \eq{CLS} for the top panels and absolute stability using \eq{ABSCOND} for the bottom panels. The $3D$ analytical lines of \eqs{q1class}{modcls} for classical stability agree with the corresponding numerical data. Above the zero-horizontal axes in the top panels, the $Q$-balls are classically unstable. Similarly, $Q$-balls above the horizontal axis, $E_Q=mQ$, are absolutely unstable. The one dimensional $Q$-balls are always classically stable. The $1D$ slopes $E_Q/m Q$ have different behaviours depending on DVP and NDVP unlike the other dimensional cases.}
  \label{fig:clsabs}
   \end{center}
\end{figure}

\fig{fig:clsabs} shows the classical and absolute stability lines for $Q$-balls. \tbl{tbl:eqwa} indicates the approximate analytical values of $\om_a$ derived by \eqs{viriwa}{ndvpwa}, which can be compared to the numerically obtained critical values $\om$ for the stabilities denoted by $\om_c,\; \om_s,\; \om_{ch},\; \om_a,$ and $\om_f$ in \tbl{tbl:dwmc}. These are defined by  $\left.\frac{dQ}{d\om}\right|_{\om_c}=\left.\frac{d^2S_\om}{d\om^2}\right|_{\om_s}=\left.\frac{d}{d\om}\bset{\frac{E_Q}{Q}}\right|_{\om_{ch}}=0$, $E_Q/Q|_{\om_a}=m$,
 and $\left.\frac{d\om}{dQ}\right|_{\om_f}=0$ respectively. The $3D$ analytical plots of $\frac{\om}{Q} \bset{\frac{dQ}{d\om}}$
in the thin- and ``thick-wall'' limits, \eqs{q1class}{modcls}, can be seen to match the corresponding numerical data in the appropriate limits of $\om$. We have confirmed numerically that  for both DVP and NDVP cases $\om_c=\om_f \simeq \om_s\simeq \om_{ch}$, see \tbl{tbl:dwmc}. This can be easily understood from \eqs{CLS}{chslope} and \eq{SAF}.

Recall \eq{modcls} with $n=4$ leads to the classical stability condition $D\le 2$ for the ``thick-wall'' case. The top panels in \fig{fig:clsabs} demonstrate that
``thick-wall'' $Q$-balls in $D \ge 3$ are classically unstable.
In \tbl{tbl:dwmc}, one can check that the absolute stability condition is more severe than the classical one. We then categorise into three types of $Q$-ball \cite{Friedberg:1976me}: \emph{absolutely stable $Q$-balls} for $\om < \om_a$, \emph{meta-stable $Q$-balls} for $\om_a \leq \om \leq \om_c$, which are not quantum-mechanically stable but classically stable, and \emph{completely unstable $Q$-balls} for $\om_c < \om$.

Both analytical values $\om_a$ in DVP and NDVP in \tbl{tbl:eqwa} agree well with the numerical ones in \tbl{tbl:dwmc}. Generally speaking, the higher dimensional $Q$-balls are more stable classically as well as quantum mechanically. Moreover, thin-wall $Q$-balls are always classically stable as demonstrated in \eq{q1class}, but the classical stability of ``thick-wall'' $Q$-balls is model- and $D$- dependent as in \eq{modcls}. The one- and two- dimensional $Q$-balls have a much richer structure than the thin- and ``thick-wall'' $Q$-balls. It is a challenging task to understand their intermediate profiles \cite{Sakai:2007ft}.

\begin{figure}[ht]
  \def\@captype{table}
    \begin{minipage}[t]{\textwidth}
    \begin{center}
      \begin{tabular}{|c||c|c|c|c|}
	\hline
		& \multicolumn{4}{|c|}{$\om_a$} \\
        \hline
$D$ &  $\mathcal{S} \gg \mathcal{U}$ & $\mathcal{S} \simeq \mathcal{U}$  or DVP &  NDVP & $\mathcal{S} \ll \mathcal{U}$ \\
        \hline
$3$ & 0.50 & 0.80 & 0.86 & 1 \\
$4$ & 0.67 & 0.86 & 0.90 & 1 \\
$5$ & 0.75 & 0.89 & 0.92 & 1 \\
        \hline
      \end{tabular}
    \end{center}
    \tblcaption{Virial relations: $\om_a$ in terms of space dimension
$D$ and ratio $\mathcal{S}/\mathcal{U}$, see \eq{viriwa}}	
  \label{tbl:eqwa}
  \end{minipage}
   \hfill
	\\
	\\
  \begin{minipage}[t]{\textwidth}
    \begin{center}
      \begin{tabular}{|c||c|c|c|c|c||c|c|c|c|c|}
	\hline
		&  \multicolumn{5}{|c||}{DVP} & \multicolumn{5}{|c|}{NDVP} \\
	\hline
	$D$ & $\omega_a$ & $\omega_c$ & $\omega_s$ & $\om_{ch}$ & $\om_f$ & $\omega_a$ & $\omega_c$ & $\omega_s$ & $\om_{ch}$  & $\om_f$ \\
    	\hline
3 & 0.82 & 0.92 & 0.92 & 0.92 & 0.92 & 0.87  & 0.94  & 0.94 & 0.94 & 0.94\\
4 & 0.86 & 0.96 & 0.96 & 0.96 & 0.96 & 0.89  & 0.97  & 0.97 & 0.97 & 0.97 \\
5 & 0.882 & 0.983 & 0.993 & 0.983 & 0.983 & 0.910 & 0.985 & 0.996 & 0.991 & 0.985 \\
\hline
      \end{tabular}
    \end{center}
    \tblcaption{The critical values for classical stability, absolute stability and stability against fission in DVP and NDVP using \eqss{ABSCOND}{CLS}{chslope} and \eq{SAF}. The critical values are defined by $\left.\frac{dQ}{d\om}\right|_{\om_c}=\left.\frac{d^2S_\om}{d\om^2}\right|_{\om_s}=\left.\frac{d}{d\om}\bset{\frac{E_Q}{Q}}\right|_{\om_{ch}}=0$, $E_Q/Q|_{\om_a}=m$,
 and $\left.\frac{d\om}{dQ}\right|_{\om_f}=0$. The numerical values of $\om_a$ coincide with the analytic ones in \tbl{tbl:eqwa}. We have confirmed numerically that  $\om_c=\om_f \simeq \om_s\simeq \om_{ch}$.}
    \label{tbl:dwmc}
  \end{minipage}
\end{figure}

\paragraph*{\underline{\bf{{Legendre relations}}}}

\begin{figure}[!ht]
    \begin{center}
    \subfigure{\label{fig:dlgrd1}\includegraphics[angle=-90,scale=0.27]{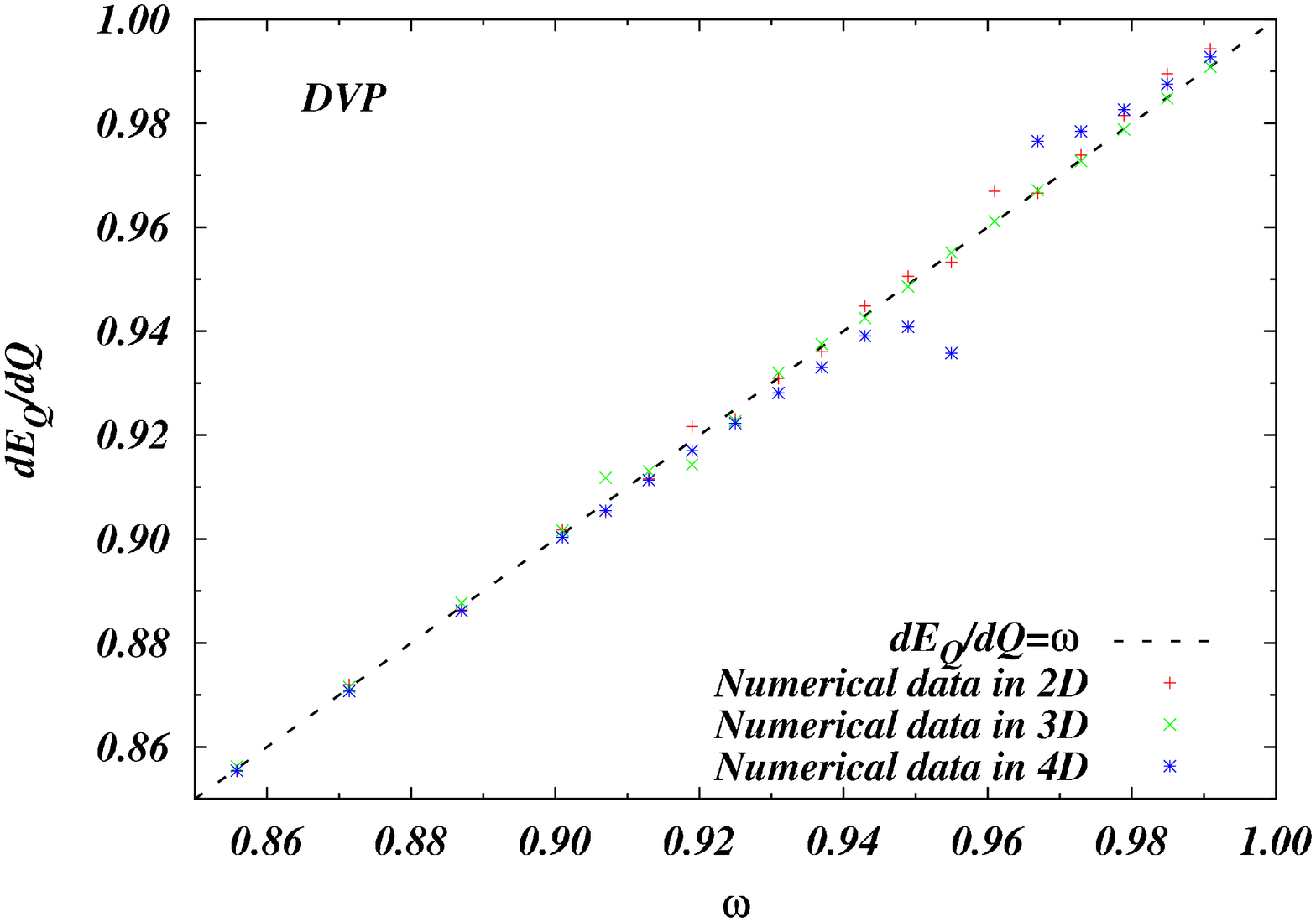}}
    \subfigure{\label{fig:ndlgrd1}\includegraphics[angle=-90,scale=0.27]{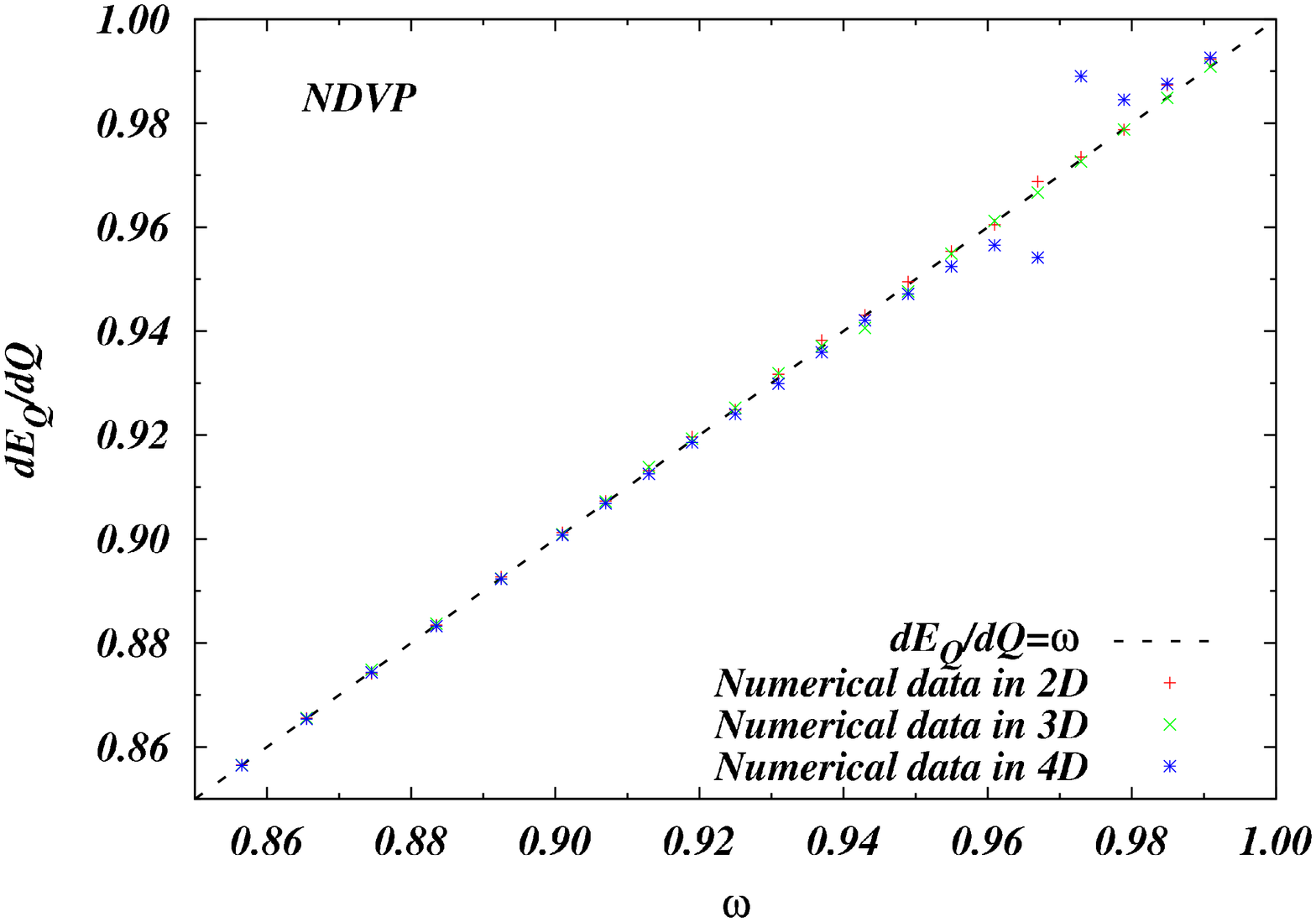}}\\
    \subfigure{\label{fig:dlgrd2}\includegraphics[angle=-90,scale=0.27]{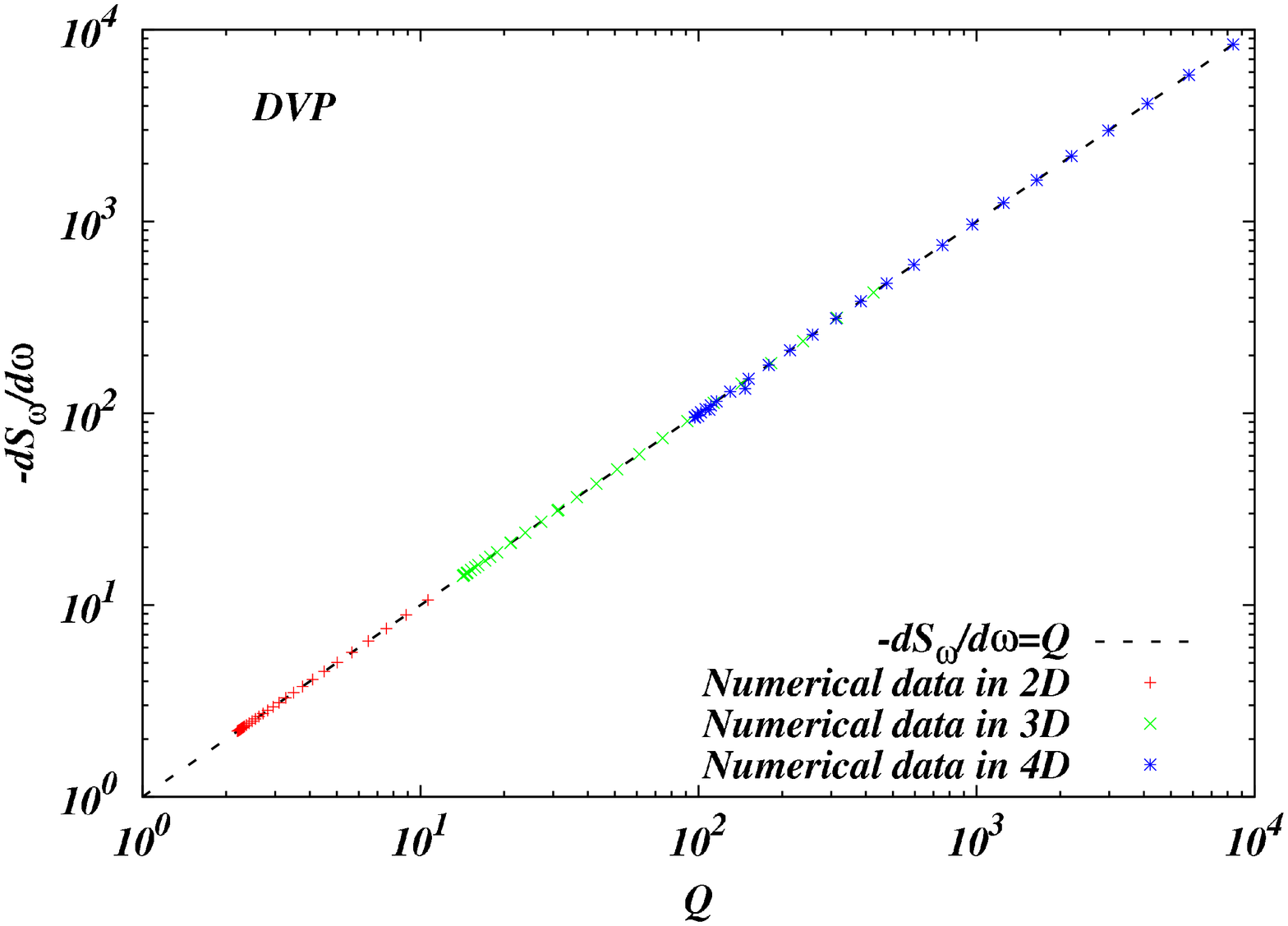}}
    \subfigure{\label{fig:ndlgrd2}\includegraphics[angle=-90,scale=0.27]{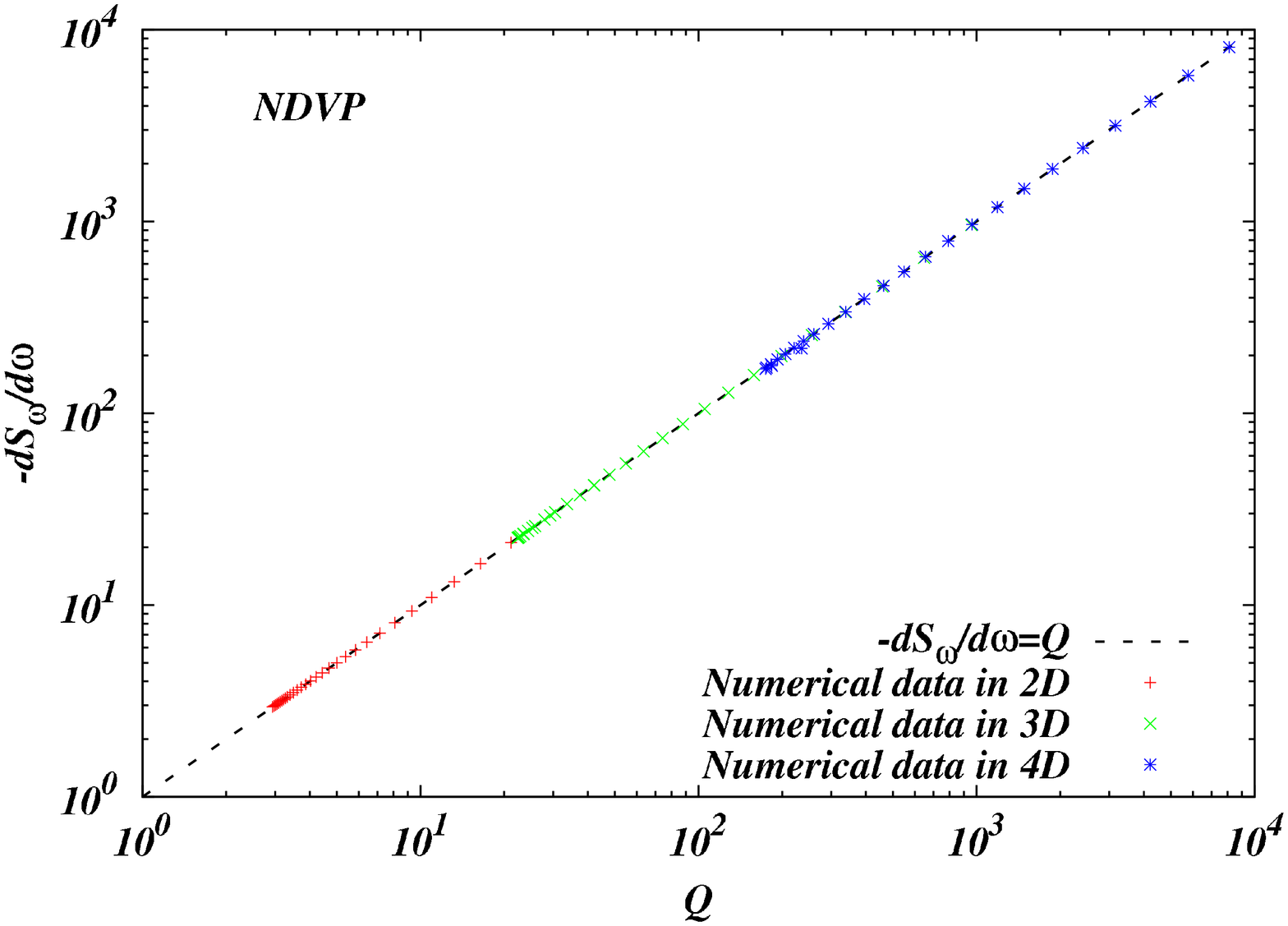}}\\
    \subfigure{\label{fig:dlgrd3}\includegraphics[angle=-90,scale=0.27]{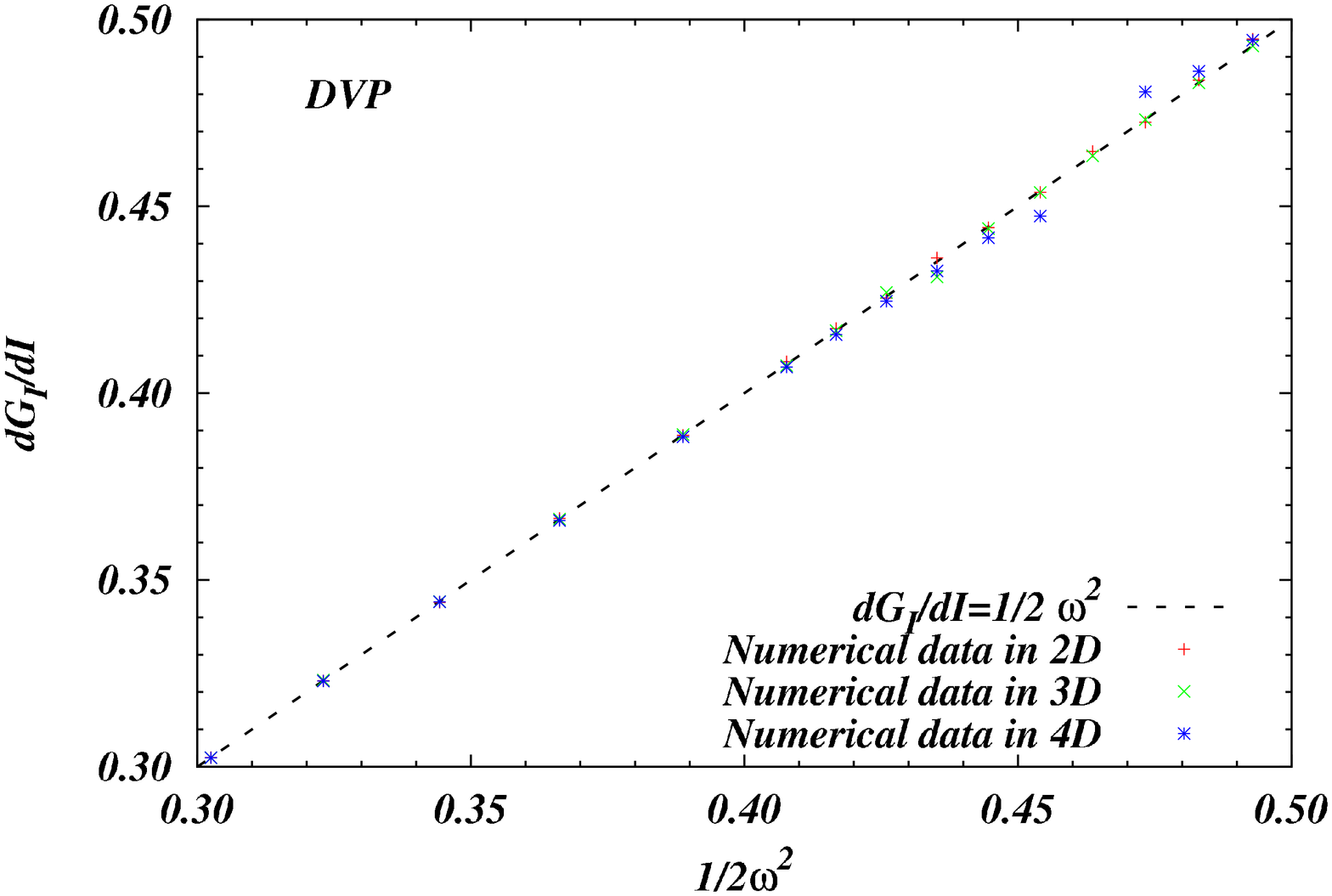}}
    \subfigure{\label{fig:ndlgrd3}\includegraphics[angle=-90,scale=0.27]{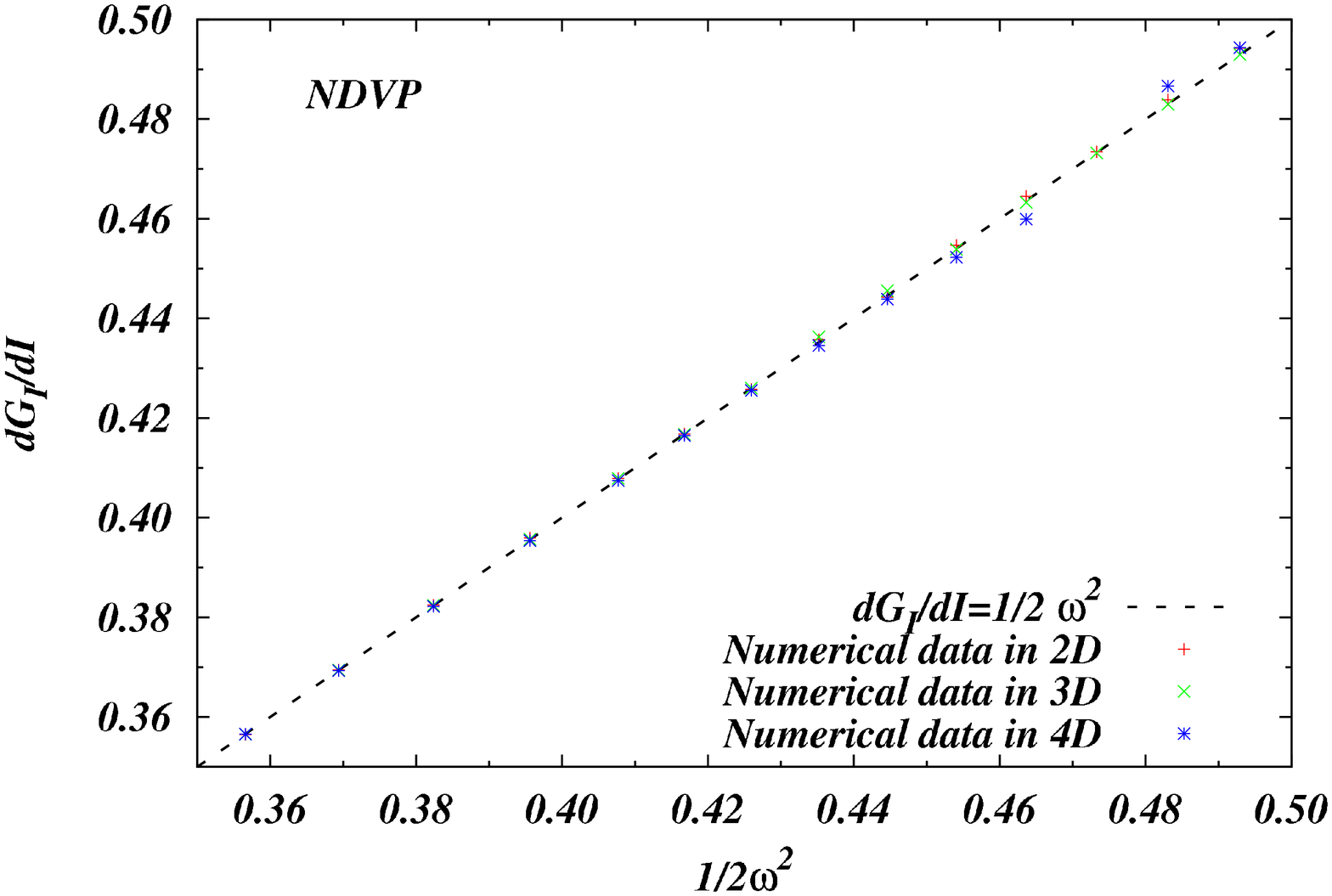}}
  \end{center}
  \caption{The Legendre relations of \eq{legendre} for both the DVP case (left panels) and the NDVP case (right panels): $\frac{dE_Q}{dQ}=\om$ (top), $-\frac{d S_\omega}{d\omega}=Q$ (middle), and $\frac{dG_I}{dI}=\half \omega^2$ (bottom). Note the excellent agreement between the analytical dotted lines and the numerical dots.}
  \label{fig:dlgrd}
\end{figure}

\fig{fig:dlgrd} shows the Legendre relations:$\frac{dE_Q}{dQ}$ v. $\om$, $-\frac{d S_\omega}{d\omega}$ v. $Q$, and $\frac{dG_I}{dI}$ v. $\half \om^2$ which can be used to check \eq{legendre}. We have also checked the validity of the Legendre transformations in \eqm{legtrns}{legtrns2}. Since the numerical results match our analytical ones, these results strengthen the validity of our analytic arguments.

\section{Conclusion}\label{ch3:conc}
We have numerically and analytically explored the stationary properties of a single $Q$-ball for an arbitrary spatial dimension $D$ in a class of polynomial potentials. By linearising the $Q$-ball \eq{QBeq} or rescaling $\So$, we have been able to consider the two limiting cases called the thin- and ``thick-wall'' $Q$-balls. The step-like ansatz of \eq{equation} can describe thin-wall $Q$-balls in the extreme limit $\om=\om_-$, whereas the modified ansatz \eq{thindanstz} is applicable to $\sigma_0(\om)\simeq \sigma_+(\om)$ which leads to wider range of parameter space $\om$ and of course  includes the previous limit. On the other hand, the limit $\om\simeq \om_+$ is used to describe ``thick-wall'' $Q$-balls in both the Gaussian ansatz \eq{gaussansatz} and our modified ansatz for the ``thick-wall'' case.

The thin-wall approximation is valid for $D \ge 2$. Since the step-like ansatz in the thin-wall approximation does not have surface effects, the characteristic slope is simply $\gamma=1$, \eq{QEGYTHIN}. With the modified ansatz including
surface effects, the classical stability for thin
wall $Q$-balls does not depend on $D$ in \eq{q1class}, but the absolute stability condition \eq{ndvpwa} does. Throughout the analysis, we have assumed \eqs{coreQ}{app2}, and imposed \eq{const} explicitly, which differs from the analysis in
\cite{Paccetti:2001uh}. Without these approximations, our calculations, in particular \eqs{shellsw}{shllsw2} and \eq{swall}, become inconsistent; similarly, the last assumption \eq{const} ensures that the shell thickness of the thin-wall $Q$-ball is real. The mechanical analogies and the numerical
results naturally explain and validate our underlying assumptions: the core sizes of the $Q$-balls are
much smaller than their corresponding thickness as seen in the middle two panels of \fig{fig:sig}, and the
surface tension depends weakly on $\om$ as seen in \tbl{tbl:ndrq}. With these
assumptions, thin-wall $Q$-balls for $\om < \om_a$ are absolutely stable. Moreover, the characteristic slopes coincide with those derived using the virial theorem. This follows from our analysis of the relative contributions between the potential and surface energies. The slopes have two types in either non-degenerate vacua potentials (NDVPs) or degenerate vacua potentials (DVPs): thin-wall $Q$-balls in NDVPs have a large energy from the charge; hence, the surface energy is less effective than the potential energy. They support the existence of $Q$-matter in the extreme limit, $\om=\om_-$. ``Thick-wall'' $Q$-balls in DVPs, however, have negligible energy from the charge compared to surface and potential energies; thus, the surface energy is well virialised with the potential energy. As seen in the left-bottom panel of \fig{fig:conf}, the configurations of energy density have peaks within the shells, which contribute to the surface energy. It would be worthwhile understanding these peaks in terms of our modified ansatz. Even in the extreme thin-wall limit, the charge and energy of the $Q$-balls in NDVPs are not proportional to the volume, \ie\ no $Q$-matter.

``Thick-wall'' $Q$-ball solutions naturally tend to the free charged and massive particle solutions \eq{freeengy}. With the simple Gaussian ansatz we have extremised $\So$ with respect to $\sigma_0$ and $R$ with fixed $\om$, while the approaches in \cite{Gleiser:2005iq} are that $E_Q$ is extremised with respect to only $R$. By extremising with respect to two degrees of freedom we are able to  recover the expected results of \eqs{apprxsigma0}{gausseq} unlike in \cite{Gleiser:2005iq}. The Gaussian ansatz, however, is valid only for $D=1$ because of \eq{gausscore}, and gives contradictory  results for the condition for classical stability. In order to remove these drawbacks in the Gaussian ansatz, we introduced another modified ansatz and used the Legendre relations to simplify the computations of $\So,\; Q$ and $E_Q$. We obtained a consistent classical stability condition \eq{modcls} which depends on $D$ and the non-linear power $n$ of the polynomial potential \eq{thckpot}. Not surprisingly, our numerical results suggest that the modified ansatz is much better than the Gaussian ansatz in the bottom two panels of \fig{fig:egy}. With the same panels, the validity condition \eq{validthck} in the modified ansatz has also been confirmed numerically.

In \eqs{viriwa}{ndvpwa} and \tbl{tbl:polres}, the analytical and numerical results found the critical value $\om_a$ subject to the assumption that the higher dimensional $Q$-balls could be treated with the thin-wall approximation over a wide range of values of $\om$. In summary, the higher dimensional $Q$-balls can be simplified into the thin- and ``thick-wall'' cases, while it is more challenging and interesting to understand stationary properties of one- and two-dimensional $Q$-balls. For example, those $Q$-balls embedded in $3D$ space (called $Q$-strings and $Q$-walls \cite{MacKenzie:2001av}) or extended $Q$-balls (nontopological strings \cite{Copeland:1988ra} and $Q$-balls with spatial spins and/or twists \cite{Radu:2008pp}) may exist in the formation of three dimensional $Q$-balls \cite{mit-web-page}.

The properties of non-thermal $Q$-balls we discussed in this chapter can lead to different consequences compared to thermal ones, \ie\ in the evolution of the Universe. The thermal effects on $Q$-balls induce subsequent radiation and evaporation. The Affleck-Dine mechanism provides a natural homogeneous condensate during an inflationary era, these fluctuations are then amplified to nonlinear objects, namely $Q$-balls if the pressure of the AD condensate is negative. The formation, dynamics, and thermalisation might have phenomenological consequences in our present universe, \eg\ gravitational waves \cite{GarciaBellido:2007af} and baryon-to-photon ratio.

\begin{figure}[!ht]
  \def\@captype{table}
  \begin{minipage}[t]{\textwidth}
   \begin{center}
      \begin{tabular}{|c||c|c|c|}
	\hline
	Model & \multicolumn{3}{|c|}{Polynomial potentials} \\
    	\hline
	$Q$-ball type & \multicolumn{2}{|c|}{Thin-wall} & ``Thick-wall''  \\ \hline
	Assumptions & \multicolumn{2}{|c|}{$R_Q \gg \delta, 1/\mu;\; \sigma_0 \simeq \sigma_+$ and $\tau$ does not depend on $\omega$} & None\\
	\hline
	Potential type & DVPs & NDVPs & Both\\
    	\hline
	& & &  \\
	$1/\gamma$ & $\frac{2D-1}{2(D-1)}$ & 1 & 1  \\
	Absolute stability & $\bigcirc$ & $\bigcirc$ & $\bigtriangleup$\\
	Classical stability & $\bigcirc$ & $\bigcirc$ & $\bigtriangleup$\\
\hline
      \end{tabular}
	\end{center}
	\tblcaption{Key analytical results for the case of polynomial potentials. Recall that the $\omega$-independent characteristic slope $\gamma \equiv E_Q/\omega Q$ leads to the proportionality relation $E_Q\propto Q^{1/\gamma}$. The symbols, $\bigcirc$ and $\bigtriangleup$, indicate that $Q$-balls are stable or can be stable subject to certain conditions, respectively. Recall that we may need the condition $\bar{\sigma(R_Q)}<\sigma_-$ in our thin-wall analysis; the readers should also note that our ``thick-wall'' analysis is valid as long as it satisfies \eq{validthck}. The $Q$-balls in the ``thick-wall'' limit are absolutely and classically stable subject to the condition \eq{modcls}.}
    \label{tbl:polres}
  \end{minipage}
\end{figure}



\chapter[$Q$-balls in MSSM flat potentials]{$Q$-balls in MSSM flat potentials}\label{ch:qbflt}

\section{Introduction}
$Q$-balls have recently attracted much attentions in cosmology \cite{Dvali:1997qv} and astrophysics \cite{Cecchini:2008su, Takenaga:2006nr, Kusenko:1997vp, Kusenko:2009iz, Shoemaker:2009ru}. A $Q$-ball is a nontopological soliton, and a number of scalar field theory models have been proposed to support the existence of nontopological solitons.

From a phenomenological point of view, the most interesting examples are probably the supersymmetric $Q$-balls arising within the framework of the Minimal Supersymmetric Standard Model (MSSM), which naturally contains a number of gauge invariant flat directions. Many of the flat directions can carry baryon (B) or/and lepton (L) number which is/are essential for Affleck-Dine (AD) baryogenesis \cite{Affleck:1984fy}. Following the AD mechanism, a complex scalar (AD) field acquires a large field value during a period of cosmic inflation and tends to form a homogeneous condensate, the AD condensate. In the presence of a negative pressure \cite{Enqvist:1997si, Lee:1994qb}, the condensate is unstable against spatial fluctuations so that it develops into nonlinear inhomogeneous lumps, namely $Q$-balls. The stationary properties and cosmological consequences of the $Q$-balls depend on how the Supersymmetry (SUSY) is broken in the hidden sector, transmitting to the observable sector through so-called messengers. In the gravity-mediated \cite{de Gouvea:1997tn} or gauge-mediated scenarios \cite{Dvali:1997qv}, the messengers correspond respectively either to supergravity fields or to some heavy particles charged under the gauge group of the standard model.

$Q$-balls can exist in scalar field potentials where SUSY is broken through effects in the supergravity hidden sector \cite{Nilles:1983ge}. This type of $Q$-balls can be unstable to decay into baryons and the lightest supersymmetric particle dark matter, such as neutralinos \cite{Fujii:2001xp}, gravitinos \cite{Ellis:1984eq, Seto:2005pj, Allahverdi:2005rh} and axinos \cite{Seto:2007ym}.  Recently, McDonald \cite{McDonald:2009cc} has argued that enhanced $Q$-ball decay in AD baryogenesis models can explain the 
observed positron and electron excesses detected by PAMELA \cite{Adriani:2008zr, Adriani:2008zq}, ATIC \cite{:2008zzr} and PPB-BETS \cite{Torii:2008xu}. By imposing an upper bound on the reheating temperature of the Universe after inflation, this mode of decay through $Q$-balls has been used to explain why the observed baryonic ($\Omega_{b}$) and dark matter ($\Omega_{DM}$) energy densities are so similar \cite{Kusenko:1997si, Enqvist:1998en}, i.e. $\Omega_{DM}/\Omega_{b}=5.65\pm0.58$ in \eq{dm/b} \cite{Spergel:2006hy}.

Scalar field potentials arising through gauge-mediated SUSY breaking \cite{de Gouvea:1997tn} tend to be extremely flat. Using one of the MSSM flat directions, namely the $QdL$  direction (where $Q$ and $d$ correspond to squark fields and $L$ to a slepton field), which has a nonzero value of $B-L$ and therefore does not spoil AD baryogenesis via the sphaleron processes that violate $B+L$ \cite{Enqvist:1998en}, Shoemaker and Kusenko recently explored the minimum energy configuration for baryo-leptonic $Q$-balls, whose scalar field consists of both squarks and sleptons \cite{Shoemaker:2008gs}. It had been assumed to that point that the lowest energy state of the scalar field corresponds to being exactly the flat direction; however in \cite{Shoemaker:2008gs, Shoemaker:2009jy}, the authors showed that the lowest energy state lies slightly away from the flat directions, and that the relic $Q$-balls, which are stable against decay into both protons/neutrons (baryons) and neutrinos/electrons (leptons) \cite{Cohen:1986ct}, may end up contributing to the energy density of dark matter \cite{Laine:1998rg, Kusenko:1997si, Enqvist:2001jd}; thus, the $Q$-balls can provide the baryon-to-photon ratio \cite{Laine:1998rg}, i.e. $n_b/n_\gamma \simeq (4.7-6.5) \times 10^{-10}$ in \eq{basym} \cite{Fields:2008zz}  where $n_b$ and $n_\gamma$ are, respectively, the baryon and photon number densities in the Universe.

In this chapter we examine analytically and numerically the classical and absolute stability of $Q$-balls using flat potentials in the two specific models mentioned above. In order to study the possible existence of lower-dimensional $Q$-balls embedded in 3+1 dimensions, we will work in arbitrary spatial dimensions $D$; although of course the $D=3$ case is of more phenomenological interest. Previous work \cite{Multamaki:1999an, Enqvist:1997si, Enqvist:1998en} on the gravity-mediated potential has used either a steplike or Gaussian ansatz to study the analytical properties of the thin and thick-wall $Q$-balls. Introducing more physically motivated ans\"{a}tze, we will show that the thin-wall $Q$-balls can be quantum mechanically stable against decay into their own free particle quanta, that both thin and thick-wall $Q$-ball solutions obtained are classically stable against linear fluctuations, and confirm that a Gaussian ansatz is a physically reasonable one for the thick-wall $Q$-ball. The one-dimensional $Q$-balls in the thin-wall limit are excluded from our analytical framework. The literature on $Q$-balls with gauge-mediated potentials has tended to use a test profile in approximately flat potentials. We will present an exact profile for a generalised gauge-mediated flat potential, and show that we naturally recover results previously published in \cite{Laine:1998rg, de Gouvea:1997tn, Enqvist:1998en}.

The rest of this chapter is organised as follows. Section \ref{gravity-mediated} provides a detailed analyses for gravity-mediated potentials, and in Sec. \ref{gauge-mediated} we investigate the case of a generalised gauge-mediated potential. We confirm the validity of our analytical approximations with complete numerical $Q$-ball solutions in Sec. \ref{numerics} before summarising in Sec. \ref{conc}. In Appendix \ref{exactsol}, we obtain an exact solution for the case of a logarithmic potential, and in Appendix \ref{appxthick}, we confirm that the adoption of a Gaussian ansatz is appropriate for the thick-wall $Q$-ball found in the gravity-mediated potentials. This chapter is published in \cite{Copeland:2009as}.

\section{Gravity-mediated potentials}\label{gravity-mediated}

The MSSM consists of a number of flat directions where SUSY is not broken. Those flat directions are, however, lifted by gauge, gravity, and/or nonrenormalisable interactions. In what follows the gravity interaction is included perturbatively via the one-loop corrections to the bare mass $m$ in \eq{Ugrav} and the nonrenormalisable interactions ($U_{NR}$), which are suppressed by high energy scales such as the grand unified theory scale $M_U\sim 10^{16}$ GeV or Planck scale $m_{pl} \sim 10^{18}$ GeV.  Here, $m$ is of order the SUSY breaking scale which could be the gravitino mass $\sim m_{3/2}$, evaluated at the renormalisation scale $M$ \cite{Nilles:1983ge}. We note that, following the majority of work in this field, we will ignore A-term contributions ( U(1) violation terms), thermal effects \cite{Allahverdi:2000zd, Anisimov:2000wx} which come from the interactions between the AD field and the decay products of the inflaton, and the Hubble-induced terms \cite{Allahverdi:2002vy} which gives a negative mass-squared contribution during inflation. It is possible that their inclusion could well change the results of the following analysis.

The scalar potential we are considering at present is \cite{Enqvist:1997si, Nilles:1983ge}
\be\label{sugra}
U=U_{grav}+U_{NR}=\half m^2 \sigma^2\bset{1+K\ln \bset{\frac{\sigma^2}{M^2}}} + \frac{|\lambda|^2}{m^{n-4}_{pl}} \sigma^{n}
\ee
where we used \eq{Ugrav}, $K$ is a factor for the gaugino correction, which depends on the flat directions, and $M$ is the renormalisation scale. Also $\lambda$ is a dimensionless coupling constant, and $U_{NR}\equiv \frac{|\lambda|^2}{m^{n-4}_{pl}} \sigma^{n}$, where $n>2$. If the MSSM flat directions include a large mass top quark, $K$ can be positive and then $Q$-balls do not exist. For flat directions that do not have a large mass top quark component, we typical find $K \simeq -[0.01-0.1]$ \cite{Enqvist:1997si, Enqvist:2000gq}. The power $n$ of the nonrenormalisable term depends on the flat directions we are choosing along which we maintain R parity. As examples of the directions involving squarks, the $u^c d^c d^c$ direction has $n=10$, whilst the $u^c u^c d^c e^c$ direction requires $n=6$. A complete list of the MSSM flat directions can be found in Table 1 of \cite{Dine:1995kz}. Since the potential in \eq{sugra} for $K<0$ could satisfy the $Q$-ball existence condition in \eq{EXIST}, where $\omega_+ \gg m$, $Q$-balls naturally exist.

In the rest of this chapter, we will focus on potentials of the form of \eq{sugra} for general $D(\ge 1)$ and $\omega$ and $n (>2)$ so that $M$ and $m_{pl}$ have the same mass dimension, $(D-1)/2$, as $\sigma$. It means that the parameters $M$ and $m_{pl}$ are only physical for $D=3$. For several cases of $n$ and $D$, the term $U_{NR}$ can be renormalisable, but we will generally call it the nonrenormalisable term for the future convenience. The readers should note that the potential \eq{sugra} has been derived only with $\mathcal{N}=1$ supergravity in $D=3$; therefore, the potential form could well be changed in other dimensions. Furthermore, the logarithmic correction breaks down for small $\sigma$ and the curvature of \eq{sugra} at $\sigma=0$ is finite due to the gaugino mass, which affects our thick-wall analysis and their dynamics. However, we concentrate our analysis on this potential form for arbitrary $D,\; n$ and any values of $\sigma$ for two main reasons. The first is that it contains a number of general semiclassical features expected of all the potentials, and the second is that it offers the opportunity to consider the lower-dimensional $Q$-balls embedded in $D=3$.

In Appendix \ref{exactsol}, we obtain the exact solution of \eq{QBeq} with the potential $U=U_{grav}$; however, exact solutions of the general potential $U$ in \eq{sugra} are fully nonlinear and can be obtained only numerically. Therefore, we will analytically examine the approximate solutions in both the thin and thick-wall limits.  Before doing so, we shall begin by imposing a restriction on $\lambda$ in \eq{sugra} in order to obtain stable $Q$-matter in NDVPs. With the further restrictions on $\lambda$ and $|K|$, we can proceed with our analytical arguments, and we will finally obtain the asymptotic $Q$-ball profile for large $r$ which will be used in the numerical section, Sec. \ref{numerics}.
 
\subsection{The existence of absolutely stable $Q$-matter}\label{exstQmat}

As we have seen, the first restriction on the parameters in \eq{sugra} is $K<0$ to satisfy \eq{EXIST}. Further, we need to restrict the allowed values of the parameter $\lambda$ to ensure that we obtain absolutely stable $Q$-matter. Notice that $Q$-matter exists in NDVPs, whilst the extreme thin-wall $Q$-balls in DVPs will not be $Q$-matter as we showed in chapter \ref{ch:qpots}.

By using the definitions of $\omega_-$ and $\sigma_+$, namely,
$\omega^2_-\equiv \left.\frac{2U}{\sigma^2}\right|_{\sigma_+}$ and $\left.\frac{d U_{\omega_-}}{d\sigma}\right|_{\sigma_+}=0$, we shall find the range of values of $\lambda$ for which absolutely stable $Q$-matter solutions exist. Moreover, we will obtain the curvature $\mu$, which is proportional to $\abs{K}$, of the effective potential $\Uo$ at $\sigma_+$.

\begin{figure}[!ht]
  \begin{center}
	\includegraphics[angle=-90, scale=0.28]{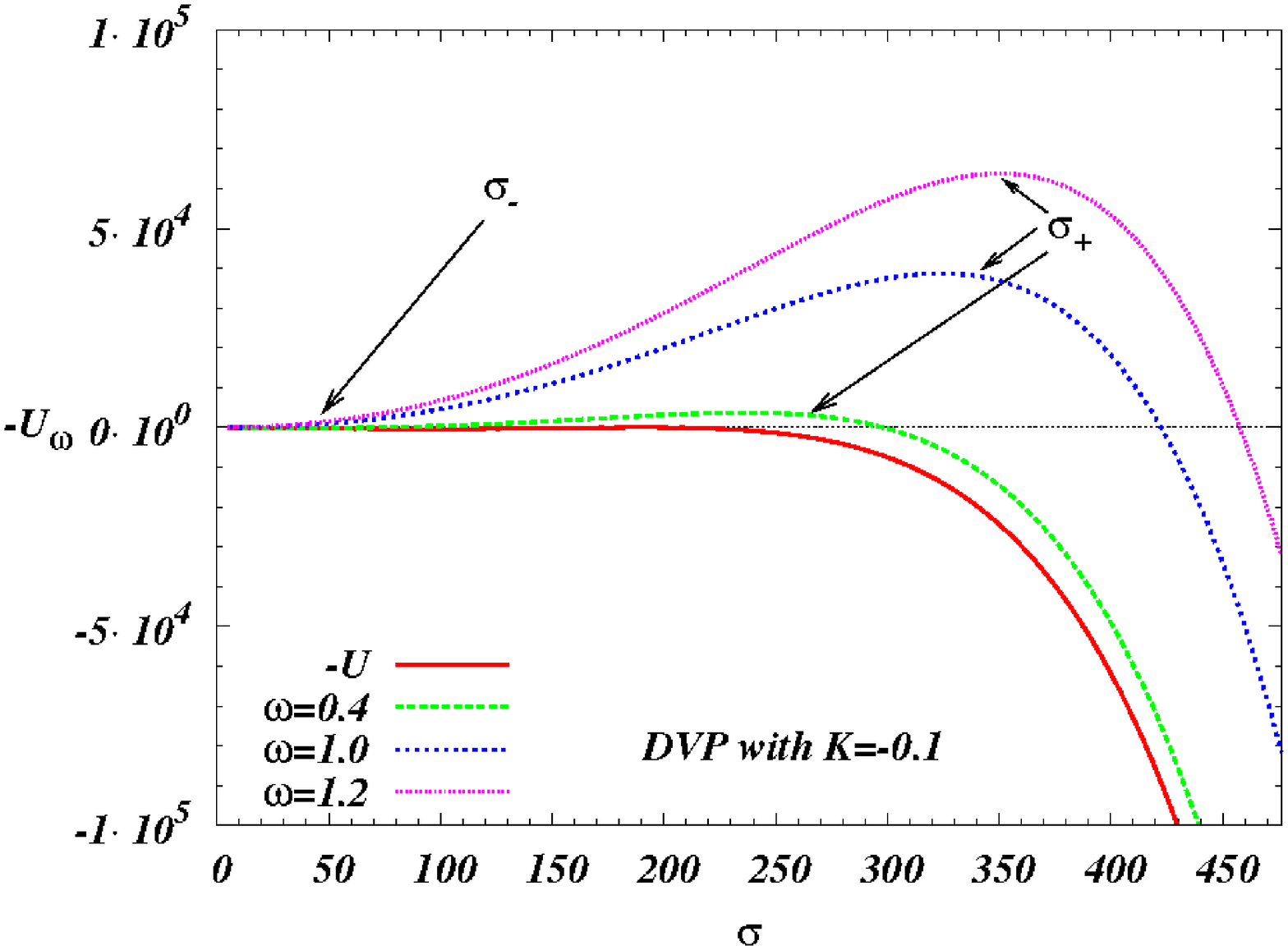}
	\includegraphics[angle=-90, scale=0.28]{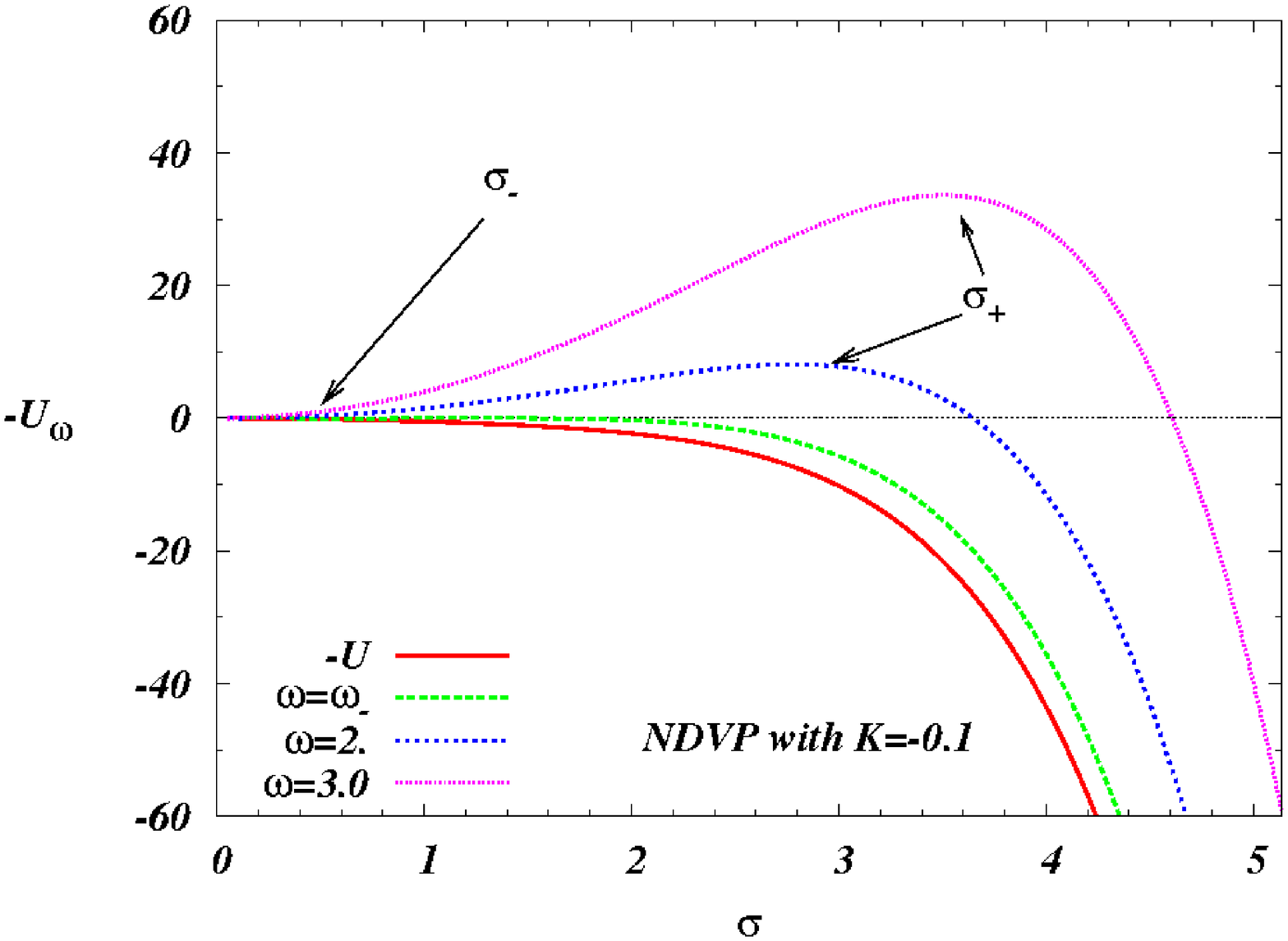}
  \end{center}
  \caption{Parameters $\sigma_{\pm}(\omega)$ for a potential of the form $U(\sigma)=\half \sigma^2\bset{1-|K|\ln\sigma^2} + \beta^2 \sigma^6$ (effective potential $\Uo=U-\half \omega^2\sigma^2$) with $|K|=0.1$. The left hand figure corresponds to the case of a DVP with $\beta^2=\frac{|K|e^{-1}}{4}\exp\bset{-\frac{2}{|K|}} \sim 1.90 \times 10^{-11}$, whilst the right hand side is the NDVP with $\beta^2=\frac{|K|e^{-1}}{4} \sim 9.20 \times 10^{-3}$, see \eq{rbeta}. The coloured lines in each plot correspond to different values of $\omega$. The variable
$\sigma_+(\omega)$ is defined as the maximum of the inverse effective potential $-\Uo$ where as  $\sigma_-(\omega)$ corresponds to $-\Uo(\sigma_-(\omega))=0$ for $\sigma_-(\omega)\neq 0$. Recalling $\omega_-=0$ in DVP, the DVP has degenerate vacua at $\sigma_+(0)=e^{1/4}\exp\bset{\frac{1}{2|K|}}\sim 1.91\times 10^2$ (red-solid line), whilst the NDVP does not. The inverse effective potential $-\Uo$ with $\omega_-=1$ in NDVP (green-dashed line), however, has degenerate vacua at $\sigma_+(\omega_-)=e^{1/4}\sim 1.28$, see the first relation in \eq{tisigom}. For the lower limit $\omega\simeq \omega_-$ (green-dashed lines), we could see $\sigma_+=e^{1/4}$, whilst the purple dotted-dashed lines show $\sigma_-(\omega) \to 0$ near the thick-wall like limit $\omega =3.0\sim \omega_+$ where $\omega_+ \gg 1$.}
  \label{fig:gravpot}
\end{figure}

The effective potential for \eq{sugra} can be rewritten in terms of new dimensionless variables $\tisig=\sigma/M,\; \tiom=\omega/m$, and 
\be\label{betani}
\beta^2=\frac{|\lambda|^2 M^{n-2}}{m^{n-4}_{pl} m^2}>0,
\ee
as
\be\label{repsugra}
 U_{\tilde{\omega}}=\half M^2 m^2 \tisig^2\bset{1-\tiom^2 - 2|K|\ln\tisig} + M^2 m^2 \beta^2 \tisig^n.
\ee
After some simple algebra and introducing $\tiom^2_-\equiv \frac{2U}{\tisig^2}|_{\tisig_+}$ and $\left.\frac{d U_{\tiom_-}}{d\tisig}\right|_{\tisig_+}=0$, we obtain 
\be\label{tisigom}
\tisig_+=\bset{\frac{|K|}{(n-2)\beta^2}}^{\frac{1}{n-2}},\hspace*{10pt} \tiom^2_-=\frac{1}{n-2}\sbset{n-2+2|K|-2|K|\ln\bset{\frac{|K|}{(n-2)\beta^2}}}.
\ee
Notice that $\tiom^2_-=0$ corresponds to DVPs where $Q$-matter solutions do not exist, whilst the extreme thin-wall $Q$-balls do exist and are absolutely stable against their own quanta as we will see. In NDVPs, $Q$-matter solutions exist and are absolutely stable when $0<\tiom^2_-< 1$, see \eq{ABSCOND}. Combining these facts and using the second relation in \eq{tisigom}, we have the constraint on $\lambda$ for stable $Q$-matter solutions to exist, namely
\bea{rbeta}
\frac{|K|e^{-1}}{n-2}\exp{\bset{-\frac{n-2}{2|K|}}}<&\beta^2&<\frac{|K|e^{-1}}{n-2},\\
\label{restbeta}\Leftrightarrow \hspace*{5pt}  \frac{|K|e^{-1}}{n-2} \frac{m^{n-4}_{pl}m^2}{M^{n-2}} \exp{\bset{-\frac{n-2}{2|K|}}} <&|\lambda|^2&<\frac{|K|e^{-1}}{n-2} \frac{m^{n-4}_{pl}m^2}{M^{n-2}},
\eea
where we have used \eq{betani} to go from \eq{rbeta} to \eq{restbeta}. Here, the lower limit of $|\lambda|^2$ corresponds to $\tiom^2_-=0$, whilst the upper limit corresponds to $\tiom^2_-=1$. The inequality in \eq{restbeta} implies that if the coupling constant $\lambda$ of the nonrenormalisable term in \eq{sugra} is too small, then it does not support the existence of $Q$-balls, whereas a large $\lambda$ coupling leads to unstable $Q$-matter. 
With the following parameter set, $m=M=1,\; |K|=0.1,\; n=6$ and the lower/upper limits of $\beta^2$ in \eq{rbeta}, \fig{fig:gravpot} shows the inverse potentials in \eq{repsugra} and their inverse effective potentials $-\Uo$ with various values of $\omega$. The lower limit, $\beta^2=\frac{|K|e^{-1}}{4}\exp{\bset{-\frac{2}{|K|}}}$, corresponds to DVPs case with $\omega_-=0$, whilst in the upper limit, $\beta^2=\frac{|K|e^{-1}}{4}$, the potentials do not have degenerate vacua with $\omega_-=1$, hence are called NDVPs. By substituting the values of $\beta^2$ into \eq{tisigom}, we obtain the values of $\sigma_+$ indicated in \fig{fig:gravpot}. Finally we can obtain the curvature, $\mu^2(\omega)\equiv\left.\frac{d^2\Uo}{d\sigma^2}\right|_{\sigma_+(\omega)}$, evaluated at $\omega_-$ , i.e.
\be\label{mucurv}
\mu^2 \equiv \mu^2(\omega_-) = m^2|K|(n-2)\propto |K|,
\ee
which implies that a small logarithmic correction $|K| \ll \order{1}$ in \eq{sugra} gives an ``extremely'' flat effective potential $\Uo$ compared to the quadratic term $m^2$ around $\sigma=\sigma_+$ for a given $n\sim \order{10^{0-1}}$.

\subsection[Thin-wall $Q$-balls]{Thin-wall $Q$-balls for $\sigma_0\simeq \sigma_+,\; R_Q \gg \delta, 1/\mu,\; D\ge 2$}\label{sect:grvthn}

For the extreme limit $\omega = \omega_-$, Coleman demonstrated that the steplike ansatz \cite{Coleman:1985ki} is applicable to the case of NDVPs because the surface effects of the thin-wall $Q$-ball in this limit are not significant. There are situations though where we would like to explore the region around $\omega = \omega_-$, corresponding to $\sigma_0 \simeq \sigma_+(\omega)$, and to do this we need to include surface effects. In chapter \ref{ch:qpots} we explained how to do this under the assumptions: $R_Q/\delta,\; \mu R_Q \gg 1,\; \sigma(R_Q)<\sigma_-(\omega),\; \sigma_+(\omega)\simeq \sigma_+(\omega_-)\equiv \sigma_+$, and that the surface tension $\tau\simeq \int^{\sigma_+}_0 d \sigma \sqrt{2 U_{\omega_-}}$ does not depend ``sensitively'' on $\omega$. Here, $R_Q,\; \delta$ are, respectively, the $Q$-ball core size and the shell thickness. We note that in \cite{Coleman:1977py}, Coleman assumed $\Uo\simeq U_{\omega_-}$ in the shell region, and this is equivalent to saying $\sigma(R_Q)<\sigma_-(\omega)$. In what follows we will be making use of Coleman's approach. By requiring this or $\sigma(R_Q)<\sigma_-(\omega)$, we can guarantee real values of shell thickness $\delta$ and surface tension $\tau$. The assumption, in which $\tau$ does not depend on $\omega$, is related to the assumptions: $\sigma_+(\omega)\simeq \sigma_+$ and $\Uo\simeq U_{\omega_-}$ in the shell region. 

Under these assumptions and for $D\ge 2$, we now apply the previous thin-wall analysis developed in chapter \ref{ch:qpots} to the present potential \eq{sugra}. The ansatz is given by \eq{thindanstz}
\be\label{thinpro}
\sigma(r) =
    \left\{
    \begin{array}{ll}
    \sigma_+ - s(r) &\; \textrm{for}\ 0 \le r < R_Q, \\
    \bar{\sigma}(r) &\; \textrm{for}\ R_Q \le r \le R_Q + \delta, \\
    0 &\; \textrm{for}\ R_Q + \delta < r,
    \end{array}
    \right.
\ee
where $R_Q,\; \delta$, the core profile $s(r)$, and the shell profile $\bar{\sigma}(r)$ will be obtained in terms of the underlying potential by extremising $\So$ with respect to $R_Q$. Each of the profile functions satisfies
\bea{}
	s^{\p\p}+\frac{D-1}{r}s^\p -\mu s &=&0,\\
	\bar{\sigma}^{\p\p}-\left.\frac{d\Uo}{d\sigma}\right|_{\bar{\sigma}}&=&0.
\eea
By recalling \eq{infqt}, we have previously found that in Eqs (\ref{RQ}-\ref{q1class}), \ie\
\bea{RQc}
R_Q&\simeq& \bset{D-1}\frac{\tau}{\eo};\; \; \So \simeq \frac{\tau}{D} \partial V_D >0;\; \; Q\simeq \omega \sigma^2_+ V_D,\\
\label{grvthnch} \frac{E_Q}{\omega Q}&\simeq& \frac{2D-1}{2(D-1)} -\frac{\omega^2_-}{2(D-1)\omega^2},\\
\label{grvthncls} \frac{\omega}{Q}\frac{dQ}{d\omega}&\simeq& 1-\frac{2D\omega^2 }{\omega^2-\omega^2_-} <0,
\eea
where we have taken the thin-wall limit $\omega\simeq \omega_-$ in the last inequality. Notice that our analytical work cannot apply for the $1D$ thin-wall $Q$-ball, see the first expression in \eq{RQc}.

\paragraph*{\underline{\bf NDVPs:}}

This type of potential supports the existence of $Q$-matter that corresponds to the regime $\mU\gg \mS$. The $Q$-matter can be absolutely as well as classically stable for the extreme limit $\omega\simeq \omega_-$, when the coupling constant $\lambda$ for the nonrenormalisable term in \eq{sugra} satisfies \eq{restbeta}. The characteristic slope is given by the first case of \eq{virieq}, and the charge and energy are linearly proportional to the volume $V_D$.

\paragraph*{\underline{\bf DVPs:}}

With the presence of degenerate minima in \eq{sugra}, in chapter \ref{ch:qpots} we obtained the ratio $\mU/\mS \sim 1$, which corresponds to the second case of \eq{virieq}. The charge and energy are not proportional to the volume $V_D$ itself in this case; hence, we cannot see the existence of $Q$-matter in the extreme limit $\omega=\omega_-=0$. Instead we can find the proportional relation simply from \eq{leg2} and \eq{grvthnch}, namely $E_Q\propto Q^{2(D-1)/(2D-1)}$.

\vspace*{15pt}

Our main approximations are based on the assumptions $\sigma_0\simeq \sigma_+, R_Q \gg \delta,\; 1/\mu,$ and $\Uo\simeq U_{\omega_-}$ in the shell region. In what follows we will see through numerical simulations that our analytic results agree well with the corresponding numerical results even in a ``flat'' potential choice $|K|=0.1,\; m=M=1,\; n=6$, which implies that $1/\mu \sim 1.58$, see \eq{mucurv}.

\subsection[``Thick-wall'' $Q$-balls]{Thick-wall $Q$-balls for $\beta^2 \lesssim |K|\lesssim \order{1}$}\label{thickgrav}

In chapter \ref{ch:qpots} we studied ``thick-wall'' $Q$-balls in general polynomial potentials, and extracted out the explicit $\omega$ dependence from the integral in $\So$ by reparameterising terms in the Euclidean action $\So$ in terms of dimensionless quantities and by neglecting higher order terms. We then made use of the technique \eq{easycalc} and obtained consistent classical and absolute stability conditions, \eqs{ABSCOND}{CLS}. For our present potential, \eq{repsugra}, which satisfies the condition, $\beta^2 \lesssim |K|\lesssim \order{1}$, we will be able to ignore the nonrenormalisable term  by introducing $\tisig=\sigma/M$ and $\beta^2$ in \eq{betani}. We can then obtain the stability conditions using the same technique as before. Indeed for the limit $\omega \gtrsim \order{m}$, we will see $\tisig(r) \sim \order{\epsilon} < \order{1}$ where $\epsilon$ is a small dimensionless constant (not $\epsilon_\omega$ in \eq{infqt}), and see $\tisig_0 \equiv \tisig(0) \geq \tisig(r)$ for any $r$ because $\tisig(r)$ is a monotically decreasing function in terms of $r$. Since the leading order of the logarithmic term, $\tisig^2 \ln{\tisig}$, in \eq{repsugra} is of $\order{\epsilon^2}$ using the L'H\^{o}pital's rules, we can ignore the nonrenormalisable term in \eq{repsugra} at the beginning of our analysis. To confirm this, in Appendix \ref{appxthick} we will keep all terms in \eq{repsugra} by introducing a Gaussian ansatz and show that the results below [\eqs{grvqeq}{grvclscond}] can also be recovered under the same assumption $\beta^2 \lesssim |K|\lesssim \order{1}$. By adapting the techniques introduced in \eq{easycalc}, in this subsection we will show how to obtain the thick-wall solutions without involving the Gaussian ansatz.

First of all we introduce two characteristic limits: the ``moderate limit'' $\omega\gtrsim \order{m}$ and the ``extreme'' limit $\omega \gg m$. We will see $\tisig_0\simeq \tisig_-(\omega)\to 0^+$ which leads to $\tisig_-(\omega) \ll \order{1}$ in the ``extreme limit'', and then even in the ``moderate limit'' we will see that the contributions from the nonrenormalisable term are negligible and that $\tisig_-(\omega)$ is a monotonically decreasing function in terms of $\omega$. Under the conditions $\beta^2 \lesssim |K|\lesssim \order{1}$ in \eq{repsugra}, we obtain
\bea{thckom1}
\bset{\frac{\omega}{m}}^2&=&1-2|K|\ln{\tisig_-(\omega)}+2\beta^2 \tisig^{n-2}_-(\omega) \sim 1-2|K|\log{\tisig_-(\omega)},\\
\label{thckom3} \frac{|K|m^2}{2\omega \tisig_-(\omega)}\frac{d\tisig_-(\omega)}{d\omega}&=&  \sbset{-1+2(n-2)\frac{\beta^2 \tisig^{n-2}_-(\omega)}{|K|}}^{-1}\sim -1 <0,\\
\label{thckom2} &\Leftrightarrow & \omega \gtrsim \order{m},\; \; \tisig_-(\omega) \sim \exp{\sbset{\frac{\mo^2}{2|K|m^2}}} \to 0,
\eea
where we used $\Uo(\tisig_-(\omega))=0$ to obtain \eq{thckom1}. It follows that $\tisig_-(\omega)\ll \order{1}$ for the thick-wall limit $\omega \gg m$, and we can ignore the nonrenormalisable term. Since \eq{thckom3} implies that $\frac{d\tisig_-(\omega)}{d\omega}<0$ in the limit $\tisig_-(\omega) < \order{1}$, $\tisig_-(\omega)$ is a monotonically decreasing function. Therefore, we can ignore the contributions from the nonrenormalisable term up to $\omega\gtrsim \order{m}$ which we call the ``moderate limit'' with the notation '$\sim$' as seen in the second relations of \eqs{thckom1}{thckom3}, instead of the ``extreme'' limit $\omega \gg m$ with the notation '$\to$'. Thus, we obtain the desired results of the second relation in \eq{thckom2}. From \eq{thckom1}, the logarithmic term may be of $\lesssim \order{1}$ for $|K|< \order{1},\; \beta^2\ll \order1$ in the ``moderate'' limit, which implies that the ``moderate limit'' is valid even when $\omega \sim \order{m}$. 

Let us define $\alpha(r)$ and $\tir$ through $\tisig(r)=a\alpha(r)$ and $r=b\tir$ where $a$ and $b$ will be obtained in terms of the underlying parameters. By substituting these reparamerised parameters $\alpha$ and $\tir$ into \eq{Uo}, and neglecting the nonrenormalisable term due to 'the L'H\^{o}pital's rules', we obtain
\bea{}
\nonumber \So &\sim& \Omega_{D-1}\int d\tir \tir^{D-1}b^D \left\{ \half \bset{\frac{aM}{b}}^2\bset{\frac{d\alpha}{d\tir}}^2 \right. \\
\label{sothick} & & - \left. \half m^2 a^2 M^2 \bset{1-\bset{\frac{\omega}{m}}^2- 2|K|\ln{a}}\alpha^2 + \half m^2|K|a^2 M^2\alpha^2\ln{\alpha^2} \right\}, \\
\label{sothick2} &=& a^2 M^2 b^{D-2}\tilde{S}(\alpha),
\eea
where $\tilde{S}\bset{\alpha(\tir/b)}\equiv \Omega_{D-1} \int d\tir \tir^{D-1} \set{\half\bset{\frac{d\alpha}{d\tir}}^2 -\half \alpha^2(1-\ln{\alpha^2})}$, which is independent of $\omega$. In going from \eq{sothick} to \eq{sothick2} we have set the coefficients of the three terms in the brackets of \eq{sothick} to be unity in order to explicitly remove the $\omega$ dependence from the integral in $\So$. In other words, we have set $a=e^{-1/2}\exp\sbset{\frac{\mo^2}{2|K|m^2}} \sim e^{-1/2}\tisig_-(\omega),\; b=\frac{1}{m\sqrt{|K|}}$. Following \eq{easycalc}, we can differentiate \eq{sothick2} with respect to $\omega$ to obtain $Q$ and then use the Legendre transformation to obtain $E_Q$. Coupled with \eqs{ABSCOND}{CLS} we obtain both the classical and absolute stability conditions. This is straightforward and yields
\bea{grvqeq}
Q&\sim& \frac{2\omega}{m^2|K|}\So,\hspace*{10pt} \frac{E_Q}{\omega Q}\sim  1+\frac{m^2|K|}{2\omega^2} \to 1,\\
\label{grvclscond}\frac{d}{d\omega}\bset{\frac{E_Q}{Q}}&\sim&1-\frac{m^2|K|}{2\omega^2}\to 1 >0,\hspace*{10pt} \frac{\omega}{Q}\frac{dQ}{d\omega}\sim 1-\frac{2\omega^2}{m^2|K|} \to - \frac{2\omega^2}{m^2|K|}<0,
\eea
where we have taken the ``extreme'' limit $\omega \gg m$ as indicated by '$\to$'. \eq{grvqeq} implies that the characteristic slope for the thick-wall $Q$-balls are tending towards the case $\mS\ll \mU$ in \eq{virieq} and \eq{grvclscond} shows that the $Q$-balls are classically stable. These results are independent of $D$. In Appendix \ref{appxthick} we will generalise the results of \eqs{grvqeq}{grvclscond} by adopting an explicit Gaussian ansatz without neglecting the nonrenormalisable term. 

Before finishing this subsection, let us comment on possibilities to have absolutely stable thick-wall $Q$-balls in the case, $|K| < \order{1},\; \beta^2 \ll \order{1}$. The results present above still hold even in the ``moderate limit'' $\omega \sim \order{m}$ for the present case. Thus, the thick-wall $Q$-balls, if they exist, can be absolutely stable when the following conditions from \eqs{ABSCOND}{grvqeq} are met:
\be\label{thckabs3}
\omega_- < m,\hspace*{10pt} \frac{\omega}{m} < \frac{1+\sqrt{1-2|K|}}{2}, \hspace*{10pt} |K| < \half,\; \beta^2\ll \order1.
\ee
It follows that for $|K| \ge 1/2$, the thick-wall $Q$-balls are always absolutely unstable. If $\omega_- \ge m$, we know $\omega > \omega_-$ in both the ``moderate'' and ``extreme'' limits, hence the thick-wall $Q$-ball is always absolutely unstable again, see \eq{ABSCOND}. Notice that the condition $\beta^2 \lesssim |K|$ implies $\omega_- \lesssim \order{m}$, see \eq{rbeta}, so the first condition in \eq{thckabs3}  can be satisfied. This then leaves only a small window of the parameter space for absolutely stable thick-wall $Q$-balls. In the numerical section, Sec. \ref{numerics}, we will confirm that the thick-wall $Q$-ball can be absolutely stable against decay into their own quanta by choosing suitable parameters, i.e. $\omega_-=0,\; \beta^2 \sim 1.90 \times 10^{-11}$, and $|K|=0.1$.

\subsection[Asymptotic profile]{Asymptotic profile for large $r$ and $\beta^2 \lesssim |K|\lesssim \order{1}$}

In order to obtain the full numerical profiles over all values of $\omega$, we should analytically determine the asymptotic profile for large $r$ in the potential \eq{sugra} which satisfies $\beta^2 \lesssim |K|\lesssim \order{1}$ as in the previous subsection. As long as the value of $r$ satisfies $r>R_\omega$ where $R_\omega$ is some large length scale and depends on $\omega$, we can assume that the friction term in \eq{QBeq} and the nonrenormalisable term in \eq{sugra} are negligible for large $r$. Hence, the $Q$-ball equation \eq{QBeq} reduces to the one-dimensional and integrable form
\be\label{asymgr}
\sigma^{\p\p}=\frac{d\Uo}{d\sigma},
\ee
where $\Uo\simeq \half m^2 \sigma^2\bset{1-\bset{\frac{\omega}{m}}^2 - |K|\log \bset{\frac{\sigma^2}{M^2}}}$. Equation (\ref{asymgr}) implies that the profile has a symmetry under the variation of $r$ because \eq{asymgr} does not depend on $r$ explicitly. Multiplying both sides of \eq{asymgr} by $\frac{d\sigma}{dr}$ leads to
\be\label{asymint}
\int^{\sigma(r)}_{\sigma(R_\omega)} \frac{d\sigma}{\sqrt{2\Uo}}=R_\omega-r,
\ee
where we have used the boundary conditions: $\sigma^\p(\infty) \to 0,\; \Uo(\sigma(\infty) \to 0)\to 0$ and $\sigma^\p(r) <0$. After some elementary algebra, the final asymptotic profile becomes 
\bea{asympro}
	\sigma(r)&=& M e^{M\mo^2/2m^2} \exp\bset{-\frac{m^2|K|M}{2} (r-r_\omega)^2},\\ 
\label{asympro2}	\frac{d}{dr}\bset{-\frac{\sigma^\p}{\sigma}}&=& m^2|K|M,
\eea
where $r_\omega\equiv R_\omega-\sqrt{\frac{\mo^2}{m^2}-\frac{2|K|}{M}\log{\bset{\frac{\sigma(R_\omega)}{M}}}}/(|K|m)$.
Equation (\ref{asympro}) is a consequence of the symmetry in \eq{asymgr} under the translation $r\to r- r_\omega$ from a Gaussian profile as seen in \eq{gauss} of Appendix \ref{exactsol}. Furthermore, \eq{asympro2} depends on the parameters $m,\; M,\; |K|$ in \eq{sugra}. We will later use the relation \eq{asympro2} as a criterion  that must be satisfied in obtaining full numerical profiles for all values of $\omega$. 

\vspace*{15pt}
We finish this section by recapping the key results we have derived for the case of the gravity-mediated potential, \eq{sugra}, in both the thin and thick-wall limits. In the thick-wall limit, we imposed the restrictions $\beta^2 \lesssim |K|\lesssim \order{1}$ on the potential to ignore the nonrenormalisable term. In both limits, we have derived the characteristic slopes in \eqs{grvthnch}{grvqeq} and the classical stability conditions in \eqs{grvthncls}{grvclscond} and shown that the $Q$ balls are classically stable in both cases. The thin-wall $Q$-balls in DVPs are always absolutely stable, and $Q$-matter in NDVPs can be absolutely stable when the coupling constant for the nonrenormalisable term satisfies \eq{restbeta}; whilst absolutely stable $Q$-balls in the thick-wall limit may exist only for \eq{thckabs3}. Finally, we obtained the general asymptotic profile, \eq{asympro}, for large $r$.

\section{Gauge-mediated potential}\label{sect:gauge}
\label{gauge-mediated}

\begin{figure}[!ht]
  \begin{center}
	\includegraphics[angle=-90, scale=0.5]{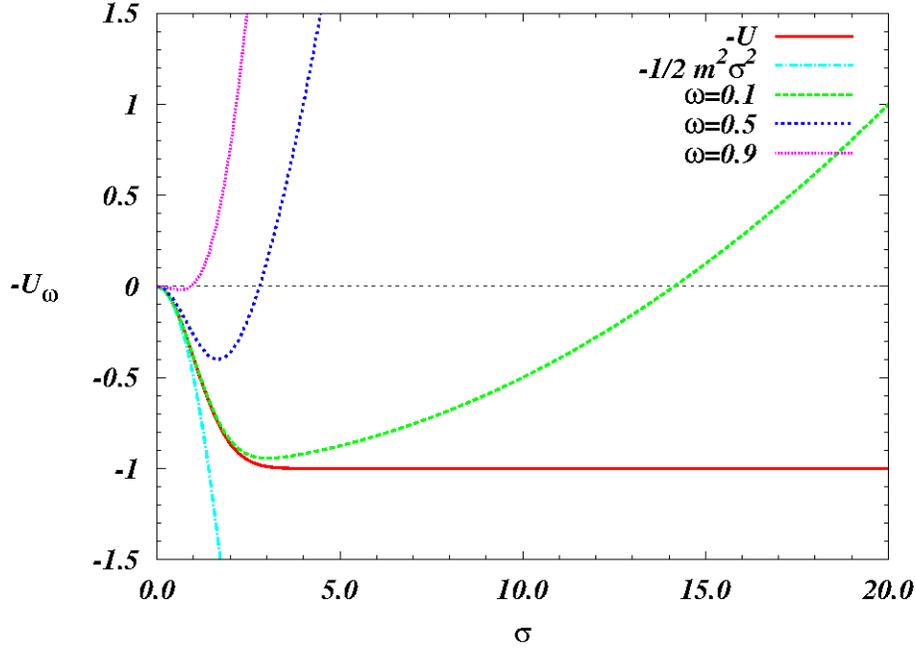}
  \end{center}
  \caption{The inverse potential $-U_{gauge}$ in \eq{apprxpot} (red-solid line) with $m=1,\; \Lambda^2=2$ which implies $U_0=1$ and the inverse effective potentials $-\Uo$ for different values of $\omega$. In order to compare between \eq{potgauge} and \eq{apprxpot}, we plot the inverse quadratic potential with the sky-blue dotted-dashed line. The two potentials are asymptotically similar, but they are different around the intermediate region of $\sigma$, where $1 \lesssim \sigma \lesssim 3$.}
  \label{fig:gaupot}
\end{figure}

The gauge-mediated scalar potential can be written in quadratic form in the low energy regime for scales up to the messenger scale $M_S$, and carries a logarithmically (extremely) flat piece in the high energy regime \cite{de Gouvea:1997tn, Dvali:1997qv}. This extreme flatness means that the thin-wall $Q$-ball we used in \eq{thinpro} cannot be applied to this situation, and so we now turn our attention to $Q$-balls in extreme flat potentials. We will generalise the results of \cite{Dvali:1997qv} to an arbitrary number of spatial dimensions and show that the known $Q$-ball profiles in \cite{Laine:1998rg, Dvali:1997qv} are naturally recovered by our more general ansatz. Moreover, we will investigate both the classical and absolute stability of these $Q$-balls. The gauge-mediated potential, which we will use in this section, is approximated by \cite{Asko:2002phd, MacKenzie:2001av}
\be\label{potgauge}
U(\sigma)=
    \left\{
    \begin{array}{ll}
    \half m^2 \sigma^2 &\; \textrm{for}\ \sigma(r) \le \sigma(R), \\
    U_0=const. &\; \textrm{for}\ \sigma(R) < \sigma(r),
    \end{array}
    \right.
\ee
where $U_0$ and $R$ are free parameters that will be determined by imposing a condition that leads to a smooth matching of the profiles at $\sigma(R),\;  U_0 =  \half m^2 \sigma^2(R)$. Notice that $Q$-balls exist within $0 < \omega < m$ in \eq{potgauge}, and the potential does not have degenerate vacua although $\omega_-\simeq 0$. Since \eq{potgauge} is not differentiable at $\sigma(R)$, we can approximate \eq{potgauge} by
\be\label{apprxpot}
U_{gauge}=\half m^2 \Lambda^2 \bset{1-e^{-\sigma^2/\Lambda^2}}
\ee
which we will use in the numerical section, Sec. \ref{numerics}. Note that $\Lambda=\sigma(R)$ corresponds to the scale below which SUSY is broken, so that $U_0=\half m^2 \Lambda^2$ in \eq{potgauge}. The potential \eq{apprxpot} differs from the one used in \cite{Campanelli:2007um}, but is similar to the potential used in \cite{Gumrukcuoglu:2008gi}. \fig{fig:gaupot} shows the inverse potential \eq{apprxpot} and the inverse effective potentials for various values of $\omega$ with $m=1,\; \Lambda^2=2$, which implies $U_0=1$. The red-solid line shows the inverse potential of \eq{apprxpot} ($- U_{gauge}$), and the sky-blue dotted-dashed line corresponds to the inverse quadratic potential of \eq{potgauge}. For sufficiently large and small $\sigma$, the two potentials in \eqs{potgauge}{apprxpot} have similar behaviour, but we can see the difference in the intermediate region of $\sigma$ where $1 \lesssim \sigma \lesssim 3$. Hence, we can expect that profiles around the thick-wall limit are different between the potentials since the thick-wall profiles are constructed in the particular region, $1 \lesssim \sigma \lesssim 3$; hence it may lead to the different stationary properties and stability conditions.

Using \eq{infqt}, the $Q$-ball equation, \eq{QBeq}, in the linearised potential \eq{potgauge} becomes
\bea{qbgau1}
	\sigma^{\p\p}_{core} + \frac{D-1}{r}\sigma^{\p}_{core}+\omega^2 \sigma_{core}&=&0, \; \textrm{for}\ 0 \le r < R,\\
\label{qbgau2}	\sigma^{\p\p}_{shell} + \frac{D-1}{r}\sigma^{\p}_{shell}-\mo^2 \sigma_{shell}&=&0, \; \textrm{for}\ R \le r,
\eea
where the profiles should be imposed to satisfy the boundary conditions, $\sigma^\p<0,\; \sigma(0)\equiv \sigma_0=finite,\; \sigma(\infty)=\sigma^\p(\infty)=0,\; \sigma^\p(0)=0$. The solutions are
\be\label{profgauge}
    \left\{
    \begin{array}{ll}
    \sigma_{core}(r)=A\ r^{1-D/2} J_{D/2-1}(\omega r) &\; \textrm{for}\ 0 \le r < R, \\
    \sigma_{shell}(r)=B\ r^{1-D/2}K_{D/2-1}(\mo r) &\; \textrm{for}\ R \le r,
    \end{array}
    \right.
\ee
where $J$ and $K$ are Bessel and modified Bessel functions respectively, with constants $A$ and $B$.
By introducing $\sigma_0$, and expanding $J_{D/2-1}(\omega r) $ for small $\omega r$ in $\sigma_{core}(r)$, and by using the condition $U_0 =  \half m^2 \sigma^2_{shell}(R)$ we obtain
\be\label{au0}
A=\sigma_0 \Gamma(D/2)\bset{\frac{2}{\omega}}^{D/2-1},\hspace*{5pt} U_0=\half m^2 B^2 R^{2-D}K^2_{D/2-1}(\mo R).
\ee
Since the energy density is smooth and finite everywhere, we have to impose a smooth continuity condition to the profiles 
$\sigma_{core}(R)=\sigma_{shell}(R)$ and $\sigma^\p_{core}(R)=\sigma^\p_{shell}(R)$, which gives
\be\label{ab}
\frac{A}{B}= \frac{K_{D/2-1}(\mo R)}{J_{D/2-1}(\omega R)}=\frac{\mo K_{D/2}(\mo R)}{\omega J_{D/2} (\omega R)}.
\ee 
We will see that the particular value of $\sigma_0$ does not change important features such as the stability condition and characteristic slope of the $Q$-ball solutions. Using \eq{ab} we obtain the following important identities, which we will make use of later 
\cite{gr}:
\bea{cont1}
 \omega \frac{J_{D/2}(\omega R)}{J_{D/2-1}(\omega R)}&=&\mo \frac{K_{D/2}(\mo R)}{K_{D/2-1}(\mo R)},\\
\label{cont2} \frac{J_{D/2}(\omega R) J_{D/2-2}(\omega R)}{J^2_{D/2-1}(\omega R)}&=&-\bset{\frac{\mo}{\omega}}^2 \frac{K_{D/2}(\mo R) K_{D/2-2}(\mo R)}{K^2_{D/2-1}(\mo R)},
\eea
where we used the recursion relations $J_{\mu -1}(z)+J_{\mu +1}(z)=\frac{2\mu}{z} J_{\mu}(z),\; K_{\mu -1}(z)- K_{\mu +1}(z)=-\frac{2\mu}{z}K_{\mu}(z)$ for any real $\mu$ and $z$.
We can easily find $\So^{core}=U_0 V_D + \bset{\half \sigma_{core} (R)\sigma^\p_{core} (R)} \partial V_D$ and $\So^{shell}=-\bset{\half \sigma_{shell}(R)\sigma^\p_{shell}(R)} \partial V_D$, and then using $\So=\So^{core}+\So^{shell}$ it follows that 
\be\label{Soth}
\So = U_0 V_D,
\ee
where we have again used the continuity relations $\sigma_{core}(R)=\sigma_{shell}(R)$ and $\sigma^\p_{core}(R)=\sigma^\p_{shell}(R)$. To find the charge $Q$, we do not make use of the Legendre relation $Q=-\frac{d\So}{d\omega}$ in \eq{easycalc}, because $R$ is a function of $\omega$, and is determined by \eq{cont1}. However, we can obtain $Q$ by substituting \eq{profgauge} directly into \eq{Uo}:
\be\label{gauq}
Q=\frac{D U_0 V_D}{\omega}  \bset{\frac{K_{D/2}(\mo R)K_{D/2-2}(\mo R)}{K^2_{D/2-1}(\mo R)}},
\ee
where we have used \eqs{au0}{ab} and \eq{cont2}, as well as the relation, $\int dy\ y Z^2_\mu(y) =$ \\ $\sbset{\frac{y^2}{2} \bset{ Z^2_{\mu}(y)-Z_{\mu-1}(y)Z_{\mu +1}(y)}}$, \cite{Asko:2002phd, gr}. Here, $\mu$ is real, and $Z$ can be either the Bessel function $J$ or the modified Bessel function $K$, and we have used the following recursion relations to obtain the indefinite integral: $z\frac{dJ_\mu}{dz}\pm \mu J_\mu=\pm z \ J_{\mu\mp 1},\; J_{\mu-1}-J_{\mu+1}=2\frac{dJ_\mu}{dz},\; z\frac{dK_\mu}{dz}\pm \mu K_\mu=-z \ K_{\mu\mp 1},\; K_{\mu-1}+K_{\mu+1}=-2\frac{dK_\mu}{dz}$.

For future reference we obtain explicit expressions for $R$ for case with an odd number of spatial dimensions. \eq{cont1} can be solved explicitly in terms of $R$ to give
\bea{gauR1}
	      \omega R&=&  \arctan \bset{\frac{\omega}{\mo}} ,\hspace*{8pt} \textrm{for} \ D=1,\\
\label{gauR3} \omega R&=& \pi-\arctan \bset{\frac{\omega}{\mo}} , \hspace*{8pt} \textrm{for} \ D=3,
\eea
where we have used $J_{3/2}(x)=\sqrt{\frac{2}{\pi x}} \bset{\frac{\sin(x)}{x} -\cos(x)},\; J_{1/2}(x)=\sqrt{\frac{2}{\pi x}}\sin(x),\; J_{-1/2}(x)=\sqrt{\frac{2}{\pi x}}\cos(x),\; K_{3/2}(x)=\sqrt{\frac{\pi}{2x}}e^{-x}\bset{1+\frac{1}{x}},\; K_{1/2}(x)=\sqrt{\frac{\pi}{2x}}e^{-x}=K_{-1/2}(x)$. 
We will discuss the classical stability for $Q$-balls in $D=1,\; 3$ in the numerical section, in which we will show stability plots arising from \eqs{gauR1}{gauR3}.

\subsection[``Thin-wall'' $Q$-balls]{``Thin-wall-like'' limit for $\mo R, \omega R \gg \order{1}$}

We now discuss both the classical and absolute stability of gauge-mediated $Q$-balls in arbitrary dimensions $D$, in the limit $\mo R,\; \omega R \gg 1$, which implies that the ``core'' size $R$ is large compared to $1/\mo,\; 1/\omega$. As we will see in the numerical section, Sec. \ref{numerics}, the limit will turn out to be equivalent to the thin-wall limit $\omega\simeq \omega_-\simeq 0$. Recall that this potential does not have degenerate vacua. Using \eqs{Soth}{gauq},
\begin{equation}
\So \simeq \frac{\omega Q}{D}\set{1+\order{(\mo R)^{-1}}},
\end{equation}
where we have used $\lim_{|z| \to \infty} K_\mu(z)\sim \sqrt{\frac{\pi}{2z}}e^{-z}\sbset{1+\frac{4\mu^2-1}{8z}+\order{z^{-2}}}$. The characteristic slope follows
\be\label{gauthnch}
\frac{E_Q}{\omega Q}\simeq \frac{D+1}{D}
\ee
from which we see immediately from \eq{leg2} that we recover the published results of \cite{Dvali:1997qv, MacKenzie:2001av}, namely $E\propto Q^{D/(D+1)}$. From \eqs{ABSCOND}{gauthnch}, the ``thin-wall-like'' $Q$-ball is absolutely stable since the present limits will cover the thin-wall limit $\omega\simeq \omega_-\simeq 0$ as we stated.

We can also obtain an explicit expression for $R(\omega)$ and $\frac{dR}{d\omega}$ in the limits $\mo R \gg 1$ and $\omega R \gg |\mu^2-\frac{1}{4}|$, where $\mu\; ( \sim \order1)$ is the argument of the Bessel function:
\bea{coregau}
\omega R &=& \bset{\frac{D+1}{4}}\pi - \arctan\bset{\frac{\omega}{\mo}},\\
\label{dRgau} \frac{dR}{d\omega}&=&-\frac{R}{\omega}\bset{1-\frac{1}{\mo R}} \simeq - \frac{R}{\omega}.
\eea
Notice that \eq{coregau} for $D=3$ reproduces the given profile in \cite{Laine:1998rg, Dvali:1997qv}, and it coincides with the exact expression derived in \eq{gauR3}. Using \eqs{gauq}{coregau} and \eq{dRgau}, we obtain
\bea{}
Q&\simeq& \frac{V_D U_0 D}{\omega},\\
\label{gaucls}\frac{\omega}{Q}\frac{dQ}{d\omega}&\simeq& -D-1 <0,
\eea
which shows that the $Q$-ball in this limit is classically stable. One can also check both $Q\simeq -\frac{d\So}{d\omega} = D U_0 V_D/\omega$ from \eq{dRgau} and  $\frac{d}{d\omega}\bset{\frac{E_Q}{Q}}\simeq\frac{D+1}{D}>0$ from \eq{gauthnch}, which are respectively consistent with \eq{gaucls} and with the result in \eq{CLS}.

\subsection[``Thick-wall'' $Q$-balls]{``Thick-wall'' limit for $D=1,\ 3,\; \dots$}

Having just discussed the ``thin-wall-like'' properties for arbitrary $D$, we turn our attention now to the  the other limit, $\omega\simeq\omega_+$. This is much more difficult to analytically explore because \eq{cont2} can only give a closed form expression for $R$ for the case where $D$ is an odd number of spatial dimensions. Therefore, we will concentrate here on the interesting cases, e.g. $D=1,\; 3$.

\paragraph*{\underline{\bf $D=3$ case:}}

From \eq{gauR3} and recalling that in the ``thick-wall'' limit, $\mo\to0,\; \omega\simeq\omega_+=m$, we obtain $R \simeq \frac{\pi}{2\omega},\hspace{10pt} \frac{dR}{d\omega}\simeq -\frac{R}{\omega}$, and by substituting these into \eq{gauq} we find 
\bea{3dcls}
	\frac{\omega}{Q}\frac{dQ}{d\omega}&\simeq& -1 + \frac{\omega^2}{\mo^2}\to \frac{\omega^2}{\mo^2} >0,\\
\label{gauthinch} \frac{E_Q}{\omega Q}&=&1+\frac{\pi \mo}{6\omega}\to 1,
\eea
which shows that the three-dimensional ``thick-wall'' $Q$-ball is classically unstable. This fact is consistent with the relation that $\frac{d}{d\omega}\bset{\frac{E_Q}{Q}}=1-\frac{\pi\omega}{6\mo}\to-\frac{\pi\omega}{6\mo}<0$ where we have used \eq{gauthinch}. It also follows that the ``thick-wall'' $Q$-ball is not absolutely stable, and the solution will decay to free particles satisfying $E_Q\to m Q$ which is the first case of \eq{virieq}.

\paragraph*{\underline{\bf $D=1$ case:}}

As in the case $D=3$, \eq{gauR1} implies $R\to 0,\; \frac{dR}{d\omega}\simeq-\frac{m^2}{\mo\omega^3}$ in the ``thick-wall'' limit. Using the above results, we obtain
\bea{1dcls}
	\frac{\omega}{Q}\frac{dQ}{d\omega} & \simeq & -1-\frac{m^2}{\omega^2}+\frac{\omega^2}{\mo^2} \to \frac{\omega^2}{\mo^2} >0,\\
\label{1dch}	\frac{E_Q}{\omega Q}&=&1+\bset{1+\frac{1}{\mo R}}^{-1}\to 1.
\eea
Note that the approximate value in \eq{1dcls} is the same as \eq{3dcls}. Then the one-dimensional ``thick-wall'' $Q$-ball is also classically unstable. This fact is again consistent with the result that\\ $\frac{d}{d\omega}\bset{\frac{E_Q}{Q}}\simeq 1+\mo R-\frac{m^2}{\omega^2} -\frac{\omega^2R}{\mo} \to -\frac{\omega^2R}{\mo} <0$. As in the three-dimensional case, the ``thick-wall'' $Q$-ball is not absolutely stable, and the solution decays into its free particles.

\subsection{Asymptotic profile}

The asymptotic profile for the large $r$ regime in this model can be described by the contribution from the quadratic term in the potential \eq{potgauge}, from which the profile is \eq{slope}, such that
\be\label{gauasym}
\sigma(r)\sim E\sqrt{\frac{\pi}{2\mo}}r^{-\frac{D-1}{2}}e^{-\mo r} \Leftrightarrow -\frac{\sigma^\p}{\sigma}\sim \frac{D-1}{2r}+\mo,
\ee
where $E$ is a constant. Note that we have used the fact that the modified Bessel function of the second kind has the relation $K_{\mu}(r)\sim \sqrt{\frac{\pi}{2r}}e^{-r}$ for large $r$ and any real $\mu$. We will use the criterion in the second expression of \eq{gauasym} in the following section.

\vspace*{15pt}

Summarising our most important results in the generalised gauge-mediated potential, the ``thin-wall-like'' $Q$-ball is classically stable for a general $D$, whilst it is absolutely stable as seen in \eqs{gauthnch}{gaucls}. On the other hand, for ``thick-wall'' $Q$-balls in $D=1,\ 3$, the $Q$-balls are both classically and absolutely unstable, as can be seen from \eqs{3dcls}{gauthinch} and \eqs{1dcls}{1dch}. Finally we obtained the general asymptotic profile \eq{gauasym} for large $r$.

\section{Numerical results}\label{numerics}

In this section, we obtain exact numerical solutions for $Q$-balls for both the gravity-mediated potential in \eq{repsugra} and the gauge-mediated potential in \eq{apprxpot} with dimensionless parameters by setting $m=M=1$ and $\Lambda^2=2$. We adopt the 4th-order Runge-Kutta algorithm and usual shooting methods to solve the second order differential equations \eq{QBeq} (for full details see the  numerical techniques developed in chapter \ref{ch:qpots}). The raw numerical data contains errors for large $r$, thus we introduce the previously obtained analytical asymptotic profiles to help control these uncertainties. In particular we use \eq{asympro2} for the gravity-mediated potential and  \eq{gauasym} for the gauge-mediated case. Using these techniques, the numerical profiles match smoothly and continuously onto the analytic ones. In order to check the previously obtained analytic results, we calculate $Q$-ball properties numerically over the whole parameter space $\omega$ except around the extreme thin-wall limit $\omega=\omega_-$, because it is difficult to obtain reliable numerical results in that limit.

\subsection{Gravity-mediated potential}

We shall investigate gravity-mediated potentials with two choices of $\lambda$ in \eq{sugra} for $|K|=0.1$ and $n=6$, which can be seen as the red solid lines in \fig{fig:gravpot}. The choice of the parameters, $|K|$ and $n$, are simply from phenomenological reasons. The degenerate vacua potential (DVP) on the left has $\omega_-=0$ ($\beta^2=\frac{|K|e^{-1}}{4}\exp\bset{-\frac{2}{|K|}} \sim 1.90 \times 10^{-11} \ll \order{1}$), and the nondegenerate vacua potential (NDVP) on the right has $\omega_-=1$ ($\beta^2=\frac{|K|e^{-1}}{4}\sim 9.20 \times 10^{-3} \ll \order{1}$), recalling \eq{rbeta}. \fig{fig:gravpot} also shows plots of the inverse effective potentials $-\Uo$ for various values of $\omega$. Because of numerical complications, we are unable to fully examine the properties in the extreme thin-wall limit; however, by solving close to this wall limit, our numerical results recover the expected analytical results we derived in \eqs{grvthnch}{grvthncls}. With the above choice of parameters, the curvature $\mu$ of $\Uo$ at $\sigma_+(\omega_-)\equiv \sigma_+$ in \eq{mucurv} is $\mu^2 \sim 0.4$ which implies that $1/\mu \sim 1.58$. From the first relation in \eq{tisigom}, we have found $\sigma_+\sim 1.28$ in NDVP and $\sigma_+\sim 1.91 \times 10^2$ in DVP. Since we have assumed $R_Q\gg 1/\mu,\; \sigma_0\simeq \sigma_+$ in our thin-wall analysis for the gravity-mediated potential, we see that it breaks down when the core size $R_Q$ becomes the same order as $1/\mu$ and/or $\sigma_0 \not\sim \sigma_+$. Although the full definition of the core size $R_Q$ is presented in chapter \ref{ch:qpots}, it is very time consuming to evaluate it properly in the simulations; hence, in this analysis we have used a more naive approach, in which we have estimated the value of $r=R_Q$ when the field profile drops quickly from its core value.  For the thick-wall limit, we required the condition $\beta^2 \lesssim |K|\lesssim \order{1}$, which is satisfied with the above chosen parameter set; hence, the analysis is valid for $\omega \gtrsim \order{1}$. Because of the choice of $|K|=0.1 < \order1$ and $\omega_-=0$ in NDVP, we will see our analysis holds even for $\omega \sim \order{1}$.

\paragraph*{\underline{\bf Hybrid profile:}}

\begin{figure}[!ht]
  \begin{center}
    \includegraphics[angle=-90, scale=0.28]{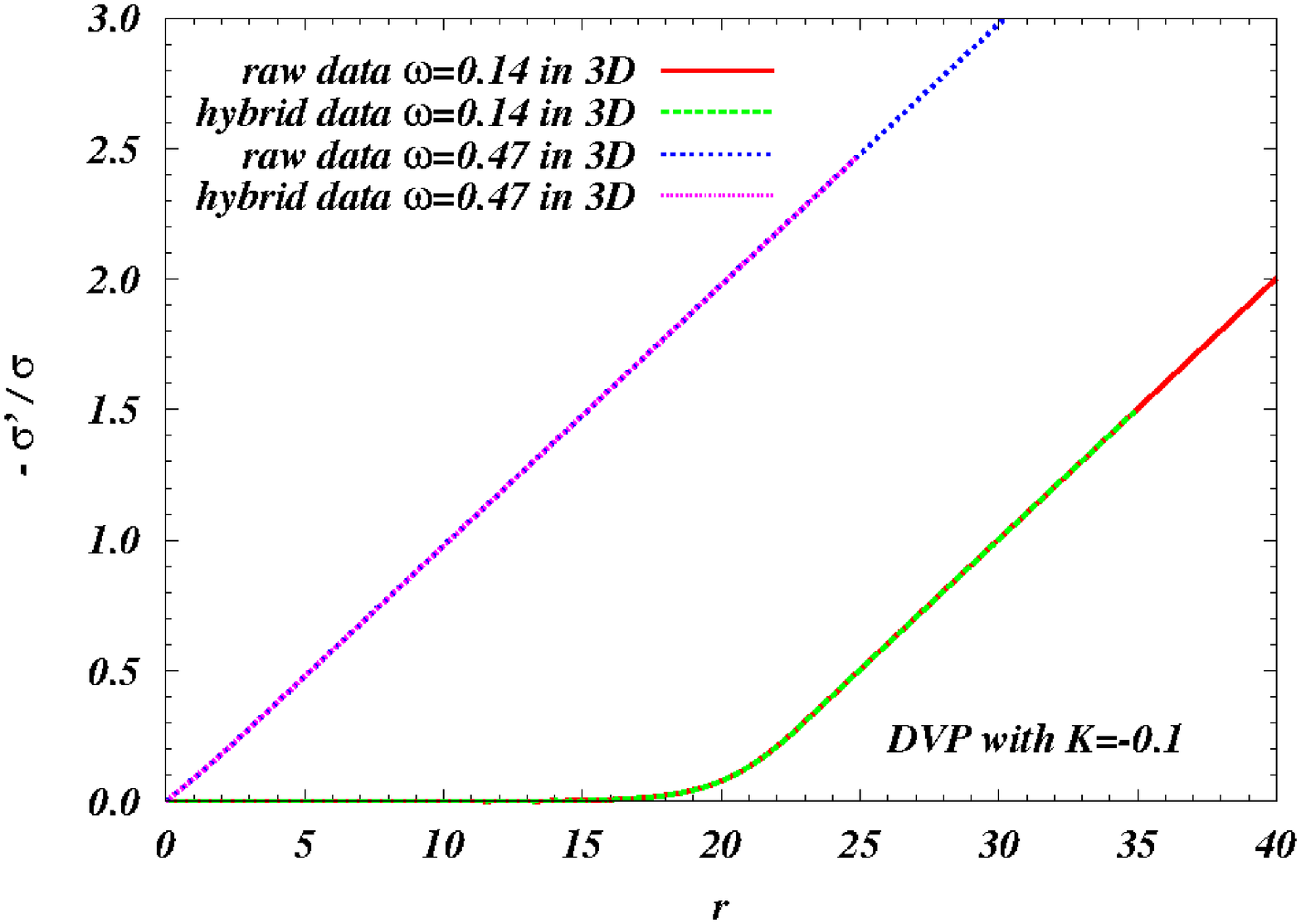} 
   \includegraphics[angle=-90, scale=0.28]{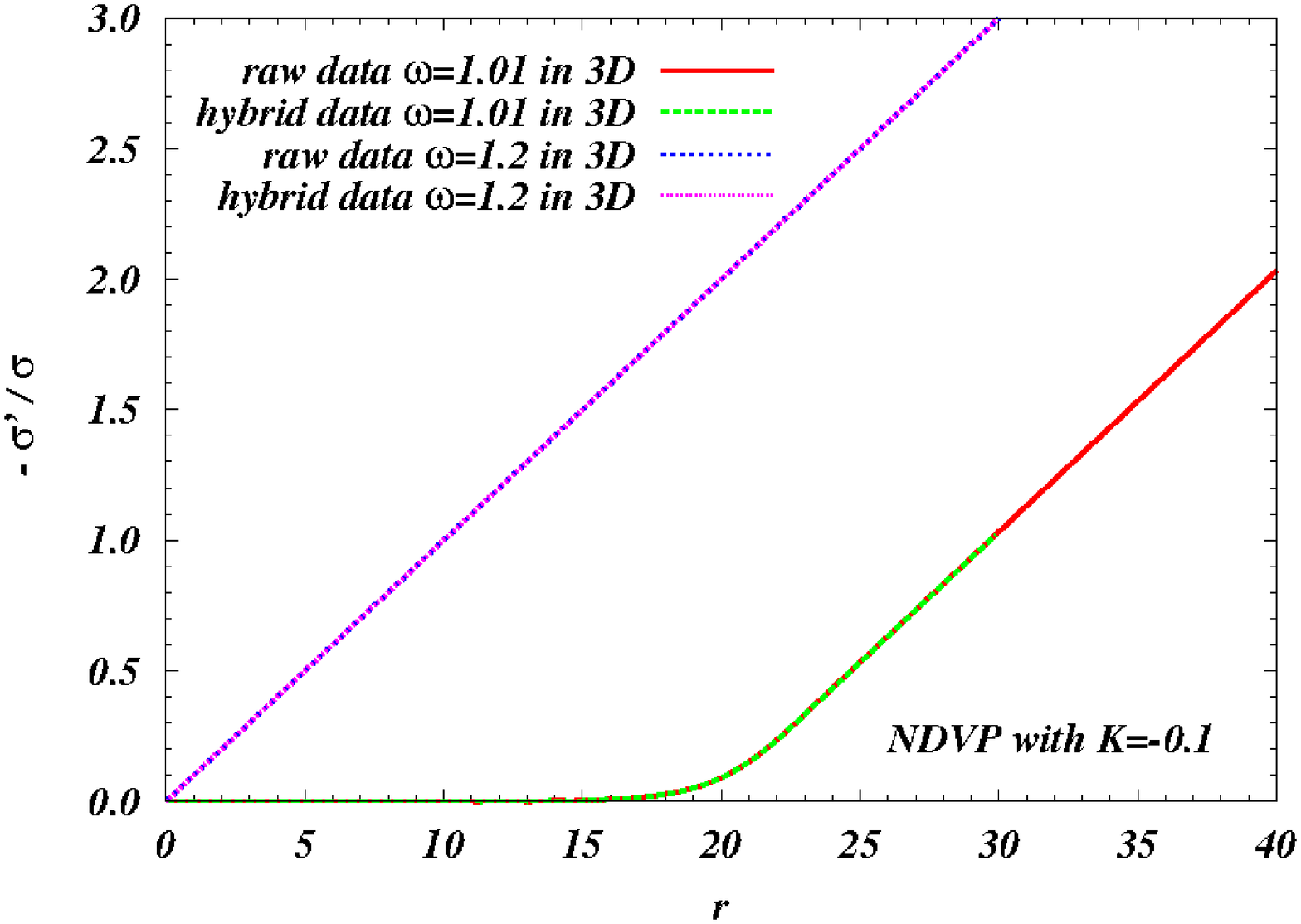} \\	
   \includegraphics[angle=-90, scale=0.28]{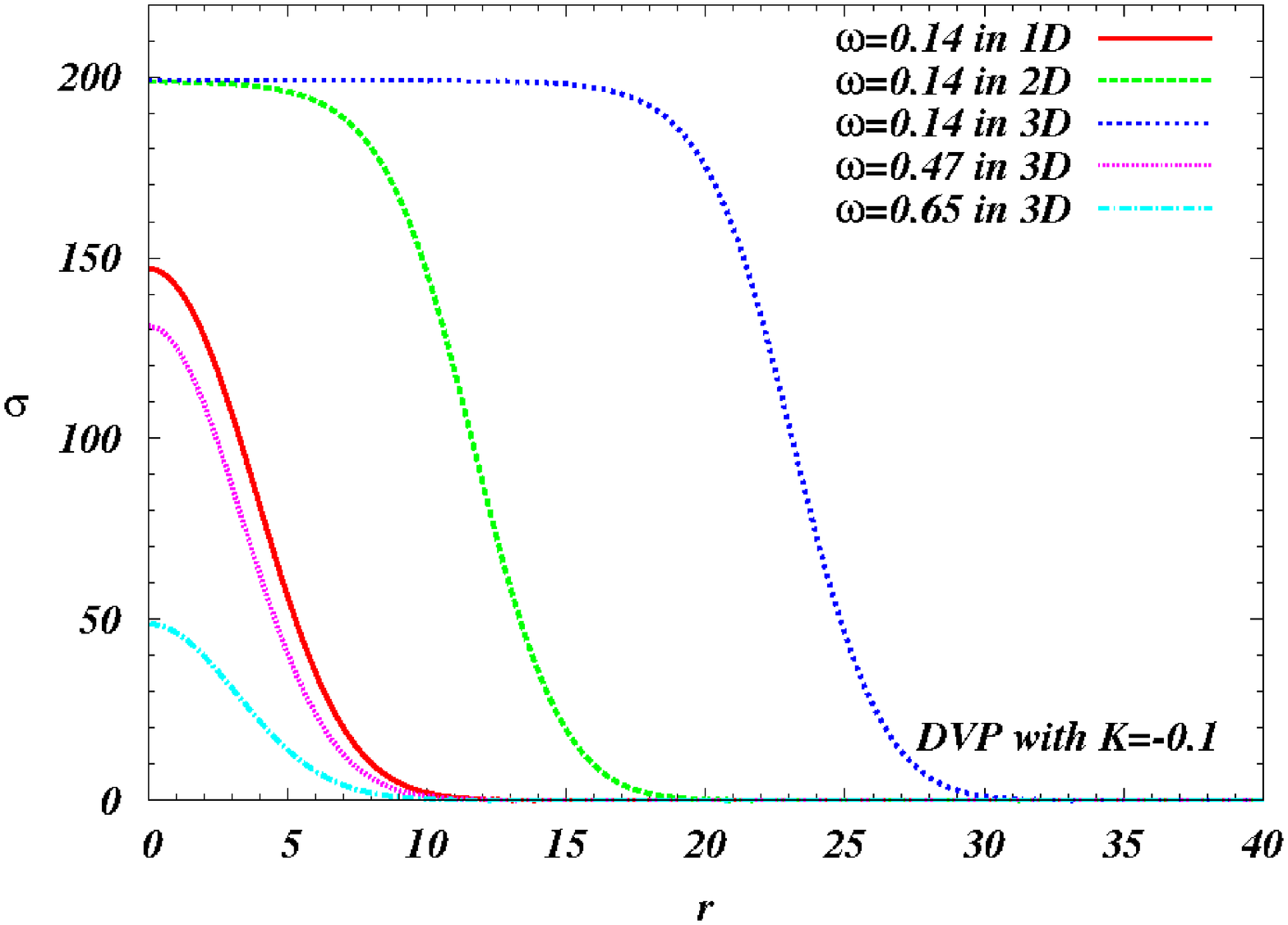} 
   \includegraphics[angle=-90, scale=0.28]{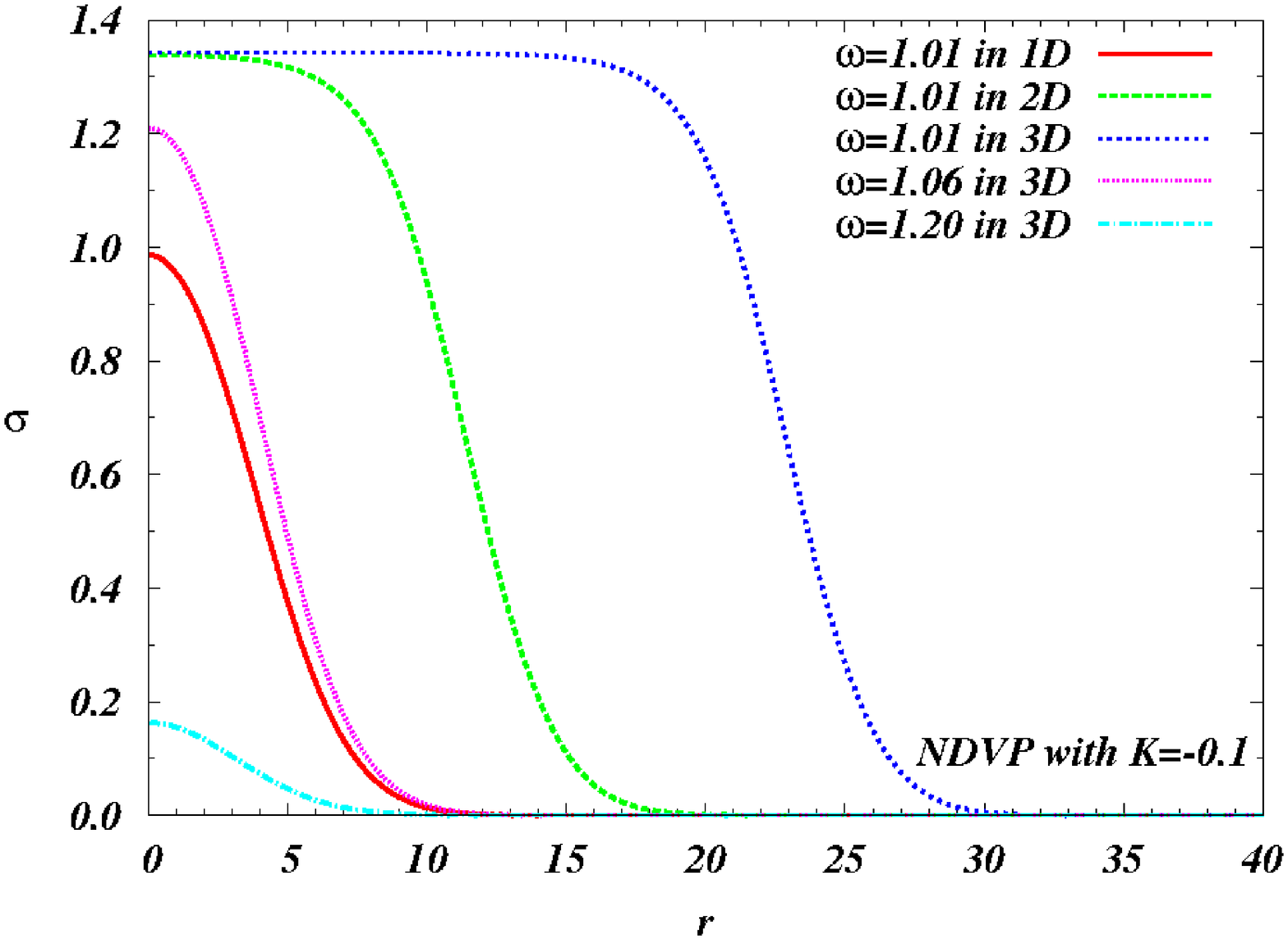} \\
   \includegraphics[angle=-90, scale=0.28]{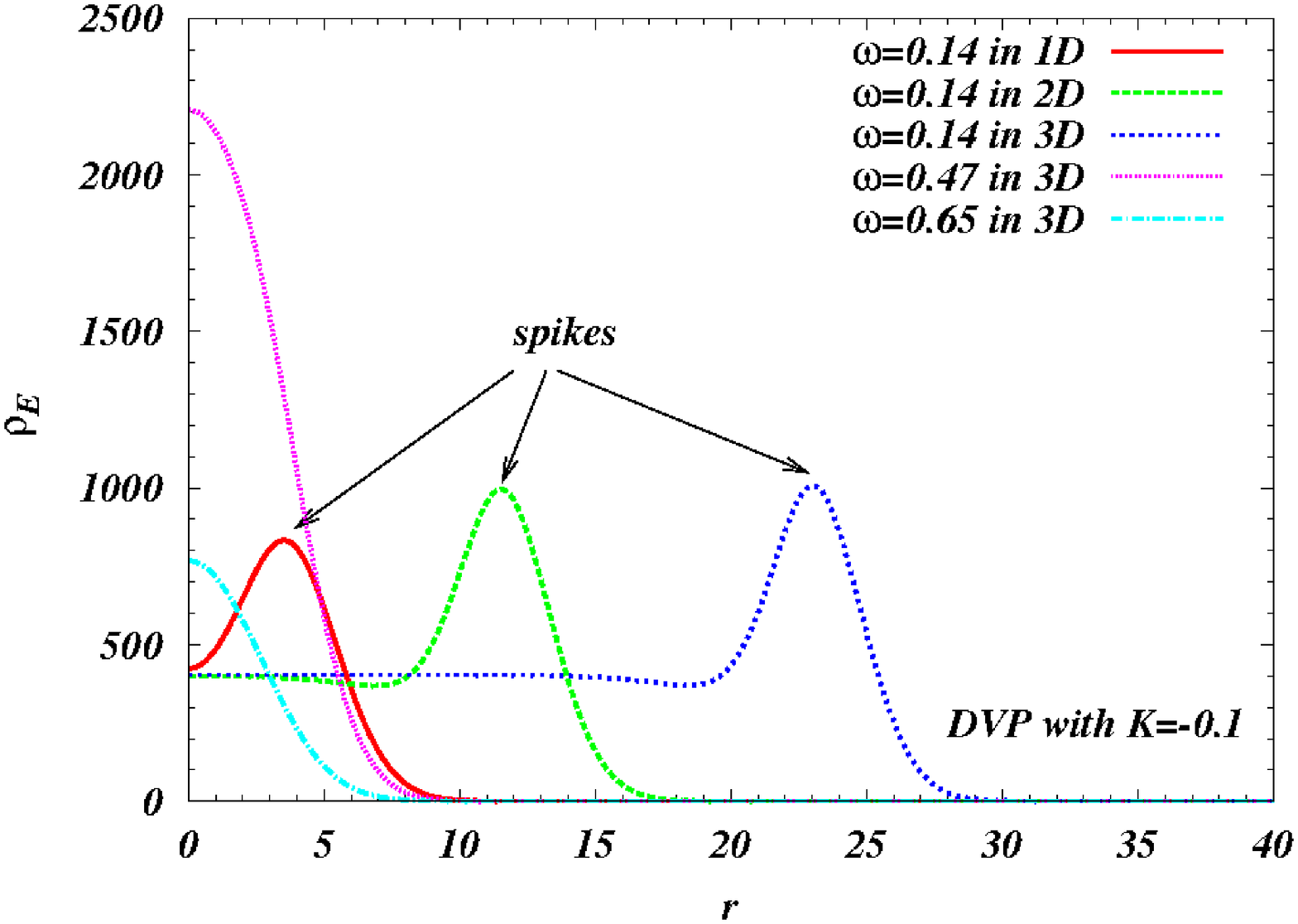}
   \includegraphics[angle=-90, scale=0.28]{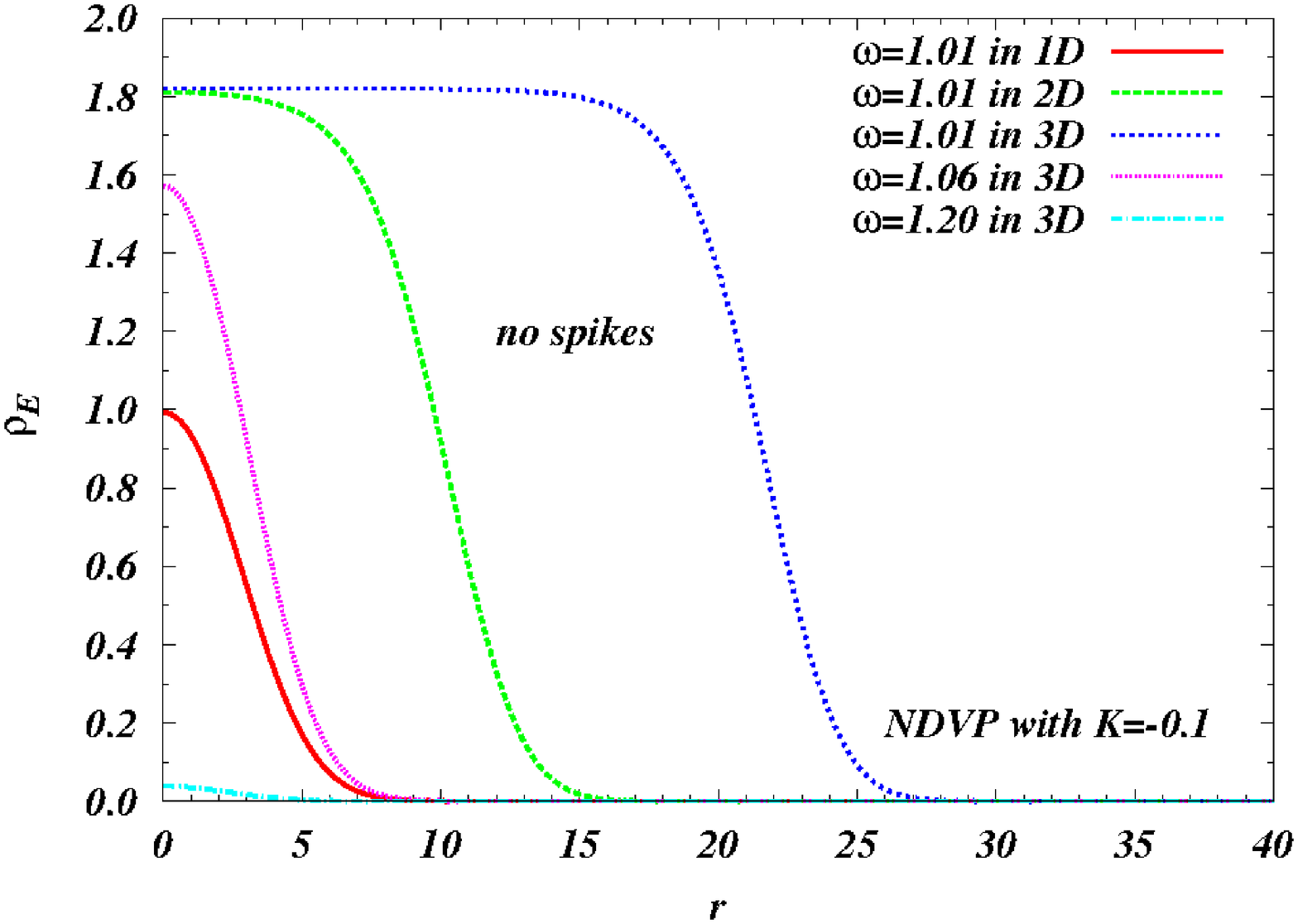}\\
  \end{center}
  \caption{The top two panels show the three-dimensional numerical slopes $-\sigma^\p/\sigma$ for two typical values of $\omega$ for both DVP (left) and NDVP (right). The raw numerical data (red-solid and blue-dotted lines) matches continuously on to the analytical asymptotic profiles for large $r$ (green-dashed and purple-dotted-dashed lines). The linear lines correspond to the Gaussian tails in \eq{asympro}, where we can see the large shifts in the thin-wall limits of $\omega$. The middle and bottom panels show, respectively, the hybrid profiles \eq{hybridprof} and the energy density configurations for the various values of $\omega$ and $D$. The spikes of the energy density configurations exist in the DVP case but not in the NDVP case.}
  \label{fig:grvpro}
\end{figure}

The numerical profiles have errors for large $r$ which correspond to either undershooting or overshooting cases; thus, to minimise the errors in the region of large $r$ we replace the numerical data by the predicted asymptotic analytical profile using the criterion \eq{asympro2} to obtain the solution for the whole range of $r$. We then have the hybrid profile which can be written as
\be\label{hybridprof}
    \sigma(r)=
    \left\{
    \begin{array}{ll}
    \sigma_{num}(r), &\ \ \textrm{for $r<R_{num}$}, \\
    \sigma_{num}(R_{num})\exp\bset{-\frac{|K|}{2}R^2_{num}-\frac{\sigma^\p_{num}(R_{num})}{\sigma_{num}(R_{num})}R_{num} }& \\
\times \exp\bset{-\frac{|K|r^2}{2}+\bset{R_{num}|K|+ \frac{\sigma^\p_{num}(R_{num})}{\sigma_{num}(R_{num})} } r} &\ \ \textrm{for $R_{num} \le r \le R_{max}$},
    \end{array}
    \right.
\ee
where $\sigma_{num}$ is the numerical raw data, $R_{num}$ is determined by $|\bset{-\sigma^\p_{num}/\sigma_{num}}^\p -1 |_{r=R_{num}} < 0.001$, and we have set $R_{max}=60$ throughout our numerical simulations in this subsection. We have calculated the following numerical properties using the above hybrid profile, \eq{hybridprof}, for $D=1,\; 2,\; 3$:

\paragraph*{\underline{\bf Profile:}}
In the top two panels of \fig{fig:grvpro} (DVP on the left and NDVP on the right), the red-solid and blue-dotted lines show the numerical slopes $-\sigma^\p/\sigma$ for two typical values of $\omega$ in $D=3$. We smoothly continue them to the corresponding analytic profiles by the methods just  described in the numerical techniques, see green-dashed and purple-dotted-dashed lines. The linear lines correspond to the Gaussian tails in \eq{asympro} and for the cases of $\omega =0.14$ (DVP) and $\omega = 1.01$ (NDVP) corresponding to the thin-wall solution we see that it is shifted from the origin to $r \simeq 21$. The middle panels show the obtained hybrid profiles of \eq{hybridprof} for the various values of $\omega$ and $D$. The higher the spatial dimension, the larger the core size $Q$-balls can  have. The energy density configurations $\rho_E(r)$ can be seen in the bottom panels of \fig{fig:grvpro}. Outside of the cores of the DVP profiles for $\omega \sim \omega_-$, we can see the same features  as we saw in the polynomial potentials we investigated in chapter \ref{ch:qpots}, namely, highly concentrated energy density spikes. In NDVP, however, the spikes cannot be seen. The presence of the spike contributes to the increase in the surface energy $\mathcal{S}$, which in turn leads to the different virialisation ratio for $\mathcal{S}/\mathcal{U}$ where $\mathcal{U}$ is the potential energy, as can be seen in \eq{virieq}.

\paragraph*{\underline{\bf Criterion for the existence of a thin-wall $Q$-ball:}}

\begin{figure}[!ht]  
  \begin{center}
	\includegraphics[angle=-90, scale=0.28]{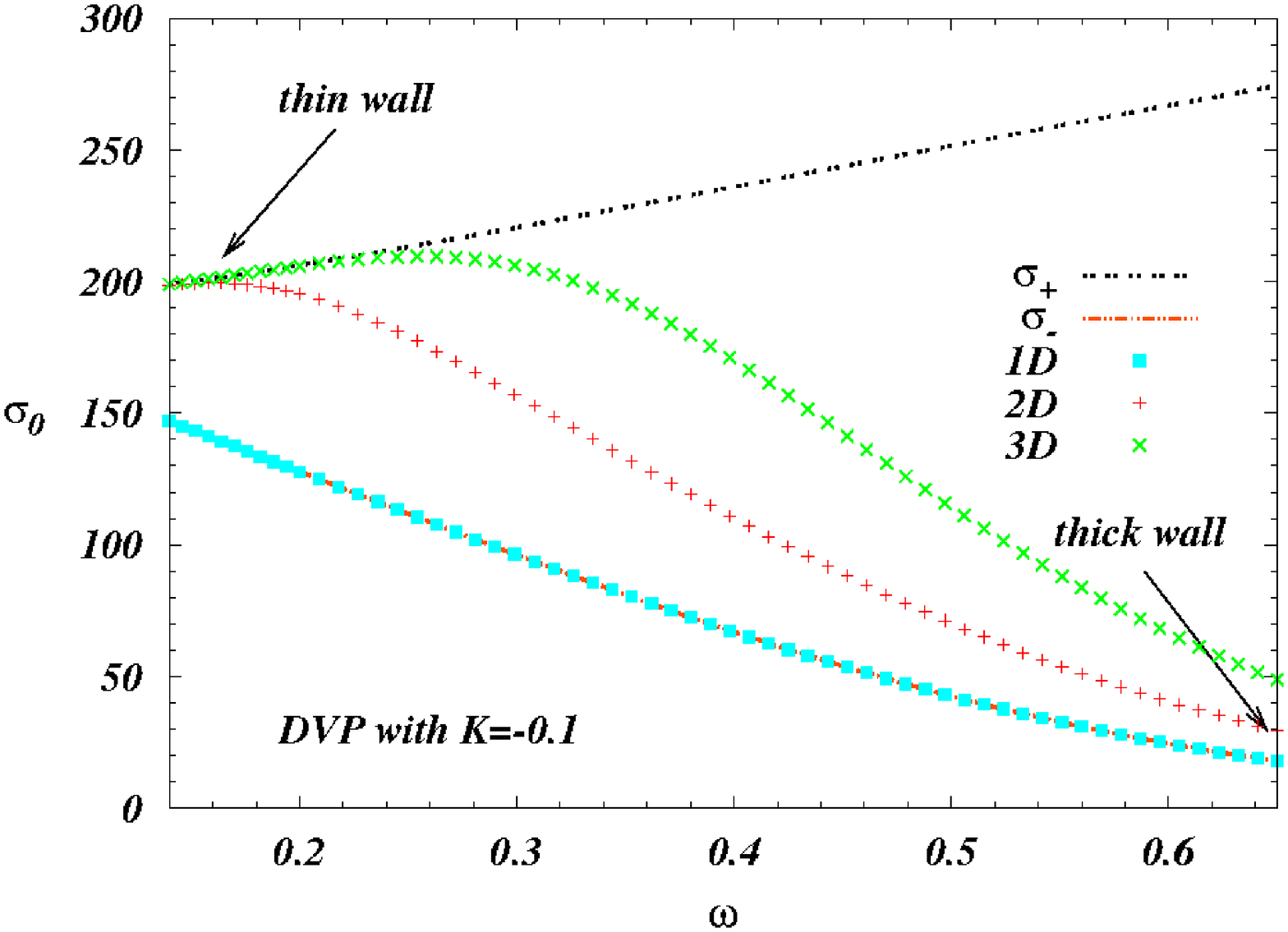}
	\includegraphics[angle=-90, scale=0.28]{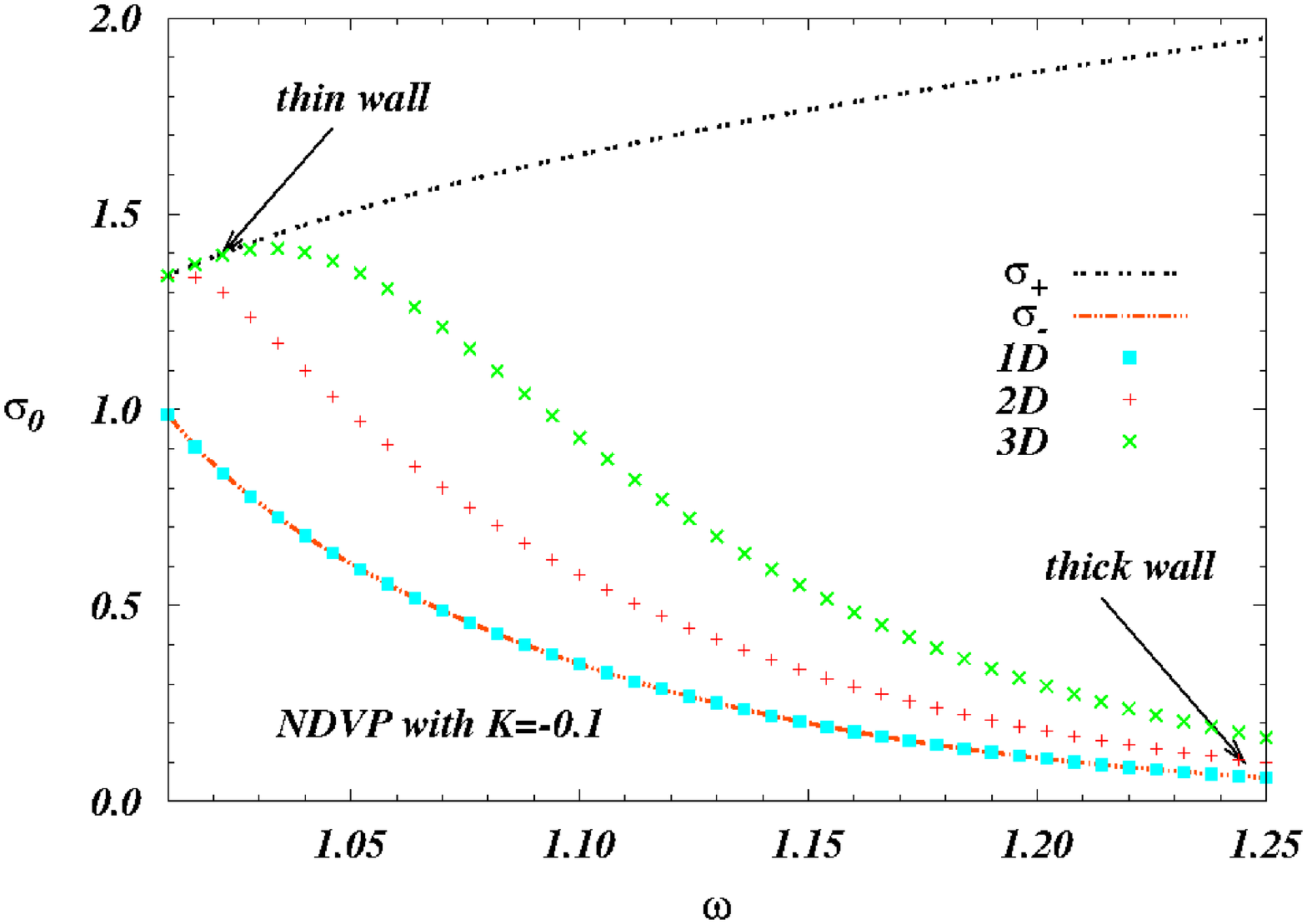}
  \end{center}
\caption{The initial value $\sigma_0(\omega)\equiv \sigma(0)$ is plotted against $\omega$. In the two panels the black-dashed and orange dotted-dashed lines show  $\sigma_\pm(\omega)$, and these lines become closer for $\omega = \omega_-$ for both types of the potentials DVP (left, $\omega_-=0$) and NDVP (right, $\omega_-=1$). Since $\sigma_0\simeq \sigma_+\equiv \sigma_+(\om_-)$ for $D=2,3$ in the region $\omega \sim \omega_-$ where $\sigma_+\sim 1.28$ in NDVP and $\sigma_+\sim 1.91\times 10^2$ in DVP, our analytical results in Sec.\ref{sect:grvthn}, are valid in this region.}
	\label{fig:grvsig}
\end{figure}

\fig{fig:grvsig} shows the numerical results for $\sigma_0(\omega)$ \\against $\omega$ for both types of potentials -- DVP (left) and NDVP (right). Our main analytical approximation relies on $\sigma_0(\omega) \simeq \sigma_+(\omega) \sim \sigma_+ \equiv \sigma_+(\omega_-)$, where we have found $\sigma_+\sim 1.28 \sim \order{1}$ in NDVP and $\sigma_+\sim 1.91 \times 10^2 \gg \order{1}$ in DVP. The $3D$ thin-wall $Q$-ball (green-crossed dots) appears for a wider range of $\omega$ than the $2D$ $Q$-ball (red-plus signs) in DVP as well as NDVP. For each case, the approximation can be valid, respectively, up to $\omega \sim 0.24$ or $\omega \sim 1.04$ with about $10\%$ errors for the $3D$ case. Near the thick-wall limit $\omega\simeq \omega_+$ for both potentials, we see $\sigma_0\simeq \sigma_- \to 0$. The one-dimensional values (skyblue-circled dots) always lie on $\sigma_-$. Note that in the $3D$ region $\omega \gtrsim 0.53$ for DVP, we can see $\sigma_0(\omega)\lesssim \order{10^2}$, which implies that the contribution from the nonrenormalisable term in \eqs{thckom1}{lam2},  i.e. $\beta^2 \tisig^4 \lesssim \order{10^{-3}} \ll \order{1},\; \order{|K|}$, is negligible compared to other terms in \eqs{thckom1}{lam2}. Hence, our analytic solution still holds in the limit $\omega \sim \order1$  as discussed in Sec.\ref{thickgrav}. 

\paragraph*{\underline{\bf Virialisation and characteristic slope:}}

\begin{figure}[!ht]
  \begin{center}
	\includegraphics[angle=-90, scale=0.28]{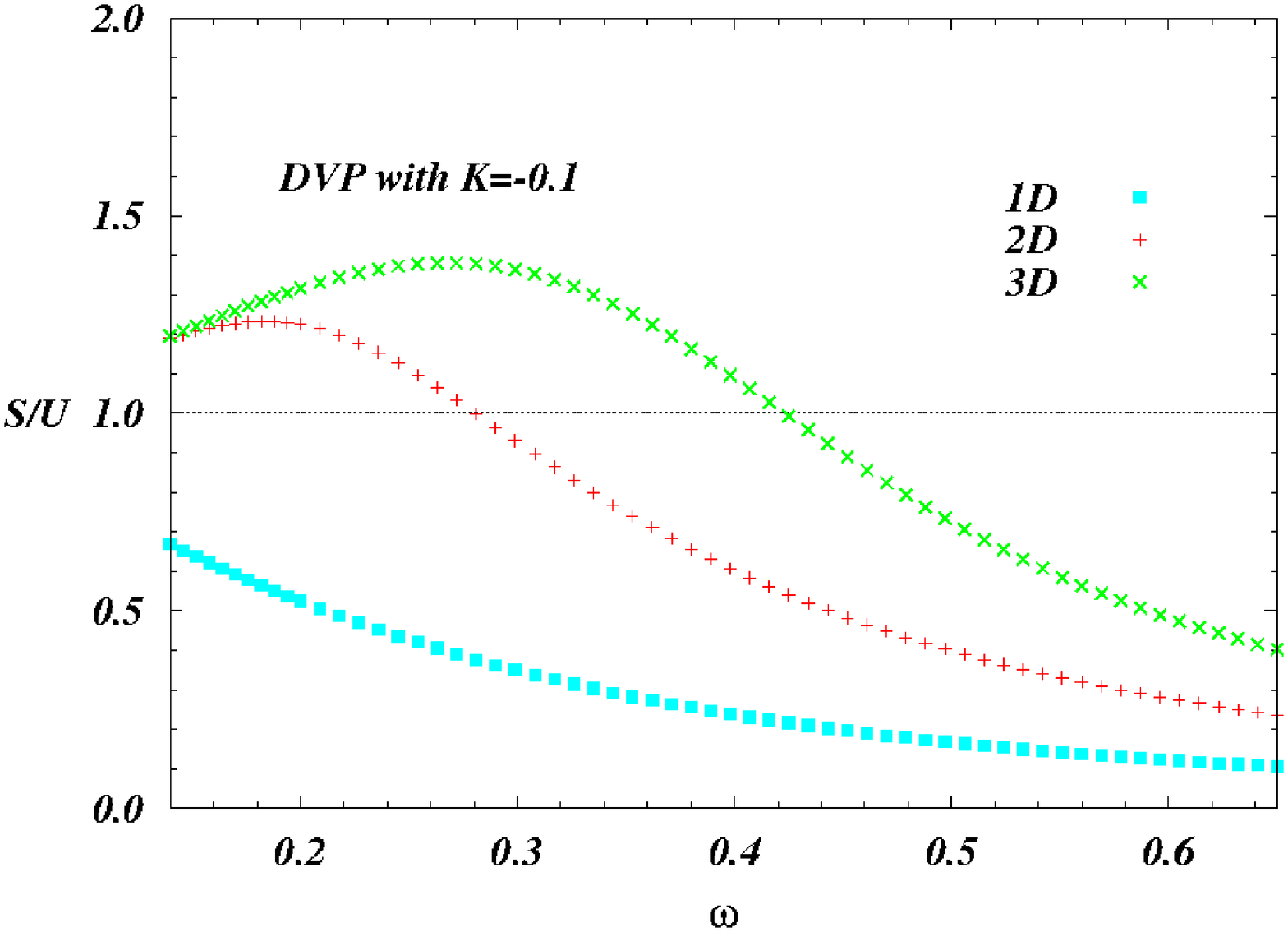}
	\includegraphics[angle=-90, scale=0.28]{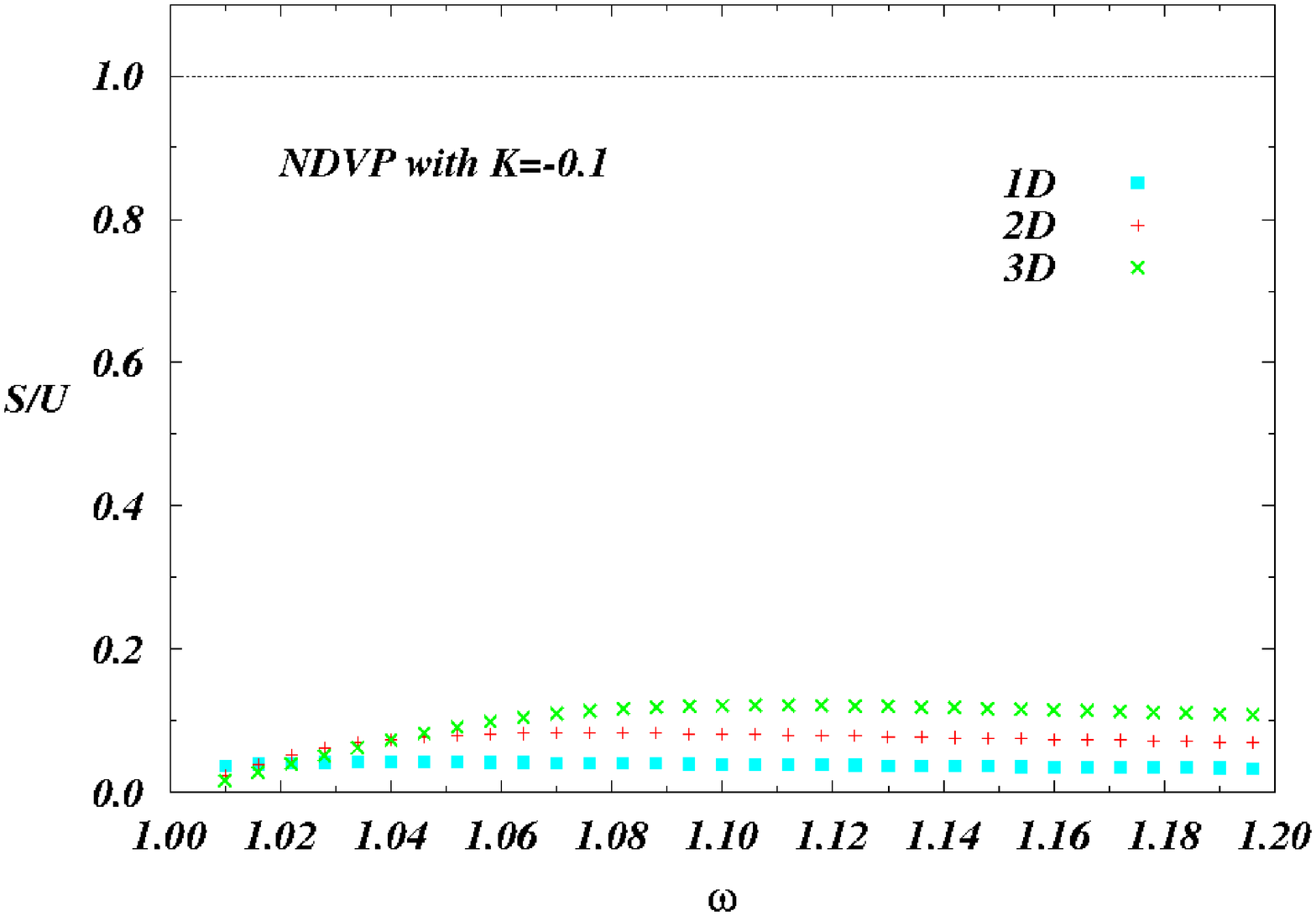}\\
	\includegraphics[angle=-90, scale=0.28]{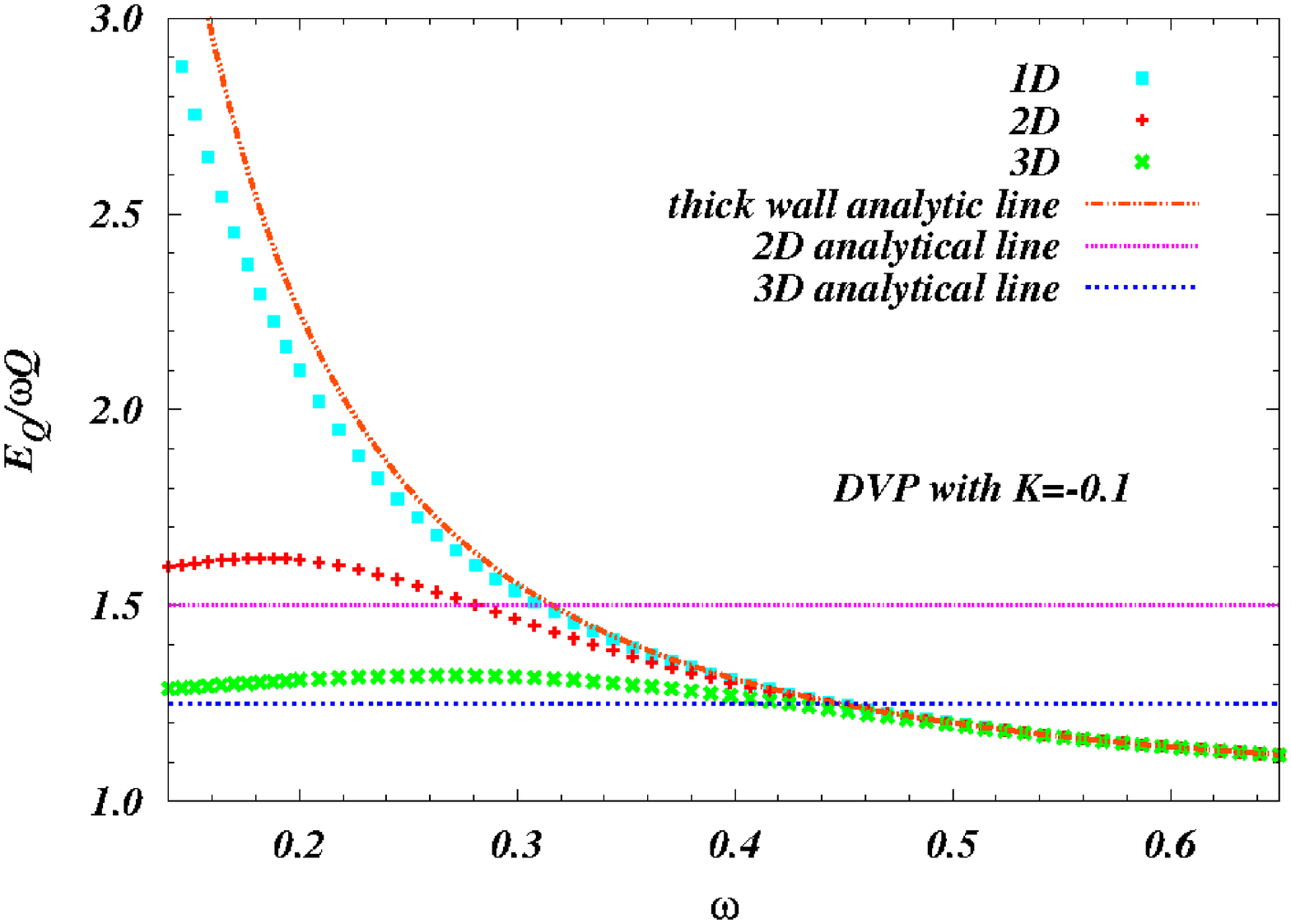}
	\includegraphics[angle=-90, scale=0.28]{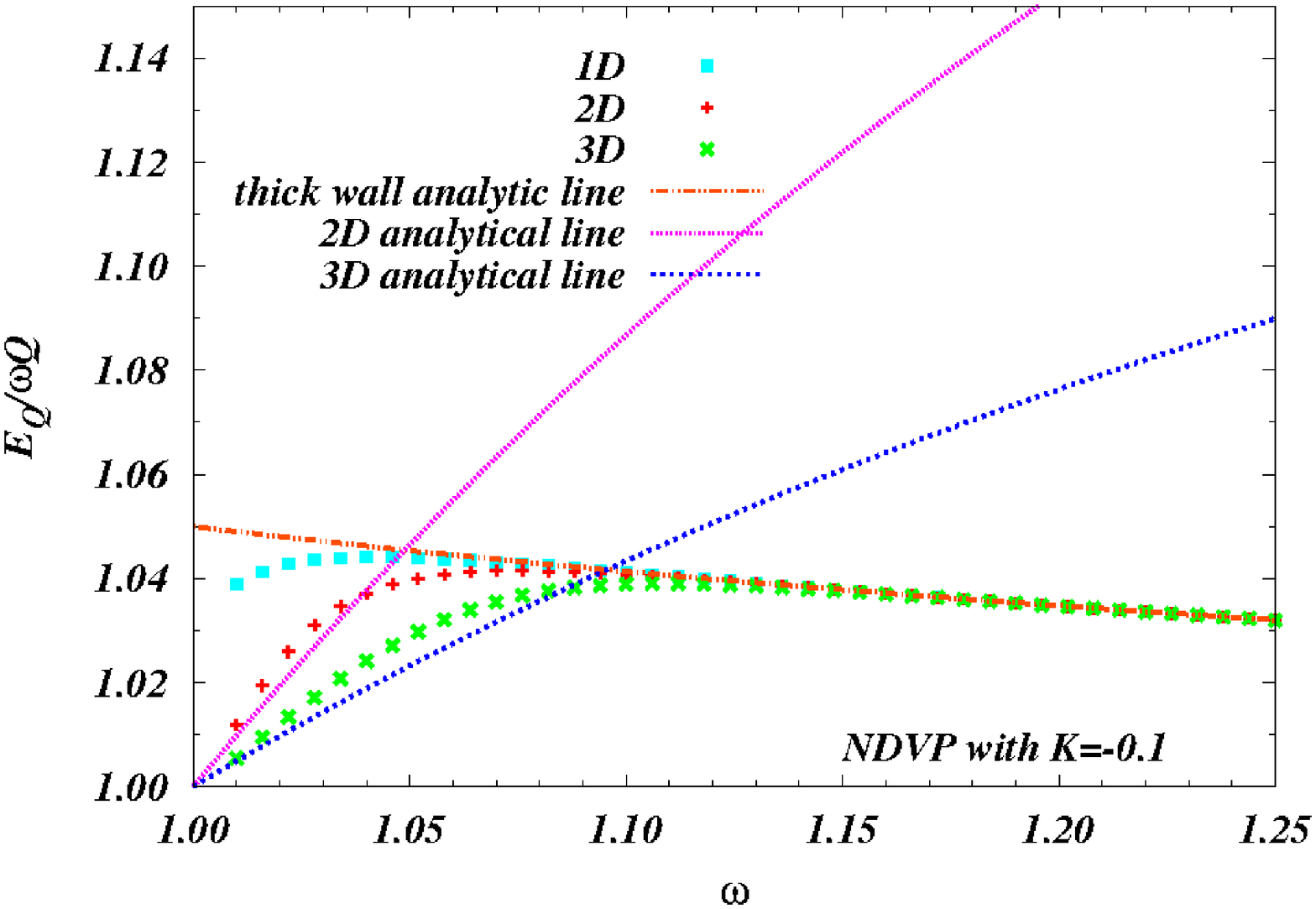}
  \end{center}
  \caption{The top panels show the ratio $\mS/\mU$ where $\mS$ and $\mU$ are the surface and potential energies, and the bottom panels show the numerically obtained  characteristic slope $E_Q/\omega Q$, in $1D$ (skyblue circled-dots), $2D$ (red plus-dots) and $3D$ (green crossed-dots). For  comparison, in the bottom panels, the thin-wall analytic lines obtained using \eq{grvthnch} are also shown (purple-dotted line for $2D$ and blue-dotted line for $3D$)  as are the thick-wall analytic lines obtained from \eqs{grvqeq}{grvch} (orange-dotted-dashed for all $D$). The analytic lines match well with the numeric data in the appropriate limits, especially for the NDVP case.}
  \label{fig:grvst}
\end{figure}

\fig{fig:grvst} shows the $Q$-ball properties plotted against the ratio of $\mS/\mU$ where $\mS$ and $\mU$ are the surface and potential energies (top panels), and the characteristic slope $E_Q/\omega Q$ (bottom panels).  For the DVP case where the thin-wall $Q$-ball satisfies $\sigma_0 \sim \sigma_+$ it appears to be heading towards  $\mS/\mU\sim 1$ as $\omega \to \omega_- = 0$ [see \eq{virieq}], in all three cases. Also we predict that the thin-wall $Q$-ball in NDVP has $\mS/\mU\sim 0$ [see \eq{virieq}], and that it is consistent with what can be seen in the  top right panel around $\omega=\omega_-=1$. The bottom panels  show analytically and numerically the characteristic slopes $E_Q/\omega Q$ in both the  thin and thick-wall limits. The analytic thin-wall lines (purple-dotted line for $2D$ and blue-dotted line for $3D$) based on \eq{grvthnch} are well fitted for the NDVP case with the corresponding numerical dots (red plus-dots for $2D$ and green crossed-dots for $3D$) as long as $\sigma_0\simeq \sigma_+$, see the criteria in \fig{fig:grvsig}. For the DVP case, our numerical data is seen to be heading in the right direction. The numerical solutions for both cases in the thick-wall region are well fitted by the analytic solution in general $D$ given by the orange-dotted-dashed lines, in the second relation of \eq{grvqeq} or \eq{grvch}. From the virial relation \eq{virieq} for $D=1$, we can only predict the extreme values of the $1D$ characteristic slope, $\gamma$, in either the DVP or NDVP case once we know what $\mS/\mU$ is. To obtain that we rely on the numerical simulations and from the top two panels in  \fig{fig:grvst}, we see that for the DVP case with $D=1$,  $\mS/\mU$ appears to be heading towards unity, implying $\gamma \gg 1$ in \eq{virieq}, whereas for the NDVP case $\mS/\mU \ll 1$, implying $\gamma \to 1$ in \eq{virieq}. Comparing these with the bottom two panels we see the behaviour for $\gamma$ appears to follow these predictions.  

\paragraph*{\underline{\bf $Q$-ball stability:}}

\begin{figure}[!ht]
  \begin{center}
	\includegraphics[angle=-90, scale=0.28]{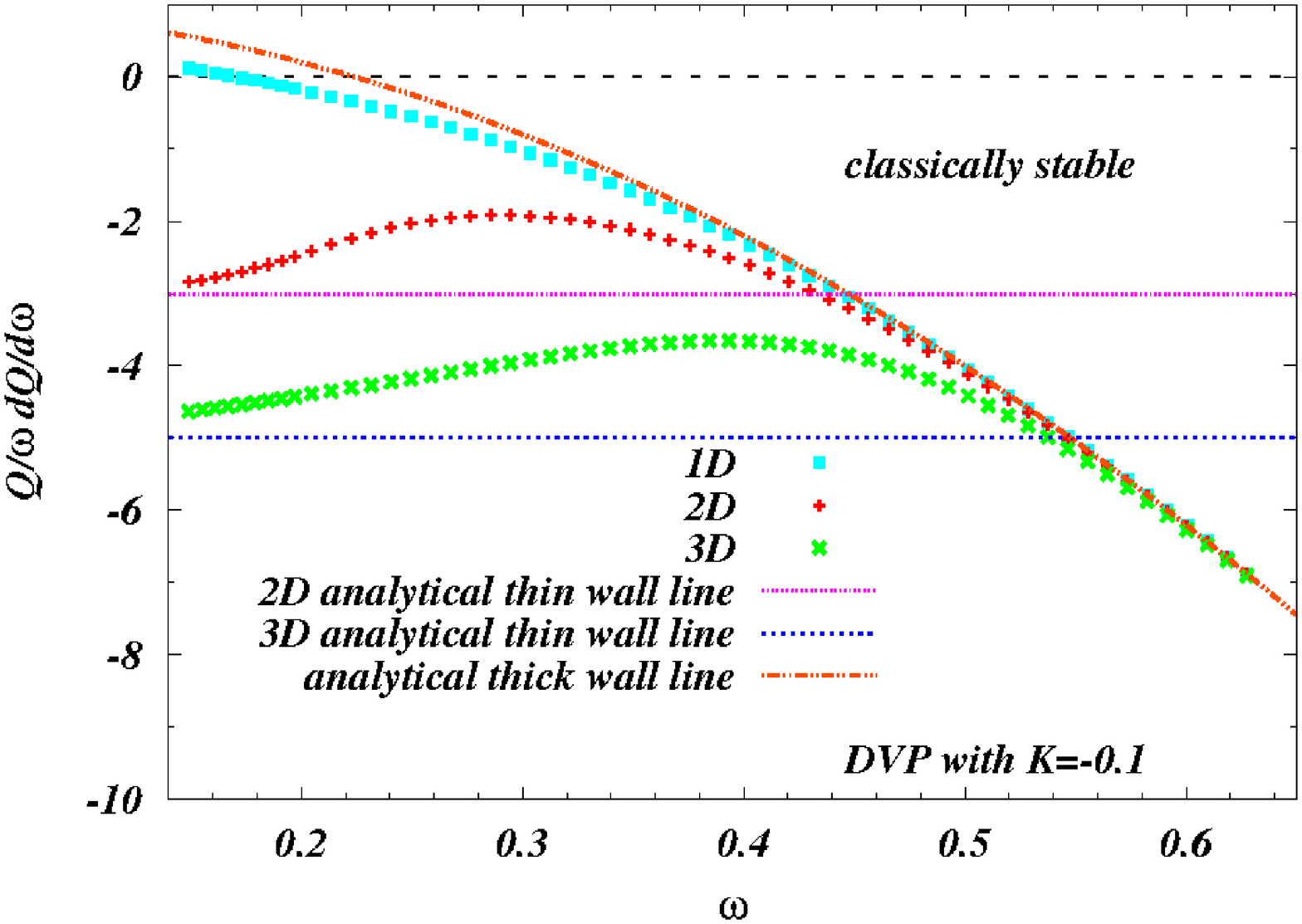} 
	\includegraphics[angle=-90, scale=0.28]{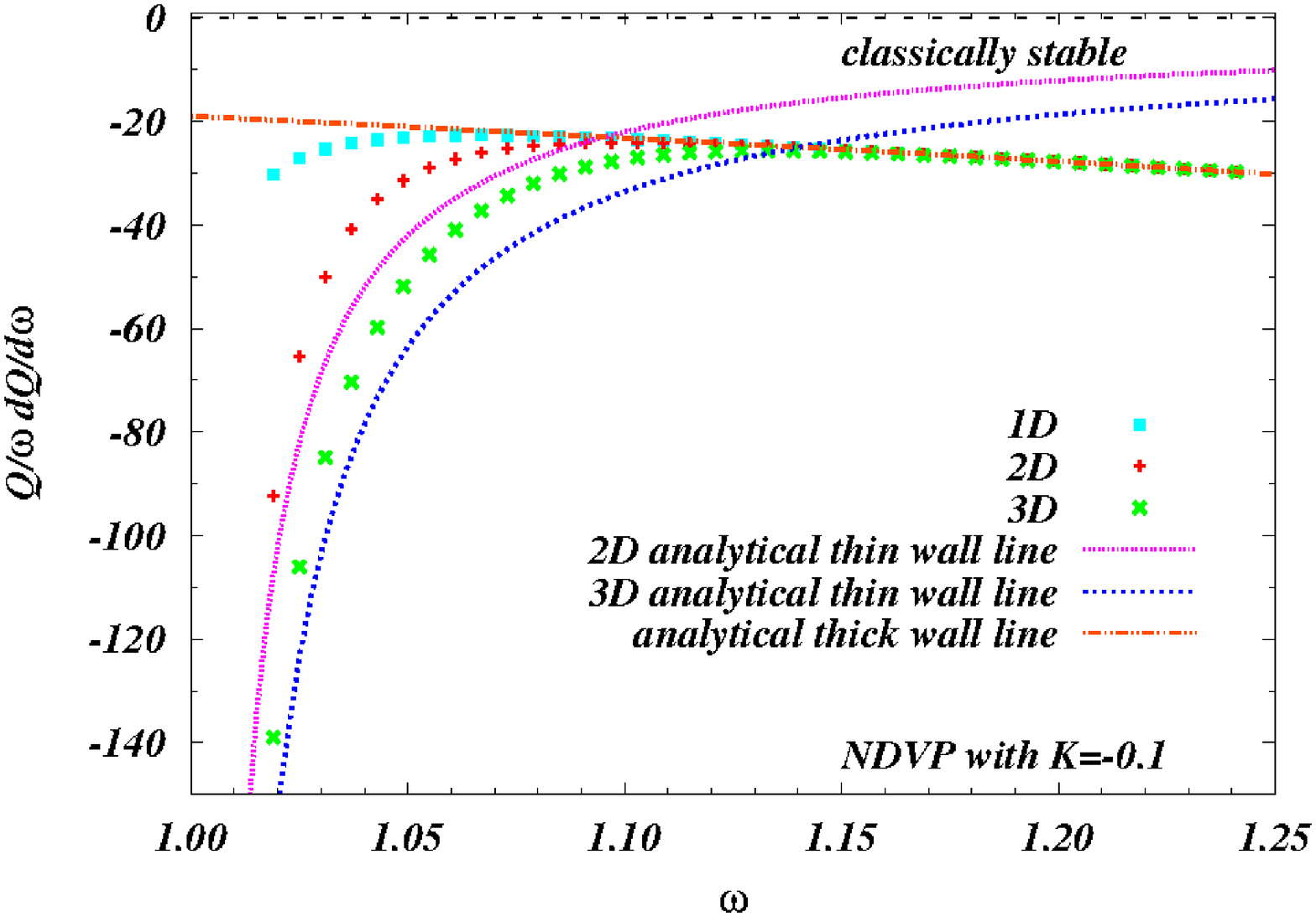} \\
	\includegraphics[angle=-90, scale=0.28]{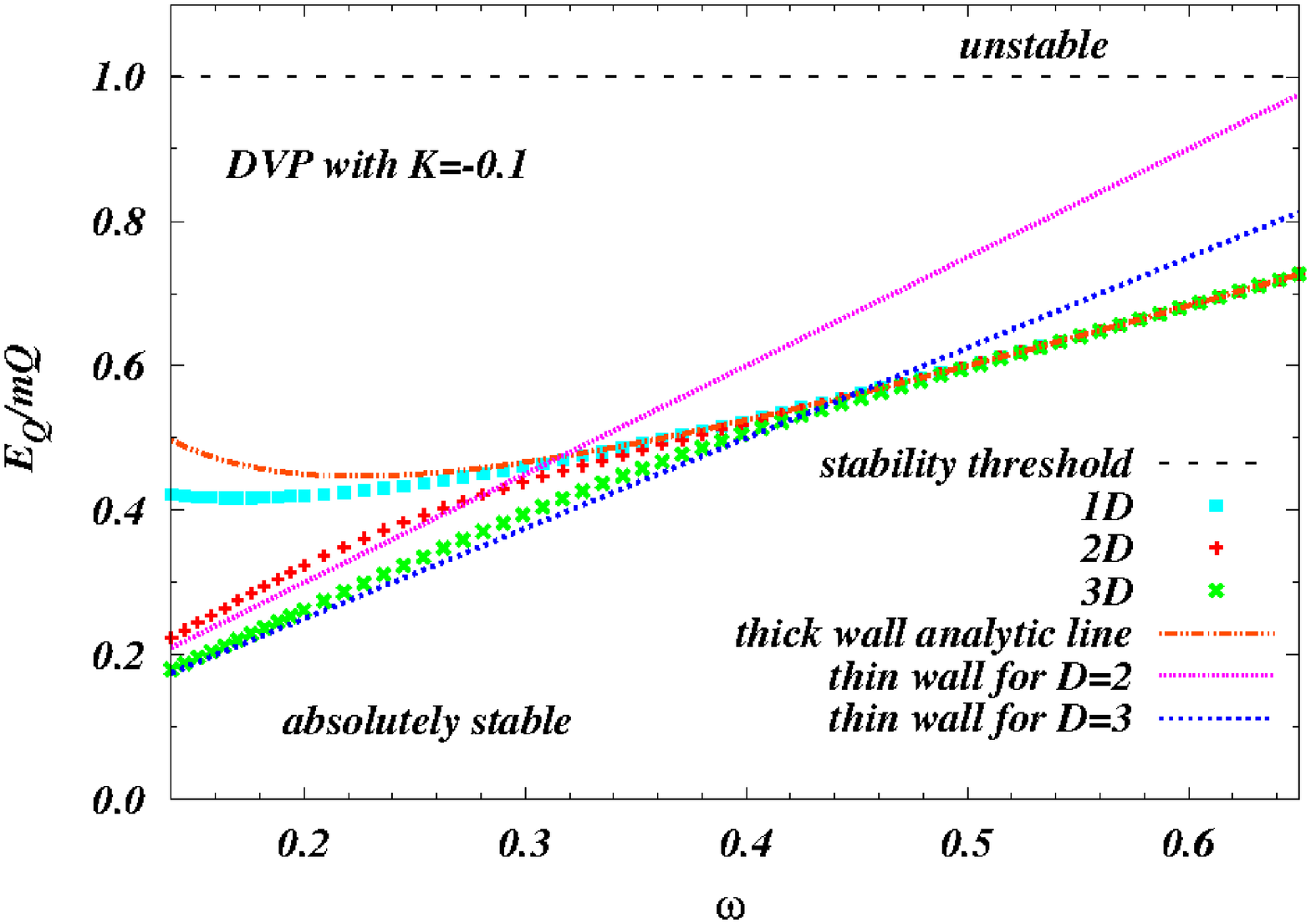} 
	\includegraphics[angle=-90, scale=0.28]{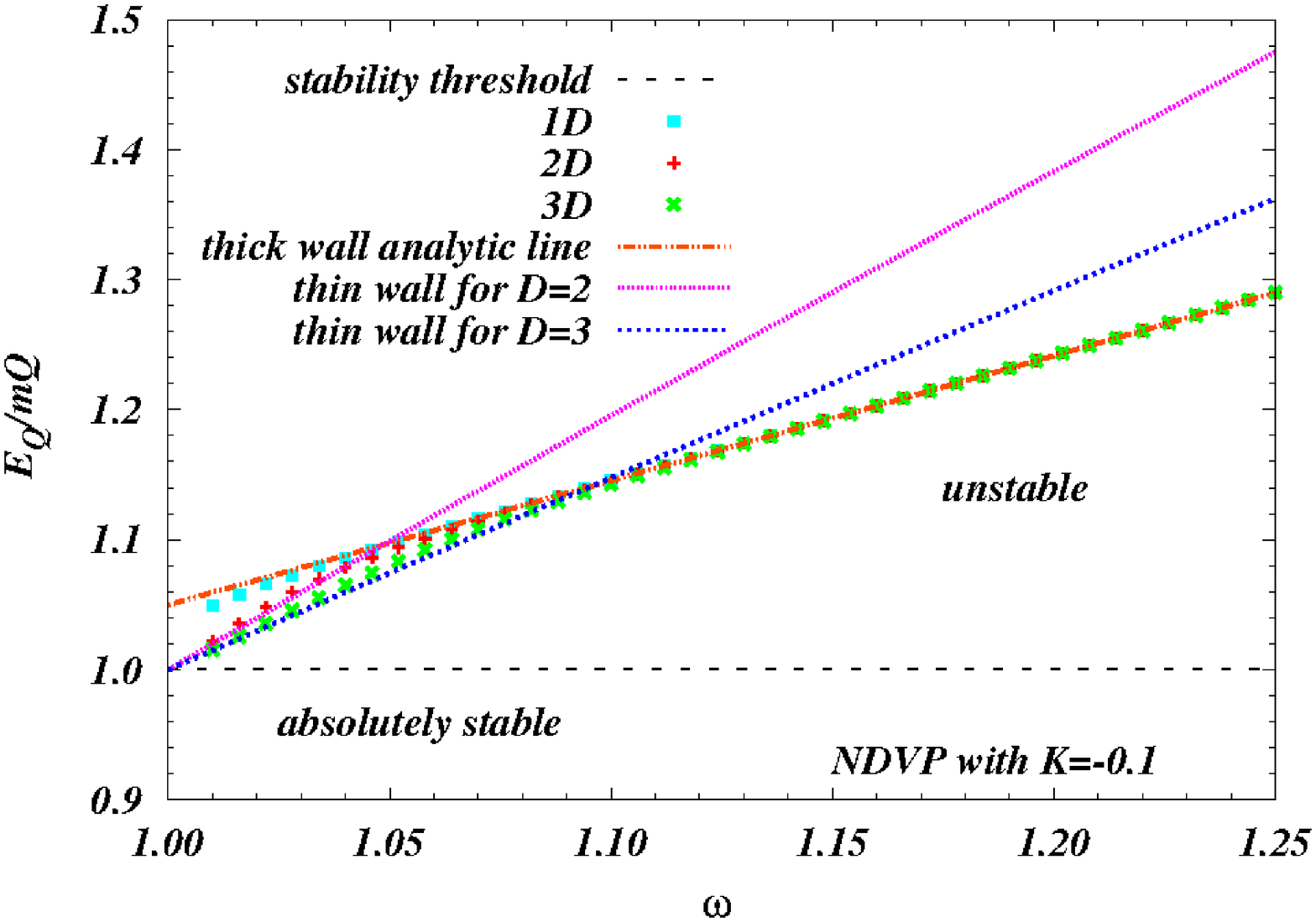}
  \end{center}
   \caption{Classical stability for the top panels and absolute stability for the bottom panels for both DVP (left) and NDVP (right). The black-dashed lines indicate the stability thresholds for both classical and absolute stability in all panels. $Q$-balls found below the lines are stable either (both) classically or (and) absolutely. In the top panels, the analytical lines using \eqs{grvthncls}{grvclscond} agree well quantitatively with the corresponding numerical data for the thick-wall regimes, but not well in the thin-wall regimes. However the numerical plots look qualitatively similar to the analytical lines in the thin-wall limit as seen with the polynomial models in the left-top panel of \fig{fig:clsabs} in chapter \ref{ch:qpots}. In addition, the analytical lines for $E_Q/mQ$ using \eqs{grvthnch}{grvch} match the numerical lines for both the thin and thick-wall limits. }
  \label{fig:grvabs}
\end{figure}

\fig{fig:grvabs} shows plots for both the classical (top panels) and absolute stability (bottom panels) with the stability threshold lines (black-dashed) for the cases of DVP (left) and NDVP (right). Let us consider the classical stability case first. For the thin-wall regime in DVP, notice that the numerical data of $\frac{\omega}{Q}\frac{dQ}{d\omega}$ (red-dot-circles for $2D$ and green-dot-crosses for $3D$) are heading towards the analytic lines of \eq{grvthncls}. For the thick-wall case, on the other hand, the analytical lines of \eq{grvclscond} (orange-dotted-dashed) fit excellently with the numerical data in all dimensions, because \eq{grvclscond} is independent of $D$. Furthermore, the $Q$-ball is classically stable over all values  of $\omega$ except for the $1D$ thin-wall case where our analytical work cannot be applied. We saw this feature of unstable $1D$ thin-wall $Q$-balls for the case of polynomial models in the left-top panel of \fig{fig:clsabs} in chapter \ref{ch:qpots}. For the absolute stability in the bottom panels, the analytical lines using \eq{grvthnch} and \eqs{grvqeq}{grvch} are matched with the numerical dots for both the thin and thick-wall limits. Here, we note how well the three-dimensional $Q$-ball (and also the higher dimensional ones as predicted in chapter \ref{ch:qpots}) can be described simply by our thin and thick-wall $Q$-balls. As our parameter set satisfies \eq{thckabs3}, we can see that absolutely stable $Q$-balls exist in DVP near the thick-wall limit. Because of the choice of $\omega_-=1$, the $Q$-ball in the NDVP case, however, is always absolutely unstable and most of the features are similar in terms of $D$. The analytical lines (top-right panel) in NDVP  agree with the corresponding numerical data qualitatively better than the lines for DVP. 

To sum up our discussion of the gravity-mediated model, our analytical estimates of the characteristic slope and other properties of the $Q$-balls are well checked against the corresponding numerical results, even though we set a ``flatter'' potential with $|K|=0.1 < \order{1}$.

\subsection{Gauge-mediated potential}

This subsection presents numerical results showing the properties of gauge-mediated $Q$-balls with $m=1,\; \Lambda^2=2$ in \eq{apprxpot}. Although we have obtained analytical results for the potential, \eq{potgauge}, the potential is neither analytic nor smooth for all $\sigma$. Therefore, we shall use the approximate potential, \eq{apprxpot}, see \fig{fig:gaupot} and we expect that \eq{apprxpot} is a suitable approximation especially for the thin-wall limit $\omega$ and large $D$. We will also see and explain the expected discrepancies that exist between the numerical and analytic results.

\paragraph*{\underline{\bf Hybrid profile:}}
As we saw in earlier examples the numerical profiles we have obtained have errors for large $r$, which correspond to either undershooting or overshooting; thus, we replace the numerical data in that regime by the exact asymptotic analytic solutions we obtained using the second relation of \eq{gauasym} to smoothly continue the numerical solutions to the corresponding analytical ones. The hybrid profile in this model is
\be\label{hybridgau}
    \sigma(r)=
    \left\{
    \begin{array}{ll}
    \sigma_{num}(r)&\ \ \textrm{for $r<R_{num}$}, \\
    \sigma_{num}(R_{num})\bset{\frac{R_{num}}{r}}^{(D-1)/2} e^{-\mo(r-R_{num})} &\ \ \textrm{for $R_{num} \le r \le R_{max}$},
    \end{array}
    \right.
\ee
where $\sigma_{num}$ is the numerical raw data, $R_{num}$ is determined by $|\frac{D-1}{2r}+\mo+\bset{\sigma^\p_{num}/\sigma_{num}}| < 0.001$, and we have again set $R_{max}=60$. We have calculated the following numerical properties using the above hybrid profile, \eq{hybridgau}, up to $D=3$.

\paragraph*{\underline{\bf Profile and energy density configuration:}}

\begin{figure}[!ht]
  \begin{center}
    \includegraphics[angle=-90, scale=0.50]{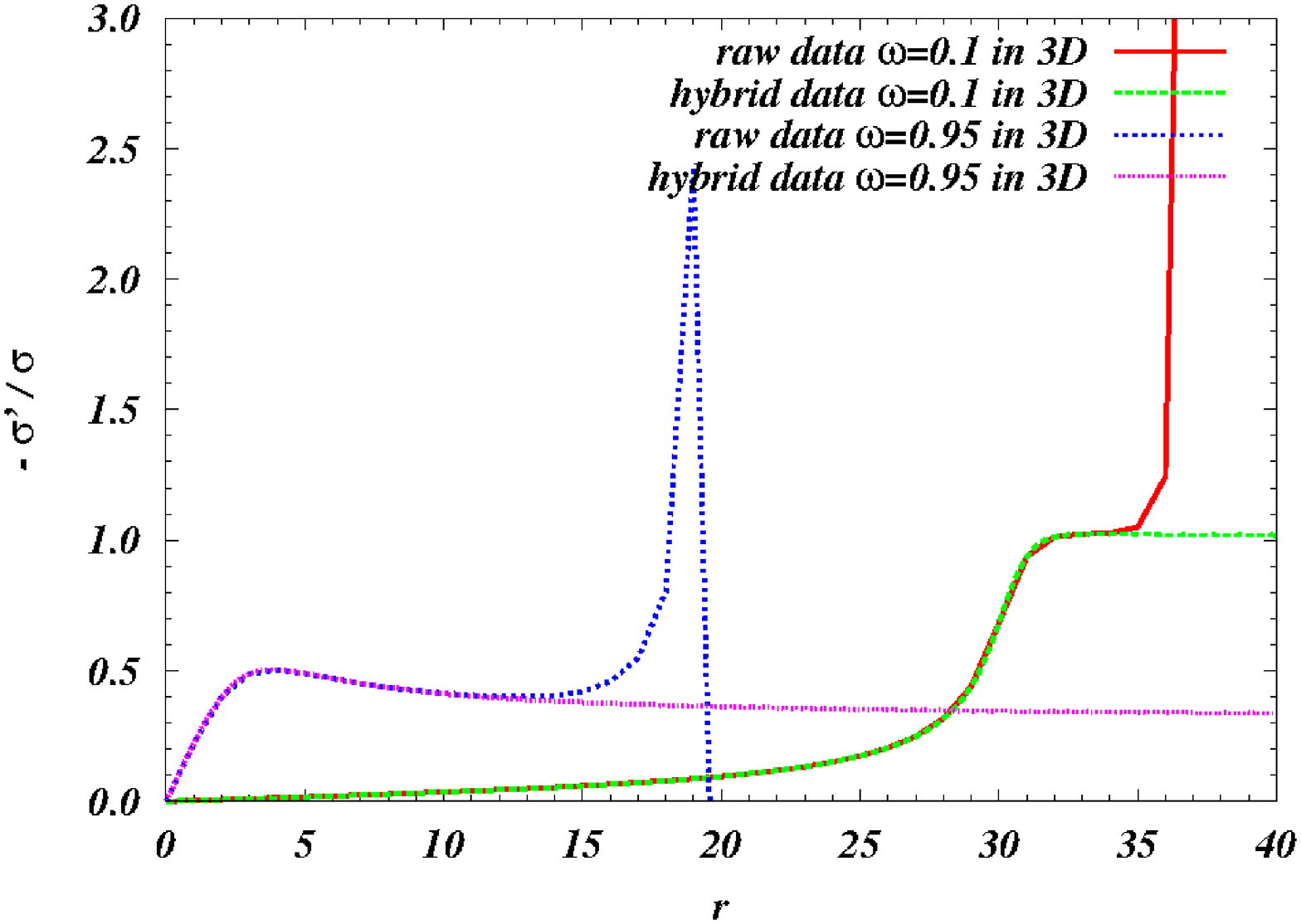}\\
   \includegraphics[angle=-90, scale=0.28]{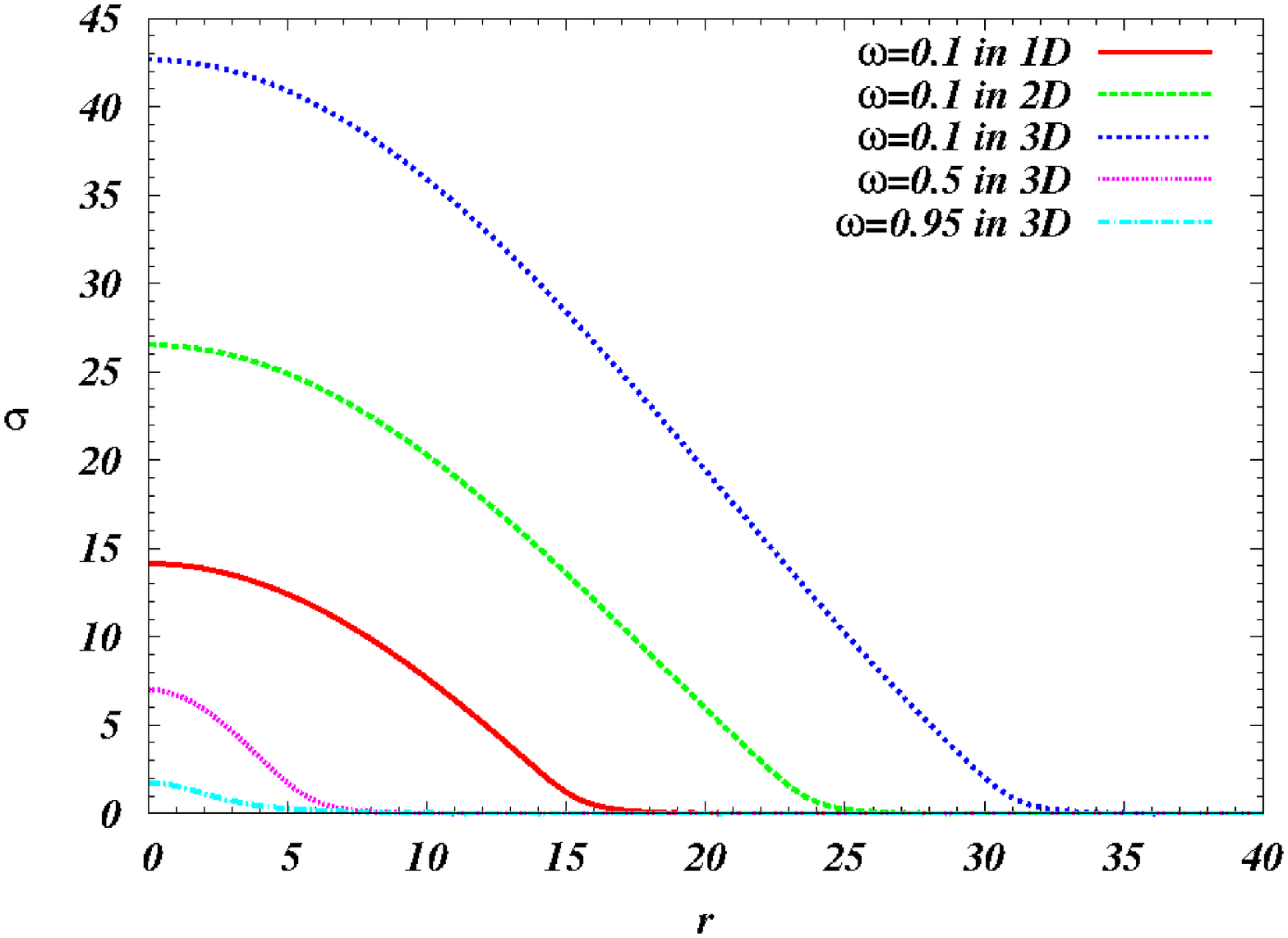} 
   \includegraphics[angle=-90, scale=0.28]{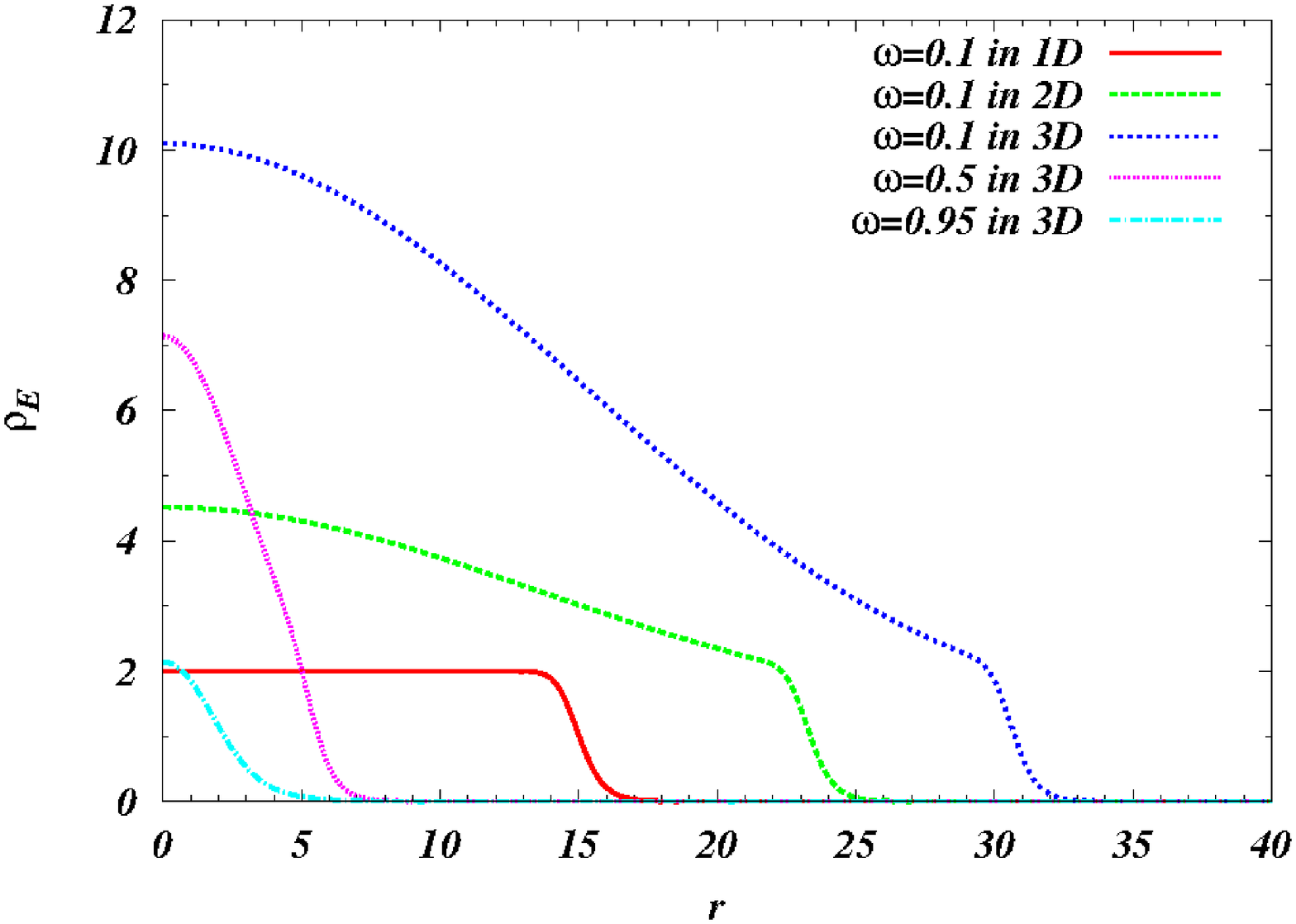}
  \end{center}
  \caption{The top panel shows the three-dimensional numerical slopes $-\sigma^\p/\sigma$ for two values of $\omega$. The raw numerical data (red-solid and blue-dotted lines) matches smoothly to the corresponding analytical asymptotic profiles for large $r$ (green-dotted and purple-dashed lines). Both the left- and right-bottom panels show, respectively, the hybrid profiles \eq{hybridgau} and the energy density configurations for the various values of $\omega$ and $D$. The spikes of energy density configurations do not exist even in the thin-wall limits. }
  \label{fig:gaupro}
\end{figure}

\fig{fig:gaupro} shows the three-dimensional numerical slopes \\$-\sigma^\p/\sigma$ for two values of $\omega$ (top), hybrid profiles (left-bottom) as in \eq{hybridgau}, and the configurations for energy density (right-bottom). In the top panel, the raw numerical data (red-solid and blue-dotted lines) is matched smoothly onto the continuous asymptotic profiles \eq{hybridgau} for large $r$ (green-dotted and purple-dashed lines). By fixing the numerical raw data using the technique \eq{hybridgau}, we show the profiles for various values of $\omega$ and $D$, see the left-bottom panel. Also the peaks of the energy density cannot be observed in the whole range of $\omega$, see the right-bottom panel.

\paragraph*{\underline{\bf Characteristic slope:}}

\begin{figure}[!ht]
  \begin{center}
	\includegraphics[angle=-90, scale=0.5]{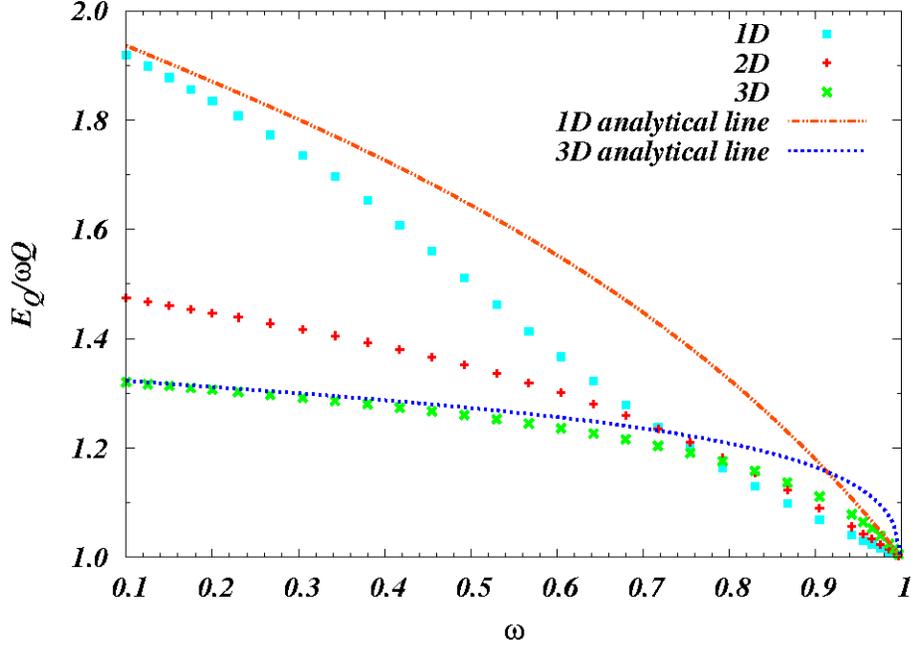}
  \end{center}
  \caption{The numeric characteristic slopes $E_Q/\omega Q$ and the analytic lines (orange-dashed line for $1D$ and blue-dotted line for $3D$) which are calculated using \eqs{gauR1}{gauR3} in the whole range of $\omega$. The $3D$ analytic line agrees with the numeric data well except near the thick-wall limit. Similarly the $1D$ analytic line agree well only in the extreme thin-wall limit.}
  \label{fig:gaueq}
\end{figure}

In \fig{fig:gaueq}, we plot both the numeric and analytic characteristic slopes $E_Q/\omega Q$ (orange-dashed line for $1D$ and blue-dotted line for $3D$). By substituting \eqs{gauR1}{gauR3} into \eq{Uo} and \eq{gauq}, we have obtained the analytic slopes covering the whole range of $\omega$. The $3D$ analytic line agrees with the numerical data except near the thick-wall limit. Similarly, the $1D$ analytic line agrees well only in the thin-wall limit. The origin of the discrepancies in the analytic versus numerical fits are the differences between the potentials themselves [\eq{potgauge} and \eq{apprxpot}]. These differences are largest between $1 \lesssim \sigma \lesssim 3$ which in turn affects the region around $0.9\lesssim \omega < 1.0$, see \fig{fig:gaupot} and \fig{fig:gaueq}. 

\paragraph*{\underline{\bf $Q$-ball stability;}}

\begin{figure}[!ht]
  \begin{center}
	\includegraphics[angle=-90, scale=0.28]{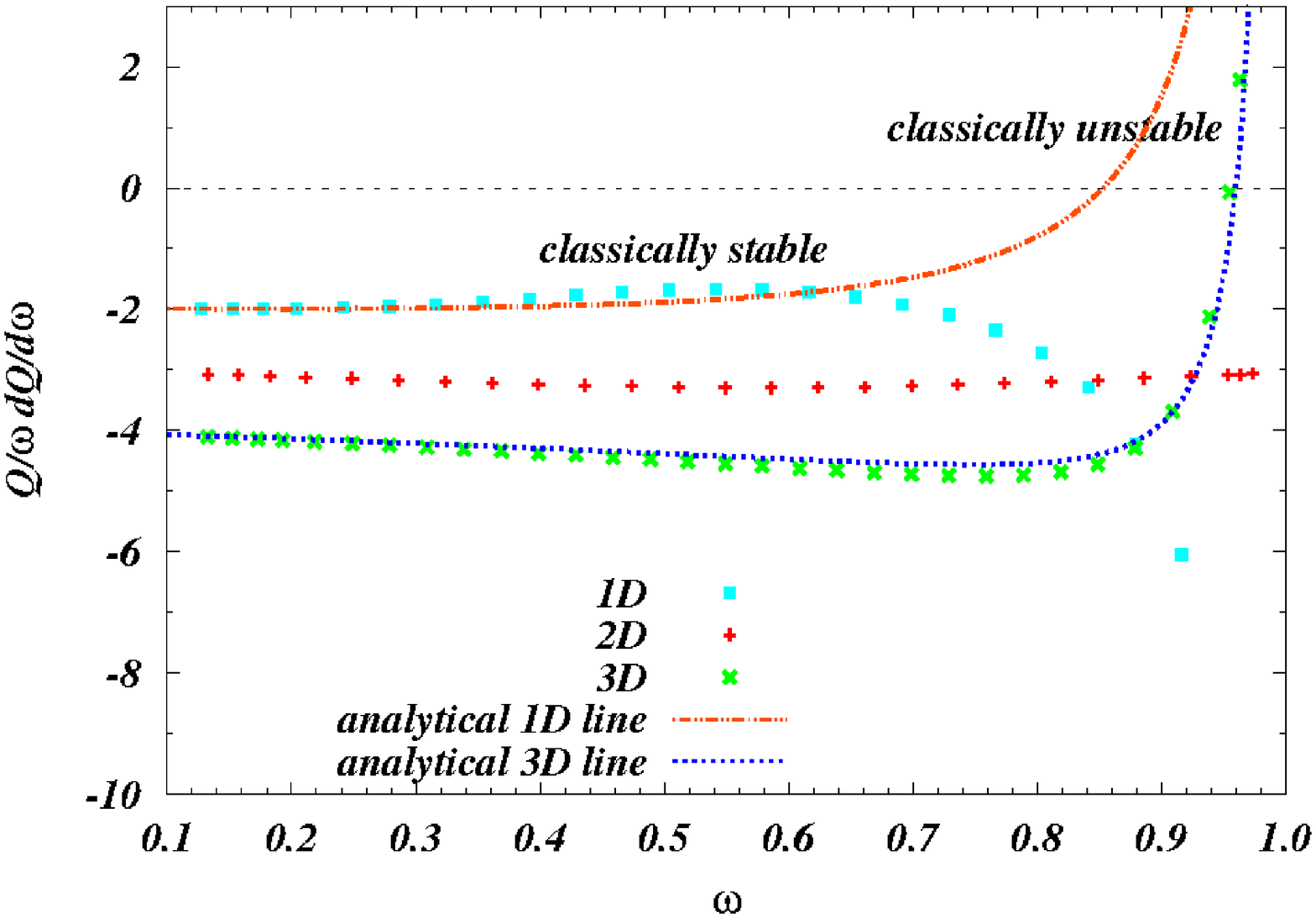} 
	\includegraphics[angle=-90, scale=0.28]{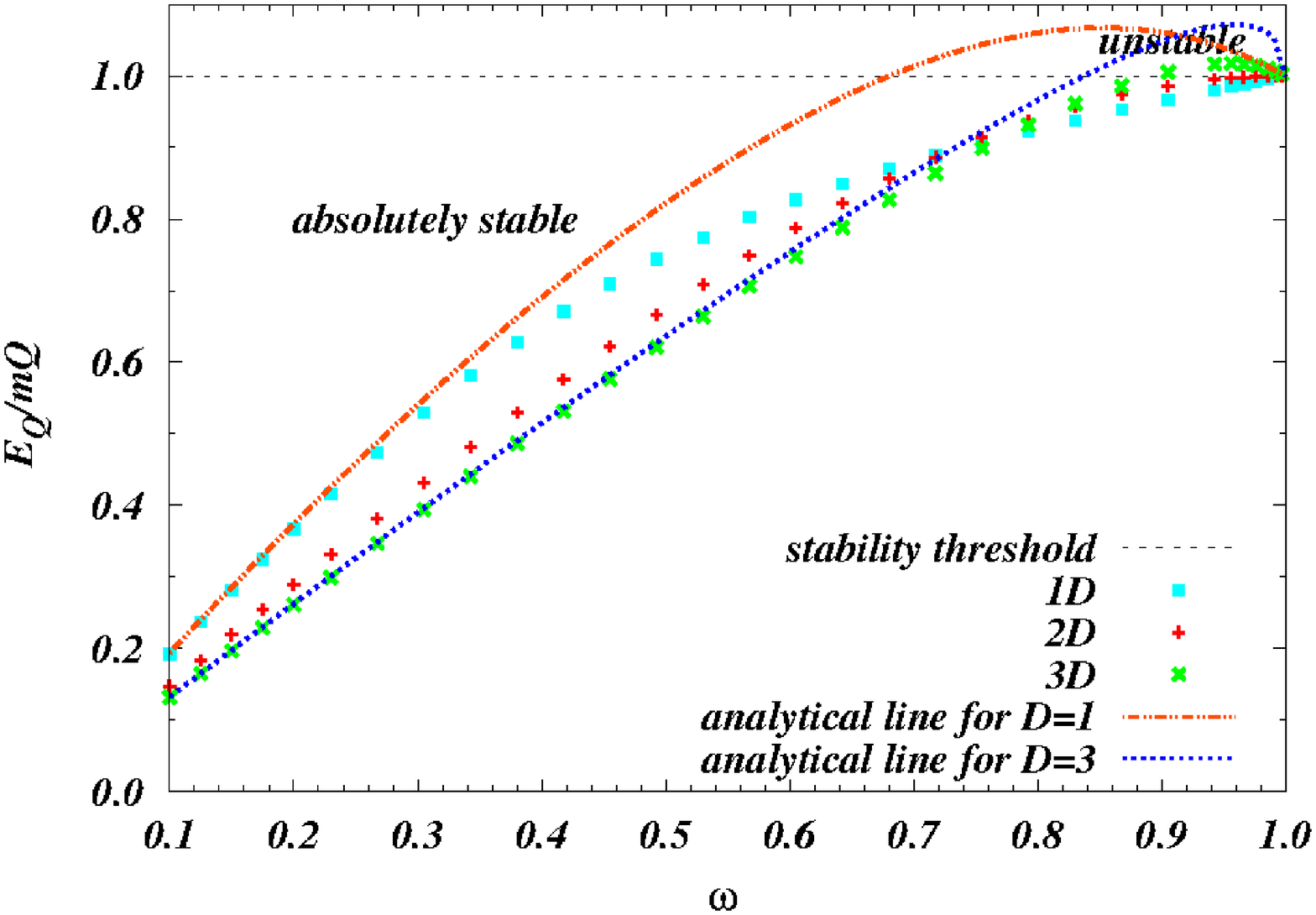} 
  \end{center}
   \caption{The stability of $Q$-balls -- Classical (left panel) and absolute (right panel). The black-dashed lines in the two panels indicate the stability thresholds for both classical and absolute stability where $Q$-balls under the lines are classically/absolutely stable. The analytic lines for $D=1,\; 3$ are calculated by substituting \eqs{gauR1}{gauR3} into \eq{gauq} and differentiating it with respect to $\omega$.}
  \label{fig:gauabs}
\end{figure}

\fig{fig:gauabs} illustrates the stability of $Q$-balls: classical stability in the left panel and absolute stability in the right panel. The black-dashed lines in both panels indicate their respective stability thresholds where $Q$-balls under the lines are stable. We calculate the analytic lines for $D=1,\; 3$ by substituting \eqs{gauR1}{gauR3} into \eq{gauq} and differentiating it with respect to $\omega$. The $3D$ numerical data can be matched with the analytic lines in both the thin and thick-wall limits. As in \eq{3dcls}, the three-dimensional $Q$-ball in the thick-wall limit is classically unstable. The numerical thick-wall $Q$-ball in $1D$ is classically stable which differs from the prediction in \eq{1dcls}. In the right panel, the analytic line for $D=3$ agrees with the numerical data except in the thick-wall limit where the analytical lines for both $1D$ and $3D$ do not match the corresponding numerical data. Furthermore, the thick-wall $Q$-ball in $1D$ is absolutely unstable as predicted analytically in \eq{1dch}, but this fact cannot be observed numerically. The reasons for this discrepancy are as before a problem with our choice of potentials. We can see that the thin-wall $Q$-balls for any $D$ are both classically and absolutely stable.

To recap, our numerical results in the gauge-mediated case are generally well fitted by our analytical estimations. Observed discrepancies between the analytical predictions and numerical data arise from the artifact of our approximated smooth potential \eq{apprxpot} for the generalised gauge-mediated potential \eq{potgauge}. We have confirmed that the thin-wall $Q$-balls for any $D$ are both absolutely and classically stable.

\section{Conclusion and discussion}
\label{conc}

We have explored stationary properties of $Q$-balls in two kinds of flat potentials, which are the gravity-mediated potential, \eq{sugra}, and the generalised gauge-mediated potential, \eq{potgauge}. Generally, the gauge-mediated potential is extremely flat compared to the gravity one; therefore, we cannot apply our thin-wall ansatz \eq{thinpro} to the gauge-mediated case. By linearising the gauge-mediated potential, we obtained the analytical properties instead. For both potential types, we both analytically and numerically examined characteristic slopes as well as the stability of the $Q$-balls in the thin and thick-wall limits. Our main analytical results are summarised in \tbl{tbl:results}.

This present chapter is of course related to chapter \ref{ch:qpots}. The key differences are that in the present work on thin-wall $Q$-balls we are assuming the value of $\sigma_+(\omega)$ for the thin-wall limit $\omega\simeq \omega_-$ depends weakly on $\omega$ and we have replaced the assumption $\sigma(R_Q)<\sigma_-(\omega)$ by the equivalent assumption (made by Coleman) $\Uo\simeq U_{\omega_-}$ in the $Q$-ball shell region \cite{Coleman:1977py}. These in turn are related to the previous requirement that the surface tension $\tau$ depends weakly on $\omega$, which can be translated into the main assumptions: $R_Q\gg \delta, 1/\mu, \sigma_0\simeq \sigma_+$, and $\Uo\simeq U_{\omega_-}$ in the shell region. Furthermore, our analytic work agrees well with the numerical results for small curvature $\mu$ with $|K|=0.1$; however, it is not clear that our analytic framework still holds even in the case of $|K|\ll \order1$, which corresponds to a case where the potential is extremely flat, see \eq{mucurv}.

\paragraph*{\underline{\bf $Q$-balls in gravity-mediated potentials:}}

It is possible to obtain absolutely stable $Q$-matter with a small coupling constant, \eq{restbeta}, for the nonrenormalisable term in \eq{sugra}. For $|K|\not\ll \order1$, a gravity-mediated potential cannot be really considered as flat, which allows us to apply our previous results, Eqs (\ref{RQ}-\ref{q1class}), in chapter \ref{ch:qpots} to describe the thin-wall $Q$-ball where $\sigma_0\simeq \sigma_+$. In the thick-wall limit by reparameterising parameters in $\So$ and neglecting the nonrenormalisable term under the conditions $\beta^2 \lesssim |K| \lesssim \order{1}$, we have obtained the stationary properties of the $Q$-ball. We showed that the ``thick-wall'' $Q$-ball is classically stable, and demonstrated that under certain conditions \eq{thckabs3} it can be absolutely stable. Although this analysis is much simpler than the analysis associated with imposing a Gaussian ansatz developed in appendix \ref{appxthick}, the former analysis assumed that the nonrenormalisable term is negligible at the beginning of the analysis. In the latter analysis, we have kept all terms in \eq{repsugra} and shown that the nonrenormalisable term is indeed negligible in the limit $\omega \gtrsim \order{m}$. Our results, \eqs{grvclscond}{grvch}, for the thick-wall $Q$-ball have recovered the previous results obtained in \cite{Enqvist:1998en, Morris:1978ca} without any contradictions for classical stability conditions as opposed to the case of using a Gaussian ansatz in a general polynomial potential in which we showed that the ansatz led to a contradiction and corrected it by introducing a physically motivated ansatz in chapter \ref{ch:qpots}. This is because the Gaussian ansatz, \eq{testgauss}, becomes the exact solution, \eq{gauss}, in the gravity-mediated potential in the limit $\omega \gtrsim \order{m}$ where the nonrenormalisable term is negligible. In \figs{fig:grvst}{fig:grvabs} the analytical lines agree well with the corresponding numerical plots in both the thin-wall and thick-wall limits. Under our numerical parameter sets, the $Q$-balls in DVP are both classically and absolutely stable up to $\omega \lesssim m$, while all of the $Q$-balls in NDVP are absolutely unstable because of our choice, $\omega_-=m$. We believe that an absolutely stable $Q$-matter exists in NDVP when we take $\omega_-<m$. Since the $Q$-balls in both potential types are always classically stable, as can be seen in the top two panels of \fig{fig:grvabs} except for the case of $1D$ $Q$-balls in the thin-wall limit to which our analytical work cannot be applied since it holds only for $D\ge 2$. We have also found the asymptotic profile \eq{asympro} for all possible values of $\omega$, see the top two panels in \fig{fig:grvpro}.

Our analytical estimations on the value of $\frac{\omega}{Q}\frac{dQ}{d\omega}$ do not agree well with the numerical results, because $\sigma_0\not\sim \sigma_+$. Nevertheless the other analytical properties are well fitted especially in NDVP, see bottom panels in \figs{fig:grvst}{fig:grvabs}. The DVP in \eq{sugra} for small $|K|$ is extremely flat as the gauge-mediated potential in \eq{potgauge}, where both of the potentials have $\omega_-\simeq 0$. Notice that the asymptotic profile for the former case has a Gaussian tail, while the latter profile is determined by the usual quadratic mass term, see \eqs{asympro}{gauasym}. By assuming that the shell effects are much smaller than the core effects in the thin-wall limit, the difference of the tails can be negligible. Indeed, we can see the thin-wall numerical lines for both the classical stability and the characteristic slope look qualitatively and quantitatively similar to each other, as can be seen in both the top/bottom left panels of \fig{fig:grvabs} and the panels of \fig{fig:gauabs}. Notice that the spikes of energy density in the gauge-mediated potential cannot be seen even though $\omega_-\simeq 0$, see \fig{fig:gaupro}.

Furthermore, we know that the potential $U_{grav}$ can be approximated by $\half m^2 M^{2|K|}\sigma^{2-2|K|}$ for small $|K| \ll \order1$, then the potential in \eq{sugra} looks similar to the confinement model in \cite{Simonov:1979rd, Mathieu:1987mr}. By neglecting the nonrenormalisable terms in the thick-wall limit, we can easily obtain the characteristic slope, $\gamma = \frac{2+|K|(D-1)}{2+|K|(D-2)}\simeq 1$, \cite{Dine:2003ax} by following the same technique as in \eq{easycalc}, which does not depend on $\omega$ but does depend on $D$ and $|K|$. It follows that $E_Q\propto Q^{1/\gamma}$ from \eq{leg2}. This result is obviously worse than our main results in \eqs{grvqeq}{grvch}, see bottom two panels in \fig{fig:grvst}, because we know that the Gaussian ansatz \eq{testgauss} can be the exact solution \eq{gauss} for $U=U_{grav}$; thus, it is not so powerful to approximate $U_{grav}$ by $\half m^2 M^{2|K|}\sigma^{2-2|K|}$ for small $|K|$.

\paragraph*{\underline{\bf $Q$-balls in gauge-mediated potentials:}}

For the gauge-mediated potential in \eq{potgauge}, we obtained the full analytic results in $D=1,\; 3$ over the whole range of $\omega$ using \eqs{gauR1}{gauR3}, see \fig{fig:gaueq} and \fig{fig:gauabs}. In the ``thin-wall'' limit for $\mo R,\; \omega R \gg \order{1}$, we reproduced the previously obtained results, \eq{gauthnch}, in \cite{Dvali:1997qv, Asko:2002phd, MacKenzie:2001av} and showed that they are both classically and absolutely stable in \eqs{gauthnch}{gaucls}. The one- and three-dimensional ``thick-wall'' $Q$-balls, on the other hand, are neither classically nor absolutely stable, see either \eqs{1dcls}{1dch} or \eqs{3dcls}{gauthinch}, respectively. Since the potential, \eq{potgauge}, is not differentiable everywhere, we have used the approximate potential, \eq{apprxpot}, instead in the numerical section, Sec. \ref{numerics}. \figs{fig:gaueq}{fig:gauabs} show that the numerical results agree with the analytical results in the thin-wall limit. The numerical data near the ``thick-wall'' limit and/or in the $1D$ case differ from the analytic lines since the profiles are computed in the region where the two potentials between \eq{apprxpot} and \eq{potgauge} are different, see \fig{fig:gaupot}. This differences come from the artifact of our approximated smooth potential \eq{apprxpot} against the generalised gauge-mediated potential \eq{potgauge}.

\paragraph*{\underline{\bf The $3D$ $Q$-balls:}}
Although we have shown $Q$-ball results for an arbitrary number of spatial dimensions $D$, only three-dimensional cases are phenomenologically interesting. $Q$-balls in flat potentials give the proportional relation $E_Q \propto Q^{1/\gamma}$, where $\gamma$ generally depends on $D$. The actual values of $1/\gamma$ for three-dimensional thin-wall $Q$-balls are $\frac{4}{5},\; 1,$ and $\frac{3}{4}$ in DVP, NDVP of gravity-mediated potentials and in gauge-mediated potentials respectively. It implies that the gauge-mediated $Q$-balls would be formed in the most energetically compact state for a large charge $Q$, so it is likely that such formed $Q$-balls would have survived any possible decay processes and thermal evaporation until the present day, and possibly become a dark matter candidate  \cite{Kusenko:1997si}.

\paragraph*{\underline{\bf Dynamics and cosmological applications:}}

The dynamics of a pair of one-dimensional $Q$-balls has been recently analysed using momentum flux \cite{Bowcock:2008dn}. 
For a large separation between the $Q$-balls, the profiles develop the usual exponential tail, $e^{-\mo r}$, in general polynomial potentials and  in \cite{Bowcock:2008dn} the authors showed that there was a solitonic force between them. Profiles in the gravity-mediated models and other confinement models, however, have different asymptotic tails, which may affect the detailed dynamics and the $Q$-ball formation \cite{Kasuya:2000wx, Multamaki:2000qb, Multamaki:2001az, Kusenko:2009cv}.

In a cosmological setting (thermal background), SUSY $Q$-balls are generally unstable via evaporation, diffusion, dissociation, and/or decay into todays baryons and lightest supersymmetric particles, if the AD field couples with the thermal plasma, which are decay products from inflaton, and/or if the field possesses a lepton number for the MSSM flat directions \cite{Enqvist:1997si, Enqvist:1998en}. Following our detailed analytical and numerical analyses of both gravity-mediated and gauge-mediated $Q$-balls, it is clear that this whole area of dynamics and cosmological implications of these $Q$-balls deserves further analyses.
\begin{figure}[!ht]
  \def\@captype{table}
  \begin{minipage}[t]{\textwidth}
   \begin{center}
      \begin{tabular}{|c||c|c|c|}
	\hline
	Model & \multicolumn{3}{|c|}{Gravity-mediated potentials} \\
    	\hline
	$Q$-ball type & \multicolumn{2}{|c|}{Thin-wall} & Thick-wall  \\ \hline
	Conditions & \multicolumn{2}{|c|}{$\blacktriangle$} & $\beta^2 \lesssim |K|\lesssim \order{1}$\\
	\hline
	Assumptions & \multicolumn{2}{|c|}{$R_Q \gg \delta, 1/\mu;\; \sigma_0 \simeq \sigma_+$ and $\Uo\simeq U_{\omega_-}$ in shell} & None\\
	\hline
	Potential type & DVPs & NDVPs & Both\\
    	\hline
	& & &  \\
	$1/\gamma$ & $\frac{2D-1}{2(D-1)}$ & 1 & 1  \\
	Absolute stability & $\bigcirc$ & $\bigtriangleup$ & $\bigtriangleup$\\
	Classical stability & $\bigcirc$ & $\bigcirc$ & $\bigtriangleup$\\
\hline
      \end{tabular}
    \end{center}
  \hfill
    \begin{center}
      \begin{tabular}{|c||c|c|}
	\hline
	Model & \multicolumn{2}{|c|}{Gauge-mediated potentials}\\
    	\hline
	$Q$-ball type &  Thin-wall & Thick-wall \\ \hline
	Conditions &  None & $D=1,3, ...$\\
	\hline
	Assumptions &  $R\gg 1/\mo, 1/\omega$ & None\\
	\hline
	Potential type & \multicolumn{2}{|c|}{NDVPs} \\
    	\hline
	& & \\
	$1/\gamma$ &  $\frac{D}{D+1}$ & 1 \\
	Absolute stability &  $\bigcirc$ & $\times$\\
	Classical stability &  $\bigcirc$ & $\times$\\
\hline
      \end{tabular}
	\end{center}
	\tblcaption{Key analytical results. Recall that the $\omega$-independent characteristic slope $\gamma \equiv E_Q/\omega Q$ leads to the proportionality relation $E_Q\propto Q^{1/\gamma}$. The symbols, $\bigcirc,\; \times,\; \bigtriangleup$, indicate that $Q$-balls are stable, unstable, or can be stable with conditions, respectively. The symbol, $\blacktriangle$, means that we may need the condition $|K|\not\ll \order{1}$. Since the Gauge-mediated potentials are extremely flat for a large field value, the potentials do not have degenerate vacua.}
    \label{tbl:results}
  \end{minipage}
\end{figure}


\chapter{Affleck-Dine dynamics, $Q$-ball formation and thermalisation}\label{ch:adqbform}

%

%
\section{Introduction}
The present baryon asymmetry in the Universe is one of the most mysterious problems in cosmology and particle physics (for a review see \cite{Dine:2003ax, Riotto:1999yt}). Within the Standard Model (SM), electroweak baryogenesis was suggested as a way to explain the inequality between the baryon and anti-baryon number, and recent developments have shifted into constructing a theory of reheating the Universe \cite{Kofman:1997yn}. Electroweak baryogenesis satisfies the well-known Sakharov's three conditions required for successful baryogenesis \cite{Sakharov:1967dj}, namely baryon number production, C and CP violation, and the process taking place out-of-equilibrium; however, the predicted CP violation in the electroweak baryogenesis is too small to explain the present observed baryon number. By satisfying the above three conditions, the Affleck-Dine (AD) baryogenesis \cite{Affleck:1984fy}, which was proposed in the theoretical framework beyond the SM, namely, the Minimal Supersymmetric Standard Model (MSSM), is a more successful scenario to tackle this puzzle, since it may solve problems of gravitino and moduli overproduction and give rise simultaneously to the ordinary matter and dark matter in the Universe. The MSSM has many gauge-invariant flat directions along which R parity is preserved. The flat directions are lifted by supersymmetry (SUSY) breaking effects arising from nonrenormalisable terms, which give a U(1) violation through A-terms. In the original scenario of the AD baryogenesis, one can parametrise one of the flat directions in terms of a complex scalar field known as an AD field (or AD condensate which consists of a combination of squarks and/or sleptons fields). The AD field evolves to a large field expectation value during an inflationary epoch in the early Universe. After inflation, the orbit of the AD field can be kicked along the phase direction due to the A-terms which generate the U(1) charge (baryon/lepton number), and then the A-terms become negligible, where the AD field rotates towards the global minimum of the scalar potential. Hence, the generated global U(1) charge is fixed and the orbit of the AD field rotates around the origin of the complex field-space, \cf\ the anomaly mediated models \cite{Randall:1998uk}. After the AD condensate decays into the usual baryons and leptons, AD baryogenesis becomes complete.

The trajectory of the AD field is identical to the planetary orbits in the well-known Kepler-problem as we will show later, replacing the Newtonian potential by an isotropic harmonic oscillator potential \cite{Tort:1989}. This coincidental classical-mechanics reduction was noted for the orbits of a probe brane in the branonium system \cite{Burgess:2003qv, Rosa:2007dr}. As general relativity predicted that planetary orbits precess by adding the relativistic correction to the Newtonian potential, we will see similar events occur for the orbits of AD fields, which are disturbed by quantum and nonrenormalisable effects instead. 

By including quantum corrections \cite{Enqvist:1997si, Nilles:1983ge} and/or thermal effects \cite{Allahverdi:2000zd} in the mass term of the standard AD scalar potentials, the AD condensate is classically unstable against spatial perturbations due to the presence of negative pressure \cite{McDonald:1993ky}, and fragments to bubble-like objects, eventually evolving into $Q$-balls \cite{Coleman:1985ki}. Lee pointed out \cite{Lee:1994qb} that $Q$-balls may form due to bubble nucleation (first order phase transition) \cite{Coleman:1977py, Callan:1977pt}, even in the case that the condensate is classically stable against the linear spatial perturbations.

We explored the complete stability analysis of $Q$-balls at zero-temperature in polynomial potentials in chapter \ref{ch:qpots} and in MSSM flat potentials in chapter \ref{ch:qbflt}. Laine \textit{et.~al.} \cite{Laine:1998rg} investigated the stability of $Q$-balls in a thermal bath. The stability of the thermal SUSY $Q$-balls is different from the one of the standard ``cold'' $Q$-balls, since they suffer from evaporation \cite{Laine:1998rg}, diffusion \cite{Banerjee:2000mb}, dissociation \cite{Enqvist:1998en}, and decays into light/massless fermions \cite{Cohen:1986ct}. Therefore, most SUSY $Q$-balls are generally not stable but long-lived, and may thermalise the Universe by decaying into baryons on their surface \cite{Enqvist:2002rj}, which could solve the gravitino and moduli over-production problems without fine-tuning. The SUSY $Q$-balls in gravity-mediated (GRV-M) models are quasi-stable decaying into the lightest SUSY particles (LSP dark matter), and the fraction of the baryons from the $Q$-balls may give the present baryon number, which can explain \eq{dm/b}, namely the similarity of the energy density between the observed baryons and dark matter \cite{Kusenko:1997si, Enqvist:1998en}. The SUSY $Q$-balls in gauge-mediated (GAU-M) models, however, can be extremely long-lived so that those $Q$-balls are candidates for cold dark matter \cite{Kusenko:1997si} and may give the present observed baryon-to-photon ratio \eq{basym} \cite{Laine:1998rg}.

The dynamics and formation of $Q$-balls have been investigated numerically. With different relative phases and initial velocities, the authors \cite{Axenides:1999hs} found a charge transfer from one $Q$-ball to the other and interesting ring formation after the collision. It has been found \cite{Radu:2008pp} that similar ring-like solutions are responsible for the excited states from the ground state ($Q$-ball) by introducing extra degrees of freedom: spatial spins \cite{Volkov:2002aj} and twists \cite{Axenides:2001pi}. The formation of $Q$-balls after inflation have been investigated in both GRV-M models \cite{Kasuya:2000wx} and GAU-M models \cite{Kasuya:1999wu, Kasuya:2001hg}, in which SUSY is broken by either gravity or gauge interactions. As we will show, the $Q$-ball formation involves nonequilibrium dynamics, which is related to reheating problem in cosmology.

The reheating process after the inflation period involves nonlinear, out-of-equilibrium, and nonperturbative phenomena so that it is extremely hard to construct a theory for the whole process, see the 2 particle irreducible effective action as a remarkable approach \cite{Berges:2004yj, Arrizabalaga:2004iw, Arrizabalaga:2005tf}. In the first stage of reheating (\emph{preheating}), it is currently well known that the fluctuations at low momenta are amplified, which leads to explosive particle production. After preheating, the subsequent stages towards equilibrium are described by the wave kinetic theory of turbulence; Micha \textit{et.~al.} \cite{Micha:2004bv} recently estimated the reheating time and temperature. These turbulent regimes appear in a large variety of nonequilibrium processes, and indeed, the evolution of $Q$-ball formation experiences the active turbulence at which stage, many bubbles collide as observed in the next stage of tachyonic preheating \cite{Felder:2000hj, GarciaBellido:2002aj}. During this bubble-collision stage within the reheating scenario, it is believed that gravitational waves may be emitted from the stochastic motion of heavy objects \cite{GarciaBellido:2002aj, Felder:2006cc, Dufaux:2007pt}. The problem of gravitational wave emissions has been discussed only in the fragmentation stage of $Q$-ball formation so far \cite{Kusenko:2009cv, Kusenko:2008zm, Dufaux:2009wn}, but not in the collision stage as opposed to the preheating cases.

In this chapter, we show analytically and numerically that in GRV-M and GAU-M models the approximate trajectory of the AD fields is, respectively, either a precessing spiral or shrinking trefoil due to quantum, nonrenormalisable, and Hubble expansion effects. Moreover, we explicitly present an exponential growth of the linear spatial perturbations in both models. By introducing $3+1$ (and $2+1$)-dimensional lattice simulations, we identify that the evolution in $Q$-ball formation involves nonequilibrium dynamics, including turbulent stages. Following the pioneering work on the turbulent thermalisation by Micha \textit{et.~al.} \cite{Micha:2004bv}, we obtain scaling laws for the evolution of variances during the $Q$-ball formation.

This chapter is divided as follows. We explore both analytically and numerically the dynamics of the AD field in Sec. \ref{sectorbit}. In Sec. \ref{sectinst}, we study the late evolution of the AD fields and the process of $Q$-ball formation, introducing detailed numerical lattice results. Finally, we conclude and discuss our results in Sec. \ref{concl}. Three appendices are included. We obtain the equations of motion and their perturbed equations for multiple scalar fields in an fixed expanding background in Appendix \ref{MULTI}. In Appendix \ref{App2}, we find elliptic forms for the orbits of AD fields. To obtain the condition of closed orbits of the AD fields, we prove Bertrand's theorem in Appendix \ref{BERT}. This chapter is based on work published in \cite{Tsumagari:2009na}, where the reader should note that we use slightly different notations from the ones introduced in chapters 2-4.

\section{The Affleck-Dine dynamics}\label{sectorbit}
In this section we investigate an equation for the orbit of an AD condensate, which coincides with the well-known orbit equation in the centre force problem in classical dynamics, \ie\  planetary motions so that we sometimes call the AD condensate, ``AD planet''. For the bound orbits, the effective potential should satisfy the condition where the curvature at the minimum of the effective potential should be positive. In the presence of the Hubble expansion, the effective potential depends on time; thus, the full solution of the orbit equations can be obtained numerically except for the case that the AD field is trapped by a quadratic potential when it can be solved analytically. In appendix \ref{App2}, we obtain the exact orbit in this exceptional case when the Hubble expansion is assumed to be small and adiabatic. The orbit of the AD planet, or more precisely an eccentricity of the elliptic motion in the complex field-space, is determined by the initial charge and energy density. In order to obtain analytic expressions of the orbit in more general potential cases in which we are more interested, we restrict ourself to work in Minkowski spacetime and on the orbit which should be nearly circular. In this case, we also obtain the perturbed orbit equation and necessary conditions for closed orbits where the orbits come back to their original positions after some rotations around the minimum of the effective potential. By approximating phenomenologically motivated models that appear in the MSSM and using the results in appendix \ref{App2}, we present, in this section, analytic motions of the nearly circular orbits and the pressure of the AD planets. Further, we check these analytic results with full numerical solutions.

\vspace*{10pt}

\begin{figure}[!ht]
  \begin{center}
	\includegraphics[angle=-90, scale=0.28]{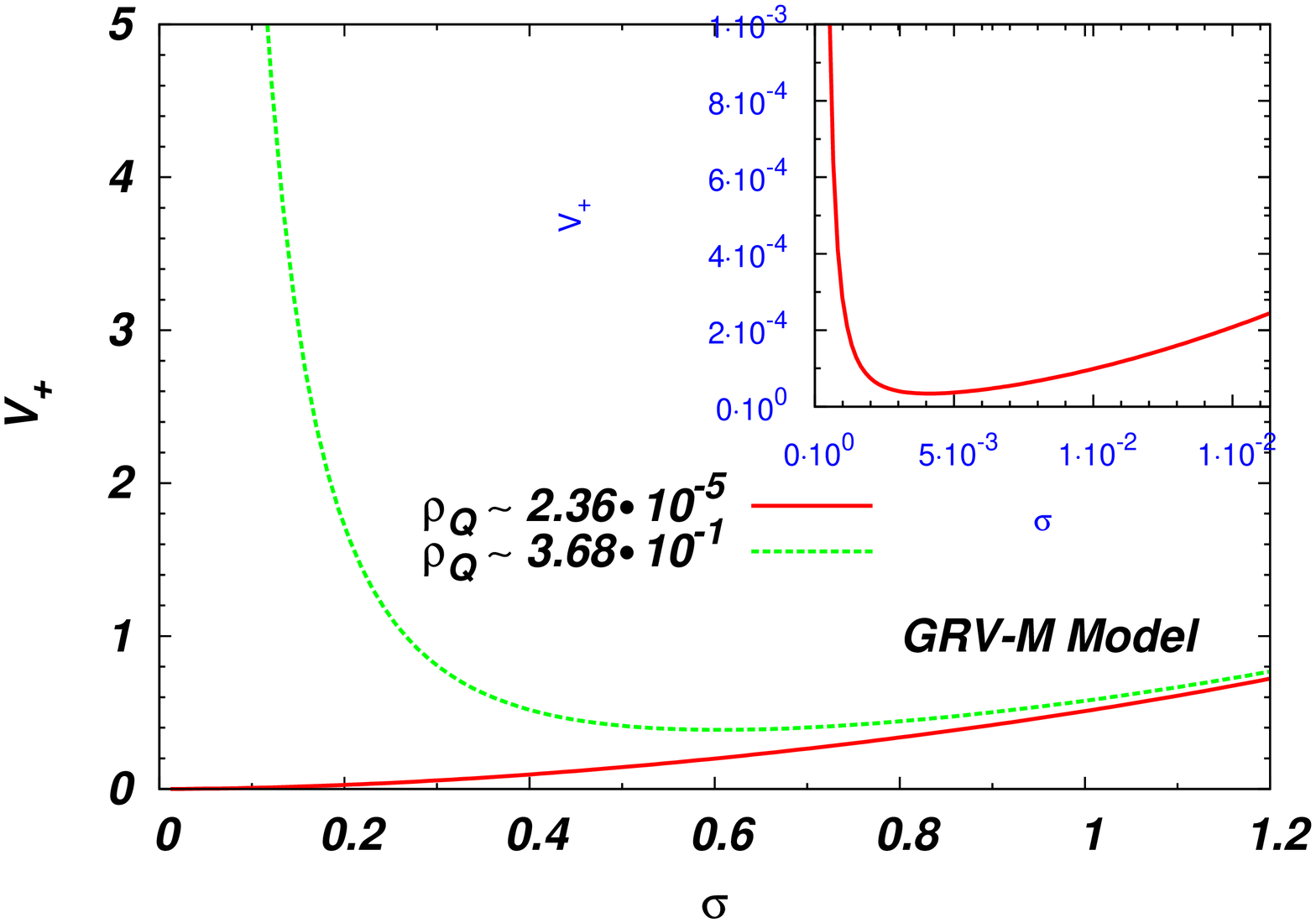}
	\includegraphics[angle=-90, scale=0.28]{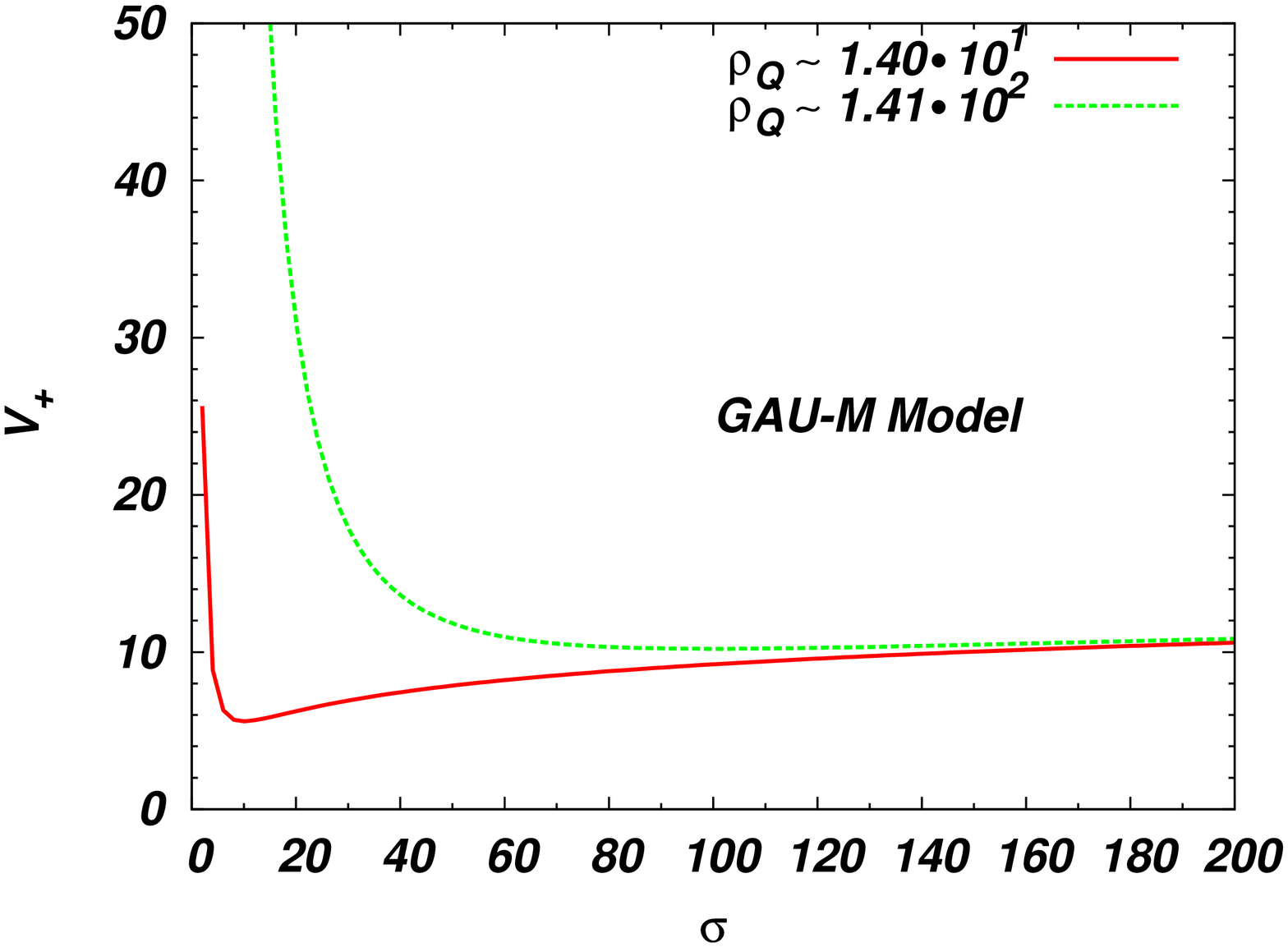}
  \end{center}
\caption{ We show the effective potentials, $V_+\equiv V(\sigma)+\frac{\rho^2_Q}{2\sigma^2}$, against $\sigma$ in two types of potentials which we call the gravity mediated model (GRV-M Model) on the left and the gauge mediated model (GAU-M Model) on the right. The potential in the GRV-M Model has the following form, $V(\sigma)=\half \sigma^2 \bset{1-|K|\ln\sigma^2}+b^2_*\sigma^6$, where, we set $|K|=0.1$ and $b^2_*=\frac{|K|}{4e}\sim 9.20 \times 10^{-3}$. The potential in the GAU-M Model is $V(\sigma)=\ln\bset{1+\sigma^2}+b^2 \sigma^6$, where we set $b^2\sim10^{-30}$. We choose the following values of $\rho_Q$: red-solid line for $\rho_Q\sim 2.36\times 10^{-5}$ and green-dashed line for $\rho_Q=1/e\sim 3.68\times 10^{-1}$ in the GRV-M Model and red-solid line for $\rho_Q\sim 1.40\times 10^1$ and green-dashed line for $\rho_Q \sim 1.41\times10^2$ in the GAU-M Model.}
  \label{fig:adpot}
\end{figure}

Let us consider a motion of AD fields in an expanding universe with scale factor $a(t)$ and Hubble parameter $H=\dot{a}/a$, where an over-dot denotes the time derivative. We investigate the AD field after they start to rotate around the origin of the effective potentials and the value of the U(1) charge $\rho_Q$ is fixed due to negligible contributions from A-terms. By decomposing the complex (AD) field $\phi$ as $\phi(t)=\sigma(t) e^{i\theta(t)}$, where $\sigma$ and $\theta$ are real scalar fields, the equations of motion for $\sigma(t)$ and $\theta(t)$ (see \eqs{sigeom}{thetaeom} in appendix \ref{MULTI}) are
\bea{radhomo}
\ddot\sigma +3H\dot\sigma+\frac{dV_+}{d\sigma}&=&0,\\
\label{phshomo}\ddot\theta+3H\dot\theta+\frac{2}{\sigma}\dot\sigma\dot\theta&=& 0\spc \lr\spc \frac{d\rho_Q}{dt}=0,
\eea
where the conserved comoving charge density is defined by $\rho_Q\equiv a^3 \sigma^2 \dot\theta$, and the effective potentials are $V_\pm=V(\sigma) \pm \frac{\rho^2_Q}{2a^{6}\sigma^2}$. Note that we will use $V_-$ shortly. From \eq{epcom}, the energy density $\rho_E$ and pressure $p$ are given by
\be\label{rhop}
\rho_E=\half {\dot{\sigma}}^2+V_+,\hspace*{30pt} p=\half {\dot{\sigma}}^2 -V_-.
\ee
With various values of the charge density $\rho_Q$, \fig{fig:adpot} shows typical effective potentials $V_+$ in Minkowski spacetime where we set $a=H=1$. The potentials shown in \fig{fig:adpot} will be used later. 

Given an initial charge and energy density (or equivalently initial momenta and position), the AD field oscillates around the value $\sigma_{cr}$, which is defined by 
\be\label{sigcr}
\left.\frac{dV_+}{d\sigma}\right|_{\sigma_{cr}}=0,
\ee
where the orbit becomes circular when it starts from there, \ie\  $\sigma(0)=\sigma_{cr},\; \dot{\sigma}(0)=0$. This orbit is bounded when the curvature is positive
\be\label{condbound}
W^2\equiv \left.\frac{d^2V_+}{d\sigma^2}\right|_{\sigma_{cr}}>0.
\ee
For example, given a power-law potential such that $V= \lambda_1\sigma^l$ where $\lambda_1$ is a dimensionful coupling constant and $l$ is the real power of the homogeneous field $\sigma$, the condition given by \eq{condbound} implies that bound orbits exist for $l<-2$ and $0<l$ if $\lambda_1>0$ and for $-2<l<0$ if $\lambda_1<0$, where we used \eq{sigcr}. Another example is the case that a scalar potential is logarithmic, \ie\  $V=\lambda_2 \ln\sigma$ where the coupling constant $\lambda_2$ is positive. In this case, \eq{condbound} is automatically satisfied. We investigate these two cases in more detail in appendix \ref{App2}.

Let us rescale the field $\sigma(t)$ as $\sigma(t)=\bset{\frac{a_0}{a(t)}}^{3/2}\tisig(t)$ where $a_0$ is the value of $a(t)$ at an initial time. It follows that the equations of motion in \eqs{radhomo}{phshomo} are
\be\label{modrad}
\ddot{\tisig}-\bset{\frac{3}{4}H^2+\frac{3}{2}\frac{\ddot{a}}{a}}\tisig-\frac{\tirhoQ^2}{\tisig^3}+\bset{\frac{a}{a_0}}^3\frac{dV(\sigma)}{d\tisig}=0, \hspace*{20pt} \frac{d\tirhoQ}{dt}=0,
\ee
where we defined $\tirhoQ\equiv \tisig^2\dot{\theta}=a^{-3}_0\rho_Q$, and the terms involving $H^2$ and $\ddot{a}/a$ are negligible under the assumption of an adiabatic Hubble expansion, \ie\  $H^2 \ll 1,\; \ddot{a}\ll a$. 

By introducing a new variable, $\tiu(t)\equiv 1/\tisig(t)$, and using the second expression in \eq{modrad}, the first expression in \eq{modrad} becomes the well-known orbit equation in the centre force problem such that
\be\label{orbiteq}
\frac{d^2\tiu}{d\theta^2}+\tiu=-\frac{1}{\tirhoQ^2}\bset{\frac{a}{a_0}}^3\frac{dV}{d\tiu}\equiv J(\tiu,t).
\ee
Notice that $J(\tiu,t)$ depends on time caused by the Hubble expansion, whereas  the time-dependence in $J$ vanishes when the potential $V$ is given by a quadratic mass term, $\half M^2 \sigma^2$, where $M$ is a mass of the AD field, $\phi$. We also discuss this case in appendix \ref{App2}.

\subsection{Model A and Model B for MSSM flat potentials}\label{MODEL-ABC}

Let us introduce two models that appear in the MSSM in which SUSY is broken due to either gravity or gauge interactions, and approximate their models in order to obtain the orbit expressions in Minkowski spacetime. The former case in the MSSM, the so-called gravity-mediated (GRV-M) model, has a scalar potential
\be\label{gravity-pot}
V=\half m^2\sigma^2\bset{1+K\ln\frac{\sigma^2}{M^2_*}} + \frac{\lambda^2}{m^{n-4}_{pl}} \sigma^n,
\ee
where $m$ is of order of the SUSY breaking scale, which could be the gravitino mass scale $m_{3/2}$ evaluated at the renormalisation scale $M_*$ \cite{Nilles:1983ge}. Also, $\lambda$ is a coupling constant for the nonrenormalisable term, which is suppressed by a high energy scale, \eg\ the Planck scale $m_{pl}\sim 10^{18}$ GeV. Here, $K$ is a factor from the gaugino-loop correction, whose value is typically $K\simeq -[0.01-0.1]$ when the flat direction does not have a large top quark component \cite{Enqvist:1997si, Enqvist:2000gq}; thus, we concentrate on the case of $K<0$ from now on. The power $n$ of the nonrenormalisable term depends on the flat directions. As examples of the directions involving squarks, the $u^cd^cd^c$ direction has $n=10$, whilst the $u^cu^cd^ce^e$ direction is $n=6$. For $|K|\ll \order{1}$, the first two terms in \eq{gravity-pot} can be approximated by $\frac{m^2M^{2|K|}_*}{2}\sigma^{2-2|K|}$, we then find that
\be\label{modelA}
V(\sigma)\simeq \frac{M^2}{2}\sigma^l+\frac{\lambda^2}{m^{n-4}_{pl}}\sigma^n \hspace*{15pt} \textrm{for}\; \; n>l
\ee
which we call 'Model A', where we set $M^2 \equiv m^2M^{2|K|}_*$ and $M$ has a mass-dimension, $\frac{4-l}{2}\simeq 1$, since $l\equiv 2-2|K|$ for $|K|\ll \order{1}$. For small values of $\sigma$, we confirm that the power $l$ is not approximately $2-2|K|$, so we will find a value of $l$ numerically in that case later.
\vspace*{5pt}

In another scenario in which SUSY is broken by gauge interactions, the so-called gauge-mediated (GAU-M) model, the scalar potential has the curvature with the electroweak mass at a low energy scale, whilst it grows logarithmically at the high energy scale (which means that the potential is nearly flat similar to the case of $l=0$ in \eq{modelA}). The scalar potential in this scenario is
\be\label{gauge-pot}
V\simeq m^4_{\phi} \ln\bset{1 + \bset{\frac{\sigma}{M_s}}^2}+\frac{\lambda^2}{m^{n-4}_{pl}}\sigma^n,
\ee
where $M_s$ is the messenger scale ($\sim 10^4$ GeV) above which the potential grows logarithmically and $m_{\phi}$ is the same scale as $M_s$. We, thus, set $M_s =m_{\phi}$ for later convenience. Then, the scalar potential at high energy scales is approximately given by \cite{Kusenko:1997si}
\be\label{modelC}
V\simeq m^4_{\phi} \ln\bset{ \frac{\sigma}{m_{\phi}}}^2+\frac{\lambda^2}{m^{n-4}_{pl}} \sigma^n.
\ee
In what follows we assume the orbit of the AD condensate is determined by the high energy scale where $\sigma_{cr}\gg m_{\phi}$, calling this case, \eq{modelC}, 'Model B'.

Using the results in Appendix \ref{App2}, we obtain the following quantities, $W,\; \Phi$ and $\state{w}$ by assuming that the dominant contribution in Model A and B is, respectively, either a power-law or logarithmic term, each of which corresponds to the first term in \eqs{modelA}{modelC}, respectively. Here, we have defined $\Phi$ as a phase difference when the radial field $\sigma$ goes from the minimum value through the maximum one and back to the same minimum point, see \eq{Phi}; in addition, $\state{w}$ is given by a value of the equation of state averaged over a rotation of the orbit, see \eq{eosg}. Note that we have defined an averaged value for a quasi-periodic quantity $X$ over an one rotation in the orbit, namely $\state{X}\equiv \frac{1}{\tau}\int^{\tau}_0 dt X(t)$.  The sub-dominant terms (nonrenormalisable terms) perturb the orbits via infinitesimally small quantities $\epsilon_A$ and $\epsilon_B$, where the subscripts correspond to the names of models introduced above. Thus, the main contributions are either \eqs{powerPhi}{powerpress} or \eqs{logPhi}{presslog}.

\vspace*{10pt}

\subsubsection{Model A -- $V(\sigma)=\frac{M^2}{2}\sigma^l+\frac{\lambda^2}{m^{n-4}_{pl}}\sigma^n$}

By recalling \eq{condbound}, we obtain the following relation for $n>l$ in Model A in Minkowski spacetime:
\be\label{modelA-W}
W^2= \frac{l(l+2)M^2\sigma^{l-2}_{cr}}{2}\bset{1+\epsilon_A},
\ee
where we have defined a positive parameter, $\epsilon_A\equiv \frac{n(n+2)}{l(l+2)}\frac{2\lambda^2}{M^2m^{n-4}_{pl}}\sigma^{n-l}_{cr}\ll 1$, which is assumed to be infinitesimally small. We also obtain $\beta^2\simeq (l+2)\left(1+\frac{n-l}{n+2}\epsilon_A\right)>0$, where $\beta$ is defined in \eq{betasq}. Substituting $\beta$ into \eqs{Phi}{eosg}, we obtain $\Phi$ and $\state{w}$:
\bea{modelA-P}
\Phi&\simeq& \frac{\pi}{\sqrt{l+2}}\bset{1+\frac{l-n}{2(n+2)}\epsilon_A}, \\ \label{modelA-w}\state{w}&=&\frac{(l-2)\bset{1+\epsilon_A\frac{l(l+2)(n-2)}{n(n+2)(l-2)}}}{(l+2)\bset{1+\epsilon_A\frac{l}{n}}}\simeq \frac{l-2}{l+2}\bset{1+\epsilon_A\frac{4l(n-l)}{n(n+2)(l-2)}}.
\eea
From \eq{modelA-P}, the orbits for $l=2-2|K|\simeq 2$ are nearly closed, but it is perturbed by the nonrenormalisable term involved with $\epsilon_A$. The result is that the periapsis appears to precess where the precession rate can be obtained from \eq{modelA-W}. The denominator of the term involving $\epsilon_A$ in the second expression of \eq{modelA-w} has $l-2\simeq -2|K|\ll \order{1}$, which implies that it would be possible to have the non-negligible contribution from the term, even though $\epsilon_A\ll \order{1}$. From now on, we restrict ourself to regions where this is not the case; therefore, the dominant contributions are the leading orders in \eqs{modelA-W}{modelA-P} and \eq{modelA-w}, which correspond to \eqs{powerPhi}{powerpress} and \eq{compress}. From \eq{modelA-w} with $\epsilon_A\simeq 0$, our results recover the result published in \cite{Enqvist:1997si}, \ie\ $\state{w}\simeq -\frac{|K|}{2}$.

\vspace*{10pt}

\subsubsection{Model B -- $V(\sigma)=m^4_{\phi} \ln\bset{\sigma/m_{\phi}}^2+\frac{\lambda^2}{m^{n-4}_{pl}}\sigma^n$}
By introducing another infinitesimally small positive parameter, $\epsilon_B\equiv \frac{n(n+2)\lambda^2\sigma^n_{cr}}{4m^4_{\phi}m^{n-4}_{pl}}\ll 1$, we obtain the following relations in Model B in Minkowski spacetime:
\bea{ModelC-W}
W^2&\simeq& \frac{4m^4_{\phi}}{\sigma^2_{cr}}\bset{1+\epsilon_B}, \spc \Phi\simeq \frac{\pi}{\sqrt{2}}\bset{1-\frac{n}{2(n+2)}\epsilon_B} \sim \frac{2\pi}{3},\\
\label{ModelC-w} \state{w}&=& \frac{ 1-2\ln\bset{\frac{\sigma_{cr}}{m_{\phi}}} +\frac{2(n-2)}{n(n+2)}\epsilon_B }{1+2\ln\bset{\frac{\sigma_{cr}}{m_{\phi}}} +\frac{2}{n}\epsilon_B} \gtrsim -1.
\eea
Since we are working in the high-energy regime, $\sigma_{cr}\gg m_{\phi}$, the pressure of the AD condensate is likely to be negative, see \eq{ModelC-w}. From the second expression of \eq{ModelC-W} for $\Phi$, the orbits are not closed and it should look like the trefoil, see \eq{logPhi}.

\vspace*{10pt}

In an expanding universe, the above orbits for Model A and B suffer from the Hubble damping so that the orbits are naively expected to be precessing spiral or shrinking trefoil in the field-space, respectively.

\subsection{Numerical results}\label{numAD}

In this subsection we present numerical results to check the analytic results, which we found in the previous subsection. To do so, we use the full potentials, \eqs{gravity-pot}{gauge-pot}, instead of \eqs{modelA}{modelC}, and then solve \eq{radhomo} numerically in Minkowski spacetime as well as in an expanding universe. We adopt the 4th order Runge-Kutta method with various sets of initial conditions, such as $\rho_Q$ and $\varepsilon^2$. Since our analytical work holds as long as $\varepsilon^2\ll \order{1}$, we are concerned with the two cases: a nearly circular orbit with $\varepsilon^2=0.1$ and a more elliptic orbit with $\varepsilon^2=0.3$. First of all, we parametrise \eqs{gravity-pot}{gauge-pot} by introducing dimensionless variables: $\mathring{\sigma}=\sigma/M_*,\; b^2_*=\frac{\lambda^2M^{n-2}_*}{m^{n-4}_{pl}m^2}=|K|e^{-1}/4,\; \mathring{t}=mt,\; \mathring{\mathbf{x}}=m\mathbf{x}$ in the GRV-M Model and $\mathring{\sigma}=\sigma/M_s,\; b^2=\frac{\lambda^2M^{n-4}_s}{m^{n-4}_{pl}},\; \mathring{t}=M_st,\; \mathring{\mathbf{x}}=M_s\mathbf{x}$ in the GAU-M Model. Since we know that $m\sim 10^2$ GeV, $M_*\sim 10^{10}$ GeV, $m_{pl}\sim 10^{18}$ GeV; hence, we can set $b^2_*\sim 9.20\times 10^{-3}\sim \order{10^{-2}}$ in the GRV-M Model, where we choose $|K|=0.1$. Notice that these choices are the same as the ones used in chapter \ref{ch:qbflt} \cite{Copeland:2009as}. On the other hand, we know that $m_{\phi}\sim M_s \sim 10^4$ GeV; hence, we can set $b^2\sim 10^{-30}$ in the GAU-M Model, where we choose $\lambda\sim 10^{-2}$ as used in the GRV-M case. Notice that we can obtain the rescaled charge density $\mathring{\rho}_Q$ and energy density $\mathring{\rho}_E$, such that $\rho_Q=mM^2_*\mathring{\rho}_Q,\; \rho_E=m^2M^2_*\mathring{\rho}_E$ in the GRV-M Model and $\rho_Q=M^3_s\mathring{\rho}_Q,\; \rho_E=M^4_s\mathring{\rho}_E$ in the GAU-M Model.

Therefore, our rescaled potentials in GRV-M and GAU-M models for a flat-direction with $n=6$ are, respectively,
\bea{numGRV}
V&=&\half \sigma^2\bset{1-2|K|\ln \sigma} +b^2_* \sigma^6,\\
\label{numGAU} V&=&\ln\bset{1+\sigma^2}+b^2\sigma^6,
\eea
where we omit over-rings for simplicity. The variables that appear within the rest of this subsection are dimensionless. We can also obtain the ratio defined by an energy density relative to (a mass multiplied by a charge density), where the mass corresponds to $m$ or $M_s$ in either GRV-M or GAU-M Model, respectively.

In order to obtain appropriate initial values of $\sigma(0),\; \dot{\sigma}(0)$ and $\dot{\theta}(0)$ satisfying the conditions $\epsilon_A,\; \epsilon_B\ll \order{1}$ and not giving too small charge densities, we shall show that we need to choose only the initial values of $\dot{\theta}(0)$ in both GRV-M and GAU-M models. First, by ignoring the nonrenormalisable term in \eq{numGRV} for the GRV-M Model, we obtain $\sigma_{cr}=\exp\bset{-\frac{1}{2|K|}\bset{\dot{\theta}^2(0)+|K|-1}}:=\sigma(0)$ from \eq{sigcr}, where we set $\sigma_{cr}:=\sigma(0)$, which implies that we are setting the initial phase to be $3\pi/2$. Since $\dot{\sigma}$ has the maximum value at $\sigma=\sigma_{cr}$, we can set $\dot{\sigma}(0):=\varepsilon^2\sigma(0)\sqrt{\dot{\theta}^2(0)-|K|/2}$ from \eq{deltaeq}, which implies that $\epsilon_A\sim 12 b^2_* \sigma^4(0)$ from the definition. We notice that $\sigma(0)\gg \order{1}$ for $\dot{\theta}(0)\ll \order{1}$; hence, it breaks the condition, $\epsilon_A \ll \order{1}$. We can also see that $\sigma(0)\ll \order{1}$ for $\dot{\theta}(0)\gg \order{1}$, so the charge density is suppressed exponentially. Therefore, we are concerned with the following two cases: $\dot{\theta}(0)=\sqrt{2}$ and $1.0$, which give, respectively, $\epsilon_A\sim 1.20\times 10^{-11},\; \rho_Q \sim 2.36\times 10^{-5}  $ and $\epsilon_A \sim 1.58\times 10^{-2},\; \rho_Q \sim 3.68\times 10^{-1}$. Similarly, in the GAU-M Model, we choose that $\sigma_{cr}=\sqrt{\frac{2}{\dot{\theta}^2(0)}-1}:=\sigma(0)$, $\dot{\sigma}(0):=\varepsilon^2\sqrt{1-\frac{3}{4}\dot{\theta}^2(0)}$ and $\epsilon_B=12b^2\sigma^6(0)$ from the definition of $\epsilon_B$. Here, we also set the initial phase is $3\pi/2$ due to $\sigma_{cr}:=\sigma(0)$. With this fact and the approximation, $\sigma_{cr}\gg \order{1}$, we need to have $\dot{\theta}(0)\ll \order{1}$. In addition, we should have $\sigma(0)< \order{10^5}$ due to the condition, $\epsilon_B < \order{1}$. Therefore, we choose $\dot{\theta}(0)=\sqrt{2}\times 10^{-1}$ and $\sqrt{2}\times 10^{-2}$ which gives, respectively, $\epsilon_B\sim 1.16\times 10^{-23},\; \rho_Q \sim 1.40\times 10^1$ and $\epsilon_B \sim 1.20\times 10^{-17},\; \rho_Q \sim 1.41\times 10^2$. 

Using the above initial conditions, we initiate the numerical simulations with 8 different sets of the initial values in the GRV-M Model and the GAU-M Model summarised in Table \ref{parameterSET}, where we call each of the parameter-sets 'SET-1, SET-2,..., and SET-8'. In \fig{fig:adpot}, we also show,  with the various charges which we introduced above, the effective potentials $V_+$ for the GRV-M potential given by \eq{numGRV} in the left panel and for the GAU-M potential given by \eq{numGAU} in the right panel.  After had carried out many trial numerical simulations, we found that the best time-step $dt$ is $dt=1.0\times 10^{-4}$ in the GRV-M case and $dt=1.0\times 10^{-3}$ in the GAU-M case.

\begin{center}
\begin{table} [!ht]
\begin{tabular} { |c|c|c|c|c|c|c|c| }

\hline
SET & Model & $\dot{\theta}(0)$ & $\sigma(0)$ & $\rho_Q$ &   $\epsilon_A$ or $\epsilon_B$ & $\varepsilon^2$ & $\rho_E/\rho_Q$ \\
\hline \hline
1 & & & & & & 0.1  & 1.46\\ 
2 & & \raisebox{1.5ex} {$\sqrt{2}$} & \raisebox{1.5ex} {$\sim 4.09\times 10^{-3}$} & \raisebox{1.5ex} {$\sim 2.36\times 10^{-5}$}&  \raisebox{1.5ex} {$\sim 1.20\times 10^{-11}$} & 0.3 & 1.51 \\
3 & \raisebox{1.5ex} {GRV-M} & & &  &  & 0.1 & 1.06 \\
4 & & \raisebox{1.5ex} {1.0} & \raisebox{1.5ex} {$\sim 6.07\times 10^{-1}$}  & \raisebox{1.5ex} {$\sim 3.68\times 10^{-1}$}&  \raisebox{1.5ex} {$\sim 1.58\times 10^{-2}$}  &  0.3 & 1.09\\ 
\hline \hline
5 & & &  & & & 0.1 & $4.00\times 10^{-1}$ \\ 
6 & & \raisebox{1.5ex} {$\sqrt{2}\times 10^{-1}$} & \raisebox{1.5ex} {$\sim9.95$} & \raisebox{1.5ex} {$\sim 1.40 \times 10^1$} &  \raisebox{1.5ex} {$\sim 1.16\times 10^{-23}$} & 0.3 & $4.03\times 10^{-1}$\\
7 & \raisebox{1.5ex} {GAU-M}  & & & &  & 0.1  & $7.22\times 10^{-2}$ \\
8 & & \raisebox{1.5ex} {$\sqrt{2}\times 10^{-2}$} &  \raisebox{1.5ex} {$\sim1.00\times 10^{2}$} & \raisebox{1.5ex} {$\sim 1.41\times 10^2$}  &  \raisebox{1.5ex} {$\sim 1.20\times 10^{-17}$}  &  0.3 & $7.25\times 10^{-2}$\\ 
\hline
\end{tabular}
\caption{We show 8 different parameter sets in both the GRV-M and GAU-M cases, where we call each of the parameter-sets 'SET-1, SET-2,..., and SET-8'. The initial parameters of $\sigma(0)$ and $\dot{\sigma}(0)$ can be obtained by the values of $\dot{\theta}(0)$. We also set $\theta(0)=\frac{3\pi}{2}$ in all cases, and show the values of $\epsilon_A$ for the GRV-M Model and the values of $\epsilon_B$ for the GAU-M Model. By substituting these values and choosing the values of the third eccentricity $\varepsilon^2=0.1$ and $0.3$, we obtain the dimensionless energy-to-(mass multiplied by charge) ratios, $\rho_E/\rho_Q$. Note we are using the dimensionless quantities.}
\label{parameterSET}
\end{table}
\end{center}

\subsubsection{The orbit of an Affleck-Dine ``planet'' in Minkowski spacetime}

First, we present numerical results in Minkowski spacetime in order to check our analytical results. We then give the ans\"{a}tze that are motivated by our analytic solutions, in an expanding universe in the next sub-subsection.

\vspace*{10pt}

\paragraph*{\underline{\bf The motion of $\sigma^2(t)$}}

\begin{figure}[!ht]
  \begin{center}
	\includegraphics[angle=-90, scale=0.28]{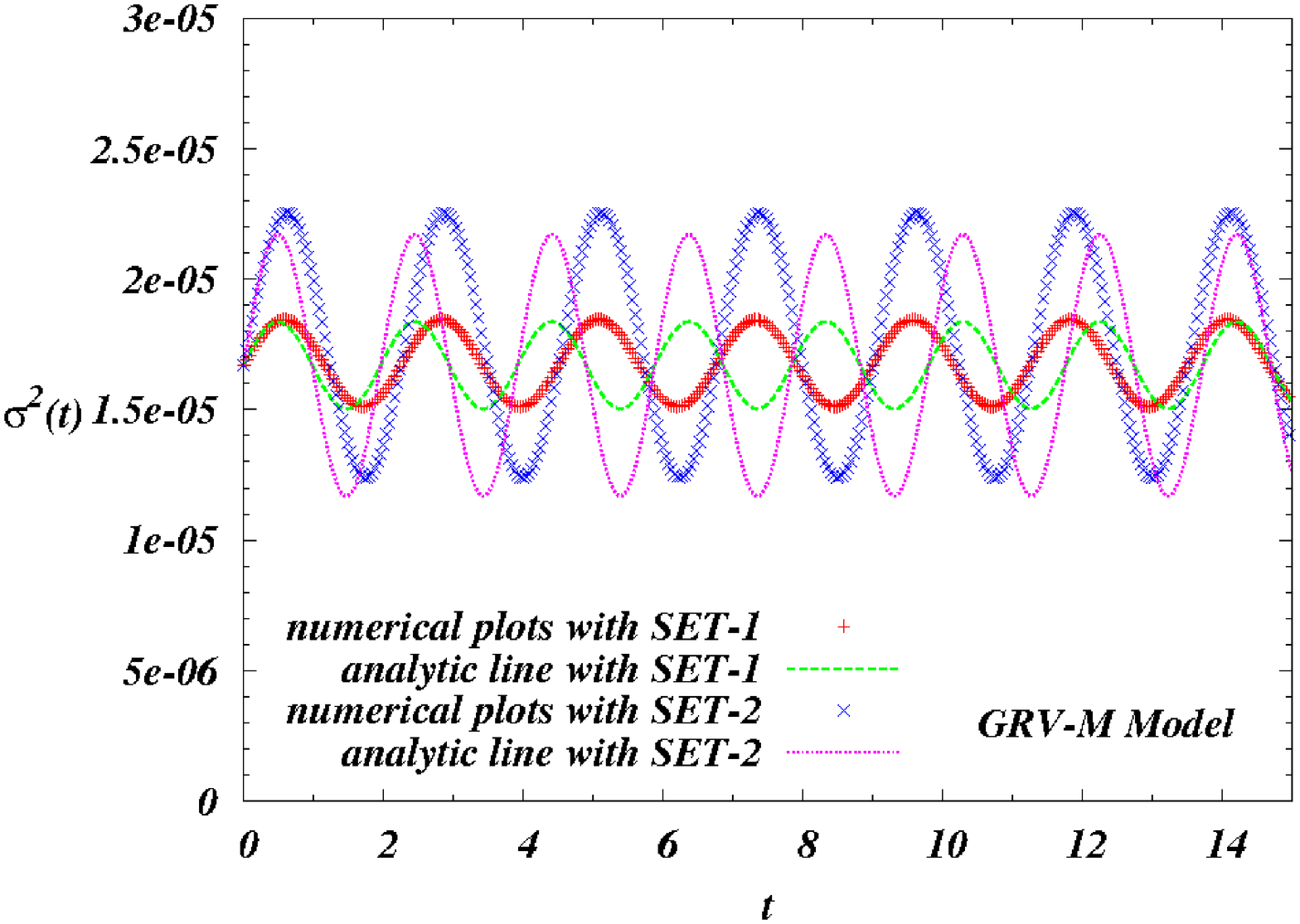}
	\includegraphics[angle=-90, scale=0.28]{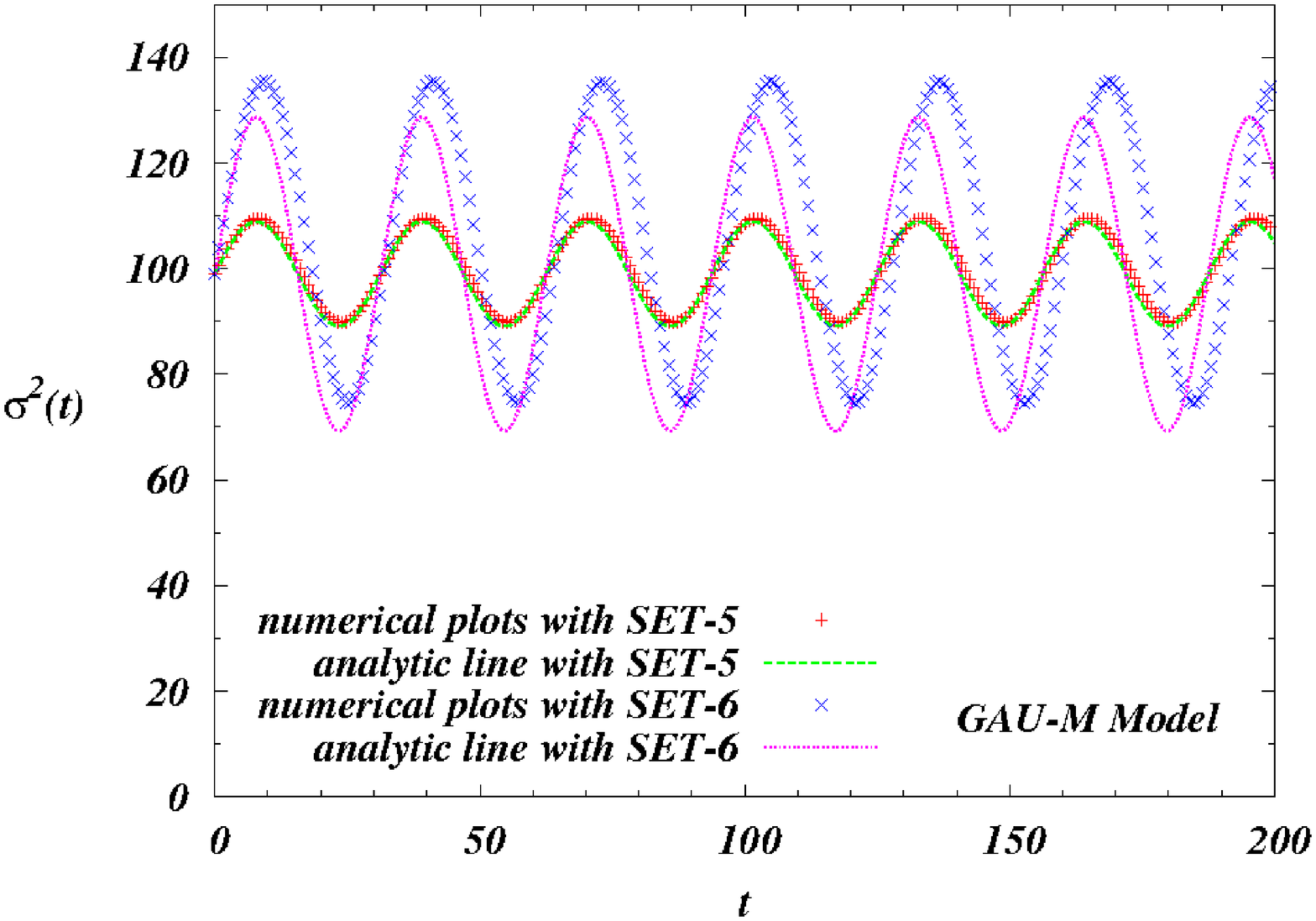}\\
	\includegraphics[angle=-90, scale=0.28]{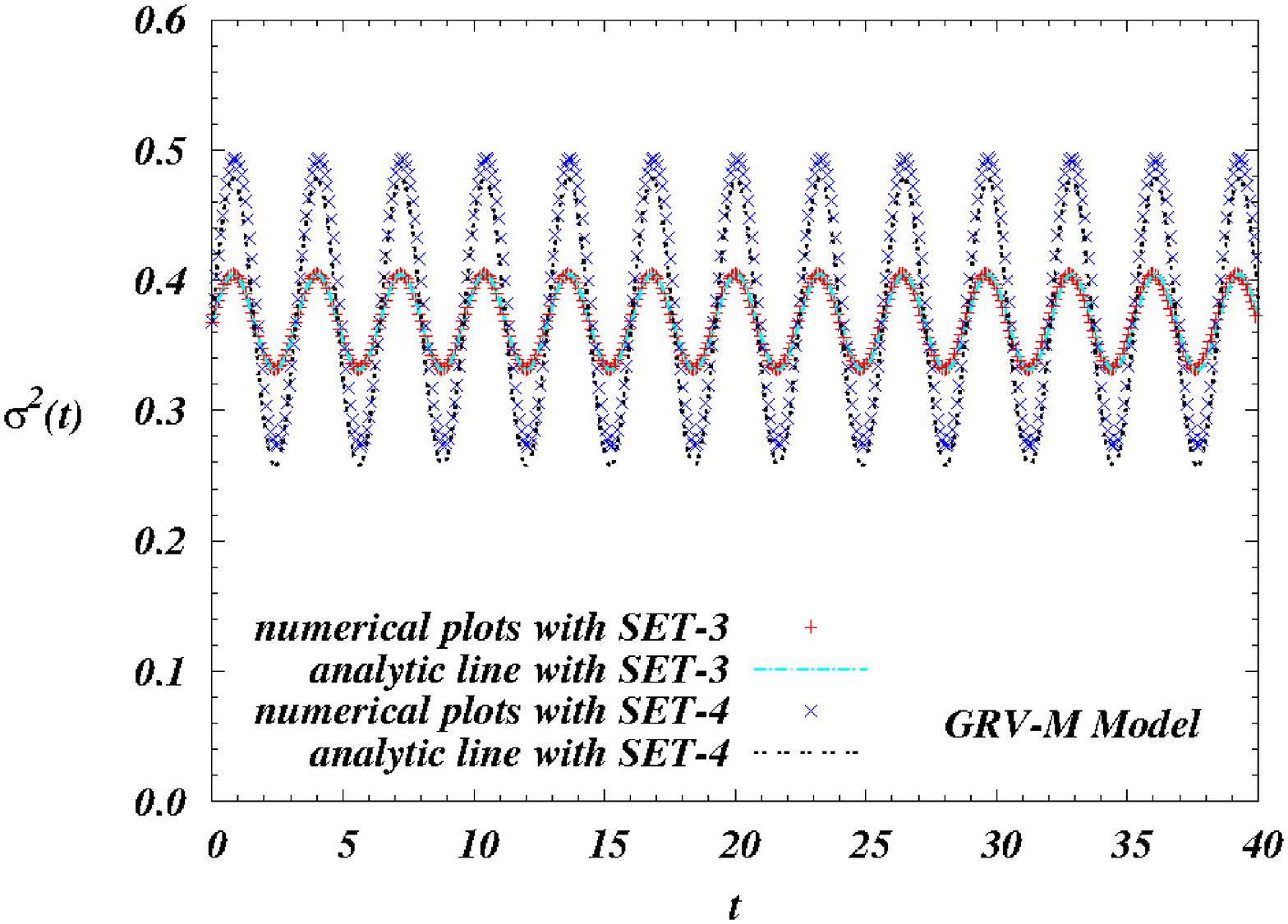}
	\includegraphics[angle=-90, scale=0.28]{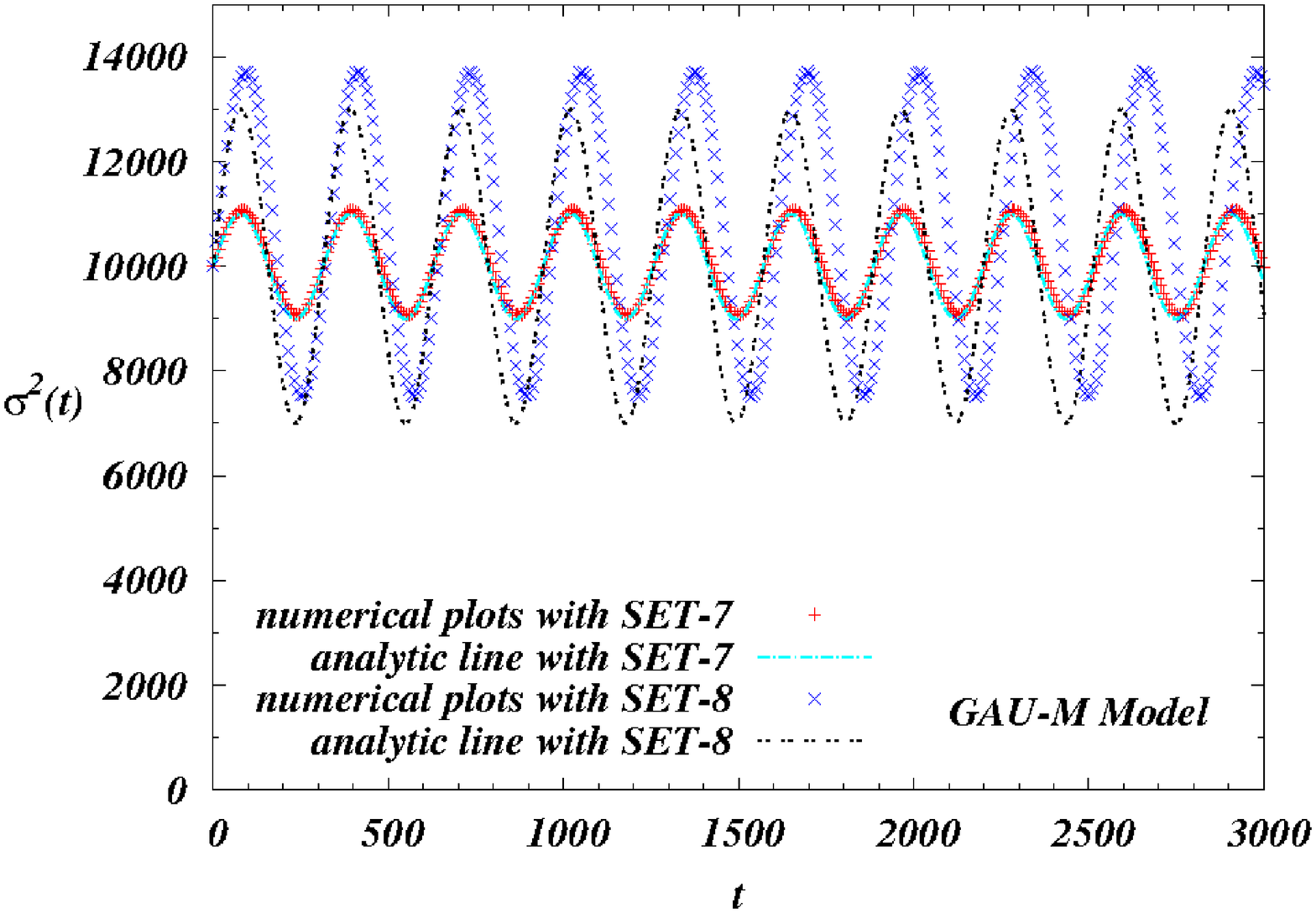}
  \end{center}
  \caption{ Using the parameter sets summarised in Table \ref{parameterSET}, we plot the numerical evolution for $\sigma^2(t)$ in both the GRV-M Model (left) and the GAU-M Model (right). In all the panels except the case for the left-top panel, the numerical plots (red-plus dots and blue-cross dots) agree well with the corresponding analytic lines, which are obtained from Sec. \ref{MODEL-ABC}. The disagreements between the numerical and analytic plots in the left-top panel come from the artifact that the analytical estimated value, $l=1.8$, in \eq{modelA}.}
  \label{fig:sigsq}
\end{figure}

In \fig{fig:sigsq}, we show the numerical solutions using the GRV-M potential with \eq{numGRV} (left) and using the GAU-M potential with \eq{numGAU} (right), and compare them with the corresponding analytic solutions which are given by \eq{sigsol}. Using the initial values whose parameter sets can be seen in Table \ref{parameterSET}, we plot the numeric and analytic solutions in \fig{fig:sigsq}. In the top-left panel, the numerical plots (red-plus dots for SET-1 and blue-cross dots for SET-2) have the same amplitudes as the analytical ones (green-dashed line for SET-1 and purple-dotted-dashed line for SET-2), we, however, can see the significant differences for the frequencies of each oscillation. We notice that these discrepancies come from the artifact of our choice with $l=2-2|K|$ in \eq{modelA}, since the choice is not appropriate for $\sigma\ll \order{1}$, recalling $\sigma(0)\sim 4.09\times 10^{-3}$ in SET-1 and SET-2. Shortly, we will obtain numerically this power $l$, and show that the semi-analytic solutions we obtained match with the numerical ones. With SET-3 and SET-4, we can see that $\sigma(0)$ is not so small as opposed to the previous cases, \ie\  $\sigma(0)\sim 6.07\times 10^{-1}$; thus, in the left-bottom panel of \fig{fig:sigsq} we can see a nice agreement between the numerical plots (red-plus dots for SET-3 and blue-cross dots for SET-4) and the analytic plots (skyblue-dotted-dashed line for SET-3 and black-dotted line for SET-4).

Similarly,  we show the numerical and analytic plots for the GAU-M potential in the right-panels of \fig{fig:sigsq} using the parameter-sets: for SET-5 and SET-6 in the right-top panel and for SET-7 and SET-8 in the right-bottom panel. By changing the values of the third eccentricity $\varepsilon^2$ (see TABLE \ref{parameterSET}), the numerical plots deviate slightly from our analytic lines in the right-top and right-bottom panels of \fig{fig:sigsq} as we can expect; in particular, we can see that our analytic values of both the frequencies and amplitudes of $\sigma^2(t)$ are larger than the numerical ones, and this difference can be significantly reduced when the orbits of the AD planets is nearly circular with $\varepsilon^2=0.1$.

\begin{figure}[!ht]
  \begin{center}
	\includegraphics[angle=-90, scale=0.4]{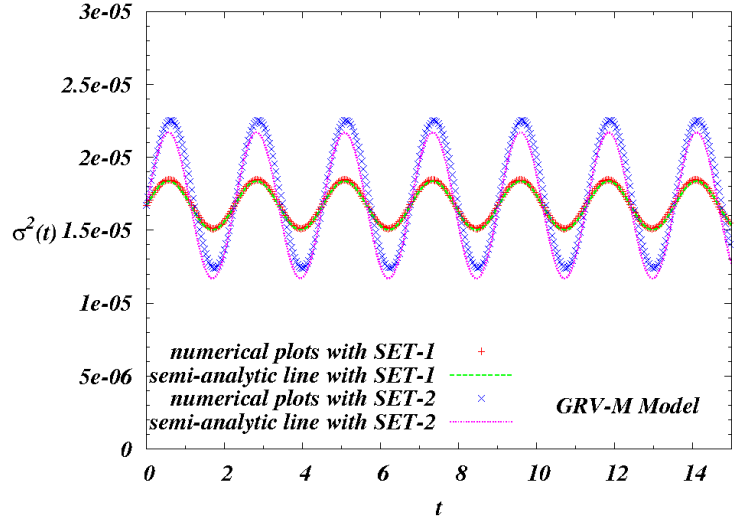}
  \end{center}
  \caption{ Substituting the numerical value, $l=1.86002$, into \eq{modelA}, we plot the semi-analytic evolution for $\sigma^2(t)$. The semi-analytic solutions agree with the numerical solutions.}
  \label{fig:sigsqSEMI}
\end{figure}

As we have seen in the left-top panel of \fig{fig:sigsq}, our analytic value, $l=1.8$, in \eq{modelA} are not good enough to reproduce the numerical solutions since $\sigma(t)\ll \order{1}$. Therefore, we set a trial function, $f(\sigma)=\half \sigma^{\alpha}+b^2_*\sigma^6$, where a numerical value $\alpha$ is found by using the 'fit' command in the numerical software called 'gnuplot'. We find that $\alpha=1.86002:=l$ is the best value of $\alpha$, where we fitted this trial function $f(\sigma)$ onto the numerical full potential in \eq{numGRV} for the range of $\sigma \in [1.0\times 10^{-2}-1.0\times 10^{-3}]$, recalling $\sigma(0)\sim 4.09\times 10^{-3}$ in SET-1 and SET-2. Using this value of $\alpha$ as the value of $l$ instead of $l=1.8$, we plot the semi-analytic evolution for $\sigma^2(t)$ in \fig{fig:sigsqSEMI} (green-dashed line for SET-1 and purple-dotted-dashed line for SET-2) against the corresponding numerical plots (red-plus dots for SET-1 and blue-cross dots for SET-2). Now, our semi-analytic solutions match with the numerical solutions.

\vspace*{5pt}

\paragraph*{\underline{\bf The average values of $w(t)$}}

Using \eqs{modelA-w}{ModelC-w}, we show both numerical values $\state{w_{num}}$ and (semi-)analytical values $\state{w_{ana}}$ of the averaged equation of state in \tbl{tbl:eos}. For all cases, the AD condensate has a negative pressure and one can say that the numerical values are of the same order as analytic values.

\begin{figure}[!ht]
  \def\@captype{table}
  \begin{minipage}[t]{\textwidth}
   \begin{center}
\begin{tabular} { |c||c|c|c|c|}

\hline 
$\state{w}$ & \multicolumn{4}{c|}{the GRV-M Model v.s. Model A} \\ 
\hline \hline
 & SET-1 & SET-2 & SET-3 & SET-4 \\
\hline 
$\state{w_{num}}$ & \multicolumn{2}{c|}{$-2.42\times 10^{-2}$} & $-4.47\times 10^{-2}$ & $-4.45\times 10^{-2}$ \\
\hline
$\state{w_{ana}}$ &  \multicolumn{2}{c|}{$-3.63\times 10^{-2}$} & \multicolumn{2}{c||}{$-5.00\times 10^{-2}$} \\
\hline
\end{tabular}
    \end{center}
  \hfill
    \begin{center}
\begin{tabular} { |c||c|c|c|c|}

\hline 
$\state{w}$ & \multicolumn{4}{c|}{the GAU-M Model v.s. Model B}\\
\hline \hline
 & SET-5 &  SET-6 & SET-7 & SET-8\\
\hline 
$\state{w_{num}}$ & $-6.43\times 10^{-1}$ & $-6.45\times 10^{-1}$ & \multicolumn{2}{c|}{$-8.00\times 10^{-1}$} \\ 
\hline
$\state{w_{ana}}$ & \multicolumn{2}{c|}{$-6.43\times 10^{-1}$} & \multicolumn{2}{c|}{$-8.04\times 10^{-1}$}\\ 
\hline
\end{tabular}
    \end{center}
\tblcaption{Using \eqs{modelA-w}{ModelC-w}, we show the both numerical values $\state{w_{num}}$ and analytical values $\state{w_{ana}}$ for the averaged equations of state. The values of $\state{w_{ana}}$ in SET-1 and SET-2 are semi-analytically obtained by substituting $l=1.86002$ into \eq{modelA}.  For all cases, the AD condensate has a negative pressure, and these analytic estimates are reasonable against the numerical values.}
\label{tbl:eos}
  \end{minipage}
\end{figure}

\vspace*{5pt}

\paragraph*{\underline{\bf The values of $\Phi$}}

In TABLE \ref{parameterPHI}, we show the numerical and (semi-)analytic values of $\Phi$ in both the GRV-M Model and GAU-M Model, which are analytically obtained in Sec. \ref{MODEL-ABC}. Our analytical values agree well with the numerical values. These values suggest that the orbits in the GRV-M Model and GAU-M Model are nearly either elliptic or trefoil, respectively.

\begin{center}
\begin{table} [!ht]
\begin{tabular} { |c||c|c|c|c||c|c|c|c| }

\hline 
$\Phi$ & \multicolumn{4}{c||}{the GRV-M Model v.s. Model A} & \multicolumn{4}{c|}{the GAU-M Model v.s. Model B}\\
\hline \hline
 & SET-1 & SET-2 & SET-3 & SET-4 & SET-5 &  SET-6 & SET-7 & SET-8 \\
\hline 
$\Phi_{num}$ & 1.591 & 1.590 & 1.605 & 1.604 & 2.210 & 2.206 & 2.221 & 2.217 \\ 
\hline
$\Phi_{ana}$ &  \multicolumn{2}{c|}{1.612 (analytic) or 1.599 (semi-analytic)} & \multicolumn{2}{c||}{1.605} & \multicolumn{2}{c|}{2.221} & \multicolumn{2}{c|}{2.221}\\ 
\hline
\end{tabular}
\caption{We show the numerical and (semi-)analytic values of $\Phi$ in both the GRV-M Model and GAU-M Model, which are analytically obtained in Section \ref{MODEL-ABC}.}
\label{parameterPHI}
\end{table}
\end{center}

\subsubsection{The orbit of an Affleck-Dine ``planet'' in an expanding universe}

We carry out our numerical simulation in an expanding universe when the inflaton field, which is trapped by a quadratic potential, starts to coherently oscillate around the vacuum during the reheating era. Then the evolution of the Hubble expansion, $H(t)$, and scale factor, $a(t)$, follows as ordinary nonrelativistic (zero-pressure) matter, see \eq{compress}. For $l=2$, we find $H=\frac{2}{3(t+t_0)}$ and $a(t)=a_0\bset{\frac{t+t_0}{t_0}}^{2/3}$, where $a_0$ is given by the value of $a(t)$ at $t=0$ and we set $a_0=0.1$. We also set the initial time as $t_0=4\times 10^2$ for the GRV-M Model and $t_0=4\times 10^4$ for the GAU-M Model. Notice that with this choice of $t_0$ our simulation starts from the same physical time because we rescaled the time by either $m\sim 10^2$ GeV or $M_s\sim 10^4$ GeV, respectively. We again solve the equation of motion, \eq{radhomo}, numerically using the 4th order Runge-Kutta method and compare them with following ans\"{a}tze we will introduce. In order to see the significant effects from Hubble expansion, we use SET-3 in the GRV-M Model and SET-7 in the GAU-M Model as the initial parameters.

In an expanding spacetime, one can guess that our analytical results in Minkowski spacetime should be changed. In particular, the amplitude of $\sigma(t)$ may decrease due to the Hubble damping as we saw in the quadratic case in appendix \ref{sec2}, and similarly the frequency $W$ in \eq{condbound} should be changed. Hence, the orbit of the AD planet can be a precessing spiral or shrinking trefoil in either GRV-M or GAU-M Model as one can see \cite{Jokinen:2002xw}. Let us give an ansatz for $\sigma^2(t)$,
\be\label{sigantz}
\sigma^2(t)=\bset{\frac{t_0}{t+t_0}}^{\alpha_1}\tisig^2 \bset{1+\varepsilon^2\cos{\bset{\widetilde{W} \cdot \bset{\frac{t_0}{t+t_0}}^{\alpha_2}\cdot t+\frac{3\pi}{2}}}}.
\ee
Here, we use the Minkowskian values of $\tisig$ and $\widetilde{W}$, and will obtain the possible values of $\alpha_{1,\ 2}$ in both models. From \eqs{sigcr}{condbound} by ignoring the nonrenormalisable term and recalling $a(t)=a_0\bset{\frac{t+t_0}{t_0}}^{2/3}$, we can find the following proportionality relations: $\sigma_{cr}(t)\propto (t+t_0)^{-4/(l+2)} \simeq (t+t_0)^{-2/(2-|K|)}$ and $W(t)\propto (t+t_0)^{-\frac{2(l-2)}{l+2}}\simeq (t+t_0)^{\frac{2|K|}{2-|K|}}$ in Model A, where we used $l=2-2|K|$. In Model B, we obtain $\sigma_{cr}\propto (t+t_0)^{-2}$ and $W(t)\propto (t+t_0)^2$. Therefore, we set $\alpha_1=\frac{4}{2-|K|},\;\alpha_2=-\frac{2-|K|}{2|K|}$ in Model A, and $\alpha_1=4,\;\alpha_2=-2$ in Model B. We believe that our ans\"{a}tze are valid as long as the nonrenormalisable term does not play a role, and the frequency of the coherent rotation, $\order{W(t)}$, is rapid compared to the Hubble expansion rate, $\order{H}$. The latter restriction implies that the rotation time scale is much shorter than the time scale of the Hubble expansion, \ie\  $W^{-1}(t)\gg H^{-1}$ \cite{Turner:1983he}.

\vspace*{10pt}

\paragraph*{\underline{\bf The motion of $\sigma^2(t)$}}

\begin{figure}[!ht]
  \begin{center}
	\includegraphics[angle=-90, scale=0.28]{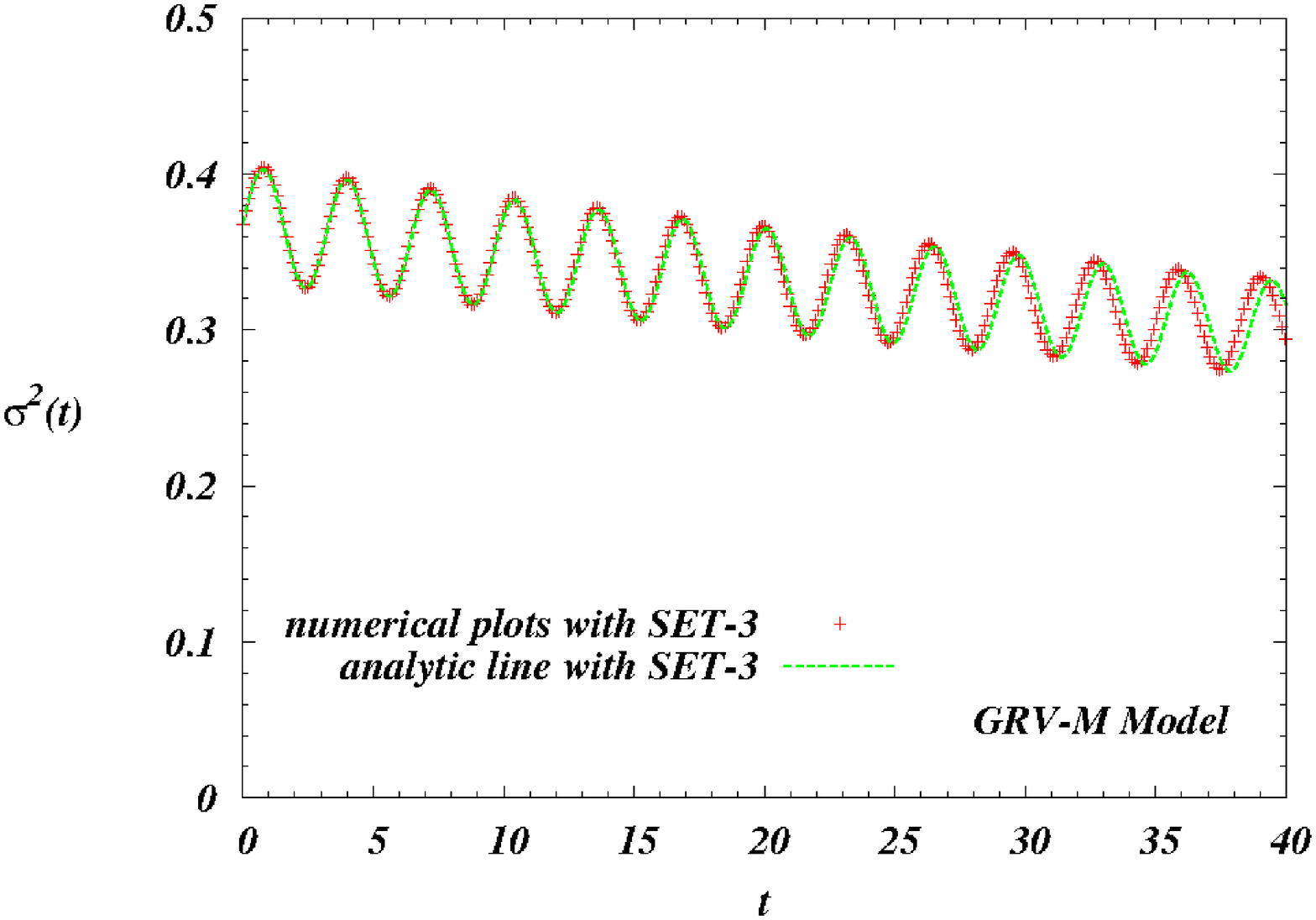}
	\includegraphics[angle=-90, scale=0.28]{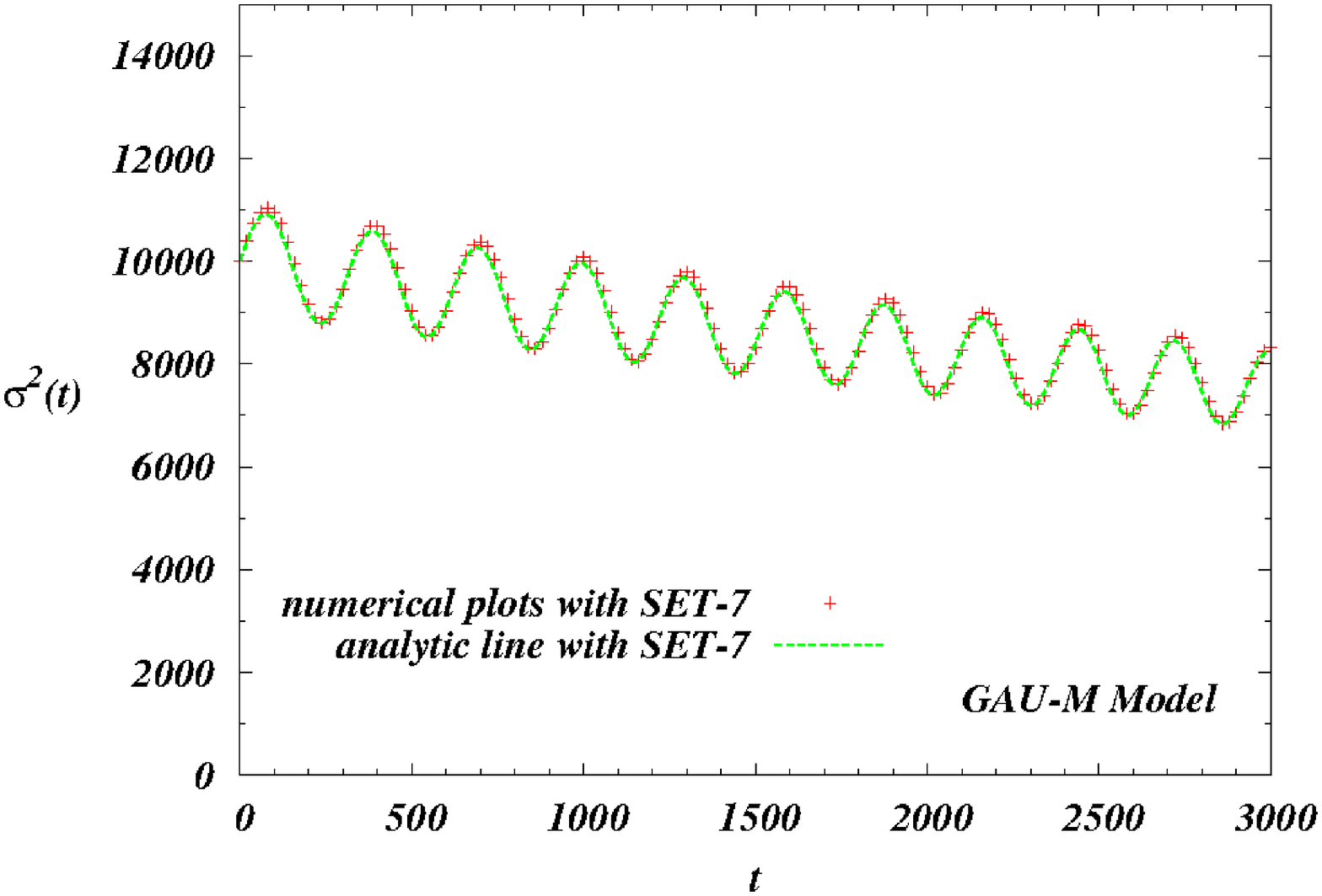}
  \end{center}
  \caption{ We plot the evolution of $\sigma^2(t)$ with the numerical data (red-plus dots) for the GRV-M Model (left) and for the GAU-M Model (right) and with the analytic data (green-dotted lines) by using our ans\"{a}tze introduced in \eq{sigantz}.}
  \label{fig:sigexp}
\end{figure}

In \fig{fig:sigexp}, we plot the evolution of $\sigma^2(t)$ with the numerical data (red-plus dots) for the GRV-M Model (left) and for the GAU-M Model (right) and with the analytic data (green-dotted lines) using our ans\"{a}tze \eq{sigantz}. The readers should compare the Minkowskian cases of SET-3 (left bottom panel) and SET-7 (right bottom panel) in \fig{fig:sigsq} with the corresponding expanding background cases. For both potential cases, the amplitudes of $\sigma^2(t)$ decrease in time as we expected, and our analytic plots excellently agree with the corresponding numerical results. In the left panel of \fig{fig:sigexp}, the difference between the analytic line and the numeric plots arises in the late time. We believe that this comes from the artifact of the approximation on $l=2-2|K|$ in the GRV-M Model, \eq{numGRV}, since the values of $\sigma^2(t)$ decrease to the region where the above approximation does not hold, \ie\  for $\sigma\ll \order{1}$ as we saw in the left-top panel of \fig{fig:sigsq}.

\vspace*{10pt}

\paragraph*{\underline{\bf The motion of the equation of state: $w(t)=p(t)/\rho_E$}}

\begin{figure}[!ht]
  \begin{center}
	\includegraphics[angle=-90, scale=0.28]{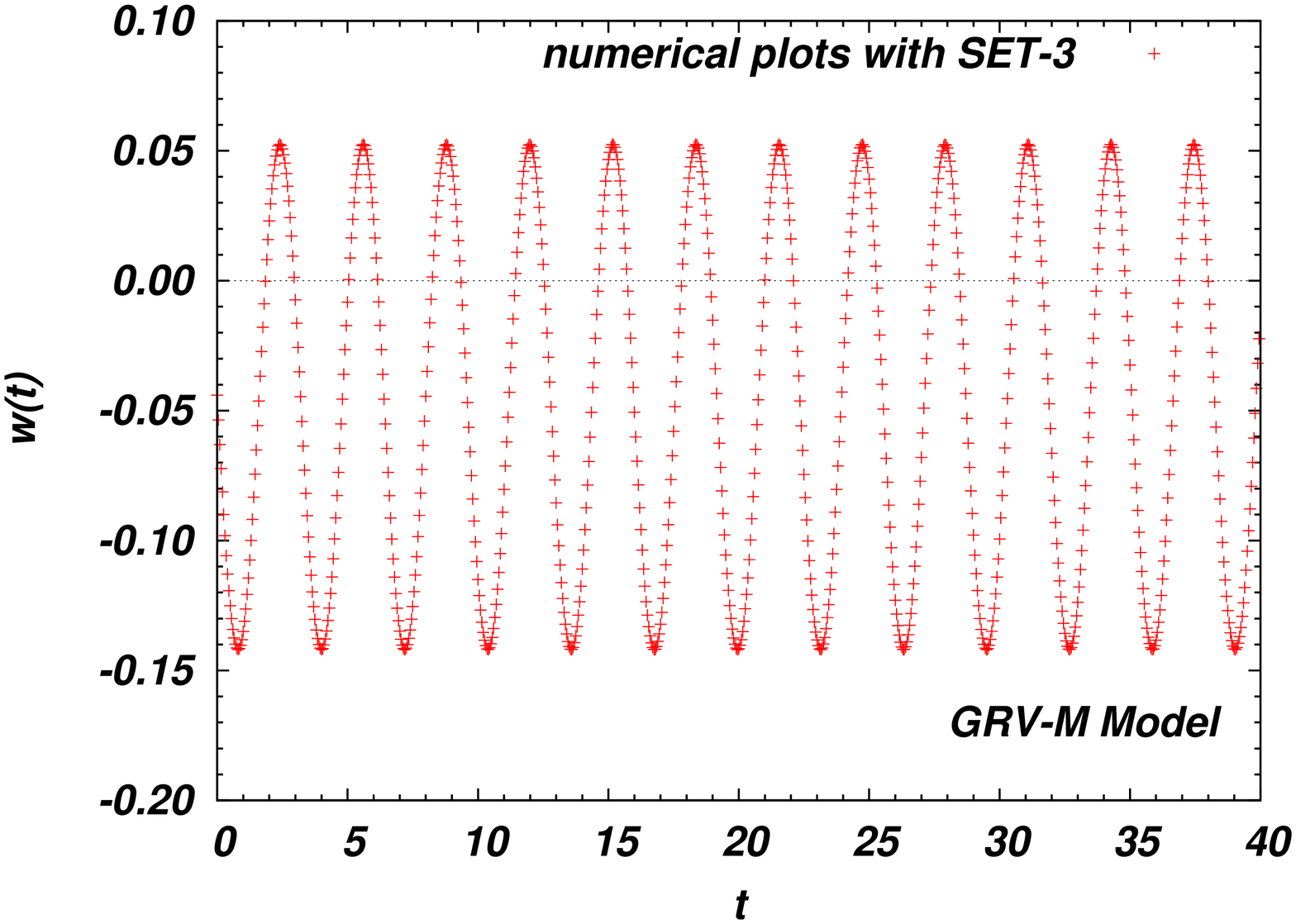}
	\includegraphics[angle=-90, scale=0.28]{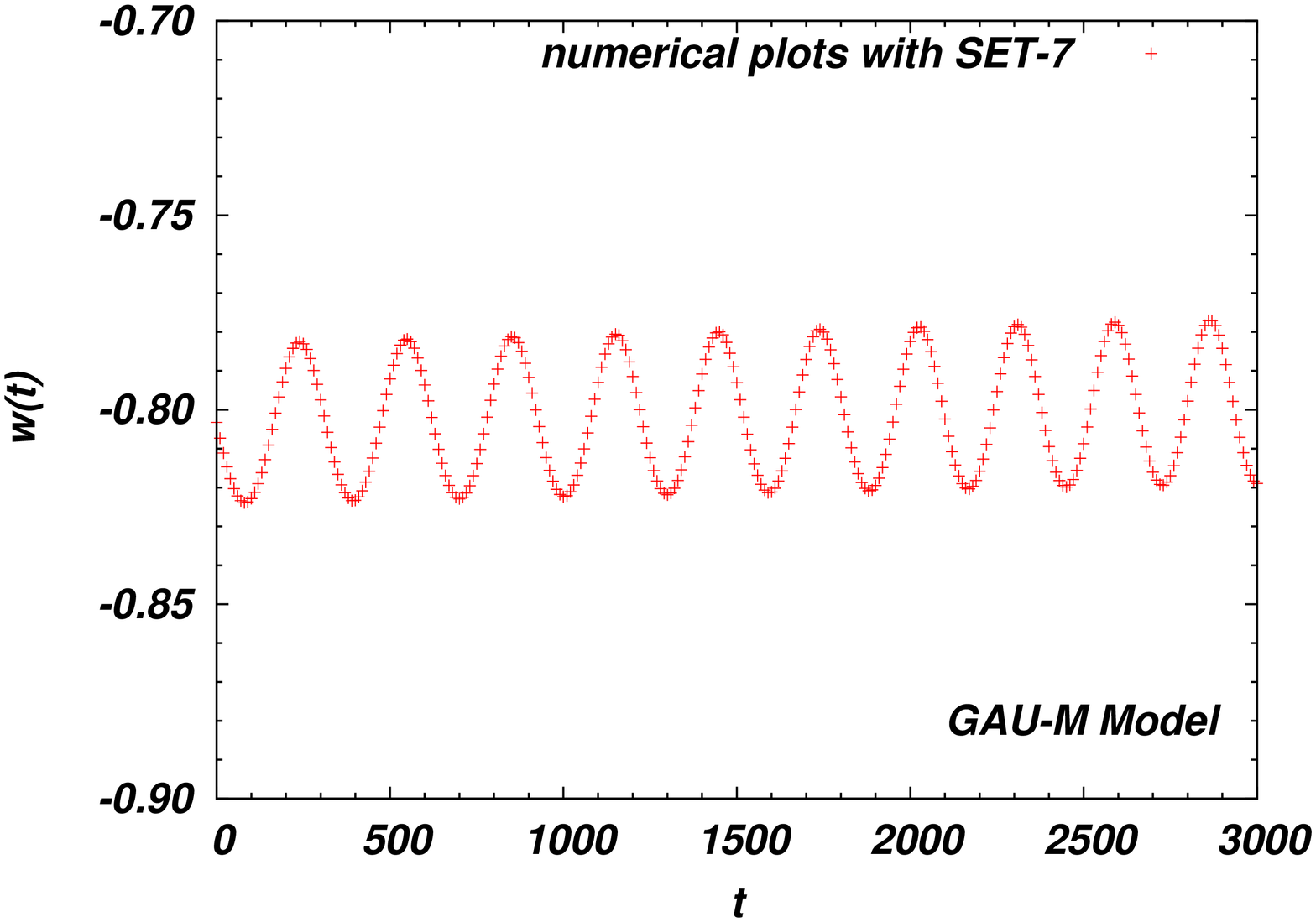}
  \end{center}
  \caption{ Using the initial conditions with SET-3 (right-panel) and SET-7 (left-panel) in Table \ref{parameterSET}, we plot the numerical values of the equation of state which are given by $w(t)\equiv p(t)/\rho_E$, where $p(t)$ and $\rho_E$ are the pressure and energy density of the AD condensate.}
  \label{fig:eqsexp}
\end{figure}

In \fig{fig:eqsexp}, we plot the numerical values of the equation of state, which is given by $w(t)\equiv p(t)/\rho_E$, where $p(t)$ and $\rho_E$ in \eq{rhop} are the pressure and energy density of the AD condensate. The averaged pressure over the rotations seems to be negative in the GRV-M Model, see the left panel; whereas, the pressure in the GAU-M Model is always negative, see the right panel. The frequencies of the rotation for $w(t)$ in both cases are, respectively, similar as the corresponding frequencies of $\sigma^2(t)$, see \fig{fig:sigexp}; however, the phases are different from the phases of $\sigma^2(t)$ approximately by $\pi$.

\vspace*{10pt}

In summary, we have analytically obtained the nearly circular orbits for both the GRV-M Model and the GAU-M Model in \eqs{numGRV}{numGAU} approximated by Model A and Model B in \eqs{modelA}{modelC}. We then checked that the semi-analytic results in \eqs{modelA-W}{modelA-P} and \eqs{ModelC-W}{ModelC-w} and our ans\"{a}tze in \eq{sigantz} agree well with the corresponding numerical results obtained by solving \eqs{radhomo}{rad} numerically. In the rest of this chapter, we investigate the late evolution for the AD condensates once the spatial perturbations generated by quantum fluctuations or thermal noise from the early oscillation \cite{Allahverdi:2000zd} become non-negligible due to the negative pressure presented in \tbl{tbl:eos} and \fig{fig:eqsexp}.

\section{$Q$-ball formation and thermalisation in Minkowski spacetime}\label{sectinst}
In this section we analyse the late evolution of the AD condensates in both the GRV-M and GAU-M models, in which we find that the spatial perturbations are amplified exponentially due to the presence of the negative pressure, and the presence of negative pressure supports the existence of nontopological solitons, \ie\  $Q$-balls. As a process of reheating the Universe, the dynamics of the $Q$-ball formation is a nonequilibrium, nonperturbative, and nonlinear process, and it includes three distinct stages: \emph{pre-thermalisation} (linear perturbation), \emph{driven turbulence} (bubble collisions), and \emph{thermalisation} towards thermal equilibrium. As opposed to the reheating process, we find that the driven turbulence stage lasts longer and the subsequent thermalisation process is different, which is caused by the presence of nontopological soliton solutions. During the turbulent stages, we find scaling laws for the variances of fields and for the spectra of the charge density. In addition, we adopt numerical lattice simulations to solve classical equations of motion in Minkowski spacetime, where our numerical code is developed from LATfield \cite{latfield-web-page}, and we present the detailed nonlinear and nonequilibrium dynamics (some videos are available \cite{mit-web-page}).

\subsection{Linear evolution -- Pre-thermalisation}

The late evolution, after the AD condensate forms, depends on the properties of the models. In the standard AD baryogenesis scenario \cite{Affleck:1984fy}, the condensate governed by the quadratic potential, \eq{quad}, decays into thermal plasma that may provide our present baryons/leptons in the Universe. By including quantum and/or thermal corrections in the mass term as in \eqs{gravity-pot}{gauge-pot}, the subsequent evolution may be different from the standard AD scenario since the AD condensate has a negative pressure. The negative pressure, which causes the attractive force among particles in the condensate, amplifies the linear spatial fluctuations exponentially. We see this exponential growth for the linear perturbations in nearly circular orbit cases with the growth rate $\dot{S}_m$, and obtain the most amplified wave-number $k_m$, which give a rough estimate on the nonlinear time $t_{NL}$ and the radii of bubbles created just after the system enters into a nonlinear regime. As long as the perturbations are much smaller than the background field values, we call this initial linear perturbation stage, '\emph{pre-thermalisation}'.

\subsubsection{Arbitrary and circular orbits}

Let us consider the linear spatial instability for an AD condensate in Minkowski spacetime. First, we perturb the AD field $\phi$ with the linear fluctuations, $\dsig$ and $\dtheta$. Equations of motion for $\dsig$ and $\dtheta$ are given by \eqs{pertsig}{pertth},
\bea{A8}
 \ddot{\dsig}-\bset{\nabla^2+\dot{\theta}^2-V^{\p\p}}\dsig-2\sigma\dot{\theta}\dot{\dtheta}=0,\\
\label{A9} \ddot{\dtheta}+\frac{2\dot{\sigma}}{\sigma}\dot{\dtheta}-\nabla^2\dtheta+\frac{2\dot{\theta}}{\sigma^2}\bset{\sigma\dot{\dsig}-\dot{\sigma}\dsig}=0.
\eea
Let us rescale $\dsig$ and $\dtheta$ in the following form
\be\label{scdsig}
\dsig\sim \dsig_0 e^{S(t)+i\mathbf{k}\cdot\mathbf{x}},\spc \dtheta \sim \dtheta_0 e^{S(t)+i\mathbf{k}\cdot\mathbf{x}}.
\ee
Notice that both of the exponents $S(t)$ should be the same in each expression for $\dsig$ and $\dtheta$ in terms of a function of the wave number $\mathbf{k}$, because we are concerned only with linear perturbations. Substituting \eq{scdsig} into \eqs{A8}{A9}, we obtain
\be\label{mat}
\left(
    \begin{array}{cc}
      \ddot{S}+\dot{S}^2+\mathbf{k}^2 -\dot\theta^2+V^{\p\p} & -2\dot\theta\dot{S} \\
      2 \dot\theta \bset{\dot{S}-\frac{\dot{\sigma}}{\sigma}} & \dot{S}^2+\frac{2\dot{\sigma} \dot{S}}{\sigma} +\mathbf{k}^2 \\
    \end{array}
  \right)
    \left(
    \begin{array}{cc}
        \dsig \\ \sigma\dtheta \\
        \end{array}
    \right) \simeq 0,
\ee
where $V^{\p\p}\equiv \frac{d^2
V}{d \sigma^2}$ and we ignore the terms $\ddot{S}$, assuming that the linear evolution is adiabatic, \ie\  $\dot{S}^2 \gg \ddot{S}$ (WKB approximation). Notice that this assumption is violated only at the beginning of this linear evolution as we will see in the numerical subsection, Sec. \ref{qb_num}. The nontrivial solution for $\dot{S}$ can be obtained by taking the determinant of the matrix in \eq{mat}, namely
\begin{eqnarray}
\nb F(\dot{S}(k),k^2)&\equiv& \dot{S}^4 + \frac{2\dot{\sigma}}{\sigma}\dot{S}^3+\bset{2\mathbf{k}^2 +3\dot{\theta}^2+V^{\p\p}}\dot{S}^2\\ \label{FS} \label{FSK} &&+\frac{2\dot{\sigma}}{\sigma}\bset{\mathbf{k}^2-3\dot{\theta}^2+V^{\p\p}}\dot{S} + \mathbf{k}^2 \bset{\mathbf{k}^2- \dot{\theta}^2+V^{\p\p} }=0.
\end{eqnarray}
Notice that the terms involving $\dot{\sigma}$ vanish if the orbit of the AD field is exactly circular. By looking for the most amplified mode $k^2_m$, which is defined by $\left.\frac{\partial F}{\partial k^2}\right|_{k^2_m}=0$ from \eq{FS}, it implies that
\be\label{kmost}
k^2_m=\frac{\dot{\theta}^2-V^{\p\p}}{2}-\dot{S}\bset{\dot{S}+\frac{\dot{\sigma}}{\sigma}}>0,
\ee
where the inequality comes from the reality condition for $k_m$.
By considering this mode in \eq{kmost} and by solving $F(\dot{S}(k),k^2_m)=0$ in \eq{FS}, the solution of the quadratic equation for $\dot{S}_m\equiv \dot{S}(k=k_m)$ is
\be\label{Sevo}
\dot{S}_m=\frac{\frac{\dot{\sigma}}{\sigma}\bset{5\dot{\theta}^2-V^{\p\p}} \pm 2 \dot{\theta}\sqrt{\bset{\dot{\theta}^2-V^{\p\p}}^2+2\bset{\frac{\dot{\sigma}}{\sigma}}^2\bset{3\dot{\theta}^2-V^{\p\p}} }}{2\bset{4\dot{\theta}^2-\bset{\frac{\dot{\sigma}}{\sigma}}^2}},
\ee
in which we are interested in the growing mode, \ie\  $Re(\dot{S}_m)>0$. Substituting \eq{Sevo} into \eq{kmost}, we may obtain the most amplified mode. Although it is rather hard to analytically solve \eq{FSK}, we know that only one instability band exists for exactly circular orbits where $\dot{\sigma}=0$;
\be\label{instability}
0<\mathbf{k}^2< \dot{\theta}^2-V^{\p\p}(\sigma),
\ee
where $\dot{\theta}$ and $\sigma=\sigma_{cr}$ are time-independent due to the circular orbits.

\vspace*{10pt}

In addition, we can estimate a possible nonlinear time $t_{NL}$ when the spatial averaged variance, Var$(\sigma)$, becomes comparable to the corresponding homogeneous mode $\sigma$. Here, we have defined Var$(\sigma)\equiv \overline{\bset{\hat{\sigma}(\mathbf{x}, t)-\overline{\sigma}}^2}$, and a hat and a bar denote an original field and a spatial average of the field, respectively. Notice that the nonlinear time in \cite{Enqvist:1998en, Pawl:2004vi} is defined by the time when the linear fluctuation $\delta\sigma$ for the most amplified mode becomes comparable to the homogeneous-mode; however, our definition is better as we will see in the numerical subsection, Sec. \ref{qb_num}. The nonlinear time with our definition can be given by
\bea{nonlintime}
\textrm{Var}(\sigma)&\sim& \delta \sigma^2_0 \exp\bset{2 N \overline{\state{\dot{S}}}\tau} \sim \delta \sigma^2_0 \exp\bset{\int^{t_{NL}}_{t_*} 2 \state{\dot{S}_m}}\sim \sigma^2_0,\\
\label{nlt}\hspace*{10pt} \lr t_{NL} & \sim & t_* + \frac{1}{\state{\dot{S}_m}}\ln\bset{\frac{\sigma_0}{\delta \sigma_0}}.
\eea
Here, we have approximated that $\overline{\state{\dot{S}}}\sim \state{\dot{S}_m}$ and that the orbits over $N$ rotations with the period $\tau$, \eq{periodgen}, can be expressed by the integral form as shown in \eq{nonlintime}. As we assumed, the spatially averaged variance of this field is not fully developed over all modes except $k=k_m$ until $t\sim t_*$, where $t_*$ is a typical time scale when the variance starts to grow with the growth rate $\overline{\state{\dot{S}_m}}$.

\vspace*{10pt}

Our main interest in this pre-thermalisation stage is the evolution of the number of particles in terms of modes, so that we consider $\rho_Q$ as the particle number here. For a free field theory, both of the positive and negative charged particle occupation numbers develop equally. The present case, however, gives different consequences due to the presences of nonlinear interactions and the initial inequality of a charge density (baryon asymmetry). Without loss of generality, we can focus on the case where the positive charge is initially present. Since the charge density is given by $\rho_Q=\hat{\sigma}^2\dot{\hat{\theta}}$, we can approximately obtain the evolution in the linear regime using \eqs{radhomo}{phshomo},
\be\label{evochrg}
\dot{\rho}_Q\simeq \sigma^2(t)\nabla^2\delta\theta.
\ee
Hence, the charge density evolves due to the linear fluctuation of the phase field. Let $n^{\pm}_k(t)$ be the amplitude of Fourier-transformed positive and negative charge density, $n^{\pm}(\mathbf{x}, t)$, which are defined through the following decomposition, $\rho_Q=n^+(\mathbf{x}, t)-n^-(\mathbf{x}, t)$. Notice that the Fourier transformed functions, $n^\pm_k$, are related to, but are potentially different from the corresponding quantum mechanical expressions, $\tilde{n}^+_k\equiv a^\dag_k a_k,\; \tilde{n}^-_k\equiv b^\dag_k b_k$ and $Q=\int d^3x \rho_Q=\int \frac{d^3k}{(2\pi)^{3/2}} \bset{\tilde{n}^+_k - \tilde{n}^-_k}$. Here, $\tilde{n}^\pm_k$ are occupation numbers for positive and negative charged particles in a free field theory, and $a_k,\; a^\dag_k,\; b_k$ and $b^\dag_k$ are the annihilation/creation operators for both of the particles, respectively. Since we are interested in the growing mode for the positive charge density $n^+_k(t)$ in \eq{evochrg} which is initially zero except the zero-momentum mode, it implies that using \eq{scdsig}
\bea{n+k0}
\nb n^+_k(t) &\simeq& k^2|\delta\theta_0|\int^{t}_{t_0} d\tilde{t} \sigma^2(\tilde{t})e^{\state{\dot{S}(k)}\tilde{t}},\\
\label{n+k} &\sim&  k^2|\delta\theta_0| \sigma^2_{cr} \frac{e^{\state{\dot{S}}(t-t_0)}}{\state{\dot{S}}} \propto e^{\state{\dot{S}}(t-t_0)},
\eea
where $t_0$ is found numerically and we assumed $\sigma^2(t)\sim \sigma^2_{cr}$, going from the first line to the second one. Therefore, the evolution of the positive charged particle number for a mode $k$ is proportional to $e^{\state{\dot{S}(k)}(t-t_0)}$.

\vspace*{10pt}

Summarising our results, \eqs{kmost}{Sevo} are generalisations of the known results \cite{Enqvist:2002si, Kasuya:2001hg, Enqvist:1999mv}, in which the orbit of the AD field was assumed to be exactly circular. We also obtained the nonlinear time $t_{NL}$ in \eq{nlt} and the exponential growth of the particle number in \eq{n+k}.

\subsubsection{Nearly circular orbits in Model A and B}\label{NCO}

Using the results obtained in the previous subsection, we can compute the most amplified mode $\state{\mathbf{k}^2_m}$ and the growing mode $\state{\dot{S}_m}$ averaged over one rotation of the nearly circular orbits for the models introduced in Section \ref{MODEL-ABC}, \ie\  Model A and Model B. We shall confirm that these values are the same as the cases when the orbits are exactly circular, which implies that the instability band, \eq{instability}, could exist even for the present nearly circular orbit cases.

\vspace*{10pt}
\paragraph*{\underline{\bf Model A:}}
Substituting the expressions, $\dot{\sigma}/\sigma,\; \dot\theta^2$ and $V^{\p\p}$ \cf\ \eqs{sigsol}{dottheta} and \eq{modelA-W}), into \eq{Sevo}, we obtain the averaged growing factor and the most amplified mode for Model A where $M^2>0$:
\bea{ext11}
\state{\dot{S}_m}&\simeq& \pm \frac{(2-l)M}{4}\sqrt{\frac{l\sigma^{l-2}_{cr}}{2}}\bset{1 + \frac{(l+2)(2n-l-2)}{2(n+2)(l-2)} \epsilon_A   },\\
\label{ext13} \state{\mathbf{k}^2_m}&\simeq& \frac{M^2l(2-l)(l+6)\sigma^{l-2}_{cr}}{32}\bset{1+\frac{(l+2)(4n-12-l^2+2nl)}{(n+2)(l-2)(l+6)}\epsilon_A},
\eea
where we substituted \eq{ext11} into \eq{kmost} to obtain $\state{\mathbf{k}^2_m}$ and these results are consistent with the case for the exactly circular orbit. In order to satisfy $\state{\mathbf{k}^2_m}>0$, we should have $l<-6,\; 0<l<2$, and \eq{ext11} implies that the condensate is unstable against spatial fluctuations when the pressure is negative with $0<l<2$, see \eq{powerpress}.

We can recover the results \cite{Enqvist:2002si} that $\state{\dot{S}_m}\simeq \frac{m|K|}{2}\bset{1+\frac{|K|}{2}}$ and $\state{\mathbf{k}^2_m}\simeq m^2|K|\bset{1-\frac{|K|}{4}}$ by setting $l=2-2|K|$ in \eqs{ext11}{ext13} and ignoring the nonrenormalisable term as done in \cite{Enqvist:2002si}, \ie\  $\state{\dot{S}_m} \simeq \frac{|K|M}{2}\bset{1-\frac{|K|}{2}}\sigma^{-|K|}_{cr}$ and $\state{\mathbf{k}^2_m}\simeq |K|M^2\bset{1-\frac{5|K|}{4}}\sigma^{-2|K|}_{cr}$. These are of the same order as their results, recalling that $\sigma^{-2|K|}_{cr}\sim \order{1}$ due to $|K|\ll \order{1}$.

\vspace*{10pt}
\paragraph*{\underline{\bf Model B:}}

Similarly, we can also obtain the averaged growing factor and the most amplified mode for Model B from \eq{ModelC-W}
\be\label{ext12}
\state{\dot{S}_m}\simeq \frac{m^2_{\phi}}{\sqrt{2}\sigma_{cr}}\bset{1-\frac{n-1}{n+2}\epsilon_B},\hspace*{5pt}
\state{\mathbf{k}^2_m}\simeq \frac{3m^4_{\phi}}{2\sigma^2_{cr}}\bset{1-\frac{2(n-3)}{3(n+2)}\epsilon_B}
\ee
which to leading order reproduces the results \cite{Kasuya:2001hg}, where the AD orbit was assumed to be exactly circular and the nonrenormalisable term was ignored.

\vspace*{10pt}
Before we finish this subsection, let us remark upon the classical and absolute stability of AD condensates. Lee found \cite{Lee:1994qb} the dispersion relation for the waves of linear fluctuations from \eq{FSK} when the orbits of the AD field are bounded. In the longwave-length limit, there exists one massive and one massless mode. The massless mode can be interpreted as the sound wave whose sound speed should be real for the classical stability reason, and the squared value of the sound speed is related to the value of $\state{w}$ in \eq{eosg}. Therefore, this stability condition for the sound waves corresponds to the sign of the pressure in the AD condensate. In other words, the AD condensate has a negative pressure if the sound speed is imaginary; equivalently, it is classically unstable against spatial fluctuations. The zero-pressure AD condensate whose energy density is minimised with respect to any degrees of freedom is equivalent to the $Q$-matter phase as Coleman discussed in \cite{Coleman:1985ki}, where the absolutely stable $Q$-matter can be excited by classically stable sound waves. 

\subsection{Non-linear evolution and nonequilibrium dynamics}

\subsubsection{Driven (Stationary) and free turbulence}

Even when the perturbations are fully developed to support the nonlinear solutions, the system is still far from thermal equilibrium. Indeed, the system enters into more stochastic stages, 'turbulence regimes', where the strength of the turbulent behaviour depends on the ``Reynolds'' number \cite{Grana:2001ms}. As a theory of reheating of the Universe, a general nonequilibrium system goes through two different turbulence stages, going from driven turbulence to free turbulence. A major energy transfer from the zero mode takes place during driven turbulence. Garcia-Bellido \textit{et. al.} \cite{GarciaBellido:2007af} observed that bubbles form and collide during this stage in tachyonic preheating, and they proposed that the bubble collisions can be an active source of gravitational waves \cite{Khlebnikov:1997di}. In the usual reheating scenarios, this stage terminates when the energy left out in the zero-mode becomes smaller than the energy stored in other modes (created particles). Since the energy exchange between zero-mode and other modes becomes negligible, the particle distribution is self-similar in time (free turbulence) and evolves towards thermal equilibrium. In the free turbulence stage, the quantum effects change the late time evolution significantly, and the created particles are distributed following Bose-Einstein statistics rather than in a classical manner. As long as an active and stable energy source exists in momentum space, we expect that the driven turbulence stage lasts for a long time. In the case of $Q$-ball formation, we expect that this active energy source corresponds to the excited states of $Q$-balls; hence, the driven turbulence stage may last longer compared to the linear perturbation regime as opposed to the usual reheating Universe scenarios. Note that during this thermalisation stage the transition from the classical to quantum regime becomes important \cite{Micha:2004bv}; in the rest of this chapter we concentrate on the case where the system is governed by classical evolution all the time.

In turbulent stages, the scaling law can be found \cite{Micha:2004bv}:
\be\label{variation}
\textrm{Var}(\sigma) \propto t^{p},
\ee
where the power $p$ depends on the parameters of the models, \eg\ the relativistic values of $p$ are $p=\frac{1}{2m-1}$ in the driven turbulence regime and $p=-\frac{2}{2m-1}$ in the free turbulence regime. Here, $m$ is the number with which particles mainly interact. For the free turbulence regime, the particle number distribution follows a scaling law from the time $t_{free}$ when the free turbulence turns on, namely
\be\label{nmbdist}
n_k(t)=t^{-\frac{4}{2m-1}}n_{k_*}(t=t_{free}),
\ee
where $k_*\equiv kt^{-\frac{1}{2m-1}}$.

\subsubsection{Thermal equilibrium state in the presence of nontopological solitons}

In this sub-subsection, we show that the condition of the negative pressure is the same as the existence condition of $Q$-balls,  \eq{EXIST}. This does not always mean that the spatially unstable condensate evolves towards $Q$-balls; with given initial conditions, the condensate may evolve into other thermo-dynamically favoured states in which the free energy is minimised. 

The ansatz of non-thermal $Q$-balls claims that $\dot{\hat{\theta}}$, which corresponds to the ``chemical potential'' $\omega$, is constant, and that the radial field $\hat\sigma$ should be time-independent and depend on the radius $r$ of the $Q$-ball, \ie\  $\hat{\phi}=\hat{\sigma}(r)e^{i\omega t}$ in \eq{qansatz}. Hence, the existence condition of $Q$-balls at zero-temperature is
\be\label{exist}
\textrm{min}\bset{\frac{2V}{\hat\sigma^2}}\leq \omega^2 < \left.\frac{d^2V}{d\hat\sigma^2}\right|_{\hat\sigma=0}.
\ee
This condition implies that the potential should grow less quickly than a quadratic term; thus, it is equivalent to the fact that the AD condensate has a negative pressure for $l<2$ in \eq{modelA}, see \eq{powerpress}. Notice that this condition only tells us that $Q$-balls may appear after an unstable AD condensate fragments. The evolution to the thermal equilibrium state is rather hard to compute analytically, and it is related to stability problems of the $Q$-balls \cite{Laine:1998rg, Copeland:2009as}. Therefore, we conduct numerical lattice simulations that give the entire processes of nonlinear as well as out-of-equilibrium evolution.

\subsection{Numerical results}\label{qb_num}

In this subsection, we present detailed numerical results from lattice simulations for both GRV-M and GAU-M models with the parameter sets, SET-3 and SET-7 shown in \tbl{parameterSET}; we then check our analytical results obtained in the previous sections. In order to solve the second-order partial differential equations, $\frac{d^2\hat{\phi}}{dt^2} -\nabla^2 \hat{\phi} + \frac{dV}{d\hat{\phi}^\dag}=0$, with the potentials introduced in \eqs{numGRV}{numGAU}, we use the following appropriate parameters: $dx=0.2,\; dt=0.02$ in the GRV-M Models and $dx=5.0,\; dt=0.2$ in the GAU-M Model, which minimise the numerical errors. Here, $dx$ is the fundamental lattice space and $dt$ is the time step. Note that the variables in this subsection are normalized by appropriate energy scales as in Sec. \ref{numAD}. We then conduct $3+1$ (and $2+1$)-dimensional lattice simulations with $512^3$ (and $512^2$) lattice units, imposing a periodic boundary condition. Our initial conditions are, $\hat\phi_0=\phi_0+\delta\phi_0$ and $\dot{\hat{\phi}}_0=\dot{\phi}_0+\delta\dot{\phi}_0$, where the initial fluctuations, $\delta\phi_0$ and $\delta\dot{\phi}_0$, are of a Gaussian noise, which are responsible for ``quantum'' fluctuations. Their fluctuations, $\delta\phi_0$ and $\delta\dot{\phi}_0$, are of order $10^{-5}$ in GRV-M case and of order $10^{-3}$ in GAU-M case. In order to visualise these detailed evolution, we use a 3D software, 'VAPOR' \cite{vapor-web-page}, and some videos of our numerical results are available in \cite{mit-web-page}.

\subsubsection{Pre-thermalisation}

\paragraph*{\underline{\bf The initial evolution --Non-adiabaticity:}}

\begin{figure}[!ht]
  \begin{center}
	\includegraphics[angle=-90, scale=0.28]{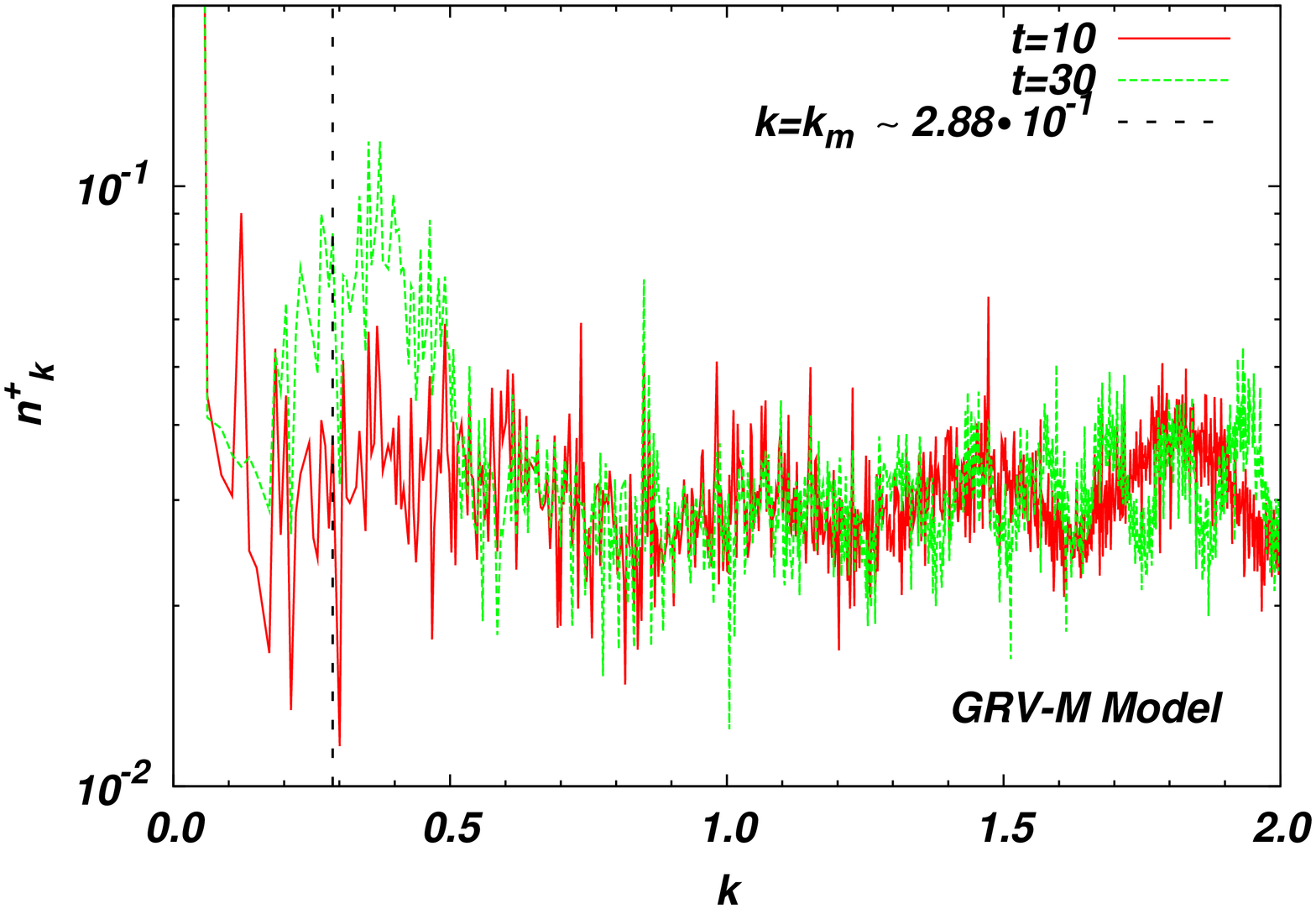}
	\includegraphics[angle=-90, scale=0.28]{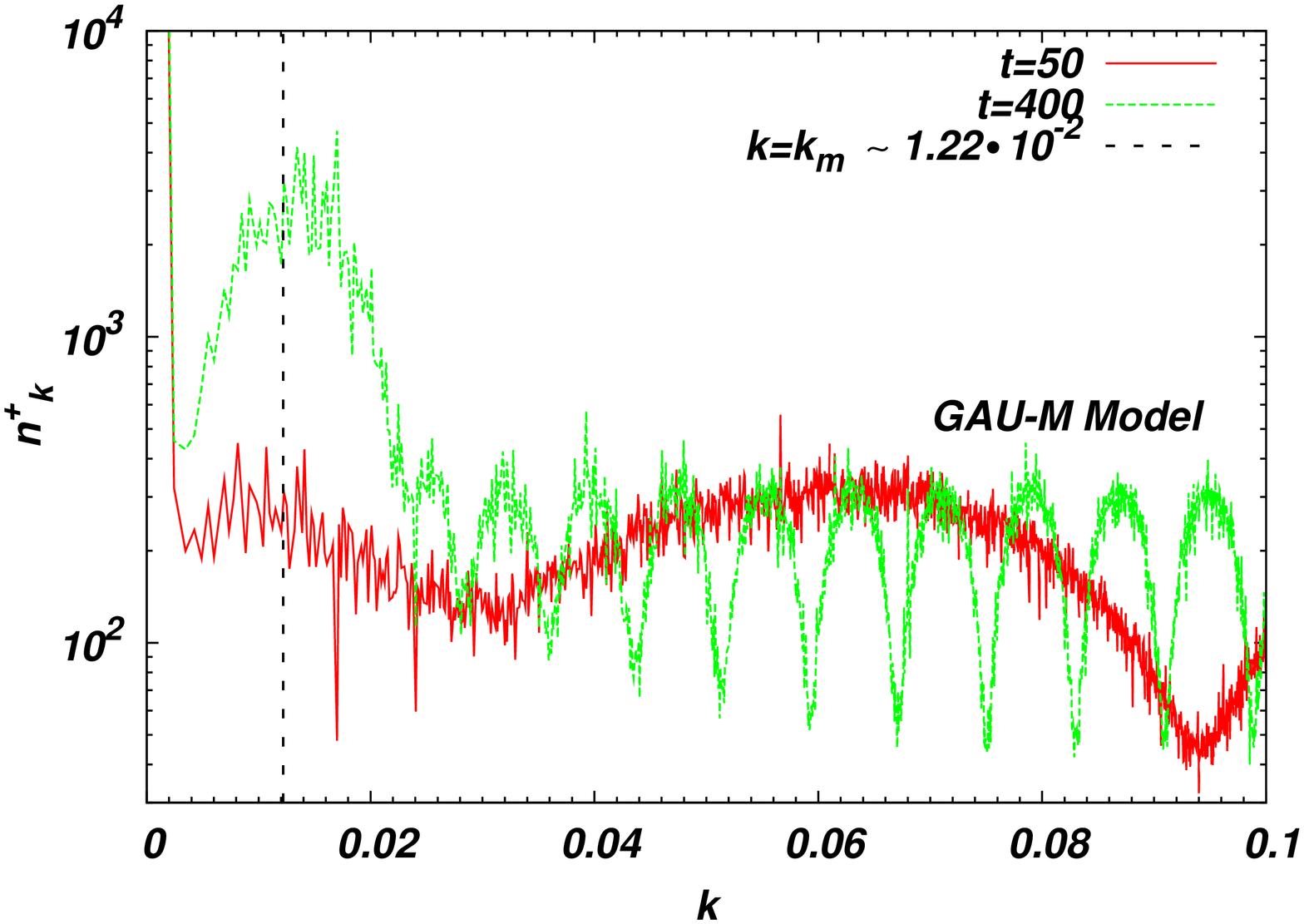}\\
	\includegraphics[angle=-90, scale=0.28]{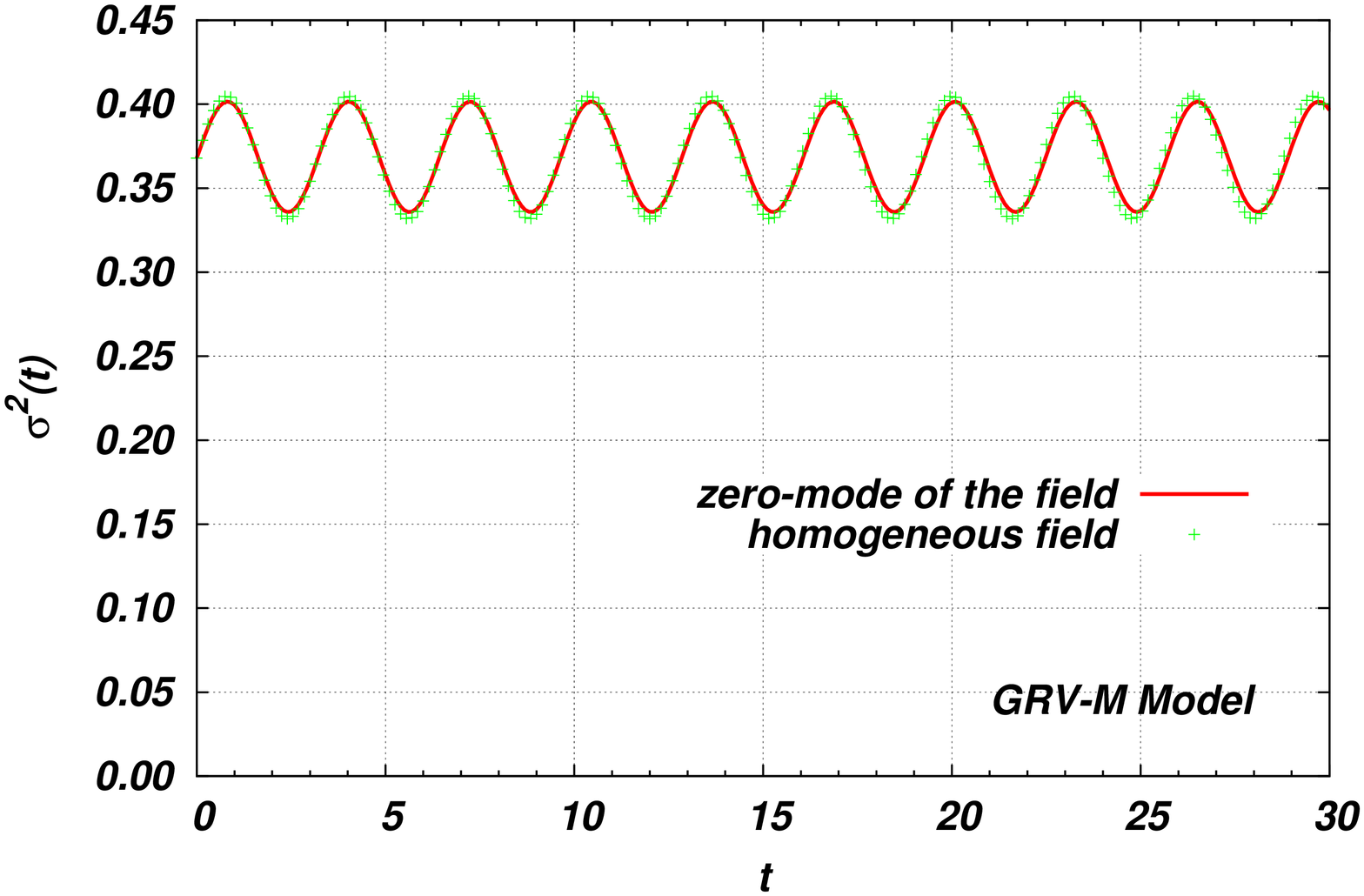}
	\includegraphics[angle=-90, scale=0.28]{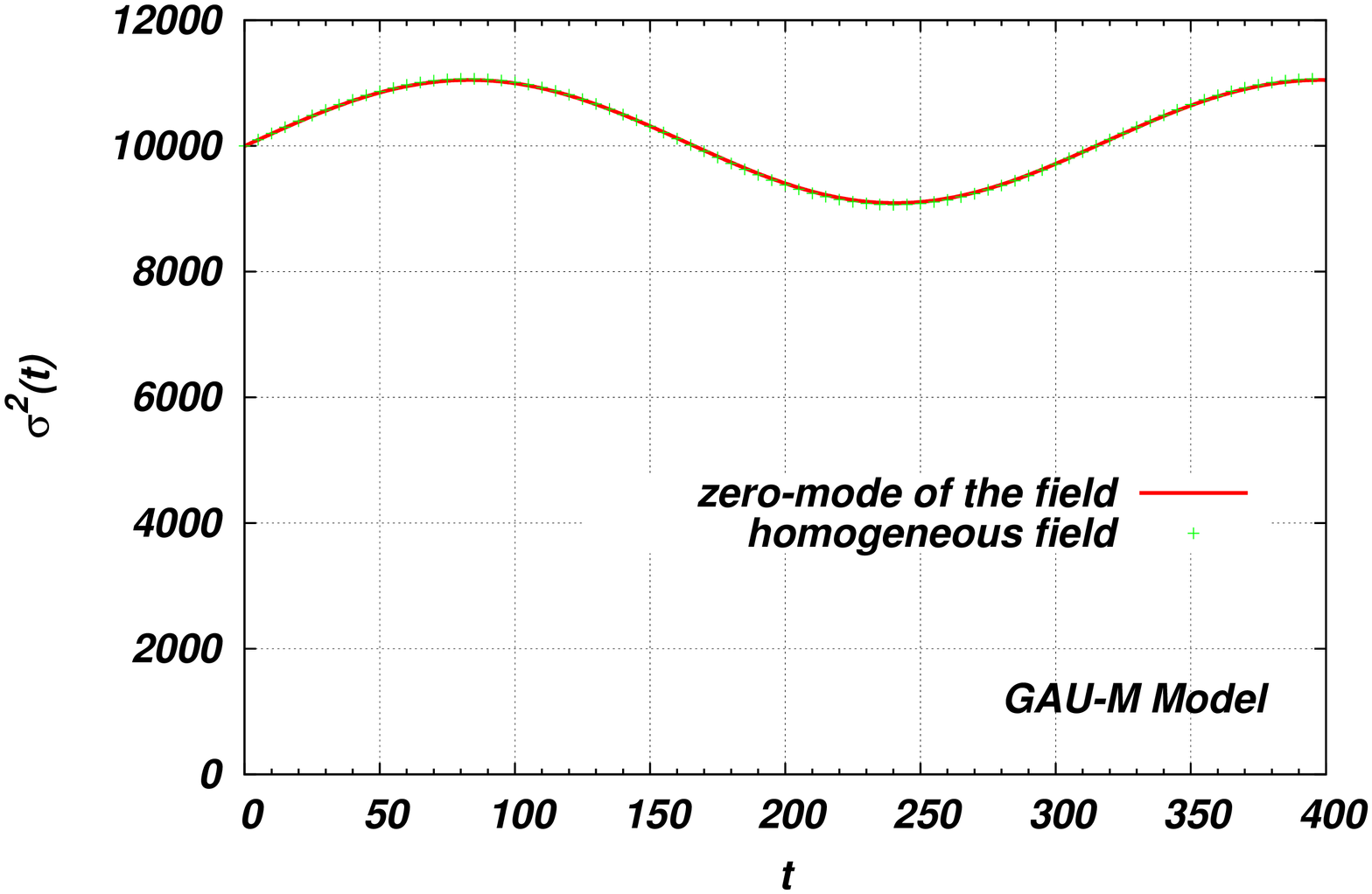}\\
	\includegraphics[angle=-90, scale=0.28]{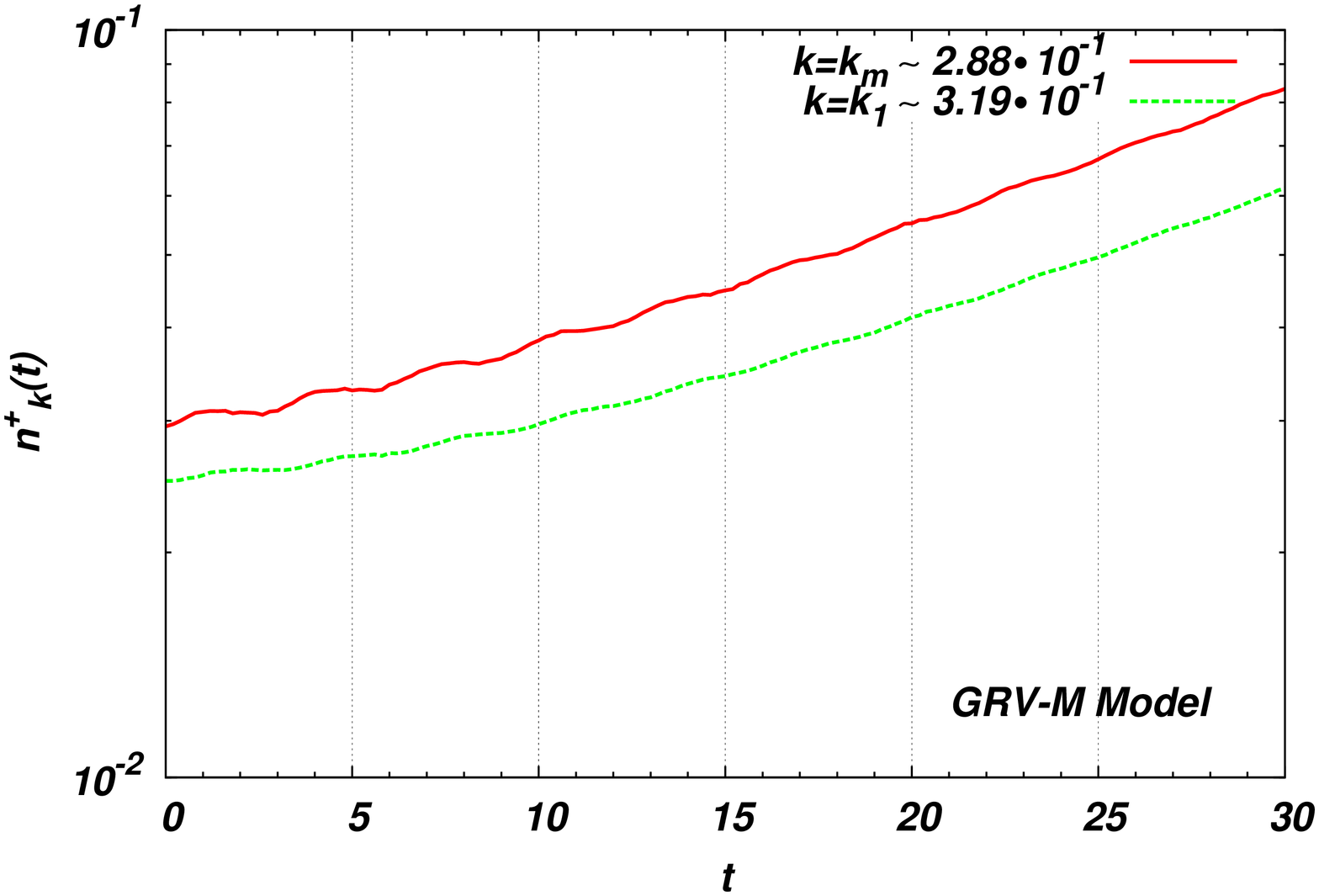}
	\includegraphics[angle=-90, scale=0.28]{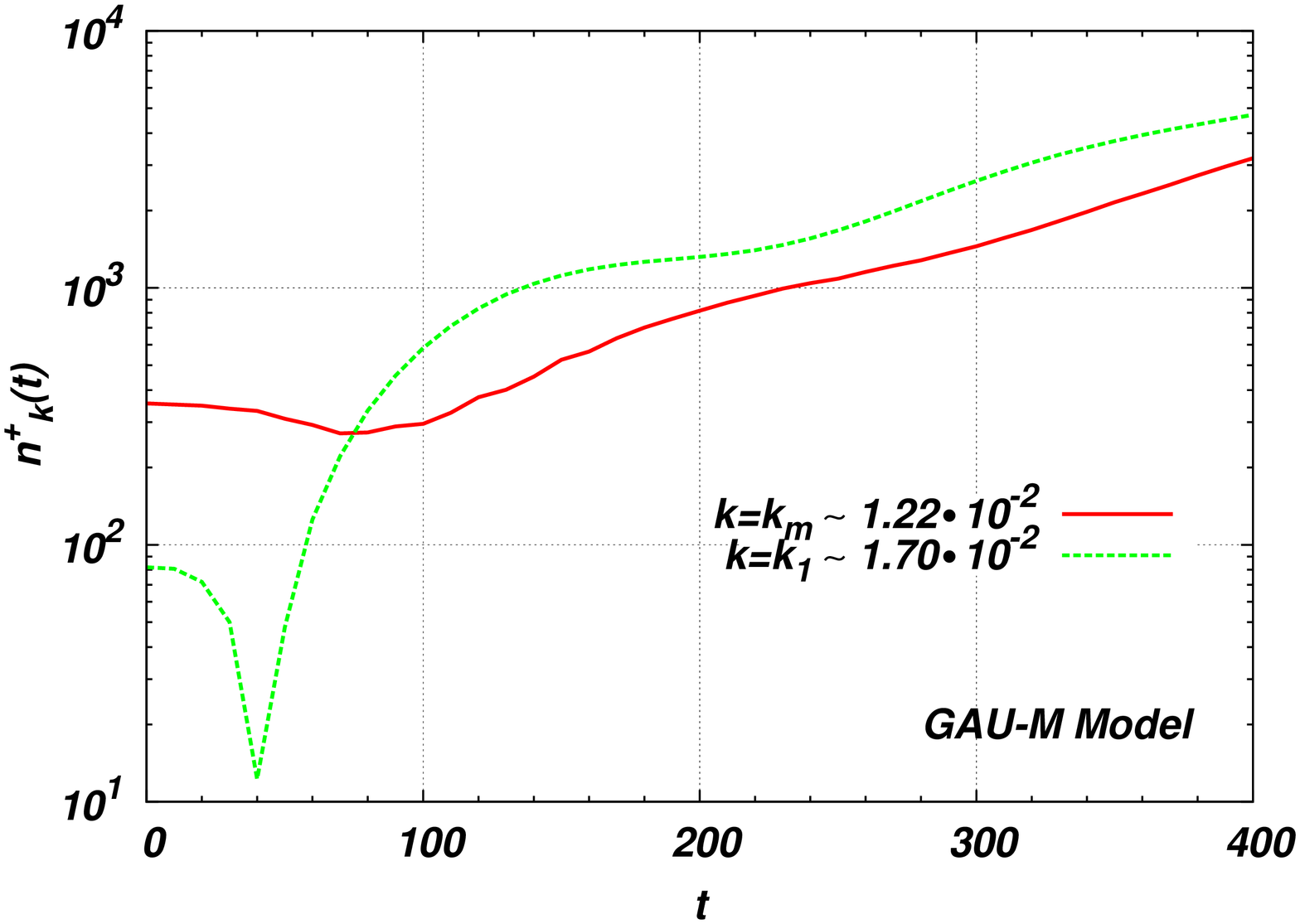}
  \end{center}
  \caption{  In the top two panels, we plot the amplitude of $n^+_k(t)$ at two different time steps for the GRV-M Model in the left panel and the GAU-M Model in the right panel, where we took the average of $n^+_\mathbf{k}(t)$ over the axes of $\mathbf{k}$. The black-dashed vertical lines indicate the analytical values of the most amplified modes $k_m$ obtained from \eqs{ext13}{ext12}. In the middle panels (GRV-M Model on left and the GAU-M Model on right), we compare the zero-mode $\overline{\sigma^2}$ (red-solid lines) and the homogeneous field $\sigma^2$ (green-plus dots) obtained in the bottom panels of \fig{fig:sigsq}. In the bottom panels of \fig{fig:nklin}, we plot the evolution of $n^+_k(t)$ for both analytic values $k_m$ (red-solid lines) and numerical values $k_1$ (green-dashed lines) of $n^+_k$ shown in the top two panels.}
  \label{fig:nklin}
\end{figure}

In the top two panels of \fig{fig:nklin}, we plot the amplitude of $n^+_k(t)$, where we took the average of $n^+_\mathbf{k}(t)$ over the axes of $\mathbf{k}$. We show the amplitudes of $n^+_k(t)$ for the GRV-M Model in the left panel and for the GAU-M Model in the right panel at two different time steps. In the panels, we indicate the analytical values of the most amplified modes $k_m$ obtained from \eqs{ext13}{ext12} with black-dashed vertical lines. In the GRV-M Model, the amplitude with $t=30$ (green-dashed line) is a little noisy to see the first peak $k_1$ in terms of $k$. Our analytical estimate, $k_m\sim 2.88\times 10^{-1}$, is located at a more infrared region than the point $k=k_1\sim 3.40\times 10^{-1}$, and the periodic structure can be seen in the higher-momentum space. In the GAU-M Model, on the other hand, we can confirm that our analytical value, $k=k_m\sim 1.22\times 10^{-2}$, agrees with the numerical value, $k_1\sim 1.70\times 10^{-2}$, in the green-dashed line; however, the analytical value appears in a slightly more infrared region. We also observe the periodic structure in the higher-momentum modes as was reported in \cite{Kasuya:2001hg}. In the middle panels (GRV-M Model on left and the GAU-M Model on right), we compare both the zero-mode, $\overline{\sigma^2}$ (red-solid lines), and the homogeneous field, $\sigma^2$ (green-plus dots), shown in the bottom panels of \fig{fig:sigsq}. The middle panels in both cases show that the zero-mode does not decay quickly, and it oscillates around $\sigma^2=\sigma^2_{cr}$. We can also check that our numerical parameters are appropriate, minimising numerical errors. In the bottom panels of \fig{fig:nklin}, we plot the evolution of $n_k(t)$ for the modes both $k_m$ (red-solid lines) and $k_1$ (green-dashed lines). In the left bottom panel, we can see the exponential growth of the amplitude in the GRV-M Model for both modes, and step-like particle production exists at the beginning of the evolution as broad resonant preheating \cite{Kofman:1997yn} (\cf\ \eq{evochrg}), and it begins to create the particles exponentially afterwards. The particles are produced quickly when the zero-mode $\sigma^2(t)$ increases in time at the beginning, see the middle panels. This is the different feature of the evolution compared to the case of resonant preheating, where particle production for the broad resonance occurs nonadiabatically when the zero mode (inflaton field) crosses the zero axis. In the right bottom panel, we can see more clearly the step-like particle creations for both modes, and then this step-like evolution smooths out, which leads to the exponential particle production as in the GRV-M case. We believe that the adiabatic condition, $\ddot{S}\ll \dot{S}^2$, is ``softly'' violated only in this initial stage since we can not see the clear exponential growth at the beginning of this evolution. In the next paragraph, we discuss the late linear evolution when this nonadiabatic evolution ceases, and show that our analytical results agree much better with our numerical ones more nicely.

\vspace*{10pt}
\paragraph*{\underline{\bf Up to the nonlinear time:}}

\begin{figure}[!ht]
  \begin{center}
	\includegraphics[angle=-90, scale=0.28]{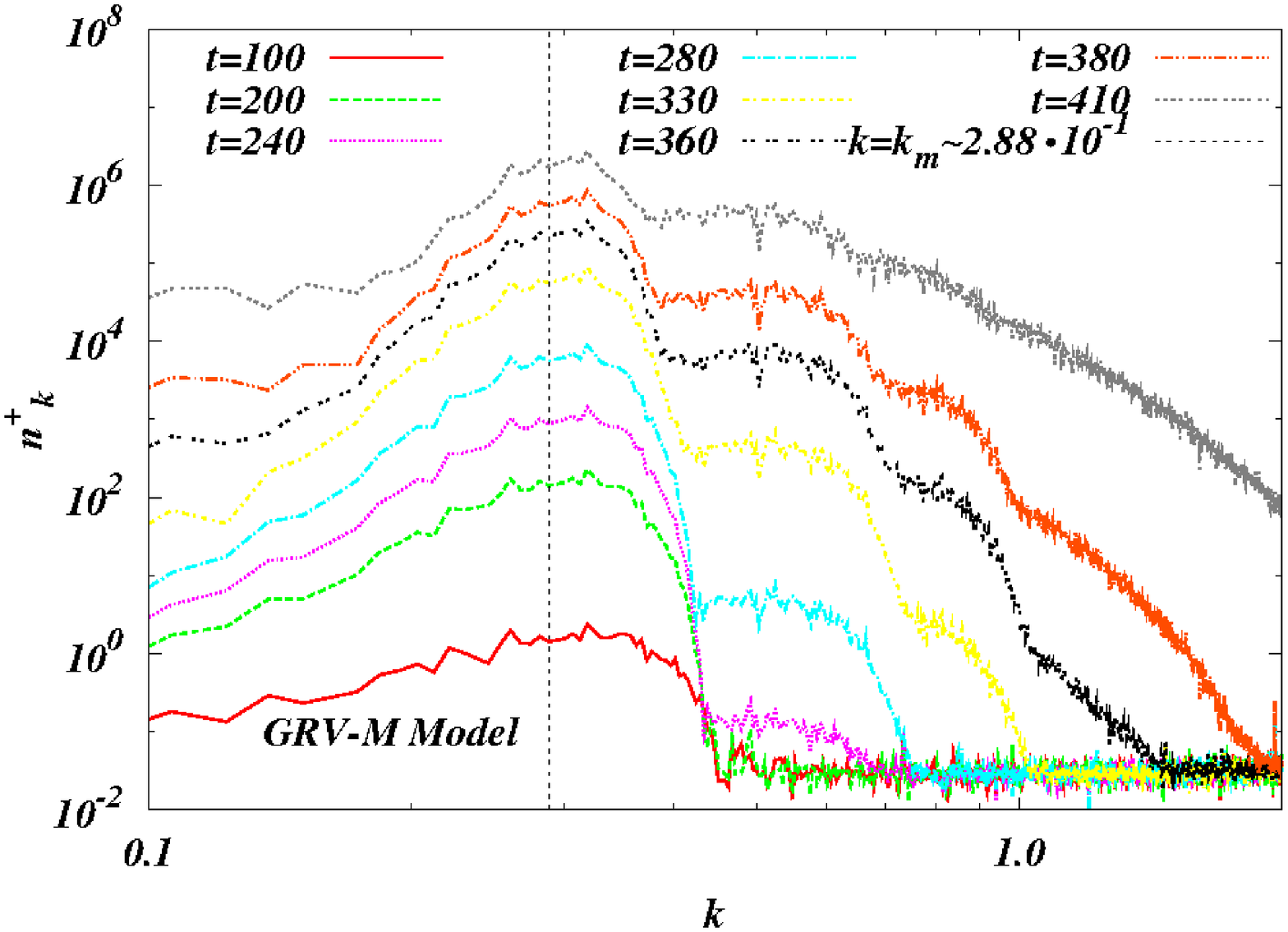}
	\includegraphics[angle=-90, scale=0.28]{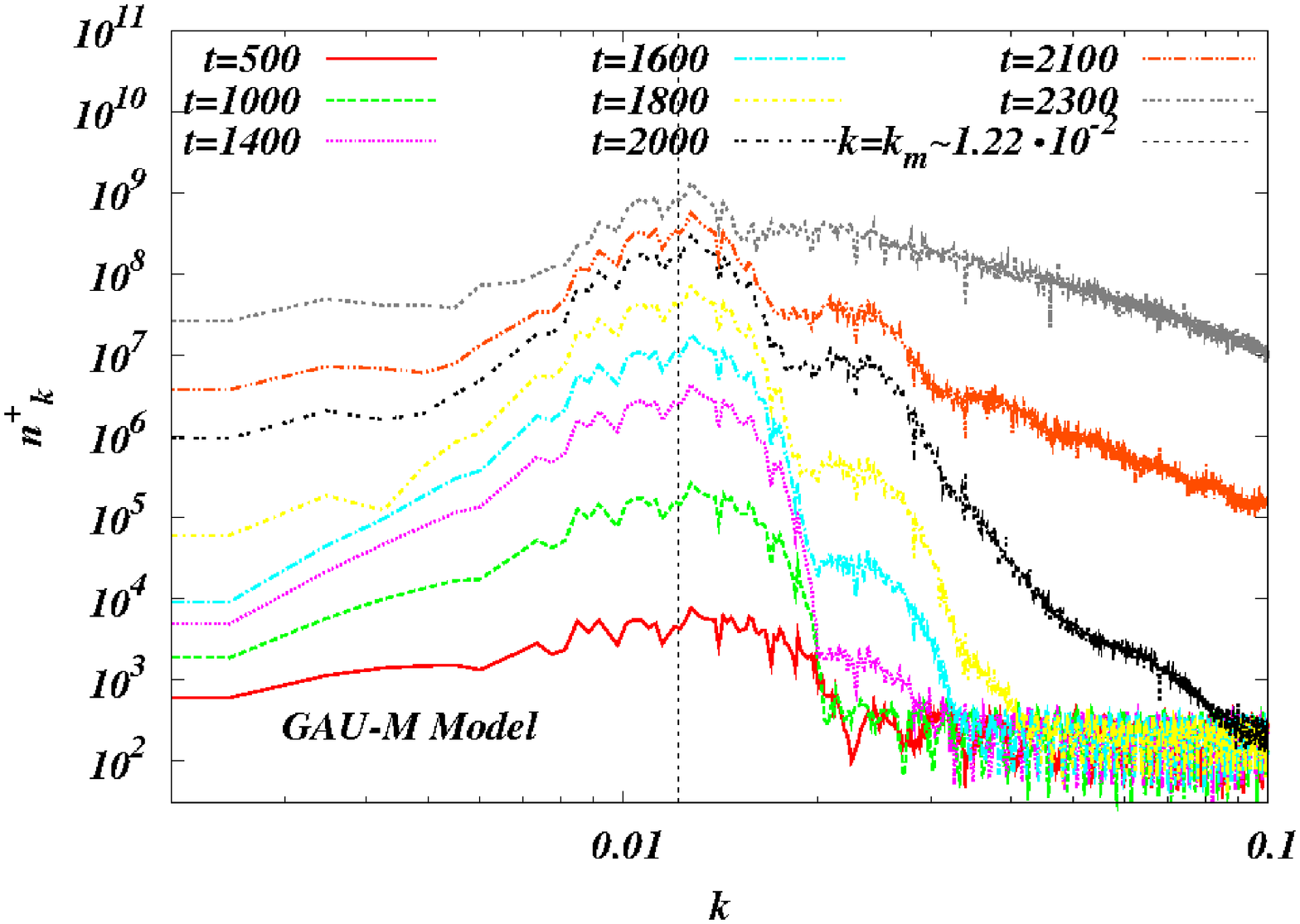}\\
	\includegraphics[angle=-90, scale=0.28]{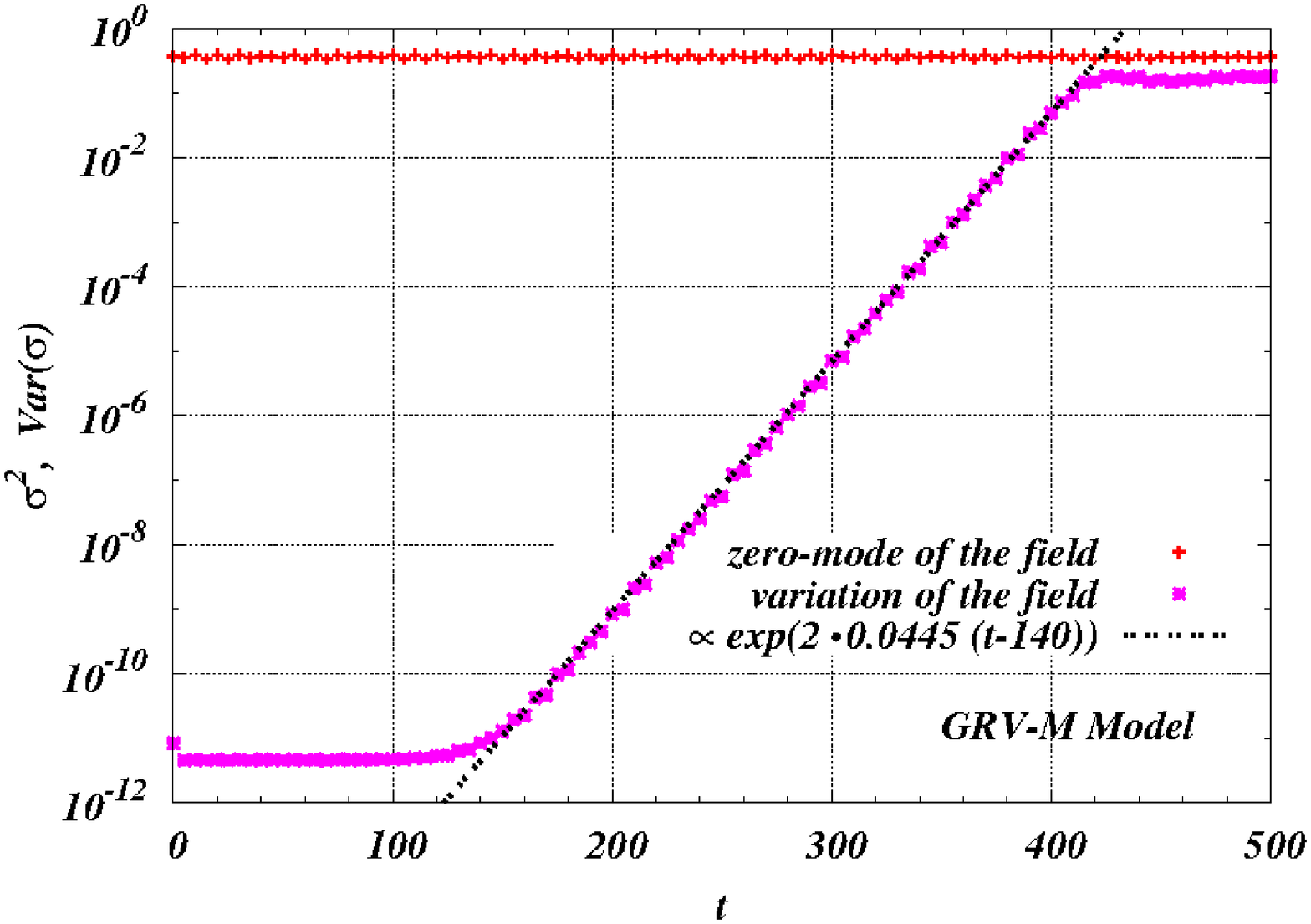}
	\includegraphics[angle=-90, scale=0.28]{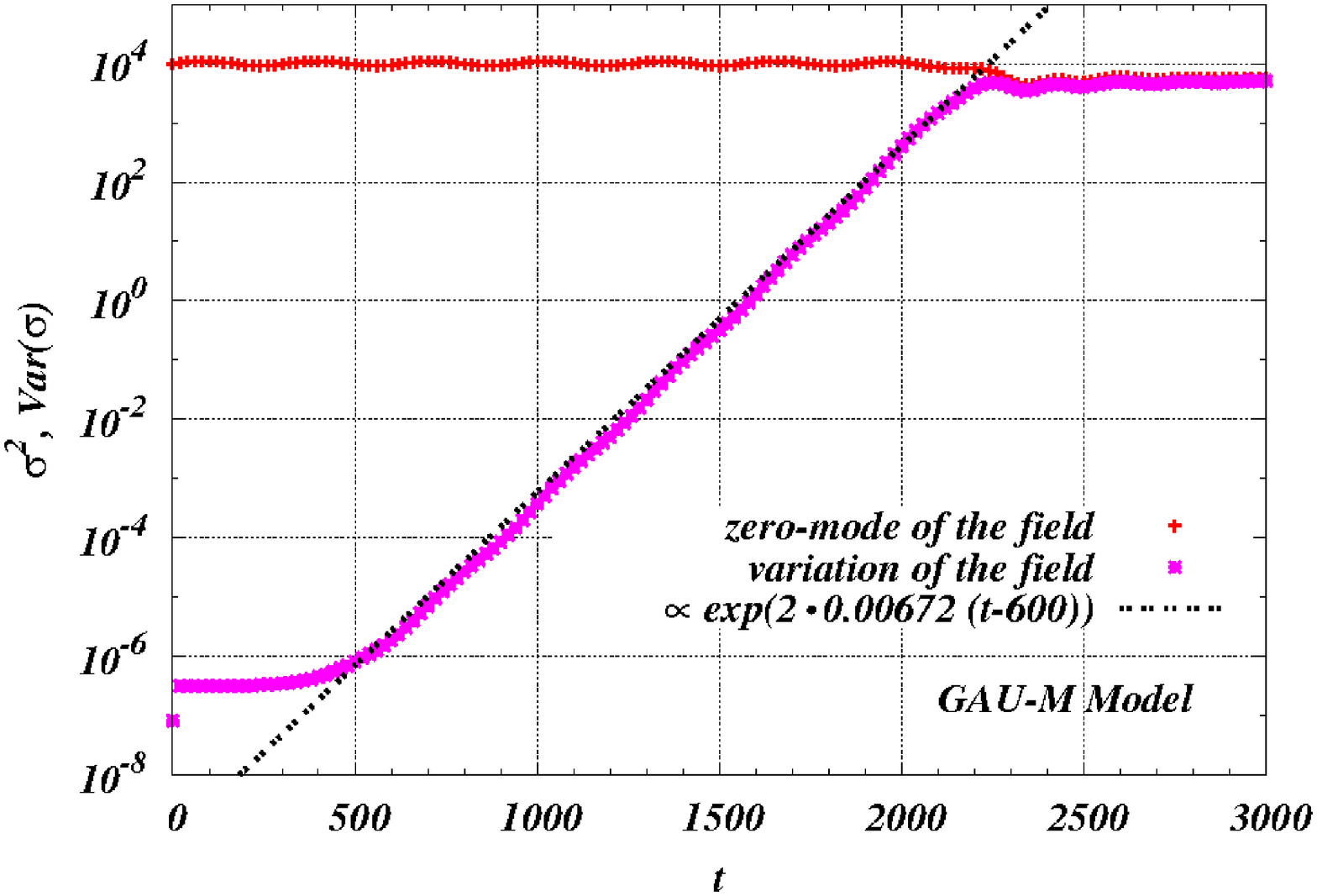}\\
	\includegraphics[angle=-90, scale=0.28]{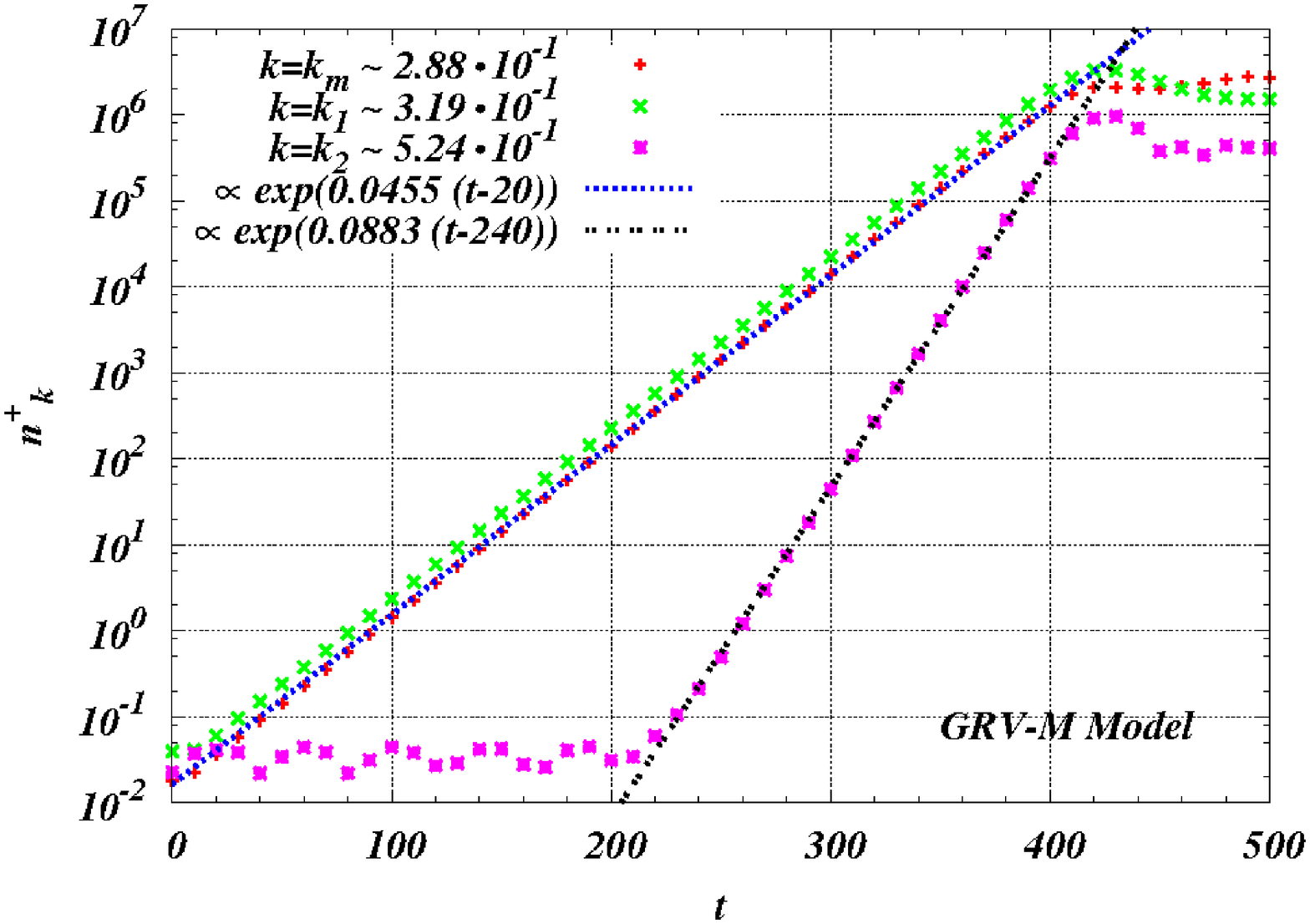}
	\includegraphics[angle=-90, scale=0.28]{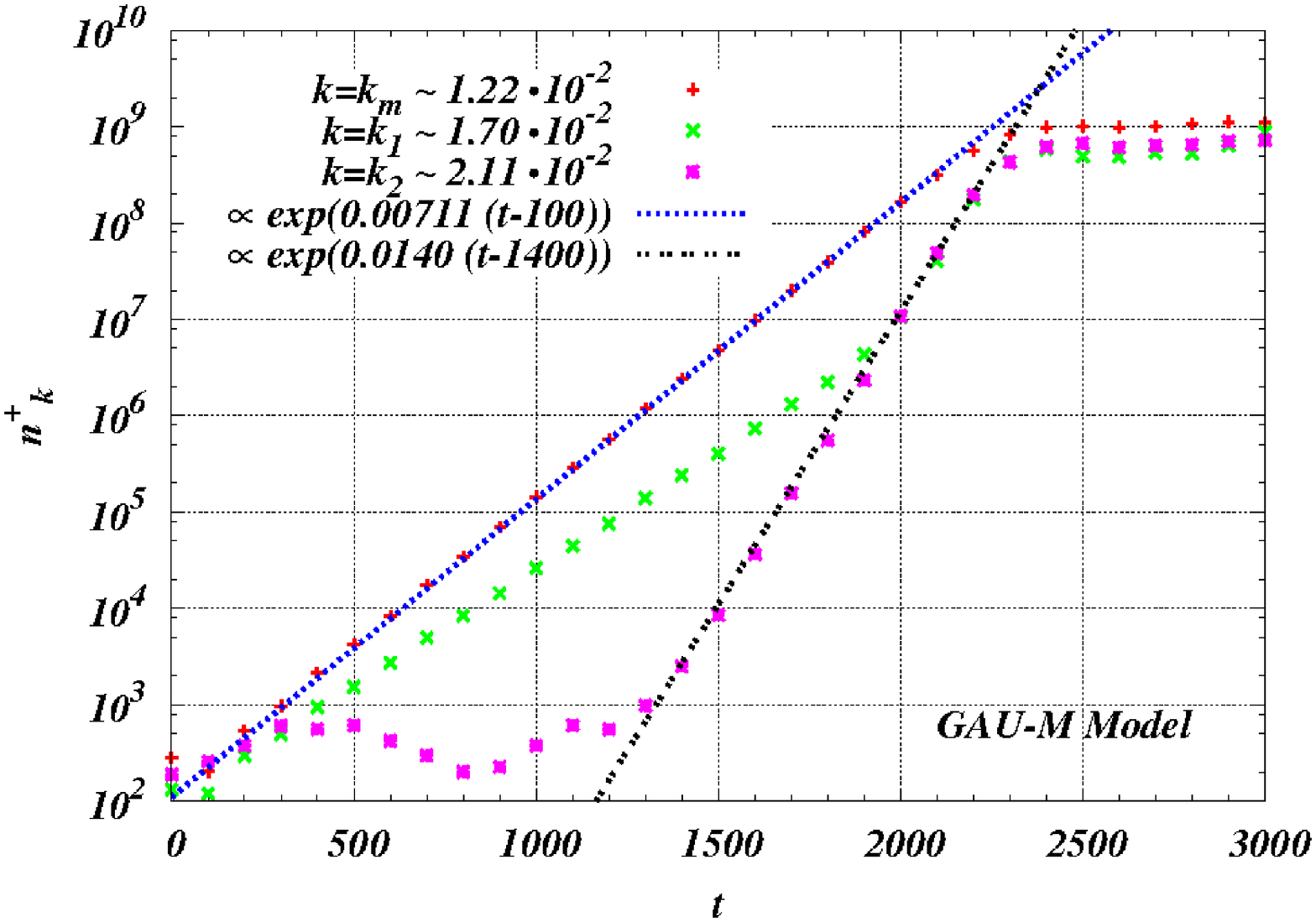}
  \end{center}
  \caption{ The top panels plot the amplitude of $n^+_k$ with various times in both the GRV-M Model (left) and the GAU-M Model (right). The analytical values of the most amplified mode $k_m$ in black-dashed vertical lines agree with the first peak, $k_1$, of the spectra in both cases. The middle panels show the evolution of zero-mode field, $\overline{\sigma^2}$ (red-plus dots), and the variance of the field, Var$(\sigma)$ (green-cross dots), up to the nonlinear time $t=t_{NL}$, where we can set $t_{NL}\sim 420$ in the GRV-M Model and $t_{NL}\sim 2200$ in the GAU-M Model. In the bottom panels, we plot the evolution of the amplitude $n^+_k$ for the first $k_1$ (red-plus dots), second peak $k_2$ (green-cross dots) modes and the analytical most amplified modes $k_m$ (purple squared-cross dots).}
  \label{fig:nknext}
\end{figure}

In \fig{fig:nknext}, we show the evolution of the various physical quantities in the late stage of linear perturbations: $n^+_k,\; \overline{\sigma^2}$ and Var$(\sigma)$. The top panels plot the amplitude of $n^+_k$ with various times in both the GRV-M Model (left) and the GAU-M Model (right). Notice that we plot them against the logarithmic scale of $k$ as opposed to the linear scale shown in the top panels of \fig{fig:nklin}. For all time steps shown there, our analytical values of $k_m$ (in black-dashed vertical lines) agree well with the first peak mode $k_1$, at which the amplitudes are most amplified. Notice that the zero-momentum mode does not decay in both cases. After the first peak of the amplitude is well developed, the second peak appears in the spectra, and later the third peak can be barely observed. Roughly speaking, the $n^{th}$ peaks appear around the values which are $k_m$ multiplied by $n$. These higher peaks are suppressed by rescattering processes in which a particle from the first peak transfers some of its momentum to a particle from the zero-momentum modes (AD condensates) \cite{Khlebnikov:1996mc}. Later, all modes of the particle spectra, $n^+_k$, develop quickly, but the first peak is still visible. The middle panels illustrate the evolution of a zero-mode field $\overline{\sigma^2}$ and the variance of the field Var$(\sigma)$ up to the nonlinear time $t=t_{NL}$. As we saw in the top panels, the zero mode does not decay even after the nonlinearity comes in, whilst the variance of the field develops exponentially from $t\sim 140$ in the GRV-M Model (left) and from $t\sim 600$ in the GAU-M Model (right). This delay of the exponential growth comes from the fact that the other modes do not evolve initially except the mode $k_m$; thus, we can set these times as $t_*$ defined in \eq{nlt}. We fit a function, $\propto exp\bset{2 \dot{S}_{num}(t-t_*)}$, against the exponential evolution for the variations, where we obtain $\dot{S}_{num} \sim 4.45\times 10^{-2}$ in the GRV-M Model and $\dot{S}_{num} \sim 6.72\times 10^{-3}$ in the GAU-M Model, which match satisfactorily with the analytical ones in \eqs{ext11}{ext12}, where we computed as $\state{\dot{S}_m}\sim 4.20\times 10^{-2}$ in the GRV-M Model and $\state{\dot{S}_m}\sim 7.07\times 10^{-3}$ in the GAU-M Model. From the middle panels, the nonlinear time is approximately both $t_{NL}\sim 420$ in the GRV-M Model and $t_{NL}\sim 2200$ in the GAU-M Model, and these values agree well with our analytic estimates in \eq{nlt}, where the analytical values are $t_{NL}\sim 262+140\sim 422$ in the GRV-M Model and $t_{NL}\sim 1628+600\sim 2228$. In the bottom panels, we plot the evolution of the amplitude $n^+_k$ for the first peak mode (red-plus dots), second peak mode (green-cross dots) and the analytical most amplified modes (purple squared-cross dots). The numerical values of the exponents for the most amplified modes $k_m$ in blue long-dotted lines, ($\dot{S}_{num} \sim 4.55\times 10^{-2}$ in the GRV-M Model and $\dot{S}_{num} \sim 7.11\times 10^{-3}$ in the GAU-M Model) match with the analytical ones in \eqs{ext11}{ext12}. The second peaks $k_2$ in black short-dotted lines start to grow at $t\sim 220$ in the GRV-M Model and at $t\sim 1300$ in the GAU-M Model, and we can set these values as $t_0$ defined in \eq{n+k}. The initial behaviour of the amplitude of second peak of $n^+_k$ seems to be quasi-periodic, which implies that $\state{S}$ for the mode, $k_2$, is pure imaginary, see \eq{scdsig} (\cf\ the bottom panels of FIG. 5 in \cite{Rosa:2007dr}). Surprisingly, the growth rates for the second peaks are about twice as large as the values of both $\state{\dot{S}_m}$ and $\dot{S}_{num}$ for Var($\sigma$) and $k_1$. Note that the initial evolution for $k_2$ is not adiabatic, so that the growth rates are not strictly exponential as we have seen in the bottom panels of \fig{fig:nklin}. For example, the growth of the first peaks, $k_m$ (or $k_1$), in the GAU-M Model is not exponential initially, but it becomes exponential as the growth of the second peak mode $k_2$.

\vspace*{10pt}
\paragraph*{\underline{\bf Bubbles pinched out of filaments:}}

\begin{figure}[!ht]
  \begin{center}
	\includegraphics[scale=0.36]{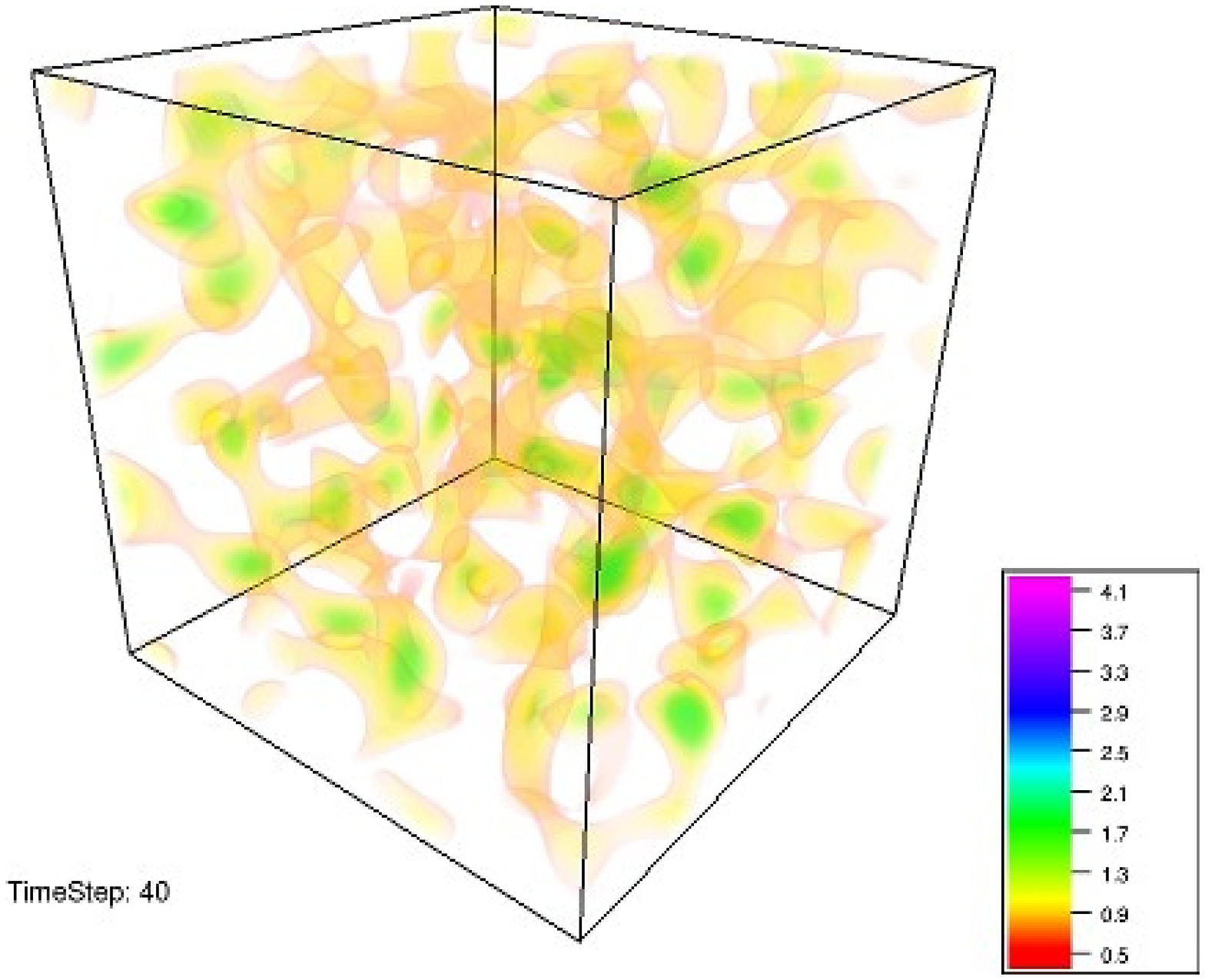}
	\includegraphics[scale=0.36]{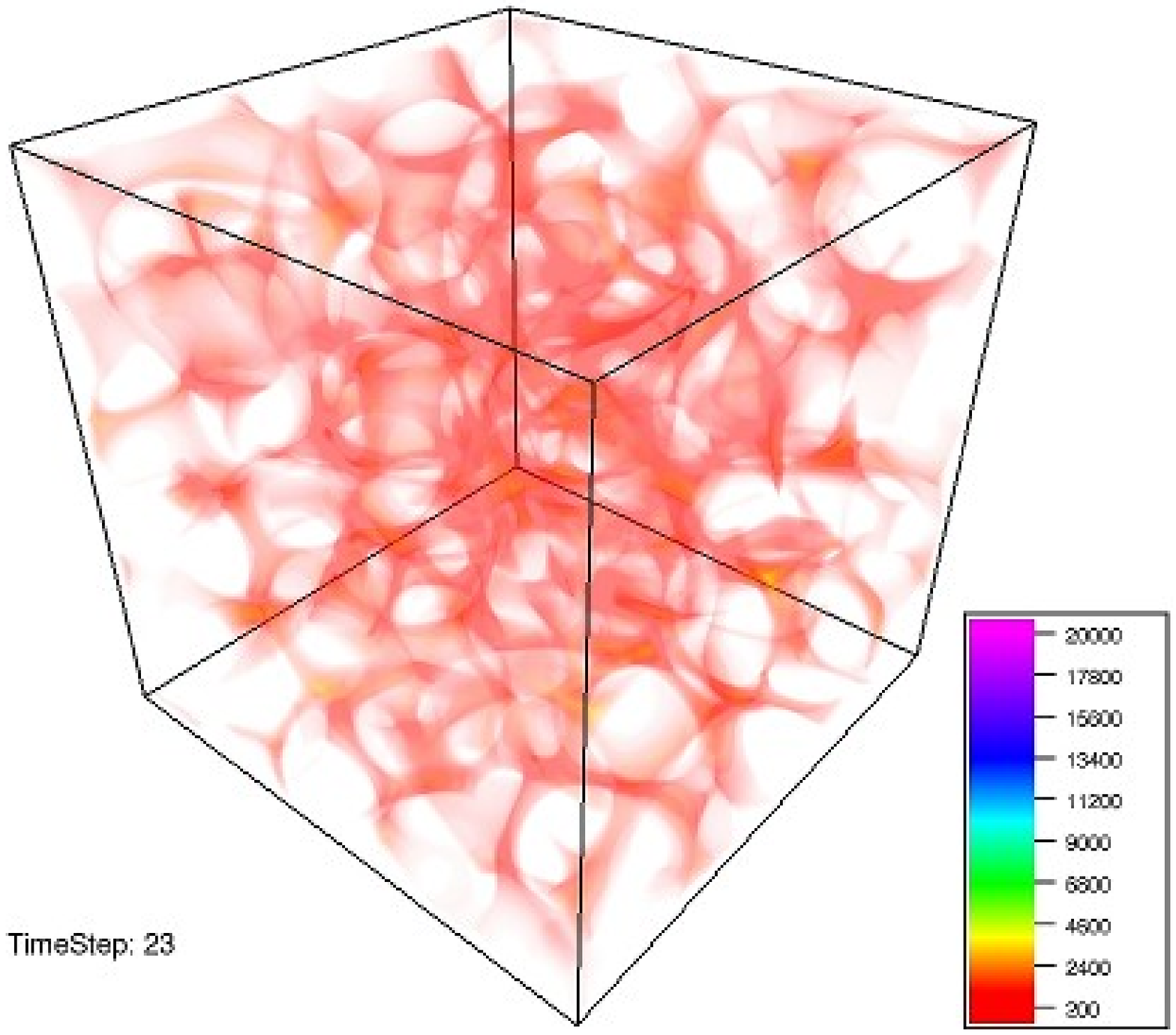}\\
	\includegraphics[scale=0.36]{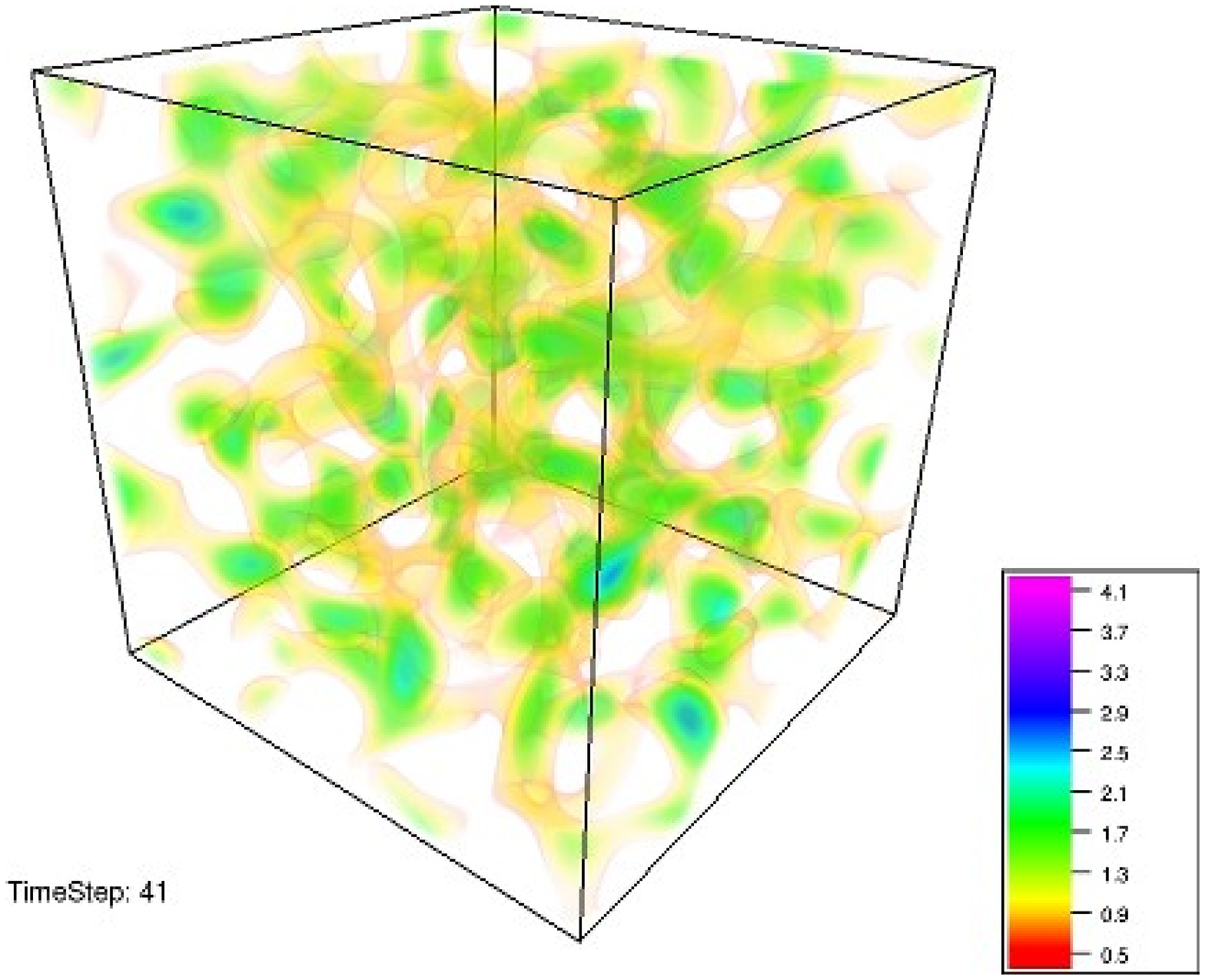}
	\includegraphics[scale=0.36]{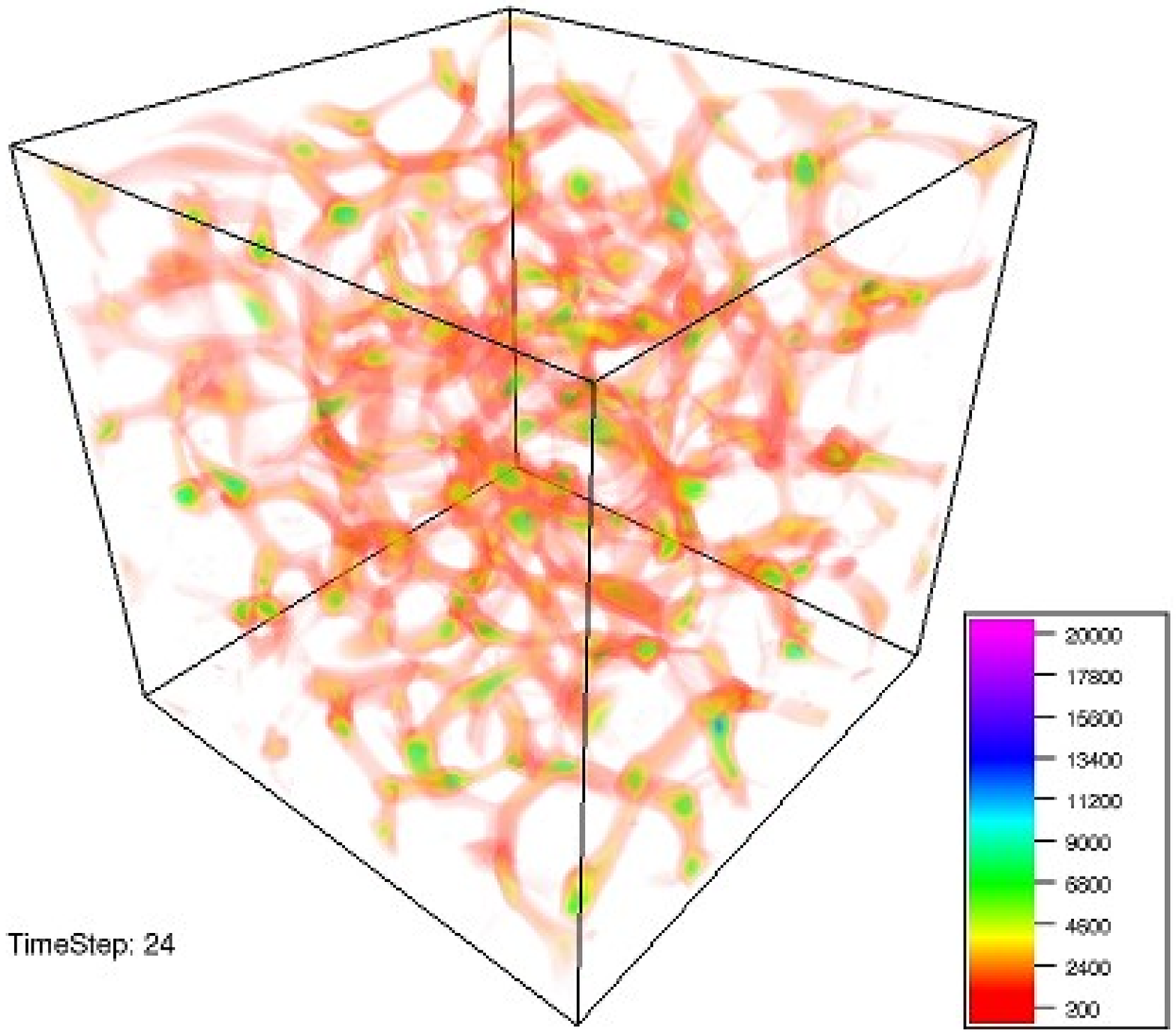}
  \end{center}
  \caption{ In the top and bottom panels, we show snapshots of the positive charge density $n^+(\mathbf{x})$ for the GRV-M Model (left panels) and the GAU-M Model (right panels) around $t\sim t_{NL}$, where 'Timestep' in the panels denotes the actual time divided by $10$ in the GRV-M Model and the actual time divided by $10^2$ in the GAU-M Model, and the colour bars illustrate the values of the positive charge density. After the nonlinearity is fully developed, many bubbles form, which are pinched out of ``highly'' concentrated charged filaments.}
  \label{fig:qb}
\end{figure}

In \fig{fig:qb}, we show snapshots of the positive charge density $n^+(\mathbf{x})$ for the GRV-M Model (left panels) and the GAU-M Model (right panels) around $t\sim t_{NL}$, where 'Timestep' in the panels denotes the actual time divided by $10$ in the GRV-M Model and the actual time divided by $10^2$ in the GAU-M Model. The colour bars illustrate the values of the positive charge density. We can see long-wavelength objects (sometimes called 'filaments') in both cases, and the charge in some regions is compactified into spheres, see bottom panels. These filaments and bubbles correspond to nonlinear solutions, which may be nontopological strings \cite{Copeland:1988ra} and the excited states of $Q$-balls, respectively. The radii of these bubbles are of the same order as the wave-length which corresponds to the most amplified modes, $k_m$. As we will see in the next subsection, these bubbles grow by colliding and merging each other. Note that this bubble creation is nothing to do with bubble nucleation in first-order phase transition as opposed to the case in \cite{Lee:1994qb}, in which case the AD condensate is classically stable against spatial perturbations, but not quantum mechanically.

\subsubsection{Nonlinear evolution}

\paragraph*{\underline{\bf Bubble collisions and mergers:}}

\begin{figure}[!ht]
  \begin{center}
	\includegraphics[scale=0.24]{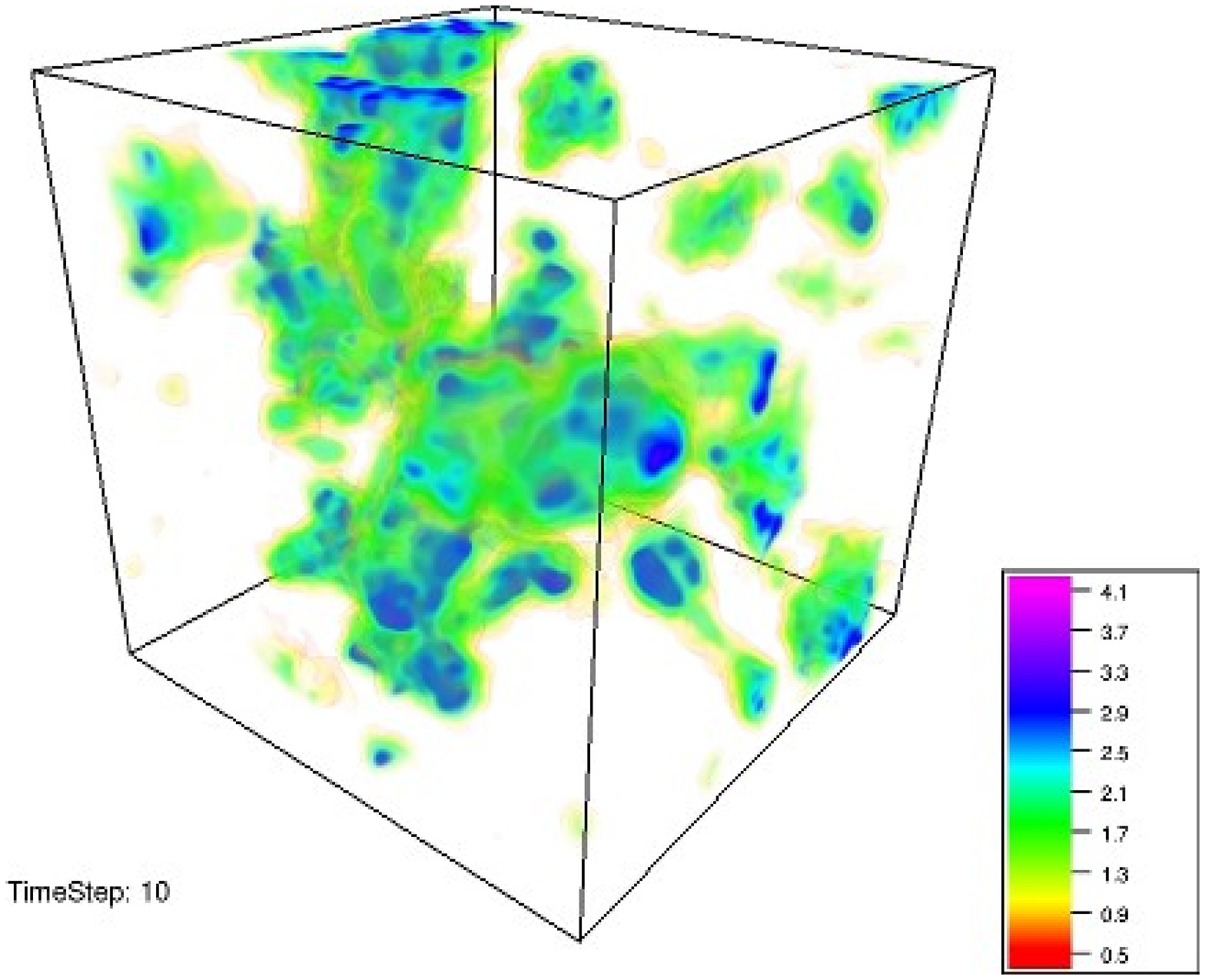}
	\includegraphics[scale=0.24]{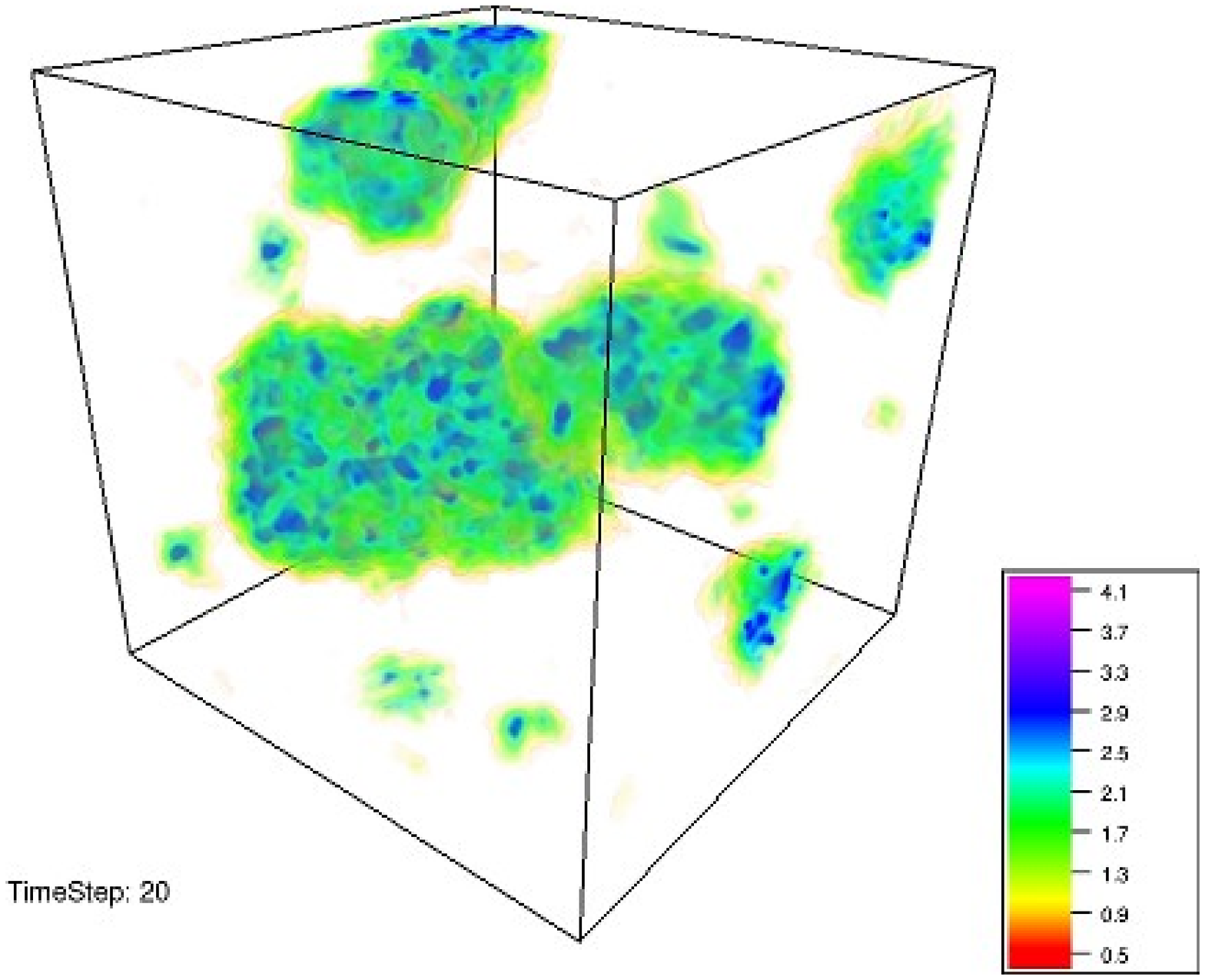}
	\includegraphics[scale=0.24]{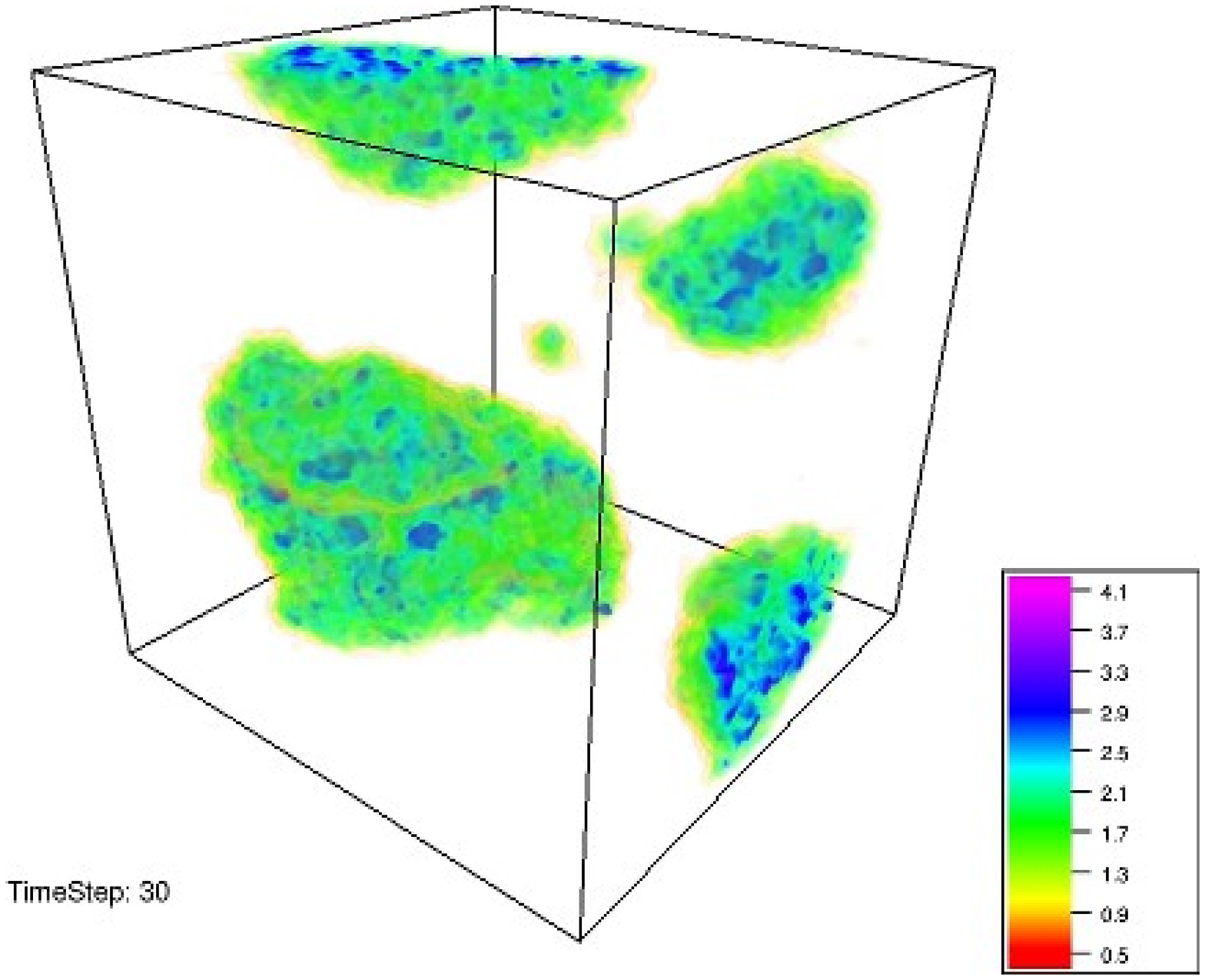}\\
	\includegraphics[scale=0.24]{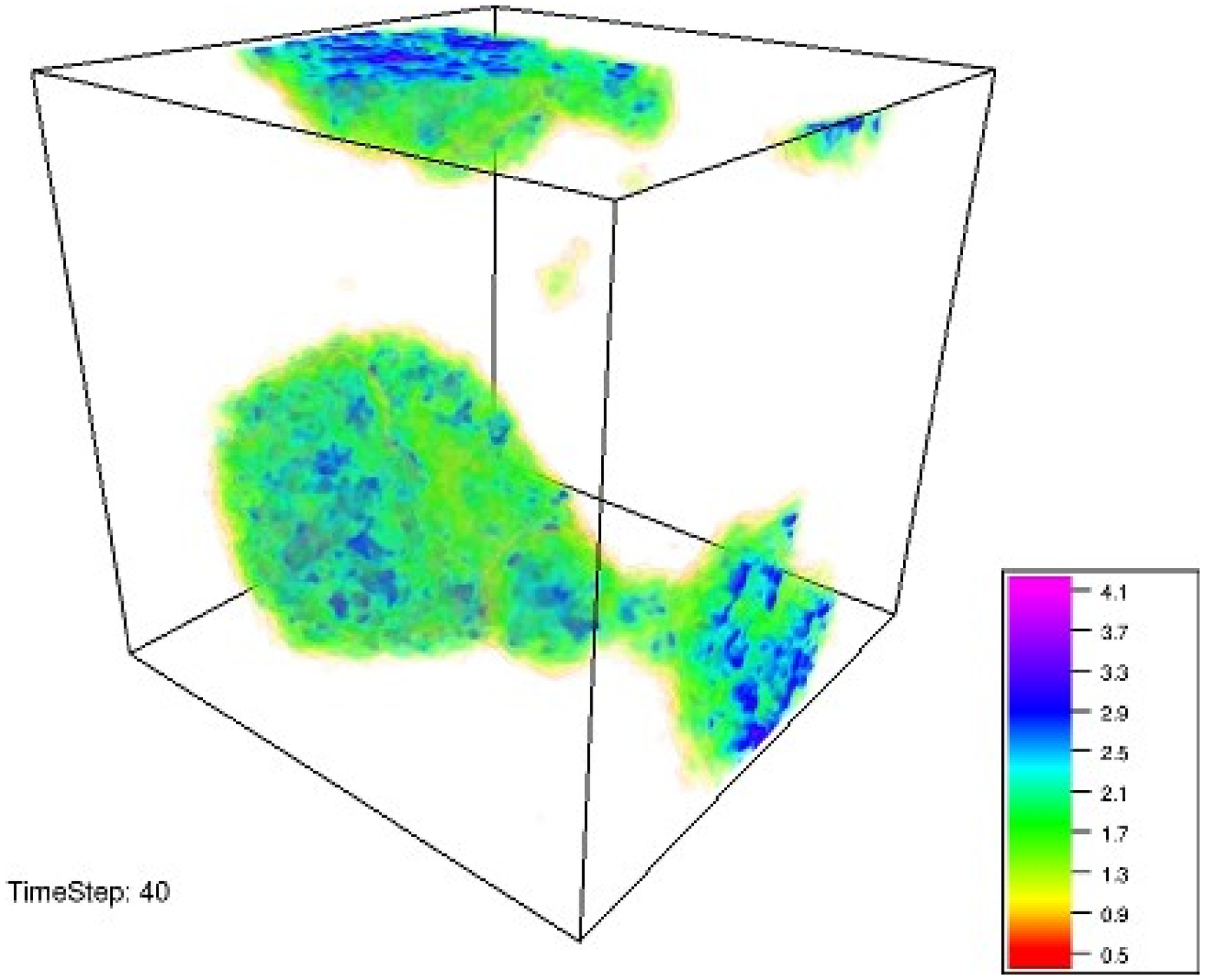}
	\includegraphics[scale=0.24]{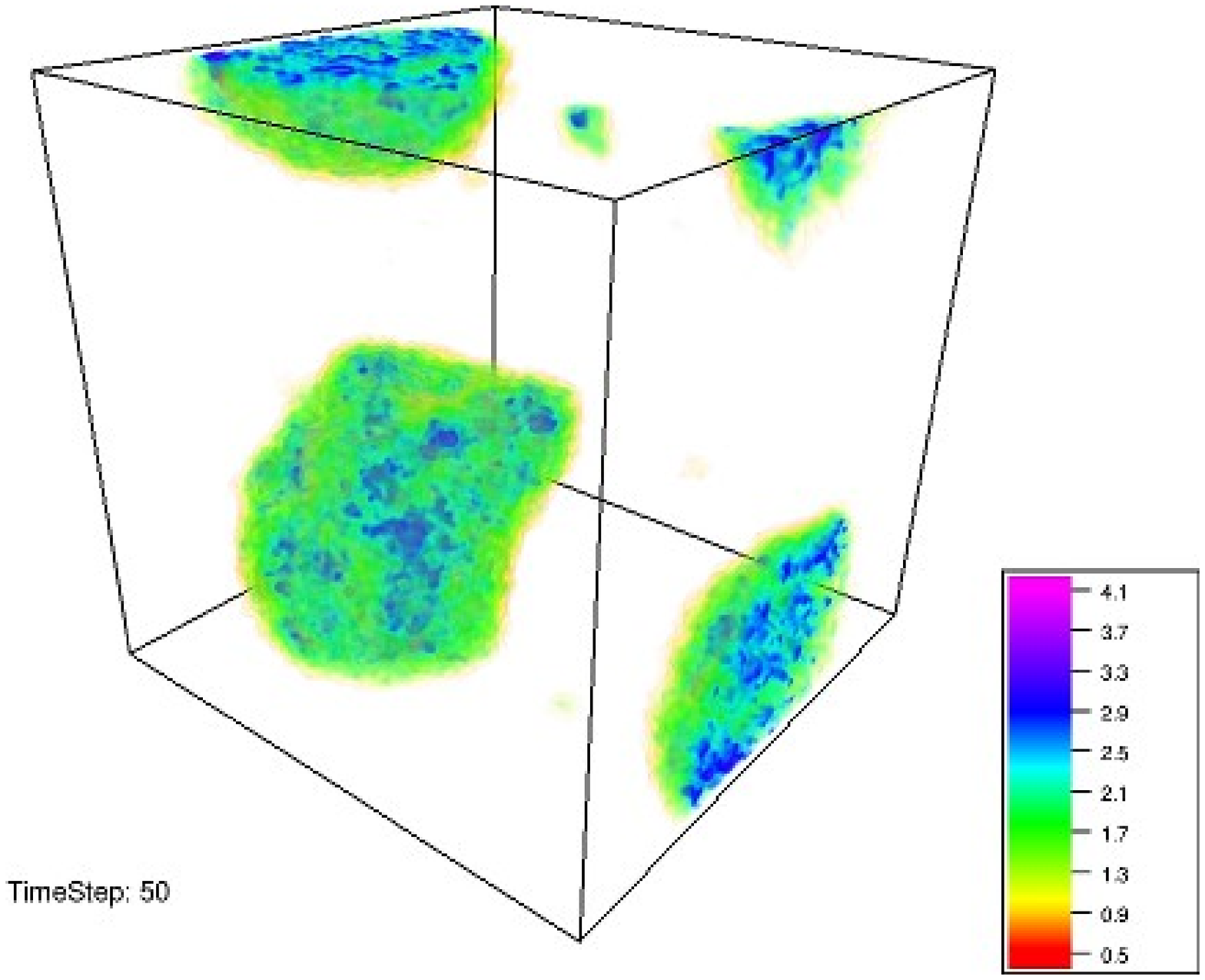}
	\includegraphics[scale=0.24]{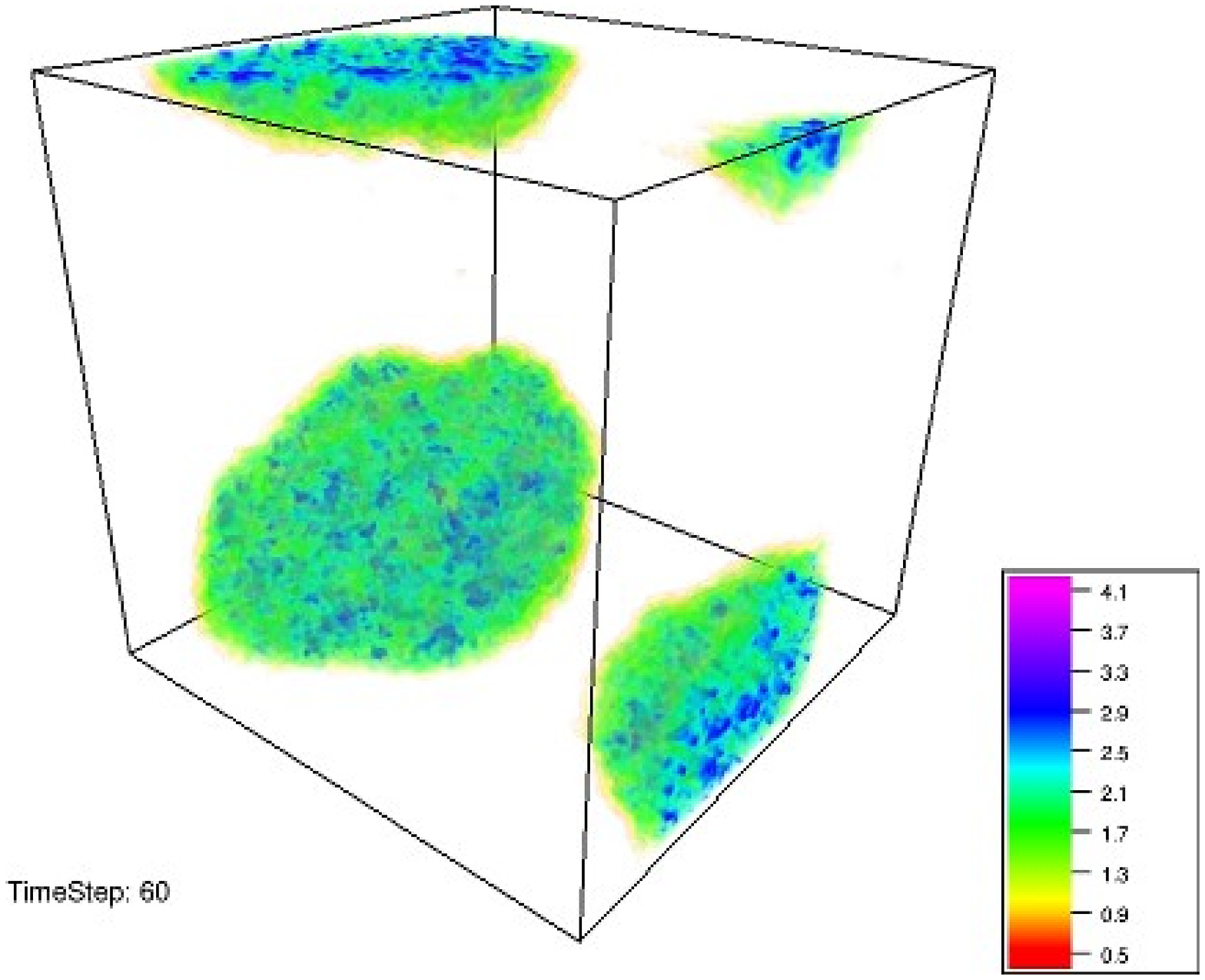}
  \end{center}
  \caption{ We show snapshots of the positive charge density for the GRV-M Model in different time steps ($t=1000,\ 2000,\ 3000,\ 4000,\ 5000$ and $6000$), where 'Timestep' in the figure denotes the actual simulation time divided by $10^2$ and the colour bars illustrate the values of the positive charge density. A few created lumps collide and merge into a large cluster by the end.}
  \label{fig:qbformgrv}
\end{figure}

In \fig{fig:qbformgrv}, we show snapshots of the positive charge density for the GRV-M Model in different time steps up to $t=6000$, where 'Timestep' in the figure denotes the actual simulation time divided by $10^2$ and the colour bars illustrate the values of the positive charge density. After the system goes into a nonlinear regime, we can see a few lumps in the first few panels of the snapshots, and those lumps merge into larger lumpy objects. Finally, we can see a large cluster, which consists of a complicated inner structure, see the last snapshot. Recall that we are using the periodic boundary condition.

\vspace*{5pt}

\begin{figure}[!ht]
  \begin{center}
	\includegraphics[scale=0.24]{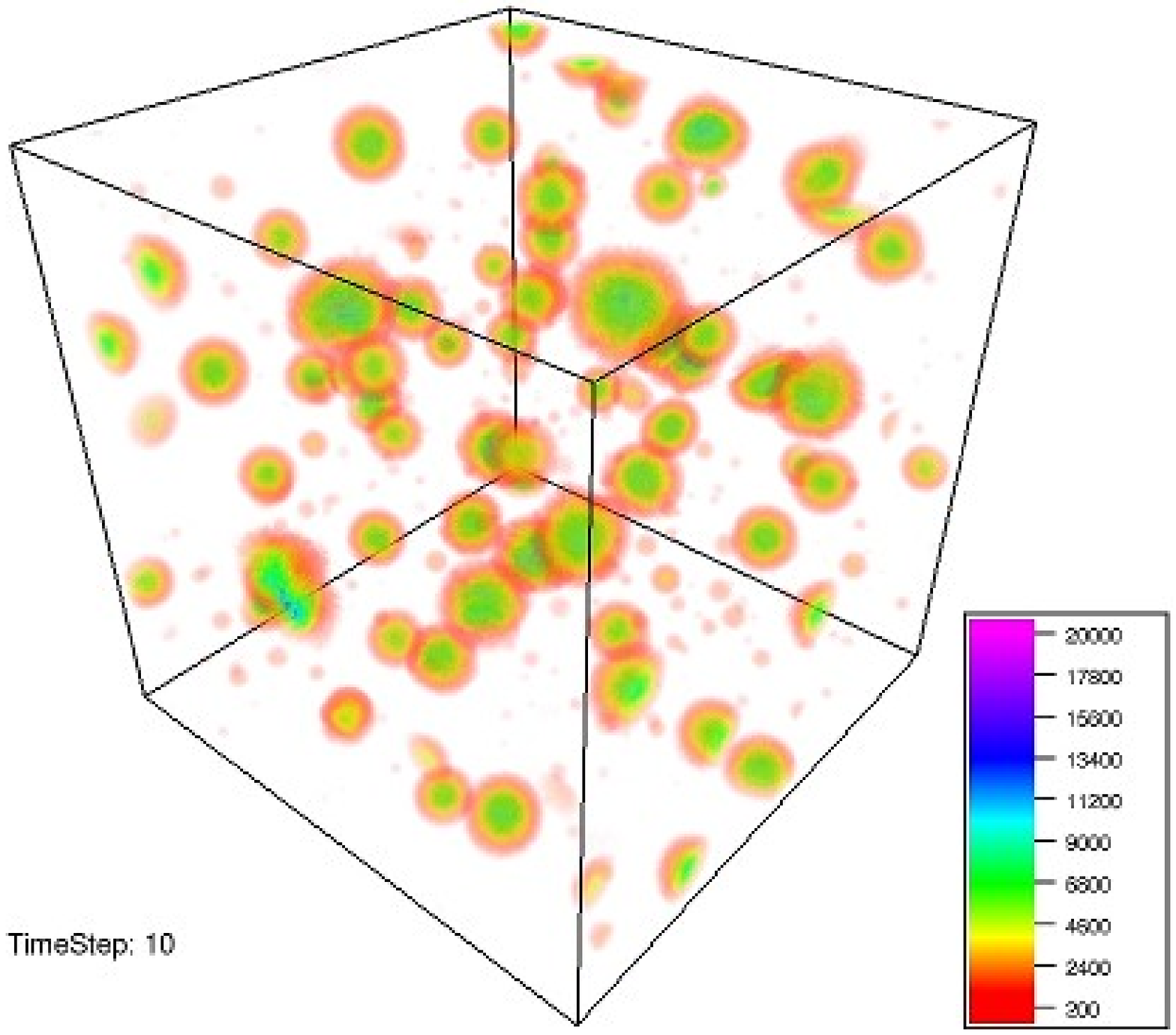}
	\includegraphics[scale=0.24]{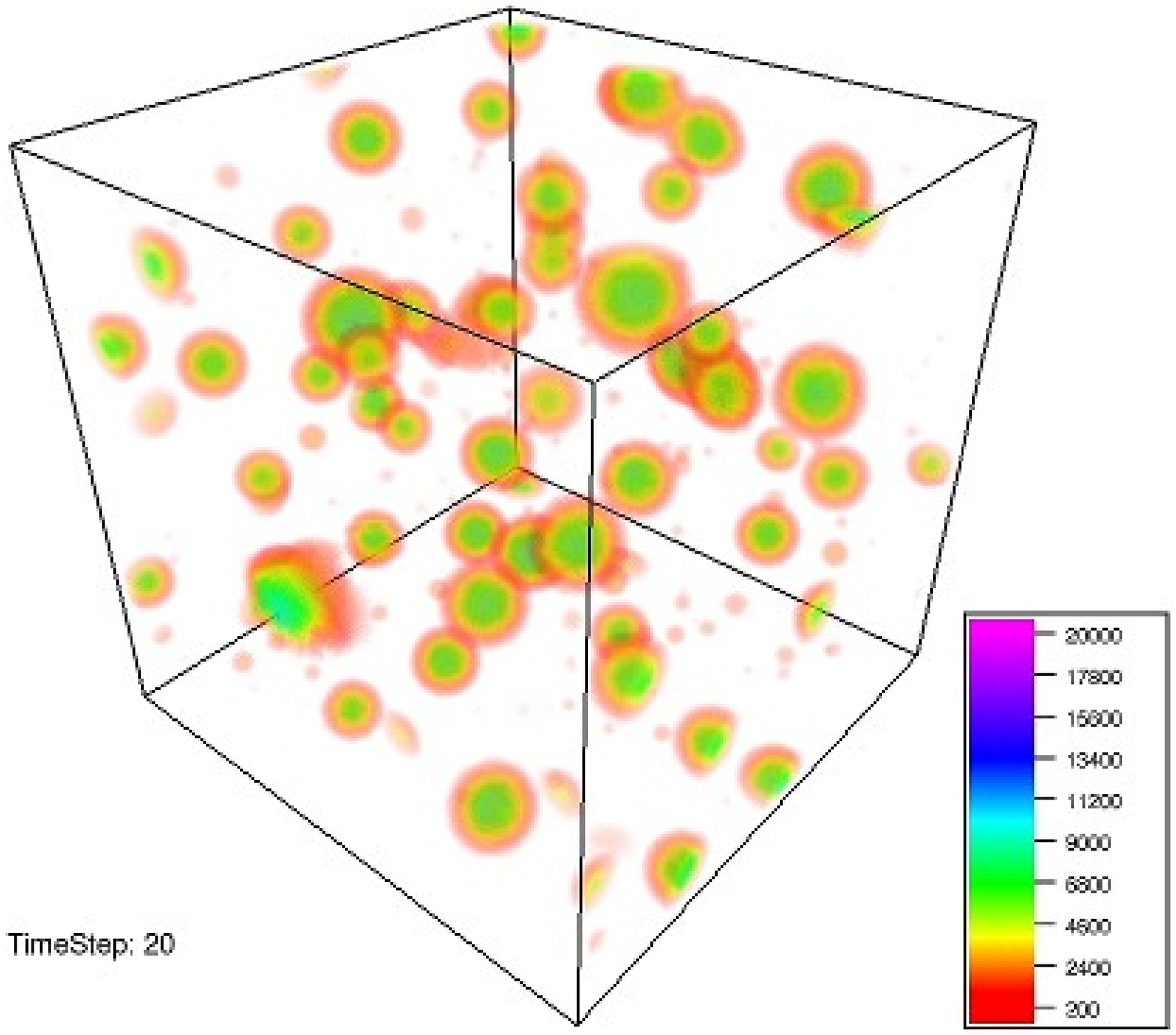}
	\includegraphics[scale=0.24]{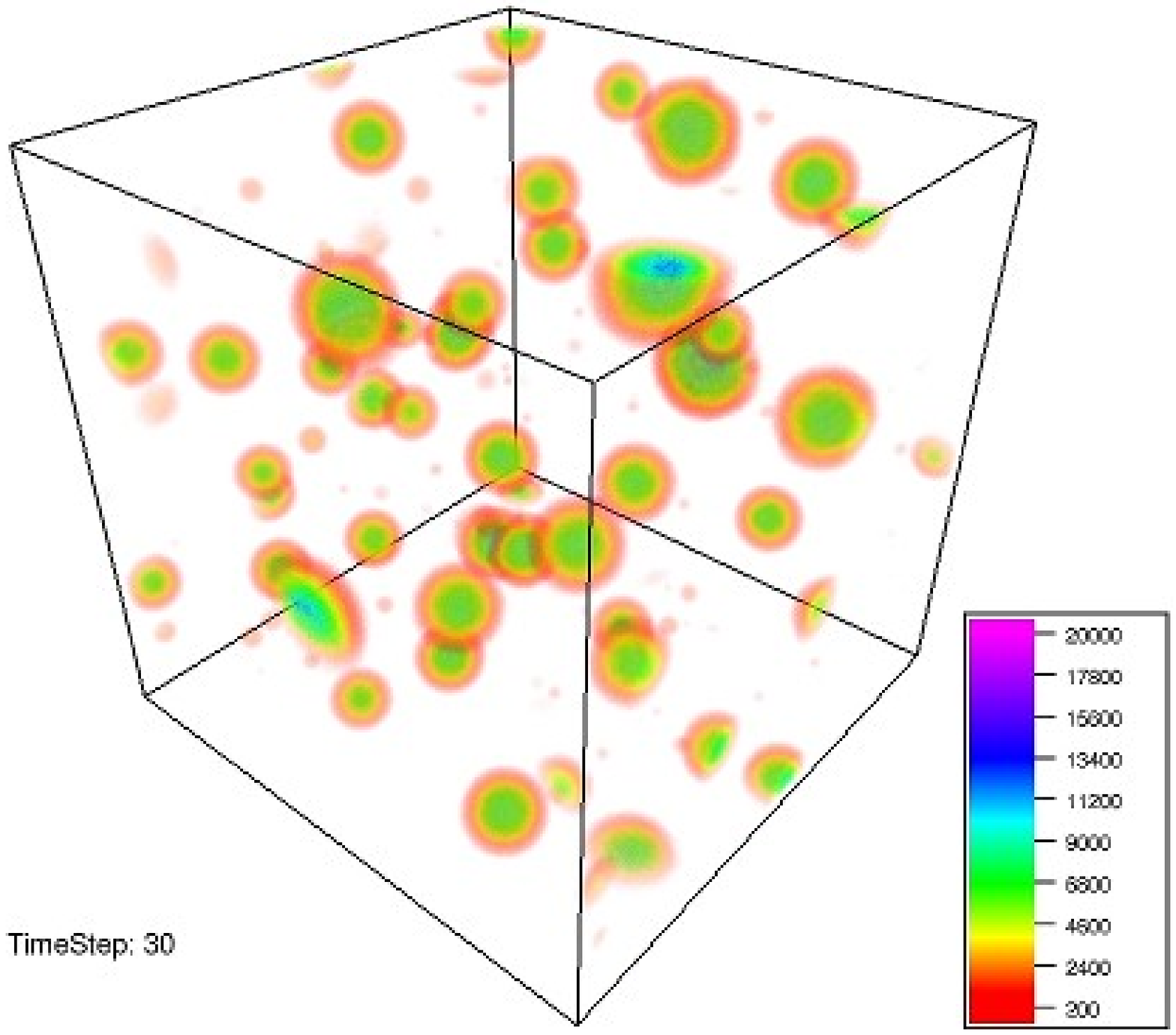}\\
	\includegraphics[scale=0.24]{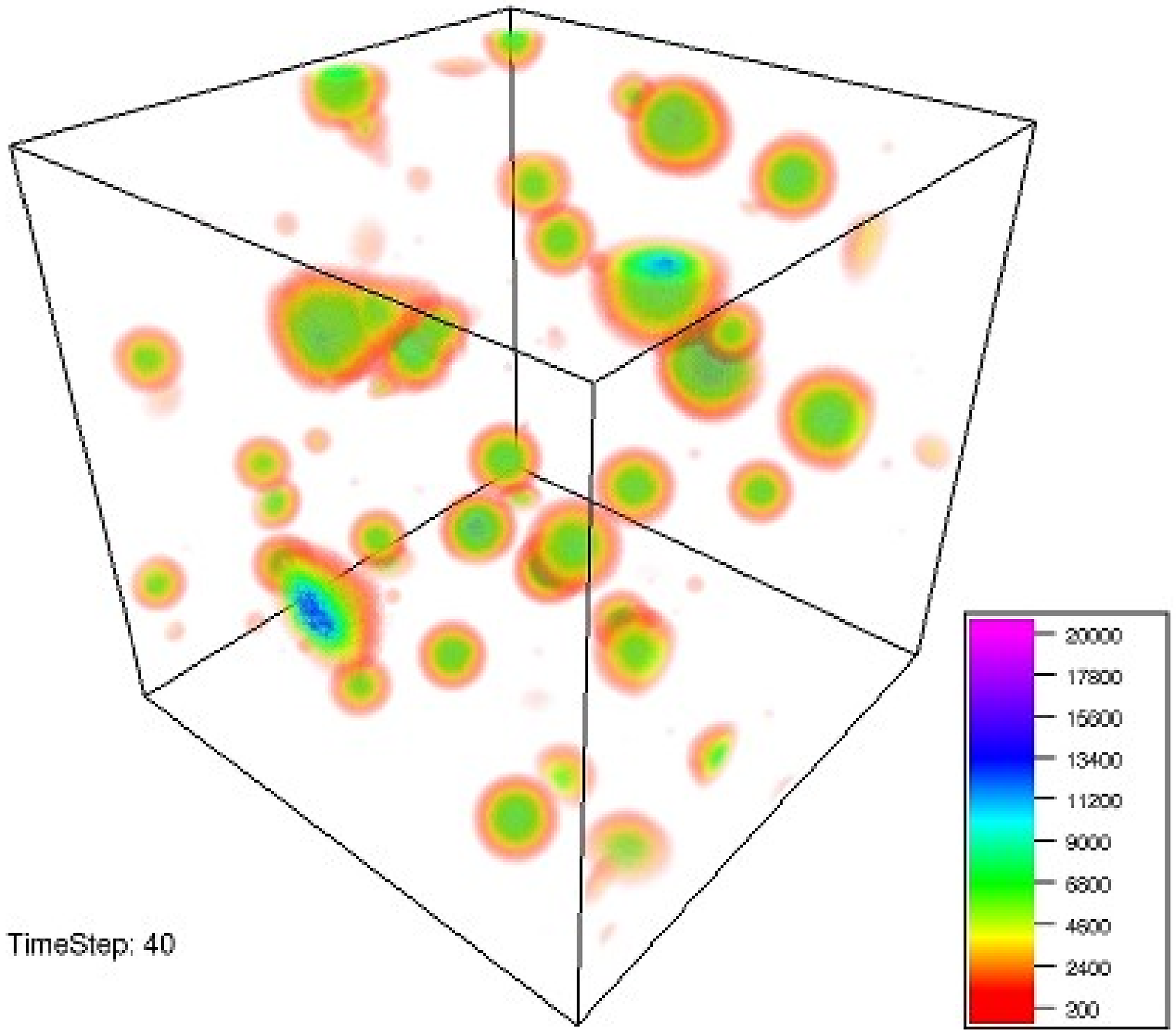}
	\includegraphics[scale=0.24]{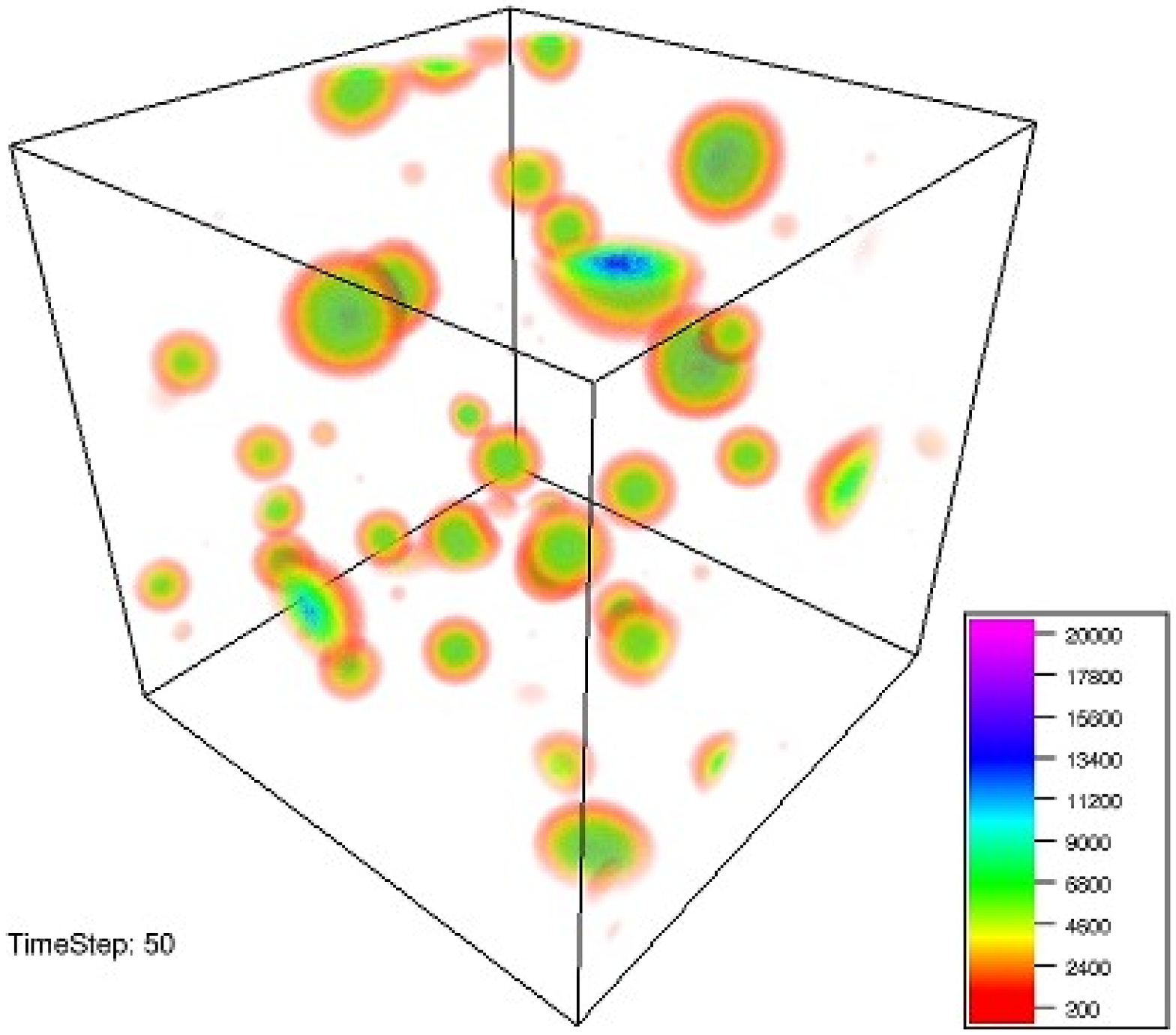}
	\includegraphics[scale=0.24]{npgau2-6.eps}
  \end{center}
  \caption{ We illustrate the detailed evolution of the positive charge density for the GAU-M Model in different time steps ($t=10000,\ 20000,\ 30000,\ 40000,\ 50000$ and $60000$), where 'Timestep' in the figure denotes the actual simulation time divided by $10^3$ and the colour bars illustrate the values of the positive charge density. There are smaller number of bubbles left by the end.}
  \label{fig:qbformgau}
\end{figure}

\fig{fig:qbformgau} shows the detailed evolution of the positive charge density for the GAU-M Model in different time steps up to $t=60000$, where 'Timestep' in the figure denotes the actual simulation time divided by $10^3$ and the colour bars illustrate the values of the positive charge density. A large number of small bubbles can be observed, and nearby bubbles collide and merge into larger bubbles. In the final panel, there are smaller number of bubbles left (compare to the first panel). We believe that this time arrow is followed because the total energy of large bubbles is smaller than the total energy of smaller bubbles, \cf\ fission stability of $Q$-balls in \eq{SAF}. These large bubbles are able to carry a large amount of charge inside of them as we saw in the left-bottom panel of \fig{fig:gauabs} in chapter \ref{ch:qbflt} in the ``thin-wall'' $Q$-ball limit. 

\vspace*{5pt}

The differences in the evolution between GRV-M and GAU-M models come from a number of facts, \eg\  different initial conditions, stability conditions and momentum fluxes due to asymptotic profiles at a large distance from the cores.

\vspace*{10pt}

\paragraph*{\underline{\bf Distributions of the negative charge density:}}

\begin{figure}[!ht]
  \begin{center}
	\includegraphics[scale=0.36]{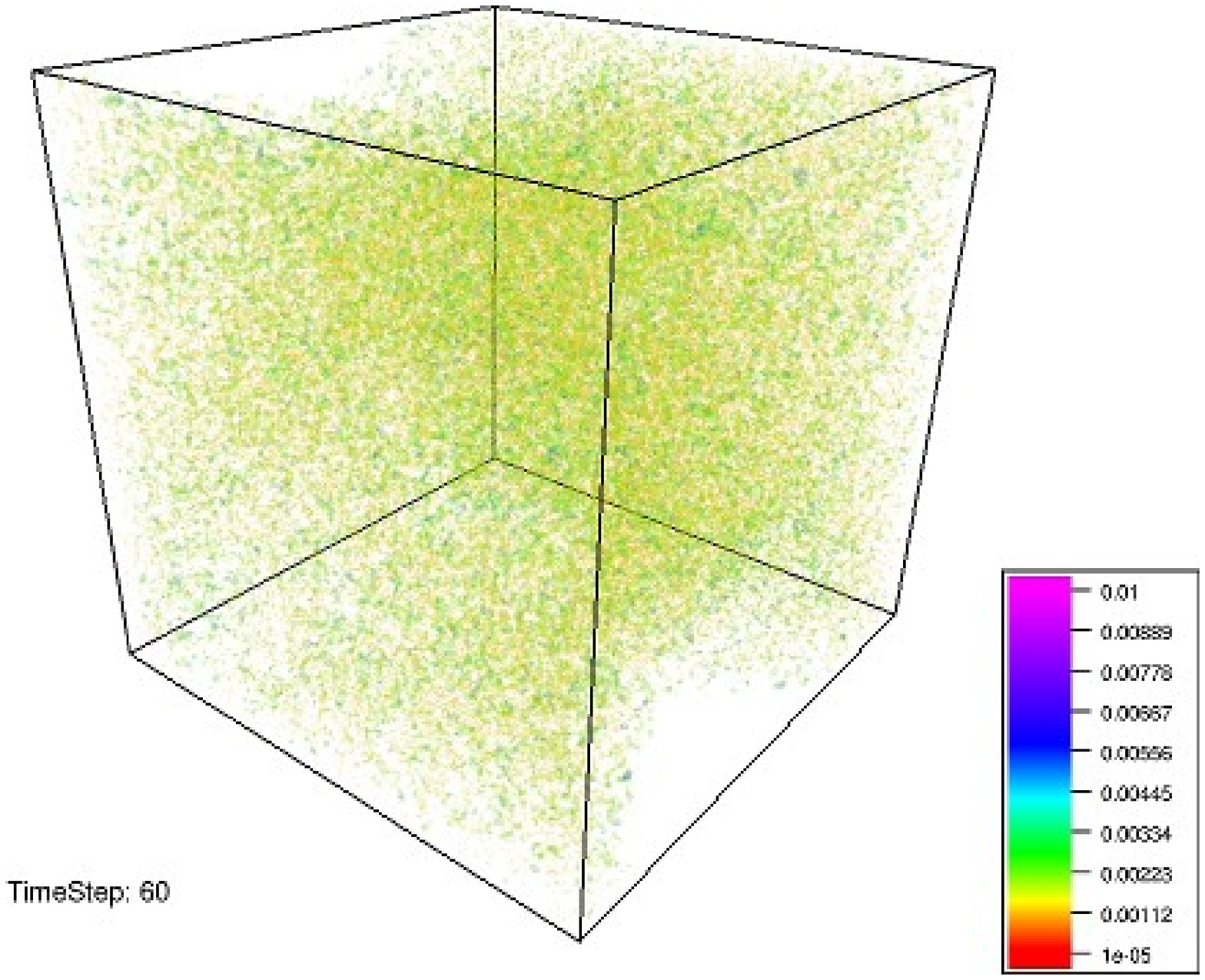}
	\includegraphics[scale=0.36]{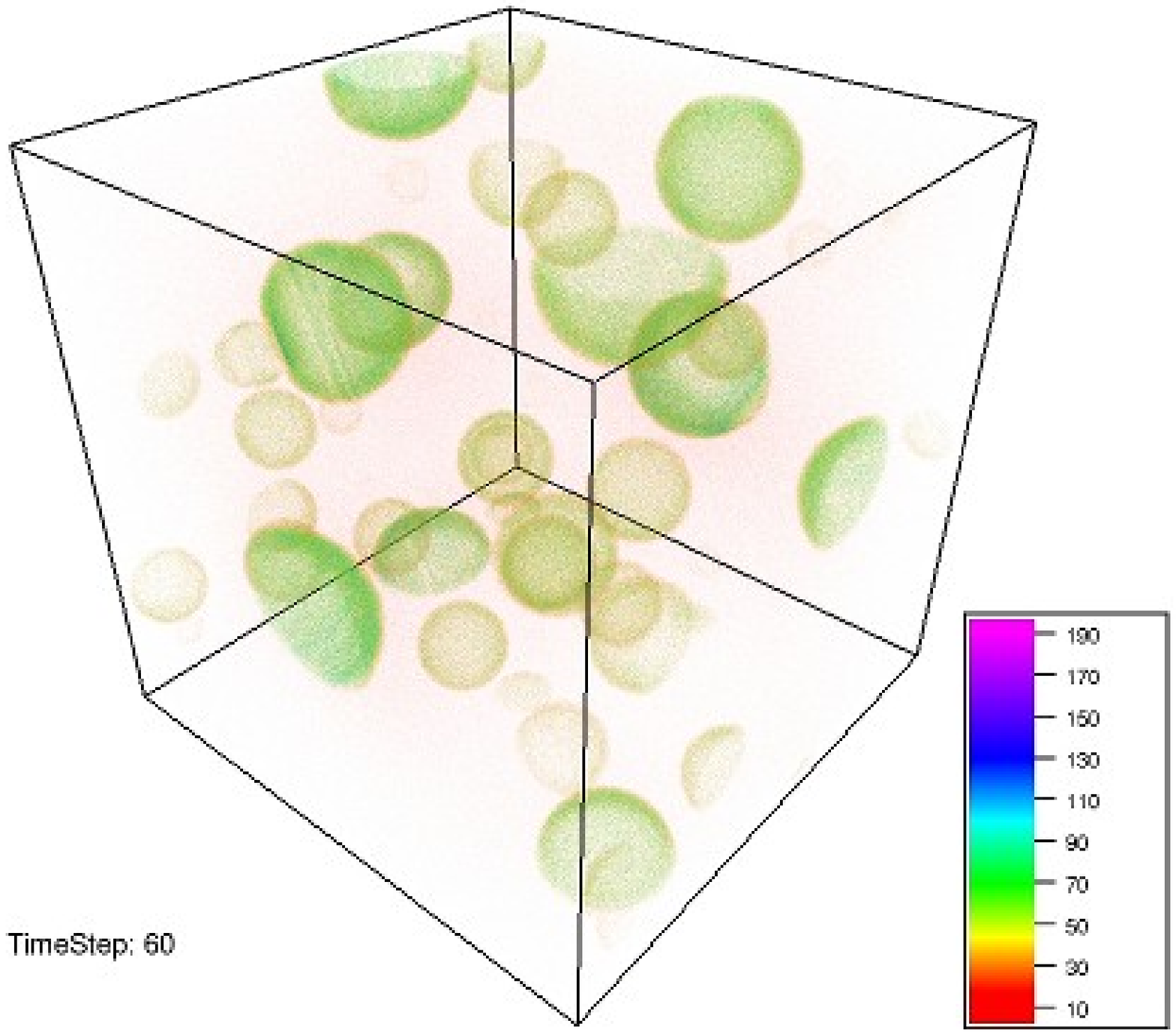}
  \end{center}
  \caption{ We present the snapshots of the negative charge density for the GRV-M Model (left panel) at $t=6.0\times 10^{3}$ and the GAU-M Model (right panel) at $t=6.0\times 10^{4}$, where the colour bars illustrate the values of the negative charge density. The negative charge for the GRV-M Model is surrounded by the large positive charged cluster; however, the distribution spreads out over the lattice space, whereas the negative charge for the GAU-M Model is concentrated around the positive charged lumps [compare them to the last panels of \figs{fig:qbformgrv}{fig:qbformgau}].}
  \label{fig:qbnm}
\end{figure}

We show snapshots of the negative charge density for the GRV-M Model (left panel) at $t=6000$ and the GAU-M Model (right panel) at $t=1.0\times 10^{5}$ in \fig{fig:qbnm}, where the colour bars illustrate the values of the negative charge density. These times correspond to the same physical times as in the final snapshots of \figs{fig:qbformgrv}{fig:qbformgau}. The values of charge density in both models are much smaller than the values of positive charge density in \figs{fig:qbformgrv}{fig:qbformgau}. This implies that we are observing the plots of thermal plasma rather than charged (nonlinear) lumps. Their distributions are quite different from each other. The negative charge density for the GRV-M Model is surrounded by the large positive charged cluster seen in the last panel of \fig{fig:qbformgrv}, and it is distributed all over the lattice; whereas, for the GAU-M Model the distributions of the negative charged plasma are highly concentrated only around the surface of the lumps (compare the last panel of \fig{fig:qbformgau}).

\vspace*{10pt}

\paragraph*{\underline{\bf Driven turbulence:}}

\begin{figure}[!ht]
  \begin{center}
	\includegraphics[angle=-90, scale=0.28]{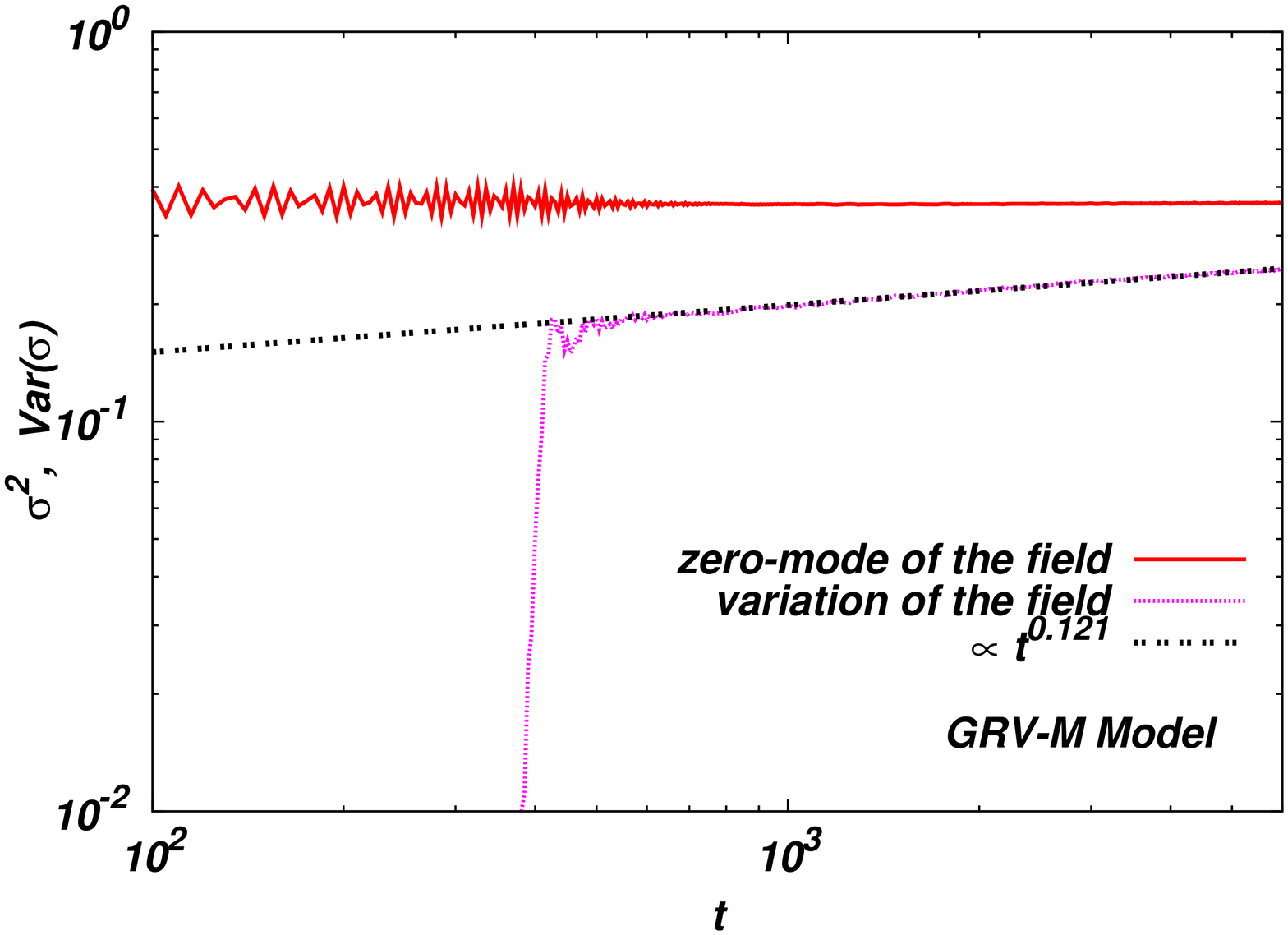}
	\includegraphics[angle=-90, scale=0.28]{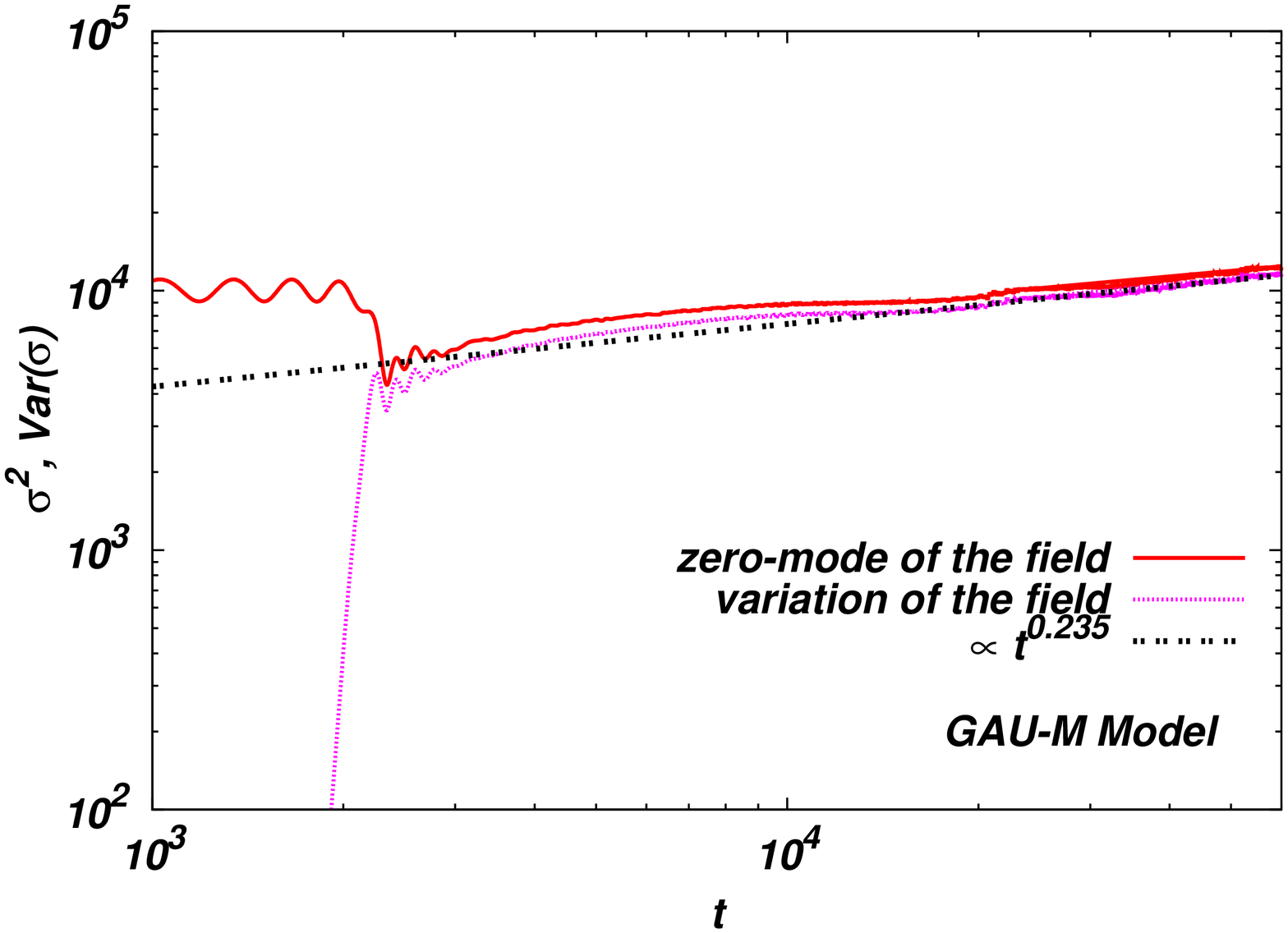}\\
	\includegraphics[angle=-90, scale=0.28]{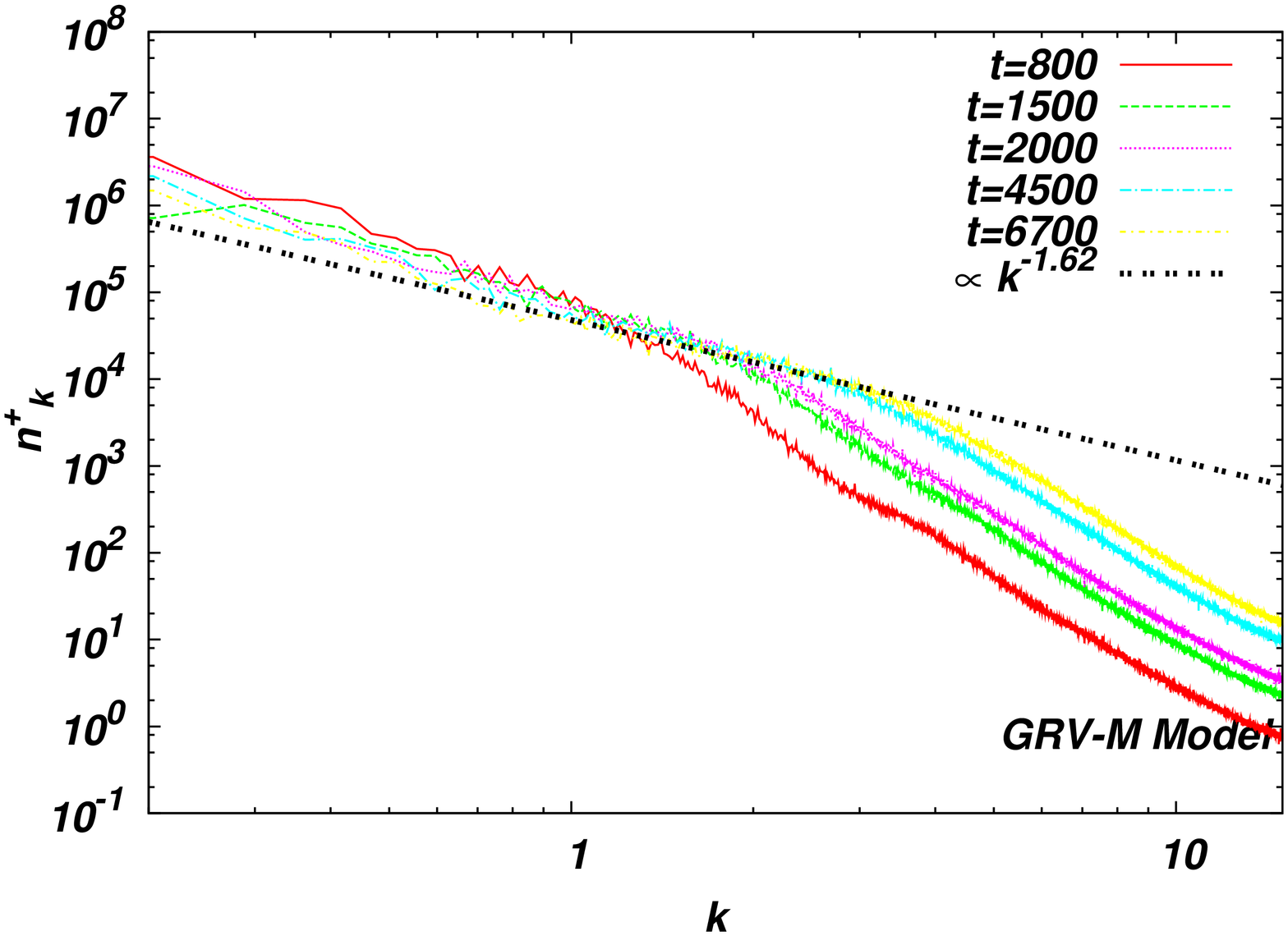}
	\includegraphics[angle=-90, scale=0.28]{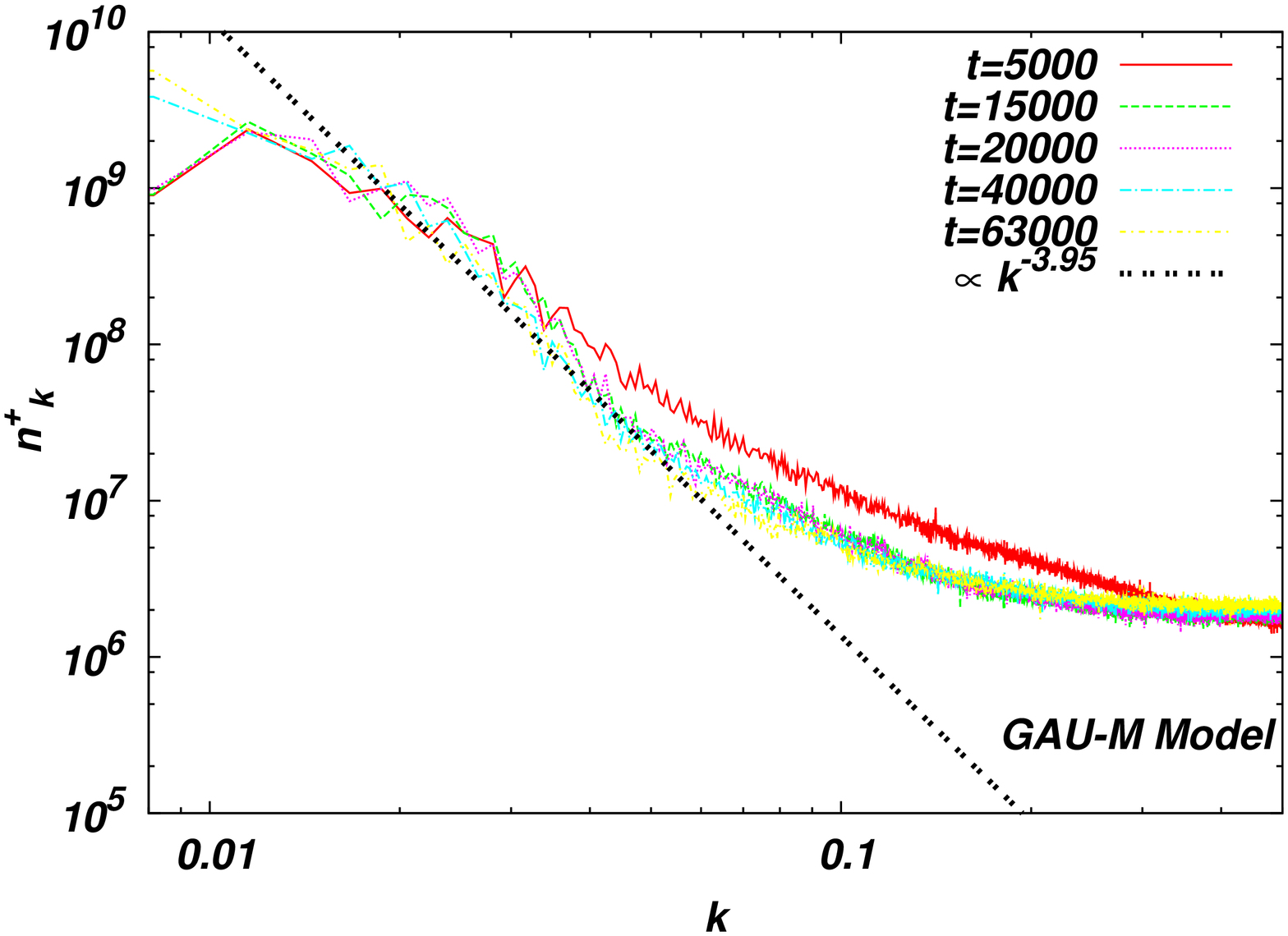}\\
	\includegraphics[angle=-90, scale=0.28]{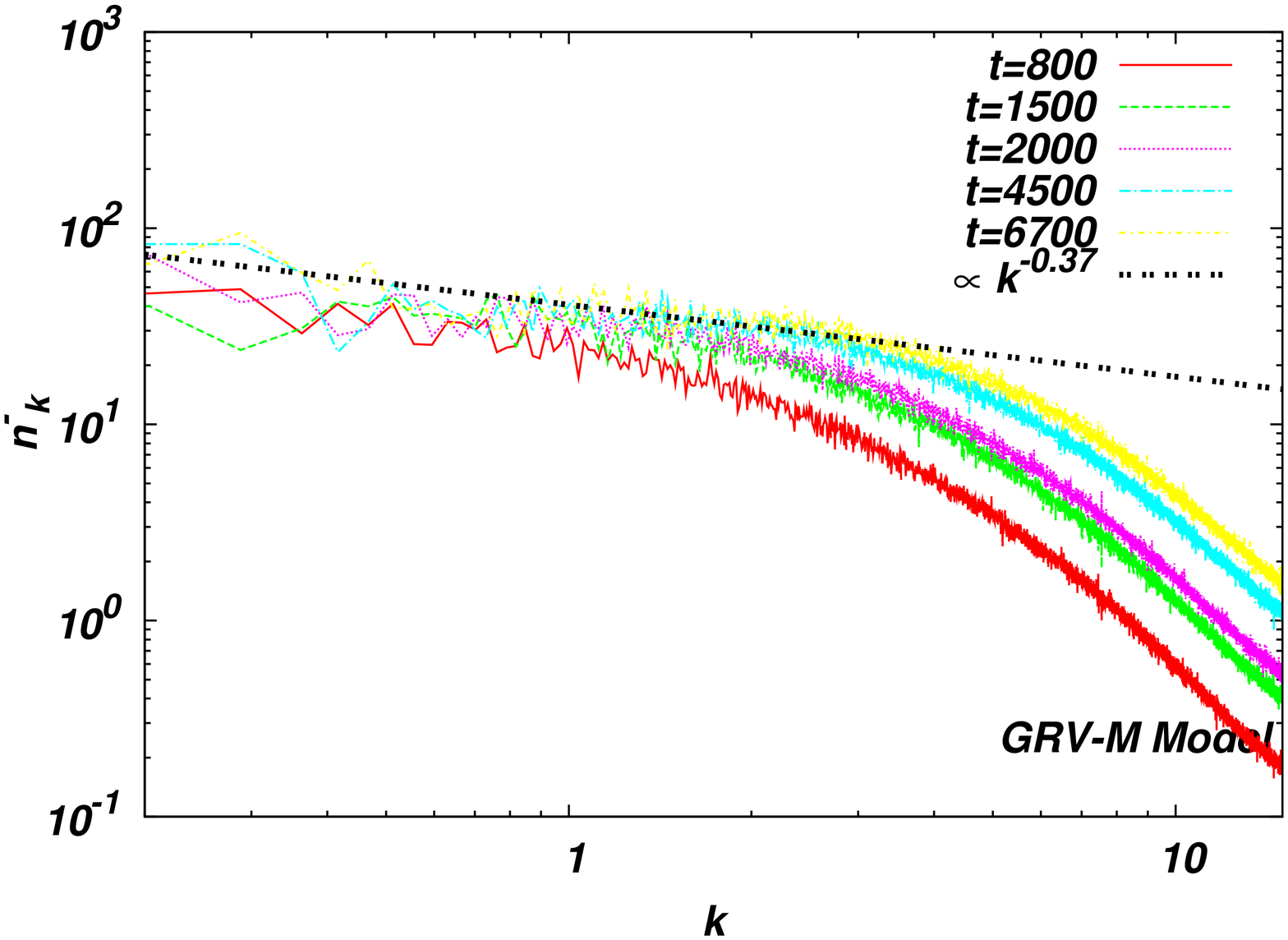}
	\includegraphics[angle=-90, scale=0.28]{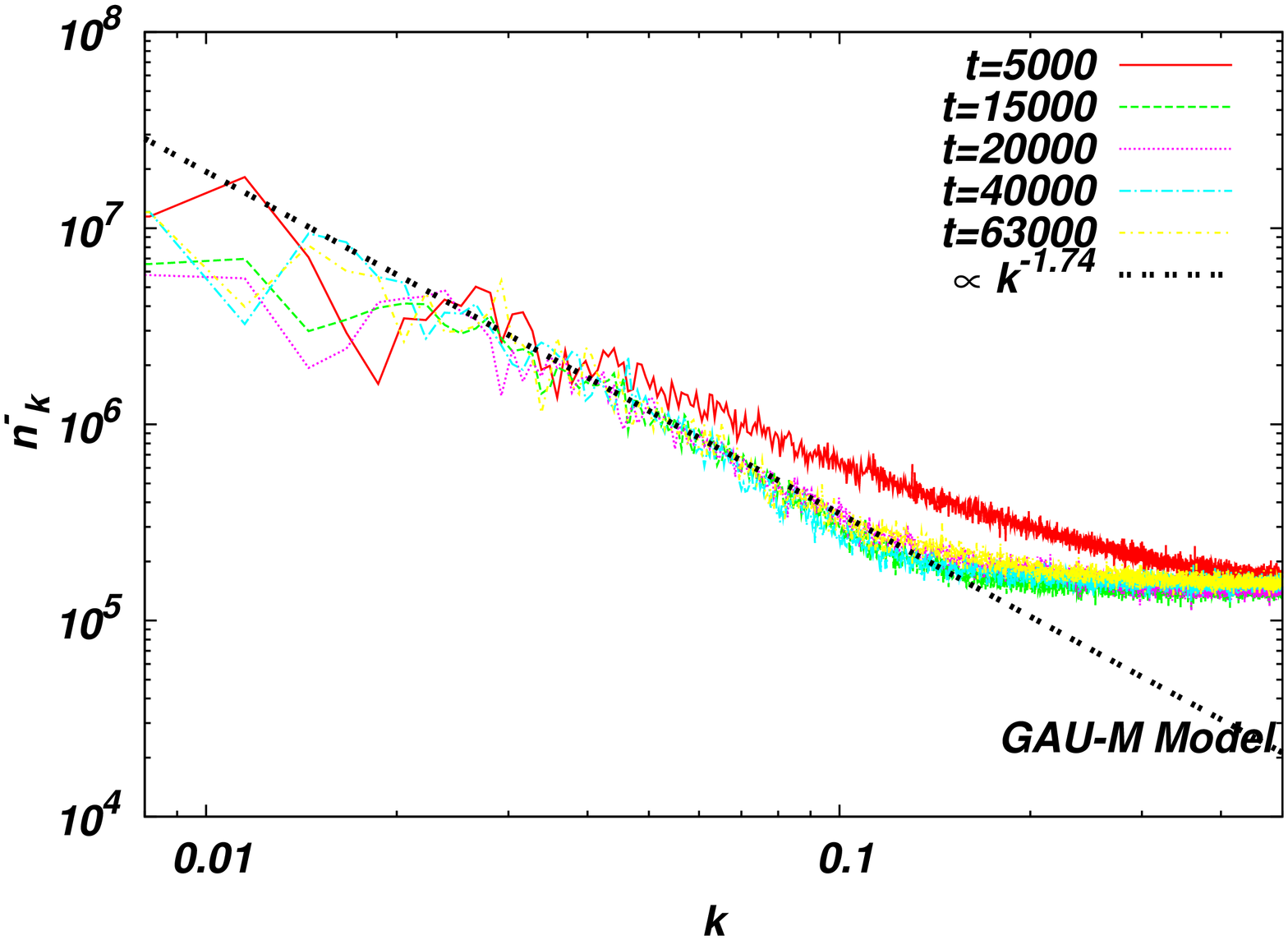}
  \end{center}
\caption{ Left panels (GRV-M Model) and right panels (GAU-M Model): the top panels show the evolution of zero-mode (red-solid lines) and the variations for $\sigma$ (dotted-dashed purple lines), whose latter evolution are fitted by a function, $\propto t^{\gamma_1}$, (black dashed lines) where we numerically obtain the value of $\gamma_1$. In the middle and bottom panels, we plot, respectively, the amplitudes of $n^+_k$ and $n^-_k$ in different times for both models, and we fit them by a function of $\propto k^{-\gamma_2}$ where $\gamma_2$ is also numerically obtained.}
  \label{fig:nknonlin}
\end{figure}

The top panels of \fig{fig:nknonlin} show the evolution of the zero-mode (red-solid lines) and the variations for $\sigma$ (dotted-dashed purple lines), whose latter evolution are fitted by a function, $\propto t^{\gamma_1}$, (black dashed lines), where $\gamma_1$ is a numerical value as the power of \eq{variation}. For both models (GRV-M Model on the left panel and the GAU-M Model on the right panel), the asymptotic evolution after the linear perturbation regime is overlapped by the function, where $\gamma_1 \sim 0.121$ for the GRV-M Model and $\gamma_1\sim 0.235$ for the GAU-M Model. Our analytic values can be matched by setting $p\sim 0.111$ with $m=5$ in the GRV-M Model and $p\sim 0.250$ with $m=3$ in the GAU-M Model, see \eq{variation}. Hence, we could identify this regime as driven (stationary) turbulence, and the main dynamics in each model is caused by either a ``five-particle'' interaction or ``three-particle'' interaction, respectively. Note that our nonrenormalisation term has a $\phi^6$ term in both models. In the middle and bottom panels of \fig{fig:nknonlin}, we plot, respectively, the amplitudes of $n^+_k$ and $n^-_k$ at different times for the GRV-M Model (left panels) and the GAU-M Model (right panels). For $n^{\pm}_k$ of the GRV-M Model, the amplitudes of the high momentum modes grow in time, whilst the lower momentum modes do not decay completely and stay for a long time. We fit a function, $\propto k^{-\gamma_2}$, (yellow dotted lines) where $\gamma_2$ is a numerical value onto the spectra at $t=6700$ for the region where the function is fitted as shown in black dashed lines. We find that $\gamma_2\sim 1.62$ for the $n^+_k$ case and $\gamma_2\sim 0.37$ for the $n^-_k$ case. In the right middle and bottom panels, we plot the amplitudes of $n^{\pm}_k$ for the GAU-M Model in various times. The amplitudes of the high momentum modes decrease as opposed to the GRV-M case, and the slopes of the spectra for $n^{\pm}_k$ at $t=63000$ in yellow-dotted lines are steeper than the GRV-M case, where we fit the numerical spectra by the following values shown in black dashed lines: $\gamma_2\sim 3.95$ for the $n^+_k$ case and $\gamma_2\sim 1.74$ for the $n^-_k$ case.

\subsubsection{From driven turbulence to near equilibrium -- Thermalisation:}

\begin{figure}[!ht]
  \begin{center}
	\includegraphics[angle=-90, scale=0.28]{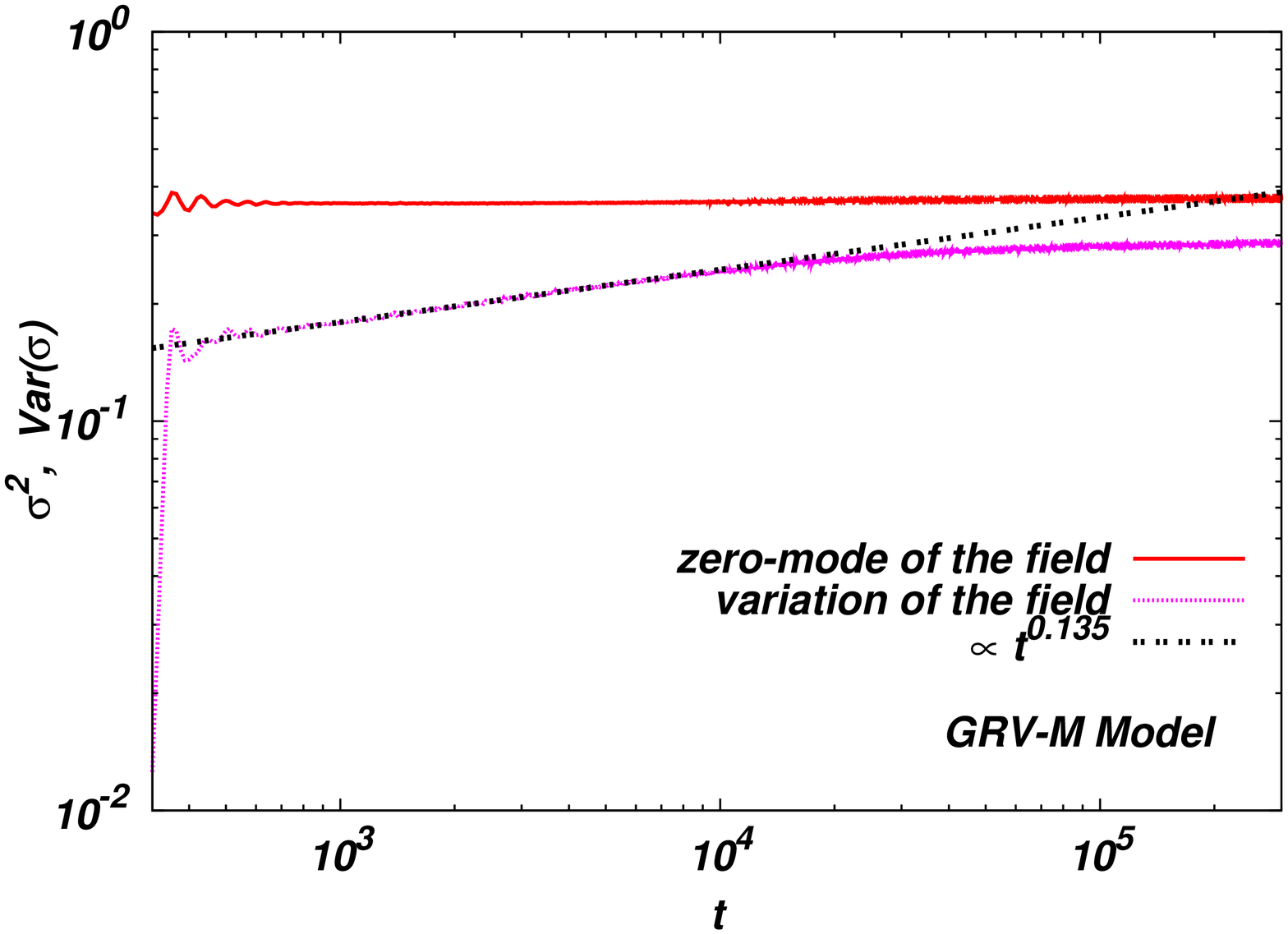}
	\includegraphics[angle=-90, scale=0.28]{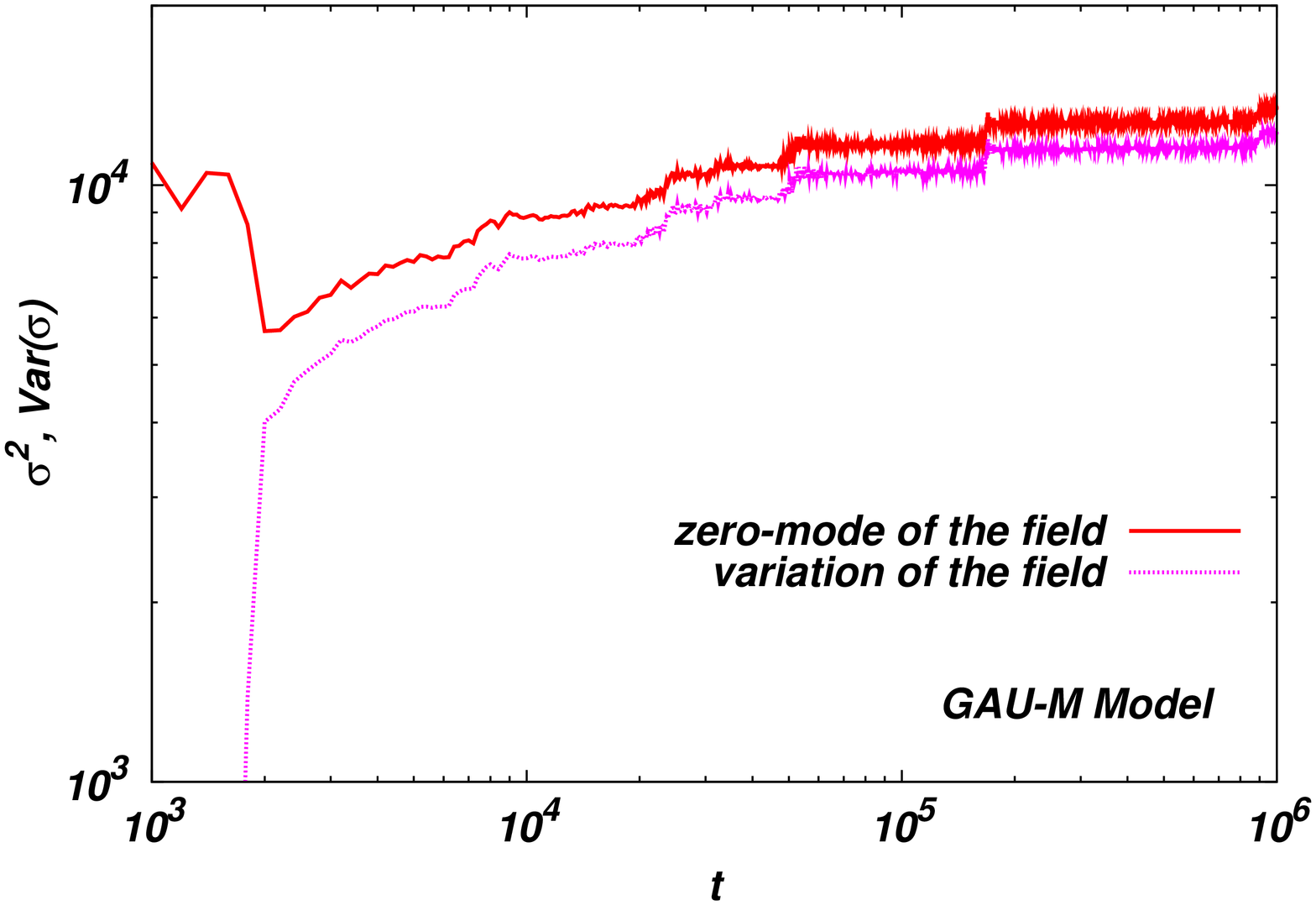}\\
	\includegraphics[scale=0.36]{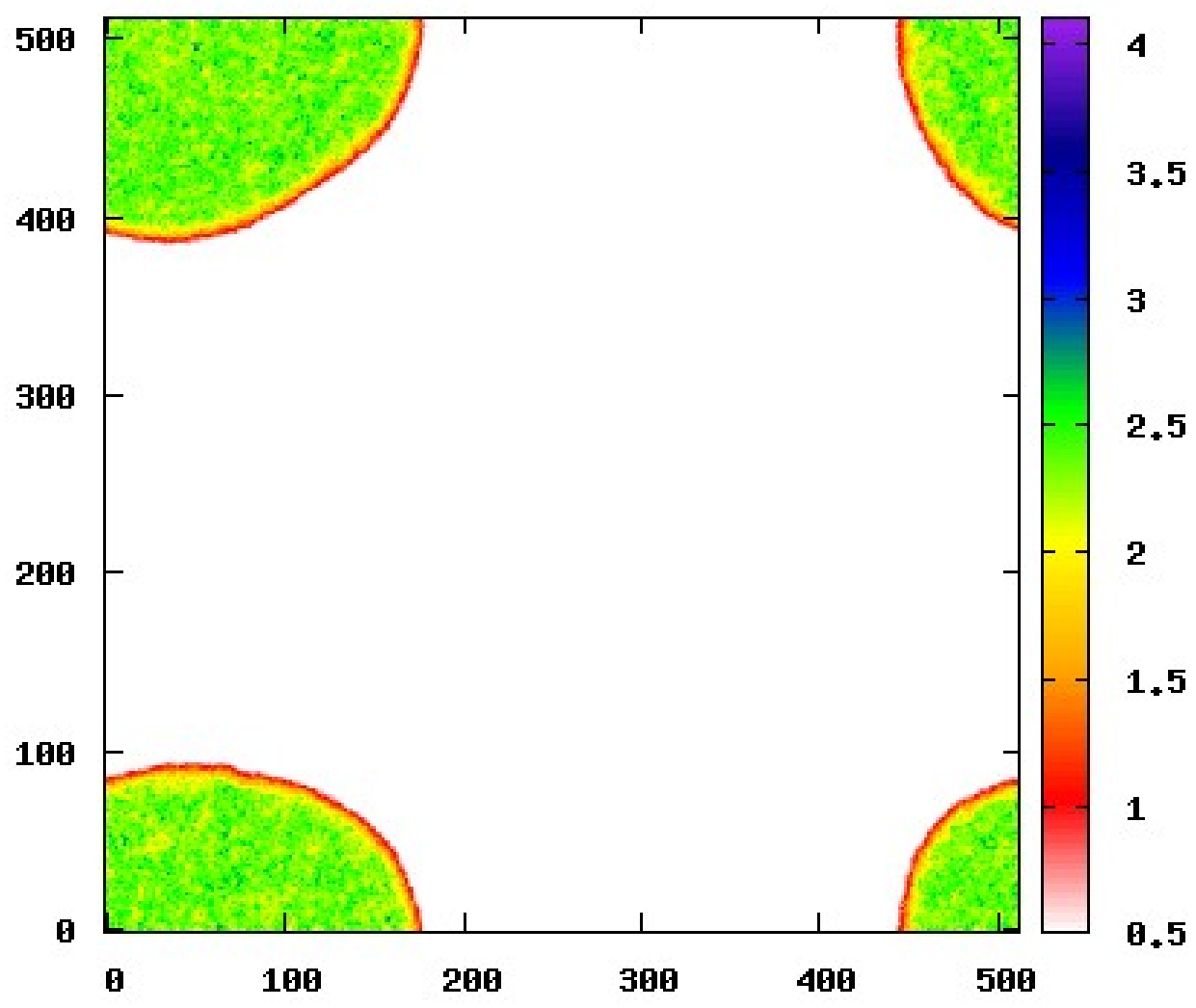}
	\includegraphics[scale=0.36]{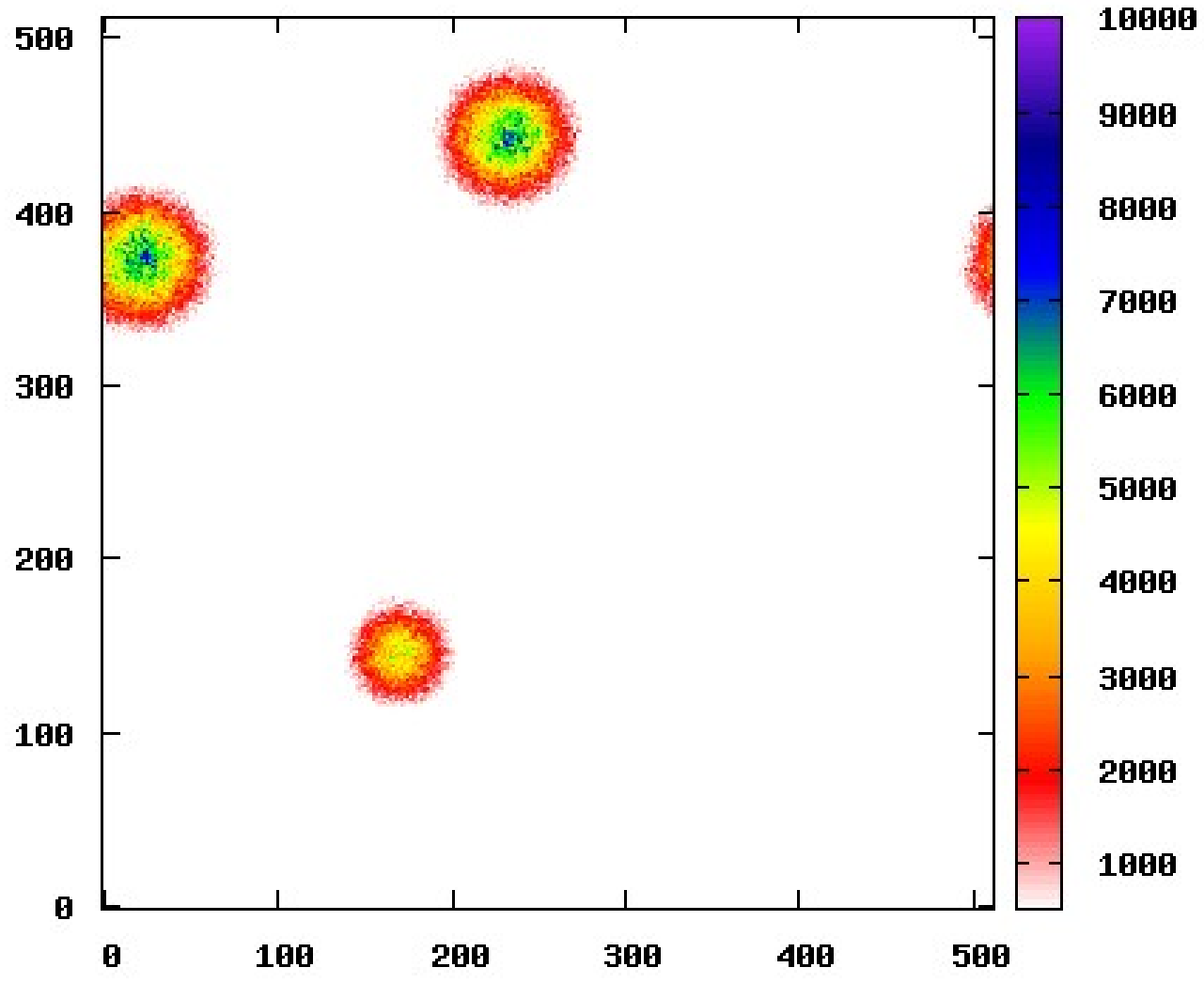}
  \end{center}
  \caption{ Left panels (GRV-M Model) and right panels (GAU-M Model) in $2+1$ dimensions: the top panels show the evolution of the zero-mode (red-solid lines) and the variations for $\sigma$ (dotted-dashed purple lines). In the bottom panels, we plot the energy density (at $t=3.5\times 10^5$ in the left-bottom panel and at $t=1.7\times 10^7$ in the right-bottom panel) instead of the charge density to compare the $Q$-ball profiles seen in \figs{fig:grvpro}{fig:gaupro} in chapter \ref{ch:qbflt}, where the colour bars illustrate the values of energy density. We can see that almost all of the charge is trapped into bubbles which may be ``thin-wall'' $Q$-balls, recall that we are imposing a periodic boundary condition.}
  \label{fig:neq}
\end{figure}

In order to significantly reduce the simulation time, we carry out $2+1$-dimensional lattice simulations with the same initial conditions as used in the $3+1$-dimensional cases, where our lattice units are reduced from $512^3$ to $512^2$. In the top panels of \fig{fig:neq} (GRV-M Model in the left panels and the GAU-M Model in the right panels), we illustrate the evolution of the zero-mode and the variances of $\sigma$, and in the bottom panels we plot the energy density (at $t=3.5\times 10^5$ in the left-bottom panel and at $t=1.7\times 10^7$ in the right-bottom panel) instead of the charge density to compare with the $Q$-ball profiles at zero-temperature, which we obtained in \figs{fig:grvpro}{fig:gaupro} in chapter \ref{ch:qbflt}. The colour bars in the bottom panels of \fig{fig:neq} illustrate the values of energy density. Note that we are using the same parameters for the GRV-M Model as the ones used in chapter \ref{ch:qbflt}, whilst the potential for the GAU-M Model used there is a generalised version of our present potential \eq{gauge-pot}, so the profiles in the GAU-M Model should look similar only qualitatively, but not quantitatively. From the top panels, we can also see, in particular the GRV-M Model, the scaling exponent evolution during the driven turbulence stage after the pre-thermalisation ends as confirmed in the top panels of \fig{fig:nknonlin}. The subsequent evolution, however, is different between each other and also unique apart from a characteristic free turbulence stage. These features of the thermalisation process are caused by stable nonlinear solutions, namely ``$Q$-balls''; in the GRV-M Model (left panels), the variance does not evolve that much after the driven turbulence stage ends and we can see thin walled like charged lumps in the end, see the left-bottom panel. In the GAU-M Model (right panels) the variance has a step-like evolution, at which stage we confirmed that two (or sometimes more) charged lumps collide and merge into a larger lump. The collision rate is very low since the motions of these ``heavy'' bubbles are nonrelativistic, but we expect that there will be only one single $Q$-ball left ultimately as similar as the GRV-M case. Generally, we observe that almost all of the total energy is trapped into these lumps, where we also confirm that the total charge is absorbed into these lumps, as reported in \cite{Kasuya:1999wu, Kasuya:2000wx}. As the ``thin-wall'' $Q$-balls in the GAU-M Model do not have an extremely thin-wall thickness \cite{Copeland:2009as}, the profiles seen in the right bottom panel do not have such a thin-shell thickness. Note that the ``thick-wall'' $Q$-balls in the GAU-M Model may suffer from classical instability and fission against spatial perturbations around the $Q$-ball solutions, and decay into smaller $Q$-balls as opposed to the case of ``thick-wall'' $Q$-balls in the GRV-M Model. The reader should also notice that the potential for the GAU-M Model in the present case is different from \eqs{potgauge}{apprxpot} in chapter \ref{ch:qbflt}, which may change the classical stability of the $Q$-balls in the ``thick-wall'' limit. Furthermore, the stability of $Q$-balls is related to their own charge $Q$ so that the initial ratio, $E/(mQ)$, can also cause the different evolution. Therefore, we believe that the evolution is very sensitive to the parameters of the models used and the initial conditions. It is worth mentioning, in the left-bottom panel, that the value of charge density within the charged cluster is slightly larger than the value of the thin-wall $Q$-balls in the zero-temperature case [compare to right bottom panel of \fig{fig:grvpro} in chapter \ref{ch:qbflt}]. We believe that this is because this charged cluster appears in the thermal background, in which the thermal effects change their profiles.

\vspace*{10pt}

Let us recap our findings in this section. We have shown in both GRV-M and GAU-M models that the AD condensate that has a negative pressure is generally unstable against linear fluctuations, and the fluctuations evolve exponentially. The condition for the presence of the negative pressure corresponds to the existence condition of $Q$-balls, and under our initial conditions shown in \tbl{parameterSET}, we observed that almost all of the total charge is trapped into a single (and a few) spherical lump(s) (``thermal $Q$-balls'') by the end of our numerical simulations. In the intermediate regions between the initial exponential amplification stage and thermalisation stage in the presence of the nonlinear solutions, we identified that the driven turbulence is active; we then found the scaling exponent evolution for the variance of $\sigma$, and we saw that this stage lasts relatively much longer than the case of tachyonic reheating.

\section{Conclusion and discussion}\label{concl}

In this chapter we have discussed both analytically and numerically two main issues: the dynamics of Affleck-Dine (AD) condensates and their subsequent nonequilibrium dynamics in the presence of nonlinear solutions. We showed that the AD dynamics has the same features as the orbital motions of planets, replacing the gravitational force by an isotropic harmonic oscillator force. As the relativistic correction to the Newtonian potential gives a precession for the planetary orbit, the orbits of AD fields are disturbed by the nonrenormalisable and quantum correction terms. Note that the essential origin of these corrections is physically different. In the presence of a negative pressure of the AD condensate, we have shown that the condensate is classically unstable, and the evolution of the system is similar to the dynamics of reheating of the Universe, \ie\  \emph{pre-thermalisation}, \emph{bubble collisions} and \emph{thermalisation}. Adopting lattice simulations, we found that the thermalisation process occurs in the presence of charged lumps, which merge into a single (or a few) ``thermal thin-walled $Q$-ball(s)'', absorbing most of the homogeneous charge distributed initially on the lattice.

\vspace*{10pt}

In Sec. \ref{sectorbit}, we introduced two phenomenological models motivated by the MSSM, \ie\  the gravity-mediated (GRV-M) model and gauge-mediated (GAU-M) model. We obtained the frequencies of the rotation for the nearly circular orbits, and showed that the condensate can have a negative pressure in both cases, see Sec. \ref{MODEL-ABC}. Furthermore, we checked numerically our analytic results with the various cases in both a non-expanding and expanding universe.

Our analytic expressions have a number of advantages. In the existing literature on preheating for complex scalar fields \cite{Postma:2003gc, Pawl:2004cs, Chacko:2002wr}, the motion of the complex scalar field is assumed to be of an elliptical form, but their ansatz does not hold [compare our expressions in \eqs{tisig}{sigsol} and \eq{sigantz} and their ansatz]. In the multi-flat direction cases, our analytic expressions of the AD field give the exact Mathieu equation if the interaction term between the AD field $\phi$ and another field $\chi$ that parametrises another flat direction, is given by $g^2 |\phi|^2|\chi|^2$, where $g$ is a coupling constant between them. The previous literature \cite{Chacko:2002wr, Allahverdi:2006xh, Allahverdi:2008pf} suggested that the resonant SUSY preheating for nearly circular orbits is not effective since the characteristic dimensionless quantity $q$ is much less than unity, recalling that broad resonant preheating (nonadiabatic evolution) occurs for $q\gg 1$. This statement also holds for our case when the orbit of the AD field is nearly circular because of $q\propto \varepsilon^2$ where $\varepsilon^2$ is the third eccentricity of the orbits, recalling that nearly circular orbits correspond to the case of $\varepsilon^2 \ll 1$. 

We obtained the successful ans\"{a}tze, \eq{sigantz}, for nearly circular orbits in an expanding universe (see also the top panels in \fig{fig:sigexp}), but our analytical expressions could be improved by the action variable technique as a real scalar field case \cite{Berkooz:2005sf}. These issues on understanding analytic forms of the orbits are related to the dynamics of spinning scalar fields, which can be responsible for the early- and late- time exponential expansions of the Universe (spinflation \cite{Easson:2007dh} and spintessence \cite{Boyle:2001du}) since the AD condensate can possess a negative pressure, which can satisfy the condition of slow-roll inflation, $w<-1/3$. In  \cite{Liddle:1998xm}, the authors discussed an oscillating field responsible for dark energy (see a recent review \cite{Copeland:2006wr}), and it gives a constraint on the power of a power-law potential in order to obtain the attractor solutions \cite{Copeland:1997et}. As in the case of real scalar fields, a complex scalar field has been investigated, see for example \cite{Kasuya:2001pr, Alimohammadi:2006qi, Wei:2005nw}. Following our analytical work, one can investigate the further analysis on dark energy for a complex scalar field and their late evolution in order to place constraints on parameters of the models, avoiding $Q$-ball formation.

\vspace*{10pt}

In Sec. \ref{sectinst}, we explored the late evolution of AD fields in Minkowski spacetime in both GRV-M and GAU-M models. As the usual nonequilibrium dynamics, we proposed that the dynamics of the $Q$-ball formation goes through three distinct regimes: \emph{pre-thermalisation}, \emph{bubble collision} (driven turbulence) and \emph{thermalisation}. We showed analytically that the AD condensate is unstable against spatial perturbations if the condensate has a negative pressure, and the perturbations grow exponentially. The presence of the negative pressure satisfies the existence condition of $Q$-balls as well as the fact that the sound wave of the perturbation has an imaginary value of the sound speed. Assuming the adiabatic linear evolution, we have analytically shown that the perturbations for the most amplified mode $k=k_m$ in \eq{kmost} grows with the exponent $\dot{S}_m$ in \eq{Sevo}, which we obtained by taking the average over one rotation of the orbits of the AD field. In the previous literature \cite{Enqvist:2002si, Kasuya:2001hg}, these values were obtained by ignoring the nonrenormalisable term and by assuming that the orbit is circular. By including the nonrenormalisable term and considering more general elliptic orbits, we recovered their results as the leading order term of our solutions in Sec. \ref{NCO}. We also showed that the nonlinear time is delayed compared to the time which the authors in \cite{Enqvist:1998en} obtained, since the other modes are not well developed when the most amplified mode starts to grow exponentially. With our $3+1$-dimensional numerical lattice simulations, which were run for a much longer time with much larger simulation sizes than the past lattice simulations in \cite{Kasuya:1999wu, Kasuya:2000wx, Enqvist:2002si, Kasuya:2001hg, Kusenko:2008zm}, our analytic results were shown to be robust. We found that the adiabatic condition is violated at the beginning stage of the linear perturbations as seen in broad resonant preheating. In the driven turbulence stage, we observed that many bubbles form and collide/merge into larger bubbles in both GRV-M and GAU-M models. Note that these bubbles are nothing to do with the bubbles due to first order phase transition. By concerning with the variance of the radial field $\sigma$, we have seen that the evolution follows a scaling exponent law as a signature of the driven turbulence \cite{Micha:2004bv}. As opposed to the case of tachyonic preheating, this driven turbulence stage, in our case, lasts for a longer time, which may be caused by the presence of classical nonlinear solutions, \ie\  ``$Q$-balls''. We saw in our $2+1$-dimensional numerical results that a thermalisation stage actually exists where the evolution for the variance of a field has a different scaling law from the one which appears in the driven (first) turbulence stage. We believe that quantum effects should be non-negligible in this late turbulence stage, and the classical thermalisation process, in our case, should be different from the corresponding quantum-mechanical thermalisation. Since the thermalisation process is generally extremely long, a lattice simulation in an expanding background encounters serious problems in the ultra-violet limits; thus, we ignored the Hubble expansion in our lattice simulations. By considering the quantum-mechanical effects as well as Hubble expansion, it is worth investigating the cosmological consequences.

In the context of a (p)reheating scenario, it has been suggested \cite{GarciaBellido:2007af} that the collision of bubbles during the driven turbulence stage can be an effective source of gravitational waves, which can be detected by LIGO \cite{ligo-web-page} and LISA \cite{lisa-web-page} in the near future. We noticed that this analysis should be applicable to the same driven turbulence stage of the $Q$-ball formation, which was initially proposed in \cite{Tsumagari:2008bv}. The problem of gravitational waves emitted in the fragmentation stage has been discussed \cite{Kusenko:2008zm}, while the analysis in the driven turbulence stage of $Q$-ball formation still remains to be done.

\vspace*{10pt}

Moreover, we assumed that the A-terms in the scalar potentials $V$, \eq{gravity-pot} and \eq{gauge-pot}, are negligible at the beginning of the analysis, where $V$ is independent of the phase field $\theta$. However, those terms are essential to generate the baryon/lepton number in the AD baryogenesis, and the dynamics of the AD field and the formation of $Q$-balls may be affected by the A-terms. Recall that the conserved global charge (baryon number) stabilises a $Q$-ball. With the inclusion of the A-term in $V$, the authors in \cite{Kawasaki:2005xc} showed that the $Q$-balls can be unstable for a strong coupling constant of the A-term, however they also claimed that the previously published stability analysis on $Q$-balls should not be affected drastically since the coupling constant of the A-term is very weak under the realistic cosmological situation. Therefore our analysis in this chapter is still valid.



\chapter{Conclusions}\label{ch:concl}

In this thesis we have studied the stability and detailed formation process of $Q$-balls in polynomial potentials and physics beyond the SM, namely the MSSM. By including quantum corrections as well as thermal corrections in the MSSM scalar potentials, the AD condensate may possess a negative pressure, whose presence implies that this condensate may fragment into nontopological solitons, known as $Q$-balls. Past work \cite{Kusenko:1997ad, Kasuya:1999wu, Kasuya:2000wx, Enqvist:1997si} has failed to convincingly demonstrated the necessary conditions to understand the stability and formation of these $Q$-balls. A proper treatment may significantly alter the existing cosmological estimates for the presence of the $Q$-balls; and this research background on $Q$-balls gave our initial motivations to explore these solutions in more detail. Our primary goal in this thesis has been to understand how $Q$-balls form and interact with each other in the very early Universe, and we have solved a number of questions related to this issue throughout this thesis.

\vspace*{10pt}

In chapter \ref{ch:chapter2}, we reviewed the fundamental aspects of standard $Q$-balls. By introducing two powerful analytical tools, the Legendre transformation and the virial theorem, we presented a remarkable way of calculating the charge $Q$ and energy $E_Q$ of a $Q$-ball, and obtained the relations of their classical and absolute stability conditions. By scaling a $Q$-ball solution and imposing the ratio between the surface and potential energy of the $Q$-ball, we obtained the virial relations and characteristic slopes $\gamma$, which give an important proportional relation, \ie\ $E\propto Q^{1/\gamma}$, see \eqs{legendre}{virieq}. We also obtained the threshold values for absolute stability of $Q$-balls in the parameter space $\omega$ in \eq{viriwa}. These values agree with the corresponding numerical results in polynomial potentials in chapter \ref{ch:qpots}, see \tbls{tbl:eqwa}{tbl:dwmc}.

\vspace*{10pt}

Following the pioneering work on nontopological solitons, \ie\ $Q$-balls, by Friedberg, Lee and Sirlin \cite{Friedberg:1976me} and Sidney Coleman \cite{Coleman:1985ki}, in chapter \ref{ch:qpots} we explored both the absolute and classical stability conditions of standard $Q$-balls in a general polynomial potential, which can appear as an effective potential with quantum/thermal corrections. In the extreme lower limit of $\omega$, namely $\om=\om_-$, we defined thin-wall $Q$-balls, which have generally an infinitesimally small thickness outside of their core. For potentials without degenerate vacua (NDVPs), we showed that a step-like profile is the appropriate ansatz in the extreme thin-wall limit, and we found that the energy of the $Q$-ball grows linearly as the charge $Q$, which is consistent with the result obtained with the virial relation in chapter \ref{ch:chapter2}. We also found that the solution is absolutely stable against their own quanta, known as $Q$-matter as ordinary matter with a zero-pressure. We noticed that this $Q$-matter phase is not generally equivalent to the state, in which the AD condensate has a negative pressure, and suffers from spatial perturbations, fragmenting into inhomogeneous states, see chapter \ref{ch:adqbform}. In order to investigate thin-wall $Q$-balls including a finite size of the shell thickness, we introduced a modified ansatz which is valid for a more wider parameter space $\omega$ in addition to $\om=\om_-$. We then recovered the solution of the $Q$-matter phase ($\om=\om_-$) as the extreme case in NDVPs, and obtained new features of the stability conditions in polynomial potentials both with and without degenerate vacua cases, \ie\ DVPs and NDVPs. With our modified ansatz for thin-wall $Q$-balls, the condition for classical stability does not depend on the number of spatial dimensions, but the absolute stability condition  does. Moreover, the characteristic slopes coincide with those derived using the virial theorem as found in the extreme thin-wall limit. The values of the characteristic slopes $\gamma$ depend on the presence of degenerate vacua in potentials, \ie\ whether NDVPs or DVPs, such that $1/\gamma=1$ in NDVPs and $1/\gamma=\frac{2D-1}{2(D-1)}$ in DVPs, recalling $E_Q\propto Q^{1/\gamma}$. On the contrary, for the upper limit of the parameter space $\omega$ we defined ``thick-wall'' $Q$-balls, which do not actually imply that the $Q$-balls have a large shell thickness compared to the core size since we cannot define explicitly both of the sizes in this limit. We confirmed that the ``thick-wall'' $Q$-ball solutions naturally tend to free-particle solutions. We also pointed out that a Gaussian ansatz in polynomial potentials has several drawbacks, whilst the other modified ansatz solved these problems and we obtained the general classical stability condition in \eq{modcls} under the validity condition \eq{validthck}. With this fact and \eq{modslope}, it implies that the ``thick-wall'' $Q$-balls are absolutely stable. We should, however, state that a Gaussian ansatz is actually valid for one of the MSSM flat scalar potentials, \ie\ gravity-mediated potentials, as shown in chapter \ref{ch:qbflt}. The key analytic results in chapter \ref{ch:qpots} were summarised in \tbl{tbl:polres}.

\vspace*{10pt}

In the late '90s, Alexander Kusenko and Mikhail Shaposhnikov \cite{Kusenko:1997ad} and Kari Enqvist and John McDonald \cite{Enqvist:1997si} discovered SUSY nontopological soliton solutions in the MSSM, which implies that these solutions may have rich cosmological consequences.  Following our analyses developed in chapter \ref{ch:qpots}, we obtained, in chapter \ref{ch:qbflt}, both analytically and numerically new stability and stationary properties of both thin- and thick-wall $Q$-balls at zero-temperature in both gravity-mediated and gauge-mediated potentials. In gravity-mediated potentials in which SUSY is broken by gravity interactions, we found that thin-wall $Q$-balls can be quantum-mechanically and classically stable against their own quanta as long as the coupling constant of the nonrenormalisable term is small enough. The values of the characteristic slopes $\gamma$ are the same as the ones computed in chapter \ref{ch:qpots} for thin-wall $Q$-balls in polynomial potentials. Further, we showed that the ``thick-wall'' $Q$-balls are classically stable against linear perturbations and may be quantum-mechanically stable under the conditions \eq{thckabs3}. As stated, a Gaussian ansatz in this model does not have any contradictions since the solution in the ``thick-wall'' limit becomes the exact Gaussian solution, \eq{gauss}, examined in appendix \ref{appxthick}. As another example of $Q$-balls in the MSSM flat potentials, we explored $Q$-balls in gauge-mediated models in which SUSY is broken by a gauge interaction. Generally speaking, the gauge-mediated potentials are extremely flat compared to the gravity-mediated and polynomial ones; therefore, we could not apply our thin-wall ansatz \eqs{thindanstz}{thinpro} to the present case. Instead, by linearising the gauge-mediated potentials, we obtained the full analytic results over the whole range of $\omega$, see \figs{fig:gaueq}{fig:gauabs}. In particular, we showed that the $Q$-balls in the ``thin-wall-like'' limit are absolutely stable in \eq{gauthnch}, while the one- and three-dimensional ``thick-wall'' $Q$-balls are completely unstable, see \eqs{1dcls}{1dch} or \eqs{3dcls}{gauthinch}, respectively. The energy ratio by unit charge for the ``thin-wall'' $Q$-balls in the gauge-mediated models is lower compared to ones computed with the other models, namely $1/\gamma=\frac{D}{D+1}$ in $E\propto Q^{1/\gamma}$. Thus, we can use this stable and energetically compact $Q$-ball solution to explain the present dimensionless energy density of dark matter, $\Omega_{DM}$, in the Universe, \ie\; $\Omega_{DM}\sim 0.23$. Our key analytic results were summarised in \tbl{tbl:results}.

\vspace*{10pt}

It has been noted \cite{Kusenko:1997ad, Enqvist:1997si} that an AD condensate with a negative pressure fragments into $Q$-balls, and Sinta Kasuya and Masahiro Kawasaki \cite{Kasuya:1999wu, Kasuya:2000wx} showed numerically that the bubble-like objects actually form from the decays of the condensate with classical lattice simulations with both gravity-mediated and gauge-mediated models. Their original work on $Q$-ball formation, however, was not done in a consistent way from the perspective of the dynamics in the AD mechanism, and their analytic results were not well checked; therefore, in chapter \ref{ch:adqbform} we reexamined the dynamics of the AD mechanism and the late evolution which includes ``$Q$-ball'' formation and the thermalisation process in both models. We identified that the dynamics for the motion of the AD field has the same properties as orbital motions of the usual planets, replacing the gravitational force by an isotropic harmonic oscillator force. By including nonrenormalisable terms and quantum corrections in the mass term of the scalar potentials, the motion of the AD fields in both models is disturbed in a similar way as the precessesion of planetary orbits occurs due to the relativistic corrections on the Newtonian potential. Furthermore, we explicitly showed that the presence of a negative pressure in the AD condensate leads to the three consequences, all of which arise from the same origin, such as the spatial instability against linear spatial perturbations, imaginary values of the sound speed, and meeting the existence condition of $Q$-balls. By adopting $3+1$-dimensional lattice simulations with more realistic initial conditions in both gravity-mediated and gauge-mediated models, we investigated both analytically and numerically the detailed processes of $Q$-ball formation, in which we found that the evolution of the system goes through the same three distinct stages as a model of reheating in the early Universe, \ie\ \emph{pre-thermalisation}, \emph{bubble collisions} (driven turbulence), and main \emph{thermalisation}. Following the wave kinetic theory of turbulence originally proposed by Raphael Micha and Igor Tkachev \cite{Micha:2004bv}, we obtained the scaling exponent law for the variance of a field during bubble collisions. Moreover, we found numerically that the classical thermalisation process is unique due to the presence of charged lumps, which merge into a single (or a few) ``thermal thin-walled $Q$-ball(s)'', absorbing most of the homogeneous charge initially distributed over the lattice space.

\vspace*{10pt}

In summary, we have explored the stability and stationary properties of $Q$-balls in polynomial potentials and the MSSM flat potentials, and the detailed formation process in the latter phenomenologically interesting potentials, adopting with lattice simulations. We showed that non-thermal ``thin-wall'' $Q$-balls, which contain a lot of charge (baryons/leptons), can be absolutely stable in any types of the above potentials, which implies that these $Q$-balls, in particular ``thin-wall'' $Q$-balls in gauge-mediated potentials, are likely to survive from thermal effects, diffusion, dissociation, and decays into fermions. Those $Q$-balls may still exist in the present Universe as invisible matter, \ie\ dark matter. This fact naturally provides the two quantities in \eq{qbasym} \cite{Laine:1998rg}.

As our future work, it is worth investigating the possibility of gravitational wave emission from the collisions of charged bubbles during the thermalisation stage as we saw in our lattice simulations. We are also interested in studying the stability and formation of $Q$-balls in hybrid inflation models, which are motivated by SUSY D-term and F-term inflation models and the original nontopological soliton model in \cite{Friedberg:1976me}. As mentioned at the end of the previous chapter, we obtained the successful ans\"{a}tze for nearly circular orbits of the AD fields in an expanding universe. But our analytic expressions could be improved by the action variable technique as a real scalar field case \cite{Berkooz:2005sf}. These issues on understanding analytic forms of the orbits are related to the dynamics of spinning scalar fields, which can be responsible for the early- and late- time exponential expansions of the Universe (spinflation and spintessence) since the AD condensate can possess a negative pressure, which can satisfy the condition of slow-roll inflation. Following our analytical work, one can investigate the further analysis on dark energy for a complex scalar field and their late evolution in order to place constraints on parameters of the inflation models, avoiding $Q$-ball formation.

Further, we assumed to ignore the effects of gauge fields on the stationary properties of $Q$-balls and the cosmological consequences in our entire thesis. However, the inclusion of the gauge fields may affect the detailed $Q$-ball profile as pointed out in \cite{Lee:1988ag}. Inside the gauged $Q$-ball, the local gauge symmetry should be broken by the non-zero field value. Then, the charge profile for the large gauged $Q$-ball has a peak around the surface due to the Coulomb repulsion, and there exists a maximum charge. These are different features from the non-gauged $Q$-balls; thus, we believe that the stability analysis should be well modified. Another question arises that 'Can we predict the observable cosmological consequences caused by the gauge field in the MSSM?'. Regarding this question, Kari Enqvist, Asko Jokinen, and Anupam Mazumdar computed the magnitude of the magnetic field, $10^{-30}$ Gauss, generated along the MSSM flat direction \cite{Enqvist:2004yy}. This value is the same order as the magnitude for the observed magnetic field in the clusters of galaxies \cite{Clarke:2000bz}. The above two ideas on both the stability of the gauged $Q$-balls and their cosmological consequences in the MSSM can be brought together; it is worth estimating the magnitude of the magnetic field in the presence of the gauged $Q$-balls.

Finally, with the forthcoming data from the high-energy experiments, such as LHC \cite{lhc-web-page}, the gravitational wave detectors, \eg\ LIGO \cite{ligo-web-page} and LISA \cite{lisa-web-page}, and the detectors of cosmic microwave background radiation, WMAP \cite{wmap-web-page} and PLANCK \cite{planck-web-page}, we believe that these experimental data may shed light on the origin of the two quantities, \eqs{basym}{dm/b}, in the future.

\chapter*{}

\begin{center}
\includegraphics[scale=0.5]{npgau2-6.eps}
\end{center}
\vspace*{\stretch{6}}
\emph{``Many of life's failures are people who did not realize \\
	\hspace*{100pt}how close they were to success when they gave up.''}
\vspace{\stretch{1}}
\begin{flushright}
{\bf -- Thomas Edison. }
\end{flushright}
\vspace{\stretch{10}}

\appendix
%
%

\chapter*{{\Huge Appendices}}
\addtocontents{toc}{\bigskip \bf Appendices \normalfont \endgraf}

\clearpage

\newpage

\chapter[Classical stability]{Classical stability}\label{ap1}

Nontopological solitons, \ie\  $Q$-balls, can be classically stable for a small nonlinear coupling constant, say $g$, when the fluctuations around the solution are of a harmonic oscillator form. It implies that the energy is  $E_Q=E^{(0)}_Q+\sum_N(\mathcal{N}_N+\half)\Omega_N+ \dots$ in the language of quantum field theory, where $E^{(0)}_Q\sim \order{1/g^2}$ is the $Q$-ball energy, $\mathcal{N}_N$ is the ``occupation number'' of the $N$th normal mode, and $\Omega_N$ is the characteristic frequency of the fluctuations. Here, the second term in $E_Q$ is of $\order{g^0}$ and the higher order terms are suppressed by the small nonlinear coupling constant $g$, \ie\ $\order{g^2,\; g^4,\ \dots}$ \cite{Lee:1991ax}. The case, $\Omega^2_N=0$, corresponds to the zero mode, \ie\  translations and phase transformations from the $Q$-ball solution. A classically stable $Q$-ball has fluctuations with $\Omega^2_N>0$, whilst if $\Omega^2_N<0$, the fluctuations exponentially grows, which means the solution is classically unstable. The canonical quantisation of the solitons is a matter of ordering the canonical variables so that one needs to additionally impose the equal time canonical commutation relations on the variables for the purpose  \cite{Friedberg:1976me, Lee:1991ax}. 

In this appendix we present the complete classical stability analysis of $Q$-balls in a number of spatial dimensions $D$, following the original work in \cite{Lee:1991ax}. In order to show the classical stability, we have to adopt the Hamiltonian formalism, starting from the Lagrange formalism. Introducing collective coordinates and concerning the zero-modes, we obtain all positive or zero eigenvalues for the fluctuations around the $Q$-ball solution subject to the condition that the charge of the $Q$-ball should decrease as a function of $\om$. Therefore, we can show that the $Q$-ball solutions are classically stable against linear fluctuations.

Let us begin with perturbing the lowest energy solution, \ie\  a soliton solution, $\sigma(\textbf{x}-\textbf{R}(t))$ with complex fluctuations $\chi(t,\textbf{x}-\textbf{R}(t))=\chi_R + i \chi_I$, where $\textbf{R}(t)$ is the location of the soliton and $|\chi|\sim \order{\epsilon}\ll \sigma$. Here, $\epsilon$ is a small quantity compared to the background field $\sigma$. Hence, the field $\phi$
\be\label{exp} 
\phi=e^{-i\theta(t)}\bset{\sigma+\chi},
\ee
where $\sigma$ satisfies the $Q$-ball equation \eq{QBeq}. We can expand $\chi$ with a complete set of complex functions $f_n(\textbf{x})$: $\chi=\sum^\infty_{n=D+2} q_n(t)f_n(\textbf{x})$ for $n \ge D+2$. Note that $q_n(t)$ is a real function due to the factor $\theta(t)$ in \eq{exp}. We shall define $f_k\propto \partial_k \sigma$ and $f_{D+1}\propto \sigma$ for $k=1,\ 2,\dots,\ D$ with $q_k=R_k(t),\ q_{D+1}=\theta(t)$, so that $f_i$ are orthonormal for $i,\ j=$1 to $\infty$, \ie\  $\int f^*_i f_j=\delta_{ij}$, where we defined $\int\equiv \int_{V_D}$, see \eq{epq}. By imposing the U(1) symmetry and the Lorentz invariance for the perturbed solution, we must have the conditions
\be\label{invp}
\int \sigma \chi_I =0,\spc \int \chi_R \nabla \sigma = 0.
\ee
\section{The second-order variations with the Lagrange formalism}
Using \eq{exp} and collective coordinates, $q_k=R_k(t)$ and $q_{D+1}=\theta(t)$, it is tedious but straightforward to express the lagrangian, $L=K(q,\dot{q})-V(q)$, up to second order, where the kinetic and potential terms are, respectively, 
\bea{kin}
K&=&\int \half |\dot{\phi}|^2=\half \tilde{\dot{q}}_i \mM_{ij}\dot{q}_j=K_0+K_1+K_2+\dots,\\
\label{vin}V&=&\int \half \abs{\nabla{\phi}}^2 + U(\abs{\phi})=V_0+V_1+V_2+\dots.
\eea
Here, a tilde denotes the inverse vectors of the original vectors. The components of $\mM_{ij}$ are
\begin{eqnarray}
\nb \mM_{D+1,D+1}&=& \int \set{\sigma^2+2\sigma \chi_R+ \chi^2_R+\chi^2_I},\\
\nb \mM_{kk^\p}=\mM_{k^\p k}&=&\int \left\{\partial_k \sigma \partial_{k^\p} \sigma + 2\partial_k \sigma \partial_{k^\p} \chi_R + \nb \partial_k\chi_R\partial_{k^\p}\chi_R + \partial_k\chi_I \partial_{k^\p}\chi_I \right\},\\
\nb \mM_{D+1,k}=\mM_{k,D+1}&=&\int \set{-2\chi_I \partial_k \sigma + \chi_R \partial_k\chi_I -\chi_I \partial_k\chi_R},\\
\nb \mM_{D+1,n}=\mM_{n,D+1}&=&\half \int \set{i(f_n-f^*_n)\chi_R + (f_n+f^*_n)\chi_I},\\
\nb \mM_{kn}=\mM_{nk}&=&-\half \int \set{(f_n+f^*_n)\partial_k\chi_R -i(f_n-f^*_n)\partial_k\chi_I},\\
\nb \mM_{n n^\p}=\mM_{n^\p n}&=&\int f^*_n f_{n^\p}=\delta_{n n^\p}.
\end{eqnarray}
The matrix $\mM$ can be expanded as a series of the infinitesimally small quantity $\epsilon$,
\be\label{Mat}
\mM=\mM_0(\epsilon^0) + \mM_1(\epsilon^1)+ \mM_2(\epsilon^2)+\order{\epsilon^3}\dots,
\ee
where 
\be
\mM_0 = \left(\begin{array}{ccc} S_0 & 0 & 0 \\ 0 & I & 0 \\ 0 & 0 & 1 \end{array} 
             \right),\ \ 
\mM_1 = \left(\begin{array}{ccc} A & B & J   \\ \tilde{B} & C & E \\ \tilde{J} & \tilde{E} & 0 \end{array} 
             \right),\ \
\mM_2 = \left(\begin{array}{ccc} F & G & 0   \\ \tilde{G} & H & 0 \\ 0 & 0 & 0 \end{array} 
             \right).
\ee
Here, we defined the matrices, column vectors, and scalars as
\bea{}
\nb S_0=M_0 \delta_{kk^\p}&,&\ I=\int \sigma^2,\\ 
\nb A=2\int \partial_k \sigma \partial_{k^\p} \chi_R&,& B= 2\int \chi_I \partial_k \sigma,\ C=2\int \sigma \chi_R,\\  
\nb E=\mM_{n,D+1}&,& \ J=\mM_{kn},\\
\nb F=\int \partial_k\chi_R\partial_{k^\p}\chi_R+\partial_k \chi_I \partial_{k^\p} \chi_I&,&\ G=\chi_I\partial_k\chi_R-\chi_R\partial_k\chi_I,\; \ H=\int \chi^2_R + \chi^2_I,
\eea
where $M_0\equiv \frac{1}{D}\int (\nabla \sigma)^2$ and the tildes are denoted as the inverse matrices (vectors) of the original matrices (vectors) again. Therefore, the components of $K$ in \eq{kin} is given by
\be\label{kin2}
K_0= \half \tilde{q}_i \mM_{0ij} q_j,\ \ K_1= \half \tilde{q}_i \mM_{1ij}q_j,\ \ K_2= \half \tilde{q}_i \mM_{2ij}q_j.
\ee

The potential terms in \eq{vin} can be also expressed by 
\bea{v0}
V_0&=&\int \half \bset{\nabla \sigma}^2 + U(\sigma),\\
V_1&=&\omega^2 \int \sigma \chi_R,\\
\label{v2} V_2&=&\half \int \set{\chi_R\bset{-\nabla^2+\frac{d^2 U}{d\sigma^2}}\chi_R + \chi_I\bset{-\nabla^2+\frac{1}{\sigma} \frac{d U}{d\sigma}}\chi_I },
\eea
where we used $\abs{\phi} \simeq \sigma+\chi_R+\frac{\chi^2_I}{2\sigma} +\order{\epsilon^3}$ and $U(\abs{\phi})\simeq U(\sigma)+\chi_R \frac{dU}{d\sigma}+\frac{\chi^2_I}{2\sigma}\frac{dU}{d\sigma}+\frac{\chi^2_R}{2} \frac{d^2 U}{d \sigma^2}+\order{\epsilon^3}+\dots$.

\section{The Hamilton formalism with canonical transformations}

In order to consider the modes of the fluctuations $\chi$, it is useful to switch the Lagrange formalism to 
the Hamiltonian formalism. Let us impose canonical transformations with \eq{kin} and \eq{vin}, we then obtain
\be
p_i=\frac{\partial L}{\partial \dot{q}_i}=\mM_{ij}\dot{q}_j \to \dot{q}_i=\mM^{-1}_{ij}p_j,
\ee
where $p_R=\dd{L}{\dot{\chi}_R}\sim\order{\epsilon}$ and $p_I=\dd{L}{\dot{\chi}_I}\sim\order{\epsilon}$ for $n,m=D+2,D+3,\dots$.
Hence, the Hamiltonian $H(q,\ p)$ is given by
\be\label{ham}
H=\tilde{p}_i \dot{q}_i -L= \half \tilde{p}_i \mM^{-1}_{ij}p_j + V(q)=H_0+H_1+H_2+...,
\ee
which is independent of $q_k$ and $q_{D+1}$; thus, the Hamiltonian equations for the soliton momenta $P_k$ and the charge $Q$
give conserved quantities: $\dot{P}_k=-\dd{H}{\dot{q}_k}=0$ and $\dot{Q}=-\dd{H}{\dot{q}_{D+1}}=0$. In the centre of mass frame \footnote{One can find the results in an arbitrary frame in \cite{Friedberg:1976me}.}, we can set
\be
P_k=\frac{\partial L}{\partial \dot{q}_k}:=0,\spc Q=\frac{\partial L}{\partial \dot{\theta}}=const.
\ee
Using \eq{Mat}, the inverse matrix of $\mM$ can be expanded by 
\be
\mM^{-1}\simeq \mM^{-1}_0 -\mM^{-1}_0\Del \mM^{-1}_0 +\mM^{-1}_0\Del \mM^{-1}_0 \Del \mM^{-1}_0 + \dots .
\ee
We obtain the kinetic terms in \eqs{kin}{kin2} as
\bea{k0}
K_0 &=& \half \bset{Q^2 I^{-1}},\spc
K_1 = -\frac{Q^2}{I^2}\int \sigma \chi_R, \\
\nb K_2 &=& \half \sum_n \set{p_n^2 -\frac{2Q}{I}\bset{\tilde{p}_n \mM_{n,D+1} + \mM_{D+1,n} p_n } +\frac{Q^2}{I^2}\abs{\mM_{D+1,n}}^2 } \\
\label{k2} &+& \frac{Q^2}{2I^2}\set{\frac{1}{M_0} \bset{2\int\chi_I \nabla \sigma}^2 
+ \frac{1}{I} \bset{2\int \sigma \chi_R}^2} -\frac{Q^2}{2I^2}\set{\int \chi^2_R + \chi^2_I}.
\eea
To ensure that the fluctuation $\chi$ is of order of $\epsilon$, we set
\be\label{h1}
H_1=\bset{\omega^2- Q^2 I^{-2}}\int \sigma \chi_R:=0 \to Q=I\omega,
\ee
where we chose the positive sign for both $Q$ and $\om$ without loss of generality.
Using $Q=I\omega$, we can reduce Eqs. (\ref{v0}-\ref{v2}) and (\ref{k0}-\ref{k2}) to
\bea{h0}
H_0&=& D M_0+I\omega,\\
\label{H2}H_2&=& \half \sum_n \abs{p_n-\omega \mM_{D+1,n}}^2 + \mV_R + \mV_I,
\eea
where 
\bea{mVR}
\mV_R &=& \half \int \chi_R \hat{h}_R \chi_R +\frac{2\omega^2}{I}\bset{\int \sigma \chi_R}^2,\\
\label{mVI}\mV_I &=& \half \int \chi_I \hat{h}_I \chi_I +\frac{2\omega^2}{M_0}\bset{\int \chi_I\nabla \sigma}^2.
\eea
Here, the differential operators $\hat{h}_R$ and $\hat{h}_I$in \eq{mVR} and \eq{mVI} are defined by
\bea{hr}
\hat{h}_R&=&-\nabla^2 + \frac{d^2 U}{d\sigma^2}-\omega^2,\\
\label{hi} \hat{h}_I&=&-\nabla^2 + \frac{1}{\sigma} \frac{d U}{d\sigma}-\omega^2.
\eea
As a result, we found the second-order Hamiltonian $H_2$ in \eq{H2}, with which we will be able to examine the stability of the perturbations using the Hamiltonian equations.

\section{Positive eigenvalues}
In order to show the classical stability of $Q$-ball solutions, we need to impose a condition for the charge of the $Q$-ball, which implies all eigenvalues, $\Lambda_{R}$ and $\Lambda_{I}$, for $\mV_R$ and $\mV_I$ in \eqs{mVR}{mVI} should be positive definite or zero. Those eigenvalues are given by
\bea{VR}
\frac{\delta \mV_R}{\delta \chi_R}&=&\hat{h}_R \chi_R + \frac{4\omega^2}{I}\sigma\bset{\int \sigma \chi_R}=\Lambda_R \chi_R,\\
\label{VI}\frac{\delta \mV_I}{\delta \chi_I}&=&\hat{h}_I \chi_I + \frac{4\omega^2}{S_0}\nabla \sigma \cdot \bset{\int \nabla \sigma \chi_I}=\Lambda_I \chi_I.
\eea 

Our first task is to show that $\hat{h}_R$ has only one negative eigenvalue, and that the rest of them are all positive or zero, whilst we will show that all eigenvalues of $\hat{h}_I$ are positive definite or zero. Each of the zero-modes should be treated with special efforts since these modes are translation and phase invariant modes of the $Q$-ball solutions. Finally, we are able to prove that the fluctuations around the $Q$-ball solution are of the usual harmonic oscillation form using the Hamiltonian equations subject to the condition that the charge $Q$ of the $Q$-ball should monotonically decrease as a function of $\om$. This is our main aim to prove from now on.

\subsection{Eigenvalues for $\hat{h}_R$ and $\hat{h}_I$}

Let $\psi_{Ri}$ and $\psi_{Ij}$ be the eigenstates of $\hat{h}_R$ and $\hat{h}_I$, such that
\be
\hat{h}_R \psi_{Ri} = \lambda_{Ri} \psi_{Ri},\hspace{10pt} \hat{h}_I \psi_{Ij} = \lambda_{Ij} \psi_{Ij},
\ee
where $\lambda_{Ri}$ and $\lambda_{Ij}$ are the corresponding eigenvalues.

\subsubsection{One negative eigenvalue for $\hat{h}_R$}\label{oneneg}

Here, we show that there is only one negative eigenvalue of $\hat{h}_R$. By differentiating \eq{QBeq} with respect to $\partial_k$, we obtain the zero-eigenfunctions of $\hat{h}_R$,
\be\label{hr}
\hat{h}_R \partial_k \sigma=0.
\ee
The eingenfunctions, $\partial_k \sigma$, come from the translational invariance, \ie\  $\sigma(\mathbf{x}+\mathbf{\eta})$, where $\eta$ is a small quantity which is responsible for the translation from the $Q$-ball solution. The eigenfunctions correspond to p-states of $\hat{h}_R$, which have a number of spatial dimensions $D$, \ie\ $\psi_{Rk}\propto \partial_k \sigma$. Since the lowest s-state eigenvalue of $\hat{h}_R$ must be lower than the lowest p-state eigenvalue, there exists at least one negative eigenvalue of s-states for $\hat{h}_R$ whose corresponding eigenfunctions are s-waves $\psi_i$. These eigenfunctions $\psi_i$ will be used to obtain the positive eigenvalues $\Lambda_R$ for $\mV_R$ in the next subsection. Before doing so, we have to be concerned with the eigenvalues for $\hat{h}_R$ and $\hat{h}_I$ in more detail.

\nt{hatr}{th} $Theorem:\ \hat{h}_R\ must\ have\ only\ one\ negative\ eigenvalue\ in\ order\ for\ \mV_R \ge 0$.\\
\textbf{Proof}

If $\hat{h}_R$ had two negative eigenvalues $\lambda_{-2}<\lambda_{-1}<0\le \lambda_0$, 
one can expand $\sigma$ as $\sigma=\sigma_{-1}\psi_{-1}+\sigma_{-2}\psi_{-2}$ and $\chi_R=c_{-1}\psi_{-1}+c_{-2}\psi_{-2}$, where the under-indices are denoted as the corresponding eigenfunctions and factors for the eigenvalues $\lambda_{-1},\; \lambda_{-2}$. Hence, $\mV_R$ in \eq{VR} becomes
\bea{}
\mV_R&=& \half \bset{c^2_{-1}\lambda_{-1}+c^2_{-2}\lambda_{-2}}+\frac{2\omega^2}{I}\bset{c^2_{-1}\sigma_{-1}+c^2_{-2}\sigma_{-2}}^2,\\
&=& \half \set{\bset{\frac{c_{-2}}{\sigma_{-1}\sigma_{-2}}}^2\lambda_{-1}+c^2_{-2}\lambda_{-2}}<0,
\eea
where without loss of generality we have set $c_{-1}=-\frac{c_{-2}}{\sigma_{-1}}\sigma_{-2}$ in the last step. The inequality holds due to $\lambda_{-2}<\lambda_{-1}<0$. The proof of the theorem is complete.

\subsubsection{Positive and zero eigenvalues for $\hat{h}_I$}

Next, we will show that all eingenvalues of $\hat{h}_I$ are positive or zero.
From \eq{QBeq} and \eq{hi}, we obtain
\be\label{hi2}
\hat{h}_I \sigma=0,
\ee
which leads to one zero s-state eigenfunction of $\hat{h}_I$: $\psi_{I0}\propto \sigma$. Since $\sigma$ has no node \cite{Lee:1991ax}, the other eigenvalues of $\hat{h}_I$ are positive definite. 

\subsection{Positive and zero eigenvalues for $\mV_R$ and $\mV_I$}

Our next task is to establish that all eigenvalues of $\mV_R$ and $\mV_I$ should be positive definite or zero, \ie
\be\label{lri}
\Lambda_R\geq 0,\spc \Lambda_I\geq 0.
\ee 
Let $\Psi_{Ri}$ and $\Psi_{Ij}$ be the real eigenfunctions for $\mV_R$ and $\mV_I$, respectively. Note that $\Psi_{Ri}$ and $\Psi_{Ij}$ are orthonormal, namely $\int \Psi_{Ri} \Psi_{Rj}=\int \Psi_{Ii} \Psi_{Ij}=\delta_{ij}$. As one can expect, we obtain the zero-eigenfunctions are $\Psi_{Rk}\propto \partial_k \sigma$ and $\Psi_{I0}\propto \sigma$. 

First of all, let us consider $\mV_I$ in \eq{mVI}. We expand $\chi_I$ with a complete set of $\Psi_{Ij}$ with $q_{Ij}\equiv Im(q_j)$, which implies that
\be
\chi_I=\sum_j q_{Ij} \Psi_{Ij}.
\ee
Since the first and second terms of the RHS in \eq{mVI} are positive definite, we obtain $\Lambda_{Ij}\geq 0$. Recalling the translational invariance \eq{invp} and \eq{hi}, we obtain the zero-eigenfunction, \ie\ $\Psi_{I0}\propto \sigma$ with $\Lambda_{I0}=0$, \cf\ \eq{hi2}.

Secondary, we will consider $\mV_R$. We expand $\chi_R$ with a complete set of $\Psi_{Ri}$ except the s-state $q_{Rj}\equiv Re(q_j)$:
\be\label{chir}
\chi_R={\sum_i}^\p q_{Ri} \Psi_{Ri}.
\ee
Here, a prime denotes the summation over $i$ except zero-eigenvalue.
Recalling the phase invariance \eq{invp} and \eq{hr}, we obtain the zero-eigenfunction, namely $\Psi_{Rk}\propto \partial_k \sigma$ with $\Lambda_{Rk}=0$, where we used \eq{invp} again, see \eq{hr}. We can then express the energy $E[f]$ with a function $f=\sigma+ \epsilon \Psi_R$,
\be
E[f]=E_0+E_2+E_3+...=E_0 + \epsilon^2 \Lambda_R +\order{\epsilon^3},
\ee
where $E_0$ corresponds to the lowest energy solution, \ie\  the $Q$-ball solution. Without s-state waves, we obtain $\Lambda_R\geq 0$.

By including the s-states of $\Psi_{Ri}$, we will show the positivity of $\Lambda_{Ri}$, where we define $z\equiv \Lambda_{Ri}$. We define $\psi_i$ as a complete set of s-state eigenfunctions, which satisfies the the orthonormal relation, $\int \psi_i \psi_j=\delta_{ij}$. Recalling that $\hat{h}_R\psi_i=\lambda_i \psi_i$ where $\lambda_1<0<\lambda_2<\dots$, we expand $\Psi_{Ri}$ and $\sigma$ in terms of $\psi_i$ with the amplitudes, $c_i$ and $\sigma_i$, \ie\
\be\label{psir}
\Psi_{Ri}=\sum_i c_i \psi_i,\hspace{5pt} \sigma=\sum_i \sigma_i \psi_i.
\ee
Using \eqs{psir}{VR}, we obtain
\be
\bset{\lambda_j -z + \frac{4\omega^2}{I}\sum_j \sigma^2_j}c_j=0,
\ee
where we multiplied $\psi_j$ and integrated over the space in \eq{VR}. Since $c_j$ is arbitrary, the solution $z$ for $G(z)$, where
\be
G(z)=1+\frac{4\omega^2}{I}\sum_j \frac{\sigma^2_j}{\lambda_j-z}=0,
\ee
corresponds to s-state eigenvalues of $\Lambda_{Ri}$.

Note that 
\be
\frac{dG}{dz}>0,\hspace{5pt} \lim_{z\to \lambda_j \pm} G(z)=\mp \infty,\hspace{5pt} \lambda_j<\Lambda_{Rj}<\lambda_{j+1}.
\ee
Since it is required to have $\Lambda_R\geq0$, we should impose that
\bea{neg0}
\lambda_1<0\leq \Lambda_1<\Lambda_2<..\spc \lr \spc G(0)\leq 0,\\
\label{g0} \textrm{where\hspace{5pt}} G(0)=1+\frac{4\omega^2}{I}\sum_j \frac{\sigma^2_j}{\lambda_j}.
\eea
We will now show that $G(0)=\frac{\om}{Q}\frac{dQ}{d\om}$, which implies that we should have a monotonically decreasing function $Q$ in terms of $\om$ due to $G(0)\leq 0$. Recalling \eq{h1}, we obtain
\bea{omegasig}
\delta Q = I \delta\omega+ 2\omega\int \sigma \delta \sigma\hspace{10pt} \lr\hspace{10pt} \frac{\omega}{Q}\frac{\partial Q }{\partial \omega}=1+\frac{2\omega}{I}\int \sigma \frac{\partial \sigma}{\partial \omega}.
\eea
Differentiating \eq{QBeq} with respect to $\omega$, we obtain
\be\label{ntr}
\hat{h}_R \frac{\partial \sigma}{\partial \omega}=2\omega \sigma.
\ee
Multiplying ${\sum_i} \sigma_i$ on the eigen-equation $\hat{h}_R\psi_i=\lambda_i\psi_i$, we then obtain
\be\label{sig}
{\sum_i}\hat{h}_R\frac{\sigma_i}{\lambda_i}\psi_i=\sigma.
\ee
By comparing \eq{ntr} with \eq{sig}, we find $\frac{1}{2\omega}\frac{\partial\sigma}{\partial \omega}\propto {\sum_i} \frac{\sigma_i}{\lambda_i}\psi_i$. By multiplying $\sum_j \sigma_j \psi_j$, integrating over space, and using the orthonormal relations, we obtain
\be\label{g1}
\int \sigma\frac{\partial \sigma}{\partial \omega}=2\omega{\sum_i} \frac{\sigma^2_i}{\lambda_i}.
\ee
Using this, we finally obtain from \eq{g0}
\bea{}
	G(0)&=&\textrm{RHS in \eq{omegasig}},\\
	&=&\frac{\omega}{Q}\frac{\partial Q}{\partial \omega},
\eea
Hence, the charges $Q(\omega)$ of $Q$-ball should decrease in terms of $\omega$ to satisfy \eq{neg0}. It follows that the condition, \eq{CLS}, namely
\be\label{cls}
\frac{\omega}{Q}\frac{dQ}{d\omega}\le 0,
\ee
ensures that the eigenvalues in \eq{VR} are all positive or zero. It will turn out that \eq{cls} corresponds to the classical stability condition for $Q$-ball solutions as we will see in the next section. 

To sum up, we showed that the eigenvalues in \eqs{VR}{VI} are all positive or zero subject to the condition \eq{cls}.

\section{Harmonic oscillations}
In this section, we show that the perturbations are of the usual harmonic oscillator form, which implies that all of the frequencies of the perturbations should be real. In order to show this, we have to be concerned with the kinetic term in the second-order Hamiltonian $H_2$ in \eq{H2} and consider the Hamiltonian equations for $H_2$. Let us construct the eigenfunctions excluding the zero eigenfunctions, which we found in the previous sections. We then obtain
\be
\chi_R={\sum_i}^\p q_{Ri}(t)\Psi_{Ri}(\mathbf{x}),\hspace{10pt} \chi_I={\sum_j}^\p q_{Ij}(t) \Psi_{Ij}(\mathbf{x}).
\ee
The canonical variables are
\be
p_R=\left( \begin{array}{c} p_{R1} \\ p_{R2} \\ \vdots \end{array}\right),\hspace{5pt} p_I=\left(\begin{array}{c} p_{I1} \\ p_{I2} \\ \vdots \end{array}\right) ,\hspace{5pt}
q_R=\left(\begin{array}{c} q_{R1} \\ q_{R2} \\ \vdots \end{array}\right),\hspace{5pt} q_I=\left(\begin{array}{c}q_{I1} \\ q_{I2} \\ \vdots \end{array}\right).
\ee
We then express \eq{mVR} and \eq{mVI} as 
\be
\mV_R=\half \tilde{q}_R \Lambda_R q_R,\hspace{20pt} \mV_I=\half \tilde{q}_I \Lambda_I q_I,
\ee
where $\Lambda_R$ and $\Lambda_I$ are diagonal matrices.
Hence, $H_2$ in \eq{H2} becomes
\bea{}
\nb H_2=\half (\tilde{p}_R p_R + \tilde{p}_I p_I) &+& \half \tilde{q}_R(\Lambda_R + \Gamma \tilde{\Gamma})q_R
+\half \tilde{q}_I(\Lambda_I +  \tilde{\Gamma} \Gamma)q_I\\
 &+& \tilde{p}_R \Gamma q_I - \tilde{p}_I \tilde{\Gamma}q_R,
\eea 
where $\Gamma$ is a real matrix whose components are
\be
\Gamma_{ij}=-\omega \int \Psi_{Ri} \Psi_{Ij}.
\ee
Introducing column vectors, $\mP \equiv \left( \begin{array}{c} p_R \\ p_I \end{array}\right)
$ and $\mQ  \equiv \left( \begin{array}{c} q_R \\ q_I \end{array}\right)$, we obtain the second-order Hamiltonian,

\bea{}
H_2&=&\half \tilde{\mP}\mP + \half \tilde{\mQ} \Delta \mQ + \tilde{\mP} \Xi \mQ
   = \half \widetilde{\left( \begin{array}{c} \mP \\ \mQ \end{array} \right)}
             \left(\begin{array}{cc} 1 & \Xi \\ \tilde{\Xi} & \Delta \end{array} 
             \right)
             \left( \begin{array}{c} \mP \\ \mQ \end{array}   \right),\\
&=&\half \tilde{\eta}\left(\begin{array}{cc} 1 & \Xi \\ \tilde{\Xi} & \Delta \end{array} 
 \right)\eta,
\eea
where we set $\Delta=\left( \begin{array}{cc} \Lambda_R+\Gamma\tilde{\Gamma} & 0 \\
0 & \Lambda_I+\tilde{\Gamma}\Gamma\\ \end{array}
 \right)$, $\Xi=-\tilde{\Xi}=\left( \begin{array}{cc} 0 & \Gamma \\ -\tilde{\Gamma} & 0\end{array}
\right)$, and $\eta=\left(\begin{array}{c} \mP \\ \mQ \end{array}\right)$. Hence, the Hamiltonian equation for $H_2$ is
\be\label{hameta}
\dd{\eta}{t}=\left(\begin{array}{cc} \Xi & -\Delta \\ 1 & \Xi \end{array}
  \right) \eta.
\ee
By imposing the normal mode solution $\eta=\eta_N$,
\be
\eta_N(t)\propto e^{-i\Omega_N t},
\ee
we will show that there exist only real solutions for $\Omega_N$ subject to the condition \eq{cls}. The solutions $\Omega_N$ in \eq{hameta} are the roots of the following quadratic equation,
\be\label{omg}
-c_1+c_2\Omega_N+\Omega^2_N=0.
\ee
Here, $c_1$ and $c_2$ are given by
\be
c_1=\mR^\dag_N \left(\begin{array}{cc}
 \Lambda_R & 0 \\ 0 & \Lambda_I
\end{array}
 \right) \mR_N, \hspace{10pt}
c_2 = \mR^\dag_N \left(\begin{array}{cc}
 0 & -2i\Gamma \\ 2i\tilde{\Gamma} & 0
\end{array}
 \right)\mR_N,
\ee
where $\mR_N$ is the coordinate column vector for $\eta_N$, which satisfies the normalisation condition, $\mR^\dag_N \mR_N=1$. Then, the solutions of \eq{omg} are
\be
\Omega_N=\half \sbset{-c_2\pm (c^2_2+4c_1)^{1/2}}.
\ee
Since $c_1$ is real and positive from \eq{lri} and $c_2$ is real, we obtain real values of $\Omega_N$. Therefore, $Q$-balls are classically stable against the spatial perturbations subject to the condition \eq{cls}.


\chapter{An exact solution}\label{exactsol}

In this appendix we will show that a Gaussian profile is an exact solution of the $Q$-ball equation in \eq{QBeq} with $\Uo=U_{grav}-\half \omega^2\sigma^2$, where $U_{grav}$ is defined in \eq{Ugrav}. Notice that the potential $U_{grav}$ becomes negative for $e^{1/2|K|} M < \sigma$; hence, the system is not bounded from below. The additional contribution from the non-renormalisable term $U_{NR}$ compensates the negative term and supports the existence of $Q$-balls in the system. Although the Gaussian exact solution is no longer a solution for the full potential $U_{grav}+U_{NR}$ in \eq{sugra}, the solution we will obtain here provides hints in suggesting a reasonable ansatz for the thick wall $Q$-ball as we will see in appendix \ref{appxthick}. 

Let us consider the following Gaussian profile:
\be\label{gauss}
\sigma_{sol}(r)=\rho_{\omega}\ \exp\bset{-\frac{|K|m^2 r^2}{2}},
\ee
where we will see that $m,\; M,$ and $|K|$ are the same parameters as in \eq{sugra} and $\rho_{\omega}$ will be shortly determined in terms of the underlying parameters. By substituting \eq{gauss} into the left-hand side of 
\eq{QBeq} it leads to 
\be\label{potgauss}
U_{grav}=\frac{m^2}{2}\sigma^2\bset{1-|K|\ln\bset{\frac{\sigma}{M}}^2}
\ee
and 
\be\label{rhom}
\rho_{\omega}= M \exp\bset{\frac{D-1}{2}+ \frac{\mo^2}{2 |K| m^2}},
\ee
where we set the integration constant as zero. Recall $\mo^2 \equiv m^2-\omega^2$. Note that the constant $M$ has the same mass dimension, $(D-1)/2$, as $\sigma$ so that the only physical case is $D=3$. The profile, \eq{gauss}, is an exact solution for $U_{grav}$ with the ``core'' radius $R_Q=\sqrt{2/m^2|K|}$ \cite{Enqvist:1998en}, which is very large compared with $m^{-1}$ for small $|K| \ll \order{1}$, and satisfies the boundary conditions for $Q$-balls, namely $\sigma^{\p}(0)=0=\sigma(\infty)=\sigma^\p(\infty)$, see chapter \ref{ch:chapter2}. In the extreme limit $\omega \gg m$, we obtain $\rho_{\omega}\to 0$ {for $|K|\lesssim \order{1}$} which implies $\sigma_0\equiv \sigma(0) \to 0$. For large $\sigma$, the potential becomes asymptotically flat, tending towards an infinite negative value. By adding the non-renormalisable term $U_{NR}$, the potential $U_{grav}$ is lifted for large $\sigma$ in \eq{sugra}, then the full potential $U_{grav}+U_{NR}$ is bounded from below, see Sec. \ref{exstQmat}. We can see the ansatz given in \cite{Enqvist:1998en} corresponds to the case where $\rho_\omega \simeq M$, which is valid only for $|K| \ll \order{1}$ and $\omega \simeq m$, see \eq{rhom}.
%
%
%

\chapter{Gaussian ansatz in gravity-mediated potentials}\label{appxthick}

In this appendix, we will investigate the thick-wall $Q$-ball in gravity-mediated models by introducing a Gaussian ansatz and keeping all terms in \eq{repsugra} as opposed to the analysis in Sec.\ref{thickgrav}. By using this profile we can perform the Gaussian integrations, and will obtain the generalised results of \eqs{grvqeq}{grvclscond} in Sec. \ref{thickgrav}. The test profile  for the case, $\omega \gtrsim \order{m}$, coincides with the solution $\sigma_{sol}$ in \eq{gauss}, which implies that the nonrenormalisable term $U_{NR}$ in \eq{sugra} is negligible.

To recap, the notation we have adopted in \eq{repsugra} is $\tisig=\sigma/M,\; \tiom=\omega/m$, $\beta^2$ is defined in \eq{betani} and we are considering the case of $n>2$. To begin with we introduce a Gaussian ansatz inspired by \eq{gauss} for the potential \eq{repsugra}
\be\label{testgauss}
\tisig(r)=\lamom \exp(-\alom^2 r^2/2),
\ee
where $\tisig_0 \equiv \tisig(0)=\lamom={\rm finite}$, and $\lamom,\; \alom$ will be functions of $\omega$ implicitly. $\lamom$ should not be confused with the coupling constant $\lambda$ in \eq{sugra}. Both $\lamom$ and $\alom$ can be determined by extremising the Euclidean action $\So$; hence, the actual free parameter here will be only $\omega$. It is crucial to note that $\lamom$ cannot be infinite in the thick-wall limit since we know that $\lamom$ is finite and tending to 0. If the nonrenormalisable term $U_{NR}$ is negligible, we can expect $\lamom \sim \rho_\omega/M \sim \tisig_-(\omega)$ and $\alom^2 \sim |K|m^2$ due to \eq{gauss}, which implies that the ``core'' radius $R_Q$ of the thick-wall $Q$-ball is $R_Q\sim\sqrt{2/m^2|K|}$. For the extreme thick-wall limit $\omega \gg m$, we shall also confirm $\lamom \to 0$, which means $\tisig_0 \to 0$.
 
By substituting \eq{testgauss} into \eq{Uo} with the potential \eq{repsugra}, we obtain $Q$ and $\So$ using the following Gaussian integrations: $\Omega_{D-1}\int^\infty_0 dr r^{D-1} e^{-k r^2}=\bset{\frac{\pi}{k}}^{D/2}$ for real $k$ where $\Omega_{D-1}\equiv \frac{2\pi^{D/2}}{\Gamma(D/2)}$. Thus,
\bea{qapx}
Q&=& M^2 \pi^{D/2} \omega \lamom^2 \alom^{-D}, \\
\label{soapx}\So&=&M^2 \pi^{D/2}\alom^{-D} \sbset{A(\alom,\; \lamom) + B(\omega,\; \lamom) + C(\lamom)},\\
\nb \textrm{where} \hspace*{10pt} A(\alom,\; \lamom)&\equiv& \frac{D\lamom^2 }{4}(\alom^2+|K|m^2),\\
\nb  B(\omega,\; \lamom)&\equiv& \frac{m^2\lamom^2}{2}\bset{1-\frac{\omega^2}{m^2}- 2|K|\ln\lamom},\\ 
\label{ABC} C(\lamom)&\equiv&m^2 \beta^2 \lamom^{n}\bset{\frac{2}{n}}^{D/2}.
\eea
Notice that $A(\alom,\; \lamom)$ comes from the gradient term and the logarithmic term in $\So$ and depends on both $\alom$ and $\lamom$. Similarly, $B(\omega,\; \lamom)$ is given by the quadratic term in the potential \eq{repsugra} and depends both on $\lamom$ and explicitly on $\omega$, whereas $C(\lamom)$ arises simply from the nonrenormalisable term in the potential. An alternative (but in this case more complicated) approach to obtain $Q$ would be the use of Legendre transformations in \eq{easycalc}.

By extremising $\So$ in terms of the two free parameters $\alom$ and $\lamom$: 
\be\label{ext}
\frac{\partial \So}{\partial \alom}=0,\hspace*{15pt} \frac{\partial \So}{\partial \lamom}=0,
\ee
we obtain 
\be\label{abc}
A+B+C=\frac{\lamom^2 \alom^2}{2},\hspace*{15pt} A+B+\frac{nC}{2}=\frac{m^2\lamom^2|K|}{2},
\ee
which implies that
\be\label{al2}
\frac{\alom^2}{m^2} = |K|-(n-2)\beta^2 \lamom^{n-2} \bset{\frac{2}{n}}^{D/2} \ge 0,
\ee
where we have eliminated the $A+B$ terms in the two expressions of \eq{abc}. Using \eq{al2} and the second expression of \eq{abc}, we obtain the relations between $\omega$ and $\lamom$
\bea{lam2}
\frac{\omega^2}{m^2}&=& 1 + |K|\bset{D-1- 2\ln\lamom} + \frac{2(n+D)-nD}{2}\beta^2 \lamom^{n-2}\bset{\frac{2}{n}}^{D/2},\\
 \label{lam4}   &\sim&
    \left\{
    \begin{array}{ll}
    1+|K|(D-1-2\ln\lamom) &\; \textrm{for}\ |K| \sim \order{1},  \\
    1-2|K|\ln\lamom  &\; \textrm{for}\ |K|<\order{1},
    \end{array}
    \right.\\
\label{lam3} \frac{d \lamom}{d\omega}&=& -\frac{\lamom \omega}{|K|m^2 F}\sim - \frac{\lamom \omega}{|K|m^2} <0,
\eea
where we have differentiated \eq{lam2} with respect to $\omega$ to obtain \eq{lam3} and have defined $F$ as $F\equiv 1 - (n-2)\frac{2(n+D)-nD}{4}\frac{\beta^2}{|K|} \lamom^{n-2} \bset{\frac{2}{n}}^{D/2}= 1+ \frac{2(n+D)-nD}{4|K|m^2}\bset{\alom^2 -m^2|K|}$. Equations (\ref{al2}, \ref{lam2}) imply that both $\alom$ and $\lamom$ are functions of $\omega$; however, these are not solvable in closed forms unless the particular limits, which were introduced in Sect.\ref{thickgrav}, are taken, as we will now show. Comparing \eqs{lam2}{lam3} with \eqs{thckom1}{thckom3}, we can see an extra contribution of $\order{|K|}$ in \eq{lam2}, which is not present in \eq{thckom1}. This difference of $(D-1)|K|$ arises because in  calculating \eq{lam2} we have used $\lamom$, whereas we have used $\tisig_-(\omega)$ in obtaining \eq{thckom1}, and although related they are not precisely the same. In the extreme thick-wall limit $\omega \gg m$, and from \eq{lam2} this implies $\lamom \to 0^+$ (recall from \eq{testgauss} that $\lamom$ has to remain finite). Considering the nonrenormalisable term in \eq{lam2}, the fact that $\beta^2 \lesssim |K|\lesssim \order{1}$ and $\lamom \to 0^+$ with $n>2$, implies that this term is subdominant and can be ignored. 
As long as $\lamom < \order{1}$, then $F\sim 1$ and the second relation of \eq{lam3} follows, which implies that $\lamom$ is a monotically decreasing function in terms of $\omega$. The limit $\lamom \sim \order{1}$ corresponds to $\omega \gtrsim \order{m}$, see \eq{lam2}. We will call this the  ``moderate limit'' and represent it by  '$\sim$'. The other case, $\omega \gg m$ (or equivalently $\lamom \ll \order{1}$), we shall call the ``extreme limit'' and represent it by '$\to$'. Depending on the logarithmic strength of $|K|$, we can obtain \eq{lam4}, which leads to the approximate expressions for $\lamom$ and can also obtain $\alom$ from \eq{al2}
\be\label{predict}
    \lamom \sim
    \left\{
    \begin{array}{ll}
    \rho_\omega/M &\; \textrm{for}\ |K| \sim \order{1} \\
    \tisig_-(\omega)  &\; \textrm{for}\ |K|<\order{1}
    \end{array}
    \right.
    \to 0;\; \;
\frac{\alom^2}{m^2} \sim |K| \to |K| \; \textrm{for}\ |K| \lesssim \order{1},
\ee
where $\alom$ is independent of $\omega$ in both the ``moderate'' and ``extreme'' limits.

Using \eqs{qapx}{soapx} and \eq{abc}, we obtain the characteristic slope in both the ``moderate'' and ``extreme'' limits,
\be\label{grvch}
 \frac{E_Q}{\omega Q}= 1+ \frac{\alom^2}{2\omega^2} \sim 1 + \frac{m^2|K|}{2\omega^2} \to 1.
\ee
In order to show their classical stability, we shall differentiate $Q$ with respect to $\omega$ using \eqs{al2}{lam2} and \eq{lam3}:
\bea{}
\nonumber\frac{\omega}{Q}\frac{dQ}{d\omega}&=&1 - \frac{2\omega^2}{m^2|K|F} \sbset{1-\frac{D(n-2)}{4\alom^2}\bset{\alom^2-m^2|K|} },\\
\label{grvclscond2} &\sim& 1- \frac{2\omega^2}{m^2|K|} \to -\frac{2\omega^2}{m^2|K|}<0,\\
\nonumber \frac{d}{d\omega}\bset{\frac{E_Q}{Q}}&=&1-\frac{1}{2\omega^2}\sbset{\alom^2 + \frac{(n-2)\omega^2}{m^2|K|F}\bset{\alom^2-m^2|K|}},\\ 
\label{grvclscond3} &\sim& 1-\frac{m^2|K|}{2\omega^2} \to 1 >0,
\eea
where we have taken the ``moderate limit'' and ``extreme limit'' and used $\alom^2 \sim m^2 |K|$ and $F=1+\frac{2(n+D)-nD}{4|K|m^2}\bset{\alom^2 -m^2|K|}\sim 1$. The classical stability condition \eq{grvclscond2} is consistent with \eq{grvclscond3}, and is consistent with \eqs{CLS}{chslope}. This is different from the result we obtained for the polynomial potentials [see \eq{gausscls} in chapter \ref{ch:qpots}], because in that case the Gaussian ansatz does not give the exact solution unlike here in  \eq{testgauss} where it does become the exact solution \eq{gauss} in both limits. The results, \eqs{grvch}{grvclscond2} and \eq{grvclscond3}, in both the ``moderate'' and ``extreme'' limits recover the key results, \eqs{grvqeq}{grvclscond}, and are independent of $D$; hence, the thick-wall $Q$-balls for all $D$ have similar properties. We can also see the small additional effects arising from the nonrenormalisable term in \eqs{grvclscond2}{grvclscond3}.

Let us summarise the important results we found in this appendix. By introducing a Gaussian test profile \eq{testgauss} inspired by the exact solution \eq{gauss} for $U_{grav}$, we computed the Euclidean action $\So$ and the charge $Q$ using Gaussian integrations. Then, we extremised $\So$ in terms of $\lamom$ and $\alom$ in \eq{ext}, which gave the relations of both $\lamom$ and $\alom$ as a function of $\omega$. By introducing two limits called ``moderate limit'' and ``extreme limit'', we confirmed that the ansatz, \eq{testgauss}, approaches \eq{gauss} in the ``moderate limit''. We established that the results \eqs{grvch}{grvclscond2} and \eq{grvclscond3} recovered the previous results in \eqs{grvqeq}{grvclscond} which are obtained simply by reparametrising in $\So$ and extracting the explicit $\omega$-dependence from the integral in $\So$ with $U=U_{grav}$ where the nonrenormalisable term was neglected at the beginning of the analysis by applying L'H\^{o}pital rules.

In addition, we would like to emphasise the main differences between our work and other earlier analyses in the literature \cite{Enqvist:1998en, Morris:1978ca}. The analytical framework adopted in  \cite{Morris:1978ca} is valid only for $|K|=1,\; D=3,\; n=4$. Our work has shown that this can be generalised to arbitrary integer values of $D$ and $n (>2)$ under the conditions $\beta^2 \lesssim |K|\lesssim \order{1}$, and that the thick-wall $Q$-ball can be classically stable. In Sect.\ref{thickgrav}, we also found that the thick-wall $Q$-ball may be absolutely stable under certain additional conditions, \eq{thckabs3}. Furthermore, Enqvist and McDonald in \cite{Enqvist:1998en} analytically obtained the same ``core'' size of thick-wall $Q$-balls, although they obtained a slightly different value for $E_Q/Q$ (see their Eq. (112)). The reason for this is because their ansatz assumed $\lamom \simeq 1$ in \eq{testgauss} by simply neglecting the nonrenormalisable term, which implies that the third term of $B(\omega,\; \lamom)$ and term $C(\lamom)$ in \eq{ABC} are absent. Hence, their analysis is valid for $|K| \ll \order{1}$ and $\omega \simeq m$, see \eq{rhom}. We, however, have kept all the terms in \eq{repsugra} and used a more general ansatz, which can be applied for $|K| \lesssim \order{1}$ and $\omega \gtrsim \order{m}$ with the restricted coupling constant of the nonrenormalisable term $\beta^2 \lesssim |K|$. In summary, in this appendix we have  
extensively investigated analytically  both the absolute and classical stability of $Q$-balls in \eq{grvch} and \eq{grvclscond2}.

%
%

\chapter{Perturbations on multiple scalar fields}\label{MULTI}

In this appendix we obtain Euler-Lagrange equations for multiple scalar fields $\hat\varphi^a$ with a symmetric field space metric $G_{ab}(\hat\varphi)=G_{ba}(\hat\varphi)$, following the notations in \cite{Sasaki:1995aw, Weinberg:2008zzc}. Our aim is to obtain equations of motion for the background homogeneous (zero-mode) fields $\varphi^{a}(t)$ and the perturbed fields $\delta\varphi^a(t,\mathbf{x})$ in a fixed unperturbed background (Friedmann-Robertson-Walker) metric, $g_{\mu\nu}=\textrm{diag}(-1,\ a(t),\ a(t),\ a(t))$, where $a(t)$ is the scale factor and $H=\dot{a}/a$ is the Hubble parameter. Here, an over-dot denotes the time-derivative. As the simplest nontrivial example of the multiple scalar fields, we find equations of motion for a complex scalar field $\hat{\phi}\equiv\hat{\sigma} e^{i\hat{\theta}}$ where $\hat{\sigma}$ and $\hat{\theta}$ are real scalar fields and the system possesses a U(1) symmetry.

Let us start off with the following action
\be\label{ext2}
S=\int d^4x \sqrt{-g}\bset{-\half g^{\mu\nu}G_{ab}(\hat{\varphi})\partial_\mu \hat\varphi^a \partial_\nu, \hat{\varphi}^b-V(\hat\varphi)},
\ee
where $g\equiv \textrm{det}\bset{g_{\mu\nu}}$ and $V(\hat\varphi)$ is a potential for $\hat{\varphi}$.
By applying the action principle, we obtain the Euler-Lagrange equation for $\hat{\varphi}$
\be\label{ELmulti}
\frac{1}{\sqrt{-g}}\partial_\rho\bset{\sqrt{-g}g^{\rho\nu}G_{cb}\partial_\nu \hat\varphi^b}=\half g^{\mu\nu}G_{ab,c}\partial_\mu \hat\varphi^a \partial_\nu \hat\varphi^b + V_{,c},
\ee
and the energy momentum tensor
\be\label{EGMmulti}
T_{\mu\nu}=G_{ab}\partial_\mu \hat\varphi^a \partial_\nu \hat\varphi^b + g_{\mu\nu}\sbset{-\half g^{\rho\sigma}G_{ab}\partial_\rho\hat\varphi^a\partial_\sigma\hat\varphi^b - V(\hat\varphi)}.
\ee
Here, we defined $G_{ab,c}\equiv \frac{dG_{ab}}{d\hat\varphi^c}$, and so on. The energy density and pressure can be given by $T_{\mu\nu}$ \cite{Weinberg:2008zzc}
\bea{energy-multi}
\rho_E&=&-\half g^{\mu\nu} G_{ab} \partial_\mu \varphi^a \partial_\nu \varphi^b + V(\varphi),\\
\label{press-multi} p&=& -\half g^{\mu\nu} G_{ab} \partial_\mu \varphi^a \partial_\nu \varphi^b - V(\varphi).
\eea
By pertubing the fields as $\hat\varphi^a=\varphi^a(t)+\delta\varphi^a(t,\mathbf{x})$ where $|\varphi|\gg |\delta\varphi|$, the homogeneous part gives, from \eq{ELmulti},
\be\label{homovp}
\frac{D}{dt}\dot{\varphi}^a+3H \dot{\varphi}^a +G^{ab}V_{,b}=0,
\ee
where the covariant derivative, $D/dt$, can be defined by the ``Christoffel symbols'' $\gamma^{a}_{bc}\equiv\half G^{ad}\times$ $\bset{G_{dc,b}+G_{db,c}-G_{bc,d}}$; thus, $\frac{D}{dt}\dot{\varphi}^a\equiv \frac{d}{dt}\dot{\varphi}^a + \gamma^{a}_{bc}\dot{\varphi}^b\dot{\varphi}^c$.
On the other hand, we can obtain the equations of motion for the pertubed fields $\delta \varphi$ from \eq{ELmulti}
\be\label{perturb}
\frac{D^2}{dt^2}\delta\varphi^a+3H\frac{D}{dt}\delta\varphi^a-\bset{\frac{\nabla}{a}}^2\delta\varphi^a-\gamma^{a}_{bcd} \dot{\varphi}^b\dot{\varphi}^c\delta\varphi^d+(V^{;a})_{;d}\delta\varphi^d=G^{ab}G_{bc,d}G^{ce}V_{,e}\delta\varphi^d,
\ee
where we used $\frac{D}{dt}\delta\varphi^a = \delta\dot{\varphi}^a+\gamma^a_{bc}\dot\varphi^b\delta\varphi^c$, defined the ``Riemann tensors'' as $\gamma^a_{bcd}\equiv \gamma^a_{bd,c}-\gamma^a_{bc,d}+\gamma^a_{ce}\gamma^e_{bd}-\gamma^a_{de}\gamma^e_{bc}$, and denoted the covariant derivative as the usual notion $';'$. Notice that we used $V_{,b}\equiv \frac{\partial V}{\partial\hat\varphi^b}(\hat\varphi)\simeq \left.\frac{\partial V}{\partial \hat\varphi^b}(\hat\varphi)\right|_{\varphi}+\delta\varphi^c \left.\frac{\partial^2 V}{\partial \hat\varphi^b\partial \hat\varphi^c}\right|_{\varphi}+\dots$.

When the system has a $O(2)\sim U(1)$ symmetry for $\hat\varphi_a=\bset{\hat\sigma, \hspace*{5pt} \hat\theta}$ and a flat field metric is $G_{ab}=\textrm{diag}(1,\; \hat\sigma^2)$, we can obtain $\gamma^1_{22}=-\hat\sigma;\, \gamma^2_{12}=\gamma^2_{21}=1/\hat\sigma$. We then induce \eq{homovp} with a potential $V(\sigma)$ to
\bea{sigeom}
\ddot{\sigma}+3H\dot{\sigma}-\sigma\dot{\theta}^2+\frac{dV}{d\sigma}&=&0,\\
\label{thetaeom} \ddot{\theta}+3H\dot{\theta}+\frac{2}{\sigma}\dot{\sigma}\dot{\theta}&=&0.
\eea
Here, the third term in \eq{sigeom} corresponds to ``centrifugal force'' due to a spin in the field space, and the third term in \eq{thetaeom} corresponds to the ``Colliori force''. In addition, the energy density and pressure can be given by from \eqs{energy-multi}{press-multi}
\be\label{epcom}
\rho_E=\half\bset{\dot{\sigma}^2+\sigma^2\dot{\theta}^2}+V,\spc p=\half\bset{\dot{\sigma}^2+\sigma^2\dot{\theta}^2}-V.
\ee
Furthermore, \eq{perturb} gives
\bea{pertsig}
&&\ddot{\dsig}+3H\dot{\dsig}-\bset{\bset{\frac{\nabla}{a}}^2+\dot{\theta}^2-\frac{d^2V}{d\sigma^2}}\dsig-2\sigma\dot{\theta}\dot{\dtheta}=0,\\
\label{pertth}&&\ddot{\dtheta}+\bset{3H+\frac{2\dot{\sigma}}{\sigma}}\dot{\dtheta}-\bset{\frac{\nabla}{a}}^2\dtheta+\frac{2\dot{\theta}}{\sigma^2}\bset{\sigma\dot{\dsig}-\dot{\sigma}\dsig}=0.
\eea
We use \eqs{sigeom}{thetaeom} to concern the orbits of AD condensates in Sec. \ref{sectorbit}, and use \eqs{pertsig}{pertth} to investigate the linear spatial instability of the condensates in Sec. \ref{sectinst}.  		
\chapter{The orbit of an Affleck-Dine ``planet''}\label{App2}

In this appendix, we obtain an exact orbit form in a quadratic potential case when the Hubble expansion is assumed to be small and adiabatic. The orbit of an AD field, or more precisely an eccentricity of the elliptic motion in the complex field-space, is determined by the initial charge and energy density. In order to obtain analytic expressions of the orbit in more general potential cases in which we are more interested, we restrict ourself to work in Minkowski spacetime and on the orbit which should be nearly circular. We then obtain the perturbed orbit equation and necessary conditions for closed orbits, where the orbits come back to their original positions after some rotations around the minimum of the effective potential. By including the effects of the Hubble expansion, in Sec. \ref{numAD} we shall introduce ans\"{a}tze, which are inspired by our solutions obtained in Minkowski spacetime. Our numerical results support the ans\"{a}tze, assuming that the rotation frequency $W$ is always much greater than the Hubble expansion $H$ \cite{Turner:1983he}.

\section{The exact orbit in an expanding universe}\label{sec2}
The exact orbit expressions of an AD field in an expanding universe can be obtained with a quadratic potential,
\be\label{quad}
V=\frac{M^2}{2}\sigma^2=\frac{M^2}{2}\bset{\frac{a_0}{a}}^3\tisig^2,
\ee
where $M$ is a mass of the field $\phi$ and we have rescaled the field $\sigma$, $\sigma(t)=\bset{\frac{a_0}{a(t)}}^{3/2}\tisig(t)$. From now on, we solve the orbit equations, \eq{modrad}, for $\tisig(t)$ at first, and then solve them for $\tiu(\theta)$, replacing the time-dependence in $\tisig(t)$ by a phase variable $\theta$. We then show that the orbits for $\tisig(t)$ and $\tiu(\theta)$ are of the usual elliptic forms with a third eccentricity $\varepsilon^2$.

\subsection{The orbit for $\tisig(t)$}

In this subsection we obtain an expression for the orbit $\tisig(t)$ with the quadratic potential \eq{quad} by solving \eq{modrad}. Substituting \eq{quad} into \eq{modrad} and ignoring the terms involving $H^2$ and $\ddot{a}/a$, we obtain
\be\label{radquad}
\ddot{\tisig} -\frac{\tirhoQ^2}{\tisig^3}+M^2\tisig=0 \spc \lr \spc \frac{d\tirhoE}{dt}=0,
\ee
where $\tirhoE\equiv \half \bset{\frac{d\tisig}{dt}}^2+\half M^2 \tisig^2+\frac{\tirhoQ^2}{2\tisig^2}\neq a^{-3}_0 \rho_E$, which is approximately conserved. Since $\half \frac{d^2}{dt^2}(\tisig^2)=\dot{\tisig}^2+\tisig\ddot{\tisig}=2\tirhoE-2M^2\tisig^2$, \eq{radquad} leads to a harmonic oscillator form,
\be\label{ext6}
\frac{d^2}{dt^2}(\tisig^2)=-4M^2\bset{\tisig^2-\frac{\tirhoE}{M^2}}
\ee
whose solution is
\bea{tisigsol}
\tisig^2(t)&=&\frac{\tirhoE}{M^2}+ A \cos\sbset{2M(t+t_0)},\\
\label{tisig}&=&\frac{\tirhoE}{M^2}\bset{1+\varepsilon^2 \cos\sbset{2M(t+t_0)} }.
\eea
Here, $B$ is some constant value and we set $t_0$ as a time when the AD field starts to rotate. We have also defined a third eccentricity $\varepsilon^2 \equiv \frac{AM^2}{\tirhoE}= \frac{\tisig^2_{max}-\tisig^2_{min}}{\tisig^2_{max}+\tisig^2_{min}}$, where the apocentral and pericentral points are, respectively, given by $\tisig^2_{max}\equiv\frac{\tirhoE}{M^2}+A$ and $\tisig^2_{min}\equiv\frac{\tirhoE}{M^2}-A$.
Notice that the circular orbit case corresponds to $\varepsilon^2=0$, which implies that $\tisig^2_{max}=\tisig^2_{min}$, and also note that the eccentricity is real and has a value between 0 and 1 in the present quadratic potential \footnote{In an inverse-squared central force, the first eccentricity can be larger than equal 1, which corresponds to the cases where the orbits are parabola or hyperbola.}.

We can obtain the period $\tau$ of this orbit,
\be\label{period}
\tau=\frac{\pi}{M}.
\ee
Substituting \eq{tisigsol} into $\tirhoE$, we obtain $A=\frac{\sqrt{\tirhoE^2-M^2\tirhoQ^2}}{M^2}$. From the above expressions for $\varepsilon^2$ and $A$, we can obtain $\frac{M\tirhoQ}{\tirhoE}=\sqrt{1-\varepsilon^4}$. Using this and \eq{tisigsol}, it ends up with
\be\label{dotth}
\dot{\theta}(t)=\frac{\tirhoQ}{\tisig^2}=\frac{M\sqrt{1-\varepsilon^4}}{1+\varepsilon^2\cos\sbset{2M(t+t_0)}}.
\ee
For the circular orbits with $\varepsilon^2=0$, $\dot{\theta}$ is time-independent as we can expect, and the ratio, $\tirhoE/(M\tirhoQ)$, is unity. While for the radial orbits with $\varepsilon^2=1$, we obtain $\dot{\theta}=0$ and $\tirhoE/(M\tirhoQ)\gg 1$ as expected.
\subsection{The orbit for $\tiu(\theta)=\tisig^{-1}(\theta)$}

What follows is that we express $\tisig(t)$ as a function of $\theta$ by using the second expression in \eq{modrad} and \eq{tisigsol}. We then obtain
\be\label{theta}
\tan(\theta-\theta_0)=\frac{\tisig_{min}}{\tisig_{max}}\tan\bset{M(t+t_0)},
\ee
where $\theta_0$ is an integration constant and we used the following integral formula, $\int \frac{dx}{a_1+a_2\cos{x}}=\frac{2}{\sqrt{a^2_1-a^2_2}}\times$\\ $\textrm{Arctan}\bset{\frac{(a_1-a_2)\tan(\frac{x}{2})}{\sqrt{a^2_1-a^2_2}}}$ with some real values $a_1$ and $a_2$. Without loss of generality, we can choose $t_0=\theta_0=0$, which implies that the orbit at $t=0,\; \tau/2$ gives, respectively, $\theta=0,\; \pi/2$, recalling \eq{period}. By comparing \eq{tisigsol} to \eq{theta}, we obtain 
\bea{tisigsol2}
\tisig^2(\theta)&=&\frac{\tisig^2_{max}\tisig^2_{min}}{\tisig^2_{min}\cos^2\theta+\tisig^2_{max}\sin^2\theta},\\
\label{uel1}\lr \tiu^2(\theta)&=&\frac{1}{\tisig^2}=\frac{\cos^2\theta}{\tisig^2_{max}}+\frac{\sin^2\theta}{\tisig^2_{min}},\\
\label{uel} &=& \frac{\tisig^2_{max}+\tisig^2_{min}}{2\tisig^2_{max}\tisig^2_{min}}\bset{1-\varepsilon^2\cos(2\theta)}.
\eea
Hence, we can see that $\theta=0$ when $\tisig=\tisig_{max}$ and $\theta=\pi/2$ when $\tisig=\tisig_{min}$. Finally, we obtain the expressions for the orbits as the usual elliptic forms in \eqs{tisig}{uel}. For the circular orbits $\varepsilon^2=0$, we can obtain $\tiu^2=const.$ from \eq{uel} as expected.
\section{The nearly circular orbits in Minkowski spacetime}
Without the Hubble expansion, we can investigate a nearly circular bounded orbit of an AD field in general potentials which satisfy \eq{condbound}. For this reason, we concentrate on the case of a non-expanding background in this section, and obtain a time-dependent expression for the nearly circular orbits as in \eq{tisig}. We then find the expression that depends on the phase $\theta$ as in \eq{uel1}. Moreover, we obtain conditions for closed orbits, in which the perturbations are expanded up to 1st order (for the complete proof of the condition up to 4th order, see appendix \ref{BERT} for Bertrand's theorem \cite{Bertrand}).

\subsection{The orbit for $\sigma(t)$}\label{sigorbitMIN}
In Minkowski spacetime, we can find an expression for the orbit $\sigma(t)$ in a general potential $V(\sigma)$ as in \eq{tisig}. Notice that the tilde variables are the same as un-tilde ones in the present non-expanding background. Recall the equation of motion, \eq{radhomo}, in Minkowski spacetime,
\be\label{rad}
\ddot{\sigma}+\frac{dV_+}{d\sigma}=0.
\ee
Suppose that the orbit that is nearly circular as $\sigma(t)=\sigma_{cr}+\delta(t)$ where $\sigma_{cr}\gg |\delta|$, recalling $\sigma_{cr}$ is defined by \eq{sigcr}. Substituting this expression for $\sigma$ into \eq{rad} and keeping $\delta$ terms up to 1st order, we obtain a harmonic oscillator form
\be\label{del}
\ddot{\delta}+W^2 \delta=0,
\ee
where the readers should recall the condition, \eq{condbound}, for the bound orbits, and $W$ is constant since we are working in Minkowski spacetime. 

Thus, the solution of \eq{del} is 
\be\label{deltaeq}
\delta(t)=\sigma_{cr} B\cos(Wt),
\ee 
where $B$ is a small positive dimensionless constant, i.e. $B\ll 1$ due to $\sigma_{cr}\gg |\delta|$, and we have set the differentiation constant to be $0$ to ensure that $\sigma(0)=\sigma_{max}$. We find that $\sigma_{max}=\sigma_{cr}(1+B)$, $\sigma_{min}=\sigma_{cr}(1-B)$, and $\sigma_{max}\sigma_{min}\simeq \sigma^2_{cr}\bset{1+\order{B^2}}$. These give $B=\frac{\sigma_{max}-\sigma_{min}}{\sigma_{max}+\sigma_{min}}$, $\sigma_{cr}=\frac{\sigma_{max}+\sigma_{min}}{2}$, and $2B\simeq \frac{\sigma^2_{max}-\sigma^2_{min}}{\sigma^2_{max}+\sigma^2_{min}}= \varepsilon^2$, where we have used the definition of the third eccentricity. We can check that the condition, $2B\simeq \varepsilon^2\ll 1$, is consistent with the fact that the orbit is nearly circular. Since $\dot{\sigma}_{max}=\dot{\sigma}_{min}=0$ and $\rho_E$ is constant, we can equate $B$ with $\rho_E$ and $\rho_Q$ using \eqs{deltaeq}{condbound}:
\be\label{Avalue}
B=\sqrt{\frac{2(\rho_E-V_+(\sigma_{cr}))}{W^2\sigma^2_{cr}}}=\sigma_{cr}\sqrt{\frac{2(\rho_E-V_+(\sigma_{cr}))}{\left.(\sigma^4V^{\p\p})\right|_{\sigma_{cr}}+3\rho^2_Q}}\simeq \frac{\varepsilon^2}{2} \ll 1,
\ee
where a prime denotes the differentiation with respect to $\sigma$. Finally, we obtain
\be\label{sigsol}
\sigma^2(t) = \sigma^2_{cr}\bset{1+\varepsilon^2\cos(Wt) + \order{\varepsilon^4}},
\ee
where $W$ is given by \eq{condbound} [compare with \eq{tisig}]. Now, we can define the period $\tau$
\be\label{periodgen}
\tau=\frac{2\pi}{W},
\ee
which reproduces \eq{period} as the case with $W=2M$. Using \eqs{phshomo}{sigcr}, we can also find
\be\label{dottheta}
\dot{\theta}\simeq\frac{\sqrt{\left.V^\p/\sigma\right|_{\sigma_{cr}}}}{1+\varepsilon^2\cos{(Wt)}}.
\ee
Using \eq{rhop}, let us compute the pressure of this AD condensate whose orbit is described by \eq{sigsol}. By expanding $V_-(\sigma)$ around $\sigma=\sigma_{cr}$ and using \eq{sigsol}, we obtain $V_-(\sigma)\simeq V_-(\sigma_{cr}) + \frac{\varepsilon^2 \rho^2_Q}{\sigma^2_{cr}} \cos{(Wt)} +\frac{\varepsilon^4\sigma^2_{cr}}{8}\bset{W^2-\frac{6\rho^2_Q}{\sigma^4_{cr}}}\cos^2{(Wt)} + \dots,$ where we have assumed that the higher order terms in $V_-$ are negligible. Therefore,
\bea{}
\nb p &\simeq& \frac{W^2\sigma^2_{cr}\varepsilon^4}{8}\bset{1-2\cos^2{(Wt)}}-V(\sigma_{cr})+\frac{\rho^2_Q}{2\sigma^2_{cr}}\bset{1-2\varepsilon^2\cos{(Wt)}+ \frac{3}{2} \varepsilon^4 \cos^2{(Wt)}  },\\
\label{pressure} \lr\hspace*{5pt} \state{p} &\simeq& -V(\sigma_{cr})+\frac{\rho^2_Q}{2\sigma^2_{cr}}.
\eea
Here, we have defined an averaged value over one rotation in the orbit, \eq{sigsol}, namely $\state{X}\equiv \frac{1}{\tau} \int^{\tau}_0 dt X(t)$ where $X$ is some quasi-periodic quantity and $\tau$ is determined by \eqs{periodgen}{condbound}. The result, \eq{pressure}, can be easily understood by the fact that the averaged pressure corresponds to the value at $\sigma=\sigma_{cr}$ since the orbit oscillate around $\sigma_{cr}$ and $\dot{\sigma}_{cr}=0$, c.f. a real scalar field case \cite{Turner:1983he}. Similarly, we can obtain the averaged energy density 
\be\label{avgengy}
\state{\rho_E} \simeq V(\sigma_{cr})+\frac{\rho^2_Q}{2\sigma^2_{cr}} + \frac{W^2\sigma^2_{cr}\varepsilon^4}{16},
\ee
where we have kept the contribution from the term involving $\varepsilon^4$.
Hence, the averaged equation of state is given by
\be\label{eosg}
\state{w}\equiv \state{\frac{p}{\rho_E}}=\frac{\frac{\rho^2_Q}{2\sigma^2_{cr}}-V(\sigma_{cr})}{\frac{\rho^2_Q}{2\sigma^2_{cr}}+V(\sigma_{cr})+W^2\sigma^2_{cr}\varepsilon^4/16}.
\ee
\subsection{The orbit for $u(\theta)=\sigma^{-1}(\theta)$}\label{uorbitMIN}

In order to obtain a $\theta$-dependent expression of the orbit as \eq{uel1}, let us switch the variable $\sigma$ to $u(\theta)\equiv1/\sigma(\theta)$. In Minkowski spacetime, where we can again drop the tilde variables here, the orbit equation \eq{orbiteq} is
\be\label{orbitMink}
\frac{d^2u}{d\theta^2}+u=-\frac{1}{\rho^2_Q}\frac{dV}{du}\equiv J(u).
\ee
Let $u_0$, which is independent of $\theta$, be the value of a circular orbit (i.e. $u_0\equiv1/\sigma_{cr}$). We then consider an orbit $u(\theta)$ that deviates slightly from $u_0$ with a fluctuation $\eta(\theta)$, i.e. $u=u_0+\eta$, where $u_0\gg |\eta|$. Since $\frac{du_0}{d\theta}=0=\frac{d^2u_0}{d\theta^2}$, \eq{orbitMink} implies that $u_0=J(u_0)$. By expanding $J(u)$ around $u=u_0$, we obtain $J(u)\simeq u_0 + \eta \left.\frac{dJ}{du}\right|_{u_0}+\dots$, where we are evaluating the derivatives at $u_0$. Hence, we can obtain the perturbed orbit equation for $\eta(\theta)$
\be\label{pertorbit}
\frac{d^2\eta}{d\theta^2}+\beta^2\eta=0,
\ee
where $\beta^2\equiv 1-\left.\frac{dJ}{du}\right|_{u_0}$ which should be positive for bounded orbits. Note that this condition, $\beta^2>0$, is equivalent to the previous condition, \eq{condbound}, since 
\be\label{betasq}
\beta^2=\frac{\sigma^4_{cr}}{\rho^2_Q}W^2=\left.\frac{3V^\p+\sigma V^{\p\p}}{V^\p}\right|_{\sigma_{cr}},
\ee
where we used the fact $V^\p=\frac{\rho^2_Q}{\sigma^3}$ at $\sigma=\sigma_{cr}$ from \eq{sigcr}.
The solution of \eq{pertorbit} is 
\be\label{etasol}
\eta=u_0C\cos(\beta\theta+\theta_0),
\ee
where $C$ and $\theta_0$ are constants, and $0<C\ll 1$ due to the fact that $u_0\gg |\eta|$. We can then show $C= B$ by equating the value of $C$ with $\rho_Q$ and $\rho_E$. Substituting $u$ into $\rho_E$ and expanding $V(u)$ around $u=u_0$ up to second order, we can find
\be\label{Cvalue}
C=\frac{1}{u_0}\sqrt{\frac{2(\rho_E-V_+(1/u_0))}{\left.\frac{d^2V(1/u)}{du^2}\right|_{u_0}+\rho^2_Q}}=B\simeq \frac{\varepsilon^2}{2},
\ee
where we used $\left.\frac{dV_+(u)}{du}\right|_{u_0}=\left.\frac{dV(u)}{du}\right|_{u_0}+\rho^2_Qu_0=0$ from \eq{sigcr}. The relation, $C=A$, is obtained by changing the variable $u$ back to $\sigma$ [compare \eq{Cvalue} with \eq{Avalue}].

Let us choose $\theta_0=\pi$ in \eq{etasol}, then we obtain
\bea{uel2}
u&=&u_0\bset{1-C\cos(\beta\theta)},\\ 
\label{uel3} u^2 &\simeq& u^2_0\sbset{1-2C\cos(\beta\theta) +\order{C^2}}.
\eea
Notice that $0<C \ll 1$ which is consistent with the condition for nearly circular orbits $\varepsilon^2\ll 1$, as we have seen in appendix \ref{sigorbitMIN} and \eq{Cvalue}. We can also find that $\sigma_{max}=\frac{\sigma_{cr}}{1-C}$ for $\beta\theta=0$ and $\sigma_{min}=\frac{\sigma_{cr}}{1+C}$ for $\beta\theta=\pi$. 

To show that the orbit $u(\theta)$ in \eq{uel3} has a similar form as \eq{uel}, let us compute the following relations: $\sigma^2_{max}+\sigma^2_{min}\simeq 2\sigma^2_{cr}\bset{1+\order{C^2}},\; \sigma^2_{max}-\sigma^2_{min}\simeq 4\sigma^2_{cr}C\bset{1+\order{C}}$ and $\sigma^2_{max}\sigma^2_{min}\simeq \sigma^4_{cr}\bset{1+\order{C^2}}$. Hence, we obtain $u^2_0\simeq \frac{\sigma^2_{max}+\sigma^2_{min}}{2\sigma^2_{max}\sigma^2_{min}}$ and $2C \simeq \varepsilon^2$, which imply that \eq{uel3} is of similar orbit form as \eq{uel}. As we computed going from \eq{uel1} to \eq{uel}, where for this case we deduce \eq{uel} from \eq{uel1}, we finally obtain
\be\label{ext7}
u^2 \simeq \frac{\cos^2\frac{\beta}{2}\theta}{\sigma^2_{max}}+\frac{\sin^2\frac{\beta}{2}\theta}{\sigma^2_{min}}.
\ee
In the next subsection, we obtain the conditions for closed orbits using \eq{ext7} \cite{Whittaker37}.

\subsection{Conditions for closed orbits and equations of state}\label{Sectclose}

Let us define an angle $\Phi$, which is the phase difference as the orbit goes from $\eta=u_0C$ to $\eta=-u_0C$,
\be\label{Phi}
\Phi \equiv \frac{\pi}{\beta}=\pi \sqrt{\left.\frac{V^\p}{3V^\p+\sigma V^{\p\p}}\right|_{\sigma_{cr}}},
\ee
where we used \eq{betasq}. For closed orbits, the angle must have the value that is $\pi$ multiplied by a rational number, i.e. $\Phi=\pi\frac{r}{q}$ where $q,\; r \in  \mathbb{Z}$; therefore, $\beta$ should be the rational number. In order to obtain the $\sigma$-independent value for $\Phi$, the potentials can be of the forms, $\frac{M^2\sigma^l}{2}(+ const.)$, $m^4_{\phi}\ln\bset{\sigma/m_{\phi}}^2(+ const.)$, and etc. Here, $M$ and $m_{\phi}$ are constant real values, and we should have $l<-2,\; 0<l$ for bound orbits, whereas we may have $-2<l<0$ for bound orbits when $M^2<0$, recalling \eq{condbound}. The constant terms in the potentials add an extra energy for the orbits, and it does not play a significant role, so that we consider the potentials without the constant terms. The former power-law potential case, $V=\frac{M^2\sigma^l}{2}$, gives 
\be\label{powerPhi}
\Phi=\frac{\pi}{\sqrt{l+2}},
\ee
which implies that the closed orbits exist for $l=(-1),\; 2,\; 7,\; \dots$. Using \eqs{pressure}{avgengy} and \eq{condbound}, we obtain 
\be\label{powerpress}
\state{p}\simeq \frac{(l-2)M^2\sigma^l_{cr}}{4}, \spc \state{\rho_E} \simeq \frac{(l+2)M^2\sigma^l_{cr}}{4}, \spc W^2=\frac{l(l+2)M^2\sigma^{l-2}_{cr}}{2}, 
\ee
which implies that the bound orbits of the AD condensate has a negative pressure for $l<2$. In the computation of $\rho_E$, \eq{avgengy}, we safely ignored the $\varepsilon^4$ term. We note that the bound orbits for $l=(-1,)\; 2$ are closed. For the quadratic potential case $l=2$, the averaged pressure is zero, in which the AD condensate corresponds to an example of nonrelativistic cold dark matter \cite{Turner:1983he}. In addition, using \eqs{eosg}{powerpress} we can find
\be\label{compress}
\state{w}\simeq \frac{l-2}{l+2}.
\ee

On the other hand, the latter logarithmic potential case, $m^4_{\phi}\ln\bset{\sigma/m_{\phi}}^2$, leads to 
\be\label{logPhi}
\Phi=\frac{\pi}{\sqrt{2}}\sim \frac{2\pi}{3},
\ee
which corresponds to the former power-law case with $l=0$. Similarly, using \eqs{pressure}{avgengy} and \eq{condbound}, we obtain 
\be\label{presslog}
\state{p}\simeq m^4_{\phi}\bset{1-2\ln{\frac{\sigma_{cr}}{m_{\phi}}}}, \spc \state{\rho_E} \simeq m^4_{\phi}\bset{1+2\ln{\frac{\sigma_{cr}}{m_{\phi}}}}, \spc W^2=\frac{4m^4_{\phi}}{\sigma^2_{cr}},
\ee
which implies that the AD condensate has a negative pressure for $\sigma_{cr}>m_{\phi}\exp{\bset{\half}}$. In the computation of $\rho_E$, \eq{avgengy}, we safely ignored the $\varepsilon^4$ term again. Using \eqs{eosg}{presslog}, we obtain
\be\label{compresslog}
\state{w}\simeq \frac{1-2\ln\bset{\frac{\sigma_{cr}}{m_\phi}}}{1+2\ln\bset{\frac{\sigma_{cr}}{m_\phi}} }.
\ee
In \eq{presslog}, we cannot clearly see the correspondence with the case for $l=0$, but we can find $W^2\simeq 0$ and $\state{w} \simeq -1$ for $m_\phi \ll \sigma_{cr}$ as the case with $l=0$.

\vspace*{10pt}

Let us comment on the pressure when the AD orbit is exactly radial, which corresponds to the zero-charge density case as for real fields \cite{Turner:1983he}. In this case, the field $\sigma(t)$ coherently oscillates around the vacuum if the potential follows a power-law, i.e. $V\propto \sigma^l$ for $l>1$, and $\state{w}$ has the same expression as \eq{compress}, but it gives a negative pressure for $1<l<2$. Note that the lower bound of $l$ ensures to be a coherent oscillation for the radially oscillating AD fields and real scalar fields.

\vspace*{10pt}

In summary, we have obtained analytically the explicit expressions, \eqs{tisig}{uel}, for the orbit of the AD fields in a quadratic potential under an expanding universe, and approximately obtained the elliptic orbit expressions, \eqs{sigsol}{ext7}, for nearly circular orbits in Minkowski spacetime in potentials which satisfy the condition \eq{condbound} for bound orbits.


\chapter{Proof of Bertrand's theorem}\label{BERT}

In Bertrand's theorem \cite{Bertrand}, there are only two allowed potential forms for closed "planetary" orbits: isotropic harmonic force and inverse-squared force. Each of the central forces gives dynamical symmetries: the Fradkin tensor \cite{Fradkin} and Runge-Lenz vector \cite{Laplace-Runge-Lenz, Laplace-Runge-Lenz2, Laplace-Runge-Lenz3, Laplace-Runge-Lenz4, Laplace-Runge-Lenz5, Laplace-Runge-Lenz6}, respectively. These dynamical charges are obtained both classically using the algebra of Poisson bracket \cite{Stehle:1967} and quantum-mechanically using the corresponding Lie algebra in the abelian case \cite{Higgs:1978yy} and the non-abelian case \cite{Leemon:1978yz}. In this appendix we prove Bertrand's theorem with the consistent notations with ones for the orbits in the AD fields obtained in Appendix \ref{App2}. 

In order to show Bertrand's theorem in the abelian case and the Minkowski spacetime \cite{tikochinsky:1073}, we expand an elliptic orbit perturbed from the circular orbit up to 4th order, and show that the allowed values of $\beta^2$ defined in \eq{betasq} are only $1$ and $4$, which correspond to the above two types of potentials, \ie\ isotropic harmonic force and inverse-squared force.

Recalling that $\rho_Q=\sigma^2\dot{\theta}$ and $\rho_E=\half \dot{\sigma}+V_+$, where $V_+(\sigma)\equiv V(\sigma)+\rho^2_Q/2\sigma^2$, we obtain $\dot{\sigma}=\sqrt{2(\rho_E-V_+)}>0$. For the motion of $\sigma(t)$ going from $\sigma_{min}$, through $\sigma_{cr}$, and to $\sigma_{max}$, by recalling \eq{Phi} we can obtain,
\bea{extB1}
2\Phi&=&2\int^{\Phi}_{0}d\theta=\int^{\sigma_{max}}_{\sigma_{min}} d\sigma \frac{2\rho_Q}{\sigma^2\sqrt{2(\rho_E-V_+)}},\\
\label{Phi2} &=& \sqrt{2}\rho_Q\int^{\rho_E}_{V_0}dV_+\frac{f(V_+)}{\sqrt{\rho_E-V_+}},
\eea
where we split the integration into two parts, i.e. $\int^{\sigma_{cr}}_{\sigma_{min}}+\int^{\sigma_{max}}_{\sigma_{cr}}$, and then changed the integration variable from $\sigma$ to $V_+$. Here, we defined $f(V_+) \equiv \frac{d}{dV_+}\bset{\frac{1}{\sigma_1}-\frac{1}{\sigma_2}}$, where $\sigma_1=\sigma_{cr}-x$ and $\sigma_2=\sigma_{cr}+y$, and we assumed that the orbit is ``nearly'' circular, i.e. $\sigma_{cr}\gg x,\ y >0$. Recall $\Phi=\pi/\beta$ in \eq{Phi} where $\beta$ should be a rational number for closed orbits as we found in the linear perturbation analysis in Appendix \ref{Sectclose}. Since $f(V_+)$ is an Abel's equation \cite{Landau}, by multiplying $1/\sqrt{\bar{V}_+-\rho_E}$ on both sides of \eq{Phi2}, where $\bar{V}_+$ is some value of $V_+$, and then by integrating it over $\rho_E$ from $V(\sigma_{cr})$ to $\bar{V}_+$, we obtain 
\be\label{bertsub}
\frac{1}{\sigma_1(V_+)}-\frac{1}{\sigma_2(V_+)}=\frac{2\sqrt{2}}{\beta \rho_Q}\sqrt{V_+-V(\sigma_{cr})},
\ee
where we changed the order of the integrations, used the formula $\int\frac{dy}{\sqrt{(y-a_1)(a_2-y)}}=2Arctan\bset{\sqrt{\frac{y-a_1}{a_2-y}}}$, and finally replaced the variable $\bar{V}_+$ by $V_+$. By taking the square of \eq{bertsub}, we obtain
\be\label{bert}
 \bset{\frac{1}{\sigma_1(V_+)}-\frac{1}{\sigma_2(V_+)}}^2=\frac{8}{\beta^2 \rho^2_Q}\bset{V_+-V_0}.
\ee

Consider the RHS of \eq{bert} by expanding $V_+$ around $\sigma=\sigma_{cr}$ up to 4th order of $x$ and $y$. Recalling that $\sigma_1=\sigma_{cr}-x$ and $\sigma_2=\sigma_{cr}+y$, where $\sigma_{cr}\gg x, y>0$, we obtain
\bea{vx}
V_+-V(\sigma_{cr})&=& \frac{x^2}{2}V^{(2)}_+(\sigma_{cr})-\frac{x^3}{6}V^{(3)}_+(\sigma_{cr})+\frac{x^4}{24}V^{(4)}_+(\sigma_{cr})+\order{x^5},\\
\label{vy}&=&  \frac{y^2}{2}V^{(2)}_+(\sigma_{cr})+\frac{y^3}{6}V^{(3)}_+(\sigma_{cr})+\frac{y^4}{24}V^{(4)}_+(\sigma_{cr})+\order{y^5},
\eea
where $V^{(2)}_+(\sigma_{cr})\equiv \left.\frac{d^2V_+}{d\sigma^2}\right|_{\sigma_{cr}},\;V^{(3)}_+(\sigma_{cr})\equiv \left.\frac{d^3V_+}{d\sigma^3}\right|_{\sigma_{cr}}$, and so on. It implies that we can equate $x$ with $y$ such that $x=y(1+cy+dy^2+\order{y^3})$ with real values, $c$ and $d$. By substituting this expression for $x$ into \eq{vx} and comparing it to \eq{vy} for each orders of $y$, it leads to
\be\label{ext3}
c=\frac{V^{(3)}_+(\sigma_{cr})}{3V^{(2)}_+(\sigma_{cr})}; \hspace*{10pt} d=c^2.
\ee
For the LHS of \eq{bert}, we again expand $\sigma_{1,\ 2}$ up to 4th order of $y$, put the results together into LHS of \eq{bert}, and then compare between the LHS and RHS of \eq{bert} for each orders of $y$. Thus, we obtain
\be\label{beta2}
\beta^2=\frac{\sigma^4}{\rho^2_Q}V^{(2)}_{+}(\sigma_{cr});\hspace*{10pt} 5c^2+8\bset{\frac{1}{\sigma^2_{cr}}+\frac{c}{\sigma_{cr}}}=\frac{V^{(4)}_+(\sigma_{cr})}{2V^{(2)}_+(\sigma_{cr})},
\ee
where the first relation corresponds to \eq{betasq}. Equation (\ref{beta2}) implies that the potentials should have the following restricted form: $V(\sigma)=\frac{M^2}{2}\sigma^{\beta^2-2}+\Lambda_1\sigma+\Lambda_0$, where $M$ and $\Lambda_{0,\; 1}$ are constants. The constraint from \eq{Phi} implies that $\Lambda_1=0$ since the angle $\Phi$ should be independent of $\sigma$ for closed orbits. Using \eq{beta2} and the fact $V^{(1)}_+(\sigma_{cr})=0$ in \eq{sigcr}, we finally obtain
\be\label{ext4}
\beta^2=1,\; 4.
\ee
Hence, the proof of the theorem is complete. We can obtain non-perturbatively the exact orbit expressions for the cases of $\beta^2=1,\; 4$, see \cite{Goldstein}.  		
\newpage
\ssp                                            


\addcontentsline{toc}{chapter}{Bibliography}


\bibliographystyle{hunsrt}

\newpage
\ssp

\end{document}